\documentclass[11pt]{article}%
\ifdefined\pdfsuppressptexinfo\pdfsuppressptexinfo=-1\fi%
\usepackage{amsmath,amssymb}
\usepackage{graphicx}
\usepackage{url}
\usepackage{xr}
\AtBeginDocument{\input{s4labels}}
\IfFileExists{tikz.sty}{\usepackage{tikz}}{%
  \usepackage{verbatim}%
  }

\ifdefined\usingsvjour
  \smartqed
\else
  \usepackage{amsthm}
  \usepackage[margin=1.15in]{geometry}
  \theoremstyle{plain}
  \newtheorem{theorem}{Theorem}[section]
  \newtheorem{proposition}[theorem]{Proposition}
  \newtheorem{lemma}[theorem]{Lemma}
  \newtheorem{corollary}[theorem]{Corollary}
  \theoremstyle{definition}
  \newtheorem{definition}[theorem]{Definition}
  \theoremstyle{remark}
  \newtheorem{remark}[theorem]{Remark}
  
  \let\qed\qedhere
  \providecommand{\smartqed}{}
  \makeatletter
  \newcommand{\institute}[1]{\g@addto@macro\@author{\\[1ex]{\normalsize
    \begin{tabular}[t]{@{}c@{}}#1\end{tabular}}}}
  \makeatother

  \newcommand{\keywords}[1]{\par\smallskip\noindent\textbf{Keywords} {\def\and{\unskip, }#1}}
  \newcommand{\subclass}[1]{\par\smallskip\noindent\textbf{Mathematics Subject Classification (2020)} {\def\and{\unskip, }#1}}
  \newcommand{\titlerunning}[1]{}
  \newcommand{\authorrunning}[1]{}
  \providecommand{\journalname}[1]{}
\fi
\numberwithin{equation}{section}
\numberwithin{table}{section}
\providecommand{\ZZ}{\mathbb{Z}}
\providecommand{\RR}{\mathbb{R}}
\providecommand{\NN}{\mathbb{N}}

\providecommand{\Zthree}{\mathbb{Z}^{3}}
\providecommand{\EE}{\mathbb{E}}
\providecommand{\PP}{\mathbb{P}}
\providecommand{\Ber}{\mathrm{Ber}}
\providecommand{\eps}{\varepsilon}
\providecommand{\Ord}{\mathcal{O}}
\providecommand{\one}{\mathbf{1}}
\providecommand{\supp}{\operatorname{supp}}
\providecommand{\dist}{\operatorname{dist}}
\providecommand{\diam}{\operatorname{diam}}
\providecommand{\Tr}{\operatorname{Tr}}
\providecommand{\avg}{\operatorname{avg}}

\providecommand{\ReP}{\operatorname{Re}}               

\providecommand{\abs}[1]{\lvert #1\rvert}
\providecommand{\norm}[1]{\lVert #1\rVert}
\providecommand{\ex}[1]{\langle #1\rangle}   

\providecommand{\Lam}{\Lambda}
\providecommand{\Lamo}{\Lambda^{\circ}}             
\providecommand{\bdtwo}{\partial_{2}}                
\providecommand{\Gp}{G'}                             

\providecommand{\XX}{\mathbb{X}}                     
\providecommand{\MM}{\mathbb{M}}                     
\providecommand{\KK}{\mathfrak{K}}                   
\providecommand{\FF}{\mathfrak{F}}                   
\providecommand{\Dep}{\mathfrak{D}}                  

\providecommand{\cU}{\mathcal{U}}                    
\providecommand{\cP}{\mathcal{P}}
\providecommand{\Grid}{\mathfrak{G}}                 

\providecommand{\Bpay}{B_{\mathrm{pay}}}
\providecommand{\Qatt}{\mathcal{Q}_{\mathrm{att}}}
\providecommand{\ncut}{n_{\mathrm{cut}}}
\providecommand{\Scut}{S_{\mathrm{cut}}}

\providecommand{\gfix}{g^{\mathrm{fix}}}             
\providecommand{\WG}{W_{G'}}                         
\providecommand{\AG}{A_{G'}}                         
\providecommand{\vt}{\tilde v}                       
\providecommand{\omb}{\omega_{\beta}}                
\providecommand{\thJ}{\vartheta_{J}}                 
\providecommand{\Rone}{R_{1}}                        

\providecommand{\bGM}{\beta_{\mathrm{GM}}}           
\providecommand{\bGMo}{\beta^{0}_{\mathrm{GM}}}     
\providecommand{\Kst}{K_{*}}                         
\providecommand{\cdel}{c_{\delta}}                   
\providecommand{\cdelsp}{c_{\delta}^{\mathrm{sp}}}
\providecommand{\astar}{a_{*}}

\providecommand{\mD}{m_{D}}
\providecommand{\mt}{\tilde m}
\providecommand{\lPV}{\lambda_{\mathrm{PV}}}
\providecommand{\MY}{M_{Y}}
\providecommand{\bD}{b_{D}}                          

\providecommand{\etaw}{\eta_{w}}

\providecommand{\rbox}{r_{\square}}

\providecommand{\RNF}{R_{\mathrm{NF}}}
\providecommand{\cNF}{c_{3}^{\mathrm{NF}}}
\providecommand{\thNF}{\theta_{\mathrm{NF}}}
\providecommand{\Funit}{F_{\mathrm{unit}}}
\providecommand{\Rresp}{R_{\mathrm{resp}}}
\providecommand{\Cresp}{C_{\mathrm{resp}}}

\providecommand{\etaT}{\eta_{T}}
\providecommand{\etaS}{\eta_{S}}
\providecommand{\epsc}{\varepsilon_{c}}
\providecommand{\epspp}{\varepsilon''}
\providecommand{\epsD}{\varepsilon_{D}}

\providecommand{\cg}{c_{g}}
\providecommand{\cgp}{c_{g}'}
\providecommand{\cL}{c_{L}}

\providecommand{\Cst}{C_{\star}'}

\providecommand{\Cfree}{C_{\mathrm{free}}}

\providecommand{\kpl}{\kappa_{\mathrm{pl}}}

\providecommand{\bC}{\mathbf{C}}                  

\providecommand{\rcov}{r_{\mathrm{cov}}}

\providecommand{\cdep}{c_{\mathrm{dep}}}           

\providecommand{\epsE}{\varepsilon_{E}}

\providecommand{\sigV}{\sigma_{V}}

\providecommand{\Kcell}{K_{\mathrm{cell}}}
\providecommand{\Kcol}{K_{\mathrm{col}}}
\providecommand{\RHK}{R_{\mathrm{HK}}}

\providecommand{\Rdep}{R_{\mathrm{dep}}}

\providecommand{\qpin}{q_{\mathrm{pin}}}
\providecommand{\Rf}{R_{f}}

\providecommand{\vCb}{v_{\mathrm{Cb}}}

\ifdefined
\fi

\begin{document}

\title{Area law for three-dimensional U(1) lattice gauge theory with the
Wilson action at every coupling}
\titlerunning{Area law for 3D U(1) Wilson lattice gauge theory}
\author{Tejasvi Singh Tomar}
\authorrunning{Tejasvi Singh Tomar}
\institute{ORCID: \url{https://orcid.org/0000-0003-4668-3639}}
\date{\normalsize Preprint. DOI: 10.5281/zenodo.23445737\\[0.5ex] October 2026}

\maketitle

\begin{abstract}
We prove that three-dimensional U(1) lattice gauge theory with the Wilson
action confines static charges at every coupling: rectangular Wilson loops
obey an area law, uniformly in the finite box, with an explicit lower bound on
the string tension at weak coupling. For the Villain action, confinement at every coupling is the theorem of G\"opfert
and Mack; for the Wilson action the question had remained open. The Wilson weight is a mixture of Villain weights, so Ginibre's inequality bounds the Wilson loop by an average of
Villain loops at a coupling larger by a fixed factor, in which a sparse random set of plaquettes is frozen at angle zero;
after this set is regularized, the multiscale analysis of G\"opfert and Mack, made uniform in it, and a Peierls estimate
give the area law.
As corollaries, the area law holds at
every coupling for all closed subgroups of U($n$) containing the scalars, also
with a Higgs field blind to the centre of U($n$), and for the Georgi\textendash Glashow
model at infinite Higgs coupling.
\subclass{81T25 \and 82B20 \and 60K35}
\end{abstract}

\section{Introduction}\label{sec:1}

\subsection{The model and the main result}\label{ssec:1.1}

Three-dimensional U(1) lattice gauge theory is the simplest gauge theory in
which static charges are expected to be confined at every coupling. By
Polyakov's mechanism \cite{Po77,BMK77}, magnetic monopoles form a plasma that
screens the magnetic field at the Debye length and squeezes the electric flux
between two static charges into a tube of finite tension. G\"opfert and Mack
proved this for the Villain action \cite{GM82}. In this paper we prove it for the
Wilson action.

\begin{definition}[Wilson and Villain measures and the Wilson loop]\label{def:1.1}
A finite box $\Lam\subset\Zthree$ is a cell complex with vertex set $\Lam$,
bonds (oriented edges) $e$ and plaquettes $p$ (free boundary conditions
throughout); for $\theta\in(\RR/2\pi\ZZ)^{E(\Lam)}$ and an oriented plaquette
$p$, $\theta_p$ is the sum of $\theta_e$ around $\partial p$.
\begin{enumerate}
\item The \emph{Wilson measure} at coupling $\beta>0$ (weak coupling means large $\beta$) and the
\emph{Villain weight} at coupling $\tau>0$ are
\begin{equation}\label{eq:1.wilson}
\mu^{W}_{\Lam,\beta}(d\theta)=\frac{1}{Z^{W}_{\Lam,\beta}}\prod_{p}e^{\beta\cos\theta_p}\prod_{e}d\theta_e ,\qquad
\vt_{\tau}(\theta)\propto\sum_{m\in\ZZ}e^{-\tau(\theta-2\pi m)^2/2};
\end{equation}
the Villain measure $\mu^{V}_{\Lam,\tau}$ is $\mu^{W}_{\Lam,\beta}$ with
$e^{\beta\cos\theta_p}$ replaced by $\vt_\tau(\theta_p)$.
\item For a rectangle $C=C_{R,T}$ of sides $R,T\in\NN$ in a coordinate plane,
an oriented closed lattice path, $\theta_C:=\sum_{e\in C}\theta_e$, and the
Wilson loop expectation is $\ex{\cos\theta_C}^{W}_{\Lam,\beta}$.
\item The model has an \emph{area law at coupling $\beta$} if there are
$\sigma>0$, $c<\infty$ and $R_1<\infty$ such that
$\ex{\cos\theta_C}^{W}_{\Lam,\beta}\le\exp(-\sigma RT+c(R+T))$ for every
rectangle $C$ with $\min(R,T)\ge R_1$ and every finite box $\Lam\supseteq C$.
\end{enumerate}
\end{definition}

This area law, uniform in the box and with an explicit perimeter term, is
stronger than positivity of the iterated-limit string tension $\alpha$ of
\cite[(1.7)]{GM82},\footnote{Equation, section and page numbers of \cite{GM82} and \cite{GM81} refer to the
preprints \cite{GM82pre} (DESY 81-036) and \cite{GM81pre} (DESY 81-014), equation numbers of \cite{Fr79} to its
preprint IHES/P/79/02, and lemma numbers of \cite{Gu80} to its preprint CLNS 79/439.} whose finite-loop estimate carries the loss term $P_F$ of
\cite[(7.7)]{GM82}. Theorem~\ref{thm:2.1}(c) shows that the infinite-volume limit exists
and that the area law passes to it. By Remark~\ref{rem:9.6}, the proof also gives the area law in this
form for the Villain action.

\begin{theorem}[Main theorem]\label{thm:1.2}
For every $\beta>0$ there are $\sigma(\beta)>0$ and $\Rone(\beta)<\infty$ such
that for every rectangle $C$ of sides $R,T$ with $\min(R,T)\ge\Rone(\beta)$
and every finite box $\Lam\supseteq C$,
\begin{equation}\label{eq:1.main}
\ex{\cos\theta_C}^{W}_{\Lam,\beta}\le\exp\bigl(-\sigma(\beta)RT+c(R+T)\bigr),
\end{equation}
where $c$ is an absolute constant. Moreover, for every $K>\Kst:=377.38$ there is
$\beta_0(K)<\infty$ such that, for every $\beta\ge\beta_0(K)$, $\sigma(\beta)$
and $\Rone(\beta)$ can be chosen, depending on $K$, so that \eqref{eq:1.main}
holds and
\begin{equation}\label{eq:1.sigma}
\sigma(\beta)\ge1.35\cdot10^{-4}\,\cg\cL\,(K\beta)^{-1/2}e^{-2.49436\,K\beta},
\end{equation}
where $\cg,\cL$ are the fixed regularization constants of
Section~\ref{app:D}; for $\beta<\beta_0(K)$, \eqref{eq:1.main} then holds with
$\sigma(\beta)$ and $\Rone(\beta)$ equal to the values so chosen at
$\beta_0(K)$.
\end{theorem}

The first assertion follows from the second with one fixed $K>\Kst$. In the first assertion the
absoluteness of $c$ is no restriction: if \eqref{eq:1.main} holds with a constant $c'=c'(\beta)$, then it
holds with $c=0$ and with $\sigma(\beta)/2$ and $\max(\Rone(\beta),4c'/\sigma(\beta))$ in place of
$\sigma(\beta)$ and $\Rone(\beta)$, since $c'(R+T)\le2c'\max(R,T)\le\frac12\sigma(\beta)RT$ when
$\min(R,T)\ge4c'/\sigma(\beta)$. The content of the first assertion is the uniformity in $\Lam$; the absolute $c$ matters
together with \eqref{eq:1.sigma}. With
$\cg=1/4$ and $\cL=1.6\cdot10^{-19}$ (Section~\ref{ssec:D.1}),
\eqref{eq:1.sigma} reads
$\sigma(\beta)\ge5.4\cdot10^{-24}(K\beta)^{-1/2}e^{-2.49436K\beta}$
(Section~\ref{ssec:D.4}).

\begin{corollary}[Groups containing the scalars, with or without Higgs fields blind to the centre of U($n$)]\label{cor:1.5}\label{cor:1.3}
Let $G\subseteq U(n)$ be a closed subgroup with $\{zI:\abs{z}=1\}\subseteq G$,
in $d=3$ with the Wilson action $\beta\ReP\Tr U_p$ (free boundary conditions),
with or without a Higgs field coupled through any action that depends on the
links only through $\rho(U_e)$, $\rho$ a representation of $G$ trivial on the
scalars. At every $\beta>0$ and all Higgs couplings, for every rectangle $C$ with
$\min(R,T)\ge\Rone(2n\beta)$ and every finite box $\Lam\supseteq C$,
$\bigl|\ex{\Tr\textstyle\prod_{e\in C}U_e}_{G,\beta}\bigr|\le2n\exp\bigl(-\sigma(2n\beta)RT+c(R+T)\bigr)$;
in particular for U(2) with an adjoint Higgs field.
\end{corollary}

\begin{corollary}[Georgi\textendash Glashow model at infinite Higgs coupling]\label{cor:1.4}
The SU(2) lattice gauge theory with the Wilson action $\beta\cdot\frac12\Tr U_p$
and a unit adjoint Higgs field at hopping coupling $\kappa=\infty$, defined in
each finite box as the limit $\kappa\to\infty$, is the U(1) Wilson model at
coupling $\beta$, and it obeys the area law \eqref{eq:1.main} at every
$\beta>0$.
\end{corollary}

\subsection{Background}\label{ssec:1.2}
\emph{The Villain action.} G\"opfert and Mack \cite{GM82} made Polyakov's
mechanism rigorous for the Villain action. By duality the model is a Coulomb
gas of monopoles, which they control by an iterated Mayer expansion down to
the Debye scale \cite{GM81} and a single-scale expansion built on the analysis
of Debye screening of Brydges and Federbush \cite{BF80,B78}. Their Theorem~1
is the large-$\beta$ bound $\alpha\ge c\,m_D\beta^{-1}$ \cite[(1.9)]{GM82}, with $m_D$ the Debye mass,
and all couplings follow \cite[Corollary~2]{GM82} by the 3-dimensional version
of Guth's inequality \cite[Lemma~4.1]{Gu80}. The expansions use the Gaussian
form of the dual weight: in \cite[Appendix~A]{GM82} the Villain action is chosen
explicitly, in place of their Wilson action (A.4).

\emph{The Wilson action.} In three dimensions the rigorous results are the area
law at strong coupling \cite{OS78}, for all compact groups, and logarithmic
confinement bounds at all couplings, for U(1) \cite{GJ77} and for closed $G\subseteq U(n)$
containing the scalars \cite{Ch26}. Fr\"ohlich's comparison \cite{Fr79} bounds
the Wilson loop of such a group by a U(1) loop with the Wilson action
$A'=-\beta'\sum_p\cos(\theta_{\partial p})$ of \cite[(6)]{Fr79}, at coupling
$2n\beta$ \cite[(23),~(24)]{Fr79}. Thus, through this comparison, an area law at every coupling for these groups
requires the Wilson case of U(1), not the Villain case.
This case is sometimes cited as known \cite{CNS,Fo,CPS25}, as is a mass gap for all $\beta$ \cite{Ni}, but no proof for the
Wilson action has been given: the expansions of \cite{GM82} do not apply to it as they stand, and whether the Villain result can be combined
with Fr\"ohlich's comparison to give an area law for U($n$) with the Wilson action was left open in \cite[v2]{Ch26}.

\emph{Related work.} The mixture of Lemma~\ref{lem:2.2} is used to prove lower bounds
for the XY model \cite[Lemma~9.6]{AHPS} and the annealed Villain model
\cite{DvEG}. Here we use it for an upper bound. The same Bessel dual weights and their
Gaussian mixture enter recent work on the critical planar XY model
\cite[\S2.1]{OAI216}. In four dimensions a Coulomb phase is
proved for the Villain \cite{Gu80,FS82} and the Wilson action \cite{Ki86}.

\subsection{Strategy}\label{ssec:1.3}
The proof has four steps.

\emph{(i) Comparison} (Section~\ref{sec:2}). Lemma~\ref{lem:2.2} expresses
$e^{\beta\cos\theta}/I_0(\beta)$ as a mixture of Villain
weights $\vt_{1/u}$, by a theorem of Hartman and
Watson \cite{HW74}, and Lemma~\ref{lem:2.3} shows, by Ginibre's inequality, that Villain loops are nondecreasing
in each coupling. We raise the couplings below
$\tau=K\beta$ to $\tau$ and the others to $\infty$, and Lemmas~\ref{lem:2.4} and~\ref{lem:2.6}
allow us to dominate the set of
the latter by a Bernoulli set of density $\delta=e^{-\cdel(K)\beta}$, where
$\cdel(K)/K\to\infty$ as $K\to\infty$. In this way Theorem~\ref{thm:2.5} bounds the
Wilson loop at $\beta$ by the average over a Bernoulli set $D$ of the Villain
loop at $\tau$ with $D$ frozen. Proposition~\ref{prop:2.9} places this Villain loop
in one box $\Lamo$ that
depends neither on $D$ nor on $\Lam$ (Lemma~\ref{lem:3.3}), where frozen plaquettes are
insulating walls (Proposition~\ref{prop:2.10}).

\emph{(ii) Regularization} (Sections~\ref{sec:3} and~\ref{sec:4}). A
deterministic construction maps each $D$ to a hole set and frozen sheets. It
freezes more plaquettes, which only increases the loop. It works on the
blocks of a loop-adapted coarse grid, and a coarse site touched by $D$ is a mark. The hole set of its first
phase is the smallest minimizer of a cut functional that penalizes leaving blocks
crowded with marks unfilled. Thus each cost bound is a comparison with an explicit
competitor. A local seal adds thin pieces of blocks to the hole set, and a second phase
enforces non-fragmentation, which is not local. The output, a slit domain $\Gp(D)$, has
uniform isoperimetry, cells of controlled size and shape with a Poincar\'e
inequality, and non-fragmentation, with constants bounded independently of $\beta$, $K$,
$C$, $D$ (Theorems~\ref{thm:3.20}, \ref{thm:3.21}).

\emph{(iii) Deterministic area law} (Sections~\ref{sec:stage1}
and~\ref{sec:stage2}, Appendices~\ref{app:A} to~\ref{app:F}). On each output $\Gp(D)$ we run the analysis of
\cite{GM82,GM81,BF80} with uniform constants, grading the outer partition
functions by the periods of their walls (the bonds where the integer field of the expansion jumps, net of
the loop current) around the handles of the insulating set (Definition~\ref{def:7.27}). We adapt the steps we need from these sources (Part~S4 proves them in full),
and Appendix~\ref{app:A} proves a tree-graph counting
step of \cite{B78} for which the deduction given there is incomplete.
Theorem~\ref{thm:7.61} states that on each output the loop decays
exponentially in the min-cut area
$\AG(C)$, in which frozen plaquettes are free.

\emph{(iv) Min-cut lower tail} (Section~\ref{sec:mincut}). A Peierls estimate
shows that $\AG(C)\ge(1-\etaT)RT$ outside an event of probability at most
$2^{4\abs{C}+5}e^{-RT}$, with $\etaT\to0$ as $K\to\infty$
(Theorem~\ref{thm:5.1}). We impose the condition $K>\Kst$ for this estimate alone
(Lemma~\ref{lem:2.6} and the remark after it).

In Section~\ref{sec:8} we combine the four steps into the following chain: for
$\beta\ge\beta_0(K)$, with $D\sim\Ber(\delta)$ on the interior plaquettes of
$\Lamo$,
\begin{align}
\ex{\cos\theta_C}^{W}_{\Lam,\beta}
&\le\EE_D\,\gfix_{\Lamo,\tau}(C;D)\label{eq:8.1}\\
&\le\EE_D\,W_{\Gp(D)}(C)\label{eq:8.2}\\
&\le\EE_D\exp\bigl(-\sigV A_{\Gp(D)}(C)+\sigV\etaS RT+c_N(R+T)\bigr)\label{eq:8.3}\\
&=\exp\bigl(-\sigV RT+\sigV\etaS RT+c_N(R+T)\bigr)\,\EE_D\exp\bigl(\sigV(RT-A_{\Gp(D)}(C))\bigr)\label{eq:8.4}\\
&\le\exp\bigl(-\sigV(1-\etaT-\etaS)RT+c_N(R+T)+4\abs{C}+4\bigr),\label{eq:8.5}
\end{align}
where $\gfix_{\Lamo,\tau}(C;D)$ is the Villain loop on $\Lamo$ with $D$ and the boundary plaquettes frozen,
$W_{\Gp(D)}(C)$ is the Villain loop on the output, and $\sigV$, $\etaS$ and $c_N$ are the string tension, the
deficit and the perimeter constant of Theorem~\ref{thm:7.61}. Theorem~\ref{thm:2.1}, by which the Wilson loop is
nondecreasing in $\beta$, extends the resulting bound to all $\beta$ (Remark~\ref{rem:8.1}).

Two natural routes fail. First, the Wilson dual weight $I_n(\beta)/I_0(\beta)$,
of order $e^{-\abs{n}\log\abs{n}(1+o(1))}$, is dominated by no Villain dual
weight $e^{-n^2/(2t)}$. Second, on the carpet graphs of \cite{CG25}, where the
Villain action is a limit of Wilson actions, the Wilson model is the limit
with the auxiliary plaquettes frozen, and hence Ginibre's inequality bounds the
Wilson loop from below, not from above.
Appendix~\ref{app:G} shows that the first obstruction persists on every finite
box (Theorem~\ref{thm:G.2}): no comparison that holds for all charges at once, such as those obtained from correlation inequalities, bounds Wilson
loops by Villain loops with bounded couplings, even with a perimeter factor (Corollary~\ref{cor:G.4}).
Theorem~\ref{thm:2.5} is such a comparison, since it holds for every integer $1$-cycle, and this is why it keeps a
random set of frozen plaquettes (Remark~\ref{rem:G.7}). Appendix~\ref{app:G} also shows that Wilson loops are not monotone in the character
coefficients of the plaquette weights (Theorem~\ref{thm:G.8}), so the Villain monotonicity of Lemma~\ref{lem:2.3}(b) is
not an instance of a general monotonicity in these coefficients (Remark~\ref{rem:G.9}).

\emph{How the paper is organized.} Sections~\ref{sec:2} and~\ref{sec:8} use one fact about the construction,
that its output is a deterministic function of $D$ whose set $\FF$ of frozen plaquettes contains $D$
(Theorem~\ref{thm:3.20}); Sections~\ref{sec:stage1}
and~\ref{sec:stage2} can be read independently of the construction, given $\AG(C)$ and the items listed after
Theorem~\ref{thm:7.61}; Section~\ref{sec:mincut} relies on the construction of Section~\ref{sec:3} and on
Theorem~\ref{thm:3.21}; and Appendix~\ref{app:G} uses only Section~\ref{sec:2}, and no other part of the paper uses it.
Part~S1 of the supplementary material gives the values of the constants with their arithmetic, and Part~S2 completes
individual proofs that the main text sketches. Part~S4, a separate file, proves in full, in the order of the main text, the
parts of the argument that the main text takes from the literature with changes or presents in compressed form. It uses the
sources only as a guide, and its introduction lists which of its sections completes which part of the main text.

Theorem~\ref{thm:1.2}, Corollaries~\ref{cor:1.3} and~\ref{cor:1.4}, Fr\"ohlich's comparison (Proposition~\ref{prop:9.1}) and all
results of Appendix~\ref{app:G} are formalized in Lean~4 with Mathlib and checked by the Lean kernel \cite{TomLean}. The
formal theorems carry no hypotheses beyond those of the statements and use only Lean's three standard axioms, so the
steps that their proofs take from the literature are formalized as well.

\subsection{Conventions}\label{ssec:1.4}
\emph{Couplings.} We write $\beta$ for the Wilson coupling, $\tau=K\beta$ for the Villain
coupling of the comparison, and $\bGM=4\pi^2\tau$ for the coupling in the
normalization of \cite{GM82}, which we use in Sections~\ref{sec:stage1} and~\ref{sec:stage2} and in
Appendices~\ref{app:A} to~\ref{app:F}.

\emph{Constants and units.} Section~\ref{app:D} fixes the constants in order. Those fixed before $K$ do not depend on $\beta$, $\tau$,
$K$, $C$ or $D$, and $c$, $C$ with or without indices denote such constants,
except where a dependence is shown or stated. This is the case for $\cdel(K)$, which also
depends on $\beta$, for $c_g'$ and $c_u$, which depend on $\beta$ through the size and the rounding
of $L_c$, and for $\cdep$, which grows like $\log\tau$ (Proposition~\ref{prop:6.18}). The letter $C$ also denotes the loop.
One site is one coarse cell of the loop-adapted grid in Sections~\ref{sec:3}
to~\ref{sec:mincut}, unless ``fine'' is written, and the lattice spacing is one
in Sections~\ref{sec:stage1} and~\ref{sec:stage2} (Section~\ref{ssec:2.5}).
}
\section{From the Wilson action to a diluted Villain model}\label{sec:2}

In this section we bound the Wilson loop by an average of Villain loops over a
Bernoulli set of frozen plaquettes, and we show that frozen plaquettes are
insulating walls.

\paragraph{Complexes and fields.} Let $X$ be a finite $2$-dimensional cell
complex with oriented edges $E(X)$, oriented plaquettes $P(X)$ and incidence
numbers $[p:e]$. For $\theta\in\mathbb{T}^{E(X)}$, $\mathbb{T}=\RR/2\pi\ZZ$, we set
$\theta_p:=\sum_e[p:e]\theta_e$, and for $n\in\ZZ^{P(X)}$, $(\partial
n)_e:=\sum_pn_p[p:e]$. An \emph{integer $1$-cycle} is a $C\in\ZZ^{E(X)}$ with zero
boundary, $\theta_C:=\sum_eC_e\theta_e$, and $Z_2(X):=\ker\partial$. The Villain
weight of Definition~\ref{def:1.1}, normalized by $\int\vt_t\,d\theta/2\pi=1$, is
$\vt_t(\theta)=\sum_{n\in\ZZ}e^{-n^2/(2t)}e^{in\theta}=(2\pi t)^{1/2}\sum_{m\in\ZZ}e^{-t(\theta-2\pi m)^2/2}>0$
by Poisson summation, and $\vt_t\to1$ uniformly as $t\to0$. Let
$\mu^{V}_{X,t}\propto\prod_p\vt_{t_p}(\theta_p)\prod_ed\theta_e$ for
$t\in(0,\infty)^{P(X)}$, and
$\mu^{W}_{X,\beta}\propto\prod_pe^{\beta\cos\theta_p}\prod_ed\theta_e$. For $\tau>0$,
$D\subseteq P(X)$ and $s>0$ let $t^{\tau,D,s}_p$ be $\tau$ off $D$ and $s$ on
$D$. Then the \emph{frozen expectation}
$g_{X,\tau}(C;D):=\lim_{s\to\infty}\ex{\cos\theta_C}^{V}_{X,t^{\tau,D,s}}$
exists by Lemma~\ref{lem:2.3}, and the plaquettes of $D$ are \emph{frozen}.

\paragraph{Ginibre's inequality.} Let $\mu\propto\exp(\sum_AJ_A\cos(A\cdot\phi))\,d\phi$
on a torus $\mathbb{T}^m$, where $A$ runs over a finite family of integer
vectors and $J_A\ge0$. Then for all integer vectors $B,B'$
\begin{equation}\label{eq:2.ginibre}
\ex{\cos(B\cdot\phi)}_\mu\ge0,\qquad
\ex{\cos(B\cdot\phi)\cos(B'\cdot\phi)}_\mu-\ex{\cos(B\cdot\phi)}_\mu\ex{\cos(B'\cdot\phi)}_\mu\ge0,
\end{equation}
by \cite[Propositions~2 and~3]{Gi70} applied to the torus of \cite[Example~4]{Gi70},
with $S$ the set of functions $\cos(m\cdot\phi)$. In particular,
$\partial\ex{\cos(B\cdot\phi)}_\mu/\partial J_A\ge0$ for all integer vectors $B$.

\subsection{Monotonicity}\label{ssec:2.1}

\begin{theorem}[Monotonicity for the Wilson action]\label{thm:2.1}
Let $C$ be an integer $1$-cycle supported in a box $\Lam\subset\Zthree$, in the
U(1) Wilson model with free boundary conditions. (a)
$\ex{e^{i\theta_C}}^{W}_{\Lam,\beta}=\ex{\cos\theta_C}^{W}_{\Lam,\beta}\in[0,1]$.
(b) $\ex{\cos\theta_C}^{W}_{\Lam,\beta}$ is nondecreasing in $\beta$, and
nondecreasing in $\Lam$ under inclusion. (c) Along every increasing sequence
of boxes $\Lam_k$ exhausting $\Zthree$ the limit
$\ex{\cos\theta_C}^{W}_{\beta}:=\lim_k\ex{\cos\theta_C}^{W}_{\Lam_k,\beta}$
exists, equals the supremum of $\ex{\cos\theta_C}^{W}_{\Lam,\beta}$ over all
boxes $\Lam$ containing the support of $C$, and is nondecreasing in $\beta$.
\end{theorem}

\begin{proof}
(a) and the monotonicity in $\beta$ are \eqref{eq:2.ginibre} with $J_p=\beta$,
together with the substitution $\theta\mapsto-\theta$. For the volume monotonicity, let
$\Lam\subseteq\Lam'$ be boxes. The expectation on
$\Lam'$ with coupling $s\in[0,\beta]$ on the plaquettes not in $\Lam$ is
nondecreasing in $s$ and equals the one on $\Lam$ at $s=0$, where the edges
outside $\Lam$ interact with nothing. (c) follows, since each box lies in
$\Lam_k$ for large $k$.
\qed
\end{proof}

\subsection{The Hartman\textendash Watson mixture and the annealed coupling field}\label{ssec:2.2}

\begin{lemma}[Hartman\textendash Watson mixture]\label{lem:2.2}
For every $\beta>0$ there is a unique probability measure $\omb$ on $(0,\infty)$ such
that, for all $\nu\ge0$ and $\theta\in\mathbb{T}$,
\begin{equation}\label{eq:2.hw}
\int_{(0,\infty)}e^{-\nu^2u/2}\,\omb(du)=\frac{I_\nu(\beta)}{I_0(\beta)},\qquad
\frac{e^{\beta\cos\theta}}{I_0(\beta)}=\int_{(0,\infty)}\vt_{1/u}(\theta)\,\omb(du),
\end{equation}
and $\omb((0,\eps))\le e^{\nu^2\eps/2}I_\nu(\beta)/I_0(\beta)$ for all $\eps>0$
and $\nu\ge0$.
\end{lemma}

\begin{proof}
By \cite[Theorem~1.3]{HW74}, $I_\nu(\beta)/I_0(\beta)=\int_0^\infty
e^{-\nu^2t}\,W(dt,\beta)$ ($0\le\nu<\infty$) with a distribution function
$W(\cdot,\beta)$ on $[0,\infty)$. We let $\omb$ be its image under $u=2t$, which has no
atom at $0$ because $I_\nu(\beta)\le(\beta/2)^\nu e^{\beta^2/4}/\Gamma(\nu+1)\to0$
as $\nu\to\infty$. The second identity follows from the first by Fourier
inversion and Fubini, as $\sum_nI_{\abs n}(\beta)/I_0(\beta)=e^\beta/I_0(\beta)<\infty$
(for the XY model it is \cite[Lemma~9.6]{AHPS}). The tail bound is the
Chernoff bound $\one_{u<\eps}\le e^{\nu^2(\eps-u)/2}$. It is unique, since a finite measure on $[0,\infty)$
is determined by its Laplace transform on $[0,\infty)$, which the first identity of \eqref{eq:2.hw} gives
at $\nu^2/2$ for every real $\nu\ge0$.
\qed
\end{proof}

\begin{lemma}\label{lem:2.3}
Let $X$ be a finite $2$-complex and $C$ an integer $1$-cycle. (a) For
$t\in(0,\infty)^{P(X)}$, $\ex{\sin\theta_C}^{V}_{X,t}=0$ and
$\ex{\cos\theta_C}^{V}_{X,t}\ge0$. (b) $\ex{\cos\theta_C}^{V}_{X,t}$ is
nondecreasing in each $t_p$. (c) (a) and (b) hold when some plaquettes are
removed (weight $1$, the limit $t_p\to0$); a removed plaquette gives a value no
larger than any $t_p>0$. (d) For fixed couplings off $D\subseteq P(X)$, the
limit as the couplings on $D$ tend to $\infty$ (jointly, or all equal to $s\to\infty$; by monotonicity the two
limits agree) exists, lies in $[0,1]$, and is
nondecreasing in each remaining coupling and in $D$; in particular $g_{X,\tau}(C;D)\in[0,1]$
exists and is nondecreasing in $D$ (for inclusion) and in $\tau$.
\end{lemma}

\begin{proof}
On a cube of $\Zthree$ with couplings in $(0,\infty)$, (b) is proved in
\cite[Corollary~1.6]{CG25}, and also in \cite[Lemma~C.1]{Gu80} for plaquette-dependent
couplings. We need the lemma for boxes and adapt these results to them as follows. Removal, $t_p\to0$,
is dominated convergence in \eqref{eq:2.dual}, whose terms decrease as
$t_p\downarrow0$ to those of the complex without $p$, and (c) follows. A box
is a cube with the outside plaquettes removed, and an integer $1$-cycle $C$ has
$\theta_C=\theta_\ell$ for one closed path $\ell$. Freezing, $t_p\to\infty$, is a
monotone limit in $[0,1]$. This yields (d), as $t^{\tau,D,s}\le t^{\tau,D',s}$
for $D\subseteq D'$, $s\ge\tau$. For (a) we use $\theta\mapsto-\theta$ and the fact that the value
with all plaquettes removed is $\int\cos\theta_C\,d\theta/(2\pi)^{\abs E}\ge0$.
This proves the lemma on boxes. In Section~\ref{S:proofsA} we prove the lemma on all finite $2$-complexes, by verifying Ginibre's
hypothesis for the Villain weights directly.
\qed
\end{proof}

\paragraph{The annealed coupling field.} Fix $X$ and $\beta>0$. Let
$Z(t):=\int\prod_p\vt_{t_p}(\theta_p)\,d\theta/(2\pi)^{\abs{E(X)}}>0$ for
$t\in(0,\infty)^{P(X)}$, and let $\mu_{\mathrm{ann}}$ be the probability measure on
$(0,\infty)^{P(X)}$ with density proportional to $Z(1/u)$, $1/u:=(1/u_p)_p$,
with respect to $\omb^{\otimes P(X)}$. The normalization is finite and positive
by \eqref{eq:2.hw} and Fubini. We also put $\thJ(u):=\sum_{n\in\ZZ}e^{-n^2u/2}$ and
$\delta_\beta(\eps):=\int_{(0,\eps)}\thJ(u)\,\omb(du)$.

\begin{lemma}[Annealed coupling field]\label{lem:2.4}
Let $X$ be a finite $2$-complex, $\beta>0$ and $\eps>0$.
(i) For every bounded measurable $F$ on $\mathbb{T}^{E(X)}$,
$\ex{F}^{W}_{X,\beta}=\int\ex{F}^{V}_{X,1/u}\,\mu_{\mathrm{ann}}(du)$.
(ii) (Tilt bound.) For every $p\in P(X)$ and every
$u_{-p}:=(u_q)_{q\ne p}$,
$\mu_{\mathrm{ann}}(u_p<\eps\mid u_{-p})\le\delta_\beta(\eps)$.
(iii) Under $\mu_{\mathrm{ann}}$ the random set $D_\eps(u):=\{p:u_p<\eps\}$ is
stochastically dominated by $\Ber(\delta')^{\otimes P(X)}$ for every
$\delta'\in[\min(\delta_\beta(\eps),1),1]$: $\int F(D_\eps(u))\,\mu_{\mathrm{ann}}(du)\le\EE_{D\sim\Ber(\delta')^{\otimes P(X)}}F(D)$
for every nondecreasing function $F$ on the subsets of $P(X)$.
(iv) $\delta_\beta(\eps)\le\inf_{\nu\ge0}e^{\nu^2\eps/2}\sum_{n\in\ZZ}I_{\sqrt{n^2+\nu^2}}(\beta)/I_0(\beta)$.
\end{lemma}

\begin{proof}
(i) We multiply the second identity of \eqref{eq:2.hw} over $p$, integrate
against $F$ and against $1$, and use Fubini.

(ii) The tilt is bounded because, as a function of one coupling, $Z$ is a combination of the functions
$e^{-n^2u_p/2}$ with nonnegative coefficients, each at most the constant
coefficient. We fix $p$ and $u_{-p}$. As
$\mu_{\mathrm{ann}}$ has density proportional to $Z(1/u)$ with respect to a
product measure, the conditional law of $u_p$ given $u_{-p}$ has density
proportional to $u_p\mapsto Z(1/u)$ with respect to $\omb$. Expanding
$\vt_{1/u_p}(\theta_p)=\sum_ne^{-n^2u_p/2}e^{in\theta_p}$, an absolutely
convergent series, we obtain $Z(1/u)=\sum_{n\in\ZZ}e^{-n^2u_p/2}A_n$,
$A_n:=\int e^{in\theta_p}\prod_{q\ne p}\vt_{1/u_q}(\theta_q)\,d\theta/(2\pi)^{\abs{E}}$,
with $A_0>0$ independent of $u_p$. Here $A_n/A_0=\ex{e^{in\theta_p}}$ in the system
with $p$ removed. Since $n\theta_p=\theta_{C'}$ for the integer $1$-cycle
$C'=n\,\partial p$, Lemma~\ref{lem:2.3}(a),(c) shows that $A_n$ is real and
$0\le A_n\le A_0$. Hence $1\le Z(1/u)/A_0\le\thJ(u_p)$, where the lower
bound comes from the term $n=0$, and $\mu_{\mathrm{ann}}(u_p<\eps\mid u_{-p})
=\int_{(0,\eps)}(Z/A_0)\,d\omb\big/\int_{(0,\infty)}(Z/A_0)\,d\omb
\le\int_{(0,\eps)}\thJ\,d\omb=\delta_\beta(\eps)$.

(iii) Let $Y_p:=\one[u_p\ge\eps]$ and fix an order $s_1,s_2,\dots$ of $P(X)$.
Since $Y_{s_1},\dots,Y_{s_j}$ are functions of $u_{-s_{j+1}}$, (ii) and the tower
property give $\PP(Y_{s_{j+1}}=1\mid Y_{s_1},\dots,Y_{s_j})\ge1-\min(\delta_\beta(\eps),1)$.
By \cite[Lemma~1.1]{LSS97} this yields the domination for $\delta'=\min(\delta_\beta(\eps),1)$,
and a monotone coupling shows that $\EE F(D)$ is nondecreasing in $\delta'$.

(iv) is the
Chernoff bound $\one_{u<\eps}\le e^{\nu^2(\eps-u)/2}$ integrated against
$\thJ(u)\,\omb(du)$, with \eqref{eq:2.hw}.
\qed
\end{proof}

\begin{theorem}[Comparison with diluted Villain loops]\label{thm:2.5}
For every finite $2$-complex $X$, integer $1$-cycle $C$, $\beta>0$,
$\eps>0$ and $\delta'\in[\min(\delta_\beta(\eps),1),1]$,
\begin{equation}\label{eq:2.W}
0\le\ex{\cos\theta_C}^{W}_{X,\beta}\le\EE_{D\sim\Ber(\delta')^{\otimes P(X)}}\,g_{X,1/\eps}(C;D).
\end{equation}
\end{theorem}

\begin{proof}
Lemma~\ref{lem:2.4}(i) represents the Wilson model as a Villain model with random
couplings $1/u_p$, and by monotonicity we round all couplings up, those at most
$1/\eps$ to $1/\eps$ and the others to $\infty$. Positivity is a case of
\eqref{eq:2.ginibre}. For fixed $u$ and $D=D_\eps(u)$ we have $1/u_p\le1/\eps$ off
$D$, and hence Lemma~\ref{lem:2.3}(b),(d) give
$\ex{\cos\theta_C}^{V}_{X,1/u}\le g_{X,1/\eps}(C;D_\eps(u))$. We integrate this against
$\mu_{\mathrm{ann}}(du)$, use Lemma~\ref{lem:2.4}(i) once more, and apply
Lemma~\ref{lem:2.4}(iii) to the function $g_{X,1/\eps}(C;\cdot)$, which is nondecreasing
by Lemma~\ref{lem:2.3}(d).
\qed
\end{proof}

\subsection{The density of frozen plaquettes}\label{ssec:2.3}

From now on we take $\eps=1/(K\beta)$, so that the Villain coupling is $\tau=K\beta$,
and set $\delta:=\delta_\beta(1/(K\beta))$ and $\cdel(K):=-(\log\delta)/\beta$, a
function of $K$ and $\beta$, although the notation shows only $K$. Moreover $\delta>0$: $\thJ\ge1$ gives
$\delta\ge\omb((0,1/(K\beta)))$, and $\omb((0,\eps))=0$ would give $I_\nu(\beta)/I_0(\beta)\le e^{-\nu^2\eps/2}$
for all $\nu\ge0$ by \eqref{eq:2.hw}, against $I_\nu(\beta)\ge(\beta/2)^\nu/\Gamma(\nu+1)$. For
$K>e/2$ let $\astar=\astar(K)>e/2$ be the solution of $\astar=K\log(2\astar)$
(unique on $(e/2,\infty)$, where $a\mapsto a/\log 2a$ increases), and put
$\cdelsp(K):=\astar\bigl(\tfrac12\log2\astar-1\bigr)+1$.

\begin{lemma}\label{lem:2.6}
Let $K>e/2$ and $\beta\ge\pi^{-2}$. Then
\begin{equation}\label{eq:2.cdelta}
\log\delta\le-\beta\Bigl(\cdelsp(K)-\frac{1}{4\astar}\Bigr)+\log\bigl((2\astar\beta+5)\pi\sqrt\beta\bigr)+\frac12 ,
\end{equation}
so that $\cdel(K)\ge\cdelsp(K)-1/(4\astar)-O_K(\log\beta/\beta)$. The function
$K\mapsto\cdelsp(K)/(2K)$ is increasing on $(e/2,\infty)$; it equals
$14.9662$ at $K=377.3635$, and it is at least $14.9663$ for $K\ge\Kst=377.38$.
Consequently, for every $K>\Kst$ there are $\epsilon_K>0$ and $\beta_1(K)$
such that
\begin{equation}\label{eq:2.threshold}
\frac{\cdel(K)}{2K}\ge14.9662+\epsilon_K\qquad(\beta\ge\beta_1(K)).
\end{equation}
\end{lemma}

The number $14.9662$ is $6\pi^2\vCb(0)=14.96613052\ldots$ rounded up. Here
$\vCb(0)=(-\Delta_{\Zthree})^{-1}(0,0)=0.2527310\ldots$ is half of Watson's integral
\cite{Wat39} in the normalization of \cite{Zu11}, whose closed form is \cite[(3.6)]{Zu11} (Section~\ref{ssec:D.1}),
and the Debye exponent $\pi^2\vCb(0)=2.4943550871\ldots$ gives
$\log(1/\mt)=\pi^2\vCb(0)\,\tau+O(\log\tau)$ (Definition~\ref{def:3.1}). Thus, for $K>\Kst$,
\eqref{eq:2.threshold} says $\delta^{1/2}\le\mt^{\,6}e^{-\epsilon_K\tau+O(\log\tau)}$.
In the Peierls count of Lemma~\ref{lem:5.16} this means that the factor $\delta^{1/2}$ that each mark costs
(Definition~\ref{def:3.6} and Lemma~\ref{lem:5.11}) is smaller than the reciprocal of the entropy $(\mt^{-1})^{6+o(1)}$ of
a root and a part.
For the first assertion of Theorem~\ref{thm:1.2} the value of $\Kst$ is not needed: with
$t:=\log2\astar(K)$ we have $K=e^t/(2t)$ and $\cdelsp(K)/(2K)=\frac14t(t-2)+te^{-t}$, which tends
to $\infty$ with $K$ (as $t>\log2K$), so every sufficiently large $K$ satisfies \eqref{eq:2.threshold}.

\begin{proof}
Heuristically, the heavy dual tail of the Wilson weight comes from $\omb$ near $u=0$.
Applying Lemma~\ref{lem:2.4}(iv) at its saddle point $\nu=\astar\beta$, with the series
of $I_\mu$, log-convexity of $\Gamma$, Stirling's inequality and
$I_0(\beta)\ge e^{\beta-1/2}/(\pi\sqrt\beta)$ for $\beta\ge\pi^{-2}$, we obtain
\eqref{eq:2.cdelta} (Section~\ref{ssec:D.3} gives the computation, the monotonicity and the two
values). For $K>\Kst$,
$\epsilon_K:=\frac12\bigl(\cdelsp(K)/(2K)-14.9662\bigr)\ge9\cdot10^{-5}$, since
$\cdelsp(K)/(2K)$ is increasing and, by the computation in Section~\ref{ssec:D.3},
equals $14.96639$ at $\Kst$. For these $K$ the value $14.9663$ of the statement gives
$\epsilon_K\ge5\cdot10^{-5}$, and as $1/(8K\astar)<10^{-7}$ for $K\ge\Kst$, either lower bound on $\epsilon_K$
shows that \eqref{eq:2.cdelta} implies \eqref{eq:2.threshold} once the
$O_K(\log\beta/\beta)$ term is below $K\epsilon_K$.
\qed
\end{proof}

\subsection{The dual representation and the frozen limit}\label{ssec:2.4}

Let $\Lam'$ be a box, regarded as a closed cubical complex with cubes
$\Lam'_3$, plaquettes $\Lam'_2$ and boundary plaquettes $\bdtwo\Lam'$, and let
$\partial:\ZZ^{\Lam'_3}\to\ZZ^{\Lam'_2}$ be the boundary map, cubes oriented
outward. Since the complex is contractible, $\partial$ is injective,
$Z_2(\Lam')=\partial\ZZ^{\Lam'_3}$, and each integer $1$-cycle $C$ in $\Lam'$
bounds an integer $2$-chain $n_0$.

\begin{lemma}[Dual representation]\label{lem:2.7}
Let $t\in(0,\infty)^{\Lam'_2}$, let $C$ be an integer $1$-cycle in $\Lam'$, and
let $n_0$ be any integer $2$-chain of $\Lam'$ with $\partial n_0=C$. Then
\begin{equation}\label{eq:2.dual}
\ex{\cos\theta_C}^{V}_{\Lam',t}=\frac{\sum_{k\in\ZZ^{\Lam'_3}}\prod_{p}e^{-(n_0+\partial k)_p^2/(2t_p)}}{\sum_{k\in\ZZ^{\Lam'_3}}\prod_pe^{-(\partial k)_p^2/(2t_p)}} .
\end{equation}
\end{lemma}

\begin{proof}
By the expansion
$\prod_p\vt_{t_p}(\theta_p)=\sum_{n\in\ZZ^{\Lam'_2}}\prod_pe^{-n_p^2/(2t_p)}e^{i(\partial n)\cdot\theta}$,
which converges absolutely and has even coefficients, $\ex{e^{i\theta_C}}=\ex{\cos\theta_C}$
is the ratio of the sums over
$\partial n=C$ and over $\partial n=0$. Since $Z_2(\Lam')=\partial\ZZ^{\Lam'_3}$ and $\partial$ is injective,
the chains $n=n_0+\partial k$,
$k\in\ZZ^{\Lam'_3}$, parametrize the first set bijectively for every choice of $n_0$ (a point that matters when
frozen plaquettes surround an edge of $C$).
\qed
\end{proof}

Dually, cubes are sites, an interior
plaquette is a bond between its two cubes, and a boundary plaquette is a bond
to an exterior ground site, where $k=0$. We now fix $\tau>0$ and a set $F\subseteq\Lam'_2$ of plaquettes to be frozen, and
let $t^{s}_p$ be $\tau$ for $p\notin F$ and $s$ for $p\in F$. The \emph{unfrozen graph} has
the cubes and the ground as vertices and the plaquettes not in $F$ as bonds, and $\mathcal K_F$ denotes
the set of its components not containing the ground. A cube
with all faces in $F$ is such a component. Fix $c_\KK\in\KK$ and put
$W:=\{\sum_{\KK\in\mathcal K_F}a_\KK\one_\KK:a\in\ZZ^{\mathcal K_F}\}$ and
$W':=\{k\in\ZZ^{\Lam'_3}:k(c_\KK)=0\ \text{for all}\ \KK\in\mathcal K_F\}$. Then
$\ZZ^{\Lam'_3}=W\oplus W'$, and $(\partial w)_p=0$ for $w\in W$, $p\notin F$.

\begin{lemma}\label{lem:2.8}
In this setting, for every integer $1$-cycle $C$ in $\Lam'$ and every integer
$2$-chain $n_0$ of $\Lam'$ with $\partial n_0=C$,
\begin{equation}\label{eq:2.frozen}
\lim_{s\to\infty}\ex{\cos\theta_C}^{V}_{\Lam',t^s}
=\frac{\sum_{k\in W'}\exp\bigl(-\frac{1}{2\tau}\sum_{p\notin F}(n_0+\partial k)_p^2\bigr)}{\sum_{k\in W'}\exp\bigl(-\frac{1}{2\tau}\sum_{p\notin F}(\partial k)_p^2\bigr)}.
\end{equation}
The quadratic form $k\mapsto\sum_{p\notin F}(\partial k)_p^2$ is positive
definite on $W'\otimes\RR$, so both sums converge; the summands are invariant
under $W$, so the sums run over $\ZZ^{\Lam'_3}/W$ ($k$ modulo one constant on
each component in $\mathcal K_F$) and do not depend on the choice of $W'$. In
particular every cube all of whose faces are frozen drops out.
\end{lemma}

\begin{proof}
Freezing deletes dual bonds; the only issue is the soft directions $W$, one
zero mode for each component in $\mathcal K_F$, isolated cubes included, such
as the cubes outside $\Lam'$ in Proposition~\ref{prop:2.9}. Along these directions only the frozen weights act, at strength $1/s$. Poisson summation over $w\in W$,
where $k=w'+w$ and the quadratic form of this summation is positive definite because $\partial$ is injective,
gives a prefactor that
cancels in the ratio times a factor in $[0,2]$ tending to $1$. A $w'$ with
$\one_{F^c}\partial w'=0$ is constant on components and vanishes on that of the
ground, which proves the positive definiteness stated in the lemma. Dominated convergence then yields
\eqref{eq:2.frozen} (see Section~\ref{S:proofsA} for the details of this proof).
\qed
\end{proof}

\subsection{One box with insulating walls}\label{ssec:2.5}

For a box $\Lam'$, $D\subseteq\Lam'_2\setminus\bdtwo\Lam'$ and $\tau>0$,
$\gfix_{\Lam',\tau}(C;D):=g_{\Lam',\tau}(C;D\cup\bdtwo\Lam')$ is the Villain
expectation on $\Lam'$ with the plaquettes of $D$ and the boundary plaquettes
frozen. We write $C\subset\Lam'\setminus\partial\Lam'$ if all edges
of $C$ lie in $\Lam'$ and none lies in the boundary of $\Lam'$.

\begin{proposition}\label{prop:2.9}\label{eq:2.onebox}
Let $C$ be an integer $1$-cycle, $\Lam'$ a box with
$C\subset\Lam'\setminus\partial\Lam'$, $K,\beta>0$, $\tau=K\beta$, $\delta$ as
in Section~\ref{ssec:2.3} and $\delta':=\min(\delta,1)$. For every finite box
$\Lam\supseteq C$,
$\ex{\cos\theta_C}^{W}_{\Lam,\beta}\le\EE_{D\sim\Ber(\delta')^{\otimes(\Lam'_2\setminus\bdtwo\Lam')}}\,\gfix_{\Lam',\tau}(C;D)$.
\end{proposition}

\begin{proof}
We use volume monotonicity and then freeze everything outside $\Lam'$. Let
$\Lam''\supseteq\Lam\cup\Lam'$ be a box and
$F_{\mathrm{out}}:=\Lam''_2\setminus(\Lam'_2\setminus\bdtwo\Lam')$. Then
Theorem~\ref{thm:2.1}(b), Theorem~\ref{thm:2.5} on $X=\Lam''$ with
$\eps=1/(K\beta)$, and Lemma~\ref{lem:2.3}(d) give
$\ex{\cos\theta_C}^{W}_{\Lam,\beta}\le\EE_{D''}\,g_{\Lam'',\tau}(C;D\cup F_{\mathrm{out}})$,
where $D''\sim\Ber(\delta')^{\otimes\Lam''_2}$ and $D:=D''\cap(\Lam'_2\setminus\bdtwo\Lam')$.
By Lemma~\ref{lem:2.8} on $\Lam''$ and on $\Lam'$ ($n_0$ in $\Lam'$), the right
side is $\gfix_{\Lam',\tau}(C;D)$, as the unfrozen plaquettes are the same, the
cubes outside $\Lam'$ are isolated, and the ground is not reached.
\qed
\end{proof}

\paragraph{Slit graphs.} Let $\Lam'$ be a box, $\FF\subseteq\Lam'_2$ with
$\bdtwo\Lam'\subseteq\FF$, and $\XX\subseteq\Lam'_3$ a set of cubes all of whose
faces lie in $\FF$. The \emph{slit graph} $\Gp=(V,E)$ has $V:=\Lam'_3\setminus\XX$ and
$E:=E(V)\setminus S$, with $E(V)$ the nearest-neighbour pairs in $V$ and $S$
those whose common plaquette lies in $\FF$. Conversely, each triple
$(\Lam',\XX,S)$, $S$ a set of bonds with both ends in $\Lam'_3\setminus\XX$,
arises in this way, with $\FF:=F(\XX)\cup F_S\cup\bdtwo\Lam'$, where $F(\XX)$ is the set of faces
of the cubes of $\XX$ and $F_S$ is the set of plaquettes dual to $S$. The phrase ``every slit graph''
means every such triple. As $E$ omits the bonds of $S$, a slit graph is in general not an induced subgraph
of $\Zthree$. Its components are denoted $\KK$ (an isolated vertex is a
component), its vertices have degree at most $6$, and
$d_{\Gp}(x,y)\ge\norm{x-y}_1$, with $d_{\Gp}=\infty$ across components. We orient
a bond $b=(x,y)\in E$ and its plaquette $p_b$ compatibly, with $p_b$ oriented as in $\partial
x$. For an integer $2$-chain $\Xi$ of $\Lam'$ with $\partial\Xi=C$ let
$\xi_b\in\ZZ$ be the coefficient of $p_b$ in $\Xi$, which for the flat rectangle
spanned by $C$ is the signed crossing number, in $\{0,\pm1\}$, and for
$k\in\ZZ$ put
\begin{equation}\label{eq:2.ZG}
Z_{\Gp}(k,\Xi):=\sum_{[m]}\exp\Bigl(-\frac{1}{2\tau}\sum_{b=(x,y)\in E}\bigl(m(y)-m(x)-k\xi_b\bigr)^2\Bigr),
\qquad \WG(C):=\frac{Z_{\Gp}(1,\Xi)}{Z_{\Gp}(0,\Xi)},
\end{equation}
where $[m]$ runs over the classes of $m\in\ZZ^V$ modulo functions constant on
each component.

\paragraph{Coarse and fine.} In Sections~\ref{sec:3} to~\ref{sec:mincut} the
hole set $\XX$ is a set of coarse sites and the frozen set $S$ is a set of coarse
bonds, as in Definition~\ref{def:3.7}, with the coarse slit graph
$(\Lamo\setminus\XX,E(\Lamo\setminus\XX)\setminus S)$. In view of Definition~\ref{def:3.4}, a coarse face carries
between $r_0^2$ and $\bar r_0^2$ fine plaquettes. On
the fine lattice the hole cubes are the fine cubes in sites of $\XX$, $F(\XX)$
is the set of their faces (Definition~\ref{def:3.6}), $F_S$ is the set of fine
plaquettes in the coarse faces dual to the bonds of $S$, and
\begin{equation}\label{eq:2.FF}
\FF:=F(\XX)\cup F_S\cup\bdtwo\Lamo .
\end{equation}
For the output $(\XX,S)=(\XX(D),S(D))$ of Theorem~\ref{thm:3.20} the fine slit
graph of this $\FF$ is the \emph{output} $\Gp=\Gp(D)$. Its vertices are the fine
cubes of $\Lamo$ outside the holes, and a bond at a vertex is missing from $E$
exactly when it leads into a hole, is frozen (dual to $F_S$), or leaves $\Lamo$.

\begin{proposition}[Frozen plaquettes as insulating walls]\label{prop:2.10}\label{eq:2.insul}
Let $\Lam'$, $\FF$, $\XX$ and $\Gp$ be as above, $\tau>0$, and $C$ an integer
$1$-cycle in $\Lam'$. For every $D\subseteq\FF\setminus\bdtwo\Lam'$,
$\gfix_{\Lam',\tau}(C;D)\le\gfix_{\Lam',\tau}(C;\FF\setminus\bdtwo\Lam')=\WG(C)$,
and $\WG(C)$ depends neither on the choice of $\Xi$ nor on the choice of $\XX$.
\end{proposition}

\begin{proof}
By Lemma~\ref{lem:2.3}(d), freezing raises the loop, and the inequality follows. Since freezing
also deletes a dual bond, holes, sheets, cuts and box walls are admissible. For
the identity we apply Lemma~\ref{lem:2.8} with $F=\FF$, $n_0=\Xi$. Here no
component reaches the ground, cubes with all faces in $\FF$ (those of $\XX$
among them) are isolated and drop out, and
$(\Xi+\partial k)_{p_b}=\xi_b+k(x)-k(y)$ gives $Z_{\Gp}(1,\Xi)/Z_{\Gp}(0,\Xi)$ with
$m=k|_V$. A change $\Xi'=\Xi+\partial U$ is absorbed by $m\mapsto m-U|_V$. On
outputs, $D\subseteq\FF(D)$ for all $D$ by Theorem~\ref{thm:3.20}, and hence
the inequality applies to every $D$.
\qed
\end{proof}

\begin{definition}[Min-cut area]\label{def:2.11}
For $\FF$ as in \eqref{eq:2.FF}, the \emph{min-cut area} of $C$ in $\Gp$ is
$\AG(C):=\min\{\#(\supp\Sigma\setminus\FF):\ \Sigma$ an integer $2$-chain of
$\Lamo$ with $\partial\Sigma=C\}$.
\end{definition}

}
\providecommand{\Lbox}{\Lambda^{\circ}}
\providecommand{\Hset}{\mathbb{X}}
\providecommand{\Mset}{\mathbb{M}}
\providecommand{\Uf}{\mathbb{U}}
\providecommand{\Fset}{\mathfrak{F}}
\providecommand{\Kc}{\mathfrak{K}}
\providecommand{\Sbar}{\bar S}
\providecommand{\Gp}{G'}
\providecommand{\Ereg}{\mathcal{E}}
\providecommand{\dist}{\operatorname{dist}}
\providecommand{\supp}{\operatorname{supp}}
\providecommand{\good}{\operatorname{good}}
\providecommand{\Qc}{\mathcal{Q}}
\providecommand{\Uc}{\mathcal{U}}

\section{The regularization: construction and main properties}\label{sec:3}

In this section we define a deterministic map $D\mapsto\Gp(D)$ from sets $D$ of frozen plaquettes to slit domains and state its main properties, Theorems~\ref{thm:3.20} (terminal state) and~\ref{thm:3.21} (history), which we prove in Section~\ref{sec:4}; the details of the proof of Lemma~\ref{lem:3.9}, including the lattice of minimizers, are in Section~\ref{S4-S:S4.1}. Throughout Sections~\ref{sec:3} to~\ref{sec:5} we fix $K>K_*$, $\beta\ge\beta_0(K)$, $\tau=K\beta$ and the constants fixed before $K$, for which we refer to Section~\ref{sec:8} and, for their values, to Section~\ref{S:constants}. The loop $C=\partial\Xi$ bounds the rectangle $\Xi:=[r_1,r_1+R]\times[t_1,t_1+T]\times\{z_0\}$ ($r_1,t_1,z_0\in\mathbb Z$, fine units), with $\min(R,T)\ge R_1$. We treat other planes by permuting the axes. All statements hold for arbitrary $D\subseteq\Lbox_2\setminus\partial_2\Lbox$. Units are coarse (one site is one coarse cell of the grid) and distances are coarse $\ell^\infty$ distances unless ``fine'' or ``Euclidean'' is written. We write $Q_\infty(u,r):=\{x:|x-u|_\infty\le r\}$.

Before the formal definitions we outline the design of the construction. The Phase-1 hole set is the smallest minimizer of a cut functional, given in Definition~\ref{def:3.forced}, and in Section~\ref{ssec:4.9} we pay its cost by comparison with competitors that replace a region, at a bounded cost per mark. Smallest minimizers are canonical, so the definition of Phase~1 involves no tie-breaking rule. Fills add whole blocks and are triggered by a count of marks. Penalizing defective interfaces or weak links instead fails by cascades, since freezing the bonds of one defect can create defects of the same kind in other blocks, cells being defined through windows, whereas filling blocks changes no bond inside other blocks. We enforce $(\mathrm{NF}_*)$ by a rule of Phase~2, the cut rule, and not by a term of the functional, since the deletion competitor used in that comparison can create a fragmented split. The local seal of Definition~\ref{def:3.seal} fills the thin pieces of blocks that minimality does not exclude.

\subsection{Scales, the box, the grid}\label{ssec:3.1}

\begin{definition}[Scales and units]\label{def:3.1}
Let $\tilde m\in[m_D/2,2m_D]$ be the screening mass of Section~\ref{sec:stage1} (fine units). The scales are: the fine side $r_0\ge1$ of a coarse cell, which is at most $\bar r_0:=2r_0-1$; the block scale $L_c$, the largest even integer $\le c_L\tilde m^{-1}/r_0$ (coarse), and $L:=r_0L_c$ (fine); $\rho_\lambda:=L_c/12$ and the hull penalty $\lambda:=1/\rho_\lambda=12/L_c$; $V_*:=(1-\eta_w)^3\rho_\lambda^3/2$ with $\eta_w:=1/10$; the window bound $k_P:=304L_c$; the removal scale $\ell:=\ell_D/r_0$ with $\ell_D:=L\lceil\tilde m^{-1}/L\rceil$; and $R_1:=(2r_\square+5)L$ (fine) with $r_\square:=(k_P+3)/L_c\le305$.
\end{definition}

The lower bounds on $\tau$ used here are $L_c\ge4$ and $\tilde mL\ge c_L/2$ (Section~\ref{S:B1.scales}). Since $c_L\le1/308$, we have
\begin{equation}\label{eq:3.Hell}
\ell\ge k_P+4L_c=308\,L_c,\qquad \tilde mL\le c_L\le\tfrac12 .
\end{equation}

\begin{definition}[The box $\Lbox$]\label{def:3.2}
Put $m:=\max(R,T)$, $M_\Lambda:=\max\bigl(3m,\ r_0(\ell+3)\bigr)$ (fine), $M'_\Lambda$ the even integer in $\{M_\Lambda,M_\Lambda+1\}$, and $o=(o_1,o_2,o_3):=(r_1+R/2,\ t_1+T/2,\ z_0)$, the centre of $\Xi$. Then $\Lbox:=\prod_{\mu=1}^3[a_\mu,b_\mu]$ with $a_\mu:=\lfloor o_\mu\rfloor-M'_\Lambda/2$ and $b_\mu:=a_\mu+M'_\Lambda$, a cube whose faces are lattice planes; $\Lbox_2$ is its set of plaquettes and $\partial_2\Lbox$ the set of those contained in $\partial\Lbox$.
\end{definition}

\begin{lemma}\label{lem:3.3}
(a) $\Lbox$ and its grid $\mathfrak G(C)$ depend only on $C$ and the constants. (b) Consecutive special planes are at fine distance $\ge L$. (c) Every coarse side of $\Lbox$ is $\ge\ell\ge k_P+4L_c$, and $\Lbox$ is a union of blocks and of removal cells. (d) The coarse sides $n_\mu$ of $\Lbox$ satisfy $\max_\mu n_\mu\le2\min_\mu n_\mu$. (e) $\Lbox$ contains the fine $\ell^\infty$-neighbourhood of radius $m$ of $\Xi$, so $\dist(C,\partial\Lbox)\ge\max(R,T)$, $C\subset\Lbox\setminus\partial\Lbox$, and the fine side of $\Lbox$ is $\ge3\max(R,T)$.
\end{lemma}

\begin{proof}
Parts (a) to (d) are proved in Section~\ref{S:B2.box}, and for (e) we use $M'_\Lambda\ge3m$ and $o_\mu-\tfrac12\le\lfloor o_\mu\rfloor\le o_\mu$. These inequalities give $r_1-a_1\ge(M'_\Lambda-R)/2\ge m$, and the integer $b_1-(r_1+R)$ is at least $o_1-\tfrac12+M'_\Lambda/2-r_1-R=(M'_\Lambda-R-1)/2\ge m-\tfrac12$, hence at least $m$. Direction~2 is the same with $T$, and $z_0-a_3=b_3-z_0=M'_\Lambda/2\ge m$.
\qed
\end{proof}

\begin{definition}[The loop-adapted grid $\mathfrak G(C)$]\label{def:3.4}
The grid $\mathfrak G(C)$, constructed in Section~\ref{S:B2.grid}, is a function of $C$ and the constants. Its \emph{special planes} are the planes of $C$ and the faces of $\Lbox$. Its \emph{sites} are boxes of fine sides in $[r_0,\bar r_0]$ forming, combinatorially, a box of $\mathbb Z^3$, with nearest-neighbour bonds $E$. The coarse face dual to a bond is the common face of its two sites. \emph{Blocks} are the boxes between consecutive \emph{block planes} of the three directions. A \emph{block face} $F'$ is the common face of two adjacent blocks, and $|F'|$ is the number of bonds across it. The \emph{block distance} of two blocks is the $\ell^\infty$ distance of their indices. For a \emph{removal cell} $\Delta$, $|\partial\Delta|$ is the number of bonds of $\mathbb Z^3$ with exactly one end in $\Delta$, bonds leaving $\Lbox$ included. The \emph{window} $\hat Q(Q)$ of a block $Q$ is the union of the blocks at block distance at most $75$ from $Q$, and $\Qc$ is the set of windows. For a block face $F'$ between blocks $Q$, $Q'$, in the plane $P'$, the \emph{tube zone} $Z(F')$ is the set of sites of $Q\cup Q'$ in the $L_c/2$ layers nearest to $P'$ on each side, and the \emph{tube} $\tau_b$ of a bond $b$ across $F'$ is the segment of its line formed by its $L_c$ sites in $Z(F')$. The grid has the following properties.
(G1) Consecutive parallel block planes are at coarse distance in $[L_c,2L_c-1]$. Blocks have coarse sides in $[L_c,2L_c-1]$, fine sides in $[L,2L-1]$, volume in $[L_c^3,8L_c^3)$, and $|F'|\in[L_c^2,(2L_c-1)^2]$. (G2) Every block is a box with six faces, each a block face or contained in $\partial\Lbox$. A line in direction $\mu$ crosses only block planes perpendicular to $e_\mu$. Every special plane is a block plane, and $\Lbox$ and every removal cell are unions of blocks. (G3) Removal cells tile $\Lbox$, have coarse sides in $[\ell,2\ell+2L_c-3]\subseteq[\ell,3\ell)$ and fine sides at most $3\ell_D$ (a block of coarse side $a\ge L_c\ge4$ has fine side at most $r_0a+r_0-1$, and $\ell\ge308L_c$; Section~\ref{S:B2.G}), and $|\partial\Delta|\in[6\ell^2,54\ell^2)$. (G4) Every window $\hat Q(Q)$ is a box of coarse sides less than $k_P$ that contains $Q_\infty(u,75L_c)\cap\Lbox$ for every site $u\in Q$, and $Q\cup Q'\subseteq\hat Q(Q)$ for every block $Q'$ adjacent to $Q$. Every site lies in at most $151^3$ windows. (G5) $Z(F')\subseteq Q\cup Q'$, tube zones of distinct parallel block faces are disjoint, and a tube crosses no block plane other than $P'$.
\end{definition}

We prove (G1) to (G5) in Section~\ref{S:B2.G} and list the sizes in Table~\ref{tab:4.1}.

\begin{lemma}[Alignment]\label{lem:3.5}
By (G2) the planes of $C$ are block planes; so, with $P:=\{x_3=z_0\}$: (A1) every plaquette of $\Xi$ lies in the block plane $P$, and no bond with both ends in one block crosses $\Xi$; (A2) every block face lies in $\Xi$ or contains no plaquette of $\Xi$; (A3) every block lies on one side of $P$, and $C$ lies on lines of block planes. Every plaquette of $\Xi$ is interior to $\Lbox$.
\end{lemma}

\begin{proof}
By the construction of $\mathfrak G(C)$ (Section~\ref{S:B2.grid}, Steps 1--3) the special planes are block planes, and in each direction the block intervals subdivide each special interval. (A1) A bond crossing $\Xi$ joins two fine cubes on opposite sides of the block plane $P$, which lie in different blocks. (A2) A block face in a plane $x_\mu=\mathrm{const}$ with $\mu\ne3$, or in a plane $x_3\ne z_0$, contains no plaquette of $\Xi$. A block face in $P$ is $I^1\times I^2\times\{z_0\}$ with block intervals $I^1$, $I^2$; since $r_1$, $r_1+R$, $t_1$, $t_1+T$ are block planes, either $I^1\times I^2\subseteq[r_1,r_1+R]\times[t_1,t_1+T]$, and the face lies in $\Xi$, or the interiors are disjoint. (A3) A block lies between consecutive block planes and $P$ is one; each side of $C$ lies on the intersection of $P$ with one of the block planes $x_1\in\{r_1,r_1+R\}$, $x_2\in\{t_1,t_1+T\}$. The last claim is Lemma~\ref{lem:3.3}(e). \qed
\end{proof}

\begin{definition}[Marks]\label{def:3.6}
For a set $Y$ of sites, $F(Y)$ is the set of fine plaquettes that are faces of fine cubes contained in sites of $Y$. A site is \emph{marked} iff one of its fine cubes has a face in $D$. The set of marked sites is $\Mset=\Mset(D)$, and $N(P):=|P|$ for $P\subseteq\Mset$. Then $D\subseteq F(\Mset)$, and a plaquette of $D$ marks at most two sites, since it is a face of at most two fine cubes.
\end{definition}

\subsection{States, functionals and steps}\label{ssec:3.2}\label{ssec:3.3}

For a set $Y$ of sites, we write $E(Y)$ for the set of bonds with both ends in $Y$, $\partial Y$ for the set of bonds of $E=E(\Lbox)$ with exactly one end in $Y$, and $V(A)$ for the set of endpoints of a bond set $A$. A bond \emph{touches} $Y$ if it has an end in $Y$. Bonds leaving $\Lbox$ are not in $E$, that is, the faces of the box are free.

\begin{definition}[States]\label{def:3.7}
A \emph{state} is a triple $\sigma=(\Hset,\Uf,S)$: the \emph{holes} $\Hset\subseteq\Lbox$, the \emph{filled blocks} $\Uf\subseteq\Hset$, a union of blocks, and the \emph{frozen bonds} $S\subseteq E(\Lbox\setminus\Hset)$, each crossing a block face and each of one \emph{kind}: $S=S_1\sqcup S_{\rm cut}$, seal bonds and cut bonds (Definitions~\ref{def:3.seal} and~\ref{def:3.17}). A block is \emph{surviving} if it is not contained in $\Uf$. A frozen bond that $\Hset$ grows to touch leaves $S$. Such a bond is \emph{converted}, and from then on it is an ordinary hole face. The \emph{slit graph} is $\Gp(\sigma):=\bigl(\Lbox\setminus\Hset,\ E(\Lbox\setminus\Hset)\setminus S\bigr)$, and $\mathcal R(\sigma):=S\cup\{b\in E:b\text{ touches }\Hset\}$ is the set of bonds missing from it. Along a history, that is, a finite sequence of states produced by the steps below (Definition~\ref{def:4.1}), the \emph{base} is $B_t:=\Mset\cup\Uf_t$, and the \emph{ever-frozen set} $\Sbar_t$ is the set of bonds frozen at some time $\le t$ that do not touch $B_t$, converted bonds included, each keeping its kind.
\end{definition}

\begin{definition}[Hull functional]\label{def:3.8}
Let $S\subseteq E$ carry weights $w_b\in[0,2]$. For $X'\subseteq\Lbox$,
\begin{equation}\label{eq:3.F}
F^S_\lambda(X'):=|\partial X'|+\sum_{b\in S\cap E(\Lbox\setminus X')}w_b+\lambda|X'| .
\end{equation}
Every frozen bond has the \emph{hull weight} $w_b:=2-\eta_w=\tfrac{19}{10}$.
\end{definition}

\begin{lemma}\label{lem:3.9}
Let $S\subseteq E$ carry weights $w_b\in[0,2]$. (a) $F^S_\lambda$ is the sum of the cut function of a graph on $\Lbox$ with nonnegative capacities and of a modular function; in particular it is submodular. (b) For every $Y\subseteq\Lbox$, the minimizers of $F^S_\lambda$ over $\{X':Y\subseteq X'\subseteq\Lbox\}$ are closed under union and intersection; the smallest, $H^S_\lambda(Y)$, is the source side of a minimum $s$\textendash$t$ cut in a network with nonnegative capacities. Bonds of $S$ that touch $Y$ may be added to $S$ or removed from it, with any weights, without changing $H^S_\lambda(Y)$. (c) $H^S_\lambda$ is extensive and idempotent: $Y\subseteq H^S_\lambda(Y)=H^S_\lambda(H^S_\lambda(Y))$. (d) Let $S$, $S'$ carry weights that agree on $S\cap S'$. If $Y\subseteq Y'$ and $\mathcal R(Y,S)\subseteq\mathcal R(Y',S')$, where $\mathcal R(Y,S):=S\cup\{b:b\text{ touches }Y\}$, then $H^S_\lambda(Y)\subseteq H^{S'}_\lambda(Y')$. In particular $H^S_\lambda$ is a closure operator for fixed $S$ and monotone in $S$.
\end{lemma}

\begin{proof}
For (a), we write a frozen bond of weight $w\le2$ as a modular term plus a cut term of capacity $\frac{2-w}2\ge0$. Indeed, with $c_b(X'):=\mathbf 1[\text{exactly one end of }b=\{u,v\}\text{ in }X']$, the three cases give $w_b\mathbf 1[u,v\notin X']+c_b(X')=w_b-\tfrac{w_b}2(\mathbf 1[u\in X']+\mathbf 1[v\in X'])+\tfrac{2-w_b}2c_b(X')$ for $b\in S$, hence
\begin{equation}\label{eq:4.Fcut}
F^S_\lambda(X')=\sum_{b\in E}\kappa_b\,c_b(X')+\sum_{x\in X'}\mu(x)+\sum_{b\in S}w_b,
\end{equation}
with $\kappa_b:=1$ for $b\notin S$, $\kappa_b:=(2-w_b)/2$ for $b\in S$, and $\mu(x):=\lambda-\frac12\sum_{b\in S,\,b\ni x}w_b$. This proves (a) and explains why frozen bonds weigh at most $2$: weights above $2$ would give negative capacities. Parts (b) and (c) are the lattice and closure properties of the minimizers of a submodular function on $\{X'\supseteq Y\}$. The network has capacities $\kappa_b$ on the bonds, $\mu(x)^\pm$ to the sink or from the source, and $\infty$ from the source to $Y$. For the last assertion of (b) we observe that, on $\{X'\supseteq Y\}$, a bond of $S$ touching $Y$ contributes $1$ or $0$ whatever its weight.

(d) By (b) we may add the bonds of $S\setminus S'$, which touch $Y'$, to $S'$ with their $S$-weights, so that $S\subseteq S'$. Then $g:=F^{S'}_\lambda-F^S_\lambda=\sum_{b\in S'\setminus S}w_b\mathbf 1[\text{no end of $b$ in }X']$ is non-increasing. With $M:=H^S_\lambda(Y)$ and $M':=H^{S'}_\lambda(Y')$, submodularity gives $F^S_\lambda(M\cap M')+F^{S'}_\lambda(M\cup M')\le F^S_\lambda(M)+F^S_\lambda(M')+g(M\cup M')\le F^S_\lambda(M)+F^{S'}_\lambda(M')$. Each term on the left is at least the corresponding one on the right. Hence $M\cap M'\supseteq Y$ is a minimizer, and $M\subseteq M'$.
\qed
\end{proof}

\begin{definition}[Hull step]\label{def:3.10}
In a history, the \emph{hull step} at time $t$ replaces the hole set by $H^{\Sbar_t}_\lambda(B_t)$, the hull from the base computed with the ever-frozen set and the hull weight, and converts the frozen bonds that touch it. A time $t$ (or its state) is \emph{hull-closed} if $\Hset_t=H^{\Sbar_t}_\lambda(B_t)$.
\end{definition}

\begin{definition}[Forced blocks and the regularization functional]\label{def:3.forced}
For a block $Q$ let $3Q$ be the union of the at most $27$ blocks at block distance at most $1$ from $Q$, and $N(3Q):=|\Mset\cap3Q|$. The block $Q$ is \emph{forced} if $N(3Q)\ge\vartheta L_c^2$, where $\vartheta:=3\cdot10^{-5}$ is the \emph{forcing threshold}, and $\mathfrak B_F=\mathfrak B_F(D)$ is the set of forced blocks. Put $\lambda_U:=2c_s/\ell$ and $\mu_{\rm B}:=25$. For a set $\Sbar$ of bonds that cross block faces, each with the hull weight, and for pairs $(\Hset,\Uf)$ with $\Mset\subseteq\Hset\subseteq\Lbox$ and $\Uf\subseteq\Hset$ a union of blocks,
\begin{equation}\label{eq:3.E}
\Ereg^{\Sbar}(\Hset,\Uf):=F^{\Sbar}_\lambda(\Hset)-(\lambda-\lambda_U)|\Uf|+\mu_{\rm B}L_c^2\,\#\{Q\in\mathfrak B_F:\ Q\not\subseteq\Uf\}.
\end{equation}
\end{definition}

Filling a block $Q$ costs at most $|\partial Q|+\lambda_U|Q|<24.6L_c^2<\mu_{\rm B}L_c^2$, filling the forced blocks costs at most $27\cdot24.6/\vartheta$ per mark, and filling a removal cell $\Delta$ costs at most $w_*$ (we give the arithmetic behind these bounds in Section~\ref{S:B1.forced}). In Lemma~\ref{lem:4.forced} we deduce from the first bound that forced blocks are filled. The last bound implies that a minimizer fills each cell whose insulation exceeds $w_*$, as Theorem~\ref{thm:4.30} shows.

\begin{lemma}\label{lem:3.minE}
Let $\Sbar$ be as in Definition~\ref{def:3.forced}. (a) $\Ereg^{\Sbar}$ is, up to a constant, the cut function of an $s$\textendash$t$ network with nonnegative capacities; its minimizers are closed under componentwise union and intersection, and the smallest one, $(\Hset^*,\Uf^*)$, is the source side of a minimum cut, a deterministic function of $\Mset$ and $\Sbar$. (b) $\Hset^*=H^{\Sbar}_\lambda(\Mset\cup\Uf^*)$. (c) The smallest minimizer is componentwise nondecreasing in $\Sbar$.
\end{lemma}

\begin{proof}
(a) We add to the network of Lemma~\ref{lem:3.9}(b) for $F^{\Sbar}_\lambda$ (with $\infty$ from the source to the marks) a node $u_Q$ per block $Q$, with capacity $\infty$ from $u_Q$ to each site of $Q$ and $(\lambda-\lambda_U)|Q|+\mu_{\rm B}L_c^2\mathbf 1[Q\in\mathfrak B_F]$ from the source to $u_Q$. Putting $u_Q$ on the source side means $Q\subseteq\Uf$. A finite cut has $\Mset\cup\Uf\subseteq\Hset$, and since the blocks tile $\Lbox$ by (G2), its capacity is $\Ereg^{\Sbar}(\Hset,\Uf)+(\lambda-\lambda_U)|\Lbox|$ up to the constant of \eqref{eq:4.Fcut}. The capacities are nonnegative, as $\lambda_U<\lambda$. Hence the minimum cuts form a lattice, and the smallest source side is the set reachable from the source in the residual network of a maximum flow. (b) For fixed $\Uf^*$, $\Ereg^{\Sbar}(\cdot,\Uf^*)-F^{\Sbar}_\lambda$ is constant on $\{\Hset\supseteq\Mset\cup\Uf^*\}$, and a smaller minimizer there would give a smaller joint minimizer. (c) Let $\Sbar\subseteq\Sbar'$, and $m$, $m'$ the smallest minimizers of $\Ereg:=\Ereg^{\Sbar}$ and $\Ereg':=\Ereg^{\Sbar'}=\Ereg+g$, $g(\Hset,\Uf):=\sum_{b\in\Sbar'\setminus\Sbar}w_b\mathbf 1[\text{no end of $b$ in }\Hset]$, which does not increase when the pair grows. Minimality of $m'$ for $\Ereg'$ and $g(m\cup m')\le g(m')$ give $\Ereg(m')\le\Ereg(m\cup m')$. Combined with submodularity, this yields $\Ereg(m\cap m')\le\Ereg(m)$. Thus $m\cap m'$ is a minimizer of $\Ereg$, and $m\subseteq m'$.
\qed
\end{proof}

For a state $(\Hset,\Uf,S)$ and a bond set $A$, we define the \emph{two-sided insulation} by $\mathcal A_2(A):=\#\{b\in A:\ b\text{ has exactly one end in }\Hset\}+2\,\#(S\cap A)$, in which a frozen bond counts $2$ whatever its weight.

\begin{definition}[Removal]\label{def:3.11}
Let $\sigma$ be a hull-closed state with base $B$. For a removal cell $\Delta\not\subseteq B$ put $w(\Delta):=\mathcal A_2(E(\Delta))$, with $E(\Delta)$ the set of bonds with both ends in $\Delta$ and $S$ the current frozen set. The \emph{removal step} removes, all at once, every removal cell $\Delta\not\subseteq B$ with $w(\Delta)>w_*:=(1+c_s)|\partial\Delta|$, where $c_s\in[3.3\cdot10^{-7},10]$ is the removal slack: $\Hset:=\Hset\cup\Delta$, $\Uf:=\Uf\cup\Delta$, and the frozen bonds touching $\Delta$ are converted. A hull step follows every removal step.
\end{definition}

\begin{definition}[Cells, interfaces]\label{def:3.12}
Let $\sigma$ be a state, $\Gp=\Gp(\sigma)$, and for $\hat Q\in\Qc$ let $\Gp\cap\hat Q$ be the subgraph of $\Gp$ induced on $\hat Q\setminus\Hset$, with components $\hat\Kc$. For a block $Q\not\subseteq\Hset$, the \emph{cells} of $Q$ are the sets $\Omega_{Q,\hat\Kc}:=Q\cap\hat\Kc$ for the components $\hat\Kc$ of $\Gp\cap\hat Q(Q)$ that meet $Q$. For a block face $F'$ between blocks $Q,Q'\not\subseteq\Hset$, an \emph{interface} $I$ is a nonempty set consisting of all $\Gp$-bonds across $F'$ that join a given cell of $Q$ to a given cell of $Q'$, and $F(I):=F'$. Frozen bonds and hole faces belong to no interface. A bond $b\in I$ is \emph{good} if its tube $\tau_b$ is a $\Gp$-path (its sites lie in $\Lbox\setminus\Hset$ and its bonds are $\Gp$-bonds), and $\good(I)$ is the number of good bonds of $I$.
\end{definition}

\begin{definition}[Weak links]\label{def:3.13}
Let $\hat Q\in\Qc$, $\hat\Kc$ a component of $\Gp\cap\hat Q$, and $f_{\Gp}(A,A')$ the number of $\Gp$-bonds between disjoint site sets $A$, $A'$. For $\chi'>0$, $A\subseteq\hat\Kc$ is a \emph{$\chi'$-weak link} of $\hat\Kc$ if $0<|A|\le|\hat\Kc|/2$ and $f_{\Gp}(A,\hat\Kc\setminus A)<\chi'|A|/k_P$. The \emph{weak-link constants} are $\chi:=2.0\cdot10^{-8}$ and $\chi_{\rm N3}:=\chi/3$.
\end{definition}

Weak links are excluded by Theorem~\ref{thm:4.rest}(c).

\begin{definition}[Auxiliary material and the seal]\label{def:3.seal}
Let $\sigma=(\Hset,\Uf,S)$ be a hull-closed state with ever-frozen set $\Sbar$, let $j$ be a surviving block, and put $V_j:=j\setminus\Hset$. The \emph{auxiliary functional} $F^{\rm aux}_j$ is $F^{\Sbar\cup\partial j}_\lambda$, every bond of $\partial j$ counted as frozen with the hull weight, on the sets $Y$ with $\Hset\subseteq Y\subseteq\Hset\cup V_j$. Its minimizers are closed under union and intersection (Lemma~\ref{lem:3.9}(a)). The smallest one is $M_j$, and $A_j:=M_j\setminus\Hset$ is the \emph{auxiliary material} of $j$. The \emph{seal step} freezes every $\Gp$-bond that crosses a face between two surviving blocks and has an end in $\bigcup_jA_j$. These bonds form $S_1$ and carry the hull weight. A hull step follows every seal step.
\end{definition}

The set $M_j$ is obtained from one maximum flow inside $j$. By Lemma~\ref{lem:4.thin}, a block with a thin piece, attached to the rest of the block by few bonds however large its contacts across the faces of $j$, has auxiliary material, and Lemma~\ref{lem:4.aux} shows that after the seal the next hull step fills the auxiliary material.

\subsection{Phase 1 and Phase 2}\label{ssec:3.4}\label{ssec:3.5}

\begin{definition}[Phase~1 and the terminal properties]\label{def:3.14}\label{def:3.15}
\emph{Phase~1} starts from the state $(\Mset,\emptyset,\emptyset)$ with $\Sbar:=\emptyset$ and repeats the following \emph{outer step} until an outer step changes neither $\Uf$ nor $\Sbar$, that is, until both loops are at rest: (1) $(\Hset,\Uf):=$ the smallest minimizer of $\Ereg^{\Sbar}$ (Lemma~\ref{lem:3.minE}); (2) with $\Uf$ fixed, as long as some surviving block has auxiliary material, perform a seal step, add its bonds to $\Sbar$ (in Phase~1, $\Sbar$ is not pruned: it keeps the seal bonds that touch the base, unlike $\Sbar_t$ of Definition~\ref{def:3.7}; the hulls are the same, as the two sets differ only by bonds touching $B_t$, but step (1) and Theorem~\ref{thm:4.30} use this unpruned set), and perform the hull step $\Hset:=H^{\Sbar}_\lambda(\Mset\cup\Uf)$. Its final state $(\Hset^1,\Uf^1,S^1)$, with $S^1:=\Sbar\cap E(\Lbox\setminus\Hset^1)$, base $B^1:=\Mset\cup\Uf^1$ and ever-frozen set $\Sbar^1:=\Sbar$, is the \emph{Phase-1 output}.

The \emph{terminal properties} of a hull-closed state with base $B$, frozen set $S$ and ever-frozen set $\Sbar$ are: (S-hull) $\Hset=H^{\Sbar}_\lambda(B)=H^S_\lambda(\Hset)$; (S-rem) no removal cell $\Delta\not\subseteq B$ has $w(\Delta)>w_*$; (S-face) every interface $I$ has $\good(I)\ge c_g|F(I)|$; (S-plug) no component of any $\Gp\cap\hat Q$, $\hat Q\in\Qc$, has a $\chi_{\rm N3}$-weak link.
\end{definition}

We show in Theorem~\ref{thm:4.30} that Phase~1 terminates and that its output is the smallest minimizer of $\Ereg^{\Sbar^1}$, has no auxiliary material and no frozen bond, and has the four terminal properties. As $\Sbar$ does not change in the last outer step, the output is the pair returned by step (1) of that outer step, and step (2) finds no auxiliary material.

\begin{definition}[Scale cost, fragmented splits, the cut rule, Phase~2]\label{def:3.17}
Let $\sigma$ be a state with slit graph $\Gp$ and cells as in Definition~\ref{def:3.12}. Every site of $\Lbox\setminus\Hset$ lies in exactly one cell, and every cell in one component of $\Gp$. A \emph{cell set} is a union $U$ of cells. Its cut $\partial_{\Gp}U$ is the set of $\Gp$-bonds with exactly one end in $U$, and $|\partial_{\Gp}U|$ is its number of (coarse) bonds. $U$ is \emph{c-connected} if it is connected in the graph on $U$ whose edges are the $\Gp$-bonds inside $U$ and the pairs of sites in a common cell. The distance of two interfaces is the least Euclidean distance (fine units) between the coarse faces dual to their bonds. For a finite set $\mathcal B$ of interfaces and $s>0$, $\varphi_s(\mathcal B)$ is the number of classes of the equivalence relation generated by ``distance $\le s$''. The \emph{scale cost} is $J(U):=\sum_{j\ge0:\ 2^jL\ge R_{\rm NF}}(\varphi_{2^jL}(\partial_{\Gp}U)-1)$, where $R_{\rm NF}:=L\,e^{c_3^{\rm NF}F_{\rm unit}}$ (fine), $F_{\rm unit}:=\sigma_FL^2$ with $\sigma_F$ the wall tension \eqref{eq:7.sigmaF}, and $\theta_{\rm NF},c_3^{\rm NF}>0$ are the cut-rule constants. A pair $(\Kc,U)$ is \emph{fragmented} if $\Kc$ is a component of $\Gp$, $U\subsetneq\Kc$ is a nonempty c-connected cell set, $\Kc\setminus U$ is c-connected, and $J(U)>\theta_{\rm NF}F_{\rm unit}|\partial_{\Gp}U|/L_c^2$.

\emph{The cut rule} acts at a hull-closed state at which no removal cell is over threshold and no surviving block has auxiliary material: a firing freezes $\partial_{\Gp}U$ for the first fragmented pair $(\Kc,U)$ in a fixed total order of the sets $U\subseteq\Lbox$ that depends only on $C$ and the constants (for instance: order the coarse sites lexicographically by their coordinates, and let $U\prec U'$ iff the least site of $U\triangle U'$ lies in $U$; any such order works, since the proofs below use only that a fragmented pair is cut, never which one; see Section~\ref{S4-S:S4.1}); these bonds form $S_{\rm cut}$, have the hull weight and count $2$ in $\mathcal A_2$ and in $w(\Delta)$; nothing is filled.

\emph{Phase~2} starts from the Phase-1 output, with base $B^1$ and ever-frozen set $\Sbar^1$ minus the bonds touching $B^1$, the ever-frozen set of Definition~\ref{def:3.7} (the unpruned $\Sbar^1$ is used only in Theorem~\ref{thm:4.30} and Section~\ref{ssec:4.9}), and repeats the following pass until a full pass changes nothing: (1) a hull step; (2) a removal step, then a hull step; (3) a seal step, then a hull step; (4) if (1) to (3) changed nothing, one firing of the cut rule, then a hull step. Its terminal state $(\Hset,\Uf,S)$ is the output of the construction.
\end{definition}

The scale cost is well defined, since $\partial_{\Gp}U$ is a union of whole interfaces (Lemma~\ref{lem:3.18}(a)). The trigger of the cut rule uses the coarse cut size, and the non-fragmentation condition that the rule enforces implies the same condition with the fine cut size. Since a coarse face carries at least $r_0^2$ fine plaquettes, we have $|\partial_{\Gp}U|_{\rm fine}/L^2\ge|\partial_{\Gp}U|/L_c^2$, and hence a split fragmented with the fine cut size is fragmented.

\begin{proposition}\label{prop:3.19}
Phase~2 terminates. Its terminal state has (S-hull), (S-rem), (S-face), (S-plug) on the global slit graph and (S-NF): for every component $\Kc$ of the final $\Gp$ and every nonempty c-connected cell set $U\subsetneq\Kc$ with $\Kc\setminus U$ c-connected,
\begin{equation}\label{eq:3.NFstar}
J(U)\le\theta_{\rm NF}\,F_{\rm unit}\,|\partial_{\Gp}U|/L_c^2 .\tag{$\mathrm{NF}_*$}
\end{equation}
If the Euclidean diameter of $\Lbox$ (fine units) is at most $R_{\rm NF}$, then Phase~2 makes no firing and the output is the Phase-1 output.
\end{proposition}

\subsection{The output and the regularization theorem}\label{ssec:3.6}

The output $(\Hset,\Uf,S)$, with base $B=\Mset\cup\Uf$ and $S=S_1\sqcup S_{\rm cut}\subseteq E(\Lbox\setminus\Hset)$, defines the slit domain $\Gp=\Gp(D):=\Gp(\Hset,S)$, the free set $\Fset$ of \eqref{eq:2.FF} and the min-cut area $A_{\Gp}(C)$ of Definition~\ref{def:2.11}. The fine slit graph is that of Section~\ref{ssec:2.5}, and cells and interfaces are those of Definition~\ref{def:3.12}. The \emph{cell graph} of a component $\Kc$ of $\Gp$ has the cells in $\Kc$ as vertices and the interfaces as edges. The weight of a cell $\Omega$ is $|\Omega|_{\rm fine}$, its number of fine cubes. The $\Uc$-cells are the parts of a partition of each cell graph given by Lemma~\ref{lem:4.19} with $V=V_{\Uc}:=\tilde m^{-3}$, one fixed choice being made for each $\Gp$. A component of weight $<V_{\Uc}$ is a single $\Uc$-cell.

\begin{theorem}[Regularization theorem: terminal form]\label{thm:3.20}
For every $D\subseteq\Lbox_2\setminus\partial_2\Lbox$:
\begin{itemize}
\item[(R1)] (\emph{Termination, determinism.}) Phase~1 and Phase~2 terminate; the output $(\Hset,S)$ is a deterministic function of $D$; and $\Mset\subseteq\Hset$, so that $D\subseteq F(\Mset)\subseteq F(\Hset)\subseteq\Fset$.
\end{itemize}
The output satisfies the hull identity and the terminal properties, read on its slit graph: (N1) (S-hull), (N2) (S-face), (N3) (S-plug) and (N4) (S-rem) of Definition~\ref{def:3.14}, and (N5) (S-NF) of Proposition~\ref{prop:3.19}. Consequently the following hold, with constants that depend only on the constants fixed before $K$ and on none of $\beta$, $\tau$, $K$, $C$, $D$, except $c_g'$ and $c_u$, which lie between two such constants.
\begin{itemize}
\item[(R2)] (\emph{Geometry.}) For every $W\subseteq\Lbox\setminus\Hset$, $|\partial_{\Lbox}W|\le a_\eta|\partial_{\Gp}W|+b_\eta\lambda|W|$ with $(a_\eta,b_\eta)=(19/9,10/9)$, where $\partial_{\Lbox}W$ is the set of bonds of $E$ with exactly one end in $W$ and $\partial_{\Gp}W$ the set of $\Gp$-bonds with exactly one end in $W$; with $c_I:=2^{2/3}$, $|\partial_{\Gp}W|\ge(9c_I/38)|W|^{2/3}$ whenever $|W|\le V_*$. No component of $\Gp$ has at most $V_*$ sites.
\item[(R5)] (\emph{Cell volume.}) Every cell satisfies $c_1L_c^3\le|\Omega|<8L_c^3$, with $c_1:=\min\bigl(c_g/2,\ ((1-\eta_w)c_I/24)^3\bigr)$.
\item[(R6)] (\emph{Poincar\'e inequality on cells.}) For every cell $\Omega$ of a block $Q$ there is a connected set $\Kc^\flat(\Omega)$ with $\Omega\subseteq\Kc^\flat(\Omega)\subseteq\hat Q(Q)$ such that every $A\subseteq\Kc^\flat(\Omega)$ with $0<|A|\le|\Kc^\flat(\Omega)|/2$ has $f_{\Gp}(A,\Kc^\flat(\Omega)\setminus A)\ge\chi|A|/(3k_P)$ (expansion form), and, for every $f:\Lbox\setminus\Hset\to\mathbb R$,
\[
\sum_{x\in\Omega}|f(x)-\bar f_\Omega|^2\le\frac{108}{\chi^2}\,k_P^2\sum_{\{x,y\}\in E_{\Gp}(\Kc^\flat(\Omega))}(f(x)-f(y))^2,
\]
where $\bar f_\Omega$ is the average of $f$ over $\Omega$ and $E_{\Gp}(\cdot)$ the set of $\Gp$-bonds with both ends in the set; each site lies in $\Kc^\flat(\Omega)$, and each $\Gp$-bond lies in $E_{\Gp}(\Kc^\flat(\Omega))$, for at most $N_{\rm ov}\le27(k_P/L_c)^3$ cells $\Omega$ (for sites: a site lies in at most $151^3$ windows, and the sets $\Kc^\flat(\Omega)$ of the cells of one block are disjoint).
\item[(R7)] (\emph{Cells and $\Uc$-cells.}) Every cell lies in one block and in one component of $\Gp$; every interface lies in one block face, and that face lies in $\Xi$ or contains no plaquette of $\Xi$. Every $\Uc$-cell is a union of cells, connected in the cell graph, of weight in $[V_{\Uc},(2+6/c_g)V_{\Uc})$, or a whole component of $\Gp$ of weight in $(c_uV_{\Uc},V_{\Uc})$, with $c_u:=r_0^3V_*\tilde m^3=((1-\eta_w)c_I\tilde mL/24)^3$; and every $\Uc$-cell has Euclidean diameter at most $C_{\Uc}\tilde m^{-1}$.
\item[(R8)] (\emph{Interfaces.}) Every interface $I$ has $\good(I)\ge c_g|F(I)|\ge c_g\max(|I|,L_c^2)$, good bonds on one line are at distance $\ge L_c$, and the tube of a good bond of $I$ lies in $Q\cup Q'\subseteq\hat Q(Q)\cap\hat Q(Q')$, with one half in each cell of $I$. A block face carries at most $1/c_g$ interfaces, and a cell has at most $6/c_g$ neighbouring cells. In fine units, $\good_{\rm fine}(I)\ge c_g'\max(|I|_{\rm fine},L^2)$ with $c_g'=c_g(1+1/L_c)^{-2}$.
\item[(R9)] (\emph{Density bound.}) Every removal cell $\Delta$ satisfies $\mathcal A_2(E(\Delta))\le w_*$ and, counting every bond of $E$ with an end in $\Delta$, $\mathcal A_2(E(\Delta)\cup(\partial\Delta\cap E))\le w_*+2|\partial\Delta|=(3+c_s)|\partial\Delta|$.
\item[(R10)] (\emph{Non-fragmentation.}) $(\mathrm{NF}_*)$ holds in the split form (S-NF) of Proposition~\ref{prop:3.19}.
\end{itemize}
\end{theorem}

Item (R1) and the terminal properties (N1) to (N5) come from the construction, and in Section~\ref{sec:4} we derive them from Propositions~\ref{prop:4.4}(b), (d) and~\ref{prop:3.19}. The items (R2) and (R5) to (R10), also proved in Section~\ref{sec:4}, then follow from the terminal state alone, through (N1) to (N5), (R10) being (N5). About cells and interfaces the construction gives more than (R5) to (R8): by Theorem~\ref{thm:4.rest} there is one cell per surviving block, of volume above $(1-3\vartheta)L_c^3$, and one interface per block face.

Below we write $P^{+R}:=\{x:\dist(x,P)\le R\}$ and $r':=2R_*+2$, where $R_*$ is the clearing-out radius of Theorem~\ref{thm:4.42}. We let $\mathcal P$ be the partition of $\Mset$ into the components of the graph joining two marks at distance at most $r'$. The constants $C_\star'$, $C_{\rm free}$ are those of Proposition~\ref{prop:4.39}.

\begin{theorem}[Regularization theorem: history form]\label{thm:3.21}
\begin{itemize}
\item[(R3)] (\emph{Block-local hull.}) At every hull-closed time of the full history (Definition~\ref{def:4.1}), in particular at the output, for every block $q$, $|(\Hset\setminus B)\cap q|<2L_cm_q$ unless it vanishes, where $m_q$ is the number of marks in $q$. So hull material lies only in blocks containing a mark: frozen bonds, filled blocks, removal cells and box faces support no hull material, alone or jointly.
\item[(a)] (\emph{Area per part.}) The Phase-1 output $(\Hset^1,\Uf^1,S^1)$ has $S^1=\emptyset$. For $P\in\mathcal P$ put $Y_P:=\Hset^1\cap P^{+R_*}$. Then $\Hset^1=\bigsqcup_{P\in\mathcal P}Y_P$, no bond joins $Y_P$ to $Y_{P'}$ for $P\neq P'$, every block of $\Uf^1$ lies in one $Y_P$, and
\[
|\partial Y_P|+\lambda|Y_P\setminus\Uf^1|+\lambda_U|Y_P\cap\Uf^1|\le C_\star'N(P),\qquad2|\partial Y_P|+|Y_P|/\ell\le C_{\rm free}N(P).
\]
\item[(b)] (\emph{Uniform reach.}) Every site of $\Hset^1$ lies within distance $R_*$ of $\Mset$, uniformly in $|\Mset|$, $\Lbox$, $C$ and $D$; so $Y_P\subseteq P^{+R_*}$.
\end{itemize}
The constants $C_\star'$, $C_{\rm free}$ and the ratio $R_*/(\ell\log\ell)$ are bounded independently of $\beta$, $\tau$, $K$, $C$ and $D$.
\end{theorem}

We prove (R3) in Corollary~\ref{cor:4.10}, (a) in Proposition~\ref{prop:4.39} and (b) in Theorem~\ref{thm:4.42}.
}
\providecommand{\Lbox}{\Lambda^{\circ}}
\providecommand{\Hset}{\mathbb{X}}
\providecommand{\Mset}{\mathbb{M}}
\providecommand{\Uf}{\mathbb{U}}
\providecommand{\Kc}{\mathfrak{K}}
\providecommand{\Sbar}{\bar S}
\providecommand{\Gp}{G'}
\providecommand{\Ereg}{\mathcal{E}}
\providecommand{\good}{\operatorname{good}}
\providecommand{\Qc}{\mathcal{Q}}
\providecommand{\Uc}{\mathcal{U}}

\section{The regularization: proofs}\label{sec:4}

In this section we prove Theorems~\ref{thm:3.20} and~\ref{thm:3.21} and Proposition~\ref{prop:3.19}. Several steps below are abbreviated, for instance in the proofs of Proposition~\ref{prop:4.4}, of Theorem~\ref{thm:4.7}(iii), (iv) and of Lemma~\ref{lem:4.17}. We first fix the class of histories to which the arguments apply.

\begin{definition}[Admissible history]\label{def:4.1}
An \emph{admissible history} is a finite sequence of states $\sigma_t=(\Hset_t,\Uf_t,S_t)$, $0\le t\le T$, such that (H1) $\sigma_0=(\Mset,\emptyset,\emptyset)$ and the first step is a hull step; (H2) every other step is a hull step or, at a hull-closed state, a \emph{fill}, which adds a union $\Uf'$ of blocks to $\Uf$ and to $\Hset$ and converts the frozen bonds touching $\Uf'$, or a \emph{freeze}, which adds to $S$ a nonempty set of $\Gp$-bonds crossing block faces, each with its kind; (H3) every fill and every freeze is followed by a hull step. The \emph{full history} is Phase~1 followed by Phase~2.
\end{definition}

Phase~1 is admissible: its first outer step is a hull step, the fill of $\Uf^*$ and a hull step, every later outer step that changes $\Uf$ is a fill and a hull step (Lemmas~\ref{lem:3.minE}(b), (c) and~\ref{lem:4.2}), and a seal step is a freeze. The full history is admissible as well, since its removal steps are fills and its seal steps and cut firings are freezes.

\subsection{The hull}\label{ssec:4.1}

We first show in Lemma~\ref{lem:4.2} that the hull steps of an admissible history only enlarge the hole set, and then derive the geometric consequences of the minimality of the hull.

\begin{lemma}\label{lem:4.2}
Let $(\sigma_t)$ satisfy (H1) to (H3) of Definition~\ref{def:4.1}. At every hull-closed time $t$, $\Hset_t=H^{\Sbar_t}_\lambda(B_t)=H^{S_t}_\lambda(\Hset_t)$ and $S_t\subseteq\Sbar_t\subseteq E(\Lbox\setminus B_t)$, and every bond of $\Sbar_t\setminus S_t$ touches $\Hset_t$. Every hull step $t\to t+1$ satisfies $\Hset_{t+1}=H^{S_t}_\lambda(\Hset_t)\supseteq\Hset_t$.
\end{lemma}

\begin{proof}
We use that $H$ is a closure operator, jointly monotone in the base and in the set of missing bonds, as stated in Lemma~\ref{lem:3.9}(d). We argue by induction over the steps. By (H1) to (H3), each time is $0$, or hull-closed, or follows a fill or a freeze at a hull-closed time $t$ and is followed by a hull step. A hull step at a hull-closed time changes nothing, and the first hull step produces $H^\emptyset_\lambda(B_0)\supseteq B_0=\Hset_0$. A freeze at $t$ of a set $A$ of $\Gp$-bonds, which do not touch $\Hset_t\supseteq B_t$, keeps $B$ and $\Hset$, while $\Sbar_{t+1}=\Sbar_t\cup A$. Thus $\Hset_t=H^{\Sbar_t}_\lambda(B_t)\subseteq H^{\Sbar_{t+1}}_\lambda(B_t)=\Hset_{t+2}$. A fill of $\Uf'$ gives $B_{t+1}=B_t\cup\Uf'$ and drops from $\Sbar$ only bonds touching $B_{t+1}$. It follows that $\mathcal R(B_t,\Sbar_t)\subseteq\mathcal R(B_{t+1},\Sbar_{t+1})$ and $\Hset_t\subseteq\Hset_{t+2}$. In this case, moreover, $\Uf'\subseteq B_{t+1}\subseteq\Hset_{t+2}$. In both cases $B_{t+1}\subseteq\Hset_{t+1}\subseteq\Hset_{t+2}$, and by monotonicity and idempotence $H^{\Sbar_{t+1}}_\lambda(\Hset_{t+1})=\Hset_{t+2}$.

We now pass from $\Sbar$ to $S$. Since a bond of $S_{t'}$ does not touch $\Hset_{t'}\supseteq B_{t'}$, we have $S_{t'}\subseteq\Sbar_{t'}$. A bond of $\Sbar_{t'}\setminus S_{t'}$ was converted, so it touches an earlier hole set and hence $\Hset_{t'}$, and by Lemma~\ref{lem:3.9}(b) it may be dropped: $H^{\Sbar_{t'}}_\lambda(\Hset_{t'})=H^{S_{t'}}_\lambda(\Hset_{t'})$. With $t'=t+1$ this is the claim on hull steps. At a hull-closed time, with $B_t\subseteq\Hset_t=H^{\Sbar_t}_\lambda(B_t)$ and idempotence, the same identity implies $\Hset_t=H^{S_t}_\lambda(\Hset_t)$. The inclusion $\Sbar_t\subseteq E(\Lbox\setminus B_t)$ holds by definition.
\qed
\end{proof}

\paragraph{Proof of (R1) and (N1) to (N5).} These properties follow from Propositions~\ref{prop:4.4}(b), (d) and~\ref{prop:3.19}.

\begin{proposition}\label{prop:4.4}
(a) Order states by $\sigma\le\sigma'$ iff $\Hset\subseteq\Hset'$, $\Uf\subseteq\Uf'$ and $\mathcal R(\sigma)\subseteq\mathcal R(\sigma')$. Under (H1) to (H3), every step that changes the state strictly increases it; in particular (H4) $\Hset_t\subseteq\Hset_{t+1}$ and $\mathcal R(\sigma_t)\subseteq\mathcal R(\sigma_{t+1})$ for every $t$, so that $\Gp$ only shrinks, and a bond is frozen at most once along an admissible history. (b) Phase~1 and Phase~2 terminate, and their outputs are deterministic functions of $D$. (c) The terminal state of Phase~2 satisfies (S-hull), (S-rem) and (S-NF), and no surviving block has auxiliary material there. (d) $\Mset\subseteq B_t\subseteq\Hset_t$ at every time; hence, for the output, $D\subseteq F(\Mset)\subseteq F(\Hset)\subseteq\Fset$.
\end{proposition}

\begin{proof}[Sketch]
Hull steps do not shrink $\Hset$ by Lemma~\ref{lem:4.2}, a fill adds blocks to $\Uf$, a freeze adds $\Gp$-bonds to $S$, and converted bonds stay missing. This proves (a). Phase~1 terminates by Theorem~\ref{thm:4.30}, and Phase~2 terminates because each effective step strictly increases the state in a finite box. Smallest minimizers are canonical and the cut rule uses a fixed order, so the outputs are deterministic. Since a terminal pass of Phase~2 changes nothing, Lemmas~\ref{lem:4.2} and~\ref{lem:4.aux}(a) imply (c). Part (d) holds because the base stays in $\Hset$.
\end{proof}

Below, we denote by $f(A,A')$ the number of bonds of $E$ between disjoint site sets $A$, $A'$, whereas $f_{\Gp}$ of Definition~\ref{def:3.13} counts $\Gp$-bonds. Similarly, $s(A,A')$ denotes the number of bonds of $S$ between $A$ and $A'$, and $s(E(A))$ the number of those with both ends in $A$. We put $N:=|\Mset|$. A state is \emph{closed} if $\Hset=H^S_\lambda(\Hset)$ for the hull weight. By Lemma~\ref{lem:4.2}, each hull-closed state of an admissible history is closed, and so is the output.

\paragraph{Proof of (R2).} We apply Lemma~\ref{lem:4.5} and Theorem~\ref{thm:4.7}(i), (ii) at the output, which is closed by (N1).

\begin{lemma}\label{lem:4.5}
Let $(\Hset,\Uf,S)$ be closed, $V:=\Lbox\setminus\Hset$ and $W\subseteq V$. Then
\begin{equation}\label{eq:4.Ra}
f(W,\Hset)+\sum_{b\in S,\ |b\cap W|=1}(w_b-1)+\sum_{b\in S,\ b\subseteq W}w_b\ \le\ |\partial_{\Gp}W|+\lambda|W| .
\end{equation}
Consequently, with $(a_\eta,b_\eta)=(1+\frac1{1-\eta_w},\frac1{1-\eta_w})=(\frac{19}9,\frac{10}9)$, $|\partial_{\Lbox}W|\le a_\eta|\partial_{\Gp}W|+b_\eta\lambda|W|$, and $N_2(W):=f(W,\Hset)+s(W,V\setminus W)+2s(E(W))\le(|\partial_{\Gp}W|+\lambda|W|)/(1-\eta_w)$.
\end{lemma}

\begin{proof}
Filling $W$ does not lower the functional: $F^S_\lambda(\Hset\cup W)\ge F^S_\lambda(\Hset)$. Since bonds of $S$ do not touch $\Hset$, passing from $\Hset$ to $\Hset\cup W$ changes a bond between $W$ and $\Hset$ from $1$ to $0$, a $\Gp$-bond between $W$ and $V\setminus W$ from $0$ to $1$, a bond of $S$ between $W$ and $V\setminus W$ from $w_b$ to $1$ and a bond of $S$ inside $W$ from $w_b$ to $0$, and adds $\lambda|W|$. This is \eqref{eq:4.Ra}. The hull weight has $w_b-1=1-\eta_w$ and $w_b\ge2(1-\eta_w)$, and the bound on $N_2(W)$ follows. For the remaining bound, note that $|\partial_{\Lbox}W|=f(W,\Hset)+s(W,V\setminus W)+|\partial_{\Gp}W|\le|\partial_{\Gp}W|+N_2(W)$.
\qed
\end{proof}

\begin{theorem}[Geometry of the slit domain]\label{thm:4.7}\label{lem:4.6}
Let $(\Hset,\Uf,S)$ be closed and $c_I:=2^{2/3}$. For $W\subseteq\Lbox\setminus\Hset$: (i) $|\partial_{\Gp}W|\ge(9c_I/38)|W|^{2/3}$ whenever $|W|\le V_*=(1-\eta_w)^3\rho_\lambda^3/2$; (ii) no component of $\Gp$ has at most $V_*$ sites; (iii) $|B_{\Gp}(x,r)|\ge\min\{(c_Ir/(48a_\eta))^3,V_*\}$ for all $x\in\Lbox\setminus\Hset$ and integers $r\ge0$; (iv) $\lambda_1^D(\Omega)\ge(c_I^2/(48a_\eta^2))|\Omega|^{-2/3}$ for $\Omega\subseteq\Lbox\setminus\Hset$ with $|\Omega|\le V_*$.
\end{theorem}

\begin{proof}
(i) In a box $\prod_{i=1}^3\{1,\dots,n_i\}$, a line parallel to $e_i$ that meets $W$ carries a bond of the box with exactly one end in $W$ unless its $n_i$ sites lie in $W$, which happens for at most $|W|/n_i$ lines, and the lines meeting $W$ number at least $3|W|^{2/3}$ by the discrete Loomis\textendash Whitney inequality \cite[Theorem~2]{LW49} and the arithmetic-geometric mean inequality. In this box, such bonds therefore number at least $3|W|^{2/3}-|W|\sum_in_i^{-1}\ge c_I|W|^{2/3}$ if $|W|^{1/3}\le0.47\min_in_i$, since $1-2^{2/3}/3=0.4708\ldots$. Lemma~\ref{lem:4.5} controls the boundary in the box up to $b_\eta\lambda|W|$, which is small below $V_*=(c_I\rho_\lambda/(2b_\eta))^3$, an identity that follows from $(c_I/2)^3=\frac12$. Since $|W|^{1/3}\le V_*^{1/3}<\rho_\lambda=L_c/12$ and each side of $\Lbox$ is at least $\ell\ge308L_c$ by Lemma~\ref{lem:3.3}(c), this and Lemma~\ref{lem:4.5} give $c_I|W|^{2/3}\le|\partial_{\Lbox}W|\le a_\eta|\partial_{\Gp}W|+b_\eta\lambda|W|^{1/3}|W|^{2/3}\le a_\eta|\partial_{\Gp}W|+\tfrac{c_I}2|W|^{2/3}$. We conclude that $|\partial_{\Gp}W|\ge(c_I/(2a_\eta))|W|^{2/3}=(9c_I/38)|W|^{2/3}$.

(ii) A component $\Kc$ of $\Gp$ has $\partial_{\Gp}\Kc=\emptyset$, so (i) excludes $0<|\Kc|\le V_*$.

(iii) (Sketch.) We put $v(r):=|B_{\Gp}(x,r)|$. Each $\Gp$-bond leaving $B_{\Gp}(x,r)$ ends at distance $r+1$. Thus $|\partial_{\Gp}B_{\Gp}(x,r)|\le6(v(r+1)-v(r))$, and by (i), $v(r+1)\ge v(r)+\alpha v(r)^{2/3}$ with $\alpha:=c_I/(12a_\eta)\le\frac14$ while $v(r)\le V_*$. Consequently, $v(r+1)^{1/3}>v(r)^{1/3}+\alpha/4$, and $v(r)^{1/3}\ge1+\alpha r/4\ge c_Ir/(48a_\eta)$ until $v$ exceeds $V_*$.

(iv) By (i), each nonempty $A\subseteq\Omega$ has $|\partial_{\Gp}A|\ge h|A|$ with $h:=(9c_I/38)|\Omega|^{-1/3}$, and the Cheeger argument in a graph of maximal degree $6$, based on the coarea formula for $f^2$ and the Cauchy\textendash Schwarz inequality, yields $\lambda_1^D(\Omega)\ge h^2/12\ge(c_I^2/(48a_\eta^2))|\Omega|^{-2/3}$.
\qed
\end{proof}

\paragraph{Segment accounting.} In Lemma~\ref{lem:4.8} and Corollary~\ref{cor:4.10} we assume that $B=\Mset'\cup U$ with $\Mset'$ a set of sites and $U$ a union of blocks and that $S\subseteq E(\Lbox\setminus B)$ carries weights $w_b\in[0,2]$, and we let $\Hset:=H^S_\lambda(B)$ and $W\subseteq\Hset\setminus B$. By Lemma~\ref{lem:4.2}, this setting holds at a hull-closed time of an admissible history, with $\Mset'=\Mset$, $U=\Uf_t$ and $S=\Sbar_t$ carrying the hull weight. A \emph{segment} $R$ of $W$ is a maximal set of consecutive sites of $W$ on an axis-parallel line, and $m(R):=|R|$. It has $m(R)-1$ \emph{interior bonds} and two \emph{ends}, the positions just beyond its extremities. An end is an \emph{$\Hset$-end} if it is a bond to $\Hset\setminus W$, a \emph{frozen end} if it is a bond of $S$ to $\Lbox\setminus\Hset$, an \emph{open end} if it is a bond of $E\setminus S$ to $\Lbox\setminus\Hset$, and a \emph{box end} if it is on $\partial\Lbox$. With $x(R)$ the number of $\Hset$-ends and $o(R)$ the number of open ends, we put $c_W(R):=x(R)+\sum_{\text{frozen ends }b}(w_b-1)+\sum_{\text{interior }b\in S}w_b-o(R)-\lambda m(R)/3$. A \emph{grid bond} has its ends in different blocks. Such a bond lies on a block plane. The \emph{grid-type terms} of $R$ are its $\Hset$-ends, frozen ends and interior frozen bonds that are grid bonds, and its box ends. The number $x_{\rm ng}(R)$ counts the $\Hset$-ends, frozen ends and interior frozen bonds that are not grid bonds. Similarly, $S^{\rm n}$ is the set of bonds of $S$ that are not grid bonds, and $f_{\rm ng}(A,A')$ counts the bonds between $A$ and $A'$ that are not grid bonds. Frozen bonds cross block faces and are therefore grid bonds, so $S^{\rm n}=\emptyset$ at each hull-closed time of an admissible history.

\begin{lemma}\label{lem:4.8}
Let $\lambda\ge12/L_c$. Every segment $R$ of $W$ satisfies $c_W(R)\le2x_{\rm ng}(R)-2m(R)/L_c$.
\end{lemma}

\begin{proof}
Insulation on block planes, which are at least $L_c$ apart, contributes at most $2$ per plane met, while the volume term is $\lambda m/3\ge4m/L_c$. Indeed, since $o=2-x-e_{\rm fr}-e_b$, where $e_{\rm fr}$ is the number of frozen ends and $e_b$ that of box ends, we have
\begin{equation}\label{eq:4.cWid}
c_W(R)=2x+\sum_{\text{frozen ends}}w_b+\sum_{\text{interior frozen}}w_b+e_b-2-\lambda m/3 .
\end{equation}
The terms that are not grid bonds contribute at most $2x_{\rm ng}(R)$, since $w_b\le2$. Each grid-type term has coefficient at most $2$ and sits at its own bond position on a block plane perpendicular to the line. The $m+1$ bond positions of $R$ lie in an interval of length $m$, which meets at most $m/L_c+1$ block planes by (G1). Therefore $c_W(R)\le2x_{\rm ng}(R)+2(m/L_c+1)-2-4m/L_c$.
\qed
\end{proof}

\begin{corollary}[Block-local hull]\label{cor:4.10}\label{lem:4.9}\label{lem:4.27}
For every union $N$ of blocks, $W:=(\Hset\setminus B)\cap N$ is empty or satisfies $|W|<\frac{L_c}3\bigl(f_{\rm ng}(W,B)+\sum_{v\in W}\deg_{S^{\rm n}}(v)\bigr)$, and every bond counted in $f_{\rm ng}(W,B)$ ends in $\Mset'\setminus U$. Consequently, for every block $q$, $|(\Hset\setminus B)\cap q|<(L_c/3)(6m_q+s_q)$ unless it vanishes, with $m_q:=|\Mset'\cap q|$ and $s_q$ the number of endpoints in $q$ of bonds of $S^{\rm n}$, and $\Hset\subseteq B\cup\bigcup\{q:\ q\cap(\Mset'\cup V(S^{\rm n}))\ne\emptyset\}$. In other words, grid bonds of $S$ of any weight in $[0,2]$, filled blocks and box faces support no hull material, alone or jointly. At a hull-closed time of an admissible history, $m_q$ is the number of marks in $q$ and $S^{\rm n}=\emptyset$.
\end{corollary}

\begin{proof}
Let $W\subseteq\Hset\setminus B$ be nonempty. Then the set $\Hset\setminus W\supseteq B$ is strictly smaller than the smallest minimizer $\Hset$. Since bonds of $S$ do not touch $B$ and each site of $W$ lies in three segments, passing from $\Hset$ to $\Hset\setminus W$ changes the contribution of an $\Hset$-end by $+1$, that of a frozen end by $w_b-1$, that of an open end by $-1$ and that of an interior frozen bond by $+w_b$, and it changes the volume term by $-\lambda|W|$, so that
\begin{equation}\label{eq:4.locrem}
F^S_\lambda(\Hset\setminus W)-F^S_\lambda(\Hset)=\sum_Rc_W(R)>0,
\end{equation}
over the segments of $W$ in the three directions. For $W=(\Hset\setminus B)\cap N$, an $\Hset$-end that is not a grid bond has both ends in one block $q\subseteq N$, so it ends in $\Hset\cap N\setminus W\subseteq B$, but not in $U$, since otherwise $q\subseteq U$. It is not frozen either, and it is therefore counted in $f_{\rm ng}(W,B)$. A frozen end that is not a grid bond is counted once in the degree sum, and an interior bond of $S^{\rm n}$ is counted twice. Combining \eqref{eq:4.locrem}, Lemma~\ref{lem:4.8} and $\sum_Rm(R)=3|W|$, we obtain $0<\sum_Rc_W(R)\le2\sum_Rx_{\rm ng}(R)-6|W|/L_c\le2(f_{\rm ng}(W,B)+\sum_{v\in W}\deg_{S^{\rm n}}(v))-6|W|/L_c$. For $N=q$, $f_{\rm ng}(W,B)\le6m_q$ and the degree sum is at most $s_q$, so the bound for a single block holds. At a hull-closed time, such as the output, the last assertion of Corollary~\ref{cor:4.10} applies, and the bound for a single block then gives (R3).
\qed
\end{proof}

For a hull-closed time $t$ and a block face $F'$ between surviving blocks $Q,Q'$, in the plane $P'$, we put $W_t:=(\Hset_t\setminus B_t)\cap(Q\cup Q')$ and $\mathrm{supp}_t(F'):=f_{\rm ng}(W_t,B_t)$. For a line $\ell\perp F'$, $b_\ell$ denotes its bond across $F'$. As $B_t\cap(Q\cup Q')$ consists of marks, $\mathrm{supp}_t(F')\le6(m_Q+m_{Q'})$.

\begin{lemma}\label{lem:4.11}
At every hull-closed time $t$ of an admissible history and for every such face $F'$: (a) the lines $\ell\perp F'$ such that $b_\ell\notin\Sbar_t$ and $\ell\cap Z(F')$ contains a site of $W_t$ and no mark, number fewer than $2\,\mathrm{supp}_t(F')$ unless there is none; (b) without the condition on $b_\ell$, these lines number fewer than $2\,\mathrm{supp}_t(F')/\eta_w$ unless there is none. A frozen bond lies on the tube of a line $\ell\perp F'$ only if it is $b_\ell$, that is, frozen bonds block no tube of another line.
\end{lemma}

\begin{proof}
A line counted in (a) or (b) carries a segment that is either long, and pays through Lemma~\ref{lem:4.8}, or short, and then has no grid-type term except $b_\ell$ and has deficit at least $\eta_w$. We put $d(R):=2x_{\rm ng}(R)-c_W(R)$ for the segments of $W_t$. By Lemma~\ref{lem:4.8}, $d(R)\ge2m(R)/L_c\ge0$, and \eqref{eq:4.locrem} implies $\sum_Rd(R)<2\sum_Rx_{\rm ng}(R)\le2\,\mathrm{supp}_t(F')$, as in the proof of Corollary~\ref{cor:4.10}. For a counted line $\ell$ let $R_\ell$ be the segment of $W_t$ along $\ell$ through a site of $\ell\cap Z(F')\cap W_t$. Distinct lines give distinct segments. If $m(R_\ell)\ge L_c/2$, then $d(R_\ell)\ge1$. Otherwise all bond positions of $R_\ell$ are at distance $<L_c$ from $P'$, so by (G1) the only possible grid-type term is $b_\ell$, which is not an $\Hset$-end, since its other end would lie in $\Hset_t\cap(Q\cup Q')\setminus W_t\subseteq B_t$, a mark of $\ell\cap Z(F')$. Thus, in this case, $d(R_\ell)=2+\lambda m/3-\beta_\ell$, with $\beta_\ell=2-\eta_w$ if $b_\ell\in\Sbar_t$ and $b_\ell$ is a bond position of $R_\ell$, and $\beta_\ell=0$ otherwise. It follows that $d(R_\ell)\ge1$ for each line counted in (a) and $d(R_\ell)\ge\eta_w$ for each line counted in (b). Summing these bounds, we obtain (a) and (b). For the last claim, recall that a frozen bond is a grid bond, and use (G5): a tube consists of $b_\ell$ and of bonds interior to $Q$ or to $Q'$, which are not grid bonds.
\qed
\end{proof}

\subsection{Cells, giants and the seal}\label{ssec:4.3}

In this subsection we control the cells by sparseness. Since forced blocks are filled (Lemma~\ref{lem:4.forced}), a surviving block has fewer than $\vartheta L_c^2$ marks near it, its hole sites lie on few lines by Corollary~\ref{cor:4.10}, and one component of its free sites, the giant, contains all but a small volume. The seal excludes thin sets (Lemma~\ref{lem:4.thin} and Theorem~\ref{thm:4.rest}), at a cost of at most $726$ frozen bonds per mark (Proposition~\ref{prop:4.sealcount}).

\begin{lemma}\label{lem:4.forced}
Let $\Sbar$ be as in Definition~\ref{def:3.forced}. The smallest minimizer $(\Hset^*,\Uf^*)$ of $\Ereg^{\Sbar}$ has $\bigcup\mathfrak B_F\subseteq\Uf^*$. Hence at every hull-closed time of the full history other than the first every forced block lies in $\Uf$, and every surviving block $q$ has $m_q\le N(3q)<\vartheta L_c^2$.
\end{lemma}

\begin{proof}
Suppose that $Q\in\mathfrak B_F$ and $Q\not\subseteq\Uf^*$. Then $Q\cap\Uf^*=\emptyset$, and we compare with $(\Hset^*\cup Q,\Uf^*\cup Q)$. Only the at most $|\partial Q|$ bonds leaving $Q$ gain weight, at most $1$ each, the volume terms change by $\lambda_U|Q|-\lambda|Q\cap\Hset^*|\le\lambda_U|Q|$, and the penalty drops by $\mu_{\rm B}L_c^2>|\partial Q|+\lambda_U|Q|$ (see Section~\ref{S:B1.forced}). This is a contradiction. For the second assertion we use that $\Uf$ never decreases along the full history, by Proposition~\ref{prop:4.4}(a), and that the second hull-closed time follows the first outer step of Phase~1. If that step fills nothing, then $\mathfrak B_F=\emptyset$.
\qed
\end{proof}

\begin{lemma}\label{lem:4.runs}
At every hull-closed time of an admissible history, every surviving block $q$ satisfies: (a) $|\Hset\cap q|<3L_cm_q$ unless $\Hset\cap q=\emptyset$; (b) unless there is none, fewer than $9m_q$ maximal runs of $\Hset$ in the three directions lie inside $q$ and are joined by both end bonds to sites of $V_q:=q\setminus\Hset$.
\end{lemma}

\begin{proof}
(a) The set $\Hset\cap q$ consists of the $m_q$ marks of $q$ and, by Corollary~\ref{cor:4.10}, fewer than $2L_cm_q$ sites of hull material. (b) At most $3m_q$ runs contain a mark. A run without marks is a segment $R$ of $W:=(\Hset\setminus B)\cap q$ with two open ends inside $q$, since in-block bonds are not frozen, and with no grid-type term. For such a segment, $d(R):=2x_{\rm ng}(R)-c_W(R)=2+\lambda m(R)/3$ by \eqref{eq:4.cWid}. As in Lemma~\ref{lem:4.11}, we have $d\ge0$ for each segment of $W$ and $\sum_Rd(R)<2f_{\rm ng}(W,B)\le12m_q$. Therefore fewer than $6m_q$ runs have no mark, which proves (b).
\qed
\end{proof}

\begin{lemma}\label{lem:4.LW}\label{lem:4.planar}
(a) Let $\mathsf Q$ be a box with sides $L_1,L_2,L_3$, $Z\subseteq\mathsf Q$, $\mathsf Q\setminus Z=A\sqcup B$, and $n_\mu$ the number of lines of $\mathsf Q$ in direction $\mu$ meeting both $A$ and $B$. Then $\min(|A|,|B|)\le4|Z|+6.4\max_\mu(n_\mu L_\mu)$. (b) Let $\mathsf R$ be a rectangle of $\mathbb Z^2$ with sides $s_1,s_2\ge s$, and $\sigma\subseteq\mathsf R$ with $|\sigma|\le s^2/4$. At least $2|\sigma|^{1/2}$ bonds of $\mathsf R$ have exactly one end in $\sigma$.
\end{lemma}

\begin{proof}
(a) Let $a_\mu$ be the number of lines in direction $\mu$ meeting $A$ but not $B$. Then $|A|\le L_\mu(a_\mu+n_\mu)$, and the lines meeting $B$ are among the other $|\mathsf Q|/L_\mu-a_\mu$. By the discrete Loomis\textendash Whitney inequality \cite[Theorem~2]{LW49}, it follows that $|B|^2\le\prod_\mu(|\mathsf Q|/L_\mu-a_\mu)\le\prod_\mu(|\mathsf Q|-|A|+n_\mu L_\mu)/L_\mu$. With $\alpha:=|A|/|\mathsf Q|$, $\beta:=1-\alpha$, $\zeta:=|Z|/|\mathsf Q|$ and $\nu:=\max_\mu n_\mu L_\mu/|\mathsf Q|$ this reads $(\beta-\zeta)^2\le(\beta+\nu)^3$. If $\nu>0.1$, then $\min(|A|,|B|)\le|\mathsf Q|/2<6.4\nu|\mathsf Q|$. If $\nu\le0.1$, then $(\beta+\nu)^{3/2}\le\beta^{3/2}+1.6\nu$, so $\beta(1-\sqrt\beta)\le\zeta+1.6\nu$. Since $1-\sqrt\beta\ge(1-\beta)/2$, in this case $\alpha(1-\alpha)\le2\zeta+3.2\nu$, and $\min(|A|,|B|)\le2|\mathsf Q|\alpha(1-\alpha)\le4|Z|+6.4\nu|\mathsf Q|$.

(b) Each row and each column that meets $\sigma$ without lying in it carries such a bond, a different one for each. If no row or column lies in $\sigma$, the $r$ rows and $c$ columns meeting $\sigma$ give $r+c\ge2(rc)^{1/2}\ge2|\sigma|^{1/2}$ bonds. If a row lies in $\sigma$ (the case of a column is symmetric), all columns meet $\sigma$, and if moreover a column lies in $\sigma$, at most $|\sigma|/s_2+|\sigma|/s_1\le s/2$ rows and columns lie in $\sigma$, so that either way there are at least $s\ge2|\sigma|^{1/2}$ bonds.
\qed
\end{proof}

\begin{proposition}[The giant]\label{prop:4.giant}
At every hull-closed time of the full history other than the first, every surviving block $q$ has $|V_q|>(1-3\vartheta)L_c^3$, and the graph $\Gp_q$ of the $\Gp$-bonds with both ends in $q$ has a unique component $C_q$, the \emph{giant}, with $|V_q\setminus C_q|<128L_cm_q$ unless $V_q=C_q$.
\end{proposition}

\begin{proof}
Since bonds inside a block are never frozen, on a line of $q$ that meets two disjoint unions $A$, $B$ of components of $\Gp_q$, two consecutive sites of $V_q$ in $A$ and in $B$ are separated by a run of Lemma~\ref{lem:4.runs}(b), a different run for each line. If $A\sqcup B=V_q$, Lemma~\ref{lem:4.LW}(a) with $Z=\Hset\cap q$, $|Z|<3L_cm_q$ and $L_\mu<2L_c$ yields $\min(|A|,|B|)<12L_cm_q+6.4\cdot9m_q\cdot2L_c<128L_cm_q$. By Lemma~\ref{lem:4.forced}, $m_q<\vartheta L_c^2$, so $|V_q|>(1-3\vartheta)L_c^3>3\cdot128L_cm_q$. If each component had fewer than $|V_q|/3$ sites, a union $A$ of components with $|V_q|/3\le|A|<2|V_q|/3$ would violate the bound. The largest component $C_q$ therefore has at least $|V_q|/3$ sites, and the bound with $A=C_q$ implies $|V_q\setminus C_q|<128L_cm_q$.
\qed
\end{proof}

A \emph{thin set} of a surviving block $j$ is a nonempty $A\subseteq V_j$ with $|A|<\varepsilon_{\rm th}L_c^3:=0.008L_c^3$ and $\mathrm{cut}_j(A)<|A|/L_c$, $\mathrm{cut}_j(A)$ being the number of $\Gp$-bonds inside $j$ between $A$ and $V_j\setminus A$.

\begin{lemma}[Thin sets]\label{lem:4.thin}
At every hull-closed time of the full history other than the first, every thin set $A$ of a surviving block $j$ satisfies $F^{\rm aux}_j(\Hset\cup A)<F^{\rm aux}_j(\Hset)$. Hence a surviving block without auxiliary material has no thin set, and $V_j=C_j$ is connected in $\Gp_j$.
\end{lemma}

\begin{proof}
We show that in each planar slice a thin set is enclosed by in-block insulation and that, by planar isoperimetry, this insulation exceeds the volume penalty. We fix a direction $\nu$. A line of $j$ in direction $\nu$ that meets $A$ either meets $V_j\setminus A$, and then carries a bond counted in $\mathrm{cut}_j(A)$ or a run of Lemma~\ref{lem:4.runs}(b), or lies in $A\cup(\Hset\cap j)$, which happens for at most $(|A|+3L_cm_j)/L_c$ lines. A slice $\sigma$ of $A$ perpendicular to $\nu$ meets each line once. Therefore, with $m_j<\vartheta L_c^2$ from Lemma~\ref{lem:4.forced}, $|\sigma|\le a:=\mathrm{cut}_j(A)+9m_j+(|A|+3L_cm_j)/L_c<(2\varepsilon_{\rm th}+12\vartheta)L_c^2\le0.0164\,L_c^2\le L_c^2/4$. The bonds of the cross-section of $j$ (sides at least $L_c$) with exactly one end in $\sigma$ end in $\Hset$ or in $V_j\setminus A$, and there are at least $2|\sigma|^{1/2}\ge2|\sigma|/a^{1/2}$ of them, where the first bound is Lemma~\ref{lem:4.LW}(b). Summing over the slices, we obtain $f_j(A,\Hset)+\mathrm{cut}_j(A)\ge2|A|/a^{1/2}\ge15.6|A|/L_c$, with $f_j(A,\Hset)$ the number of bonds inside $j$ from $A$ to $\Hset$.

Passing from $\Hset$ to $\Hset\cup A$ changes $F^{\rm aux}_j$ by $-f_j(A,\Hset)$ on these hole faces, by at most $0$ on the bonds of $\partial j$ at $A$ (from $w_b$ to $1$, or from $1$ to $0$), by $+\mathrm{cut}_j(A)$ on the in-block bonds to $V_j\setminus A$, and by $+\lambda|A|$. In total the change is at most $-15.6|A|/L_c+2\,\mathrm{cut}_j(A)+12|A|/L_c<0$. If $j$ has no auxiliary material, $\Hset$ minimizes $F^{\rm aux}_j$, so $j$ has no thin set. Moreover, a component of $\Gp_j$ other than $C_j$ has $\mathrm{cut}_j=0$ and, by Proposition~\ref{prop:4.giant}, fewer than $128L_cm_j<\varepsilon_{\rm th}L_c^3$ sites, so it would be thin.
\qed
\end{proof}

\begin{lemma}[The seal]\label{lem:4.aux}
Let $t$ be a hull-closed time of an admissible history and $j$ a surviving block with $A_j\neq\emptyset$. (a) Some $\Gp$-bond that crosses a face between $j$ and a surviving block has an end in $A_j$. (b) If a seal step at $t$ is followed by a hull step with the same $\Uf$, then $A_j\subseteq\Hset_{t+2}$.
\end{lemma}

\begin{proof}
On the sets between $\Hset:=\Hset_t$ and $\Hset\cup A_j$, the functionals $F:=F^{\Sbar_t}_\lambda$ and $F^{\rm aux}_j$ differ by a constant unless a $\Gp$-bond leaves $A_j$ across a face of $j$. Indeed, since a block adjacent to $j$ that is not surviving lies in $\Hset$, the other bonds of $\partial j$ at $A_j$ touch $\Hset$ or are frozen. Those at $V_j\setminus A_j$ keep their values. (a) Otherwise $F(\Hset\cup A_j)-F(\Hset)=F^{\rm aux}_j(M_j)-F^{\rm aux}_j(\Hset)<0$, since $M_j\ne\Hset$ is the smallest minimizer. This contradicts $\Hset=H^{\Sbar_t}_\lambda(B_t)$, as $\Hset\cup A_j\supseteq B_t$.

(b) Let $X':=\Hset_{t+2}\supseteq\Hset$ (Lemma~\ref{lem:4.2}), $F':=F^{\Sbar_{t+2}}_\lambda$, $A':=A_j\setminus X'$ and $C:=\Hset\cup(A_j\cap X')\subseteq X'$. On the sets between $C$ and $M_j=C\cup A'$ only the sites of $A'$ vary. The bonds inside $j$ at these sites count alike in $F'$ and $F^{\rm aux}_j$, and after the seal each bond of $\partial j$ at them is frozen or touches $\Hset$, in both. Consequently, $F'-F^{\rm aux}_j$ is constant on these sets. If $A'\neq\emptyset$, then $C\subsetneq M_j$, and submodularity of $F'$ implies $F'(X'\cup A')-F'(X')\le F'(C\cup A')-F'(C)=F^{\rm aux}_j(M_j)-F^{\rm aux}_j(C)<0$, contradicting the minimality of $X'=H^{\Sbar_{t+2}}_\lambda(B_t)$.
\qed
\end{proof}

\begin{proposition}\label{prop:4.sealcount}
Along the full history, a block face $F'$ between blocks $Q$, $Q'$ carries at most $121(m_Q+m_{Q'})$ bonds frozen by seal steps, converted ones included. Hence seal steps freeze at most $C_{\rm seal}N$ bonds, all on faces $F'$ with $m_Q+m_{Q'}>0$, where $C_{\rm seal}:=6\cdot121=726$.
\end{proposition}

\begin{proof}
Each seal bond on $F'$ has an end in auxiliary material, which the next hull step fills while $Q$ and $Q'$ still survive. This follows from Lemma~\ref{lem:4.aux}(b), since in Phase~1 $\Uf$ is fixed during the inner loop and in Phase~2 the hull step follows the seal before any removal. Consequently, at the last hull-closed time $T$ at which $Q$ and $Q'$ both survive, each seal bond on $F'$ has an end in $W_T$ in the layer next to $F'$, hence in $Z(F')$, and distinct bonds lie on distinct lines $\ell\perp F'$. At most $m_Q+m_{Q'}$ of these lines carry a mark in $Z(F')$, and fewer than $2\,\mathrm{supp}_T(F')/\eta_w\le120(m_Q+m_{Q'})$ of the others meet $W_T$ in $Z(F')$, the first inequality being Lemma~\ref{lem:4.11}(b). Since a block has six faces, summing over all block faces yields the bound on the total number of seal bonds.
\qed
\end{proof}

\begin{theorem}[Cells at rest]\label{thm:4.rest}
Let $t$ be a hull-closed time of the full history other than the first at which no surviving block has auxiliary material. Then: (a) every surviving block $q$ carries exactly one cell, $\Omega_q=V_q$, connected in $q$, with $(1-3\vartheta)L_c^3<|\Omega_q|<8L_c^3$, and every other block lies in $\Hset$; (b) every block face $F'$ between surviving blocks contains no $\Gp$-bond, or carries exactly one interface $I$; in the latter case $F'$ contains no cut bond and $\good(I)\ge(1-134\vartheta)|F'|$; in particular (S-face) holds; (c) every component $\Kc$ of $\Gp\cap\hat Q$, $\hat Q\in\Qc$, satisfies $f_{\Gp}(A,\Kc\setminus A)\ge h_\star|A|$ for $0<|A|\le|\Kc|/2$, with $h_\star:=2.2\cdot10^{-11}/L_c$, and (S-plug) holds.
\end{theorem}

\begin{proof}
(a) By Lemma~\ref{lem:4.thin}, $V_q=C_q$ is connected in $q$, and therefore lies in one component of $\Gp\cap\hat Q(q)$ and is the only cell of $q$. The volume bounds follow from Proposition~\ref{prop:4.giant} and (G1). Finally, a block that is not surviving lies in $\Uf\subseteq\Hset$, which is the last assertion of (a).

(b) By (a), the $\Gp$-bonds across $F'$, if any, join the cell of $Q$ to the cell of $Q'$ and form one interface. Cuts fire only at such times, where by (a) a cut, which is a union of whole interfaces by Lemma~\ref{lem:3.18}(a), freezes all $\Gp$-bonds of each face its bonds cross. As $\Gp$ only shrinks, a face with a $\Gp$-bond carries no cut bond. A line $\ell\perp F'$ is good unless $\ell\cap Z(F')$ contains a mark, or $b_\ell$ is a seal bond, or else $b_\ell\notin\Sbar_t$ and $\ell\cap Z(F')$ contains a site of $W_t$. Indeed, on each remaining line the tube has no hole site, and its bonds are $\Gp$-bonds, since bonds inside a block are never frozen. The first case holds for at most $m_Q+m_{Q'}$ lines and, by Proposition~\ref{prop:4.sealcount}, the second holds for at most $121(m_Q+m_{Q'})$ lines. The third holds for fewer than $2\,\mathrm{supp}_t(F')\le12(m_Q+m_{Q'})$ lines, where the first inequality comes from Lemma~\ref{lem:4.11}(a). With $m_Q+m_{Q'}\le N(3Q)<\vartheta L_c^2\le\vartheta|F'|$, we obtain $\good(I)\ge(1-134\vartheta)|F'|\ge c_g|F'|$.

(c) We show that expansion can fail only through a thin set, which Lemma~\ref{lem:4.thin} excludes. Let $A\subseteq\Kc$, $0<|A|\le|\Kc|/2$, and $\mathrm{cut}:=f_{\Gp}(A,\Kc\setminus A)$. Since windows are unions of blocks, by (a) $\Kc$ is a union of sets $V_j$ of surviving blocks $j\subseteq\hat Q$, connected through faces that contain $\Gp$-bonds. We put $A_j:=A\cap V_j$ and $B_j:=V_j\setminus A_j$. A line of $j$ meeting $A_j$ and $B_j$ carries a bond counted in $\mathrm{cut}$ or a run of Lemma~\ref{lem:4.runs}(b). If $\min(|A_j|,|B_j|)\ge\varepsilon_{\rm th}L_c^3$ for some $j$, then $\varepsilon_{\rm th}L_c^3\le12L_cm_j+12.8L_c(\mathrm{cut}+9m_j)$ by Lemma~\ref{lem:4.LW}(a), so $\mathrm{cut}\ge(\varepsilon_{\rm th}-127.2\vartheta)L_c^2/12.8\ge\varepsilon_{\rm th}L_c^2/25.6$. Otherwise each block of $\Kc$ is either \emph{in} ($|B_j|<\varepsilon_{\rm th}L_c^3$) or \emph{out} ($|A_j|<\varepsilon_{\rm th}L_c^3$), but not both, since $|V_j|>2\varepsilon_{\rm th}L_c^3$. If an in-block and an out-block share a face $F'$ that contains $\Gp$-bonds, we follow each line $\perp F'$ from the face into either block until it reaches the majority side of that block through a bond counted in $\mathrm{cut}$ or a run, or stays in the minority side and $\Hset$ for the whole block. In this case the minority side of each block meets the face layer in at most $\mathrm{cut}+9\vartheta L_c^2+(\varepsilon_{\rm th}+3\vartheta)L_c^2$ sites, all other bonds among the $\Gp$-bonds across $F'$, of which there are at least $(1-134\vartheta)L_c^2$ by (b), join $A$ to $\Kc\setminus A$, and $3\,\mathrm{cut}\ge(1-158\vartheta-2\varepsilon_{\rm th})L_c^2$. Otherwise all blocks are in, which is impossible since then $|A|\ge\sum_j(|V_j|-\varepsilon_{\rm th}L_c^3)>|\Kc|/2$, or all are out. In the latter case each nonempty $A_j$ has fewer than $\varepsilon_{\rm th}L_c^3$ sites and is not thin, so $\mathrm{cut}\ge\sum_j\mathrm{cut}_j(A_j)\ge|A|/L_c$. In all cases $\mathrm{cut}\ge\min(3.1\cdot10^{-4}L_c^2,|A|/L_c)\ge h_\star|A|$, where we use that $|A|<k_P^3/2$ by (G4), together with the arithmetic of Section~\ref{S:B1.cells}. As $3k_Ph_\star\ge\chi$, we have $\chi_{\rm N3}/k_P=\chi/(3k_P)\le h_\star$, and (S-plug) holds.
\qed
\end{proof}

\begin{theorem}[The Phase-1 output]\label{thm:4.30}\label{prop:3.16}
Phase~1 terminates. Its output $(\Hset^1,\Uf^1,S^1)$ is the smallest minimizer of $\Ereg^{\Sbar^1}$, no surviving block has auxiliary material there, and: (i) every bond of $\Sbar^1$ touches $\Hset^1$, so $S^1=\emptyset$, and crosses a face between two blocks that together contain a mark; (ii) the four terminal properties (S-hull), (S-rem), (S-face), (S-plug) hold.
\end{theorem}

\begin{proof}
By Lemma~\ref{lem:3.minE}(c), each outer step returns $\Uf^*\supseteq\Uf$, since $\Sbar$ only grows, and by Lemma~\ref{lem:4.aux}(a) each inner round freezes at least one bond. Each outer step except the last enlarges $\Uf$ or $\Sbar$, which only grow in a finite box. Phase~1 thus terminates. In the last outer step the inner loop performs no seal step, so $\Sbar=\Sbar^1$ throughout that step. Thus step (1) of the last outer step returns the smallest minimizer of $\Ereg^{\Sbar^1}$, with hole set $H^{\Sbar^1}_\lambda(\Mset\cup\Uf^1)$ by Lemma~\ref{lem:3.minE}(b). This pair is the output, and step (2) finds no auxiliary material at it.

(i) Each bond of $\Sbar^1$ has an end in auxiliary material, which the next hull step fills by Lemma~\ref{lem:4.aux}(b). For the faces crossed by these bonds we use Proposition~\ref{prop:4.sealcount}.

(ii) Property (S-hull) follows from Lemma~\ref{lem:4.2}, and (S-face), (S-plug) follow from Theorem~\ref{thm:4.rest}. For (S-rem), let $\Delta\not\subseteq B^1$ be a removal cell with $\Delta\not\subseteq\Hset^1$ (otherwise $w(\Delta)=0$). We compare with $(\Hset^1\cup\Delta,\Uf^1\cup\Delta)$, admissible since $\Delta$ is a union of blocks. The bonds of $E(\Delta)$ with one end in $\Hset^1$ lose their weight $1$, since by (i) no bond is frozen. At most $|\partial\Delta|$ bonds gain at most $1$, the volume terms grow by at most $\lambda_U|\Delta|$, and the penalty does not grow. Therefore $w(\Delta)\le|\partial\Delta|+\lambda_U|\Delta|\le(1+c_s)|\partial\Delta|=w_*$, since $|\Delta|/|\partial\Delta|<\ell/2$ for a box with sides in $[\ell,3\ell)$ (see Section~\ref{S:B1.forced}).
\qed
\end{proof}

\begin{lemma}\label{lem:4.16}
Let the state satisfy (S-plug), $\hat Q\in\Qc$, and $\Kc$ a component of $\Gp\cap\hat Q$. Then $f_{\Gp}(A,\Kc\setminus A)\ge\chi|A|/(3k_P)$ for $A\subseteq\Kc$ with $0<|A|\le|\Kc|/2$, and $\sum_{x\in\Kc}|f(x)-\bar f_{\Kc}|^2\le(108/\chi^2)k_P^2\sum_{\{x,y\}\in E_{\Gp}(\Kc)}(f(x)-f(y))^2$ for every $f$.
\end{lemma}

\begin{proof}
The first claim is (S-plug), since $\chi_{\rm N3}/k_P=\chi/(3k_P)$. The Poincar\'e inequality follows from the Cheeger argument and a median. Let $\mu$ be a median of $f$ on $\Kc$. Since each of $(f-\mu)_\pm$ is supported on at most $|\Kc|/2$ sites, the argument of Theorem~\ref{thm:4.7}(iv) in the graph $(\Kc,E_{\Gp}(\Kc))$, with $h:=\chi/(3k_P)$, bounds $\sum(f-\mu)_\pm^2$ by $12h^{-2}=108k_P^2/\chi^2$ times the Dirichlet form of $(f-\mu)_\pm$. We add the two bounds, using $(a-b)^2\ge(a_+-b_+)^2+(a_--b_-)^2$, and replace $\mu$ by the mean.
\qed
\end{proof}

\begin{lemma}[Good bonds]\label{lem:4.17}
At every state satisfying (S-face), (R8) holds, a fine bond across $F(I)$ being good if the segment of $L$ fine sites of its line centred at it is a path of the fine slit graph.
\end{lemma}

\begin{proof}[Sketch]
The first claim follows from (S-face) and (G1), and the second holds because bonds across block faces lie on block planes, at least $L_c$ apart. A good tube is a $\Gp$-path in $Z(F')\subseteq Q\cup Q'\subseteq\hat Q(Q)\cap\hat Q(Q')$, by (G5) and (G4). Thus its half in $Q$ lies in the component of $\Gp\cap\hat Q(Q)$ of its end, the cell of $I$ in $Q$. The interfaces of one face are disjoint with at least $c_g|F'|$ bonds each, and a block has six faces. Together, these facts bound the number of interfaces of a face and the number of neighbouring cells of a cell. In fine units, a block face meets at most one exceptional layer on each side, so $|F'|_{\rm fine}<r_0^2|F'|(1+1/L_c)^2$, while each coarse good bond carries at least $r_0^2$ fine good bonds.
\qed
\end{proof}

\begin{theorem}[Cells]\label{thm:4.18}
At every closed state with (S-face) and (S-plug), such as the output, (R5), (R6) and the cell part of (R7) hold, with $\Kc^\flat(\Omega):=\hat\Kc$ for the cell $\Omega=Q\cap\hat\Kc$ of a block $Q$.
\end{theorem}

\begin{proof}
For (R5), $|\Omega|\le|Q|<8L_c^3$ by (G1). An interface on a face $F'$ of $Q$ provides, by Lemma~\ref{lem:4.17}, at least $c_gL_c^2$ good half-tubes of $L_c/2$ sites in $\Omega$, on distinct lines. Without an interface no $\Gp$-bond leaves $\Omega$, since inside $Q$ a $\Gp$-bond joins two sites of one cell and across a face it belongs to an interface. In that case $\Omega$ is therefore a union of components of $\Gp$, each with more than $V_*=((1-\eta_w)c_I/24)^3L_c^3$ sites by Theorem~\ref{thm:4.7}(ii). For (R6), we apply Lemma~\ref{lem:4.16} in $\hat\Kc\supseteq\Omega$ and use that the mean minimizes the sum of squares. The overlap bound holds because a bond lies in at most $151^3\le27(k_P/L_c)^3$ windows (G4) and the components $\hat\Kc$ of distinct cells of one block are distinct. For (R7), we use that $\hat\Kc$ is connected in $\Gp$, together with Lemma~\ref{lem:3.5}(A2).
\qed
\end{proof}

\begin{lemma}[$\Uc$-cells]\label{lem:4.19}
Let $\mathcal T$ be a connected graph of maximal degree at most $D$, with vertex weights $w(v)\in(0,V)$ and total weight at least $V$. Then the vertex set of $\mathcal T$ splits into parts, each connected in $\mathcal T$, of weight in $[V,(2+D)V)$. Applied to the cell graphs of the output with $D=6/c_g$, cell weights $|\Omega|_{\rm fine}$ and $V=V_{\Uc}=\tilde m^{-3}$, it gives the $\Uc$-cell part of (R7), with $C_{\Uc}$ of Section~\ref{S:B1.U}, provided $\tau$ is so large that $\tilde mL\ge c_L/2$.
\end{lemma}

\begin{proof}
We cut a rooted spanning tree from the leaves upward, cutting off a vertex with its uncut descendants once their weight reaches $V$. A cut part weighs less than $w(v)+DV<(1+D)V$, and the light remainder at the root joins an adjacent cut part. In the application, the cell graph is connected, since $\Gp$-bonds between cells are interface bonds, and Lemma~\ref{lem:4.17} bounds its degree by $6/c_g$. A cell weighs less than $8L^3\le\tilde m^{-3}$ by \eqref{eq:3.Hell}. By Theorem~\ref{thm:4.7}(ii), a light component has more than $V_*$ sites, and thus weight more than $r_0^3V_*=c_uV_{\Uc}$. Finally, a $\Uc$-cell holds fewer than $(2+6/c_g)\tilde m^{-3}/(c_1L^3)$ connected cells by (R5), in blocks of Euclidean diameter less than $2\sqrt3L$.
\qed
\end{proof}
}

\providecommand{\good}{\operatorname{good}}
\providecommand{\Kc}{\mathfrak{K}}
\providecommand{\Uf}{\mathbb{U}}
\providecommand{\Ereg}{\mathcal{E}}

\subsection{Payment and reach of the Phase-1 output}\label{ssec:4.9}

In this subsection we bound the cost of the Phase-1 output by means of the three comparisons of Lemma~\ref{lem:4.density}. By filling everything near a point we bound the cost of a ball; by deleting a region without marks we bound its cost by the bonds of a sphere around it; and with a good sphere, chosen by averaging over the radii, the cost decays geometrically away from the marks. Consequently, no hole lies far from a mark, as Theorem~\ref{thm:4.42} states, and in Proposition~\ref{prop:4.39} we pay for the pieces near the clusters of marks by comparison with the cheapest state that covers the marks and fills the forced blocks of the cluster.

Throughout, $(\XX^1,\Uf^1)$ is the Phase-1 output and $\Ereg:=\Ereg^{\bar S^1}$. By Theorem~\ref{thm:4.30}, $(\XX^1,\Uf^1)$ minimizes $\Ereg$, and each bond of $\bar S^1$ touches $\XX^1$ and has both ends within distance $4L_c$ of a mark. Thus, at $(\XX^1,\Uf^1)$, the frozen bonds count as ordinary bonds, and the penalty vanishes by Lemma~\ref{lem:4.forced}. For a set $A$ of sites we put $\Ereg_A:=\#\{b\in\partial\XX^1:\ b\subseteq A\}+\lambda|(\XX^1\setminus\Uf^1)\cap A|+\lambda_U|\Uf^1\cap A|$, let $B_r(x):=Q_\infty(x,r)\cap\Lamo$, and for a fixed site $x$ let $f(r):=\Ereg_{B_r(x)}$, nondecreasing in $r$. Each competitor $(\XX',\Uf')$ below has $\MM\subseteq\XX'$ and $\Uf'\subseteq\XX'$ a union of blocks, and so satisfies $\Ereg(\XX',\Uf')\ge\Ereg(\XX^1,\Uf^1)$.

\begin{lemma}\label{lem:4.density}\label{lem:4.deletion}\label{lem:4.decay}
(a) For every site $x$ and every $r\ge0$, $f(r)\le g(r):=6(2r+4L_c)^2+\lambda_U(2r+4L_c)^3$. (b) Let $B$ be a set of sites with $\dist(B,\MM)\ge6L_c$, and $\hat B$ the union of the blocks meeting $B$. Then $\Ereg_B\le\mathrm{sl}(B):=f(B,\XX^1\setminus B)+\lambda|\Uf^1\cap(\hat B\setminus B)|$. (c) Put $w:=\lceil120\ell/c_s\rceil$ and $w':=w+2L_c$. If $\dist(B_{r+w'}(x),\MM)\ge6L_c$, then $f(r)\le f(r+w')/5$.
\end{lemma}

\begin{proof}
(a) Let $\hat B$ be the union of the blocks meeting $B_r(x)$, a box of sides less than $2r+4L_c$. We compare with $(\XX^1\cup\hat B,\Uf^1\cup\hat B)$. In this competitor the bonds of $\partial\XX^1$ with both ends in $\hat B$ disappear, at most $|\partial\hat B|\le6(2r+4L_c)^2$ bonds appear, the frozen bonds still touch the hole set, each site of $\hat B$ ends with the weight $\lambda_U$, and the penalty does not grow. It follows that $f(r)\le\Ereg_{\hat B}\le|\partial\hat B|+\lambda_U|\hat B|$.

(b) A forced block has a mark within distance $4L_c$ of each of its sites. Consequently, no block meeting $B$ is forced, and no bond of $\bar S^1$ has an end in $B$. We compare with $(\XX^1\setminus B,\Uf')$, where $\Uf'$ is the union of the blocks of $\Uf^1$ that do not meet $B$. This competitor contains the marks and all forced blocks. In passing to it, the bonds of $\partial\XX^1$ with both ends in $B$ disappear, at most $f(B,\XX^1\setminus B)$ bonds appear, the frozen bonds still touch the hole set, the sites of $B\cap\XX^1$ lose their weights, and those of $\Uf^1\cap(\hat B\setminus B)$ change from $\lambda_U$ to $\lambda$.

(c) We put $A:=B_{r+w'}(x)\setminus B_r(x)$ and apply (b) to $B_s:=B_s(x)$ for $r\le s<r+w$. Since a site of $\XX^1$ at distance exactly $s+1$ from $x$ has at most six bonds into $B_s$, we have $\sum_sf(B_s,\XX^1\setminus B_s)\le6|\XX^1\cap A|$. A site of $\Uf^1$ outside $B_s$ lies in a block meeting $B_s$ for fewer than $2L_c$ values of $s$, and therefore $\sum_s\lambda|\Uf^1\cap(\hat B_s\setminus B_s)|\le2\lambda L_c|\Uf^1\cap A|=24|\Uf^1\cap A|$. Combining the two bounds, we obtain $\sum_s\mathrm{sl}(B_s)\le30|\XX^1\cap A|\le30(\lambda^{-1}+\lambda_U^{-1})\Ereg_A\le(30\ell/c_s)\Ereg_A$, using $\lambda^{-1}+\lambda_U^{-1}=L_c/12+\ell/(2c_s)\le\ell/c_s$. Thus some $s$ has $\mathrm{sl}(B_s)\le\Ereg_A/4$. The bonds and sites counted in $\Ereg_A$ are not counted in $f(r)$. Therefore $\Ereg_A\le f(r+w')-f(r)$, and $f(r)\le f(s)\le\mathrm{sl}(B_s)\le(f(r+w')-f(r))/4$.
\qed
\end{proof}

\begin{theorem}\label{thm:4.42}
Let $J$ be the least integer with $5^J>g((J+1)w')$, and $R_*:=(J+1)w'+6L_c$. Every site of $\XX^1$ lies within distance $R_*$ of a mark, and $R_*=O((\ell/c_s)\log(\ell/c_s))$, uniformly in $|\MM|$, $\Lamo$, $C$ and $D$.
\end{theorem}

\begin{proof}
Let $\dist(x,\MM)>R_*$. Then $\dist(B_{(J+1)w'}(x),\MM)>6L_c$, and applying Lemma~\ref{lem:4.density}(c) $J$ times from $r=w'$, together with Lemma~\ref{lem:4.density}(a), we obtain $f(w')\le5^{-J}f((J+1)w')\le5^{-J}g((J+1)w')<1$. So no bond of $\partial\XX^1$ has both ends in the box $B_{w'}(x)$, which therefore lies in $\XX^1$ or misses it. In the first case $f(w')\ge\lambda_U|B_{w'}(x)|>1$. Therefore $x\notin\XX^1$. With the arithmetic of Section~\ref{S:C1}, this proves Theorem~\ref{thm:3.21}(b).
\qed
\end{proof}

\begin{proposition}\label{prop:4.39}
Theorem~\ref{thm:3.21}(a) holds with $\Cst:=C_1+2C_{\rm seal}$, $C_1:=7+27\cdot24.6/\vartheta$, and $\Cfree:=\Cst(2+1/(2c_s)+1/3696)$.
\end{proposition}

\begin{proof}
By Theorem~\ref{thm:4.30}(i), $S^1=\emptyset$. We compare with the competitor that replaces $Y_P$ by the marks and the forced blocks of $P$. The sets $Y_P$ cover $\XX^1$ by Theorem~\ref{thm:4.42}. Marks of distinct parts are more than $2R_*+2$ apart, and hence no bond joins $P^{+R_*}$ to $P'^{+R_*}$. Moreover, a block of $\Uf^1$ is connected and lies in $\XX^1$. The marks of a set $3Q$ are within distance $6L_c\le r'$ of each other, and hence those of a forced block lie in one part. The union $U_F(P)$ of the forced blocks with marks in $P$ lies in $Y_P$, by Lemma~\ref{lem:4.forced} and because a forced block lies within $4L_c$ of its marks. It has at most $27N(P)/(\vartheta L_c^2)$ blocks, since a mark lies in at most $27$ sets $3Q$. The competitor $\XX':=(\XX^1\setminus Y_P)\cup P\cup U_F(P)$, $\Uf':=(\Uf^1\setminus Y_P)\cup U_F(P)$ contains the marks, $\Uf'$ is a union of blocks containing all forced blocks, and the change is confined to $Y_P$, which no bond joins to the rest of $\XX^1$. A bond of $\bar S^1$ that loses contact with the hole set touches $Y_P$, so the marks of its two blocks lie within $R_*+4L_c<r'$ of $P$ and therefore in $P$. There are at most $C_{\rm seal}N(P)$ such bonds by Proposition~\ref{prop:4.sealcount}, each costing at most $2$. Since $|\partial Q|+\lambda_U|Q|<24.6L_c^2$ and $\lambda=12/L_c\le1$ ($L_c\ge35$ by item~(e) of Section~\ref{ssec:D.5}), minimality implies $|\partial Y_P|+\lambda|Y_P\setminus\Uf^1|+\lambda_U|Y_P\cap\Uf^1|\le(6+\lambda)N(P)+24.6\cdot27N(P)/\vartheta+2C_{\rm seal}N(P)\le\Cst N(P)$.
The second inequality follows from $|Y_P|/\ell\le(\lambda\ell)^{-1}\lambda|Y_P\setminus\Uf^1|+(\lambda_U\ell)^{-1}\lambda_U|Y_P\cap\Uf^1|$, with $(\lambda\ell)^{-1}\le1/3696$ and $(\lambda_U\ell)^{-1}=1/(2c_s)$.
\qed
\end{proof}

\subsection{Phase 2: insulation, density bound and cuts}\label{ssec:4.4}\label{ssec:4.7}

In this subsection we first bound the insulation in Phase~2, in Lemma~\ref{lem:4.21}, and then prove the density bound (R9), Lemma~\ref{lem:3.18} and Proposition~\ref{prop:3.19}. Each hole site added in Phase~2 outside the removed cells is hull material, whose removal does not lower the hull functional, while the Phase-1 insulation $\mathrm{ins}^1:=\partial\XX^1$ is fixed in advance and bounded in Theorem~\ref{thm:3.21}(a). Throughout, $R_t$ is the union of the cells removed in Phase~2 before time $t$. A cell $\Delta$ removed in Phase~2 is removed at a hull-closed time $t_\Delta$, at which $w(\Delta)$ is evaluated. We denote by $s_\Delta$ the number of bonds of $E(\Delta)$ frozen in Phase~2 before $t_\Delta$, converted ones included, with no end in $\XX^1$.

\begin{lemma}\label{lem:4.21}
(a) Let $t$ be a hull-closed time of Phase~2, $N'$ a box, $W':=(\XX_t\setminus(\XX^1\cup R_t))\cap N'$, $\mathrm{Ins}_{N'}(Y):=\#\{b\in E(N'):\text{exactly one end in }Y\}+2\#\{b\in\bar S_t\cap E(N'):\text{no end in }Y\}$, and $\mathrm{cut}_{N'}$ the number of bonds across $\partial N'$ with inner end in $W'$ whose outer end lies in $\XX_t$ or which are frozen. Then
\begin{equation}\label{eq:4.I5}
\mathcal A_2(E(N');t)\le\mathrm{Ins}_{N'}(\XX^1\cup R_t)+\mathrm{cut}_{N'}.
\end{equation} (b) Every cell $\Delta$ removed in Phase~2 satisfies $w(\Delta)\le|\mathrm{ins}^1\cap E(\Delta)|+2s_\Delta+|\partial\Delta|$, hence $c_s|\partial\Delta|<|\mathrm{ins}^1\cap E(\Delta)|+2s_\Delta$.
\end{lemma}

\begin{proof}
(a) We put $\XX:=\XX_t$ and $V:=\Lamo\setminus\XX$. Since $B_t=\MM\cup\Uf^1\cup R_t\subseteq\XX^1\cup R_t\subseteq\XX$ by Proposition~\ref{prop:4.4}(a), $W'$ misses $B_t$ and $\mathrm{Ins}_{N'}(\XX\setminus W')=\mathrm{Ins}_{N'}(\XX^1\cup R_t)$, while $\mathrm{Ins}_{N'}(\XX)=\mathcal A_2(E(N');t)$, since by Lemma~\ref{lem:4.2} the bonds of $\bar S_t$ with no end in $\XX$ are those of $S_t$. This proves (a) if $W'=\emptyset$. Otherwise $F^{\bar S_t}_\lambda(\XX\setminus W')>F^{\bar S_t}_\lambda(\XX)$, because $\XX=H^{\bar S_t}_\lambda(B_t)$. Written bond by bond as in \eqref{eq:4.locrem}, this inequality reads $f_{E\setminus\bar S_t}(W',V)+\lambda|W'|<f(W',\XX\setminus W')+\sum_{b\in\bar S_t\text{ between }W',V}(w_b-1)+\sum_{b\in\bar S_t,\,b\subseteq W'}w_b$, with $f_{E\setminus\bar S_t}$ counting unfrozen bonds. When we pass from $\mathrm{Ins}_{N'}(\XX\setminus W')$ to $\mathrm{Ins}_{N'}(\XX)$, the contribution of a bond of $E(N')$ between $W'$ and $\XX\setminus W'$ changes by $-1$, that of a bond of $\bar S_t$ inside $W'$ by $-2$, that of a bond between $W'$ and $V$ by $+1$ if unfrozen and by $-1$ if frozen, and that of every other bond by $0$. Inserting the inequality, we see that the bonds between $W'$ and $\XX\setminus W'$ inside $N'$ cancel, frozen bonds inside $N'$ with an end in $W'$ get the coefficient $w_b-2\le0$, and the bonds counted in $\mathrm{cut}_{N'}$ remain, each with coefficient at most $1$. We conclude that $\mathcal A_2(E(N');t)<\mathrm{Ins}_{N'}(\XX^1\cup R_t)+\mathrm{cut}_{N'}-\lambda|W'|$.

(b) We take $N':=\Delta$ and $t:=t_\Delta$. Since $\Delta\cap R_t=\emptyset$ and, by Theorem~\ref{thm:4.30}, each bond of $\bar S^1$ touches $\XX^1$, $\mathrm{Ins}_\Delta(\XX^1\cup R_t)=|\partial\XX^1\cap E(\Delta)|+2\#\{b\in\bar S_t\cap E(\Delta):\text{no end in }\XX^1\}\le|\mathrm{ins}^1\cap E(\Delta)|+2s_\Delta$, and $\mathrm{cut}_\Delta\le|\partial\Delta|$. The last claim follows from the trigger.
\qed
\end{proof}

\begin{proposition}[Density bound]\label{prop:4.22}
At every state satisfying (S-rem), in particular at the output, (R9) holds, and $(3+c_s)|\partial\Delta|<54(3+c_s)\ell^2$.
\end{proposition}

\begin{proof}
Since a frozen bond is a missing pair of neighbours counted at both ends, the density count of Section~\ref{sec:stage1} is two-sided, namely the hull's own count $\mathcal A_2$. If $\Delta\not\subseteq B$, then $\mathcal A_2(E(\Delta))=w(\Delta)\le w_*$ by (S-rem). If $\Delta\subseteq B\subseteq\XX$, this count is $0$. A bond of $\partial\Delta\cap E$ weighs at most $2$. This yields the bound in which these bonds are counted as well, and the last assertion holds as $|\partial\Delta|<54\ell^2$ by (G3). For the constant $c_0$ of the conversion into fine plaquettes we refer to Section~\ref{S:B1.c0}.
\qed
\end{proof}

We recall that Phase~2 fills only by removal and freezes by the seal and the cut rule, and that the cut rule fires only at hull-closed states without auxiliary material, where Theorem~\ref{thm:4.rest} holds.

\begin{lemma}\label{lem:3.18}
Let the firings $j=1,2,\dots$ of the cut rule have times $t_j$, components $\Kc_j$ and sides $U_j$, $\Kc_j\setminus U_j$. (a) Every bond of $\partial_{\Gp}U$ joins two different cells, so it crosses a block face; $\partial_{\Gp}U$ is a union of whole interfaces. (b) After the cut $j$ no $\Gp$-bond joins $U_j$ to $\Kc_j\setminus U_j$; every later component of $\Gp$ lies in $U_j$, in $\Kc_j\setminus U_j$, or outside $\Kc_j$. The sides of the cuts form a laminar family. (c) $|\partial_{\Gp}U_j|\ge c_gL_c^2$ for every cut. (d) Cut bonds support no hull material, alone or with seal bonds, filled blocks and box faces, and cut bonds block no tube of another interface. (e) If $J(U)>0$, then $\partial_{\Gp}U$ contains two interfaces at distance $>R_{\rm NF}$.
\end{lemma}

\begin{proof}
(a) A $\Gp$-bond with both ends in a block $Q$ lies in $\Gp\cap\hat Q(Q)$, and hence its ends lie in one cell. Therefore a bond of $\partial_{\Gp}U$ joins cells of different blocks, and all bonds across $F'$ between a cell of $U$ and a cell outside $U$ lie in $\partial_{\Gp}U$. (b) The cut consists of all $\Gp$-bonds between $U_j$ and $\Kc_j\setminus U_j$, where $\Kc_j$ is a component of $\Gp$ at time $t_j$, and $\Gp$ only shrinks by (H4) of Proposition~\ref{prop:4.4}. Consequently, each later component lies in $U_j$, in $\Kc_j\setminus U_j$, or outside $\Kc_j$. If $j_a$ fires before $j_b$, the sides of $j_b$ lie in $\Kc_{j_b}$, a component at time $t_{j_b}$, so each of them lies in a side of $j_a$ or misses both. As the two sides of one cut are disjoint, any two sides are disjoint or nested. (c) By (a) the cut contains a whole interface $I$, and (S-face) holds at $t_j$ by Theorem~\ref{thm:4.rest}(b). Thus $|\partial_{G'}U_j|\ge\good(I)\ge c_gL_c^2$, where the second inequality comes from Lemma~\ref{lem:4.17}.
(d) Cut bonds are grid bonds with the hull weight, and they therefore enter Corollary~\ref{cor:4.10} and Lemma~\ref{lem:4.11} as grid-type terms, as seal bonds do. A tube crosses no other block plane, by (G5), and this proves the last assertion of (d).
(e) If $J(U)>0$, then $\varphi_s(\partial_{G'}U)\ge2$ for some $s=2^jL\ge R_{\mathrm{NF}}$, and interfaces in different $s$-clusters are more than $s$ apart.
\qed
\end{proof}

\begin{proof}[Proof of Proposition~\ref{prop:3.19}]
Termination is Proposition~\ref{prop:4.4}(b). At the terminal state a full pass changes nothing. Thus the state is hull-closed, which by Lemma~\ref{lem:4.2} implies (S-hull), and no removal cell is over threshold, which is (S-rem). For the same reason, no surviving block has auxiliary material, by Lemma~\ref{lem:4.aux}(a), so (S-face) and (S-plug) hold by Theorem~\ref{thm:4.rest}. Finally, the cut rule has no candidate at the terminal state, which is (S-NF) and contains the form used in Section~\ref{ssec:7.8}.

We now assume that the Euclidean diameter of $\Lamo$ is at most $R_{\mathrm{NF}}$. Then no pair is fragmented, since the right side of \eqref{eq:3.NFstar} is positive and, by Lemma~\ref{lem:3.18}(e), $J(U)>0$ would give two interfaces more than $R_{\mathrm{NF}}$ apart. Since, moreover, by Theorem~\ref{thm:4.30} the Phase-1 output is hull-closed, satisfies (S-rem) and has no auxiliary material, Phase~2 changes nothing in this case.
\qed
\end{proof}
}
\providecommand{\Gp}{G'}
\providecommand{\XX}{\mathbb{X}}
\providecommand{\KK}{\mathfrak{K}}
\providecommand{\mt}{\tilde m}
\providecommand{\bGM}{\beta_{\mathrm{GM}}}
\providecommand{\eps}{\varepsilon}
\providecommand{\Lc}{\Lambda^{\circ}}
\providecommand{\AGp}{A_{\Gp(D)}}
\providecommand{\PD}{\mathbb{P}_{D}}
\providecommand{\ED}{\mathbb{E}_{D}}
\providecommand{\etaT}{\eta_{T}}
\providecommand{\rbar}{\bar r_{0}}
\providecommand{\Cfree}{C_{\mathrm{free}}}
\providecommand{\Cresp}{C_{\mathrm{resp}}}
\providecommand{\Rresp}{R_{\mathrm{resp}}}
\providecommand{\RNF}{R_{\mathrm{NF}}}
\providecommand{\thetaNF}{\theta_{\mathrm{NF}}}
\providecommand{\Funit}{F_{\mathrm{unit}}}
\providecommand{\kpl}{\kappa_{\mathrm{pl}}}
\providecommand{\epsc}{\varepsilon_{c}}
\providecommand{\epsD}{\varepsilon_{D}}
\providecommand{\Ffr}{\mathfrak{F}}
\providecommand{\Mm}{\mathbb{M}}
\providecommand{\Sep}{\mathfrak{S}}
\providecommand{\Bfr}{\mathfrak{B}}
\providecommand{\Bpay}{B_{\mathrm{pay}}}
\providecommand{\Scut}{S_{\mathrm{cut}}}
\providecommand{\ncut}{n_{\mathrm{cut}}}
\providecommand{\Qatt}{\mathcal{Q}_{\mathrm{att}}}
\providecommand{\Uf}{\mathbb{U}}
\providecommand{\supp}{\operatorname{supp}}
\providecommand{\diam}{\operatorname{diam}}
\providecommand{\dist}{\operatorname{dist}}
\providecommand{\ZZ}{\mathbb{Z}}
\providecommand{\RR}{\mathbb{R}}

\section{The min-cut lower tail}
\label{sec:5}\label{sec:mincut}

In this section we show that the frozen set of the construction of Section~\ref{sec:3} lowers the min-cut area of Definition~\ref{def:2.11} by at most $\etaT RT$, except with probability at most $2^{4|C|+5}e^{-RT}$. The proof has two ingredients. Section~\ref{ssec:5.2} shows that, as $\Lc$ is acyclic, a minimizer can be taken constant on each piece of the hole set. In Sections~\ref{ssec:5.3} to~\ref{ssec:5.7} we then prove that the frozen plaquettes which a part of the marks attaches are paid for by the marks of that part.

The units, $\Xi$, $\Lc_2$, $\partial_2\Lc$, $F(Y)$, $F_S$, $\Ffr$, $\Gp(D)$, $\XX$ and $S=S_1\sqcup\Scut$ are those of Sections~\ref{ssec:2.5}, \ref{sec:3} and~\ref{ssec:3.6}. A coarse face carries at most $\rbar^{2}$ fine plaquettes, $\rbar=2r_0-1$, and $\Xi$ is a union of coarse faces, since the plane of $C$ is a special plane of $\mathfrak G(C)$. We write $\PD:=\mathrm{Ber}(\delta)^{\otimes(\Lc_2\setminus\partial_2\Lc)}$ for the law of $D$ and $\ED$ for its expectation.

\subsection{Statement}
\label{ssec:5.1}

\begin{theorem}[Min-cut lower tail]\label{thm:5.1}
Fix $K>K_*$ and $\beta\ge\beta_0(K)$. For every rectangle $C$ with $\min(R,T)\ge R_1(\beta)$ and every $\sigma\in(0,1]$,
\begin{gather}\label{eq:5.1}
\PD\bigl(\AGp(C)\le(1-\etaT)RT\bigr)\le 2^{4|C|+5}e^{-RT},\\
\ED\exp\bigl(\sigma(RT-\AGp(C))\bigr)\le\exp\bigl(\sigma\etaT RT+4|C|+4\bigr),\notag
\end{gather}
where
\begin{equation}\label{eq:5.2}
\etaT:=\frac{3c_T''\rbar^{2}C_3}{|\log\eps''|},
\end{equation}
with $\eps''$ of \eqref{eq:5.epspp} and the constants $c_T''$ and $C_3$ of Section~\ref{S:const5}. Moreover, there is $\tau_0(K)$ such that for $\tau\ge\tau_0(K)$ we have $\etaT\le6c_T''\rbar^{2}C_3/(\gamma_T(K)\tau)$, with $\gamma_T(K)$ of Lemma~\ref{lem:5.16}, and $\etaT=O\bigl(\rbar^{2}(C_\star'+\Cresp)(1+1/c_s)/(\tau\log^2K)\bigr)$ with an absolute implied constant, uniformly in $K>K_*$; so $\etaT\to0$ as $K\to\infty$, and as $\beta\to\infty$ at fixed $K>K_*$.
\end{theorem}

The conditions that define $\beta_0(K)$ in item~(a) of Section~\ref{ssec:D.5} are lower bounds on $\tau$, while the constants $c_T''$, $C_3$, $\Cfree$, $C_\star'$, $\Cresp$ are free of $\beta$, $\tau$, $K$, $C$, $D$. The condition $\tau\ge\tau_0(K)$ is used only for the order of $\etaT$, not for \eqref{eq:5.1}.

\subsection{The deterministic side}
\label{ssec:5.2}

In this subsection we let $(\XX,S)$ be any state with $S\subseteq E(\Lc\setminus\XX)$ and compute $\AGp(C)$ with $\Ffr=F(\XX)\cup F_S\cup\partial_2\Lc$. Then $F_S\cap F(\XX)=\emptyset$.

\emph{Box topology.} Let $B$ be a box of $\Zthree$ with its cubical complex. (i) Over $\ZZ$ and over $\ZZ_2$ this complex is the tensor product of the complexes of three intervals, each chain-homotopy equivalent to the coefficients in degree $0$; hence every $k$-cycle with $k\ge1$ is a boundary. (ii) $\partial$ is injective on $3$-chains: in a nonzero $3$-chain, a cube that is extreme in the lexicographic order has a face shared with no other cube of the chain, and the boundary has a nonzero coefficient there. (iii) $H_1(\partial B;\ZZ)=0$: if $z$ is an integer $1$-cycle on the boundary bonds, then $z=\partial c$ for a $2$-chain $c$ of $B$ by (i); the part of $c$ on the interior plaquettes is a closed $1$-cochain on the dual box, hence the coboundary of a $0$-cochain $f$ on the cubes, and $c-\partial f$ is a $2$-chain of boundary plaquettes with boundary $z$. (iv) By (i) over $\ZZ_2$, the boundaries of the unit squares generate the $\ZZ_2$ cycle space of the box graph.

\begin{lemma}\label{lem:5.3}
Every integer $2$-chain $\Sigma$ in $\Lc$ with $\partial\Sigma=C$ satisfies $\#(\supp\Sigma\setminus\partial_2\Lc)\ge RT$. Moreover $\AGp(C)\le RT$.
\end{lemma}

\begin{proof}
There are $RT$ disjoint closed dual paths that link $C$, namely nested closed paths of fine cubes in the planes perpendicular to $\Xi$ through its plaquette rows. They lie inside $\Lc$ off $\partial_2\Lc$ by Lemma~\ref{lem:3.3}(e), and each of them crosses $\Xi$ once (Section~\ref{S:proofs5}). As $\Lc$ is acyclic (Box topology (i)), $\Sigma-\Xi$ is a boundary. Each path therefore meets $\supp\Sigma$, with intersection number $\pm1$, and different paths meet it in different plaquettes. The second claim holds because $\AGp(C)\le|\supp\Xi|=RT$. \qed
\end{proof}

A \emph{piece partition} is a partition $\mathcal Y$ of the sites of $\XX$ into pieces, which need not be connected. For a site set $Y$, let $\partial Y$ be the set of coarse bonds with exactly one end in $Y$, $\operatorname{int}F(Y)$ the set of fine plaquettes whose two fine cubes lie in sites of $Y$, and $\partial^fY$ the set of those with one fine cube in a site of $Y$ and the other in a site of $\Lc\setminus Y$. Then
\begin{equation}\label{eq:5.3}
F(Y)=\operatorname{int}F(Y)\sqcup\partial^fY\sqcup\bigl(F(Y)\cap\partial_2\Lc\bigr),\qquad|\partial^fY|\le\rbar^{2}|\partial Y|.
\end{equation}
Indeed, a plaquette of $F(Y)$ lies in $\partial_2\Lc$ iff it is a face of only one fine cube of $\Lc$, and otherwise its second cube lies in a site of $Y$ or of $\Lc\setminus Y$; a plaquette of $\partial^fY$ lies in the common coarse face of two adjacent sites $y\in Y$, $y'\notin Y$, which is dual to the bond $\{y,y'\}\in\partial Y$ and carries at most $\rbar^{2}$ fine plaquettes.

\begin{lemma}\label{lem:5.4}
There is a minimizer $\tilde\Sigma$ of Definition~\ref{def:2.11} ($\partial\tilde\Sigma=C$, $\#(\supp\tilde\Sigma\setminus\Ffr)=\AGp(C)$), determined by $(\XX,S)$, such that for every $Y$ of every piece partition
\begin{equation}\label{eq:5.4}
\supp\tilde\Sigma\cap F(Y)\subseteq\partial^fY\cup\bigl(\supp\Xi\cap\operatorname{int}F(Y)\bigr)\cup\partial_2\Lc .
\end{equation}
\end{lemma}

\begin{proof}
By acyclicity a minimizer is the flat rectangle plus a boundary, and we may change the boundary freely on hole cubes, where all plaquettes are free. Let $\Sigma$ be the first minimizer in a fixed order and $\Sigma-\Xi=\partial k$, $k$ an integer function on the fine cubes of $\Lc$ (as in Lemma~\ref{lem:5.3}; $k$ is unique by Box topology (ii)). We define $\tilde k:=k$ on the fine cubes of sites of $\Lc\setminus\XX$, $\tilde k:=0$ on those of sites of $\XX$, and $\tilde\Sigma:=\Xi+\partial\tilde k$. Then $\partial\tilde\Sigma=C$, and $\tilde\Sigma$, $\Sigma$ differ only on faces of fine cubes of sites of $\XX$, which lie in $F(\XX)\subseteq\Ffr$. It follows that $\#(\supp\tilde\Sigma\setminus\Ffr)=\AGp(C)$. For $p\in\operatorname{int}F(Y)$ both cubes of $p$ carry $\tilde k=0$, and therefore $\tilde\Sigma_p=\Xi_p$. Together with \eqref{eq:5.3}, this is \eqref{eq:5.4}, and $\tilde\Sigma$ does not involve $\mathcal Y$. \qed
\end{proof}

\begin{lemma}[$C$-rooted minimizer]\label{lem:5.5}
There is a minimizer $\Sigma^\circ$, determined by $(\XX,S)$ alone, with the following properties, in which $\supp\Sigma^\circ$ denotes its support in $\Lc_2\setminus\partial_2\Lc$ and $a_\Sigma:=|\supp\Sigma^\circ|$. We have $\supp\Sigma^\circ\subseteq\supp\tilde\Sigma$, and every component of $\supp\Sigma^\circ$ for edge adjacency contains a plaquette incident to an edge of $C$. Moreover, \eqref{eq:5.4} holds for $\Sigma^\circ$ for every piece partition, and $a_\Sigma\ge RT$.
\end{lemma}

\begin{proof}
In a fixed order, we replace each component of $\supp\tilde\Sigma\setminus\partial_2\Lc$ with no plaquette at an edge of $C$ by a free $2$-chain on $\partial_2\Lc$ with the same boundary, which exists because that boundary lies on $\partial\Lc$ (Box topology (iii)) (the plaquettes of $\supp\tilde\Sigma$ at an edge off $C\cup\partial\Lc$ are pairwise adjacent). Then \eqref{eq:5.4} passes to subsets, and $a_\Sigma\ge RT$ follows from Lemma~\ref{lem:5.3}. Since $(\XX,S)$ is a function of $D$ by Theorem~\ref{thm:3.20}(R1), so is $\Sigma^\circ$. \qed
\end{proof}

\begin{proposition}\label{prop:5.7}\label{cor:5.8}\label{lem:5.6}
For a piece partition $\mathcal Y$, call $Y\in\mathcal Y$ \emph{touched} if $\supp\Sigma^\circ\cap F(Y)\ne\emptyset$, and $b\in S$ \emph{touched} if $\supp\Sigma^\circ$ meets the coarse face dual to $b$. Then
\begin{equation}\label{eq:5.5}
a_\Sigma-\AGp(C)=\#(\supp\Sigma^\circ\cap\Ffr)\le\rbar^{2}\Bigl[\sum_{Y\ \mathrm{touched}}\bigl(2|\partial Y|+|Y|/\ell\bigr)+\#\{b\in S\ \mathrm{touched}\}\Bigr].
\end{equation}
In particular, if $\eta\in(0,1]$ and $\AGp(C)\le(1-\eta)RT$, then the right side of \eqref{eq:5.5} is at least $a_\Sigma-(1-\eta)RT\ge\eta RT$.
\end{proposition}

\begin{proof}
The identity holds as $\partial_2\Lc\subseteq\Ffr$ and $\Sigma^\circ$ is a minimizer, and $a_\Sigma\ge RT$ by Lemma~\ref{lem:5.5}. Off $\partial_2\Lc$ the free set is $F(\XX)\sqcup F_S$, and we bound the contributions of the two sets in turn. Each touched bond of $S$ contributes at most $\rbar^{2}$ plaquettes. A face of $\Xi$ carrying a plaquette of $\operatorname{int}F(Y)$ has both its sites in $Y$. Its coarse column perpendicular to the plane of $C$, which is disjoint from those of the other faces, therefore either lies in $Y$ or contains a bond of $\partial Y$, and in the first case it has $n_3\ge\ell$ sites, in view of Lemma~\ref{lem:3.3}(c). Consequently, $|\supp\Xi\cap\operatorname{int}F(Y)|\le\rbar^{2}(|\partial Y|+|Y|/\ell)$, and \eqref{eq:5.4} and \eqref{eq:5.3} show that a touched $Y$ contributes at most $\rbar^{2}(2|\partial Y|+|Y|/\ell)$. \qed
\end{proof}

\subsection{The Phase-1 Peierls estimate}
\label{ssec:5.3}

In this subsection we work with the Phase-1 output $(\XX^1,\Uf^1)$, which has no frozen bond, take $\cP$, $r'=2R_*+2$ and the pieces $Y_P$, $P\in\cP$, from Theorem~\ref{thm:3.21}, and compute $\Ffr^1$, $A^1:=\AGp(C)$ and $\Sigma^\circ$ for $(\XX^1,\emptyset)$. For a plaquette set $\Xi'$, let $Z(\Xi')$ be the set of sites with a fine cube having a face in $\Xi'$. Then $|Z(\Xi')|\le2|\Xi'|$, as a plaquette is a face of at most two fine cubes. We also write $Z(\Sigma^\circ):=Z(\supp\Sigma^\circ)$. A part $P$ is \emph{touched} if $Y_P$ is.

\begin{lemma}\label{lem:5.10}\label{lem:5.9}\label{eq:5.6}
$\#(\supp\Sigma^\circ\cap\Ffr^1)\le\rbar^{2}\sum_{P\ \mathrm{touched}}(2|\partial Y_P|+|Y_P|/\ell)\le\rbar^{2}\Cfree\sum_{P\ \mathrm{touched}}N(P)$, and every touched part satisfies $\dist(P,Z(\Sigma^\circ))\le R_*$.
\end{lemma}

\begin{proof}
The two inequalities follow from Proposition~\ref{prop:5.7}, applied to $\mathcal Y^1:=\{Y_P\}$, and from Theorem~\ref{thm:3.21}(a). For the last assertion, note that a touched plaquette is a face of a fine cube of a site of $Y_P$. Thus $Z(\Sigma^\circ)$ meets $Y_P\subseteq P^{+R_*}$ (Theorem~\ref{thm:3.21}(b)). \qed
\end{proof}

\begin{lemma}[Marks]\label{lem:5.11}
Put $q_m:=3(\rbar+1)^3\delta^{1/2}\le24r_0^3\delta^{1/2}$. For every finite set $W$ of sites, $\PD(W\subseteq\Mm)\le q_m^{|W|}$.
\end{lemma}

\begin{proof}
A plaquette marks at most two sites, and a site has at most $3(\rbar+1)\rbar^2\le3(\rbar+1)^3$ plaquettes on its fine cubes. If $W\subseteq\Mm$, a choice for each $x\in W$ of such a plaquette in $D$ takes at least $|W|/2$ distinct values, and there are at most $(3(\rbar+1)^3)^{|W|}$ choices. \qed
\end{proof}

\begin{lemma}[Animals]\label{lem:5.12}
In a graph of maximal degree $D_{\max}$, the number of connected vertex sets of size $n$ containing a given vertex is at most $(eD_{\max})^{n-1}$. The number of plaquette sets $\Xi'\subseteq\Lc_2$ with $|\Xi'|=a$ each of whose components (for edge adjacency) contains a plaquette incident to an edge of $C$ is at most $2^{4|C|}(2\kpl)^a$, with $\kpl:=12e$.
\end{lemma}

\begin{proof}
Section~\ref{S:proofs5} gives both counts in more detail. By \cite[Lemma~2.1 and the sentence after it, p.~141]{GK04} the first number is at most $\binom{Dn}{n-1}/n\le(eD)^{n-1}$, where $D:=D_{\max}$ denotes the maximal degree of the graph. For the second, we choose the first plaquette at an edge of $C$ of each component, the sizes and the components. There are $2^{4|C|}$ choices for the first plaquettes and $2^a$ for the sizes, and the components are counted by the first count with $D_{\max}=12$. \qed
\end{proof}

We put $\epsc:=8e\,(2R_*+1)^3(2r'+1)^3q_m$.

\begin{proposition}\label{prop:5.13}\label{eq:5.7}\label{eq:5.8}
Assume $\epsc\le\frac12$. For every finite set $Z$ of sites and every $u_0>0$, $\PD\bigl(\sum_{P\in\cP:\ \dist(P,Z)\le R_*}N(P)\ge u_0\bigr)\le2^{|Z|+1}\epsc^{u_0}$.
\end{proposition}

\begin{proof}
Since the partition is random, we take the union bound over all candidate families of mark sets, not over the partition, and we use only that parts are pairwise disjoint subsets of $\Mm$, $r'$-connected by definition (the graph on $P$ joining sites at distance at most $r'$ is connected, see Theorem~\ref{thm:3.21}). Thus the event lies in the union, over $u\ge u_0$ and all families $(P_1,\dots,P_m)$ of pairwise disjoint $r'$-connected site sets with $\dist(P_i,Z)\le R_*$ and $\sum_i|P_i|=u$, of the deterministic events $\{P_1\cup\dots\cup P_m\subseteq\Mm\}$, each of which has probability at most $q_m^u$ by Lemma~\ref{lem:5.11}. A family is fixed by the multiset of roots $z_i\in Z$ within $R_*$ of $P_i$, the sizes, marks $y_i\in P_i$ within $R_*$ of $z_i$, and the sets $P_i\ni y_i$. These sets are connected in a graph of degree less than $(2r'+1)^3$, and Lemma~\ref{lem:5.12} bounds their number. There are $2^{|Z|+m}$ values for the multiset of roots, $2^u$ for the sizes and $(2R_*+1)^3$ for each mark. Summing over $m\le u$, we bound the probability by $2^{|Z|}\sum_{u\ge u_0}\epsc^{\,u}\le2^{|Z|+1}\epsc^{u_0}$. \qed
\end{proof}

\begin{lemma}\label{lem:5.14}
Let $k\ge1$ and $j\in\{0,1,2,3\}$ be integers, put $c_T(k):=1+\log(2^{k+1}\kpl)$ and $c_T:=c_T(2)=2+\log96$, and let $\eta\in(0,1]$, $\eps\in(0,\frac12]$ and $F_*\ge1$. Let $E$ be an event with $E\subseteq\bigcup_{\Xi'}E_{\Xi'}$, the union over the plaquette sets $\Xi'$ of Lemma~\ref{lem:5.12} with $|\Xi'|\ge RT$, where $\PD(E_{\Xi'})\le2^{k|\Xi'|+j}\eps^{(|\Xi'|-(1-\eta)RT)/F_*}$. If $\eta|\log\eps|/F_*\ge c_T(k)$, then $\PD(E)\le2^{4|C|+j+2}e^{-RT}$, and every $X$ with $0\le X\le RT$ and $X<\eta RT$ off $E$ satisfies $\ED e^{\sigma X}\le e^{\sigma\eta RT+4|C|+4}$ for every $\sigma\in(0,1]$.
\end{lemma}

\begin{proof}
With $\lambda:=\log(2^{k+1}\kpl)$ and $\omega:=|\log\eps|/F_*$, Lemma~\ref{lem:5.12} bounds $\PD(E)$ by $2^{4|C|+j}$ times a geometric series in $a\ge RT$ of ratio $e^{-(\omega-\lambda)}\le e^{-1}$, and the hypothesis reads $\eta\omega\ge1+\lambda$. The moment bound then follows from $0\le X\le RT$, as we show in Section~\ref{S:proofs5}. \qed
\end{proof}

\begin{proposition}[Phase-1 output]\label{prop:5.15}
Assume $\epsc\le\frac12$ and $\etaT^{(1)}:=c_T\rbar^{2}\Cfree/|\log\epsc|\le1$. Then for every rectangle $C$ with $\min(R,T)\ge R_1(\beta)$ and every $\sigma\in(0,1]$, the two bounds of \eqref{eq:5.1} hold with $A^1$, $\etaT^{(1)}$ and $2^{4|C|+3}$ in place of $\AGp(C)$, $\etaT$ and $2^{4|C|+5}$.
\end{proposition}

\begin{proof}
We argue as in the proof of Theorem~\ref{thm:5.1}, without block families. On $E:=\{A^1\le(1-\etaT^{(1)})RT\}$, Proposition~\ref{prop:5.7} with $\mathcal Y^1$ and Lemma~\ref{lem:5.10} give $\sum N(P)\ge u_0(|\Xi'|):=(|\Xi'|-(1-\etaT^{(1)})RT)/(\rbar^{2}\Cfree)$ over the parts within $R_*$ of $Z(\Xi')$, where $\Xi':=\supp\Sigma^\circ$ is a function of $D$ by Lemma~\ref{lem:5.5}. For each deterministic $\Xi'$, Proposition~\ref{prop:5.13} bounds the probability of this event by $2^{2|\Xi'|+1}\epsc^{u_0(|\Xi'|)}$. Lemma~\ref{lem:5.14}, applied with $k=2$, $j=1$, $\eps=\epsc$, $F_*=\rbar^{2}\Cfree$ and $X:=RT-A^1\in[0,RT]$ (by Lemma~\ref{lem:5.3}), yields both bounds. \qed
\end{proof}

\begin{lemma}[Threshold $K_*$]\label{lem:5.16}
For $K>K_*=377.38$ there is a $\beta$-independent $\gamma_T(K)>0$ with
\begin{equation}\label{eq:5.9}
|\log\epsc|\ge\tau\Bigl(\frac{c_\delta(K)}{2K}-14.9662\Bigr)-O(\log\tau+\log r_0)\ge\gamma_T(K)\,\tau-O(\log\tau+\log r_0),
\end{equation}
and \eqref{eq:5.9} holds with $\eps''$ of \eqref{eq:5.epspp} in place of $\epsc$.
\end{lemma}

\begin{proof}
One mark costs $\delta^{1/2}=e^{-\tau c_\delta(K)/(2K)}$, and the root and part entropies of Proposition~\ref{prop:5.13} cost $e^{6\log R_*+O(\log r_0)}\le e^{14.9662\,\tau+O(\log\tau+\log r_0)}$, since $6\pi^2\vCb(0)<14.9662$. For the second inequality we use the lower bound on $c_\delta(K)$ of Lemma~\ref{lem:2.6} rather than its asymptotic form, and we take $\gamma_T(K):=(c_\delta^{\mathrm{sp}}(K)-1/(4a_*))/(2K)-14.9662$, which vanishes at $K=377.3635$ by the computation in Section~\ref{S:const5}. \qed
\end{proof}

\subsection{Phase 2: separators and leaves}
\label{ssec:5.4}

In Sections~\ref{ssec:5.4} to~\ref{ssec:5.6} we bound the Phase-2 cut bonds by the insulation around them. Since a bond can lie next to many cuts, the main tool is the Leaf Lemma below: $n$ cuts at one place and scale have $n+1$ pairwise disjoint sides, each of which crosses an annulus around that place and carries insulation of its own there.

\emph{Setting.} The cuts $j$, their times $t_j$, components $\KK_j$ and sides $U_j$ are those of Section~\ref{ssec:4.4}. We put $U_j':=\KK_j\setminus U_j$, $\mathrm{cut}_j:=\partial_{\Gp_{t_j}}U_j$ and $A_j:=|\mathrm{cut}_j|$. At a cut time the state is hull-closed without auxiliary material, so (S-face) and (S-plug) hold (Theorem~\ref{thm:4.rest}) and each cell at time $t_j$ lies in one block and has at least $c_1L_c^3$ sites (Theorem~\ref{thm:4.18}). A \emph{c-path} is a sequence of sites, consecutive ones joined by a $\Gp$-bond or lying in a common cell. A step of a c-path moves less than $2L_c$ in each coordinate. The \emph{box graph} is $\Lc$ with all coarse nearest-neighbour bonds (holes and frozen bonds ignored), with box-components and box paths, and $\partial(Y,W)$ denotes the set of box bonds from $Y$ to $W$. Bonds are \emph{square-adjacent} if they are edges of a common unit square. The distance of two bonds is the least $\ell^\infty$ distance of their ends, and a family of bond sets is \emph{$k_P$-linked} if the graph joining members with bonds at distance at most $k_P$ is connected.

\begin{definition}[Separators]\label{def:5.17}
Fix a cut $j$ and write $U:=U_j$, $U':=U_j'$. Let $U^\alpha$ be the box-components of $U$, and $\mathcal W(\alpha)$ the set of box-components $W$ of $\Lc\setminus U^\alpha$ with $W\cap U'\ne\emptyset$. Put $\Sep^*_j:=\bigcup_\alpha\bigcup_{W\in\mathcal W(\alpha)}\partial(U^\alpha,W)$, and call each $\partial(U^\alpha,W)$ a \emph{piece} of $\Sep^*_j$. Define $\Sep^{**}_j$ and its pieces in the same way with $U$ and $U'$ exchanged, and put $\Sep_j:=\Sep^*_j\cup\Sep^{**}_j$.
\end{definition}

\begin{lemma}[Separators]\label{lem:5.18}
For every cut $j$: (a) every bond of $\Sep^*_j$ joins a site of $U$ to a site outside $U$ and is a cut bond of $j$ or insulating at time $t_j$ (a bond of $\partial\XX_{t_j}$ or of $S_{t_j}$); the same holds for $\Sep^{**}_j$ with $U'$; (b) every piece is square-connected; (c) every piece contains a cut bond of $j$, distinct pieces of $\Sep^*_j$ are disjoint, and in particular $\Sep^*_j$ has at most $A_j$ pieces; the same holds for $\Sep^{**}_j$; (d) every cut bond of $j$ lies in a piece of $\Sep^*_j$ and in a piece of $\Sep^{**}_j$; (e) every box path from $U$ to $U'$ contains a bond of $\Sep^*_j$ and a bond of $\Sep^{**}_j$; (f) the pieces of $\Sep_j$ are $k_P$-linked.
\end{lemma}

\begin{proof}
Items (a) and (c) to (e) are elementary and are proved in Section~\ref{S:proofs5}. The separators are defined in the box graph, and the argument stays in the box graph except for (a) and the paths of (c) and (f).

(b) The set $\Pi:=\partial(U^\alpha,W)$ is a minimal set of box bonds separating a site $y\in U^\alpha$ from a site $x\in W$. It separates, because each box neighbour of $W$ outside $W$ lies in $U^\alpha$. It is minimal, because for $e=(u,v)\in\Pi$, $u\in U^\alpha$, $v\in W$, a box path inside $U^\alpha$ from $y$ to $u$, then $e$, then a box path inside $W$ from $v$ to $x$, meets $\Pi$ only in $e$. The unit squares generate the cycle space of the box graph over $\ZZ_2$ (Box topology (iv), Section~\ref{ssec:5.2}). We use Timár's lemma only in the box graph, since in the slit graph the unit squares need not generate the cycle space. By \cite[Lemma~1]{Ti}, for a minimal separating edge set $\Pi$ and a generating set $\mathcal C$ of the cycle space, each partition $(\Pi_1,\Pi_2)$ of $\Pi$ admits a cycle of $\mathcal C$ meeting both $\Pi_1$ and $\Pi_2$. Taking for $\mathcal C$ the unit squares, we obtain the square-connectedness of $\Pi$.

(f) A cell is $\Omega=Q\cap\hat\KK$ with $\hat\KK$ a component of $\Gp_{t_j}\cap\hat Q(Q)$. In particular, two sites of one cell are joined by a $\Gp$-path inside $\hat Q(Q)$, which by (G4) is a box of sides less than $k_P$, and this path lies in $\KK_j$ if the cell does. Bonds with all ends in $\hat Q(Q)$ are at distance at most $k_P-1$. A $\Gp_{t_j}$-path in $\KK_j$ enters or leaves $U^\alpha$ only through cut bonds, and such a bond with outer end $v$ lies in the piece $\partial(U^\alpha,W)$, where $W\ni v$ is a box-component of $\Lc\setminus U^\alpha$. Here $v\in\KK_j\setminus U=U'$, as a site of $U$ box-adjacent to $U^\alpha$ lies in $U^\alpha$.

We first link the pieces with the same $\alpha$ and different $W$. A $\Gp$-bond inside $U'$ joins box-adjacent sites outside $U^\alpha$ and therefore stays in one box-component of $\Lc\setminus U^\alpha$. For $W_1\ne W_2$ in $\mathcal W(\alpha)$ we take a c-path in $U'$ from $W_1\cap U'$ to $W_2\cap U'$, which exists because $U'$ is c-connected. Along it the box-component of $\Lc\setminus U^\alpha$ changes only at same-cell steps, between two sites of a cell $\Omega'\subseteq U'$ lying in components $W\ne\tilde W$. The $\Gp$-path inside $\hat Q$ joining them is a box path from $W$ to $\tilde W$ and hence meets $U^\alpha$. Its first entry crosses a cut bond of $\partial(U^\alpha,W)$ and its last exit one of $\partial(U^\alpha,\tilde W)$, since the path stays in $W$ before the entry and in $\tilde W$ after the exit. As both bonds lie in $\hat Q$, these two pieces contain bonds at distance at most $k_P$. Following the c-path, we conclude that all pieces $\partial(U^\alpha,W)$, $W\in\mathcal W(\alpha)$, are linked.

We now turn to pieces with different $\alpha$. A $\Gp$-bond inside $U$ stays in one $U^\alpha$. Thus, along a c-path in $U$, the box-component changes only at same-cell steps between sites of a cell $\Omega\subseteq U$ lying in $U^{\alpha_1}\ne U^{\alpha_2}$. The $\Gp$-path inside $\hat Q$ joining them leaves $U^{\alpha_1}$ and enters $U^{\alpha_2}$ through cut bonds of pieces of $\alpha_1$ and of $\alpha_2$, both in $\hat Q$. Each $\alpha$ has a piece, since the box-component of $\Lc\setminus U^\alpha$ containing a site of $U'$ lies in $\mathcal W(\alpha)$. Combining this with the first case, we find that all pieces of $\Sep^*_j$ are linked. The same holds for the pieces of $\Sep^{**}_j$, with $U$ and $U'$ exchanged, and by (d) a cut bond lies in a piece of each. Hence all pieces of $\Sep_j$ are linked. \qed
\end{proof}

For a site $x$ and $r>0$, $B_\infty(x,r)$ is the set of bonds with both ends within $\ell^\infty$ distance $r$ of $x$. Put $c_{\mathrm{sh}}:=c_1/216$, and let $\mathrm{side}_{\min}$ be the least coarse side of $\Lc$.

\begin{lemma}[Admissible cuts]\label{lem:5.22}\label{def:5.21}\label{lem:5.20}\label{lem:5.19}
For a site $x$ of $\Lc$ and $27L_c\le\rho\le\mathrm{side}_{\min}/24$, a cut $j$ is \emph{admissible at $(x,\rho)$} if (i) a cut bond of $j$ has both ends within distance $\rho/2+2$ of $x$; (ii) $U_j$ and $U_j'$ (at time $t_j$) each contain a site farther than $9\rho$ from $x$; (iii) $|\mathrm{cut}_j\cap B_\infty(x,9\rho)|\le c_{\mathrm{sh}}\rho L_c/2$. If $j$ is admissible at $(x,\rho)$, then $B_\infty(x,9\rho)$ contains at least $c_{\mathrm{sh}}\rho L_c$ bonds of each of $\Sep^*_j$ and $\Sep^{**}_j$.
\end{lemma}

\begin{proof}
Both sides of the cut cross about $\rho/(9L_c)$ disjoint shells around $x$, and in each shell the separator, which by Lemma~\ref{lem:5.18}(e) contains all box bonds between the sides it separates, is large by a box Cheeger inequality. We carry this out briefly here, and in full in Section~\ref{S:proofs5}. In a box $B=[n_1]\times[n_2]\times[n_3]\subset\ZZ^3$ (shell boxes need not be cubes), with $n:=\max_in_i$, each $A\subseteq B$ with $|A|\le|B|/2$ has at least $|A|/(2n)$ bonds to $B\setminus A$. To see this, we route all ordered pairs along coordinate-ordered paths. At most $n|B|$ of them use a given bond, and at least $|A|\,|B|/2$ join $A$ to $B\setminus A$. The \emph{shell boxes} form a tensor-product partition of $\Lc$ with sides in $[6L_c,12L_c)$, and the shell $\mathrm{Sh}_i$ is the union of the boxes at index distance $3i$, $3i+1$ or $3i+2$ from the box of $x$. A c-path moves less than $2L_c$ per step and cells lie in one block. It follows that each of the at least $\rho/(9L_c)$ disjoint, face-connected shells $\mathrm{Sh}_i$ with $\rho/(36L_c)+1\le i\le\rho/(4L_c)-1$ lies within $9\rho$ of $x$ and contains a whole cell of $U_j$ and one of $U_j'$, by (i) and (ii). In each shell, the Cheeger inequality in one shell box or in the union of two adjacent ones gives $(c_1/24)L_c^2$ bonds of $\Sep^*_j$ and as many of $\Sep^{**}_j$. Summing over the shells, we obtain $(\rho/(9L_c))(c_1/24)L_c^2=c_{\mathrm{sh}}\rho L_c$ bonds of each. \qed
\end{proof}

\begin{lemma}[Leaf Lemma]\label{lem:5.24}\label{lem:5.23}
Let $n\ge1$ and let $j_1,\dots,j_n$ be distinct cuts admissible at $(x,\rho)$. Let $I$ be the set of bonds $b\in B_\infty(x,9\rho)$ such that for some $i$, $b\in\Sep_{j_i}$, $b$ is not a cut bond of $j_i$, and $b$ is insulating at time $t_{j_i}$. Then $|I|\ge(n+1)\,c_{\mathrm{sh}}\rho L_c/4$.
\end{lemma}

\begin{proof}
The $2n$ sides $U_{j_i}$, $U_{j_i}'$ (at the respective times) are pairwise disjoint or nested (Lemma~\ref{lem:3.18}(b)) and distinct, since the component of a later cut lies in one side of an earlier cut or outside its component. Hence the minimal sides, the \emph{leaves}, are pairwise disjoint. We show that there are at least $n+1$ of them by adding the cuts in firing order. When a cut $j$ is added, its two sides are leaves, since an earlier side contains $\KK_j$ or is disjoint from it. An earlier leaf stops being one only if it contains $\KK_j$, while the earlier sides containing $\KK_j$ form a chain, with at most one minimal element. Thus each cut adds two leaves and removes at most one. For a leaf $L=U_j$ (resp.\ $U_j'$), Lemma~\ref{lem:5.22} gives at least $c_{\mathrm{sh}}\rho L_c$ bonds of $\Sep^*_j$ (resp.\ $\Sep^{**}_j$) in $B_\infty(x,9\rho)$, of which at most $c_{\mathrm{sh}}\rho L_c/2$ are cut bonds of $j$ by Lemma~\ref{lem:5.22}(iii), while Lemma~\ref{lem:5.18}(a) shows that the others are insulating at $t_j$ and have an end in $L$, so that they lie in $I$. Since a bond has two ends and the leaves are disjoint, a bond of $I$ is counted for at most two leaves. \qed
\end{proof}

\subsection{Phase 2: charges and domination}
\label{ssec:5.5}

We use that a cut fires only on a fragmented split, whose scale cost exceeds its area. We spread the area of the cut over its thin clusters and, using the Leaf Lemma, charge each thin cluster to insulation of its own, at a rate $\epsD$ that is doubly exponentially small.

\emph{Scales.} For the scales $s=2^jL\ge\RNF$ of Definition~\ref{def:3.17}, which are lattice lengths, we put $\rho(s):=s/(128\rbar)$, a length in coarse units. Points of two sites at coarse distance $d$ are at lattice Euclidean distance at most $\sqrt3\,\rbar(d+1)$. Consequently, (U1) two cut bonds whose ends are pairwise within coarse distance $d$ lie in one $s$-cluster if $\sqrt3\,\rbar(d+1)\le s$, and (U2) cut bonds of different $s$-clusters have all their ends at coarse distance greater than $s/(\sqrt3\rbar)-1>73\rho(s)-1$. The \emph{charge boxes} at scale $s$ partition the coarse lattice into cubes $Q$ of side $\lceil\rho(s)\rceil$ with centres $x_Q$.

\emph{Thin clusters, good scales.} Fix a cut $j$ and a scale $s$. An $s$-cluster $c$ of $\mathrm{cut}_j$ is \emph{thin} if it contains a bond with an end $x_c$ such that $|c\cap B_\infty(x_c,10\rho(s))|\le c_{\mathrm{sh}}\rho(s)L_c/2$, and \emph{fat} otherwise. For a thin cluster we fix such a bond and end $x_c$. The scale $s$ is \emph{good} for $j$ if $\varphi_s\ge2$ and at least half of the $s$-clusters are thin. Let $\varphi_s^{\mathrm{thin}}$ be the number of thin $s$-clusters.

\begin{lemma}[Charges]\label{lem:5.25}
Assume $\RNF\ge\max\bigl(3456\rbar L_c,\ 2048\rbar L_c/(c_{\mathrm{sh}}\thetaNF\Funit)\bigr)$. For every cut $j$: (a) $J'_j:=\sum_{s\ \mathrm{good}}(\varphi_s-1)\ge J(U_j)/2$; (b) with $q_j(s,c):=A_j(\varphi_s-1)/(J'_j\varphi_s^{\mathrm{thin}})$ ($s$ good, $c$ a thin $s$-cluster), $\sum_{s\ \mathrm{good}}\sum_{c\ \mathrm{thin}}q_j(s,c)=A_j$ and $q_j(s,c)<4L_c^2/(\thetaNF\Funit)$; (c) for good $s$ and thin $c$, with $Q\ni x_c$ the charge box, $j$ is admissible at $(x_Q,\rho(s))$ and $x_{c'}\notin Q$ for every other thin $s$-cluster $c'$.
\end{lemma}

\begin{proof}
(a) A fat cluster has more than $c_{\mathrm{sh}}\rho L_c/2$ bonds, $\rho=\rho(s)$, and clusters are disjoint subsets of $\mathrm{cut}_j$. Consequently, at a scale with $\varphi_s\ge2$ that is not good, $\varphi_s-1<2\varphi_s^{\mathrm{fat}}<4A_j/(c_{\mathrm{sh}}\rho L_c)=512\rbar A_j/(c_{\mathrm{sh}}sL_c)$. Summing over dyadic $s\ge\RNF$ gives less than $1024\rbar A_j/(c_{\mathrm{sh}}\RNF L_c)\le J(U_j)/2$, by the assumption on $\RNF$ and the inequality $J(U_j)>\thetaNF\Funit A_j/L_c^2$, which holds because the split is fragmented.

(b) We have $\sum_cq_j(s,c)=A_j(\varphi_s-1)/J'_j$, and these sums add up to $A_j$ over good $s$. At a good scale $\varphi_s-1<\varphi_s\le2\varphi_s^{\mathrm{thin}}$, so $q_j(s,c)<2A_j/J'_j\le4A_j/J(U_j)<4L_c^2/(\thetaNF\Funit)$.

(c) Take for $x_Q$ a central site of $Q\cap\Lc$ (nonempty, as it contains $x_c$; for instance the centre of $Q$ moved coordinatewise into $\Lc$), so that $x_Q\in Q$ and $|x_c-x_Q|_\infty\le\lceil\rho\rceil/2\le\rho/2+1$, $\rho=\rho(s)$. The numerical inequalities used in (c), and the range $27L_c\le\rho\le\mathrm{side}_{\min}/24$ of Lemma~\ref{lem:5.22}, are established in Section~\ref{S:const5}. (i) The chosen bond of $c$ has both ends within $\rho/2+2$ of $x_Q$. (ii) Since $\varphi_s\ge2$, there is a cut bond in another $s$-cluster, with one end in $U_j$ and one in $U_j'$. By (U2) its ends are farther than $9\rho$ from $x_Q$. (iii) $B_\infty(x_Q,9\rho)\subseteq B_\infty(x_c,10\rho)$, and by (U1) all cut bonds of $j$ in $B_\infty(x_c,10\rho)$ lie in $c$, which is thin. Finally, if $x_c,x_{c'}\in Q$, the two chosen bonds lie in one cluster by (U1). \qed
\end{proof}

\emph{Classes.} For a bond set $Y$, $\Bfr(Y)$ is the set of blocks containing an end of a bond of $Y$. The \emph{classes} $\mathcal J$ are the connected components of the graph on cuts with $j\sim j'$ if $\Bfr(\Sep_j)\cap\Bfr(\Sep_{j'})\ne\emptyset$. We put $\mathfrak X(\mathcal J):=\bigcup_{j\in\mathcal J}\Sep_j$, $\Bfr(\mathcal J):=\Bfr(\mathfrak X(\mathcal J))$ and $A(\mathcal J):=\sum_{j\in\mathcal J}A_j$. Let $\Scut$ be the set of all bonds frozen by the cut rule and $S_1^{(2)}$ the set of bonds frozen by seal steps in Phase 2, and put $\mathrm{ins}^1:=\partial\XX^1$. The \emph{payable bonds} form $\Bpay:=\mathrm{ins}^1\cup\Scut\cup S_1^{(2)}\cup\partial\XX_{\mathrm{final}}\cup\bigcup_\Delta\partial\Delta$, the last union over the cells removed in Phase 2. We denote by $m_q$ the number of marks of a block $q$, and by $h(\Bfr)$ the number of sites in the blocks of $\Bfr$ that are hull material at some time of Phase 2. By Proposition~\ref{prop:4.4} a bond is frozen at most once, so each cut bond belongs to exactly one cut.

\begin{lemma}[Classes]\label{lem:5.26}
(a) Distinct classes have disjoint $\mathfrak X$'s and disjoint $\Bfr$'s. (b) The cut bonds of every $j\in\mathcal J$ lie in $\mathfrak X(\mathcal J)$, and every cut bond lying in $\mathfrak X(\mathcal J)$ belongs to a cut of $\mathcal J$. Moreover, $|\mathfrak X(\mathcal J)\cap\Scut|=A(\mathcal J)$. (c) $\Bfr(\mathcal J)$ is a union of at most $2A(\mathcal J)$ sets, each $*$-connected in the block lattice ($26$-adjacency), which are linked at block distance at most $k_b:=\lceil k_P/L_c\rceil+2$. (d) Every bond of $\mathfrak X(\mathcal J)\setminus\Bpay$ is, at some time of Phase 2, a hole face of a site $y\notin\XX^1$ that is hull material at that time and lies in a block of $\Bfr(\mathcal J)$. So $|\mathfrak X(\mathcal J)\setminus\Bpay|\le6\,h(\Bfr(\mathcal J))\le12L_c\sum_{q\in\Bfr(\mathcal J)}m_q$.
\end{lemma}

\begin{proof}
(a) A common bond or block relates two cuts.

(b) The first claim follows from Lemma~\ref{lem:5.18}(d). For the second, a cut bond of $j''$ in $\Sep_j$ lies in $\Sep_{j''}$, so $j''\sim j$. The equality in (b) follows from the two claims.

(c) In view of Lemma~\ref{lem:5.18}(c), the set $\Sep_j$ has at most $2A_j$ pieces, and Lemma~\ref{lem:5.18}(b), (f) shows that they are square-connected, hence with $*$-connected blocks, and $k_P$-linked. Bonds at distance $\le k_P$ have blocks at block distance $\le k_P/L_c+1\le k_b$, and related cuts share a block, which gives the linking asserted in (c).

(d) Let $b\in\Sep_j\setminus\Bpay$, $j\in\mathcal J$. Cut bonds and bonds frozen at $t_j$ lie in $\Scut\cup S_1^{(2)}\subseteq\Bpay$, since by Theorem~\ref{thm:4.30} the bonds frozen in Phase 1 are converted. It follows by Lemma~\ref{lem:5.18}(a) that $b=(y,u)$ is a hole face at $t_j$, with $y\in\XX_{t_j}$ and $u\in\KK_j\subseteq\Gp_{t_j}$. If $y\in\XX^1$, then $b\in\partial\XX^1=\mathrm{ins}^1$, as $u\in\Gp_{t_j}\subseteq\Gp_{\XX^1}$, where the inclusion holds because $\Gp$ only shrinks. If $y$ lies in a cell $\Delta$ removed in Phase 2 before $t_j$, then $b\in\partial\Delta$, as $u\notin\Delta\subseteq\XX_{t_j}$. Otherwise $y\notin\XX^1\cup R_{t_j}\supseteq\Mm\cup\Uf^1\cup R_{t_j}=B_{t_j}$, where $R_t$ is as in Section~\ref{ssec:4.4} and we use that Phase 2 fills only by removal. Hence, in this case, $y$ is hull material at $t_j$, also if it lies in a cell removed after $t_j$. Its block contains an end of $b$ and hence lies in $\Bfr(\mathcal J)$. Since a site carries six bonds, this gives the first inequality in (d). For the last bound, let $q\not\subseteq\Uf^1$ be a block. If its removal cell is never removed, the hull material $(\XX_t\setminus B_t)\cap q=(\XX_t\cap q)\setminus\Mm$ only grows in $t$. If the cell is removed at time $t_\Delta$, all hull material of $q$ before $t_\Delta$ is hull material just before $t_\Delta$, because removal cells are unions of blocks by Lemma~\ref{lem:3.3}(c). Corollary~\ref{cor:4.10} at the final time, or just before the removal, gives fewer than $2L_cm_q$ sites per block. \qed
\end{proof}

We call a hole face of (d) \emph{transient} if its other end is absorbed later. There can be $L_c$ times as many transient faces as local sources. We absorb them into $\epsD$ and attach them to the contour only through blocks.

\begin{proposition}\label{prop:5.27}\label{eq:5.10}
Under the assumption on $\RNF$ of Lemma~\ref{lem:5.25}, for every set $\mathcal J$ of cuts, $A(\mathcal J)\le\epsD|\mathfrak X(\mathcal J)|$ with $\epsD:=3.2768\cdot10^{7}\rbar L_c/(c_{\mathrm{sh}}\thetaNF\Funit\RNF)$, and if $\mathcal J$ is a class and $\epsD\le\frac12$,
\begin{equation}\label{eq:5.11}
A(\mathcal J)\le2\epsD\Bigl(|\mathfrak X(\mathcal J)\cap\Bpay\setminus\Scut|+12L_c\sum_{q\in\Bfr(\mathcal J)}m_q\Bigr).
\end{equation}
\end{proposition}

\begin{proof}
We assign the charge $q_j(s,c)$ of Lemma~\ref{lem:5.25} to the pair $(Q,s)$ with $Q\ni x_c$. Fix $(Q,s)$ and let $j_1,\dots,j_n$ be the cuts of $\mathcal J$ with a charge at $(Q,s)$. By Lemma~\ref{lem:5.25}(b), (c), each of them has exactly one charge there, of size less than $4L_c^2/(\thetaNF\Funit)$, and is admissible at $(x_Q,\rho(s))$. The Leaf Lemma gives $I_Q\subseteq B_\infty(x_Q,9\rho(s))\cap\mathfrak X(\mathcal J)$ with $|I_Q|\ge(n+1)c_{\mathrm{sh}}\rho(s)L_c/4$. Hence the charge at $(Q,s)$ is at most $4nL_c^2/(\thetaNF\Funit)\le16L_c\,|\mathfrak X(\mathcal J)\cap B_\infty(x_Q,9\rho(s))|/(c_{\mathrm{sh}}\thetaNF\Funit\rho(s))$. A bond lies in $B_\infty(x_Q,9\rho(s))$ for at most $20^3$ charge boxes $Q$ of one scale, and $\sum_s1/\rho(s)\le256\rbar/\RNF$ over dyadic $s\ge\RNF$. Summing over $(Q,s)$, we obtain $A(\mathcal J)\le20^3\cdot16\cdot256\,\rbar L_c|\mathfrak X(\mathcal J)|/(c_{\mathrm{sh}}\thetaNF\Funit\RNF)=\epsD|\mathfrak X(\mathcal J)|$. For \eqref{eq:5.11}, we write $\mathfrak X(\mathcal J)$ as the disjoint union of $\mathfrak X\cap\Scut$, $\mathfrak X\cap\Bpay\setminus\Scut$ and $\mathfrak X\setminus\Bpay$. The first has $A(\mathcal J)$ elements and the last is bounded by Lemma~\ref{lem:5.26}(b), (d). Thus $A\le\epsD(A+R)$ with $R$ the bracket of \eqref{eq:5.11}, and $A\le\epsD R/(1-\epsD)\le2\epsD R$. \qed
\end{proof}

\subsection{Phase 2: responses}
\label{ssec:5.6}

Cuts change $\Gp$ only on their cut bonds, which cross block faces and carry the hull weight $2-\eta_w$. The hull, removal and seal steps therefore respond to cuts as they respond to marks, and we bound the responses locally by the marks, the cut bonds and the bonds of $\mathrm{ins}^1=\partial\XX^1$. Since each frozen bond is a grid bond and supports no hull material (Corollary~\ref{cor:4.10}), a response creates no new source, so the bound on the responses need not be iterated. Section~\ref{S:const5} collects the constants of this subsection and the next, with the arithmetic behind them. Let $m(Z)$ be the number of marks in $Z$, and let $\ncut(Z)$ and $n_{\mathrm{seal}}(Z)$ be the numbers of cut bonds and of bonds frozen by seal steps in Phase~2 (converted ones included) with both ends in $Z$. For a removal cell $\Delta_0$, $\mathcal N(\Delta_0)$ is the set of the at most $27$ removal cells within index distance $1$ of $\Delta_0$, with union $U(\Delta_0)$, and $\mathcal N_2(\Delta_0)$ is the set of the at most $125$ removal cells within index distance $2$ of $\Delta_0$, with union $U_2(\Delta_0)$.

\emph{Response units.} These are (1) the bonds of $S_1^{(2)}$, (2) the bonds of $\partial\XX_{\mathrm{final}}\setminus\mathrm{ins}^1$, that is, the hole faces created in Phase 2 and present at the end, and (3) the pairs $(\Delta,b)$ with $\Delta$ a cell removed in Phase 2 and $b\in\partial\Delta$, and transient hole faces are not units. Each unit is placed at its bond, and $\mathrm{units}(Z)$ denotes the number of units with an end in $Z$. We assign each unit to the removal cell of the first of its ends in $\Lc$, in a fixed order (a bond of a unit of type (3) may leave $\Lc$). Each bond of $\Bpay\setminus\Scut$ lies in $\mathrm{ins}^1$ or is the bond of a unit.

\begin{lemma}[Units]\label{lem:5.31}
For every removal cell $\Delta_0$, $\mathrm{units}(\Delta_0)\le C_a\,m(U_2(\Delta_0))+C_b\,\ncut(U_2(\Delta_0))+C_c\,|\mathrm{ins}^1\cap E(U_2(\Delta_0))|$, with $C_a:=4362+5808/c_s$, $C_b:=2+4/c_s$ and $C_c:=1+2/c_s$.
\end{lemma}

\begin{proof}
We bound the units near $\Delta_0$ by the seal count or, through Lemma~\ref{lem:4.21}, by comparison with the Phase-1 output. Let $U:=U(\Delta_0)$, $m:=m(U_2(\Delta_0))$, $n:=\ncut(U)$ and $i:=|\mathrm{ins}^1\cap E(U)|$. A seal bond with an end in a block $q$ lies on one of the six faces of $q$, and by Proposition~\ref{prop:4.sealcount} the face shared with $q'$ carries at most $121(m_q+m_{q'})$ of them. Summing over the blocks of $\Delta_0$, resp.\ of $U$, we find that at most $1452\,m$ units of type (1) have an end in $\Delta_0$ and that $n_{\mathrm{seal}}(U)\le1452\,m$. A unit $(\Delta,b)$ with an end in $\Delta_0$ has $\Delta\in\mathcal N(\Delta_0)$, and by Lemma~\ref{lem:4.21}(b) the units of type (3) of the cells of $\mathcal N(\Delta_0)$ number less than $u_3:=c_s^{-1}(i+2n_{\mathrm{seal}}(U)+2n)$. We turn to the units of type (2). For $1\le r\le L_c$ the boxes $N_r:=\{x:\dist(x,\Delta_0)\le r\}$ lie in $U$ and have disjoint outer layers. A final hole face with an end in $\Delta_0$ lies in $E(N_r)$ and has exactly one end in $\XX_\infty$, so it is counted in $\mathcal A_2(E(N_r);\infty)$, which \eqref{eq:4.I5} bounds by $\mathrm{Ins}_{N_r}(\XX^1\cup R_\infty)+\mathrm{cut}_{N_r}$. A bond of $E(N_r)$ with exactly one end in $\XX^1\cup R_\infty$ is a bond of $\mathrm{ins}^1$ or the bond of a unit of type (3) of a cell of $\mathcal N(\Delta_0)$, and a bond of $\bar S_\infty\cap E(N_r)$ with no end in $\XX^1$ was frozen in Phase~2. Hence $\mathrm{Ins}_{N_r}(\XX^1\cup R_\infty)\le i+u_3+2n_{\mathrm{seal}}(U)+2n$. The sites counted in $\mathrm{cut}_{N_r}$ are hull material in the outer layer of $N_r$, each with at most three bonds leaving $N_r$, and the blocks of $U$ hold fewer than $2L_cm$ sites of hull material (Corollary~\ref{cor:4.10}). Consequently, $\mathrm{cut}_{N_r}\le6m$ for some $r$. Adding these bounds, we obtain $\mathrm{units}(\Delta_0)\le1452m+2u_3+i+2n_{\mathrm{seal}}(U)+2n+6m$. Inserting $n_{\mathrm{seal}}(U)\le1452\,m$ here and in the bound on the units of type (3) gives the claim. \qed
\end{proof}

\begin{lemma}\label{lem:5.32}
There are weights $\varpi(u,z)\ge0$, for $u$ a response unit and $z$ a \emph{source} (a mark, a cut bond or a bond of $\mathrm{ins}^1$), with $\varpi(u,z)=0$ if $\dist(u,z)>\Rresp:=10\ell$ and $\sum_z\varpi(u,z)=1$ for every $u$, the weights being functions of $D$, such that every mark receives $\sum_u\varpi(u,z)\le125C_a$, every cut bond at most $125C_b$, and every bond of $\mathrm{ins}^1$ at most $125C_c$.
\end{lemma}

\begin{proof}
Since Lemma~\ref{lem:5.31} bounds the units of a cell by the sources around it, we spread each unit over those sources. For a unit $u$ assigned to $\Delta_0$ put $\varpi(u,z):=c_z/D(\Delta_0)$ for the sources $z$ of $U_2(\Delta_0)$, that is, the marks in $U_2(\Delta_0)$, the cut bonds and the bonds of $\mathrm{ins}^1$ in $E(U_2(\Delta_0))$. Here $c_z=C_a$, $C_b$ or $C_c$ according to the kind of $z$, and $D(\Delta_0)$ is the right side of Lemma~\ref{lem:5.31}. Then $\sum_z\varpi(u,z)=1$, and $D(\Delta_0)\ge\mathrm{units}(\Delta_0)>0$ if a unit is assigned to $\Delta_0$. A source lies in $U_2(\Delta)$ for at most $125$ cells $\Delta$ and therefore receives at most $125c_z$. A unit assigned to $\Delta_0$ has an end in $\Delta_0$, a source of $U_2(\Delta_0)$ lies in a cell at index distance at most $2$, and cells have sides less than $3\ell$. Hence $\dist(u,z)\le9\ell+1\le\Rresp$. \qed
\end{proof}

The constant $\Cresp:=375C_a+125C_c\Cst$ is fixed before $K$. A part $P$ owns the bonds of $\partial Y_P$, which by Theorem~\ref{thm:3.21}(a) are at most $\Cst N(P)$ bonds of $\mathrm{ins}^1$. The marks and the $\mathrm{ins}^1$ bonds of $P$ therefore receive at most $(125C_a+125C_c\Cst)N(P)\le\Cresp N(P)$, and a cut bond receives at most $125C_b\le\Cresp$.

\subsection{Counting and the proof of Theorem~\ref{thm:5.1}}
\label{ssec:5.7}

In this subsection $(\XX,S)$ is the Phase-2 output, with $\Ffr$ and the minimizer $\Sigma^\circ$ of Lemma~\ref{lem:5.5} computed for it, and we write $\Xi':=\supp\Sigma^\circ$ and $a:=a_\Sigma=|\Xi'|\ge RT$. Let $R':=R_*+2\Rresp$ and
\begin{equation}\label{eq:5.epspp}
\eps'':=8e\,(2R'+4L_c+1)^3\,(2r'+1)^3\,q_m .
\end{equation}
The hole set only grows, as Proposition~\ref{prop:4.4} shows, and hence $\XX^1\subseteq\XX$.

As pieces we take the sets $Y_P$, $P\in\cP$, of Theorem~\ref{thm:3.21}(a), the sets $Y_\Delta:=\Delta\setminus\XX^1$ for the cells $\Delta$ removed in Phase 2, which lie in $\XX$, and the sets $Y_{\Delta'}:=(\XX\setminus\XX^1)\cap\Delta'$ for the removal cells $\Delta'$ not removed in Phase 2. These sets partition $\XX$. A piece is touched if $\Xi'$ meets its $F(\cdot)$, and a part $P$ is touched if $Y_P$ is. The cell of a touched $Y_\Delta$ or $Y_{\Delta'}$ is a \emph{touched cell}. Let $u_{\mathrm{touch}}$ be the number of response units that have an end in a touched cell or are of type (1) with a coarse face containing a plaquette of $\Xi'$ (each unit counted once), $c_{\mathrm{touch}}$ the number of bonds of $S\cap\Scut$ whose coarse face contains a plaquette of $\Xi'$, and $i_{\mathrm{touch}}$ the number of bonds of $\mathrm{ins}^1$ with an end in a touched cell.

\begin{proposition}\label{prop:5.34}
$\#(\Xi'\cap\Ffr)\le\rbar^{2}\bigl[\Cfree\sum_{P\ \mathrm{touched}}N(P)+20(u_{\mathrm{touch}}+c_{\mathrm{touch}}+i_{\mathrm{touch}})+(26L_c/\ell)\sum_{\Delta'\ \mathrm{touched}}m(\Delta')\bigr]$, the last sum over the touched cells $\Delta'$.
\end{proposition}

\begin{proof}
We apply Proposition~\ref{prop:5.7} to the pieces defined above. The removal cells cut the Phase-2 hole material into pieces of diameter less than $9\ell$, each entering with its volume divided by $\ell$, and we charge the boundary of each piece to Phase-1 insulation, to response units and to its volume. A touched part costs at most $\Cfree N(P)$, as in Lemma~\ref{lem:5.10}. A touched $Y_\Delta$ costs $2|\partial Y_\Delta|+|Y_\Delta|/\ell$, with $|Y_\Delta|/\ell<27\ell^2\le4.5|\partial\Delta|$. A bond of $\partial Y_\Delta$ either lies in $\partial\Delta$, in which case it is the bond of a unit $(\Delta,b)$ of type (3), or joins $\XX^1\cap\Delta$ to $\Delta\setminus\XX^1$, in which case it is a bond of $\mathrm{ins}^1$ with both ends in $\Delta$.

A touched $Y:=Y_{\Delta'}$ costs $2|\partial Y|+|Y|/\ell$. A bond $(y,v)$ of $\partial Y$, $y\in Y$, is of one of the following four types. (a) $v\in\Gp$, and then the bond is a final hole face with no end in $\XX^1$, the bond of a unit of type (2). (b) $v\in\XX^1$, so the bond lies in $\mathrm{ins}^1$. (c) $v$ lies in a cell $\Delta$ removed in Phase 2, so $(\Delta,(y,v))$ is a unit of type (3). (d) $v\in(\XX\setminus\XX^1)\cap\Delta''$ for another cell $\Delta''$ not removed in Phase 2. For a bond of type (d), we walk from $y$ along its coordinate line into $\Delta'$. If the walk leaves $Y$ inside $\Delta'$, it leaves through a bond from $Y$ to $\Delta'\setminus Y\subseteq\Gp\cup\XX^1$, of type (a) or (b) with both ends in $\Delta'$, and distinct walks end at distinct bonds. Otherwise the walk crosses $\Delta'$ inside $Y$, through at least $\ell$ sites, and we charge $1/\ell$ to each of them. Since a site lies on three lines, each with at most two type-(d) bonds at its ends, it receives at most $6/\ell$. Hence $|\partial Y|$ is at most the number of bonds of types (a) to (c) with an end in $Y$, plus those of types (a), (b) inside $\Delta'$, plus $6|Y|/\ell$. Phase 2 fills only by removal, so $Y$ is hull material and Corollary~\ref{cor:4.10} at the final time gives $|Y|<2L_cm(\Delta')$.

Since the Phase-1 output has no frozen bond, a touched bond of $S$ lies in $S_1^{(2)}$ or in $S\cap\Scut$. In the first case it is a unit of type (1), counted in $u_{\mathrm{touch}}$, and in the second it is counted in $c_{\mathrm{touch}}$. Each unit or bond of $\mathrm{ins}^1$ charged above has an end in the touched cell charging it, so it is counted in $u_{\mathrm{touch}}$ or $i_{\mathrm{touch}}$, and the coefficients add up to at most $20$. \qed
\end{proof}

\emph{The attached family.} Let $\mathcal T$ be the set of classes with a cut bond counted in $c_{\mathrm{touch}}$, and $\mathcal F$ the smallest set of classes containing $\mathcal T$ with the closure property: if $\varpi(u,z)>0$ for a cut bond $z$ of a class $\mathcal J'$, where (i) $u$ is a unit counted in $u_{\mathrm{touch}}$, or (ii) the bond of the unit $u$ lies in $\mathfrak X(\mathcal J)$ for some $\mathcal J\in\mathcal F$, then $\mathcal J'\in\mathcal F$. We put $\Bfr_{\mathcal F}:=\bigsqcup_{\mathcal J\in\mathcal F}\Bfr(\mathcal J)$ and $A_{\mathcal F}:=\sum_{\mathcal J\in\mathcal F}A(\mathcal J)$. Let $Z_{\mathcal F}$ be the set of sites of the blocks of $\Bfr_{\mathcal F}^{+1}$ (the blocks at block distance at most $1$ from $\Bfr_{\mathcal F}$) and $\Qatt$ the set of parts $P$ with $\dist(P,Z(\Xi')\cup Z_{\mathcal F})\le R'$, and put $N_{\mathcal Q}:=\sum_{P\in\Qatt}N(P)$. In Lemma~\ref{lem:5.35} and Proposition~\ref{prop:5.36} we assume the condition
\begin{equation}\label{eq:5.16}
8\epsD(1+8L_c)(C_\star'+3\Cresp+6)(48\Cresp+2)\le1,
\end{equation}
which holds for $\beta\ge\beta_0(K)$, as $\epsD$ is doubly exponentially small.

\begin{lemma}[Attached family]\label{lem:5.35}
(a) the units counted in $u_{\mathrm{touch}}$ and the units whose bond lies in $\bigcup_{\mathcal F}\mathfrak X(\mathcal J)$ number at most $\Cresp(N_{\mathcal Q}+A_{\mathcal F})$ each, so at most $2\Cresp(N_{\mathcal Q}+A_{\mathcal F})$ together; (b) $A_{\mathcal F}\le8\epsD(1+8L_c)(C_\star'+3\Cresp+6)\,N_{\mathcal Q}\le N_{\mathcal Q}/(48\Cresp+2)$; (c) $c_{\mathrm{touch}}\le A_{\mathcal F}$, $i_{\mathrm{touch}}\le C_\star'N_{\mathcal Q}$ and $|\Bfr_{\mathcal F}|\le C_{\Bfr}N_{\mathcal Q}$ ($C_{\Bfr}$ of Section~\ref{S:const5}); (d) $\Bfr_{\mathcal F}$ has $m\le2A_{\mathcal F}$ $*$-components.
\end{lemma}

\begin{proof}
(a) A unit of the first family has an end within $3\ell$ of $Z(\Xi')$, since a touched cell meets $Z(\Xi')$ and has sides less than $3\ell$, and since a unit of type (1) whose face meets $\Xi'$ has both ends in $Z(\Xi')$. A unit of the second family has an end in $Z_{\mathcal F}$. Thus, by Lemma~\ref{lem:5.32}, a source $z$ with $\varpi(u,z)>0$ lies within $\Rresp+3\ell+1$ of $Z(\Xi')\cup Z_{\mathcal F}$. If such a source is a mark, it is a mark of a part of $\Qatt$. If it is a cut bond, it belongs to a class of $\mathcal F$ by the closure property, and there are at most $A_{\mathcal F}$ such bonds, by Lemma~\ref{lem:5.26}(b). If it is a bond of $\mathrm{ins}^1=\bigsqcup_P\partial Y_P$, its owner $P$ has a site of $Y_P\subseteq P^{+R_*}$ within $\Rresp+3\ell+2$ of $Z(\Xi')\cup Z_{\mathcal F}$ and so lies in $\Qatt$. As each unit has total weight at least $1$, the loads stated after Lemma~\ref{lem:5.32} imply that each family has at most $\Cresp N_{\mathcal Q}+125C_bA_{\mathcal F}\le\Cresp(N_{\mathcal Q}+A_{\mathcal F})$ elements.

(b) We sum \eqref{eq:5.11} over $\mathcal J\in\mathcal F$. The bonds of $\mathfrak X(\mathcal J)\cap\Bpay\setminus\Scut$ are bonds of $\mathrm{ins}^1$ or of units. A bond of $\mathrm{ins}^1$ lies in $\partial Y_P$ for one part $P$, which has an end of it in $Y_P\subseteq P^{+R_*}$, so $P$ lies within $R_*+1$ of $Z_{\mathcal F}$ and belongs to $\Qatt$. A part owns at most $C_\star'N(P)$ such bonds, by Theorem~\ref{thm:3.21}(a). The units are bounded by (a), and the marks in blocks of $\Bfr_{\mathcal F}$ belong to parts of $\Qatt$. Therefore
\[
A_{\mathcal F}\le2\epsD\bigl[C_\star'N_{\mathcal Q}+2\Cresp(N_{\mathcal Q}+A_{\mathcal F})+12L_cN_{\mathcal Q}\bigr].
\]
The coefficient of $A_{\mathcal F}$ on the right is $4\epsD\Cresp\le\frac12$ by \eqref{eq:5.16}. Hence $A_{\mathcal F}\le4\epsD(C_\star'+2\Cresp+12L_c)N_{\mathcal Q}$, which gives the first inequality. The second inequality is \eqref{eq:5.16}.

(c) Lemma~\ref{lem:5.26}(b) implies that a touched cut bond belongs to a cut of a class in $\mathcal T$, and thus $c_{\mathrm{touch}}\le A_{\mathcal F}$. A bond counted in $i_{\mathrm{touch}}$ has an end within $3\ell$ of $Z(\Xi')$, so its owner lies within $R_*+3\ell+1\le R'$ of $Z(\Xi')$, in $\Qatt$, and Theorem~\ref{thm:3.21}(a) gives $i_{\mathrm{touch}}\le C_\star'N_{\mathcal Q}$. For $|\Bfr_{\mathcal F}|$, we charge each block of $\Bfr(\mathcal J)$ to a bond of $\mathfrak X(\mathcal J)$ with an end in it if that bond lies in $\Bpay$, with at most two blocks per bond. Otherwise, by Lemma~\ref{lem:5.26}(d), the block is that of a site $y$ of hull material or one of the six face neighbours of the block of that site, and we charge it to the first mark of the block of $y$, which exists by Corollary~\ref{cor:4.10}. In this way each mark receives at most seven blocks. Counting as in (a) and (b) gives $|\Bfr_{\mathcal F}|\le C_{\Bfr}N_{\mathcal Q}$.

(d) By Lemma~\ref{lem:5.26}(c), $\Bfr(\mathcal J)$ is a union of at most $2A(\mathcal J)$ $*$-connected sets. \qed
\end{proof}

\begin{proposition}\label{prop:5.36}
$\#(\Xi'\cap\Ffr)\le\rbar^{2}C_3N_{\mathcal Q}$, $|\Bfr_{\mathcal F}|\le C_{\Bfr}N_{\mathcal Q}$ and $m\le M_mN_{\mathcal Q}$, with $C_3$, $C_{\Bfr}$ free of $\beta$, $\tau$, $K$, $C$, $D$ and $M_m=O(\epsD L_c\Cresp)$ (Section~\ref{S:const5}).
\end{proposition}

\begin{proof}
We insert Lemma~\ref{lem:5.35} into Proposition~\ref{prop:5.34}. Touched parts lie in $\Qatt$ by Lemma~\ref{lem:5.10}. By Lemma~\ref{lem:5.35}(c) and the proof of (a), $u_{\mathrm{touch}}+c_{\mathrm{touch}}+i_{\mathrm{touch}}\le(\Cresp+C_\star')N_{\mathcal Q}+(\Cresp+1)A_{\mathcal F}$. The marks of the touched cells belong to parts of $\Qatt$, so $\sum_{\Delta'}m(\Delta')\le N_{\mathcal Q}$, and $26L_c/\ell\le26/308$. Together with $A_{\mathcal F}\le N_{\mathcal Q}/(48\Cresp+2)$ from Lemma~\ref{lem:5.35}(b), these estimates give the first bound, and the other two follow from Lemma~\ref{lem:5.35}(c), (b), (d). \qed
\end{proof}

\begin{lemma}\label{lem:5.37}
Assume $\eps''\le\frac12$. For a fixed set $Z$ of sites, a fixed set $\Bfr'$ of blocks and $u_0>0$, $\PD\bigl(\sum_{P\in\cP:\ \dist(P,Z\cup\bigcup\Bfr')\le R'}N(P)\ge u_0\bigr)\le2^{|Z|+|\Bfr'|+1}(\eps'')^{u_0}$.
\end{lemma}

\begin{proof}
We argue as in the proof of Proposition~\ref{prop:5.13}, with roots in $Z\cup\Bfr'$, which give at most $2^{|Z|+|\Bfr'|+m}$ multisets of roots, and with the mark $y_i$ within $R'$ of its root. As block sides are less than $2L_c$, there are $(2R'+4L_c+1)^3$ choices for each mark. \qed
\end{proof}

For Lemma~\ref{lem:5.38} we set $d_0:=\lceil R'/L_c\rceil+2$. The block distance is the $\ell^\infty$ distance of block indices, and $q(x)$ is the block of a site $x$. Since block sides are at least $L_c$, sites at distance $r$ lie in blocks at block distance at most $r/L_c+1$. A block set $\Bfr'$ is \emph{anchored to} a plaquette set $\Xi'$ if the graph on $\{o\}\cup\{*\text{-components of }\Bfr'\}$, in which $o$ is joined to a component $C$ when the component $C$ has a block within block distance $d_0$ of $q(x)$ for some $x\in Z(\Xi')$, and two components are joined when they have blocks within block distance $d_0$, is connected.

\begin{lemma}\label{lem:5.38}
(a) $\Bfr_{\mathcal F}$ is anchored to $\Xi'=\supp\Sigma^\circ$. (b) For every plaquette set $\Xi'$ with $|\Xi'|=a$, the number of block sets anchored to $\Xi'$ with $b$ blocks and $m$ $*$-components is at most $2^{2a+2b+m}(2d_0+1)^{3m}(26e)^b$.
\end{lemma}

\begin{proof}
(a) We build $\mathcal F$ by the closure, starting from $\mathcal F_0:=\mathcal T$ and letting $\mathcal F_{t+1}$ be $\mathcal F_t$ with the classes forced by $\mathcal F_t$. A class of $\mathcal T$ has a touched cut bond, whose ends lie in $Z(\Xi')$ and whose blocks lie in its $\Bfr$. A class $\mathcal J'\in\mathcal F_{t+1}\setminus\mathcal F_t$ has a cut bond $z$ with $\varpi(u,z)>0$, hence $\dist(u,z)\le\Rresp$, for a unit $u$ of type (i) or (ii) relative to $\mathcal F_t$. In case (i), $u$ has an end within $3\ell$ of $Z(\Xi')$, so the block of an end of $z$ is within block distance $(\Rresp+3\ell+1)/L_c+1\le d_0$ of $q(x)$ for some $x\in Z(\Xi')$. In case (ii), the block of an end of $z$ is within block distance $(\Rresp+1)/L_c+2\le d_0$ of $\Bfr(\mathcal J)$, $\mathcal J\in\mathcal F_t$, as $u$ has an end in a block of $\Bfr(\mathcal J)$. In both cases this block lies in $\Bfr(\mathcal J')$, as $z\in\mathfrak X(\mathcal J')$, and $\Bfr(\mathcal J')$ is a union of $*$-connected sets linked at block distance $k_b\le d_0$ by Lemma~\ref{lem:5.26}(c), each of these sets lying in one component of $\Bfr_{\mathcal F}$. By induction on $t$, each component meeting $\Bfr(\mathcal J)$, $\mathcal J\in\mathcal F_t$, is connected to $o$, and each component of $\Bfr_{\mathcal F}$ meets some $\Bfr(\mathcal J)$.

(b) Along a breadth-first spanning tree from $o$, built with fixed orders, the block set is fixed by the following data, which Section~\ref{S:proofs5} lists in more detail:
\begin{itemize}
\item the component sizes, with $2^b$ choices;
\item the anchors of the children of $o$ in $Z(\Xi')$, where $|Z(\Xi')|\le2a$, and of each component among its blocks, with $2^{2a+b+m}$ choices in all;
\item the root positions relative to the anchors, with $(2d_0+1)^{3m}$ choices;
\item the components, for which Lemma~\ref{lem:5.12} with $D_{\max}=26$ gives $(26e)^b$ choices. \qed
\end{itemize}
\end{proof}

\begin{proof}[Proof of Theorem~\ref{thm:5.1}]
The conditions of item~(a) of Section~\ref{ssec:D.5} include \eqref{eq:5.16} and
\begin{equation}\label{eq:5.17}
2^{29}\cdot26e\cdot(\eps'')^{1/(3C_{\Bfr})}\le\tfrac12,\qquad2(2d_0+1)^3(\eps'')^{1/(3M_m)}\le\tfrac12 .
\end{equation}

On $E:=\{\AGp(C)\le(1-\etaT)RT\}$, as $a=\AGp(C)+\#(\Xi'\cap\Ffr)$, Proposition~\ref{prop:5.36} gives $N_{\mathcal Q}\ge\frac13(u_1(a)+b/C_{\Bfr}+m/M_m)$ with $b:=|\Bfr_{\mathcal F}|$ and $u_1(a):=(a-(1-\etaT)RT)/(\rbar^{2}C_3)$.

We first reduce the estimate to deterministic pairs. The minimizer $\Sigma^\circ$ and the block family $\Bfr_{\mathcal F}$ are functions of $D$, defined through fixed orders. Indeed, this holds for the output $(\XX,S)$ by Theorem~\ref{thm:3.20}(R1), for the minimizer by Lemma~\ref{lem:5.5}, for the classes, for the weights $\varpi$ by Lemma~\ref{lem:5.32}, and for the smallest closed family $\mathcal F$. Thus, on $E$, the pair $(\supp\Sigma^\circ,\Bfr_{\mathcal F})$ takes one deterministic value $(\Xi',\Bfr')$, with $\Xi'$ a plaquette set as in Lemma~\ref{lem:5.12} and with $\Bfr'$ anchored to $\Xi'$ by Lemma~\ref{lem:5.38}(a). Hence $E\subseteq\bigcup_{(\Xi',\Bfr')}E(\Xi',\Bfr')$, the union over all such deterministic pairs, where $E(\Xi',\Bfr')$ is the event that the parts $P$ within $R'$ of $Z(\Xi')\cup\bigcup\Bfr'^{+1}$ have $\sum N(P)\ge\frac13(u_1(|\Xi'|)+|\Bfr'|/C_{\Bfr}+m(\Bfr')/M_m)$. In $E(\Xi',\Bfr')$ the random partition appears only as a family of disjoint $r'$-connected mark sets, by Theorem~\ref{thm:3.21}, and Lemma~\ref{lem:5.37} sums over all such families.

It remains to sum over these pairs. Lemma~\ref{lem:5.37}, applied with $Z=Z(\Xi')$, where $|Z|\le2a$, and with the block set $\Bfr'^{+1}$ of at most $27b$ blocks, gives $\PD(E(\Xi',\Bfr'))\le2^{2a+27b+1}(\eps'')^{(u_1+b/C_{\Bfr}+m/M_m)/3}$. Summing over the anchored $\Bfr'$ with Lemma~\ref{lem:5.38}(b), we obtain the factors $(2^{29}\cdot26e\,(\eps'')^{1/(3C_{\Bfr})})^b$ and $(2(2d_0+1)^3(\eps'')^{1/(3M_m)})^m$, each at most $2^{-b}$, resp.\ $2^{-m}$, by \eqref{eq:5.17}. The union over $\Bfr'$ therefore has probability at most $2^{4a+3}(\eps'')^{u_1(a)/3}$. Finally, Lemma~\ref{lem:5.14} with $k=4$, $j=3$, $\eps=\eps''$, $F_*=3\rbar^{2}C_3$ and $X:=RT-\AGp(C)$, for which $\etaT|\log\eps''|/F_*=c_T''\ge c_T(4)$ and, by Lemma~\ref{lem:5.3}, $0\le X\le RT$, gives both bounds of \eqref{eq:5.1}.

\emph{The order of $\etaT$.} By Lemma~\ref{lem:5.16} with $\eps''$, $|\log\eps''|\ge\gamma_T(K)\tau/2$ for $\tau\ge\tau_0(K)$, where $\tau_0(K)$ is chosen so that the error term of \eqref{eq:5.9} is at most $\gamma_T(K)\tau/2$ (it grows logarithmically in $\tau$). This gives the first bound, and the second follows from $C_3=O((C_\star'+\Cresp)(1+1/c_s))$ (Section~\ref{S:const5}) and $\gamma_T(K)\ge1.4\cdot10^{-6}\log^2K$ for $K>K_*$ (Section~\ref{S4-S:S4.2}). \qed
\end{proof}

}
\providecommand{\bGM}{\beta_{\mathrm{GM}}}
\providecommand{\mt}{\tilde m}
\providecommand{\mD}{m_{D}}
\providecommand{\MY}{M_{Y}}
\providecommand{\lPV}{\lambda_{\mathrm{PV}}}
\providecommand{\Gp}{G'}
\providecommand{\XX}{\mathbb{X}}
\providecommand{\KK}{\mathfrak{K}}
\providecommand{\Dep}{\mathfrak{D}}
\providecommand{\RHK}{R_{\mathrm{HK}}}
\providecommand{\rcov}{r_{\mathrm{cov}}}
\providecommand{\Rdep}{R_{\mathrm{dep}}}
\providecommand{\etaw}{\eta_{w}}
\providecommand{\Zthree}{\mathbb{Z}^{3}}
\providecommand{\RR}{\mathbb{R}}
\providecommand{\ZZ}{\mathbb{Z}}
\providecommand{\rhoref}{\rho^{\mathrm{ref}}}
\providecommand{\vCb}{v_{\mathrm{Cb}}}
\providecommand{\eps}{\varepsilon}
\providecommand{\Ord}{\mathcal{O}}

\section{Stage 1: the Mayer expansion to the Debye scale on slit domains}
\label{sec:stage1}

In this section we prove \cite[Prop.~3]{GM82}, with the iterated Mayer expansion of \cite{GM81}, for the gas whose kernel is the Neumann Green function of a slit graph (Theorem~\ref{thm:6.5}), and we derive heat-kernel, covariance and depletion bounds on outputs. Since the Neumann field of a distant source is enhanced near a free edge of a frozen sheet, the only pointwise bounds on a Neumann kernel or its gradient that Sections~\ref{sec:stage1} and~\ref{sec:stage2} use near such an edge are the diagonal lower bounds of Lemma~\ref{lem:6.1} and heat-kernel decay from the isoperimetry (R2), in the form of Theorem~\ref{thm:6.9} and Lemma~\ref{lem:6.15}. The remaining inputs are the exact $L^{1}$ sums of Lemma~\ref{lem:6.2} and, in Lemma~\ref{lem:7.34}, the free-space solid angle.

\paragraph{Conventions.} The lattice is the fine lattice (spacing $1$), $\bGM=4\pi^{2}\tau$, and distances are Euclidean unless $d_{\Gp}$ is written. Slit graphs $\Gp=(V,E)$, $V=\Lambda'\setminus\XX$, $E=E(V)\setminus S$, their components $\KK$, the bound $d_{\Gp}(x,y)\ge\|x-y\|_{1}$ and the outputs $\Gp=\Gp(D)$ are those of \S\ref{ssec:2.5}. We define $(-\Delta_{\Gp}f)(y):=\sum_{y'\sim y}(f(y)-f(y'))$ over the $\Gp$-neighbours of $y$. For $m>0$, $v^{(m)}_{\Gp}:=(-\Delta_{\Gp}+m^{2})^{-1}$, $v_{\Gp}:=v^{(\MY)}_{\Gp}$, $v:=v^{(\MY)}_{\Zthree}$, $v_{\Lambda'}$ is $v^{(\MY)}$ of the box $\Lambda'$ with all its bonds, and $\vCb$ is the Coulomb kernel of $\Zthree$. On $\RR^{V}_{0}:=\{\varphi:\sum_{x\in\KK}\varphi(x)=0$ for all $\KK\}$, $G_{N}:=(-\Delta_{\Gp})^{+}$ is the inverse of $-\Delta_{\Gp}$. The \emph{image energies} are $\Phi(x)=\Phi_{\Gp}(x):=\frac12\bGM(v_{\Gp}(x,x)-v(0))$ and $I_{\Gp}(x):=v_{\Gp}(x,x)-v_{\Lambda'}(x,x)$, and the \emph{depleted set} is $\Dep:=\{x\in V:\Phi_{\Gp}(x)>\theta_{0}\}$, $\theta_{0}=0.01$. For a matrix $K$ on $V$ and $r\ge0$, $\|K\|_{r}$ is the larger of $\sup_{x}\sum_{y}|K(x,y)|e^{r\,d_{\Gp}(x,y)}$ and the same expression with $x$ and $y$ exchanged. This norm is submultiplicative, and $\|K\|_{2\to2}\le\|K\|_{0}\le\|K\|_{r}$ for all such $K$ and $r$. The activity of a unit monopole is $z:=e^{-\bGM\vCb(0)/2}$, and $\mD^{2}:=2\bGM z$ \cite[(1.8a)]{GM82}. Hence $\mD=(8\pi^{2}\tau)^{1/2}e^{-\pi^{2}\vCb(0)\tau}$. Numerical values are given in \S\ref{S:E1.mass}. The mass $\mt\in[\mD/2,2\mD]$ and the Pauli\textendash Villars mass $\MY=\mt/\lPV$ are fixed in \eqref{eq:6.mass}. Until then $\MY$ is any mass in $I_{Y}:=[\mD/(2\lPV),2\mD/\lPV]$, and all statements hold uniformly in that range. Besides the loop, the letter $C$ denotes, depending on the context, the parameter $C\in(0,1)$ of Theorem~\ref{thm:6.5} (the notation of \cite{GM82}), a generic constant, such as those of (S5), \eqref{eq:6.lead} and Proposition~\ref{prop:6.6}, and, in Theorem~\ref{thm:6.12} and Appendix~\ref{app:C}, the covariance $(C_{0}^{-1}+\nu)^{-1}$.

\subsection{Kernel facts on every slit graph}
\label{ssec:6.1}

\begin{lemma}[Thomson monotonicity]\label{lem:6.1}
Let $H\subseteq H'$ be graphs with $V(H)\subseteq V(H')$, $E(H)\subseteq E(H')$ and $V(H)$ finite; $H'=\Zthree$ is allowed. Let $f$ be supported on $V(H)$. (i) For every $m>0$, $\bigl(f,v^{(m)}_{H}f\bigr)\ge\bigl(f,v^{(m)}_{H'}f\bigr)$. (ii) If $f$ has zero sum on every component of $H$, then it has zero sum on every component of $H'$, and $\bigl(f,(-\Delta_{H})^{+}f\bigr)\ge\bigl(f,(-\Delta_{H'})^{+}f\bigr)$; for $H'=\Zthree$ the right side is $(f,\vCb f)$. (iii) For every slit graph $\Gp\subseteq\Lambda'$ and $x\in V$: $v^{(m)}_{\Gp}(x,x)\ge v^{(m)}_{\Zthree}(0)$ and $v_{\Gp}(x,x)\ge v_{\Lambda'}(x,x)$. Hence $\Phi_{\Gp}\ge0$ and $I_{\Gp}\ge0$.
\end{lemma}

\begin{proof}
In the variational formula $(f,v^{(m)}_{H}f)=\sup_{\varphi}[2(f,\varphi)-\sum_{\{y,y'\}\in E(H)}(\varphi(y)-\varphi(y'))^{2}-m^{2}\sum_{y\in V(H)}\varphi(y)^{2}]$, restricting $\varphi$ from $V(H')$ to $V(H)$ keeps $(f,\varphi)$ and does not raise the energy, which gives (i). In (ii), $f$ is orthogonal to the kernels of both Laplacians, and the formula holds at $m=0$; for $H'=\Zthree$ let $m\downarrow0$ in (i): the left side tends to $(f,(-\Delta_H)^{+}f)$, by the spectral theorem on $f\perp\ker(-\Delta_H)$, and the right side increases to $(f,\vCb f)$, by monotone convergence in the Fourier formula. We obtain (iii) by taking $f=\delta_{x}$ in (i).
\qed
\end{proof}

\begin{lemma}\label{lem:6.2}
Let $\Gp$ be a slit graph and $0<m<m'$. Then: (i) $v^{(m)}_{\Gp}$ is symmetric and entrywise nonnegative, and $v^{(m)}_{\Gp}(x,y)=0$ when $x$ and $y$ lie in different components; (ii) $\sum_{y\in V}v^{(m)}_{\Gp}(x,y)=m^{-2}$ for every $x\in V$; (iii) $v^{(m)}_{\Gp}-v^{(m')}_{\Gp}=(m'^{2}-m^{2})\,v^{(m)}_{\Gp}v^{(m')}_{\Gp}$, and this matrix is entrywise nonnegative and positive semidefinite, with row sums $m^{-2}-m'^{-2}$; (iv) $\operatorname{spec}(-\Delta_{\Gp})\subseteq[0,12]$.
\end{lemma}

\begin{proof}
Since $\deg_{\Gp}\le6$, the matrix $P:=I-(-\Delta_{\Gp})/6$ is symmetric, nonnegative and stochastic, with no entry across components, and the series $v^{(m)}_{\Gp}=(6+m^{2})^{-1}\sum_{n\ge0}(6/(6+m^{2}))^{n}P^{n}$ gives (i) and (ii). Part (iii) is the resolvent identity, and (iv) is Gershgorin's theorem.
\qed
\end{proof}

\begin{proposition}\label{prop:6.3}
Let $\Gp$ be a slit graph, $\bGM>0$, $k\in\RR$ and $f:V\to\RR$. Put $\tilde Z(k,f):=\sum_{[n]}\exp\bigl(-\frac{1}{2\bGM}\sum_{b\in E}(\nabla_{b}(n-kf))^{2}\bigr)$, where $n$ runs over $(2\pi\ZZ)^{V}$ modulo the addition of a constant in $2\pi\ZZ$ on each component. Let $u:=G_{N}-v_{\Gp}$ on $\RR^{V}_{0}$ and let $d\mu_{u}$ be the centred Gaussian measure on $\RR^{V}_{0}$ with covariance $u$. Then: (a) $u=\MY^{2}(-\Delta_{\Gp})^{+}(-\Delta_{\Gp}+\MY^{2})^{-1}$ on $\RR^{V}_{0}$, and $u$ is positive definite there. (b) $\tilde Z(k,f)=\mathcal{N}_{0}\sum_{q}e^{ik(q,f)}e^{-\bGM(q,G_{N}q)/2}$, the sum over $q\in\ZZ^{V}$ with zero sum on every component, where $\mathcal{N}_{0}=\prod_{\KK}\mathcal{N}_{\KK}>0$ depends only on $\Gp$ and $\bGM$. An isolated vertex carries only $q=0$ and has $\mathcal{N}_{\KK}=1$. (c) With $\theta:=\sum_{\KK}\theta_{\KK}1_{\KK}$,
\begin{equation}\label{eq:6.SG}
\sum_{q\ \mathrm{neutral}}e^{ik(q,f)}e^{-\bGM(q,G_{N}q)/2}=\int_{[0,2\pi)^{\mathcal{K}}}\prod_{\KK}\frac{d\theta_{\KK}}{2\pi}\int d\mu_{u}(\varphi)\,Z^{Y}\bigl(\varphi+k\bGM^{-1/2}f+\bGM^{-1/2}\theta\bigr),
\end{equation}
where $Z^{Y}(\psi):=\sum_{q\in\ZZ^{V}}e^{i\bGM^{1/2}(q,\psi)}e^{-\bGM(q,v_{\Gp}q)/2}$ is the Yukawa gas on $\Gp$, with no neutrality constraint. (d) $Z^{Y}(\psi)>0$ for real $\psi$, and $Z^{Y}=\prod_{\KK}Z^{Y}_{\KK}(\psi|_{\KK})$. Consequently the coefficients $\rho_{s}$ of the expansion of $\ln Z^{Y}$ in Theorem~\ref{thm:6.5} vanish unless all their points lie in one component.
\end{proposition}

\begin{proof}
On each component $\KK$ we carry out the argument of \cite[(2.4) to (2.12)]{GM82}, which is written there for a measure on $\Zthree$ with charges confined to $\Lambda$ and no zero mode (change (c) of Table~\ref{tab:6.subst}). We root $\KK$ at a point $x_{0}\in\KK$. Poisson summation over $(2\pi\ZZ)^{|\KK|-1}$ then gives the neutral charges $\hat q=(q,-\sum q)$, with $\langle q,A^{-1}q\rangle=\langle\hat q,G_{N}\hat q\rangle$ for the restriction $A$ of $-\Delta_{\KK}$ to $\KK\setminus\{x_{0}\}$. One variable $\theta_{\KK}$ for each component imposes neutrality on every component, for any number of components of any sizes. Finally, $G_{N}=v_{\Gp}+u$, and (a) follows from $(-\Delta_{\Gp}+\MY^{2})^{-1}1_{\KK}=\MY^{-2}1_{\KK}$, which holds by Lemma~\ref{lem:6.2}(ii). The positivity $Z^{Y}>0$ is cited in \cite[p.~21]{GM82}. We obtain it by Poisson summation over $\ZZ^{V}$, which writes $Z^{Y}(\psi)$ as a sum of positive Gaussians in $n-\bGM^{1/2}\psi$, $n\in(2\pi\ZZ)^{V}$, with covariance $\bGM v_{\Gp}$. The factorization and (d) hold because $v_{\Gp}$ and the potentials $v^{\ell}_{\Gp}$, defined as in \cite[(2.5), p.~6, and (4.1), p.~17]{GM81}, vanish across components by Lemma~\ref{lem:6.2}(i), and \cite[\S3]{GM82} builds $\rho_{s}$ from these potentials through the tree formula \cite[(3.7)]{GM81}.
\qed
\end{proof}

\begin{proposition}\label{prop:6.4}
Let $M_{1}\ge\MY$ be the mass of \cite[(3.20)]{GM82}, and on a slit graph $\Gp$ let $v^{1}_{\Gp}:=v^{(M_{1})}_{\Gp}$, with the coincident-site rule \cite[(5.1a)]{GM81}, and $v^{2}_{\Gp}:=v_{\Gp}-v^{1}_{\Gp}$. Then the stability bounds \cite[(4.2)]{GM81} hold on $\Gp$ with $\gamma_{1}=\frac12v^{1}_{\Zthree}(0)$, with the same $\eps_{1}>0$ as on $\Zthree$ \cite[Prop.~8]{GM81}, and with $\eps_{2}=\delta_{2}=0$. Moreover $v^{\ell}_{\Gp}(x,y)\ge0$, and
\begin{equation}\label{eq:6.vtilde}
\tilde v^{1}:=\sup_{y}\sum_{x}v^{1}_{\Gp}(x,y)=M_{1}^{-2},\qquad\tilde v^{2}:=\sup_{y}\sum_{x}v^{2}_{\Gp}(x,y)=\MY^{-2}-M_{1}^{-2}
\end{equation}
exactly, as on $\Zthree$ \cite[(5.3)]{GM81}.
\end{proposition}

\begin{proof}
By Lemma~\ref{lem:6.1} each Yukawa energy on $\Gp$ is at least its $\Zthree$ value for the same configuration. Hence the margins of \cite[Prop.~8]{GM81} transfer. At level one, with the rule (5.1a) and $m(x):=\sum_{i}m_{i}\delta_{x_{i}x}$, $E_{\Gp}=\frac12(m,v^{1}_{\Gp}m)+\frac12\sum_{x_{i}=x_{j},\ \mathrm{sign}\,m_{i}=-\mathrm{sign}\,m_{j}}|m_{i}m_{j}|\,v^{1}_{\Gp}(x_{i},x_{i})\ge E_{\Zthree}$ by Lemma~\ref{lem:6.1}(i), (iii), and $E_{\Zthree}\ge\gamma_{1}+\eps_{1}(-1+\sum_{i}m_{i}^{2})$, coincident positions included. At level two, $\frac12(m,v^{2}_{\Gp}m)\ge0$ by Lemma~\ref{lem:6.2}(iii), that is (4.2) with $\eps_{2}=\delta_{2}=0$. Positivity and \eqref{eq:6.vtilde} follow from Lemma~\ref{lem:6.2}(i) to (iii).
\qed
\end{proof}

\subsection{The Stage 1 theorem}
\label{ssec:6.2}

\begin{theorem}[Stage 1 on slit domains]\label{thm:6.5}
Fix $\kappa_{S},\mu<\infty$ and $C,\delta_{\mathrm{GM}}\in(0,1)$. There are $\lambda_{0}=\lambda_{0}(\kappa_{S},\mu,C,\delta_{\mathrm{GM}})>0$ and, for $\lPV\le\lambda_{0}$, $\bGMo=\bGMo(\lPV,\kappa_{S},\mu,C,\delta_{\mathrm{GM}})$ such that for $\bGM\ge\bGMo$ and every $\MY\in I_{Y}$ the following hold on every slit graph $\Gp$, with constants independent of $\Gp$ and $\MY$. Write $\xi=(m,x)$ with $m\in\ZZ\setminus\{0\}$ and $x\in V$, $-\xi:=(-m,x)$, and $\int d\xi:=\sum_{m}\sum_{x\in V}$.
\par(S1) \emph{Expansion.} For real $\psi$ on $V$, $Z^{Y}(\psi)=\exp(-V_{\mathrm{eff}}(\psi))$ with
\begin{equation}\label{eq:6.Veff}
-V_{\mathrm{eff}}(\psi)=\rho_{0}+\sum_{s\ge1}\frac{1}{s!}\int d\xi_{1}\cdots d\xi_{s}\,\rho_{s}(\xi_{1},\dots,\xi_{s})\,\eps(\xi_{1})\cdots\eps(\xi_{s}),
\end{equation}
where $\eps(\xi):=e^{im\bGM^{1/2}\psi(x)}-1$,
the coefficients $\rho_{s}=\rho_{s}^{\Gp}$ are real and symmetric, $\rho_{0}$ is a constant, and the series converges absolutely, uniformly in real $\psi$. With Proposition~\ref{prop:6.3}, for every $k$ and $f$,
\begin{equation}\label{eq:6.S1}
\tilde Z(k,f)=\mathcal{N}_{0}\int\prod_{\KK}\frac{d\theta_{\KK}}{2\pi}\int d\mu_{u}(\varphi)\,\exp\bigl[-V_{\mathrm{eff}}\bigl(\varphi+k\bGM^{-1/2}f+\bGM^{-1/2}\theta\bigr)\bigr].
\end{equation}
\par(S2) \emph{Coefficient bounds.} For $s\ge1$ and every $a_{1}\subseteq V$,
\begin{multline}\label{eq:6.S2}
\int_{x_{1}\in a_{1}}\!\!d\xi_{1}\int d\xi_{2}\cdots\int d\xi_{s}\,|\rho_{s}(\xi_{1},\dots,\xi_{s})|\,e^{2\kappa_{S}(-1+\sum_{i}|m_{i}|)}\,e^{(1-\delta_{\mathrm{GM}})A\,L_{\Gp}(x_{1},\dots,x_{s})}\\
\le s!\,|a_{1}|\,\frac{\mD^{2}}{\bGM}\,e^{-\mu(s-1)}(1-C)^{-s-10}
\end{multline}
with $A=0$, where $L_{\Gp}(a_{1},\dots,a_{n})$ is the tree length of \cite[(1.11a), (1.11b)]{GM82} computed with the distance $3^{-1/2}d_{\Gp}$ and auxiliary points in $V$, $L_{\Gp}(x_{1},\dots,x_{s}):=L_{\Gp}(\{x_{1}\},\dots,\{x_{s}\})$, and $0\cdot\infty:=0$. Moreover, for all $a_{1},\dots,a_{s}\subseteq V$
\begin{multline}\label{eq:6.S2sets}
\int_{x_{1}\in a_{1}}\!\!d\xi_{1}\cdots\int_{x_{s}\in a_{s}}\!\!d\xi_{s}\,|\rho_{s}(\xi_{1},\dots,\xi_{s})|\,e^{2\kappa_{S}(-1+\sum_{i}|m_{i}|)}\\
\le s!\,|a_{1}|\,\frac{\mD^{2}}{\bGM}\,e^{-(1-\delta_{\mathrm{GM}})A\,L_{\Gp}(a_{1},\dots,a_{s})-\mu(s-1)}(1-C)^{-s-10}.
\end{multline}
\par(S2$^+$) \emph{Charge moments.} If $\mu\ge\ln3$, then for every $a_{1}\subseteq V$
\[
\sum_{s\ge1}\frac{2^{s}}{s!}\int_{x_{1}\in a_{1}}\!\!d\xi_{1}\int d\xi_{2}\cdots\int d\xi_{s}\,|\rho_{s}(\xi_{1},\dots,\xi_{s})|\,e^{2\kappa_{S}\sum_{i}|m_{i}|}
\le|a_{1}|\,\frac{\mD^{2}}{\bGM}\,e^{2\kappa_{S}+\mu}(1-C)^{-9};
\]
in particular $\sum_{s\ge1}\frac{2^{s}}{s!}\int|\rho_{s}|(\sum_{i}|m_{i}|)^{n}\le n!(2\kappa_{S})^{-n}|V|\frac{\mD^{2}}{\bGM}e^{2\kappa_{S}+\mu}(1-C)^{-9}$ for every $n\ge0$.
\par(S3) \emph{Symmetries.} $\rho_{s}(-\xi_{1},\dots,-\xi_{s})=\rho_{s}(\xi_{1},\dots,\xi_{s})$; in particular $\sum_{m}m\,\rho_{1}(m,x)=0$ at every $x\in V$. Moreover $V_{\mathrm{eff}}(\psi+2\pi\bGM^{-1/2}I)=V_{\mathrm{eff}}(\psi)$ for every $I\in\ZZ^{V}$.
\par(S4) \emph{One-body term.} The bounds of Proposition~\ref{prop:6.6} hold.
\par(S5) \emph{Dipole kernel.} $\nu(x_{1},x_{2}):=\bGM\sum_{m_{1},m_{2}}m_{1}m_{2}\,\rho_{2}((m_{1},x_{1}),(m_{2},x_{2}))$ is real and symmetric, and $\|\nu\|_{2\to2}\le\|\nu\|_{0}\le Ce^{-\mu}\mD^{2}$. If $Ce^{-\mu}\le\frac18$, the right side is at most $\frac12\mt^{2}$.
\par With $\MY$ fixed in \eqref{eq:6.mass}, on every output (S2) holds also with $A=\sqrt3\,\kappa_{Y}\MY$ (Proposition~\ref{prop:6.11}), and, if $\delta_{\mathrm{GM}}\le\frac12$, (S5) in the form $\|\nu\|_{\rcov}\le Ce^{-\mu}\mD^{2}$ ($\rcov$ of Theorem~\ref{thm:6.12}).
\end{theorem}

\begin{proof}[Proof of Theorem~\ref{thm:6.5}]
By Lemma~\ref{lem:6.1}, removing bonds raises all quadratic energies, and by Lemma~\ref{lem:6.2} it conserves the mass of the walk. The stability margins of \cite{GM81} and the pinned $L^{1}$ norms of the links are therefore the same as on $\Zthree$, without image factors. Hence the proof of \cite[Prop.~3]{GM82}, together with the results of \cite{GM81} that it uses, runs on $\Gp$ with the changes (a) to (h) of Table~\ref{tab:6.subst}, and its other steps carry over without change, for the same reason. Numbers refer to the DESY preprints identified in the footnote in \S\ref{ssec:1.1}.

\begin{center}
\footnotesize
\begin{tabular}{@{}p{0.1\textwidth}p{0.85\textwidth}@{}}
\hline\noalign{\smallskip}
& other steps\\
\noalign{\smallskip}\hline\noalign{\smallskip}
\cite{GM81} & \S2, (2.5) to (2.25), Prop.~2, (2.26) (its finite pieces are left arbitrary until the estimates are made, p.~6), Prop.~3, (2.30), App.~A (only (2.27) is used), Prop.~4, (3.7), App.~B, (B.4) (stated on p.~32 for any real symmetric pair array $w(ij)$, and used here for the $v^{\ell}_{\Gp}$, $\ell\ge1$), (3.8) to (3.12), (4.3) (with (4.2) for all subconfigurations, by Proposition~\ref{prop:6.4}), Lemma~5, (4.8), App.~C (combinatorial), Prop.~6, Cor.~7, Lemma~9 (through the pinned norms of change (a) only)\\
\noalign{\smallskip}
\cite{GM82} & (3.1) to (3.18) (for real $\psi$, (3.14) uses only $|1+\eps(\xi)|=1$, p.~25), (4.1) to (4.11), Lemma~6 (any metric), (4.14) to (4.22), Prop.~7, Cor.~8 (through $\|v^{\ell}\|_{A}$ only), (4.25) ($\gamma_{1}$ is the constant of $\Zthree$), Prop.~9, (4.26) to (4.30), (4.32) (through the preceding steps, with $L_{\Gp}$ and the rate $(1-\delta_{\mathrm{GM}})A$)\\
\noalign{\smallskip}\hline
\end{tabular}
\end{center}

We read the statement of \cite[Prop.~3]{GM82} with two corrections: $\eps(\xi)=-1+\exp[im\bGM^{1/2}\varphi(x)]$, as in its (3.4), and $2\bGM\mD^{-2}\rho_{1}(\pm1,x)\to1$ at its end. The constants depend only on $(\lPV,\kappa_{S},\mu,C,\delta_{\mathrm{GM}})$ and, on outputs, on those of Theorem~\ref{thm:6.9}.

A complete proof is given in Section~\ref{S4-S:S4.3}. There the tree formula \cite[Prop.~4, (3.7), App.~B]{GM81} and the tree estimate \cite[Lemma~5, App.~C]{GM81} are replaced by the forest formula of Brydges, Kennedy, Abdesselam and Rivasseau \cite{BK87,AR95}, its decoupling property and Cayley's formula; the recursive bound obtained there is slightly stronger than \cite[(4.16)]{GM82}, so that the analogues of \cite[Cor.~8, Prop.~9]{GM82} follow by their printed arguments.

\begin{table}[!htb]
\caption{Changes to \cite[Prop.~3]{GM82} and the results of \cite{GM81} it uses, on a slit graph $\Gp$.}\label{tab:6.subst}
\footnotesize
\begin{tabular}{@{}p{0.135\textwidth}p{0.31\textwidth}p{0.525\textwidth}@{}}
\hline\noalign{\smallskip}
change & steps & use of $\Zthree$ $\to$ on $\Gp$\\
\noalign{\smallskip}\hline\noalign{\smallskip}
(a) pinning & \cite[(4.5a), (4.6), pp.~18 to 20]{GM81}; \cite[(4.12), (4.13), p.~32]{GM82} & translation invariance $\to$ pinned norms in max form, re-rooted sums\\
(b) stability & \cite[Prop.~8, (5.1a) to (5.13), App.~D]{GM81}; \cite[Lemma~5, (3.25)]{GM82} & Fourier analysis $\to$ Proposition~\ref{prop:6.4}: same $\gamma_{1}$, $\eps_{1}$; $\eps_{2}=\delta_{2}=0$\\
(c) Gaussian & \cite[(2.4) to (2.12), pp.~20, 21]{GM82} & measure on $\Zthree$, no zero mode, $Z_{\Lambda}(\phi)>0$ cited $\to$ Proposition~\ref{prop:6.3}: one zero mode per component, $Z^{Y}>0$ proved\\
(d) resolvents & \cite[(3.19) to (3.22)]{GM82} & resolvents of $\Zthree$ $\to$ of $\Gp$; rule (3.19) ($=$ \cite[(5.1a)]{GM81}), same $M_{1}$ (3.20)\\
(e) metric & \cite[(3.23), (3.24)]{GM82} & $3^{-1/2}\|x-y\|_{1}$ $\to$ $\|x-y\|_{\Gp}:=3^{-1/2}d_{\Gp}(x,y)$, in $\|\cdot\|_{A}$ and $L_{\Gp}$\\
(f) link norms & \cite[Lemma~5, (3.26); (4.23), (4.24), p.~34]{GM82} & momentum space $\to$ \eqref{eq:6.vtilde}, Proposition~\ref{prop:6.11}; Debye condition $\lPV^{2}\le C^{3}(1-C)^{8}e^{-2\kappa_{S}-\mu}/24$\\
(g) $\sigma^{R}$ & \cite[(4.33), (4.34), p.~36]{GM82} & $v(0)$ $\to$ $v_{\Gp}(x,x)$ in $\sigma^{R}=\exp[-\bGM m^{2}v(0)/2]$\\
(h) tree length & \cite[(4.30) to (4.32), pp.~35, 36; (4.12), p.~32]{GM82} & $L(a_{1},\dots,a_{s})$ on the right of (4.31) $\to$ $L_{\Gp}(x_{1},\dots,x_{s})$ inside the sum over all positions\\
\noalign{\smallskip}\hline
\end{tabular}
\end{table}

(a) \emph{Pinned norms.} Where \cite[p.~18]{GM81} and \cite[p.~32]{GM82} use translation invariance, the former after a preparatory majorization by functions of differences of the arguments, we take instead, on $\Gp$, the norms in the max form \cite[(4.5b)]{GM81}, $\|\sigma_{\alpha}\|(m_{\alpha}):=\max_{i\in\alpha}\sup_{x\in V}\sum_{x_{\alpha}\in V^{n}:\,x_{i}=x}|\sigma_{\alpha}(\xi_{\alpha})|$, with the weights of \cite[(4.12)]{GM82} inserted in the same way, and $\tilde v^{\ell}:=\sup_{y}\sum_{x}|v^{\ell}_{\Gp}(x,y)|$. We sum the positions as in the trimming of the tree in \cite[p.~18]{GM81}, after re-rooting the tree at the vertex containing the pinned constituent, in order of decreasing tree distance from it. Each summed vertex is then a leaf attached by one link $(j,k)$, and $\sum_{x_{j}}|v^{\ell}_{\Gp}(x_{j},x_{k})|\le\tilde v^{\ell}$ uniformly in $x_{k}$. The $s$-integrals and tree sums are bounded by \cite[Lemma~5]{GM81} independently of the positions. Hence \cite[(4.7), (4.10), (4.13), (4.16), (4.17)]{GM81} hold without change, and $\int d\xi_{\alpha}|\sigma_{\alpha}|\le|V|\,\|\sigma_{\alpha}\|$.

(f) \emph{Link norms.} For $A\ge0$ we put $\|v^{\ell}_{\Gp}\|_{A}:=\sup_{y}\sum_{x}|v^{\ell}_{\Gp}(x,y)|\,e^{A\|x-y\|_{\Gp}}$. At $A=0$ the values are \eqref{eq:6.vtilde}, on all slit graphs. On an output, Proposition~\ref{prop:6.11} gives $\|v_{\Gp}\|_{A},\|v^{2}_{\Gp}\|_{A}\le3\MY^{-2}$ and $\|v^{1}_{\Gp}\|_{A}\le3M_{1}^{-2}$ for $A\le\sqrt3\,\kappa_{Y}\MY$, since $\|x-y\|_{\Gp}=3^{-1/2}d_{\Gp}(x,y)$, the rate of \eqref{eq:6.LW} is nondecreasing in the mass, and $\MY\le M_{1}\le m_{W}$. These bounds replace \cite[(3.26)]{GM82} in \cite[(4.23)]{GM82} and in \cite[(4.24)]{GM82}. In \cite[(4.23)]{GM82}, $3M_{1}^{-2}$ is a power of $\bGM$, absorbed by $e^{\bGM\eps_{1}}$. In \cite[(4.24)]{GM82}, $3\MY^{-2}$ replaces $2(\delta M)^{-2}$ at $A=(1-\delta)M$, so the factor $\frac14\delta^{2}$ of the Debye condition \cite[p.~34]{GM82} becomes $\frac16$. The condition then holds on $I_{Y}$ once $\lPV^{2}\le C^{3}(1-C)^{8}e^{-2\kappa_{S}-\mu}/24$ (\S\ref{S:E1.debye}). This threshold does not depend on $\Gp$, and all other thresholds of the proof are those of $\Zthree$.

(g) \emph{The leading one-body term.} For a single constituent the recursion \cite[(3.12)]{GM82} gives $\sigma^{\ell+1}=\sigma^{\ell}\exp[-\bGM m^{2}v^{\ell}_{\Gp}(x,x)/2]$, hence $\sigma^{R}(m,x)=\exp[-\bGM m^{2}v_{\Gp}(x,x)/2]$ in place of \cite[(4.33)]{GM82}. The argument on \cite[p.~36]{GM82}, through the pinned norm of \cite[Prop.~9]{GM82}, bounds the remainder in \cite[(4.34)]{GM82} by $e^{-\mu}(1-e^{-\mu})^{-2}\mD^{2}\bGM^{-1}(1-C)^{-9}$, uniformly in $x$. With the weights $e^{2\kappa_{S}(|m|-1)}$ the same argument gives this bound up to a constant factor, since we use the pinned norm with weight $e^{(2\kappa_{S}+\mu)\sum_{j}|m_{j}|}$ and $\sum_{j\ge2}|m_{j}|\ge t-1$. Thus, with $A=0$ and uniformly in $x$ and $\Gp$,
\begin{equation}\label{eq:6.lead}
\rho_{1}(m,x)=e^{-\bGM m^{2}v_{\Gp}(x,x)/2}+R(m,x),\qquad\sum_{m}|R(m,x)|\,e^{2\kappa_{S}(|m|-1)}\le Ce^{-\mu}\mD^{2}/\bGM .
\end{equation}
It replaces the final assertion of \cite[Prop.~3]{GM82}, whose error at fixed $\mu$ is of order $e^{-\mu}$.

(h) \emph{The tree length inside the sum.} We state the bound (S2) with the tree length of the points inside the integral, which is what the chain \cite[(4.30) to (4.32)]{GM82} gives. By \cite[(4.30)]{GM82}, $|\rho_{s}(\xi_{1},\dots,\xi_{s})|\le s!\sum_{t\ge s}\binom ts\sum_{[\alpha]:\,C(\alpha)=\{1,\dots,t\}}\int d\xi_{s+1}\cdots d\xi_{t}\,|\sigma^{R}_{\alpha}(\xi_{\alpha})|$. Applied to the sets $\{x_{i}\}$, \cite[(4.9)]{GM82} gives $L_{\Gp}(x_{1},\dots,x_{s})\le\tilde L(x_{1},\dots,x_{t})$ for all $t\ge s$ and $x_{s+1},\dots,x_{t}\in V$, where $\tilde L$ is the least length of a tree on exactly the vertices $x_{1},\dots,x_{t}$ \cite[(4.8)]{GM82}, computed with $3^{-1/2}d_{\Gp}$. In \cite[(4.31)]{GM82} the restriction $x_{i}\in a_{i}$ ($i\ge2$) enters only through $L(a_{1},\dots,a_{s})\le\tilde L(x_{1},\dots,x_{t})$, since the pinned norm \cite[(4.12)]{GM82}, taken in the max form of change (a), sums all positions with the weight $e^{A\tilde L(x_{\alpha})}$. Hence (4.31) holds with $x_{2},\dots,x_{s}$ summed over $V$ and the factor $e^{(1-\delta_{\mathrm{GM}})AL_{\Gp}(x_{1},\dots,x_{s})}$ on the left in place of $e^{-(1-\delta_{\mathrm{GM}})AL(a_{1},\dots,a_{s})}$ on the right. Combining this with (4.32), \cite[Prop.~9]{GM82} and the sum over $x_{1}\in a_{1}$, we obtain \eqref{eq:6.S2}. The bound \eqref{eq:6.S2sets}, which has the form of \cite[(1.18)]{GM82}, follows from it, since $L_{\Gp}(a_{1},\dots,a_{s})\le L_{\Gp}(x_{1},\dots,x_{s})$ whenever $x_{i}\in a_{i}$.

With these changes the proof of \cite[Prop.~3]{GM82} gives (S1) and (S2). Since the Debye condition is the only threshold that involves $\MY$, they hold uniformly in $\MY\in I_{Y}$. By Proposition~\ref{prop:6.3}(d), $\rho_{s}=0$ in (S2) unless $x_{1},\dots,x_{s}$ lie in one component. Together with Proposition~\ref{prop:6.3}, (S1) gives \eqref{eq:6.S1}. For (S3), the activities are equal, the potentials depend on the charges through $m_{i}m_{j}$ and the rule (3.19), both invariant under $m\mapsto-m$, and $\eps(\xi)$ is $2\pi\bGM^{-1/2}$-periodic in $\psi(x)$. (S4) is Proposition~\ref{prop:6.6}. For (S5), $\nu$ is real and symmetric because $\rho_{2}$ is, and it vanishes across components by Proposition~\ref{prop:6.3}(d). Since $|m_{1}m_{2}|\le Ce^{2\kappa_{S}(|m_{1}|+|m_{2}|-1)}$, \eqref{eq:6.S2sets} at $s=2$ with $a_{1}=\{x_{1}\}$ and any $a_{2}\subseteq V$ gives $\sum_{x_{2}\in a_{2}}|\nu(x_{1},x_{2})|\le Ce^{-\mu}\mD^{2}e^{-(1-\delta_{\mathrm{GM}})A\,L_{\Gp}(\{x_{1}\},a_{2})}$. With $A=0$ and $a_{2}=V$ this is $\|\nu\|_{0}\le Ce^{-\mu}\mD^{2}$, at most $\frac18\mD^{2}\le\frac12\mt^{2}$ once $Ce^{-\mu}\le\frac18$. On outputs we take $A=\sqrt3\,\kappa_{Y}\MY$ and, for an integer $k\ge0$, $a_{2}=\{x_{2}:d_{\Gp}(x_{1},x_{2})\ge k\}$, whose tree length with $\{x_{1}\}$ is at least $3^{-1/2}k$. With these choices and $\gamma_{\nu}:=(1-\delta_{\mathrm{GM}})\kappa_{Y}\MY$, the tail sums $T_{k}:=\sum_{x_{2}:\,d_{\Gp}(x_{1},x_{2})\ge k}|\nu(x_{1},x_{2})|$ satisfy $T_{k}\le Ce^{-\mu}\mD^{2}e^{-\gamma_{\nu}k}$. Summation by parts, $e^{rd}=1+\sum_{1\le k\le d}(e^{rk}-e^{r(k-1)})$, gives for $0\le r<\gamma_{\nu}$
\begin{align*}
\sum_{x_{2}}|\nu(x_{1},x_{2})|\,e^{r\,d_{\Gp}(x_{1},x_{2})}&=T_{0}+\sum_{k\ge1}(e^{rk}-e^{r(k-1)})T_{k}\\
&\le\Bigl(1+\frac{1-e^{-r}}{e^{\gamma_{\nu}-r}-1}\Bigr)Ce^{-\mu}\mD^{2}\le\Bigl(1+\frac{r}{\gamma_{\nu}-r}\Bigr)Ce^{-\mu}\mD^{2},
\end{align*}
which is at most $2Ce^{-\mu}\mD^{2}$ for $r\le\gamma_{\nu}/2$. For $\delta_{\mathrm{GM}}\le\frac12$ we have $\gamma_{\nu}\ge\frac12\kappa_{Y}\MY\ge2\kappa_{0}\mt\ge2\rcov$, where the middle inequality follows from the last assertion of Proposition~\ref{prop:6.11}, and hence $\|\nu\|_{\rcov}\le2Ce^{-\mu}\mD^{2}$. This is (S5) on outputs, with the constant $C$ doubled. Finally, (S2$^+$) follows from the pinned-norm bound of \cite[Prop.~9]{GM82} at $\kappa_{S}+\frac12\ln3$, since $\sum_{s=1}^{t}\binom ts2^{s}\le3^{t}\le\prod_{i\le t}3^{|m_{i}|}$, and the moments from $(\sum_{i}|m_{i}|)^{n}\le n!(2\kappa_{S})^{-n}e^{2\kappa_{S}\sum_{i}|m_{i}|}$ (Section~\ref{S4-S:S4.3}).
\qed
\end{proof}

\begin{proposition}[Additive one-body bound]\label{prop:6.6}
Let $\rhoref(m):=\lim_{\Lambda\uparrow\Zthree}\rho^{\Lambda}_{1}(m,0)$, the one-point coefficient of the Yukawa gas on $\Zthree$: $\rho^{\Lambda}_{1}(m,x)$ is the series of \cite[Prop.~3]{GM82} over the configurations in $\Lambda$, whose terms do not depend on $\Lambda$, so the limit exists by dominated convergence, does not depend on the base point, and satisfies \eqref{eq:6.lead} and \eqref{eq:6.S4c} with the same constants (Section~\ref{S4-S:S4.3}). Let $\Phi=\Phi_{\Gp}$. Uniformly in $x\in V$ and $\Gp$,
\begin{align}
\bigl|\rho_{1}(\pm1,x)-\rhoref(1)e^{-\Phi(x)}\bigr|&\le Ce^{-\mu}\rhoref(1),\label{eq:6.S4a}\\
\rho_{1}(\pm1,x)&\le\rhoref(1)\bigl(1+Ce^{-\mu}\bigr),\label{eq:6.S4b}\\
\sum_{|m|\ge2}|\rho_{1}(m,x)|\,m^{2}e^{2\kappa_{S}(|m|-1)}&\le C\mD^{2}/\bGM .\label{eq:6.S4c}
\end{align}
Consequently $|\rho_{1}(\pm1,x)-\rhoref(1)|\le\rhoref(1)\bigl(1-e^{-\Phi(x)}+Ce^{-\mu}\bigr)$.
\end{proposition}

\begin{proof}
Subtracting from \eqref{eq:6.lead} its form on $\Zthree$, $\rhoref(1)=e^{-\bGM v(0)/2}+R^{\infty}$, $|R^{\infty}|\le Ce^{-\mu}\mD^{2}/\bGM$, gives \eqref{eq:6.S4a} and \eqref{eq:6.S4b}. Here we use $e^{-\bGM v_{\Gp}(x,x)/2}=e^{-\bGM v(0)/2}e^{-\Phi(x)}$: the coefficient $\rhoref(1)$ carries the same self-energy factor as $\rho_{1}$ up to the image term $e^{-\Phi}$, so the subtraction involves no error of order $\bGM\MY$. We also use the bound $\Phi\ge0$ of Lemma~\ref{lem:6.1} and $\mD^{2}/\bGM\le2\rhoref(1)(1-2Ce^{-\mu})^{-1}$, which holds because $e^{-\bGM v(0)/2}\ge z$. In \eqref{eq:6.S4c} the leading terms are at most $e^{-2\bGM v(0)}\le e^{-\mu}\mD^{2}/\bGM$, and in the remainders $m^{2}$ is absorbed by the $\mu$-part of the weight $e^{(2\kappa_{S}+\mu)\sum_{j}|m_{j}|}$ of change (g), since $m^{2}e^{-\mu|m|}\le4e^{-2\mu}$ for $|m|\ge2$ and $\mu\ge1$, so that the Debye condition of change (f) suffices at the same $\kappa_{S}$.
\qed
\end{proof}

\begin{remark}\label{rem:6.7}
We state Proposition~\ref{prop:6.6} in additive form only, because the multiplicative form $\rho_{1}(\pm1,x)=\rhoref(1)e^{-\Phi(x)}(1+\Ord(e^{-\mu}))$ fails at a first-layer corner site of a box. There a tight neutral pair, which feels only its dipole image, makes the one-site density exponentially larger in $\bGM$.
\end{remark}

\paragraph{The masses $\mt$ and $\MY$.} For $M\in I_{Y}$ let $\rho^{\mathrm{ref}}_{M}$ be the one-point coefficient of the Yukawa gas on $\Zthree$ with kernel $(-\Delta_{\Zthree}+M^{2})^{-1}$, and $\mt(M)^{2}:=\bGM\sum_{m}m^{2}\rho^{\mathrm{ref}}_{M}(m)$. By \eqref{eq:6.lead} and \eqref{eq:6.S4c} on $\Zthree$, in which the terms with $|m|\ge2$ carry the weight $e^{2\kappa_{S}(|m|-1)}\ge e^{2\kappa_{S}}$, we have $\mt(M)^{2}=2\bGM e^{-\bGM v^{(M)}_{\Zthree}(0)/2}+\Ord(e^{-\mu}+e^{-2\kappa_{S}})\mD^{2}=\mD^{2}(1+\Ord(\bGM M))+\Ord(e^{-\mu}+e^{-2\kappa_{S}})\mD^{2}$, as $v^{(M)}_{\Zthree}(0)=\vCb(0)-\Ord(M)$. Since $\bGM\mD\to0$, $\mt(M)\in[\mD/2,2\mD]$ once $\mu$ and $\kappa_{S}$ exceed a threshold that depends on $C$ through the factor $(1-C)^{-9}$ in the remainder of \eqref{eq:6.lead} \cite[p.~36]{GM82}, and $\bGM\ge\bGMo$. Since $M_{1}$ and the rule (3.19) do not depend on $M$, the mass $M$ appears only in $v^{2}_{M}=v^{(M)}-v^{(M_{1})}$, so each term of the expansion of $\rho^{\mathrm{ref}}_{M}$ is continuous in $M$. The $t$-th term is $\Ord(te^{-\mu(t-1)})$ uniformly in $M\in I_{Y}$ \cite[p.~36]{GM82}. With $M_{-}:=\mD/(2\lPV)$, we have $0\le v^{2}_{M}\le v^{2}_{M_{-}}$ entrywise by the resolvent identity, so at fixed $t$ the terms are dominated by those with $v^{2}_{M_{-}}$ in place of $v^{2}_{M}$, whose bound in \cite{GM82} needs exactly the Debye condition of change (f) at $M=M_{-}$. By dominated convergence, $M\mapsto\mt(M)/\lPV$ is a continuous map of $I_{Y}$ into itself. Its fixed points form a nonempty closed set. We let $\MY$ be the smallest fixed point and put
\begin{equation}\label{eq:6.mass}
\mt:=\mt(\MY)=\lPV\MY\in[\mD/2,2\mD],\qquad\rhoref:=\rho^{\mathrm{ref}}_{\MY},\qquad\mt^{2}=\bGM\sum_{m}m^{2}\rhoref(m).
\end{equation}
The inverse mass $\mt^{-1}$, which like $\rhoref$ is determined by the gas on $\Zthree$ alone, before any construction, is the unit of length of Sections~\ref{sec:3} to~\ref{sec:mincut}, and $\rhoref$ is the reference density of Proposition~\ref{prop:6.6} and of Section~\ref{sec:stage2}.

\subsection{Heat kernel and covariance on outputs of the construction}
\label{ssec:6.3}

We prove the statements of this subsection in Appendix~\ref{app:F}, from (R2) and, for Proposition~\ref{prop:6.10}, from (R5) and (R6), without any Poincar\'e inequality or volume-growth hypothesis. Full proofs of Proposition~\ref{prop:6.11} and Theorem~\ref{thm:6.12} are given in Section~\ref{S:E2}. The \emph{coarse slit graph} of an output is $\Gamma=(V_{c},E_{c})$, $V_{c}:=\Lambda^{\circ}\setminus\XX$, $E_{c}:=E(V_{c})\setminus S$, the graph $\Gp$ of Sections~\ref{sec:3} to~\ref{sec:mincut} (\S\ref{ssec:2.5}). For $W\subseteq V_{c}$, $\partial_{\Gamma}W$ is the set of bonds of $E_{c}$ with exactly one end in $W$. A \emph{coarse cell} is a site of the grid of Definition~\ref{def:3.4}, a box of fine sides in $[r_{0},\bar r_{0}]$, not a cell of Definition~\ref{def:3.12}. In fine units we put
\begin{equation}\label{eq:6.scales}
V_{*}^{f}:=r_{0}^{3}V_{*}/2,\qquad\RHK:=(1-\etaw)c_{I}c_{\rho}L/3,\qquad K_{0}:=(1-\etaw)c_{I}c_{\rho}c_{L}/6\le\mt\RHK\le2K_{0},
\end{equation}
with $c_{I}=2^{2/3}$, $c_{\rho}=1/12$, $V_{*}=(1-\etaw)^{3}\rho_{\lambda}^{3}/2$ (coarse units). The bounds on $\mt\RHK$ follow from Definition~\ref{def:3.1} (\S\ref{S:E1.scales}). Let $p_{t}(x,y)$ be the kernel of $e^{t\Delta_{\Gp}}$, the transition probability of the walk $X_{t}$ that jumps across each bond of $\Gp$ at rate $1$, and let $\tau(x,\rho):=\inf\{t:d_{\Gp}(x,X_{t})>\rho\}$.

\begin{theorem}[Heat kernel at scale $\RHK$]\label{thm:6.9}
Let $\Gp$ be an output, and let $C_{H}$ (a function of $c_{F}$ of Proposition~\ref{prop:6.10}) and $\eps_{1}^{\mathrm{HK}}$ be the absolute constants of \S\ref{S:E1.HK}. There are $C_{\mathrm{mo}},c_{\mathrm{ex}},C_{\mathrm{gs}},c_{\mathrm{gs}},C_{M}>0$, depending only on $C_{H}$, such that for $\bGM\ge\bGMo$, $x,y\in V$ and $d:=d_{\Gp}(x,y)$: (HK0) $p_{t}(x,y)\le\exp(-\alpha d+6(\cosh\alpha-1)t)$ for all $\alpha,t\ge0$; in particular $p_{t}(x,y)\le e^{-d/17}$ for $t\le d$; (HK1) $p_{t}(x,y)\le\max\bigl(C_{H}t^{-3/2},4/V_{*}^{f}\bigr)\le C_{H}\max\bigl(t^{-3/2},\RHK^{-3}\bigr)$ for all $t>0$; (HK1$'$) $\mathbb{E}_{x}d_{\Gp}(x,X_{t})\le C_{\mathrm{mo}}t^{1/2}$ for $0\le t\le\RHK^{2}$; (HK2) $\mathbb{P}_{x}(\tau(x,\rho)<t)\le\sqrt2\exp(-c_{\mathrm{ex}}\min(\rho^{2}/t,\rho))$ for $\rho\ge1$ and $0<t\le\eps_{1}^{\mathrm{HK}}\RHK^{2}$; (HK3) $p_{t}(x,y)\le C_{\mathrm{gs}}t^{-3/2}\exp(-c_{\mathrm{gs}}\min(d^{2}/t,d))$ for $1\le t\le2\eps_{1}^{\mathrm{HK}}\RHK^{2}$; (HK4) $\mathbb{E}_{x}e^{\theta d_{\Gp}(x,X_{s})}\le1+C_{M}\theta s^{1/2}$ for $0<s\le\eps_{1}^{\mathrm{HK}}\RHK^{2}$ and $0\le\theta\le\min(s^{-1/2},c_{\mathrm{ex}}/2)$.
\end{theorem}

\begin{proposition}[Coarse-to-fine transfer]\label{prop:6.10}
Let $\Gp$ be an output, read on the fine lattice. (i) \emph{Isoperimetry.} $|\partial_{\Gp}A|\ge c_{F}|A|^{2/3}$ for every $A\subseteq V$ with $|A|\le V_{*}^{f}$, where the absolute constant $c_{F}$ (\S\ref{S:E1.ctf}) does not depend on $r_{0}$. (ii) \emph{Faber\textendash Krahn, components.} For $\Omega\subseteq V$ with $|\Omega|\le V_{*}^{f}$, $\lambda_{1}^{D}(\Omega)\ge(c_{F}^{2}/12)|\Omega|^{-2/3}$, where $\lambda^{D}_{1}(\Omega)$ is the smallest eigenvalue of $-\Delta_{\Gp}$ on functions vanishing off $\Omega$; and every component of $\Gp$ has more than $2V_{*}^{f}$ vertices. (iii) \emph{Cheeger transfer.} If a set $K$ of coarse cells satisfies $|\partial_{K}W|\ge h_{c}\min(|W|,|K\setminus W|)$ for every $W\subseteq K$, then its blow-up $K^{f}$ satisfies $|\partial_{K^{f}}A|\ge r_{0}^{-1}\min\{\frac12,h_{c}/C_{\mathrm{bl}}\}\min(|A|,|K^{f}\setminus A|)$ for every $A\subseteq K^{f}$, with the absolute constant $C_{\mathrm{bl}}$ of \S\ref{S:E1.ctf}. (iv) \emph{Cell Poincar\'e inequality.} For every fine cell $\Omega$ of a block $Q$ there is a connected set $\KK^{\flat}$ with $\Omega\subseteq\KK^{\flat}\subseteq V\cap\hat Q(Q)$ such that $\sum_{x\in\Omega}|\varphi(x)-\bar\varphi_{\Omega}|^{2}\le C_{W}^{f}k_{\square}^{2}\sum_{b\in E(\KK^{\flat})}(\nabla_{b}\varphi)^{2}$ for every $\varphi:V\to\RR$, with $C_{W}^{f}:=108C_{\mathrm{bl}}^{2}/\chi^{2}$ and $k_{\square}:=r_{0}(k_{P}+3)$; every fine site and every fine bond lies in at most $N_{\mathrm{ov}}\le27(k_{P}/L_{c})^{3}$ of the sets $\KK^{\flat}$, resp.\ $E(\KK^{\flat})$; and $|\Omega|\ge c_{1}L^{3}$. (v) \emph{Diameter.} Under (L8-2) of \S\ref{sec:7.6}, every graph $(\KK^{\flat},E(\KK^{\flat}))$ is connected, of diameter $d_\flat\le2+0.469\,\bGM/\mt=2+0.235L'$ in its own graph distance; in particular every $\KK^{\flat}$ has $d_{\Gp}$-diameter less than $L'$.
\end{proposition}

\begin{proposition}\label{prop:6.11}
Let $\Gp$ be an output, $\kappa_{W}:=\min(1,1/(2C_{M}))$ and $m_{W}:=\min(1,c_{\mathrm{ex}}/2)$. For every mass $0<M\le m_{W}$,
\begin{equation}\label{eq:6.LW}
\sup_{x}\sum_{y}v^{(M)}_{\Gp}(x,y)e^{\theta_{M}d_{\Gp}(x,y)}\le3M^{-2},\qquad\theta_{M}:=\kappa_{W}M\min\bigl(1,(\eps_{1}^{\mathrm{HK}})^{1/2}M\RHK\bigr).
\end{equation}
In particular, if $\MY\le m_{W}$,
\begin{equation}\label{eq:6.CY}
\sup_{x}\sum_{y}v_{\Gp}(x,y)e^{\kappa_{Y}\MY d_{\Gp}(x,y)}\le3\MY^{-2},\qquad\kappa_{Y}:=\kappa_{W}\min\bigl(1,(\eps_{1}^{\mathrm{HK}})^{1/2}K_{0}/\lPV\bigr),
\end{equation}
and the same bound holds for $v^{2}_{\Gp}$ (since $0\le v^{2}_{\Gp}\le v_{\Gp}$ entrywise) and for $v^{1}_{\Gp}$ at mass $M_{1}\le m_{W}$. Moreover $\kappa_{Y}\MY\ge4\kappa_{0}\mt$, where $\kappa_{0}:=\kappa_{W}(\eps_{1}^{\mathrm{HK}})^{1/2}K_{0}$.
\end{proposition}

We use only one property of the constant $\kappa_{W}$: since $C_{M}$ depends only on the absolute constant $C_{H}$ (Theorem~\ref{thm:6.9}), $\kappa_{W}$ does not depend on $\tau$, $K$, $C$ or $D$.

\begin{theorem}[Covariance bounds]\label{thm:6.12}
Let $\Gp$ be an output and $C_{0}:=(\MY^{-2}\Delta_{\Gp}^{2}-\Delta_{\Gp}+\mt^{2})^{-1}$ on $\RR^{V}$. Assume $\lPV\le\frac14$, $\MY\le m_{W}$ and $\mt\le\min(1/68,m_{W}/2)$ (all true for $\bGM\ge\bGMo$). Put $\kappa_{0}':=\min(\kappa_{0}/2,(2c_{\mathrm{gs}})^{1/2})$ and $\rcov:=\min(\kappa_{0},\kappa_{0}')\mt$, and let $K_{1},K_{1}',K_{1}''$ and $K_{3}\le3.5$ be the constants of \S\ref{S:E1.cov}, which depend only on $(c_{F},c_{\rho},c_{L},\lPV)$. Then for all $x,y\in V$: (H1) $C_{0}(x,x)\le K_{1}\MY$ (the zero modes included); (H1$'$) $0\le C_{0}(x,y)\le K_{1}'\MY e^{-\kappa_{0}'\mt d_{\Gp}(x,y)}$; (H2) $\|C_{0}\|_{\kappa_{0}\mt}\le K_{3}\mt^{-2}$; (H$_{C}$) if $\nu$ is symmetric with $\|\nu\|_{\rcov}\le\eps_{\nu}\mt^{2}$ and $2K_{3}\eps_{\nu}\le1$, then $C:=(C_{0}^{-1}+\nu)^{-1}$ exists and is positive definite, $C(x,x)\le2K_{1}\MY$, $|C(x,y)|\le2K_{1}''\MY e^{-\rcov d_{\Gp}(x,y)}$ and $\|C\|_{\rcov}\le2K_{3}\mt^{-2}$; the same bounds hold for every matrix $C_{s}$ with $|C_{s}(x,y)|\le|C(x,y)|$ entrywise and $C_{s}(x,x)\le C(x,x)$, such as the interpolated covariances of Appendix~\ref{app:C}.
\end{theorem}

\subsection{Depletion}
\label{ssec:6.4}

In this subsection we bound the number of depleted sites per cell. The face kernel of Theorem~\ref{thm:6.16} contains all collective effects of the walls. By Chebyshev's inequality, (R9) bounds the number of depleted sites; sparse insulation costs a factor of order $\log\tau$ in this bound, and the estimates of Section~\ref{sec:stage2} allow such a factor. Throughout, $\Lambda'=\Lambda^{\circ}$, $M:=\MY$, and $u_{x}:=v_{\Lambda'}(\cdot,x)$ is the kernel of the box without holes. The \emph{missing pairs} are the $(y,y')$ with $y\in V$, $|y-y'|=1$, and $y'\in\XX$ (the set $\mathfrak{M}_{\XX}$) or $\{y,y'\}\in S$ (the set $\mathfrak{M}_{S}$). Their union is $\mathfrak{M}:=\mathfrak{M}_{\XX}\cup\mathfrak{M}_{S}$. A frozen bond gives one pair at each end (frozen bonds are \emph{two-sided}), and $N_{2}(U):=\#\{(y,y')\in\mathfrak{M}:y\in U\}$. For $x,y\in V$ we put $g_{x}(y):=\sum_{y':\,(y,y')\in\mathfrak{M}}(u_{x}(y)-u_{x}(y'))$.

\begin{lemma}\label{lem:6.13}
For every slit graph $\Gp\subseteq\Lambda'$ and $x\in V$: (a) for $y\in V$,
\[
\bigl((-\Delta_{\Gp}+M^{2})u_{x}|_{V}\bigr)(y)=\delta_{x}(y)-g_{x}(y);
\]
(b) $I_{\Gp}(x)=\sum_{(y,y')\in\mathfrak{M}}v_{\Gp}(x,y)\bigl[u_{x}(y)-u_{x}(y')\bigr]$.
\end{lemma}

\begin{proof}
For (a), at $y\in V$ the operator $-\Delta_{\Lambda'}+M^{2}$ is $-\Delta_{\Gp}+M^{2}$ plus the terms of the pairs of $\mathfrak{M}$ at $y$, and $(-\Delta_{\Lambda'}+M^{2})u_{x}=\delta_{x}$. For (b) we apply $v_{\Gp}$ to (a) at $x$.
\qed
\end{proof}

\begin{lemma}\label{lem:6.14}
There is an absolute constant $A_{0}^{\mathrm{AD}}$ such that on every output, for every $z$ and every $r\ge1$ (fine units), $N_{2}(V\cap B(z,r))\le A_{0}^{\mathrm{AD}}r^{2}+A_{0}^{\mathrm{AD}}r^{3}/(r_{0}\rho_{\lambda})$.
\end{lemma}

\begin{proof}
At the final state Lemma~\ref{lem:4.5} applies to all sets $W\subseteq V_{c}$ of coarse cells, with $\partial_{\Gamma}W$ as the coarse cut $\partial_{\Gp}W$. In \eqref{eq:4.Ra} a frozen bond with one end in $W$ carries $w_{b}-1=1-\etaw$ and one inside $W$ carries $w_{b}\ge2(1-\etaw)$, where $w_{b}=2-\etaw$ is the hull weight. Hence the coarse count $N_{2}^{c}(W)$ of missing pairs with first entry in $W$ is at most $(|\partial_{\Gamma}W|+\lambda|W|)/(1-\etaw)$. Here a frozen bond inside $W$ gives two missing pairs with first entry in $W$, one for each orientation, and its weight pays for both; this reduces the density bound to the minimality of the hull on a ball. We first consider $r\ge r_{0}$, and let $W$ be the set of coarse cells of $V_{c}$ that meet $B(z,r)$. Then $|W|\le C(r/r_{0})^{3}$, and each bond of $\partial_{\Gamma}W$ joins a coarse cell meeting $B(z,r)$ to a coarse cell of $V_{c}$ that does not, so $|\partial_{\Gamma}W|\le C(r/r_{0})^{2}$. Since a coarse missing pair carries fewer than $4r_{0}^{2}$ fine ones, $N_{2}(V\cap B(z,r))\le4r_{0}^{2}\cdot\frac{10}{9}\bigl(C(r/r_{0})^{2}+C(r/r_{0})^{3}/\rho_{\lambda}\bigr)$. For $1\le r<r_{0}$, the fine missing pairs with first entry in $B(z,r)$ cross at most six coarse grid planes, each in at most $Cr^{2}$ bonds, counted with both orientations.
\qed
\end{proof}

\begin{lemma}\label{lem:6.15}
On every output, for all $x,y\in V$, $v_{\Gp}(x,y)\le C_{\mathrm{Gr}}\bigl[(1+|x-y|)^{-1}+\eta_{G}\bigr]$ with $\eta_{G}:=\mt(K_{0}^{-1}+\lPV^{2}K_{0}^{-3})$ and $C_{\mathrm{Gr}}$ depending only on the constants of Theorem~\ref{thm:6.9}; and $v_{\Gp}(x,y)=0$ when $x$ and $y$ lie in different components.
\end{lemma}

\begin{proof}
We split the time integral in $v_{\Gp}=\int_{0}^{\infty}e^{-M^{2}t}p_{t}\,dt$, with $R:=\RHK$. On $[0,1]$ we use $p_{t}\le1$ and (HK0). On $[1,2\eps_{1}^{\mathrm{HK}}R^{2}]$, (HK3), split at $t=d_{\Gp}(x,y)$, gives $C/(1+|x-y|)$. Beyond this interval, (HK1) gives $C(R^{-1}+M^{-2}R^{-3})\le C\eta_{G}$, since $R\ge K_{0}/\mt$ and $M=\mt/\lPV$. Across components $v_{\Gp}=0$ by Lemma~\ref{lem:6.2}(i). The zero mode of a component $\KK$ contributes $M^{-2}/|\KK|\le1/(2M^{2}V_{*}^{f})$, by Proposition~\ref{prop:6.10}(ii), and this lies inside $M^{-2}R^{-3}$.
\qed
\end{proof}

\begin{theorem}[Face-kernel bound]\label{thm:6.16}
On every output, for every $x\in V$, with $r(y):=|x-y|$, $I_{\Gp}(x)\le C_{F}S_{3}(x)+C_{F}\eta_{G}S_{2}(x)$, where $S_{3}(x):=\sum_{(y,y')\in\mathfrak{M}}(1+r(y))^{-3}e^{-\MY r(y)/2}$, $S_{2}(x):=\sum_{(y,y')\in\mathfrak{M}}(1+r(y))^{-2}e^{-\MY r(y)/2}$, $C_{F}:=C_{\mathrm{Gr}}C_{L}'$ with $C_{L}'$ the constant of Lemma~\ref{lem:F.7}, and $S_{2}(x)\le C_{S}\log(2/\MY)$ with $C_{S}:=CA_{0}^{\mathrm{AD}}(1+\lPV/(c_{\rho}c_{L}))$ (\S\ref{S:E1.dep}).
\end{theorem}

\begin{proof}
We take absolute values in Lemma~\ref{lem:6.13}(b), so no cancellation across a sheet is used and the Neumann kernel appears only through its values. For each pair, $0\le v_{\Gp}(x,y)\le C_{\mathrm{Gr}}[(1+r(y))^{-1}+\eta_{G}]$ by Lemma~\ref{lem:6.15}, and $|u_{x}(y)-u_{x}(y')|\le C_{L}'(1+r(y))^{-2}e^{-\MY r(y)/2}$ by (L2) of Lemma~\ref{lem:F.7}. That lemma applies with $M=\MY$, because $|y-y'|=1$, the sides of $\Lambda^{\circ}$ are, by Lemma~\ref{lem:3.3}(c), at least $\ell_{D}\ge\frac12\mt^{-1}\ge\MY^{-1}$, and $\MY\le\frac15$ for $\bGM\ge\bGMo$. A frozen bond appears twice, each time with its own first entry. For the bound on $S_{2}$ we sum Lemma~\ref{lem:6.14} over dyadic shells $2^{k}\le r<2^{k+1}$. Each shell inside $r_{0}\rho_{\lambda}$ gives at most $8A_{0}^{\mathrm{AD}}$, and up to $r\le\MY^{-1}$ these contributions yield the logarithm. For the shells beyond we use $e^{-\MY r/2}$, and the growth term gives at most $CA_{0}^{\mathrm{AD}}/(\MY r_{0}\rho_{\lambda})\le2CA_{0}^{\mathrm{AD}}\lPV/(c_{\rho}c_{L})$ (\S\ref{S:E1.dep}(f)).
\qed
\end{proof}

We put $\eps_{0}:=\theta_{0}/(2\pi^{2}\tau)$, so that $\Phi(x)>\theta_{0}$ if and only if $v_{\Gp}(x,x)-v(0)>\eps_{0}$, and $\eps_{F}:=\eps_{0}/(4C_{F})$. With the constants $L_{g}$, $C_{R}$ of \S\ref{S:E1.dep}, let $\Rdep:=\lceil32A_{0}^{\mathrm{AD}}L_{g}/\eps_{F}\rceil=\lceil C_{R}\tau/\theta_{0}\rceil$, and let (F) be the condition $C_{F}\eta_{G}C_{S}\log(2/\MY)\le\eps_{0}/4$.
The constant $c_{0}:=z\tau\,\bar r_{0}^{\,2}(w_{*}+2|\partial\Delta|)$, maximized over removal cells $\Delta$, comes from the two-sided bound (R9), proved in Proposition~\ref{prop:4.22}. It satisfies $c_{0}<25(3+c_{s})$ (\S\ref{S:B1.c0}).

\begin{theorem}[Depletion per removal cell]\label{thm:6.17}
Let $\Gp$ be an output and assume $\Rdep\le r_{0}\rho_{\lambda}$, $\Rdep\le\ell_{D}$ and (F). Then for every removal cell $Q$, $\#(\Dep\cap Q)\le C(z\theta_{0})^{-1}[c_{0}+C_{\mathrm{box}}+c_{0}^{\mathrm{sp}}\log(C'\tau/\theta_{0})]$, where $c_{0}^{\mathrm{sp}}\le c_{0}$ is the part of the insulation lying in sparse blocks (defined in the proof), $C_{\mathrm{box}}$ is the box-face term, and $C,C'$ are absolute up to $C_{R}$ and $C_{F}$.
\end{theorem}

\begin{proof}
(i) \emph{Reduction to $S_{3}$.} Since $v_{\Gp}(x,x)-v(0)=I_{\Gp}(x)+[v_{\Lambda'}(x,x)-v(0)]$, we have $\Dep\subseteq\{I_{\Gp}>\eps_{0}/2\}\cup\{\Phi_{\Lambda'}>\theta_{0}/2\}$, where $\Phi_{\Lambda'}(x):=\frac12\bGM(v_{\Lambda'}(x,x)-v(0))$. On the first set Theorem~\ref{thm:6.16} and (F) give $S_{3}(x)>\eps_{F}$.

(ii) \emph{Far insulation.} The pairs at distance more than $R:=\Rdep$ from $x$ contribute to $S_{3}(x)$ at most $\sum_{k\ge0}(2^{k}R)^{-3}N_{2}(V\cap B(x,2^{k+1}R))e^{-\MY2^{k}R/2}$. By Lemma~\ref{lem:6.14}, with $R\le r_{0}\rho_{\lambda}$ and $\MY r_{0}\rho_{\lambda}\ge c_{\rho}c_{L}/(2\lPV)$, this is at most $16A_{0}^{\mathrm{AD}}L_{g}/R\le\eps_{F}/2$ (\S\ref{S:E1.dep}(g)). There the shells with $2^{k+1}R\le r_{0}\rho_{\lambda}$ give $16A_{0}^{\mathrm{AD}}/R$, and the others give a logarithm. Hence each $x\in\Dep$ outside the box set has $S_{3}^{\mathrm{near}}(x)>\eps_{F}/2$, where $S_{3}^{\mathrm{near}}$ counts the pairs with first entry within $R$ of $x$.

(iii) \emph{Blocks.} We cover $Q$ by grid cubes $q$ of side $R$. Let $3q$ be the block of $3\times3\times3$ such cubes centred at $q$, put $N_{3q}:=N_{2}(V\cap3q)$, and call $q$ dense if $N_{3q}\ge R^{2}$, sparse otherwise. A dense $q$ has $\#(\Dep\cap q)\le R^{3}\le RN_{3q}$. For a sparse $q$, Chebyshev's inequality gives $\#(\Dep\cap q)\le(2/\eps_{F})\sum_{x\in q}S_{3}^{\mathrm{near}}(x)\le(2/\eps_{F})N_{3q}(4\pi\log(4R)+C_{3})$ plus the box set. Each pair lies in at most $27$ blocks $3q$, and the pairs counted for $q\cap Q\neq\emptyset$ lie within $2R\le2r_{0}\rho_{\lambda}<\ell_{D}$ of $Q$, hence in the $27$ removal cells $Q'$ around $Q$, as the removal cells form a product grid with sides at least $\ell_{D}$. Each of these cells has $N_{2}(Q')\le4r_{0}^{2}(w_{*}+2|\partial\Delta|)\le4c_{0}/(z\tau)$ by (R9), since a coarse pair carries fewer than $4r_{0}^{2}$ fine ones and removed cells contain no site of $V$. Hence $A_{\mathrm{dense}}+A_{\mathrm{sparse}}\le4\cdot27^{2}c_{0}/(z\tau)$, where $A_{\bullet}:=\sum_{q\ \bullet}N_{3q}$. We put $c_{0}^{\mathrm{sp}}:=z\tau A_{\mathrm{sparse}}/(4\cdot27^{2})\le c_{0}$ and substitute $R\le2C_{R}\tau/\theta_{0}$ and $2/\eps_{F}=16\pi^{2}C_{F}\tau/\theta_{0}$.

(iv) \emph{Box faces.} By the image representation of the box kernel (proof of Lemma~\ref{lem:F.7}) and (L1), $v_{\Lambda'}(x,x)-v(0)\le C\sum_{F}d_{F}(x)^{-1}+3C_{\rm im}/\ell_{D}$, the sum over the faces $F$ of $\Lambda'$, $d_{F}(x)$ the layer distance of $x$ from $F$, and the last term from the images other than the $26$ nearest ones, which have a coordinate at distance at least $2n_\mu\ge2\ell_D$ from $x$ ($n_\mu$ the sides of $\Lambda'$). As $\Rdep\le\ell_{D}$ and $C_{R}\ge24\pi^{2}C_{\rm im}$ (\S\ref{S:E1.dep}), $3C_{\rm im}/\ell_{D}\le\theta_{0}/(2\bGM)$. Hence the set $\{\Phi_{\Lambda'}>\theta_{0}/2\}$ lies within distance $C\tau/\theta_{0}$ of $\partial\Lambda'$ and meets a removal cell, of side at most $3\ell_{D}$, in at most $C\ell_{D}^{2}\tau/\theta_{0}\le C_{\mathrm{box}}/(z\theta_{0})$ sites, since $\ell_{D}^{2}\tau\le C\tau/\mt^{2}\le C/z$.
\qed
\end{proof}

\begin{proposition}\label{prop:6.18}
Under the hypotheses of Theorem~\ref{thm:6.17}, for every $\mathcal{U}$-cell $\Delta_{\mathcal{U}}$ of the output, $(\mt^{2}/\bGM)\,\#(\Dep\cap\Delta_{\mathcal{U}})\le \cdep:=A_{0}^{D}+A_{1}^{D}\log(C'\tau/\theta_{0})$, with $A_{0}^{D}:=CN_{\mathrm{cov}}(c_{0}+C_{\mathrm{box}})/\theta_{0}$, $A_{1}^{D}:=CN_{\mathrm{cov}}c_{0}^{\mathrm{sp}}/\theta_{0}$ and $N_{\mathrm{cov}}\le(C_{\mathcal{U}}\mt^{-1}/\ell_{D}+6)^{3}$.
\end{proposition}

\begin{proof}
Steps (i) to (iv) of the proof of Theorem~\ref{thm:6.17} apply to any region $U$ in place of $Q$, the pairs being counted in the union of the blocks $3q$ over the cubes $q$ meeting $U$. By (R7) the cell $\Delta_{\mathcal{U}}$ has diameter at most $C_{\mathcal{U}}\mt^{-1}$. Hence that union lies in a cube of side $C_{\mathcal{U}}\mt^{-1}+4R$, which meets at most $N_{\mathrm{cov}}$ removal cells, each with sides $\ge\ell_{D}\ge R$ and with $N_{2}\le4c_{0}/(z\tau)$. Finally $\mt^{2}/\bGM\le4\mD^{2}/\bGM=8z$. Since a $\mathcal{U}$-cell lies in one component and depletion is counted site by site, the bound does not depend on the number of components.
\qed
\end{proof}

\begin{lemma}\label{lem:6.19}
Let $\theta_{0}=0.01$ and
\[
C_{*}^{\mathrm{dep}}:=\max\{17.8\,C_{R}/(c_{\rho}c_{L}),\,3.6\cdot10^{3}C_{F}C_{S}(K_{0}^{-1}+\lPV^{2}K_{0}^{-3})\}.
\]
If $\tau\ge2$, $2.494\,\tau\ge\ln(C_{*}^{\mathrm{dep}}/\theta_{0})+2.5\ln\tau$ and $\mD/2\le\mt\le2\mD$, then $\Rdep\le r_{0}\rho_{\lambda}$, $\Rdep\le\ell_{D}$ and (F) hold.
\end{lemma}

\begin{proof}
By hypothesis, $C_{*}^{\mathrm{dep}}\tau^{5/2}e^{-2.494\tau}\le\theta_{0}$. With $\mt\le2\mD$, $r_{0}\rho_{\lambda}=c_{\rho}L\ge c_{\rho}c_{L}/(2\mt)$ and $\log(2/\MY)\le\ln(1/\mD)\le2.4944\,\tau$, the three conclusions follow by arithmetic (\S\ref{S:E1.619}).
\qed
\end{proof}

The value $\theta_{0}=0.01$ is dictated by Section~\ref{sec:stage2}, which needs $\eps_{\nu}+\theta'\le1/7$ with $\theta'\ge1-e^{-\theta_{0}}$. The condition of Lemma~\ref{lem:6.19} is a lower bound on $\tau$.
}

\providecommand{\bGM}{\beta_{\mathrm{GM}}}
\providecommand{\mt}{\tilde m}
\providecommand{\MY}{M_Y}
\providecommand{\lPV}{\lambda_{\mathrm{PV}}}
\providecommand{\Lo}{\Lambda^{\circ}}
\providecommand{\Kc}{\mathfrak{K}}
\providecommand{\Df}{\mathfrak{D}}
\providecommand{\Ff}{\mathfrak{F}}
\providecommand{\Xh}{\mathbb{X}}
\providecommand{\EG}{E_{G'}}
\providecommand{\AG}{A_{G'}}
\providecommand{\Floc}{F_1^{\mathrm{loc}}}
\providecommand{\UU}{\mathcal{U}}
\providecommand{\sF}{\sigma_F}
\providecommand{\sV}{\sigma_V}
\providecommand{\Funit}{F_{\mathrm{unit}}}
\providecommand{\Kcol}{K_{\mathrm{col}}}
\providecommand{\Rf}{R_f}
\providecommand{\good}{\operatorname{good}}
\providecommand{\sat}{\operatorname{sat}}
\providecommand{\supp}{\operatorname{supp}}
\providecommand{\lk}{\operatorname{lk}}
\providecommand{\dGp}{d_{G'}}
\providecommand{\one}{\mathbf{1}}
\providecommand{\Z}{\mathbb{Z}}
\providecommand{\R}{\mathbb{R}}
\providecommand{\Glong}{\Gamma_{\mathrm{long}}}
\providecommand{\Nlong}{\mathcal N^{\mathrm{long}}}

\section{Stage 2: the single-scale expansion and the deterministic area law}
\label{sec:stage2}\label{sec:7}

In this section we prove Theorem~\ref{thm:7.61}: for each output $G'=G'(D)$ of the
construction and each rectangle $C$ with $\min(R,T)\ge R_1(\beta)$,
$W_{G'}(C)\le\exp(-\sV[\AG(C)-\eta_SRT]+c_N(R+T))$. We carry out the
single-scale analysis of \cite[Section~6, Appendices~C to E]{GM82}, with the lemmas
of \cite[Section~9]{BF80} that it uses, on $G'$. The structural changes are the following. First, the outer
partition functions are graded by the class of their walls (their periods around
the handles of the insulating set), and the Kirkwood\textendash Salsburg system of
\cite[Appendix~4]{BF80} is solved for all classes at once in a weighted $\ell^1$ algebra.
Second, the handles enter the bounds in a single place: the area weight of a class
comes from one bound at fixed class (Proposition~\ref{prop:7.57}).
In Subsections~\ref{ssec:7.1} to \ref{ssec:7.5} we derive the class-graded polymer expansion
(Corollary~\ref{cor:7.30}), and in Subsections~\ref{ssec:7.6} to \ref{ssec:7.9} we bound it. Propositions~\ref{prop:7.3}, \ref{prop:7.4} and~\ref{prop:7.8} are proved in full with Proposition~\ref{prop:7.25}, in Section~\ref{S4-S:S4.5}.

\setcounter{subsection}{-1}
\subsection{Standing conventions}
\label{ssec:7.0}\label{sec:7.0}

We use the following conventions in Section~\ref{sec:stage2} and Appendices~\ref{app:A} to \ref{app:C}. The \emph{cells}
of Definition~\ref{def:3.12} may be disconnected. Interfaces, walls and tubes are defined on
cells, and polymers, seeds and the Kirkwood\textendash Salsburg system on the $\UU$-cells
of Theorem~\ref{thm:3.20}\,(R7). By $|X|$ we mean the number of $\UU$-cells of $X$, not a Debye
volume, and the constants of Appendix~\ref{app:B} are per $\UU$-cell. We work in fine lattice units,
$a=1$. Sums over sites and bonds are plain sums, each bond counted once.
The collar scale $L'$ is fixed by $\mt L'=\Kcol\bGM$, $\Kcol=2$, where $\bGM=4\pi^2\tau$. The min-cut
area $\AG(C)$ of Definition~\ref{def:2.11} (a support count) and the forced
wall content $A(p)$ of Lemma~\ref{lem:7.46} (an $\ell^1$ quantity) are related by
Theorem~\ref{thm:7.17} and Lemma~\ref{lem:7.59}. Constants $C$, $c$ are absolute or depend
only on the constants fixed before $K$ (Section~\ref{sec:8}), and change from line to
line. We write $\langle\cdot,\cdot\rangle$ and $\|\cdot\|$ for the $\ell^2$ inner product and norm on $V$ or $E$, and
$\|\varphi\|_Y^2:=\sum_{x\in Y}\varphi(x)^2$.

\subsection{Representation}
\label{ssec:7.1}\label{sec:7.1}

Throughout, $G'=(V,E)$ is the fine slit graph of Subsection~\ref{ssec:2.5} for the set
\eqref{eq:2.FF}, with set of components $\mathrm{Comp}(G')\ni\Kc$. From
Subsection~\ref{ssec:7.2} on, $G'$ will be an output of the construction.

\begin{definition}[Operators and the loop on $G'$]\label{def:7.1}
(a) For an oriented bond $b=(x,y)\in E$ and $\varphi:V\to\R$, $\nabla\varphi(b):=\varphi(y)-\varphi(x)$, and
$\Delta:=-\nabla^*\nabla$ is the Neumann Laplacian of $G'$. (b) On $\R^V$,
$u^{-1}:=\MY^{-2}(-\Delta)(-\Delta+\MY^2)$ and $C_0:=(u^{-1}+\mt^2)^{-1}=F(-\Delta)$, $F(t):=(\MY^{-2}t^2+t+\mt^2)^{-1}$.
(c) For an integer $2$-chain $\Xi$ in $\Lo$ with $\partial\Xi=C$ and $b=(x,y)\in E$, $\xi_b$ is the
coefficient in $\Xi$ of the plaquette $p_b$ dual to $b$, oriented as $\partial c_y$ ($c_y$ the
cube of $y$), and $j(b):=2\pi\xi_b$ (the opposite orientation of Subsection~\ref{ssec:2.5}
amounts to $\Xi\mapsto-\Xi$ and leaves \eqref{eq:2.ZG} unchanged, $m\mapsto-m$). With $n=2\pi m$,
\eqref{eq:2.ZG} reads $Z(k,\Xi):=Z_{G'}(k,\Xi)=\sum_{[n]}\exp(-\frac1{2\bGM}\sum_{b\in E}(\nabla n(b)-kj(b))^2)$, $[n]$ running over $(2\pi\Z)^V$ modulo functions constant on each component;
$W_{G'}(C)=Z(1,\Xi)/Z(0,\Xi)$ does not depend on $\Xi$ (Proposition~\ref{prop:2.10}). (d) The
Neumann solid angle $f=f_{G'}$ solves $-\Delta f=\nabla^*j$; $\Rf:=\nabla f-j$ and $\EG:=\|\Rf\|^2$.
\end{definition}

The operator $C_0$ is positive definite, since $u^{-1}\ge0$. As $C_0$ and $u^{-1}$ are both functions
of $-\Delta$ and $\Delta\one_\Kc=0$, we have
\begin{equation}\label{eq:7.F2}
C_0\one_\Kc=\mt^{-2}\one_\Kc\quad(\Kc\in\mathrm{Comp}(G')),\qquad
\mt^2C_0-1=-C_0u^{-1}=-u^{-1}C_0 .
\end{equation}
Since both ends of each bond lie in one component, $\nabla^*j\perp\ker\Delta$. Hence $f$ exists, unique
up to one constant per component, and $\Rf$, $\EG$ do not depend on the choice.

\begin{lemma}\label{lem:7.2}\label{prop:7.15}
Let $\Pi$ be the orthogonal projection of $\ell^2(E)$ onto $\operatorname{ran}\nabla$. Then
$\nabla f=\Pi j$ and $\Rf=-(1-\Pi)j$; hence $\nabla^*\Rf=0$, $\langle\nabla\varphi,\Rf\rangle=0$ for every
$\varphi:V\to\R$, $\langle j,\Rf\rangle=-\EG$, and $\EG=\min_\varphi\|\nabla\varphi-j\|^2$. For every $n:V\to\R$ and
$k\in\Z$, $\sum_{E}(\nabla n-kj)^2=\sum_{E}(\nabla(n-kf))^2+k^2\EG$, so
$Z(k,\Xi)=e^{-k^2\EG/2\bGM}\,\tilde Z(k,\Xi)$ with
$\tilde Z(k,\Xi):=\sum_{[n]}\exp\bigl(-\frac1{2\bGM}\sum_E(\nabla(n-kf))^2\bigr)$.
\end{lemma}

\begin{proof}
The solid angle is the gradient part of the loop current, since $f$ solves the
Euler\textendash Lagrange equation $\nabla^*(\nabla\varphi-j)=0$ of $\|\nabla\varphi-j\|^2$. Hence
$\nabla f=\Pi j$ and $\Rf\in\operatorname{ran}(1-\Pi)=\ker\nabla^*$, and the identities follow. For the
last one we write $\nabla n-kj=\nabla(n-kf)+k\Rf$ and use $\Rf\perp\operatorname{ran}\nabla$.
\end{proof}

\paragraph{The input from Stage 1.}
We use (S1) to (S5) of Theorem~\ref{thm:6.5}, with $\xi=(m,x)$ and $\int d\xi$ as there, $u$
and $d\mu_u$ of Proposition~\ref{prop:6.3} (on $\R^V_0$, where $u$ inverts $u^{-1}$), and $P:=2\pi\bGM^{-1/2}$. By \eqref{eq:6.S1} with
$\theta_\Kc=\bGM^{1/2}c_\Kc$ there is $\mathcal N_0>0$, independent of $k$ and $\Xi$, such that
\begin{equation}\label{eq:7.S1}
\tilde Z(k,\Xi)=\mathcal N_0\int_{[0,P)^{\mathrm{Comp}(G')}}\prod_\Kc\frac{dc_\Kc}{P}
\int d\mu_u(\varphi)\,e^{-V_{\mathrm{eff}}(\varphi+c\mid k\Xi)},\qquad
c:=\sum_\Kc c_\Kc\one_\Kc,
\end{equation}
\begin{equation}\label{eq:7.Veff}
-V_{\mathrm{eff}}(\varphi\mid k\Xi)=\rho_0+\sum_{s\ge1}\frac1{s!}\int d\xi_1\cdots d\xi_s\,
\rho_s(\xi_1,\dots,\xi_s)\,\varepsilon_k(\xi_1)\cdots\varepsilon_k(\xi_s),
\end{equation}
with $\varepsilon_k(\xi):=e^{im(\bGM^{1/2}\varphi(x)+kf(x))}-1$ and the real coefficients $\rho_s=\rho_s^{G'}$ of
(S1), and the series converges absolutely, uniformly in real $\varphi$. By (S3),
$V_{\mathrm{eff}}(\varphi+PI\mid k\Xi)=V_{\mathrm{eff}}(\varphi\mid k\Xi)$ for $I\in\Z^V$. On outputs the dipole kernel $\nu$
of (S5) has $\|\nu\|_{r_{\mathrm{cov}}}\le\varepsilon_\nu\mt^2$, with $2K_3\varepsilon_\nu\le1$. The space $\R^V_0$, the norms $\|\cdot\|_r$, the
image energy $\Phi=\Phi_{G'}\ge0$ and the depleted set $\Df$ are those of the conventions of
Section~\ref{sec:stage1}. The quantities $\rho^{\mathrm{ref}}$, $\mt$ and $\MY=\mt/\lPV$ are as in \eqref{eq:6.mass}, and (S4$'$)
denotes Proposition~\ref{prop:6.6} together with \eqref{eq:6.S4c} for $\rho^{\mathrm{ref}}$ in place of
$\rho_1(\cdot,x)$ (proved in Section~\ref{S4-S:S4.3}).

\begin{proposition}[{\cite[(6.1) to (6.4)]{GM82} on $G'$}]\label{prop:7.3}
Assume (S1) to (S3). (a) $Z(k,\Xi)=e^{-k^2\EG/2\bGM}\tilde Z(k,\Xi)$, with $\tilde Z$ given by
\eqref{eq:7.S1}: one zero-mode variable $c_\Kc\in[0,P)$ per component of $G'$. (b) The
integrand of \eqref{eq:7.S1} is $P$-periodic in each $c_\Kc$; hence (a) does not change
when $f$ is changed by a constant on each component. (c) $\rho_0$ cancels in $W_{G'}(C)$.
\end{proposition}

Lemma~\ref{lem:7.2} and \eqref{eq:7.S1} give part (a) as in \cite[(6.1) to (6.4)]{GM82}, except that there is
one zero mode per component (GM's Gaussian measure is defined on $(a\Z)^3$ and has
none), the loop entering only through the phases $\varepsilon_k$, and that
$e^{-k^2\EG/2\bGM}$ replaces $e^{-k^2(J,v_{Cb}J)/2\beta}$. Part (b) is (S3) with $I=\one_\Kc$. Part (c) holds as in
\cite[(6.5)]{GM82}, because $\rho_0$ gives the same factor in $Z(1,\Xi)$ and $Z(0,\Xi)$. By (c), we set $\rho_0=0$ from now on.

\paragraph{Cells, cell averages and block spins.}
We write $-V_{\mathrm{eff}}(\varphi\mid k\Xi)=\int d\xi\,\rho_1(\xi)\varepsilon_k(\xi)+E'(\varphi)$, where $E'$
collects the terms $s\ge2$ of \eqref{eq:7.Veff}. The cells $\Omega=\Omega_{Q,\hat\Kc}$ of
Definition~\ref{def:3.12} partition $V$, each in one component. For a cell $\Omega$ and
$x\in\Omega$ let $A(x)=A_\Omega$ and $\bar f(x)=\bar f_\Omega$ be the averages of $\varphi$ and $f$ over $\Omega$,
$\delta:=\varphi-A$, $\delta f:=f-\bar f$ and $\tilde\delta:=\delta+k\bGM^{-1/2}\delta f$. Then $\sum_{x\in\Omega}\tilde\delta(x)=0$ for
each cell, and $\bGM^{1/2}\varphi+kf=(\bGM^{1/2}A+k\bar f)+\bGM^{1/2}\tilde\delta$. The block spins are the
elements of $\mathcal H:=\{h:V\to P\Z\text{ constant on every cell}\}$.

\begin{proposition}\label{prop:7.4}
Assume (S3) and $\sum_m m^2|\rho^{\mathrm{ref}}(m)|<\infty$. (a) For every real $\varphi$ on $V$,
$e^{\int d\xi\,\rho_1(\xi)\varepsilon_k(\xi)}=\Theta(\varphi)\,e^{G(\varphi)}$ with
$\Theta(\varphi):=\sum_{h\in\mathcal H}e^{-\frac12\mt^2\|\varphi+k\bGM^{-1/2}f-h\|^2}$, $G=G_1+G_2$, $e^{G_1}=\prod_\Omega r_\Omega(A_\Omega)$, $G_2=G_{2,1}+G_{2,2}+G_{2,3}$,
\begin{equation}\label{eq:7.rOmega}
r_\Omega(A):=\frac{\exp\bigl\{|\Omega|\sum_m\rho^{\mathrm{ref}}(m)\bigl[e^{im(\bGM^{1/2}A+k\bar f_\Omega)}-1\bigr]\bigr\}}
{\sum_{n\in\Z}\exp\bigl\{-\frac12\mt^2|\Omega|\,(A+k\bGM^{-1/2}\bar f_\Omega-nP)^2\bigr\}},
\end{equation}
\begin{align}
G_{2,1}&:=\int d\xi\,\rho^{\mathrm{ref}}(m)\bigl[e^{im\bGM^{1/2}\tilde\delta(x)}-1-im\bGM^{1/2}\tilde\delta(x)+\tfrac12m^2\bGM\tilde\delta(x)^2\bigr],\notag\\
G_{2,2}&:=\int d\xi\,\rho^{\mathrm{ref}}(m)\bigl[e^{im(\bGM^{1/2}A(x)+k\bar f(x))}-1\bigr]\bigl[e^{im\bGM^{1/2}\tilde\delta(x)}-1-im\bGM^{1/2}\tilde\delta(x)\bigr],\label{eq:7.G2}\\
G_{2,3}&:=\int d\xi\,\bigl(\rho_1(m,x)-\rho^{\mathrm{ref}}(m)\bigr)\bigl[e^{im(\bGM^{1/2}\varphi(x)+kf(x))}-1\bigr].\notag
\end{align}
$\Theta$, $r_\Omega$, $G_1$ and the $G_{2,i}$ are real, $\Theta>0$, $r_\Omega>0$, and $e^G$ does not
depend on $h$. The identity holds for every partition of $V$ in place of the cells.
(b) Under (S1) to (S3),
\begin{equation}\label{eq:7.GM612}
\tilde Z(k,\Xi)=\mathcal N_0\sum_{h\in\mathcal H}\int\prod_\Kc\frac{dc_\Kc}{P}\int
d\mu_u(\varphi)\,e^{-\frac12\mt^2\|\varphi+c+k\bGM^{-1/2}f-h\|^2}\,e^{G(\varphi+c)}\,e^{E'(\varphi+c)}.
\end{equation}
\end{proposition}

We obtain part (a) as in \cite[(6.6), (6.8) to (6.11)]{GM82} and part (b) as in \cite[(6.12)]{GM82}, with
cells in place of blocks, $|\Omega|$ in place of $L^3$ and $\rho^{\mathrm{ref}}(m)$ in place of
$\rho_1(m,0)$ in \cite[(6.7), (6.9), (6.10)]{GM82}, since on $G'$ the one-body coefficient
depends on the site. The difference is the term $G_{2,3}$. Also, (6.12) needs a sign
correction. In (b) we move the sum over $h$ out of the integrals by Tonelli's theorem, as the
moduli of the $h$-terms sum to $e^{\int\rho_1\varepsilon_k}|e^{E'}|$.

\paragraph{The deficit term.}
Since $e^{im\bGM^{1/2}h}=1$ for $h\in\mathcal H$, (S3) for $\rho_1$ and $\rho^{\mathrm{ref}}$ gives, for each
$h\in\mathcal H$ and with $u_h:=\bGM^{1/2}(\varphi+k\bGM^{-1/2}f-h)$,
$G_{2,3}=\sum_m\sum_{x\in V}(\rho_1(m,x)-\rho^{\mathrm{ref}}(m))(\cos(m\,u_h(x))-1)$, and $G_{2,3}(Y)$ is the same
sum over $x\in Y$. For a union $Y$ of $\UU$-cells, $N_\Df(Y)$ denotes the number of its $\UU$-cells that
meet the depleted set $\Df=\{\Phi>\theta_0\}$, $\theta_0=0.01$.

\begin{proposition}\label{prop:7.5}
Assume (S3) and (S4$'$), and put
\begin{equation}\label{eq:7.thetaprime}
\theta':=1-e^{-\theta_0}+C\bigl(e^{-\mu}+e^{-2\kappa_S}\bigr).
\end{equation}
Uniformly in real $\varphi$, in $h\in\mathcal H$ and in $G'$: (a)
$|G_{2,3}(V\setminus\Df)|\le\theta'\cdot\frac12\mt^2\|\varphi+k\bGM^{-1/2}f-h\|^2_{V\setminus\Df}$; (b) for every $Y\subseteq V$,
$-C(e^{-\mu}+e^{-2\kappa_S})\frac{\mt^2}{\bGM}\#(\Df\cap Y)\le G_{2,3}(\Df\cap Y)\le C\frac{\mt^2}{\bGM}\#(\Df\cap Y)$; (c) if $G'$ is an
output of the construction, then $e^{-C\cdep N_\Df(Y)}\le e^{G_{2,3}(\Df\cap Y)}\le e^{C\cdep N_\Df(Y)}$ for every
union $Y$ of $\UU$-cells, with $\cdep$ the constant of Proposition~\ref{prop:6.18}.
\end{proposition}

\begin{proof}[Sketch]
Off $\Df$, $\Phi\le\theta_0$ and (S4$'$) give
$\sum_m|\rho_1(m,x)-\rho^{\mathrm{ref}}(m)|m^2\le\theta'\mt^2/\bGM$ (Section~\ref{S:constF}), and
$|\cos(mu)-1|\le\frac12m^2u^2$ gives (a). On $\Df$ the deviation is at most the density,
and its negative part is at most $Ce^{-\mu}\rho^{\mathrm{ref}}(1)$ plus the tails, as $e^{-\Phi}\le1$. This proves (b).
Part (c) follows from (b) and Proposition~\ref{prop:6.18}, which gives $(\mt^2/\bGM)\#(\Df\cap\Delta_\UU)\le\cdep$ for each $\UU$-cell $\Delta_\UU$.
\end{proof}

In Lemma~\ref{lem:7.31} we will move the quadratic part of $G_{2,3}(V\setminus\Df)$ into the
covariance. Throughout Section~\ref{sec:stage2} we assume the \emph{standing
hypothesis}
\begin{equation}\label{eq:7.standing}
\varepsilon_\nu+\theta'\le\tfrac17 ,
\end{equation}
which, by Lemma~\ref{lem:6.19} and Section~\ref{sec:8}, holds for $\theta_0=0.01$, $\mu$ and $\kappa_S$ large, and $\bGM\ge\bGMo$.

\paragraph{Walls.}
For $h\in\mathcal H$ we put $n:=(\bGM^{1/2}/2\pi)h\in\Z^V$, define the \emph{wall field}
$w:=\nabla h-k\bGM^{-1/2}j$ on $E$, and put $N:=(\bGM^{1/2}/2\pi)w$, that is,
$N(b)=n(y)-n(x)-k\xi_b\in\Z$ for $b=(x,y)$. The wall bonds are $\Sigma_E:=\{b\in E:N(b)\ne0\}$, the
wall sites $\Sigma_V$ are their endpoints, and we put $N^\vee:=\sum_{b\in E}N(b)p_b$, with $p_b$ oriented as in
Definition~\ref{def:7.1}\,(c). A wall is where the cell-constant field fails to absorb the jump of the loop
current, which is constant on interfaces because they are aligned with $\Xi$.

\begin{lemma}[Walls]\label{lem:7.6}\label{lem:7.7}
Let $h\in\mathcal H$ and $k\in\Z$. (a) $N(b)\ne0$ only on interface bonds, and $N$ is
constant on each interface $I$; we write $N_I$. (b) Frozen bonds, bonds into $\Xh$ and
bonds leaving $\Lo$ are not in $E$; they carry no wall. (c) $\partial N^\vee=-kC+\partial Y$ for an
integer $2$-chain $Y$ supported in $\Ff=F(\Xh)\cup F_S\cup\partial_2\Lo$.
(d) For every cycle $\gamma$ of $G'$, $\sum_{b\in\gamma}N(b)=-k\,\lk(\gamma,C)$. Moreover $N(h)=N(h')$ if and
only if $h-h'$ is constant on every component of $G'$; so $N$ determines the class $[h]$
of $h$ modulo $\mathcal C_P:=\{\sum_\Kc Pn_\Kc\one_\Kc:n_\Kc\in\Z\}$. (e) \emph{(Relatively closed walls.)} For an
integer field $M:E\to\Z$ with $M^\vee:=\sum_{b\in E}M(b)p_b$, the following are equivalent: (i) $M$
is relatively closed: $\partial M^\vee=\partial Y$ for an integer $2$-chain $Y$ supported in $\Ff$; (ii) $M$
has zero periods: $\sum_{b\in\gamma}M(b)=0$ for every cycle $\gamma$ of $G'$; (iii) $M=\nabla n'$ for some
$n':V\to\Z$.
\end{lemma}

\begin{proof}
(a) We have $\nabla h=0$ on bonds inside a cell, which do not cross $\Xi$ by Lemma~\ref{lem:3.5}\,(A1).
Other bonds join two blocks, and by Theorem~\ref{thm:3.20}\,(R7) an interface lies in one block face, which lies in
$\Xi$ or misses it. (b) is the definition of $E$. (c) Extend $n$
by $0$ to $\Xh$ and let $U:=\sum_{x\in V}n(x)c_x$. A plaquette of a cube of $V$ not dual to a bond of
$E$ is a face of a hole cube, a frozen face or in $\partial_2\Lo$, so $\partial U=\sum_{b\in E}\nabla n(b)p_b+Y_1$
with $\supp Y_1\subseteq\Ff$. Likewise, $\sum_{b\in E}\xi_bp_b=\Xi-Y_2$, with $Y_2$ the part of $\Xi$ in $\Ff$, since $\Xi$ is
interior to $\Lo$ by Lemma~\ref{lem:3.5} and $F(\Xh)$ contains all faces of hole cubes. So
$N^\vee=\partial U-Y_1-k\Xi+kY_2$ and $\partial N^\vee=-kC+\partial(kY_2-Y_1)$. Thus the wall spans $C$ relative to the insulating set and may end on faces of hole cubes, on frozen faces and on $\partial\Lo$. (d) $\sum_{b\in\gamma}\nabla n(b)=0$, and
$\sum_{b\in\gamma}\xi_b$ counts the crossings of $\gamma$ with $\Xi$, that is, $\lk(\gamma,C)$. If $N(h)=N(h')$,
then $\nabla(n-n')=0$. (e) Since the box is acyclic, the relative cycle $M^\vee-Y$ is a boundary
$\partial W$, which a cycle of $G'$ crosses as often inwards as outwards. As plaquettes dual to
bonds of $E$ are not in $\Ff$, this proves (i)$\Rightarrow$(ii). Integrating along paths, we obtain (ii)$\Leftrightarrow$(iii),
and the computation of (c) with $k=0$ gives (iii)$\Rightarrow$(i).
\end{proof}

\paragraph{Handles.} In $\Z^3$ the period condition of Lemma~\ref{lem:7.6}\,(d) is local, and
by Lemma~\ref{lem:7.6}\,(e) walls with zero periods are removed by $h\mapsto h-Pn'$. In $G'$ a cycle $\gamma$
around a handle of the insulating set with $\lk(\gamma,C)=0$ may be crossed once each, with
opposite signs, by two separated wall pieces. We therefore write the shift in the source form of
Proposition~\ref{prop:7.9} and, in Definition~\ref{def:7.27}, grade the outer partition functions
by the class of their walls.

\begin{proposition}\label{prop:7.8}
Assume (S1) to (S3) and (S5). For each $h\in\mathcal H$ let $g=g_h:V\to\R$ be
\emph{equivariant}: $g_{h+c}=g_h+c$ for $c\in\mathcal C_P$. Put $H:=h-k\bGM^{-1/2}f$,
$C^{-1}:=C_0^{-1}+\nu$, $E(\psi):=E'(\psi+g)+\tfrac12\langle\psi,\nu\psi\rangle$, $g_c:=\mt^2C_0H$,
$F_1:=\tfrac12\mt^2\|g-H\|^2+\tfrac12\langle g,u^{-1}g\rangle$ and $F_2:=\langle\psi,C_0^{-1}(g-g_c)\rangle$. Then $C^{-1}\ge\frac12C_0^{-1}>0$, and with $d\mu_C$ the centred Gaussian measure
on $\R^V$ with covariance $C$,
\begin{equation}\label{eq:7.translated}
\tilde Z(k,\Xi)=\sum_{[h]\in\mathcal H/\mathcal C_P}\mathcal N\int d\mu_C(\psi)\,
e^{E(\psi)}\,e^{G(\psi+g)}\,e^{-F_1}\,e^{-F_2},
\end{equation}
where $\mathcal N>0$ depends on $G'$, $\MY$, $\mt$ and $\nu$, and not on $k$, $\Xi$ or
$h$. Each term of \eqref{eq:7.translated} does not depend on the representative of $[h]$.
\end{proposition}

The proof follows \cite[(6.13) to (6.20)]{GM82}, with $\chi_\Lambda\equiv1$ on $V$ and $E$ the $E$ of
\cite[(6.14c)]{GM82}, except that the sum over $h$ is unfolded over
$\mathcal H/\mathcal C_P$ and the mass acts at every site. The integrand of \eqref{eq:7.GM612} is
invariant under $(\varphi+c,h)\mapsto(\varphi+c+c',h+c')$, $c'\in\mathcal C_P$, since each cell lies in one
component, $r_\Omega$ is $P$-periodic, and the phases of $E'$ shift by multiples of $2\pi$.
Hence the sum over $c'$ turns the zero-mode integrals into Lebesgue measure on
$\R^{\mathrm{Comp}(G')}$. As $u^{-1}\one_\Kc=0$, the measure becomes
$e^{-\frac12\langle\hat\varphi,u^{-1}\hat\varphi\rangle}d\hat\varphi$ on $\R^V$, and $C_0=(u^{-1}+\mt^2)^{-1}$ is one operator on
$\R^V$, whereas in \cite[(6.15)]{GM82} the mass term is $\mt^2\chi_\Lambda$. By (S5), $C^{-1}\ge\frac12C_0^{-1}$,
and equivariance with \eqref{eq:7.F2} gives the independence of the representative.

\subsection{The shift and the localized $F_1$}
\label{ssec:7.2}\label{sec:7.2}

From now on $G'$ is an output of the construction and $\bGM\ge\bGMo$, so that $L'\ge16$
and $L/2<L'$. We write $d=\dGp$ for the graph distance of $G'$ ($d(x,y)=\infty$ for sites in
different components) and $N_r(A):=\{x\in V:d(x,A)\le r\}$. By Theorem~\ref{thm:3.20}\,(R7), each cell lies in one
$\UU$-cell and each $\UU$-cell lies in one component. A good tube of
an interface bond $b$ (Subsection~\ref{ssec:7.3}) is a straight path of $G'$ of $L$ sites
with middle bond $b$.

\paragraph{The wall neighbourhood.} The set $\hat\Sigma$ is the union of the $\UU$-cells $\Delta$
with $d(\Delta,\Sigma_V)<L'$, as on \cite[p.~47]{GM82} with $\UU$-cells for unit cubes and $d$ for
GM's distance $3^{-1/2}\|x-y\|_{1}$ of \cite[(1.11a), (3.23)]{GM82}. The \emph{wall components} $S_\beta$ are the unions of the
$\UU$-cells of the connected components of the graph on the $\UU$-cells in $\hat\Sigma$,
two being adjacent when a bond of $G'$ joins them. We put $\Sigma_\beta:=\Sigma_V\cap S_\beta$ and $w_\beta:=w\one_{E(S_\beta)}$, where $E(S_\beta)$
is the set of bonds of $E$ with both ends in $S_\beta$. The
\emph{wall sources} and the \emph{standard cutoff} are $s_\beta:=\MY^{-2}(-\Delta+\MY^2)\nabla^*w_\beta$ and
$\chi_\beta:=\zeta(d(\cdot,\Sigma_\beta))$, with $\zeta=1$ on $[0,L'/2]$, $\zeta(t)=3-4t/L'$ on $[L'/2,3L'/4]$
and $\zeta=0$ on $[3L'/4,\infty)$. The transition set $A_\beta$ is the set of endpoints of bonds
$\{x,y\}\in E$ with $\chi_\beta(x)\ne\chi_\beta(y)$. The shift, which near each wall component subtracts the response $C_0s_\beta$ to the wall source of that component, is
\begin{equation}\label{eq:7.shift}
g:=H-\sum_\beta\chi_\beta C_0s_\beta,\qquad D:=g-H=\sum_\beta D_\beta,\quad D_\beta:=-\chi_\beta C_0s_\beta .
\end{equation}

\begin{proposition}\label{prop:7.9}
For every $h\in\mathcal H$ and $k\in\{0,1\}$: (a) every wall bond lies in $E(S_\beta)$ for
exactly one $\beta$, so $w=\sum_\beta w_\beta$ with disjoint supports; $\{x:d(x,\Sigma_\beta)<L'\}\subseteq S_\beta$;
every interface carrying a wall lies in one $E(S_\beta)$, and the good tubes of its bonds
lie in $S_\beta$; (b) $\chi_\beta=0$ off $S_\beta$, $\chi_\beta=1$ on $\Sigma_\beta$, $d(A_\beta,\Sigma_\beta)\ge L'/2-1\ge7$,
$N_1(A_\beta)\subseteq S_\beta$, and $\chi_\beta$ is $(4/L')$-Lipschitz for $d$; (c) $u^{-1}H=\sum_\beta s_\beta$ and
$\supp s_\beta\subseteq N_1(\Sigma_\beta)$; (d) $g=H$ off $\bigcup_\beta\supp\chi_\beta$, in particular $g+k\bGM^{-1/2}f=h$
on $V\setminus\hat\Sigma$; on $S_\beta$, $g=H-\chi_\beta C_0s_\beta$; and $g$ is equivariant; (e) $F_2':=C_0^{-1}(g-g_c)$
satisfies
\begin{equation}\label{eq:7.F2prime}
F_2'=\sum_\beta F'_{2,\beta},\qquad F'_{2,\beta}:=-[C_0^{-1},\chi_\beta]\,C_0s_\beta,\qquad
\supp F'_{2,\beta}\subseteq N_1(A_\beta)\subseteq S_\beta,
\end{equation}
$C_0^{-1}D_\beta+s_\beta=F'_{2,\beta}$ (as $\chi_\beta=1$ on $\supp s_\beta$), and on $N_1(A_\beta)$ one has $s_\beta=0$ and $F_2'=-C_0^{-1}\chi_\beta C_0s_\beta$; (f) given $S_\beta$, the
functions $s_\beta$, $D|_{S_\beta}$ and $F'_{2,\beta}$ depend on $h$ only through $N|_{E(S_\beta)}$.
\end{proposition}

\begin{proof}
Parts (a), (b), (d) and (f) are support checks. A geodesic of length less than $L'$ to $\Sigma_\beta$
stays in $\hat\Sigma$, hence in one wall component, and good tubes lie within $L/2<L'$ of their
middle bond. The function $\zeta$ is $(4/L')$-Lipschitz, which gives the Lipschitz bound of (b). Since $A_\beta\subseteq\{L'/2-1<d(\cdot,\Sigma_\beta)\le3L'/4+1\}$
and $L'\ge16$, we have $N_1(A_\beta)\subseteq\{d(\cdot,\Sigma_\beta)<L'\}\subseteq S_\beta$. Finally, $\nabla c=0$ for $c\in\mathcal C_P$, and
$w_\beta=2\pi\bGM^{-1/2}N$ on $E(S_\beta)$, whose wall bonds have endpoint set $\Sigma_\beta$.

(c) Inserting $-\Delta f=\nabla^*j$, we find $-\Delta H=\nabla^*\nabla h-k\bGM^{-1/2}\nabla^*j=\nabla^*w$, so $u^{-1}H=\MY^{-2}(-\Delta+\MY^2)\nabla^*w=\sum_\beta s_\beta$ by (a).
That is, on each bond with
$N=0$ the jumps of $h$ across $\Xi$ cancel $j$, and hence $u^{-1}H$ is a pure wall source. Moreover, $\nabla^*w_\beta$ is supported in $\Sigma_\beta$, and $-\Delta+\MY^2$ enlarges supports by at most one step.

(e) By \eqref{eq:7.F2}, $C_0^{-1}g_c=\mt^2H$ and $C_0^{-1}H-\mt^2H=u^{-1}H$, and by (c) we obtain
$F_2'=\sum_\beta(s_\beta-C_0^{-1}\chi_\beta C_0s_\beta)$. A site $x\in N_1(\Sigma_\beta)\setminus\Sigma_\beta$ is adjacent to some
$z\in\Sigma_\beta$, and $\chi_\beta(x)=1$, since otherwise $z\in A_\beta$, against (b). Thus $\chi_\beta=1$ on
$N_1(\Sigma_\beta)\supseteq\supp s_\beta$, and hence $s_\beta-C_0^{-1}\chi_\beta C_0s_\beta=-[C_0^{-1},\chi_\beta]C_0s_\beta$. The kernel
$C_0^{-1}(x,y)=(\MY^{-2}\Delta^2-\Delta+\mt^2)(x,y)$ vanishes unless $d(x,y)\le2$. Hence
$([C_0^{-1},\chi_\beta]v)(x)=\sum_yC_0^{-1}(x,y)(\chi_\beta(y)-\chi_\beta(x))v(y)$ vanishes unless $\chi_\beta$ changes
along a path of at most two bonds from $x$, which puts $x$ in $N_1(A_\beta)\subseteq S_\beta$. Since
$d(A_\beta,\Sigma_\beta)\ge7\ge3$, the sets $N_1(A_\beta)$ and $N_1(\Sigma_\beta)$ are disjoint, so $s_\beta=0$ on $N_1(A_\beta)$, and
there $F_2'=-C_0^{-1}\chi_\beta C_0s_\beta$.
\end{proof}

The construction above uses no extension of $h|_{S_\beta}$ to $V$. A construction through an extension $h^e_\beta$ of $h|_{S_\beta}$ that is
constant on the components of $V\setminus S_\beta$, as in \cite[(6.24)]{GM82}, is not available
on $G'$: such an extension exists only if $w_\beta$ has zero periods on all cycles of
$G'$, and a wall spanning the central hole of a solid torus of $\Xh$ has period $\pm1$
on a meridian of the torus. When an extension exists, \eqref{eq:7.shift} agrees with
\cite[(6.24), (6.25)]{GM82} up to the choice of cutoff and $C_0$ for $C_{00}$.

\begin{proposition}\label{prop:7.10}
For every $h\in\mathcal H$, $k\in\Z$ and $D:V\to\R$, with $g:=H+D$,
\begin{equation}\label{eq:7.exactF1}
2F_1=\mt^2\|D\|^2+\|w+\nabla D\|^2+\MY^{-2}\|\Delta D-\nabla^*w\|^2-k^2\bGM^{-1}\EG .
\end{equation}
\end{proposition}

\begin{proof}
As $u^{-1}=-\Delta+\MY^{-2}\Delta^2$ and $-\Delta=\nabla^*\nabla$, we have $\langle g,u^{-1}g\rangle=\|\nabla g\|^2+\MY^{-2}\|\Delta g\|^2$.
Here the gradient of $g$ differs from the wall field plus $\nabla D$ by a multiple of the
Biot\textendash Savart field, which is orthogonal to gradients and has inner product
$-\EG$ with $j$. Now $\nabla g=\nabla h-k\bGM^{-1/2}(j+\Rf)+\nabla D=(w+\nabla D)-k\bGM^{-1/2}\Rf$ and, by
Lemma~\ref{lem:7.2}, $\langle w+\nabla D,\Rf\rangle=-k\bGM^{-1/2}\langle j,\Rf\rangle=k\bGM^{-1/2}\EG$, so
$\|\nabla g\|^2=\|w+\nabla D\|^2-k^2\bGM^{-1}\EG$, and the perimeter energy separates exactly. Next $\Delta g=\Delta h+k\bGM^{-1/2}\nabla^*j+\Delta D=\Delta D-\nabla^*w$.
Adding $\mt^2\|g-H\|^2=\mt^2\|D\|^2$ gives \eqref{eq:7.exactF1}.
\end{proof}

\begin{corollary}[Localization]\label{cor:7.11}
Let $D$ be the shift \eqref{eq:7.shift}. For each wall component put
\begin{equation}\label{eq:7.F1Sbeta}
2F_1(S_\beta):=\mt^2\|D_\beta\|^2+\|w_\beta+\nabla D_\beta\|^2+\MY^{-2}\|\Delta D_\beta-\nabla^*w_\beta\|^2,
\end{equation}
and for every union $Y$ of wall components and of $\UU$-cells outside $\hat\Sigma$,
$\Floc(Y):=\sum_{\beta:S_\beta\subseteq Y}F_1(S_\beta)$. Then: (a) $F_1=\Floc(V)-\frac12k^2\bGM^{-1}\EG$;
(b) $F_1(S_\beta)\ge0$, and given $S_\beta$ it is a function of $N|_{E(S_\beta)}$; if
$E(S_\beta)$ contains no $\Xi$-bond, it does not depend on $(k,\Xi)$; (c) for every such $Y$,
\begin{equation}\label{eq:7.H6p}
\Floc(Y)\ge\tfrac12\mt^2\sum_{x\in Y}(g-H)^2+\tfrac12\sum_{b\in E(Y)}\bigl|\nabla g+k\bGM^{-1/2}\Rf\bigr|^2;
\end{equation}
(d) $Z(k,\Xi)$ is the right side of \eqref{eq:7.translated} with $e^{-F_1}$ replaced by
$e^{-\Floc(V)}$, and the prefactor $e^{-k^2\EG/2\bGM}$ of Lemma~\ref{lem:7.2} cancels.
\end{corollary}

\begin{proof}
Each term of \eqref{eq:7.exactF1} but the last is a sum of squares of functions
supported in the wall components: $\nabla D_\beta$ is supported on $E(S_\beta)$ and $\Delta D_\beta$ on $S_\beta$, since a bond
leaving $S_\beta$ from $\supp\chi_\beta$ would end in $A_\beta\subseteq S_\beta$. This gives (a). Part (b) is
Proposition~\ref{prop:7.9}\,(f), and off $\Xi$-bonds $N=\nabla n$ does not involve $k$. For (c) we drop
the terms with $\MY^{-2}$, as $\nabla g+k\bGM^{-1/2}\Rf=w+\nabla D$. Part (d) follows from Lemma~\ref{lem:7.2},
Proposition~\ref{prop:7.8} and (a), and since the prefactor cancels, $\EG$ does not enter the perimeter constant.
\end{proof}

\subsection{The wall tension and the spanning area}
\label{ssec:7.3}\label{sec:7.3}

\paragraph{Tubes on the fine lattice.} Let $b=(x,y)$ be a bond of an interface
$I=I(\Omega,\Omega')$, oriented from $\Omega$ to $\Omega'$, with $y=x+\sigma e_\mu$, $\sigma=\pm1$. We put
$\tau_b^-:=\{x-i\sigma e_\mu:0\le i<L/2\}$, $\tau_b^+:=\{y+i\sigma e_\mu:0\le i<L/2\}$ and
$\tau_b:=\tau_b^-\cup\tau_b^+$, a segment of $L$ sites ($L$ is even). The bond $b$ is
\emph{good} if $\tau_b\subseteq V$ and the $L-1$ consecutive bonds of $\tau_b$ lie in $E$.
We write $\good(I)$ for the number of good bonds of $I$. On the fine lattice the face rule gives, by
Theorem~\ref{thm:3.20}\,(R8) and Lemma~\ref{lem:4.17},
$\good(I)\ge c_g'\max(|I|,L^2)$ for each interface, with $c_g':=c_g(1+1/L_c)^{-2}\ge\frac9{16}c_g$.

\begin{lemma}\label{lem:7.12}
Let $z_1,\dots,z_L$ be a path, let $\tilde H=c_-$ on $\{z_1,\dots,z_{L/2}\}$ and
$\tilde H=c_+$ on $\{z_{L/2+1},\dots,z_L\}$, $J:=c_+-c_-$, and let
$m\,(L/2)\le1/\sqrt2$. Then for every $g$,
$\frac12m^2\sum_{i\le L}(g(z_i)-\tilde H(z_i))^2+\frac12\sum_{i<L}(g(z_{i+1})-g(z_i))^2\ge\frac16\,m^2\,(L/2)\,J^2$.
\end{lemma}

\begin{proof}[Sketch]
With $\bar g_\pm$ the averages of $g$ on the two halves, we write
$J=(c_+-\bar g_+)+(\bar g_+-\bar g_-)+(\bar g_--c_-)$ and apply Cauchy\textendash Schwarz to each
term. Heuristically, the mass term supplies the lower bound if $g$ does not follow the jump, and the gradient term
supplies it if $g$ does.
\end{proof}

\begin{lemma}[Tube geometry]\label{lem:7.13}
(a) If $b\in I(\Omega,\Omega')$ is good, then $\tau_b^-\subseteq\Omega$ and $\tau_b^+\subseteq\Omega'$. (b) Good tubes on
a common lattice line are site-disjoint. Hence every site lies in at most $3$ good
tubes and every bond in at most one. (c) No bond of $\tau_b$ other than $b$ crosses $\Xi$,
and $\xi$, $\nabla h$, $N$ are constant on every interface.
\end{lemma}

\begin{proof}[Sketch]
A good tube is a straight path of $G'$ crossing one block face at its middle bond,
and by (G1) of Definition~\ref{def:3.4} blocks are at least $L$ long. The chain $\Xi$ lies in grid planes
by Lemma~\ref{lem:3.5}\,(A1), and for (c) we use Lemma~\ref{lem:7.6}\,(a).
\end{proof}

\begin{theorem}[Wall tension]\label{thm:7.14}
Let $h\in\mathcal H$, $k\in\{0,1\}$, and let $Y$ be a union of wall components and
of $\UU$-cells outside $\hat\Sigma$. With $J_I:=2\pi\bGM^{-1/2}N_I$,
\begin{equation}\label{eq:7.sigmaF}
\Floc(Y)\ge\frac1{18}\,\mt^2\,\frac L2\sum_{I\subseteq Y}\good(I)\,J_I^2\ge2\sF\sum_{b\in E(Y)}N(b)^2,
\qquad\sF:=\frac{\pi^2c_g'}{18}\,\frac{\mt^2L}{\bGM},
\end{equation}
where $I\subseteq Y$ means that both cells of $I$ lie in $Y$. We put $\Funit:=\sF L^2$.
\end{theorem}

\begin{proof}
For a good bond $b$ of an interface $I\subseteq Y$, Lemma~\ref{lem:7.13}\,(a), (c) shows that $\tau_b\subseteq Y$ and, with
$\vartheta_b:=D+J_b\one_{\tau_b^+}$, that $w+\nabla D=\nabla\vartheta_b$ on the bonds of $\tau_b$ and
$g-H=\vartheta_b-J_b\one_{\tau_b^+}$ on $\tau_b$. Hence, as $\mt L/2\le c_L/2\le1/\sqrt2$,
Lemma~\ref{lem:7.12} with $m=\mt$, $c_-=0$ and $c_+=J_b$ bounds the
part of the right side of \eqref{eq:7.H6p} on $\tau_b$ below by $\frac16\mt^2\frac L2J_b^2$.
Summing over good bonds, we count each site at most three times and each bond at
most once, by Lemma~\ref{lem:7.13}\,(b). The second inequality follows from
$\sum_{E(Y)}N^2=\sum_{I\subseteq Y}|I|N_I^2$ and $\good(I)\ge c_g'|I|$.
The constant $\frac1{18}$ is the $\frac16$ of Lemma~\ref{lem:7.12} divided by the overlap
$3$ of good tubes, and the remaining factor $2$ in front of $\sF$ is used in
Subsection~\ref{ssec:7.8}. By comparison, \cite[Lemma~12, (C.9)]{GM82} gives $\frac1{60}$ together with a perimeter
term, which the bound of the theorem does not contain.
\end{proof}

\begin{theorem}[Walls spanning the loop]\label{thm:7.17}
Let $k=1$, let $n:V\to\Z$ be arbitrary, and $N(b):=\nabla n(b)-\xi_b$ on $E$. Then
$\sum_{b\in E}N(b)^2\ge\#\{b\in E:N(b)\ne0\}\ge\AG(C)$.
\end{theorem}

\begin{proof}
With $U$ as in the proof of Lemma~\ref{lem:7.6}\,(c), we correct $\Xi$ by the boundary of the integer field
and obtain the integer $2$-chain $\Sigma:=\Xi-\partial U$
in $\Lo$, with $\partial\Sigma=C$. The coefficient of $p_b$ in $\Sigma$ is $\xi_b-\nabla n(b)=-N(b)$ for
$b\in E$, and all other plaquettes of $\Lo$ are faces of hole cubes, frozen faces or
faces in $\partial_2\Lo$, hence lie in $\Ff$. So the support of this chain outside the free set $\Ff$ is exactly the
set of wall bonds, $\supp\Sigma\setminus\Ff=\{p_b:b\in E,\ N(b)\ne0\}$, and
Definition~\ref{def:2.11}, applied to this $2$-chain spanning $C$, gives $\AG(C)\le\#\{N\ne0\}\le\sum_EN^2$, since $N$ is integer valued.
\end{proof}

\subsection{Smallness of the fluctuation term $F_2$}
\label{ssec:7.4}\label{sec:7.4}

Throughout this subsection we assume $\lPV\le\frac14$ and $\mt\le\MY\le1$, which hold for $\bGM\ge\bGMo$,
and put $R_s:=(-\Delta+s)^{-1}$ for $s>0$ and
\begin{equation}\label{eq:7.partialfrac}
r_\pm:=\tfrac12\MY^2\bigl(1\pm(1-4\lPV^2)^{1/2}\bigr),\qquad
\mathcal G:=C_0\MY^{-2}(-\Delta+\MY^2).
\end{equation}
As $r_++r_-=\MY^2$ and $r_+r_-=\mt^2\MY^2$, partial fractions in $t=-\Delta$ give
$C_0=\varpi(R_{r_-}-R_{r_+})$ and $\mathcal G=c_+R_{r_-}-c_-R_{r_+}$, with $\varpi:=\MY^2/(r_+-r_-)$ and
$c_\pm:=r_\pm/(r_+-r_-)$. Bounds on these quantities, the derivation of the numerical constants of
Lemma~\ref{lem:7.18} and the arithmetic in the proofs of Theorem~\ref{thm:7.19} and
Corollary~\ref{cor:7.21} are given in Section~\ref{S:constF}. For $X$ a union of wall
components and of $\UU$-cells outside $\hat\Sigma$ we put
$F_2(X):=\langle\one_X\psi,F_2'\rangle=\sum_{\beta:S_\beta\subseteq X}\langle\psi,F'_{2,\beta}\rangle$, where the last equality holds by Proposition~\ref{prop:7.9}\,(e).

\begin{lemma}\label{lem:7.18}
(a) \emph{(Combes\textendash Thomas.)} Let $\rho:V\to\R$ satisfy $|\rho(x)-\rho(y)|\le\alpha$ on every bond
of $G'$, and let $\kappa_{\mathrm{CT}}:=s-6(\cosh\alpha-1)>0$. Then $\|e^\rho R_se^{-\rho}\|\le\kappa_{\mathrm{CT}}^{-1}$.
(b) Let $B\subseteq E$ have endpoint set $V_B$, and let $A\subseteq V$ with $d(A,V_B)\ge r\ge0$. Then
$\|\one_A\mathcal G\nabla^*\one_B\|\le8.71\,\mt^{-2}e^{-\mt r/3}$. (c) If $\chi:V\to[0,1]$ is
$\Lambda_\chi$-Lipschitz for $d$, then $\|[C_0^{-1},\chi]\|\le\Lambda_\chi(144\MY^{-2}+6)\le150\,\Lambda_\chi\MY^{-2}$.
\end{lemma}

\begin{proof}
All three are operator norms from the Schur test and the degree bound $6$, so missing
bonds enter only through the degree. (a) The bound follows by the method of \cite[Theorem~1, p.~257]{CT73},
in a lattice form not stated there. With $e^\rho(-\Delta+s)e^{-\rho}=-\Delta+s+\mathfrak B$, pairing the two
orientations of each bond gives $\mathrm{Re}\langle f,\mathfrak Bf\rangle=-\sum_{\text{bonds}}2(\cosh t-1)\mathrm{Re}(\bar f(x)f(y))$,
$t:=\rho(x)-\rho(y)$, which is at least $-6(\cosh\alpha-1)\|f\|^2$. As $-\Delta\ge0$, it follows that
$\mathrm{Re}\langle f,(-\Delta+s+\mathfrak B)f\rangle\ge\kappa_{\mathrm{CT}}\|f\|^2$. The antisymmetric part of
$\mathfrak B$ drops out, so $\rho$ may have kinks, such as those of $d$ at slit edges and hole corners, and
the second-order form matters: in (b), at $s=r_-\le1.08\,\mt^2$ and $\alpha=\mt/3$, the first-order
quantity $s-6(e^\alpha-1)$ is negative. (b) We apply (a) with $\rho:=\frac{\mt}3\min\{d(\cdot,V_B),r\}$ to the two
resolvents of $\mathcal G$, and the Schur test to $e^\rho\nabla^*e^{-\rho}$. (c) We use
$[C_0^{-1},\chi]=\MY^{-2}(\Delta[\Delta,\chi]+[\Delta,\chi]\Delta)-[\Delta,\chi]$.
\end{proof}

\begin{theorem}[{Control of $F_2$ by $F_1$; \cite[Lemma~14]{GM82} on $G'$}]\label{thm:7.19}
For every $h\in\mathcal H$, $k\in\{0,1\}$ and every union $X$ of wall components and of
$\UU$-cells outside $\hat\Sigma$,
$\sum_{x\in X}F_2'(x)^2=\sum_{\beta:S_\beta\subseteq X}\|F'_{2,\beta}\|^2\le\mt^2\,\varepsilon(L')\,\Floc(X)$, where
\begin{equation}\label{eq:7.epsL}
\varepsilon(L'):=C_\varepsilon\,c_g^{-1}\,\lPV^4\,\mt^{-9}\,(\mt L/2)^{-1}(\mt L')^{-2}e^{-\mt L'/3},
\end{equation}
with the absolute constant $C_\varepsilon$ of Section~\ref{S:constF}.
\end{theorem}

\begin{proof}
By Proposition~\ref{prop:7.9}\,(e),
$F'_{2,\beta}=-[C_0^{-1},\chi_\beta]\one_{A'}\mathcal G\nabla^*\one_Bw_\beta$ with $A':=N_1(A_\beta)$, $B:=\supp w_\beta$,
$d(A',V_B)\ge L'/2-2$ and $\Lambda_\chi=4/L'$. The commutator is localized where $\chi_\beta$ changes, at distance at least
$L'/2-1$ from the walls, and at that distance the Combes\textendash Thomas decay dominates the polynomial loss, because $\mt L'\propto\bGM$.
We apply Lemma~\ref{lem:7.18}\,(b), (c), and bound
$\sum_\beta\|w_\beta\|^2$ by Theorem~\ref{thm:7.14} with $\good(I)\ge c_g'|I|$.
\end{proof}

This proof follows \cite[Appendix~D, (D.3) to (D.16)]{GM82}, with the following changes.
\begin{itemize}
\item[(1)] The operator
norms of Lemma~\ref{lem:7.18} replace GM's pointwise bounds \cite[(D.10)]{GM82}, which come from
\cite[Lemma~B.3]{GM82} and fail within distance $2$ of a missing bond. This loses a
factor $\MY^{-4}\mt^{-4}$.
\item[(2)] The source form of Proposition~\ref{prop:7.9}\,(e) replaces
\cite[(D.3) to (D.5)]{GM82}, since $h^e_\beta$ need not exist on $G'$.
\item[(3)]
Theorem~\ref{thm:7.14} replaces \cite[Lemma~12, (C.21)]{GM82}, so the bound has no
perimeter term $P_F$ and is uniform in $\Lo$, $C$, $\Xi$, the hole set $\Xh$, $S$, $h$, $k$
and $X$.
\end{itemize}

\begin{corollary}[Gaussian $F_2$ factors]\label{cor:7.20}
Let $C'$ be positive semidefinite on $\R^V$ with $\|C'\|\le t_C\mt^{-2}$, and let
$d\mu_{C'}$ be the centred Gaussian measure with covariance $C'$. For every
$p\in\R$ and every $X$ as in Theorem~\ref{thm:7.19},
\[
\int d\mu_{C'}(\psi)\,e^{-pF_2(X)}=\exp(\tfrac12p^2\langle\one_XF_2',C'\one_XF_2'\rangle)\le\exp(\tfrac12p^2t_C\,\varepsilon(L')\,\Floc(X)).
\]
\end{corollary}

\begin{proof}
Since $F_2(X)=\langle\psi,\one_XF_2'\rangle$ is linear in $\psi$, its Laplace transform is Gaussian.
The bound then follows from $\langle F,C'F\rangle\le\|C'\|\|F\|^2$ and Theorem~\ref{thm:7.19}. We will apply the corollary with the $C'$ of
Lemma~\ref{lem:7.31}, for which $C'\le\alpha_DC_0\le\frac76\mt^{-2}$, so that $t_C=\frac76<2$, and with block-diagonal
restrictions $\sum_j\one_{Y_j}C_0\one_{Y_j}$, for which $t_C=1$.
\end{proof}

\begin{corollary}[Collar]\label{cor:7.21}
Let $p_2$, $p_3$, $p'$, $q'$ and $c_\lambda$ be the Hölder exponents and the constant of
Subsection~\ref{ssec:7.6}, and $\varepsilon_0:=\min\{(1-1/p_3-c_\lambda)/(2p_2),\,2(p'/q')^2\}$
(values in Section~\ref{S:constF}). There is $\bGMo$ such that $\varepsilon(L')\le\varepsilon_0$ for
$\bGM\ge\bGMo$.
\end{corollary}

\begin{proof}
We have $\mt L'=2\bGM$, while for $\varepsilon(L')\le\varepsilon_0$ we need only
$\mt L'\ge27\log(1/\mt)+O(1)$ (Section~\ref{S:constF}), with $\log(1/\mt)\le2.4944\tau$ and
$\Kcol=2>27\cdot2.4944/(4\pi^2)=1.706$. A collar with $\mt L'$
fixed, as in \cite[p.~47]{GM82}, does not suffice on $G'$.
\end{proof}

\begin{proposition}\label{prop:7.22}
Let $D_E:=C_\UU\mt^{-1}$ bound the Euclidean diameter of $\UU$-cells
(Theorem~\ref{thm:3.20}\,(R7)) and $\eta_{\mathrm{col}}:=(2\sqrt2L+D_E+3)/L'$. For every wall
component $S_\beta$, with $n_I(\beta)$ the number of interfaces of $S_\beta$ carrying a wall:
(a) $F_1(S_\beta)\ge\frac{\pi^2c_g'}9\frac{\mt^2L^3}{\bGM}n_I(\beta)=2\Funit\,n_I(\beta)$;
(b) $S_\beta$ has at most $n_I(\beta)\frac{4\pi}{3c_u}(\mt L')^3(1+\eta_{\mathrm{col}})^3$ $\UU$-cells;
(c) for $\bGM\ge\bGMo$, every $\UU$-cell of $\hat\Sigma$ receives a share at least $\Phi_\UU$ of
$\Floc$, that is, $F_1(S_\beta)\ge\Phi_\UU\cdot\#\{\UU\text{-cells of }S_\beta\}$, where the constant
$\Phi_\UU$ (Section~\ref{S:constF}) satisfies $\Phi_\UU\to\infty$ as $\bGM\to\infty$.
\end{proposition}

\begin{proof}[Sketch]
(a) is Theorem~\ref{thm:7.14} with $Y=S_\beta$, $J_I^2\ge4\pi^2/\bGM$ and $\good(I)\ge c_g'L^2$. (b) Each
$\UU$-cell of $S_\beta$ lies within $d$-distance $L'$ of a wall interface, and by (R7) this interface lies in one
block face and $\UU$-cells are disjoint with at least $c_u\mt^{-3}$ sites. (c) We divide
(a) by (b). Then $\Phi_\UU\to\infty$, because the energy is exponential in $\bGM$, whereas the collar
is polynomial in $\mt L'\propto\bGM$.
\end{proof}

\subsection{The class-graded expansion}
\label{ssec:7.5}\label{sec:7.5}

By Corollary~\ref{cor:7.11}\,(d), $Z(k,\Xi)=\sum_{[h]}\mathcal N\int d\mu_C\,e^{E}Q_h$ with
$Q_h:=e^{G(\psi+g)}e^{-\Floc(V)}e^{-F_2}$. We expand the Gaussian integral at fixed $h$
and then resum the walls, graded by their class.

\begin{definition}[Closures, wall polymers, seeds]\label{def:7.23}
Fix $h\in\mathcal H$. (a) The closure $\operatorname{cl}(S)$ of a union $S$ of $\UU$-cells is $S$
together with the $\UU$-cells joined to $S$ by a bond of $G'$. Closures of wall
components are \emph{linked} if they share a $\UU$-cell; a \emph{wall polymer} $\mathcal S_\gamma(h)$
is the union of a maximal linked family of closures. (b) The wall polymers and the
$\UU$-cells contained in no wall polymer form a partition $\tilde{\mathcal Y}(h)$ of $V$, whose
members are the \emph{elements}. A set is \emph{admissible} (for $h$) if it is a union of
elements; $\sat_h(S)$ is the union of the elements that meet $S$. (c) $X_1$ is the
union of the $\UU$-cells that contain an endpoint of a bond of $G'$ crossing $\Xi$, and
$\Delta_1$ is a fixed $\UU$-cell. The \emph{seeds} are $Y_1(h):=\sat_h(X_1)$ in the expansion of
$Z(1,\Xi)$ and $Y_1(h):=\sat_h(\Delta_1)$ in that of $Z(0,\Xi)$. (d)
$\mathfrak M(h):=\{X\supseteq Y_1(h):X\text{ admissible}\}$. (e) For admissible $Y$,
$G(Y):=\sum_{\Omega\subseteq Y}\ln r_\Omega(A_\Omega)+\sum_iG_{2,i}(Y)$ (the site sums of \eqref{eq:7.G2}
restricted to $Y$) and $Q_h(Y):=e^{G(Y)}e^{-\Floc(Y)}e^{-F_2(Y)}$ ($F_2(Y)$ as in Subsection~\ref{ssec:7.4});
$E(Y)$ and $E(Y,s)$ are defined in Appendix~\ref{app:C}.
\end{definition}

An admissible set is a union of wall components and of $\UU$-cells outside $\hat\Sigma$,
as a $\UU$-cell of $\hat\Sigma$ joined by a bond to $S_\beta$ belongs to $S_\beta$. Hence $\Floc(Y)$ and
$F_2(Y)$ are defined. An element that meets $X_1$ lies in each admissible $X\supseteq X_1$, since the elements partition $V$ and admissible sets are unions of elements.
Therefore $\mathfrak M(h)=\{X\supseteq X_1:X\text{ admissible}\}$, as in the rule of
\cite[p.~50]{GM82}, and with $\Delta_1$ for $k=0$, as on \cite[p.~51]{GM82}.

\begin{lemma}\label{lem:7.24}
Let $Y,Y'$ be disjoint admissible sets. Then
$Q_h(Y\cup Y')=Q_h(Y)Q_h(Y')$; in particular $Q_h=Q_h(X)Q_h(V\setminus X)$ for
admissible $X$. Moreover $Q_h(Y)$, and every term of $E(Y,s)$, are functions of
$\psi|_Y$ and $N|_{E(Y)}$. The same holds with $G^\natural$
of Lemma~\ref{lem:7.31} in place of $G$.
\end{lemma}

\begin{proof}[Sketch]
We follow \cite[(E.4)]{GM82}, with admissible sets in place of unions of unit cubes. The action is
a sum of cell terms ($\ln r_\Omega$), site terms ($G_{2,i}$, the terms of $G^\natural$ and of $E$) and
wall-component terms ($F_1(S_\beta)$, $F'_{2,\beta}$), and an admissible set contains each cell and
wall component entirely or not at all. As $g+k\bGM^{-1/2}f=h+D$ with $h$ cell-constant and $P\Z$-valued, each phase is $e^{im\bGM^{1/2}(\psi+D)}$, and $r_\Omega$ is $P$-periodic. By Proposition~\ref{prop:7.9}\,(f) and Corollary~\ref{cor:7.11}\,(b), $D|_Y$, $\Floc(Y)$ and
$F_2(Y)$ depend on $h$ only through $N|_{E(Y)}$.
\end{proof}

Since $\Floc$ does not split inside a wall polymer, this factorization holds for admissible sets
only, so we saturate the seed, as $X_1$ itself need not be admissible.

\begin{proposition}[{Cluster expansion at fixed $h$; \cite[Lemma~E.1, Proposition~E.2]{GM82} on $G'$}]\label{prop:7.25}\label{prop:C.8}\label{lem:C.1}\label{lem:C.2}\label{lem:C.6}\label{lem:C.7}
Assume (S1) to (S5) and the hypotheses of Lemma~\ref{lem:C.4}, and let $\mathbf C\in\{C,C'\}$
($C'$ of Lemma~\ref{lem:7.31}). Let $U=V$, or $U=V\setminus X_0$ for a union $X_0$ of $\UU$-cells;
let $N$ be a wall field on $E(U)$ (see the paragraph on wall fields before Lemma~\ref{lem:7.wall}, where $E(Y,s)$ and $Q_N$
are defined for wall fields) whose wall polymers have their closures in $U$ (for $U=V$,
$N=N(h)$ with any $h\in\mathcal H$), with $D$, $S_\beta$, $F_1(S_\beta)$, $F'_{2,\beta}$ and the elements defined from $N$
(Proposition~\ref{prop:7.9}\,(f), Corollary~\ref{cor:7.11}\,(b)); and let $Y_1\subseteq U$ be
nonempty and admissible for $N$ on $U$ (it may depend on $N$ and need not be
connected). With $\mathbf C_U:=\one_U\mathbf C\one_U$ on $\R^U$,
\begin{equation}\label{eq:7.clusterexp}
\int d\mu_{\mathbf C_U}(\psi)\,e^{E(U)}Q_N(U)=\sum_{X}\mathcal K_N(X)\int d\mu_{\mathbf C_U}(\psi)\,
e^{E(U\setminus X)}Q_N(U\setminus X),
\end{equation}
where $X$ runs over the admissible subsets of $U$ containing $Y_1$,
$\mathcal K_N(X):=\sum_{\bar y:\,X(\bar y)=X}\mathcal K_{\bar y}$ over the sequences $\bar y=(Y_1,\dots,Y_n)$ of pairwise disjoint
nonempty admissible sets with union $X(\bar y)=X$, and
\begin{equation}\label{eq:7.activity}
\mathcal K_{\bar y}:=\int_{[0,1]^{n-1}}ds\int d\mu^{\bar y}_s(\psi)\,
e^{E^{\bar y}(X(\bar y),s)}\,\kappa(\bar y,s)\,Q_N(X(\bar y)),
\end{equation}
with $d\mu^{\bar y}_s$ (covariance \eqref{eq:C.E3} at $s_n=0$), $E^{\bar y}$ and the operator $\kappa(\bar y,s)$ of
Appendix~\ref{app:C}, built from $\mathbf C_U$.
Every sequence has length at most $e_N+1$, $e_N$ the number of elements of $U$ not
contained in $Y_1$; all sums are finite, all integrals converge absolutely, and
$\int d\mu_{\mathbf C_U}e^{E(Y)}Q_N(Y)\in(0,\infty)$ for every admissible $Y\subseteq U$. $\mathcal K_{\bar y}$ depends only
on $\bar y$, the $X(\bar y)\times X(\bar y)$ block of $\mathbf C$ and $N|_{E(X(\bar y))}$, not on $U$.
\end{proposition}

We give the proof in Appendix~\ref{app:C}.

\paragraph{Admissibility in terms of walls.} For a set $W$ of wall bonds with endpoint
set $V(W)$, $\hat\Sigma(W)$ is the union of the $\UU$-cells $\Delta$ with $d(\Delta,V(W))<L'$. For a
union $X$ of $\UU$-cells, $\partial X$ is the set of bonds of $G'$ joining $X$ to $X^c:=V\setminus X$,
and we put $W_X:=\Sigma_E\cap E(X)$ and $W_{X^c}:=\Sigma_E\cap E(X^c)$. We consider the conditions (in)
$\operatorname{cl}(\hat\Sigma(W_X))\subseteq X$, (out) $\operatorname{cl}(\hat\Sigma(W_{X^c}))\subseteq X^c$, and ($\partial$) $N=0$ on $\partial X$.

\begin{lemma}\label{lem:7.26}
Let $k\in\{0,1\}$, $h\in\mathcal H$, and let $X$ be a union of $\UU$-cells with
$X\supseteq X_1$ if $k=1$ and $X\supseteq\Delta_1$ if $k=0$. Then
$X\in\mathfrak M(h)$ if and only if (in), (out) and ($\partial$) hold. In that case
the elements of $\tilde{\mathcal Y}(h)$ inside $X$ are determined by $X$ and $N|_{E(X)}$,
and those inside $X^c$ by $X$ and $N|_{E(X^c)}$.
\end{lemma}

\begin{proof}[Sketch]
By Proposition~\ref{prop:7.9}\,(a), each wall component, with its closure, belongs to the side that contains its walls.
No element straddles $\partial X$ when $\partial X$ carries no wall, since
under ($\partial$) we have $\hat\Sigma=\hat\Sigma(W_X)\cup\hat\Sigma(W_{X^c})$, no $\UU$-cell of one being adjacent to one of the
other by (in) and (out), and the wall components inside $X$ are the components of
$\hat\Sigma(W_X)$, with $W_X=\supp N|_{E(X)}$.
\end{proof}

At $k=1$, since both ends of each $\Xi$-bond lie in $X_1\subseteq X$, the set $\partial X$ contains no $\Xi$-bond,
and ($\partial$) is equivalent to $h(x)=h(y)$ for all $\{x,y\}\in\partial X$, as at $k=0$. By Lemmas~\ref{lem:7.24}
and \ref{lem:7.26}, for admissible $X\supseteq X_1$ the activity $\mathcal K_h(X)$ with first set
$\sat_h(X_1)$ depends on $h$ only through $N|_{E(X)}$.

\paragraph{Wall fields.} A \emph{wall field} on a union $U$ of $\UU$-cells is a function
$N:E(U)\to\Z$, extended by $0$ to $E$, that vanishes on the bonds inside cells and is
constant on each interface. By Lemma~\ref{lem:7.6}\,(a), the restriction $N(h)|_{E(U)}$ is a
wall field for $h\in\mathcal H$ and $k\in\{0,1\}$. The objects $S_\beta$, $\chi_\beta$, $s_\beta$, $D$,
$F_1(S_\beta)$, $\Floc$, $F'_{2,\beta}$ and $F_2(Y)$ of Subsection~\ref{ssec:7.2} are functions of $w:=2\pi\bGM^{-1/2}N$ alone,
as Proposition~\ref{prop:7.9}\,(f) and Corollary~\ref{cor:7.11}\,(b) show. For a wall field $N$ we define $G(Y)$,
$G^\natural(Y)$ (Lemma~\ref{lem:7.31}), $E(Y)$ and $E(Y,s)$ by their formulas with
$\psi+g+k\bGM^{-1/2}f$ replaced by $\psi+D$ in all phases and all cell averages, and put $Q_N(Y):=e^{G(Y)}e^{-\Floc(Y)}e^{-F_2(Y)}$. Under this replacement,
$A_\Omega+k\bGM^{-1/2}\bar f_\Omega$ becomes $\operatorname{avg}_\Omega(\psi+D)$, and $\tilde\delta$ becomes
$(\psi+D)-\operatorname{avg}_\Omega(\psi+D)$.

\begin{lemma}[Wall form]\label{lem:7.wall}
(a) If $U$ is admissible for $h\in\mathcal H$ ($U=V$, or $U=X^c$ with $X\in\mathfrak M(h)$), $k\in\{0,1\}$ and $N=N(h)|_{E(U)}$, then the elements of $N$ are the elements of $h$ inside $U$, $Q_N(Y)=Q_h(Y)$, and
$E(Y,s)$ is the same in both definitions, for every admissible $Y\subseteq U$.
(b) Definition~\ref{def:7.23} applies, and Lemma~\ref{lem:7.24}, Proposition~\ref{prop:7.25} and
Lemma~\ref{lem:7.26} hold, for every wall field $N$ on $E(U)$ whose wall polymers have their
closures in $U$, with $V$ replaced by $U$ and without the condition on the seed in
Lemma~\ref{lem:7.26}; Theorem~\ref{thm:7.14} holds for every wall field; and
Proposition~\ref{prop:7.5} holds with $\|\varphi+k\bGM^{-1/2}f-h\|$ replaced by $\|\psi+D\|$.
\end{lemma}

\begin{proof}[Sketch]
(a) As in the proof of Lemma~\ref{lem:7.24}, $h$ drops out. (b) Since the action consists of cell, site and wall-component
terms, these proofs use $h$ only
through (a) and the walls. In Theorem~\ref{thm:7.14}, \eqref{eq:7.H6p} becomes
\eqref{eq:7.F1Sbeta} without its last term, and $w=0$ inside cells. The proof of
Proposition~\ref{prop:7.5} is pointwise in $\bGM^{1/2}(\psi+D)(x)$, and Appendix~\ref{app:C} uses Lemma~\ref{lem:C.3}
and the Poincaré inequality on cells, pointwise in $(\psi,D)$ (details in Section~\ref{S:F2}).
\end{proof}

\begin{definition}[Classes and outer partition functions]\label{def:7.27}
(a) The \emph{m3-graph} has the sites of $V$ as vertices, joined by the bonds of $E$ and
by the pairs of sites in a common cell; a closed walk $\gamma$ in it has the length $|\gamma|$,
the sum of the $d_{G'}$-distances of its steps. The \emph{class} of a function $N$ on $E$
(extended by $0$ to the same-cell pairs) is $[N]:\gamma\mapsto\sum_{e\in\gamma}N(e)$ (oriented sum) on
closed m3-walks. (b) A wall field $N$ is \emph{long} if $[N](\gamma)=0$ for every closed
m3-walk $\gamma$ with $|\gamma|<2L'$. $\Nlong(U)$ is the set of long wall fields on $E(U)$, and
$\Glong:=\{[N]:N\in\Nlong(V)\}$; it is a group. (c) For a union $X$ of $\UU$-cells and
$p\in\Glong$,
\begin{equation}\label{eq:7.Zprime}
\mathcal Z_p(X):=\sum_N\mathcal N\int d\mu_C(\psi)\,e^{E(X^c)}\,Q_N(X^c),
\end{equation}
the sum over the $N\in\Nlong(X^c)$ that satisfy (out) and have $[N]=p$, with the
integrand of Lemma~\ref{lem:7.wall}; $\mathcal Z(X):=(\mathcal Z_p(X))_{p\in\Glong}$, $\mathcal Z_p(X):=0$
for $p\notin\Glong$, $Z_p:=\mathcal Z_p(\emptyset)$ and $Z:=Z_0$. (d) For $k\in\{0,1\}$ and a union $X$ of
$\UU$-cells, with $X\supseteq X_1$ if $k=1$, an \emph{inside field} at coupling $k$ is a wall field
$N_X$ on $E(X)$ that satisfies (in) and for which $[(N_X+k\xi)\one_{E(X)}]$ vanishes on the
closed m3-walks in $X$; its \emph{outer class} is $p_k(N_X):=-[(N_X+k\xi)\one_{E(X)}]$.
\end{definition}

Integrating along m3-paths, we see that, for $N$ vanishing on the bonds inside cells (for instance a wall field), $[N]=0$ if and only if $N=\nabla n$ with $n:V\to\Z$ constant
on each cell (on the cycles of $G'$ alone $n$ need not be constant on cells, which
may be disconnected). With Lemmas~\ref{lem:7.6}\,(d) and~\ref{lem:7.wall}\,(a), this characterization gives
$Z=Z(0,\Xi)$: the configurations of $Z_0$ are the walls of the $[h]\in\mathcal H/\mathcal C_P$ at $k=0$.
A long wall field has period $0$ around each plaquette of $G'$, which is a closed walk of
length $4$. For $X\supseteq X_1$, $\xi$ vanishes off $E(X)$, so by Lemma~\ref{lem:7.6}\,(d) an inside
field at $k=1$ has the periods $-\lk(\cdot,C)$ on the cycles of $G'$ in $X$, and $p_1(N_X)(\gamma)=-\lk(\gamma,C)-\sum_{e\in\gamma\cap E(X)}N_X(e)$ on the cycles of $G'$.
In general, such a field is not long.

\begin{lemma}[Long fields]\label{lem:7.45}
(a) A closed m3-walk $\gamma$ with $|\gamma|<2L'$ that contains a wall bond of a wall
component $S_\beta$ has all its sites in $S_\beta$. Hence, if $N\in\Nlong(U)$ has its wall
polymers' closures in $U$ and $Y\subseteq U$ is a union of elements of $N$, then $N\one_{E(Y)}$
and $N\one_{E(U\setminus Y)}$ are long, and their classes add up to $[N]$. (b) Let $h\in\mathcal H$,
$k=1$ and $X\in\mathfrak M(h)$. Then $N(h)|_{E(X)}$ is an inside field at $k=1$, $N(h)|_{E(X^c)}$ is
long and satisfies (out), and $[N(h)|_{E(X^c)}]=p_1(N(h)|_{E(X)})\in\Glong$.
\end{lemma}

\begin{proof}
A closed walk of length less than $2L'$ stays at $d_{G'}$-distance less than $L'$
from each of its sites. (a) So all sites of $\gamma$ lie in $S_\beta$
(Proposition~\ref{prop:7.9}\,(a)), all wall bonds of $\gamma$ lie in $E(S_\beta)$, and $S_\beta$ lies in
one element. Thus the period of $N\one_{E(Y)}$ on $\gamma$ is $0$ or that of $N$, and $N$
vanishes on the bonds between $Y$ and $U\setminus Y$ by Lemma~\ref{lem:7.26}. (b) We put
$N:=N(h)=\nabla n-\xi$, $N_X:=N|_{E(X)}$ and $N':=N|_{E(X^c)}$. By Lemma~\ref{lem:7.26}, $N_X$
satisfies (in), $N'$ satisfies (out), and $N=0$ on $\partial X$. Since $\xi$ vanishes off $E(X)$, we have
$N_X+\xi\one_{E(X)}=\nabla n$ on $E(X)$ and $[N']=-[N_X+\xi\one_{E(X)}]=p_1(N_X)$, as $[N+\xi]=0$. A short
closed walk through a site of $X$ carries no wall of $N'$, by (out), and on one inside
$X^c$ the period of $N'=\nabla n$ is $0$. Hence $N'$ is long.
\end{proof}

For unions $Z\subseteq Y$ of $\UU$-cells and a wall field $N_Y$ on $E(Y)$ of which $Z$ is a union
of elements, we write $\mathcal K^{(Z)}_{N_Y}(Y)$ for the activity of Proposition~\ref{prop:7.25} at $N_Y$ with
first set $Z$, and $\tilde{\mathcal K}(\Delta)$ for that of a wall-free $\UU$-cell $\Delta$ ($n=1$). The finiteness
of $\sum_{N_Y}|\mathcal K^{(Z)}_{N_Y}(Y)|$ and of $\sup_p\mathcal Z_p(X)$ used below is Proposition~\ref{prop:7.51},
proved without Theorem~\ref{thm:7.29} for the activities and outer partition functions built with $(C',G^\natural,\mathfrak c\mathcal N)$
(Lemma~\ref{lem:7.31}\,(f)), the only ones used from Lemma~\ref{lem:7.31} on.

\begin{theorem}[Resummation]\label{thm:7.29}\label{lem:7.49}
Assume the hypotheses of Proposition~\ref{prop:7.25}, and that the sums below converge absolutely; for the objects built with
$(C',G^\natural,\mathfrak c\mathcal N)$ of Lemma~\ref{lem:7.31}\,(f) they do, by Proposition~\ref{prop:7.51}. (a) For $X\supseteq X_1$ and an inside field $N_X$ at $k=1$, let
$\mathcal K_{N_X}(X):=\mathcal K^{(\sat(X_1))}_{N_X}(X)$. Then
\[
Z(1,\Xi)=\sum_{X\supseteq X_1}\sum_{N_X}\mathcal K_{N_X}(X)\,\mathcal Z_{p_1(N_X)}(X),
\]
the inner sum over the inside
fields at $k=1$. (b) \emph{(Kirkwood\textendash Salsburg
identity.)} Let $X_0\subsetneq V$ be a union of
$\UU$-cells and $\Delta\subseteq X_0^c$ a $\UU$-cell. For every $p\in\Glong$,
\begin{equation}\label{eq:7.7.KS}
\mathcal Z_p(X_0)=\sum_{Y\ni\Delta}\ \sum_{N_Y}\mathcal K^{(\sat(\Delta))}_{N_Y}(Y)\,
\mathcal Z_{p-[N_Y]}(X_0\cup Y),
\end{equation}
the sums over the unions $Y\subseteq X_0^c$ of $\UU$-cells containing $\Delta$ and the
$N_Y\in\Nlong(Y)$ satisfying (in), with first set $\sat_{N_Y}(\Delta)$. The term $Y=\Delta$, $N_Y=0$ is $\tilde{\mathcal K}(\Delta)\mathcal Z_p(X_0\cup\Delta)$. (c) \emph{(Prescribed
walls.)} Let $R$ be a family of disjoint unions $\mathcal S$ of
$\UU$-cells, each with a long wall field $N_{\mathcal S}$ on $E(\mathcal S)$ of which it is a wall
polymer, $\sigma:=(\bigcup R,\sum_{\mathcal S}N_{\mathcal S})$, and for $r\in\Glong$ let $Z_r(R)$ be the part of
$Z_r$ from the configurations $N$ in which every $\mathcal S\in R$, with $N|_{E(\mathcal S)}=N_{\mathcal S}$,
is an element. Then
$Z_r(R)=\sum_{Y\supseteq\bigcup R}\sum_{N_Y}\mathcal K^{(\bigcup R)}_{N_Y}(Y)\,\mathcal Z_{r-[N_Y]}(Y)$, the sum over
the $N_Y\in\Nlong(Y)$ satisfying (in) with $N_Y|_{E(\bigcup R)}=\sum_{\mathcal S}N_{\mathcal S}$ of which $\bigcup R$
is a union of elements.
\end{theorem}

\begin{proof}
(a) By Corollary~\ref{cor:7.11}\,(d) and Proposition~\ref{prop:7.25} with $U=V$ and first set
$\sat_h(X_1)$, $Z(1,\Xi)=\sum_{[h]}\mathcal N\sum_{X\in\mathfrak M(h)}\mathcal K_h(X)\int d\mu_C\,e^{E(X^c)}Q_h(X^c)$.
Fix $X\supseteq X_1$. The map $[h]\mapsto(N(h)|_{E(X)},N(h)|_{E(X^c)})$ is a bijection from
$\{[h]:X\in\mathfrak M(h)\}$ onto the pairs $(N_X,N')$ of an inside field at $k=1$ and a
configuration $N'$ of $\mathcal Z_{p_1(N_X)}(X)$. It maps into this set by Lemma~\ref{lem:7.45}\,(b), and it is
injective by Lemma~\ref{lem:7.6}\,(d). It is onto, since $N:=N_X+N'$ has $[N+\xi]=[(N_X+\xi)\one_{E(X)}]+[N']=0$,
hence $N=N(Pn)$ with $n$ cell-constant, and (in), (out), ($\partial$) give $X\in\mathfrak M(Pn)$
(Lemma~\ref{lem:7.26}). The activity depends only on $N_X$ (Lemmas~\ref{lem:7.24},
\ref{lem:7.26}). Since both ends of each $\Xi$-bond lie in $X_1\subseteq X$, the outer integral
depends only on $N'$ and is the summand of \eqref{eq:7.Zprime} (Lemma~\ref{lem:7.wall}\,(a)).
The outer integrals are $\ge0$ (Lemma~\ref{lem:7.44}), and the absolute convergence allows
the sums to be exchanged. Only inside fields with $p_1(N_X)\in\Glong$ occur, by Lemma~\ref{lem:7.45}\,(b),
and for them $\sum_{N_X}|\mathcal K_{N_X}(X)|<\infty$ (Proposition~\ref{prop:7.51}, for the objects built with $(C',G^\natural,\mathfrak c\mathcal N)$). (b), (c) In (b) we fix a configuration $N$ of $\mathcal Z_p(X_0)$
and expand on $U=X_0^c$ with first set $\sat_N(\Delta)$ by Proposition~\ref{prop:7.25} in wall form (Lemma~\ref{lem:7.wall}\,(b)).
In (c) we fix a configuration $N$ of $Z_r(R)$ and expand in the same way on $U=V$ with first set $\bigcup R$, possibly
disconnected. The index sets agree: if each $\mathcal S\in R$ is an element of $N$, then $\bigcup R$ is admissible;
conversely, every site within distance $L'$ of a wall of $N_{\mathcal S}$ lies in a wall component of
$N_{\mathcal S}$, hence in $\mathcal S$ (Proposition~\ref{prop:7.9}\,(a)), so the wall components of
$\sum_{\mathcal S}N_{\mathcal S}$ are those of the $N_{\mathcal S}$, closures from different $\mathcal S$ are
disjoint, and the elements of $N_Y$ in $\bigcup R$ are the $\mathcal S$ (Lemma~\ref{lem:7.26}). By
Lemma~\ref{lem:7.26} in wall form and Lemma~\ref{lem:7.45}\,(a), the map
$N\mapsto(Y,N\one_{E(Y)},N\one_{E(U\setminus Y)})$ is a bijection onto the triples of $Y$, $N_Y$ as stated
and a configuration of $\mathcal Z_{p-[N_Y]}(X_0\cup Y)$ in (b), and of $\mathcal Z_{r-[N_Y]}(Y)$ in (c). The summand
factorizes (Lemma~\ref{lem:7.24}), and the absolute convergence allows the sums to be
exchanged. For $Y=\Delta$ and $N_Y=0$, the $\UU$-cell $\Delta$ is a wall-free element.
\end{proof}

At fixed walls the expansion factorizes over the elements, admissibility splits
into conditions on the two sides (Lemma~\ref{lem:7.26}), and the class of the walls
is additive. This class is therefore all that the inside constrains on the outer side.
In (b) the removal of $Y$ moves the class by $[N_Y]$, and each pair $(X,p)$, $X\ne\emptyset$,
enters the identity for $(X\setminus\Delta,p)$ through the wall-free term $Y=\Delta$, the form of
\cite[(A4.3), (A4.5)]{BF80}. The class cannot be dropped: removing the polymer of one
of the two walls in the paragraph headed \emph{Handles} (Section~\ref{sec:7.1}) leaves outer walls of nonzero class, which
are not the walls of a height on the complement.

\begin{corollary}[Loop expansion]\label{cor:7.30}
Under the hypotheses of Theorem~\ref{thm:7.29}, in the finite box,
\begin{equation}\label{eq:7.loopexp}
W_{G'}(C)=\sum_{X\supseteq X_1}\ \sum_{N_X}\mathcal K_{N_X}(X)\,\frac{\mathcal Z_{p_1(N_X)}(X)}Z,
\end{equation}
the inner sum over the inside fields at $k=1$, and $\mathcal Z_{p_1(N_X)}(X)\ge0$.
\end{corollary}

\begin{proof}
We divide Theorem~\ref{thm:7.29}\,(a) by $Z=Z(0,\Xi)>0$ and use Lemma~\ref{lem:7.44}. Here
$\mathcal Z_p(X)/Z$ is a ratio of two absolutely convergent
sums in the finite box, which Theorem~\ref{thm:7.52} only bounds.
\end{proof}
}
%
\providecommand{\bGM}{\beta_{\mathrm{GM}}}
\providecommand{\mt}{\tilde m}
\providecommand{\Gp}{G'}
\providecommand{\Floc}{F_1^{\mathrm{loc}}}
\providecommand{\Dep}{\mathfrak D}
\providecommand{\cU}{\mathcal U}
\providecommand{\cK}{\mathcal K}
\providecommand{\fK}{\mathfrak K}
\providecommand{\fc}{\mathfrak c}
\providecommand{\fn}{\mathfrak n}
\providecommand{\cN}{\mathcal N}
\providecommand{\cP}{\mathcal P}
\providecommand{\bD}{b_D}
\providecommand{\lPV}{\lambda_{\mathrm{PV}}}
\providecommand{\MY}{M_Y}
\providecommand{\sF}{\sigma_F}
\providecommand{\sV}{\sigma_V}
\providecommand{\epsE}{\varepsilon_E}
\providecommand{\Funit}{F_{\mathrm{unit}}}
\providecommand{\Kcell}{K_{\mathrm{cell}}}
\providecommand{\Kcol}{K_{\mathrm{col}}}
\providecommand{\cnear}{c_{\mathrm{near}}}
\providecommand{\CR}{C_{\mathrm{BS}}}
\providecommand{\kbox}{k_{\square}}
\providecommand{\rbox}{r_{\square}}
\providecommand{\Kflat}{\mathfrak K^{\flat}}
\providecommand{\Ncal}{\mathcal N}
\providecommand{\eps}{\varepsilon}
\providecommand{\dist}{\operatorname{dist}}
\providecommand{\supp}{\operatorname{supp}}
\providecommand{\sat}{\operatorname{sat}}
\providecommand{\avg}{\operatorname{avg}}
\providecommand{\Ad}{\mathrm{A}}

\subsection{Bounds on the activities}
\label{sec:7.6}\label{ssec:7.6}

In this subsection we bound the activities of Section~\ref{sec:7.5} on $\Gp$ by
\cite[Section~9]{BF80} and \cite[Section~7]{GM82}. On $\Gp$ the argument
differs from these references in the following points. There is a
one-body deficit, and Lemma~\ref{lem:7.31} moves its quadratic part into the covariance.
Lemmas~\ref{lem:7.34} and~\ref{lem:7.35} avoid both an extension of the wall field across $\Xi$ and a
Poincar\'e inequality across a slit.
Finally, the covariance rate is small and independent of $\tau$ (Proposition~\ref{prop:7.38}).
Lemma~\ref{lem:7.31} is proved in full in Section~\ref{S4-S:S4.6}, which uses Section~\ref{S4-S:S4.4} for its one-body bounds and for Theorem~\ref{thm:7.29} and Corollary~\ref{cor:7.30} with $(C',G^\natural,\fc\cN)$. Lemmas~\ref{lem:7.39} and~\ref{lem:7.47} and Propositions~\ref{prop:7.long}, \ref{prop:7.51} and~\ref{prop:7.56} are proved in full in Section~\ref{S4-S:S4.8}.

Throughout, $X$ is a polymer (a union of elements of $\tilde Y(h)$ in the sense of
Definition~\ref{def:7.23}), $|X|$ is its number of $\cU$-cells,
$\vartheta:=\psi+D$, where $D:=g-H$ is the field in the phases, and
$\tilde\delta:=P_X\vartheta$, where $P_X\vartheta:=\vartheta-\avg_\Omega\vartheta$
on each cell $\Omega\subseteq X$ and $P_X\vartheta:=0$ off $X$. We put
$C_u:=2+6/c_g$, so that $\#\Delta<C_u\mt^{-3}$ for each $\cU$-cell $\Delta$
by (R7), and $\|A\|_r:=\sup_x\sum_y|A(x,y)|e^{r\,d_{\Gp}(x,y)}$ for a matrix
$A$ on $V$ and $r\ge0$. Constants $C,C_n$ depend only on those of
Theorem~\ref{thm:6.5} and Proposition~\ref{prop:6.6}.

\paragraph{The deficit decomposition.}
By Proposition~\ref{prop:7.4} the inhomogeneity is confined to the term
$G_{2,3}$ of \eqref{eq:7.G2}, which by (S3) is
$\sum_{m\neq0}\sum_{x\in V}(\rho_1(m,x)-\rho^{\mathrm{ref}}(m))(\cos(m\bGM^{1/2}\vartheta(x))-1)$.
Let $G_{2,3}^{\Dep}$ be its part with $x\in\Dep$, and put
\begin{align}
\mu_{\Dep}(x)&:=\bGM\sum_{m\neq0}m^2\bigl(\rho_1(m,x)-\rho^{\mathrm{ref}}(m)\bigr)
\,1_{V\setminus\Dep}(x),\label{eq:7.6.mu}\\
G_{2,3}^{\natural}&:=\sum_{m\neq0}\sum_{x\in V\setminus\Dep}
\bigl(\rho_1(m,x)-\rho^{\mathrm{ref}}(m)\bigr)
\bigl(\cos(m\bGM^{1/2}\vartheta(x))-1+\tfrac12m^2\bGM\psi(x)^2\bigr).\notag
\end{align}
Then, identically in $\psi$,
\begin{equation}\label{eq:7.6.split}
G_{2,3}=G_{2,3}^{\Dep}-\tfrac12\langle\psi,\mu_{\Dep}\psi\rangle+G_{2,3}^{\natural}.
\end{equation}
We write $G_{2,3}^{\Dep}(Y)$, $G_{2,3}^\natural(Y)$ for the sums over $x\in Y$. Finally, let
$G^\natural:=G_1+G_{2,1}+G_{2,2}+G_{2,3}^{\Dep}+G_{2,3}^{\natural}$ and
$C':=(C_0^{-1}+\nu+\mu_{\Dep})^{-1}$, where the
operator $\mu_{\Dep}$ depends only on $\Gp$ (through $\rho_1$).
At a site with $D(x)=0$ we have $\partial_{\psi(x)}G_{2,3}
=-\mu_{\Dep}(x)\psi(x)+O(\theta'\mt^2\bGM|\psi(x)|^3)$, with
$\theta'\ge1-e^{-\theta_0}$ fixed, that is, with no power of $\bGM$, whereas
\cite[(9.813) to (9.819)]{BF80} need $\bGM^{1/2}$ per derivative of a
one-body coefficient of order $\mt^2/\bGM$. For this reason we move the quadratic part into
the covariance, as $\nu$ is, and $G_{2,3}^\natural$ is of fourth order
(type vi of \cite[(9.811)]{BF80}).

\begin{lemma}\label{lem:7.31}
Assume the standing hypothesis \eqref{eq:7.standing}, $\eps_\nu+\theta'\le1/7$,
with $\eps_\nu$ of Section~\ref{ssec:7.1} (the output form of (S5)) and $\theta'$ from \eqref{eq:7.thetaprime}. Then,
uniformly in $\psi$, $h$, $k$, $\Xi$ and $\Gp$, with $C_{\Dep}$ the constant $C$ of
Proposition~\ref{prop:7.5}(b):
(a) $|\mu_{\Dep}(x)|\le\theta'\mt^2$ for every $x$; for every union $Y$
of $\cU$-cells,
$-C_{\Dep}(e^{-\mu}+e^{-2\kappa_S})\,\cdep N_{\Dep}(Y)\le G_{2,3}^{\Dep}(Y)\le
C_{\Dep}\cdep N_{\Dep}(Y)$ and
\[
|G_{2,3}^\natural(Y)|\le\tfrac12\theta'\mt^2\textstyle\sum_{x\in Y\setminus\Dep}
\bigl(\vartheta(x)^2+\psi(x)^2\bigr).
\] (b) At a site $x\notin\Dep$ with $D(x)=0$, the $n$-th derivative in
$\psi(x)$ of the $x$-term of $G_{2,3}^\natural$ is at most
$C_n\theta'\mt^2\bGM|\psi(x)|^{4-n}$ for $0\le n\le4$ and at most
$C_n\theta'\mt^2\bGM^{n/2-1}$ for $n\ge4$. At $x\in\Dep$, the $n$-th derivative
($n\ge1$) of the $x$-term of $G_{2,3}^{\Dep}$ is at most
$C_n(\mt^2/\bGM)\bGM^{n/2}$. (c) $C'$ exists, is positive definite, and $C'\le\alpha_DC_0$ as
forms, with $\alpha_D:=(1-\eps_\nu-\theta')^{-1}\le 7/6$. (d) The bounds of Theorem~\ref{thm:6.12} hold for $C'$ and for every
interpolated covariance $C'_s$ of \eqref{eq:C.decomp}:
$C'_s(x,x)=C'(x,x)\le\alpha_DK_1M_Y$,
$|C'_s(x,y)|\le 2K_1''M_Ye^{-r_{\mathrm{cov}}d_{\Gp}(x,y)}$,
$\|C'_s\|_{r_{\mathrm{cov}}}\le 2K_3\mt^{-2}$ and
$\|C'_s\|_{2\to2}\le\alpha_D\mt^{-2}$. (e) As measures on $\mathbb R^V$,
\begin{equation}\label{eq:7.6.cov}
e^{G(\psi+g)}\,d\mu_C(\psi)=\fc\,e^{G^\natural}\,d\mu_{C'}(\psi),\qquad
\fc:=\Bigl(\frac{\det C'}{\det C}\Bigr)^{1/2},
\end{equation}
and $\fc$ does not depend on $h$, $k$ or $\Xi$. (f) The representation \eqref{eq:7.translated}, Corollary~\ref{cor:7.11}\,(d) and Proposition~\ref{prop:7.25} hold with $(C,G,\cN)$ replaced by
$(C',G^\natural,\fc\cN)$; Theorem~\ref{thm:7.29} and Corollary~\ref{cor:7.30} hold for the activities
and outer partition functions built with $(C',G^\natural,\fc\cN)$, with the same left sides
$Z(1,\Xi)$, $Z_r(R)$ and $W_{\Gp}(C)$; and $W_{\Gp}(C)$, every $Z_p$ and every $Z_r(R)$ are unchanged.
\end{lemma}

\begin{proof}
\emph{(a)} Arguing as in the proof of Proposition~\ref{prop:7.5}(a), with $m^2$
replaced by $|m|^n$, we obtain, for $x\notin\Dep$ and $n\ge2$,
\begin{equation}\label{eq:7.6.mom}
\bGM\sum_{m\neq0}|m|^n\,|\rho_1(m,x)-\rho^{\mathrm{ref}}(m)|\le C_n\theta'\mt^2,
\end{equation}
with $C_2=1$ once $C$ in \eqref{eq:7.thetaprime} is large, and at each $x$
the left side is at most $C_n\mt^2$. For $n=2$ this bounds $\mu_{\Dep}$.
In the same case, together with $|\cos y-1+\frac12z^2|\le\frac12y^2+\frac12z^2$, applied with
$y=m\bGM^{1/2}\vartheta(x)$ and $z=m\bGM^{1/2}\psi(x)$, it also bounds
$G_{2,3}^\natural$. The bound on $G_{2,3}^{\Dep}(Y)$ follows from Proposition~\ref{prop:7.5}(b) together with
Proposition~\ref{prop:6.18}.

\emph{(b)} If $D(x)=0$, then
$\vartheta(x)=\psi(x)$ and $f(\psi):=\cos(m\bGM^{1/2}\psi)-1+\frac12m^2\bGM\psi^2$
vanishes to fourth order with $|f^{(n)}|\le|m|^n\bGM^{n/2}$. At $x\in\Dep$,
$|\partial_\psi^n\cos(m\bGM^{1/2}\vartheta)|\le|m|^n\bGM^{n/2}$. We then sum with
\eqref{eq:7.6.mom}.

\emph{(c)} The Schur test gives
$\|\nu\|_{2\to2}\le\|\nu\|_{r_{\mathrm{cov}}}\le\eps_\nu\mt^2$, part (a) gives
$\|\mu_{\Dep}\|_{2\to2}\le\theta'\mt^2$, and
$C_0^{-1}=u^{-1}+\mt^2\ge\mt^2$ by Definition~\ref{def:7.1}. Hence
$C_0^{-1}+\nu+\mu_{\Dep}\ge(1-\eps_\nu-\theta')C_0^{-1}\ge\frac67C_0^{-1}>0$,
and (c) follows, since inversion reverses the order of positive definite forms.

\emph{(d)} Theorem~\ref{thm:6.12} applies to $C'=(C_0^{-1}+\nu')^{-1}$,
$\nu':=\nu+\mu_{\Dep}$, as
$2K_3\|\nu'\|_{r_{\mathrm{cov}}}\le2K_3(\eps_\nu+\theta')\mt^2\le\mt^2$,
where we used $K_3\le3.5$. The diagonal bound follows from (c). Block compressions keep the
diagonal and positivity and increase no other bound, which extends these bounds to the
interpolated covariances.

\emph{(e)} By \eqref{eq:7.6.split},
$e^{G(\psi+g)}=e^{G^\natural}e^{-\frac12\langle\psi,\mu_{\Dep}\psi\rangle}$. Since
$C^{-1}=C_0^{-1}+\nu$ by Proposition~\ref{prop:7.8}, we have
$e^{-\frac12\langle\psi,\mu_{\Dep}\psi\rangle}d\mu_C(\psi)
=(\det2\pi C)^{-1/2}e^{-\frac12\langle\psi,C'^{-1}\psi\rangle}d\psi=\fc\,d\mu_{C'}(\psi)$.
The operators $C_0$, $\nu$, $\mu_{\Dep}$ depend only on $\Gp$, $M_Y$ and $\mt$, so the factor $\fc$ does not depend on $h$, $k$ or $\Xi$, and since \eqref{eq:7.6.cov} is an identity of measures, it also preserves absolute integrability.

\emph{(f)} We insert \eqref{eq:7.6.cov} in Proposition~\ref{prop:7.8}. Those
proofs use only factorization over cells and sites, the margin of
Lemma~\ref{lem:C.4} ($q_M\le1/6$ for $C'$) and the quadratic bound on
$|E(X,s)|$, and, as $G-G^\natural=-\frac12\langle\psi,\mu_{\Dep}\psi\rangle$ does not depend on the wall field, \eqref{eq:7.6.cov} applies to each configuration of $Z(k,\Xi)$, $Z_p$ and $Z_r(R)$, so these are unchanged; the outer partition functions with $X\ne\emptyset$ are not the same for the two choices, and from here on only those built with $(C',G^\natural,\fc\cN)$ are used.
\end{proof}

Like the $\mathcal Z_p(X)$, the activities are now built with
$(C',G^\natural,\fc\cN)$. For a polymer $X$, $\cN(X)$ is the set of wall
configurations $N_X:=N|_{E(X)}$ of cell-constant labels $h_X$ with values in
$P\mathbb Z$ ($P=2\pi\bGM^{-1/2}$) whose walls lie in $E(X)$. At $k=0$ the
statements of this subsection and Lemma~\ref{lem:7.53} hold for the long
wall fields of Definition~\ref{def:7.27}, by Proposition~\ref{prop:7.long}.
We write $n_I(X)$ for the number of interfaces with both cells in $X$, and put $A_w:=2\Funit$.

\begin{lemma}[Wall entropy]\label{lem:7.32}
(a) $\Floc(Y)\ge A_w\sum_{I\subseteq Y}N_I^2$ for every union $Y$ of elements
of $\tilde Y(h)$, where $N_I$ is the constant value of $N$ on the interface
$I$ (Lemma~\ref{lem:7.6}(a)). (b) For every polymer $X$ and every $c_4>0$ with
$e^{-c_4A_w}\le\frac13$,
$\sum_{N_X\in\cN(X)}e^{-c_4\Floc(X)}\le(1+3e^{-c_4A_w})^{n_I(X)}$
$\le e^{c_4'|X|}$ with $c_4':=9C_ue^{-c_4A_w}/(c_gc_1(\mt L)^3)$.
\end{lemma}

\begin{proof}
We adapt the proof of \cite[Lemmas~9.2, 9.3]{BF80} to $\Gp$ as follows.
(1) The face sum of \cite[(9.117)]{BF80} becomes the interface sum. Indeed, walls
are supported on interfaces by Lemma~\ref{lem:7.6}(a), and by Theorem~\ref{thm:7.14}
with $\mathrm{good}(I)\ge c_g'L^2$, which holds by (R8) in fine units and Lemma~\ref{lem:4.17},
an interface $I\subseteq Y$ gives at least
$\frac1{18}\mt^2\frac L2c_g'L^2\cdot4\pi^2N_I^2/\bGM=2\Funit N_I^2$. (2) The
sum over $h$ becomes the injection $N_X\mapsto(N_I)_I\in\mathbb Z^{n_I(X)}$,
each interface giving $\sum_ne^{-c_4A_wn^2}\le1+3e^{-c_4A_w}$. (3) The cube
count becomes $n_I(X)\le3C_u|X|/(c_gc_1(\mt L)^3)$, since a cell has at least
$c_1L^3$ sites by (R5) and at most $6/c_g$ neighbours by (R8), and a $\cU$-cell has
fewer than $C_u\mt^{-3}$ sites. With $\log(1+3t)\le3t$ this gives the second inequality of (b),
and by Corollary~\ref{cor:7.11}(d) no perimeter term arises. For the value $c_4=c'/4$ of Remark~\ref{rem:A.10}, $c_4'$
is doubly exponentially small in $\tau$, as
$A_w\ge(\pi^2c_g'/72)c_L^3(\mt\bGM)^{-1}$ by $\mt L\ge c_L/2$.
\end{proof}

In Debye units, for the Peierls factor $r_\Omega$ of \eqref{eq:7.rOmega},
we put $v_\Omega:=\mt^3|\Omega|$, $\hat\rho(m):=\mt^{-3}\rho^{\mathrm{ref}}(m)$,
$\bD:=\bGM\mt$, $\hat P:=2\pi\bD^{-1/2}$, and
\begin{equation}\label{eq:7.6.Rhat}
\hat R_v(x):=\frac{\exp\bigl\{v\sum_m\hat\rho(m)\bigl[e^{im\bD^{1/2}x}-1\bigr]\bigr\}}
{\sum_{n\in\mathbb Z}\exp\bigl\{-\frac12v(x-n\hat P)^2\bigr\}},
\end{equation}
so that $r_\Omega(A)=\hat R_{v_\Omega}(\mt^{-1/2}(A+k\bGM^{-1/2}\bar f_\Omega))$.
By (R5) in fine units and $c_L/2\le\mt L\le c_L$, we have
$v_\Omega\in[v_-,v_+]:=[c_1c_L^3/8,\,8c_L^3]$. Let
$\eta_*:=\sup_{0<s\le\pi}[1-2\bD\sum_m\hat\rho(m)(1-\cos ms)/s^2]
\le\bar\eta:=1-\frac4{\pi^2}\bigl(1-C_{\hat\rho}e^{-2\kappa_S}\bigr)+C_{\hat\rho}e^{-2\kappa_S}\le1-\frac4{\pi^2}+2C_{\hat\rho}e^{-2\kappa_S}$, where
$C_{\hat\rho}:=4C\ge\bD\sum_{|m|\ge2}|\hat\rho(m)|\,m^2e^{2\kappa_S(|m|-1)}$ with $C$ of \eqref{eq:6.S4c}
($\mt\ge\mD/2$); unlike $\eta_*$, the bound $\bar\eta$ does not depend on $\bD$.

\begin{lemma}\label{lem:7.33}
Fix $\eta_H\in(\bar\eta,1)$. There are $b_0>0$ and $c_7,c_7',c_7''$, depending only
on $v_\pm$, $\eta_H$, $C_{\hat\rho}$ and $\kappa_S$, such that for
$\bD\le b_0$, $v\in[v_-,v_+]$, all real $x$ and all integers $N\ge0$,
\[
|\partial_x^N\hat R_v(x)|\le c_7\,(c_7'\bD^{1/6})^Ne^{c_7''N\log N}
e^{\frac12\eta_Hvx^2}.
\]
Consequently, uniformly in $h$, with $\partial A_\Omega/\partial\psi(x)=1/|\Omega|$
($x\in\Omega$),
\[
|\partial_A^Nr_\Omega(A_\Omega)|\le c_7\,(c_7'\bD^{1/6}\mt^{-1/2})^N
e^{c_7''N\log N}\exp\Bigl(\tfrac12\eta_H\mt^2\sum_{x\in\Omega}\vartheta(x)^2\Bigr).
\]
\end{lemma}

The first bound is proved as \cite[Lemma~9.7, (9.51)]{BF80}, with $\tilde l_D=1$,
$L^3\to v$, $\beta\to\bD$, $\rho_i\to\hat\rho(m)$, $\tau\to\hat P$, except for the following
points. (1) There are infinitely many charges. The Taylor expansions of the factors
I, II of \cite[(9.53)]{BF80} converge for $|\mathrm{Im}\,x|\le\bD^{-1/6}$ by
the tails of (S4$'$), as $e^{|m|\bD^{1/3}}e^{-2\kappa_S|m|}$ is summable.
(2) The constants are uniform in $v\in[v_-,v_+]\subset(0,\infty)$, not for a
fixed volume $L^3$. (3) We need an explicit exponent for
$\alpha_Dp'\eta_H<1$ in (L8-1). The inequalities $1-\cos s\ge2s^2/\pi^2$ on $[0,\pi]$ and
$2\bGM\rho^{\mathrm{ref}}(1)\ge(1-C_{\hat\rho}e^{-2\kappa_S})\mt^2$ bound $\eta_*$, and
each $\eta_H\in(\bar\eta,1)$ is admissible in \cite[(9.59)]{BF80}. The second
bound follows from the $\hat P$-periodicity, from
$A_\Omega+k\bGM^{-1/2}\bar f_\Omega=h_\Omega+\avg_\Omega\vartheta$ with
$h_\Omega\in P\mathbb Z$, and from
$|\Omega|(\avg_\Omega\vartheta)^2\le\sum_{x\in\Omega}\vartheta(x)^2$. With
$\gamma:=\eta_H\mt^2$, the second bound is the analogue on $\Gp$ of \cite[(7.9), (7.10)]{GM82}: each
derivative of $G_1$ contributes a factor $\bD^{1/6}$, times $e^{\frac12\gamma\sum_X\vartheta^2}$.

\paragraph{H\"older exponents and the hypotheses (L8-1) to (L8-4).}
We fix $\eta_H\in(\bar\eta,1)$, H\"older exponents $p_1,\dots,p_4>1$ with
$\sum_i1/p_i=1$ and $p_1$ even, and $p'>p_3$, $q'$ with $1/p'+1/q'=1/p_3$,
at the values given in Section~\ref{S:G.L8}. Let $C_W^f$, $\kbox=\rbox L$,
$N_{\mathrm{ov}}$ and $\Kflat_\Omega\supseteq\Omega$, with sites in the box
$\hat Q(Q)$ of fine side at most $\kbox$, be those of the cell Poincar\'e
inequality of Proposition~\ref{prop:6.10}(iv), and let
$C_P':=C_W^f\rbox^2N_{\mathrm{ov}}$, so that $C_P'L^2=C_W^f\kbox^2N_{\mathrm{ov}}$.
Hypothesis (L8-1) is the standing hypothesis \eqref{eq:7.standing} together with
$\hat\eta:=\alpha_Dp'\eta_H<1$, which implies $\alpha_Dp_3\eta_H<1$ (in \cite[(9.638)]{BF80} the constant $\hat\eta$
is denoted $\hat\beta<1$). (L8-2) requires
$C_P'(\mt L)^2\le\eta_H/8$, so that
$\theta_P:=(4\hat\eta/\eta_H)C_P'(\mt L)^2\le\frac12$ and
$\nu_0:=\max(\hat\eta,\theta_P)<1$. (L8-3) requires $\eps(L')\le 2(p'/q')^2$, with
$\eps(L')$ of Theorem~\ref{thm:7.19}, and (L8-4) states that each $\Kflat_\Omega$ has
$d_{\Gp}$-diameter less than $L'$. At the values fixed above, (L8-1) holds. Hypothesis (L8-2) holds by the choice of
$c_L$, since $\mt L\le c_L$, (L8-3) holds by Corollary~\ref{cor:7.21}, and (L8-4)
follows from (L8-2) by Proposition~\ref{prop:6.10}(v). A cell
$\Omega\subseteq Q$ is \emph{near} if the $\ell^\infty$-distance from the
sites of $\hat Q(Q)$ to $C$ is at most $\frac12$, and \emph{far} otherwise.
For a union $Y$ of elements, $E(\cN(Y))$ is the union of the $E(\Kflat_\Omega)$
over the near cells $\Omega\subseteq Y$.

\paragraph{The free-space solid angle.}
Let $\Omega_\Xi(y)\in(-2\pi,2\pi)$ be the signed solid angle of the flat
rectangle $\Xi$ seen from a point $y$ off its plane, increasing by $4\pi$
when $y$ crosses $\Xi$ in the direction in which $\xi_b=+1$. For a site $x$
of $\Gp$ (a dual site of $\mathbb Z^3$, never in the plane of $\Xi$) we put
$f^C(x):=\frac12\Omega_\Xi(x)$ and, for each dual bond $b=(x,x')$ of
$\mathbb Z^3$, $R^C(b):=f^C(x')-f^C(x)-j(b)$, where $j(b)=2\pi\xi_b$, which on the
bonds of $\Gp$ is the jump field of $\Xi$ (Section~\ref{sec:7.1}). The field
$f^C$ depends only on $C$, not on $\Gp$, unlike the Neumann solid angle
$f_{\Gp}$ of Lemma~\ref{lem:7.2}, and $R^C$ differs from $R_f$. Let
$S(D):=\sum_{b:\,d(b,C)\le D}|R^C(b)|^2$, the sum running over dual bonds of
$\mathbb Z^3$, and $D_{\cN}:=\sqrt3\,\kbox+1$.

\begin{lemma}\label{lem:7.34}
(a) Assume (L8-4). Let $Y$ be a union of elements of $\tilde Y(h)$ and
$\Omega\subseteq Y$ a cell. For every $\vartheta\in\mathbb R^V$, every
function $a$ on the sites of $\Kflat_\Omega$ and every $\eps\in(0,1]$,
\begin{gather*}
\|P_\Omega\vartheta\|^2\le(1+\eps)C_W^f\kbox^2\|\nabla\vartheta+w_Y\|^2_{E(\Kflat_\Omega)}
+(1+1/\eps)k^2\bGM^{-1}\delta_\Omega(a)^2,\\
\delta_\Omega(a):=(C_W^f)^{1/2}\kbox\|\nabla a-j\|_{E(\Kflat_\Omega)}+\|P_\Omega a\|_\Omega ,
\end{gather*}
where $w_Y:=\sum_{S_\beta\subseteq Y}w_\beta$. If $\Omega$ is far, there is an
$a$ with $\delta_\Omega(a)=0$. (b) For every dual bond $b$ of $\mathbb Z^3$, $d(b,C)\ge\frac12$ and
$|R^C(b)|\le 2\pi/d(b,C)$; at every dual site, $|f^C|<\pi$. (c) On every output $\Gp$,
$\sum_{\Omega\text{ near}}\|R^C\|^2_{E(\Kflat_\Omega)}\le N_{\mathrm{ov}}\|R^C\|^2_{E(\cN(V))}
\le N_{\mathrm{ov}}S(D_{\cN})$, and if $|C|\ge8\kbox$, then
\[
S(D_{\cN})\le C_{\mathrm{BS}}|C|\,(\log\kbox+C_{\mathrm{sh}}),\qquad C_{\mathrm{BS}}:=24\pi^3,\quad
C_{\mathrm{sh}}:=92 .
\]
\end{lemma}

\begin{proof}
Heuristically, on a box crossed by $C$ the trivial primitive leaves a flat sheet of
jumps, of energy $\asymp\kbox^2$, and with the solid angle only the
Biot\textendash Savart residual remains, of energy $\asymp\kbox\log\kbox$ per unit length
of $C$.

\emph{(a)} We first show that $w=w_Y$ on $E(\Kflat_\Omega)$. Let
$b'\in E(\Kflat_\Omega)$ be a wall bond. It belongs to one component $w_{\beta'}$ and has a wall
site $z$ as an endpoint. Fix $x\in\Omega$. Then $d_{\Gp}(x,z)<L'$ by (L8-4). Hence
each $\cU$-cell met by a $\Gp$-path of length $d_{\Gp}(x,z)$ from $x$ to $z$
lies in $\hat\Sigma$, consecutive ones being joined by a bond, and by Proposition~\ref{prop:7.9} the $\cU$-cell of
$x$ lies in $S_{\beta'}$. This $\cU$-cell also lies in the
element of $Y$ containing $\Omega$, and each $S_\beta$ lies in one element, in view of
Definition~\ref{def:7.23}. Hence $S_{\beta'}\subseteq Y$. Other bonds carry
$w=0=w_Y$. Next we define $a_\Omega:=h-k\bGM^{-1/2}a$ on the sites of $\Kflat_\Omega$.
Here, at $k=0$, for long wall fields, the label is replaced by the primitive of
Lemma~\ref{lem:7.prim} (Proposition~\ref{prop:7.long}), and at $k=1$, for the inside fields
of outer class in $\Glong$, by the primitive of Lemma~\ref{lem:7.insideprim}\,(b). As $h$ is constant on cells, $P_\Omega h=0$ and
$P_\Omega\vartheta=P_\Omega(\vartheta+a_\Omega)+k\bGM^{-1/2}P_\Omega a$. On
$E(\Kflat_\Omega)$, since $w=\nabla h-k\bGM^{-1/2}j$, we have
$\nabla(\vartheta+a_\Omega)=\nabla\vartheta+w_Y+k\bGM^{-1/2}(j-\nabla a)$. The
Poincar\'e inequality of Proposition~\ref{prop:6.10}(iv) for
$\vartheta+a_\Omega$ on $\Kflat_\Omega$ and the triangle inequality give
$\|P_\Omega\vartheta\|\le(C_W^f)^{1/2}\kbox\|\nabla\vartheta+w_Y\|_{E(\Kflat_\Omega)}
+k\bGM^{-1/2}\delta_\Omega(a)$. We then square, using
$(x+y)^2\le(1+\eps)x^2+(1+1/\eps)y^2$.

Now let $\Omega\subseteq Q$ be far, and let $\Pi$ be the plane of $\Xi$. A dual bond
crossing $\Pi$ crosses it at the centre of a plaquette, and $j(b)\neq0$ iff
that plaquette lies in $\Xi$. The plaquettes crossed by bonds of $\hat Q(Q)$
form a rectangle. If this rectangle contained plaquettes in and outside $\Xi$, it would
contain two sharing an edge $e\in\partial\Xi=C$ at $\ell^\infty$-distance
$\frac12$ from a site of $\hat Q(Q)$, which is excluded for a far cell. Hence
$j=2\pi\sigma\nabla1_{\Pi^+}$ on $E(\hat Q(Q))$ for some $\sigma\in\{-1,0,1\}$,
where $\Pi^+$ is the set of sites on one side of $\Pi$. With this $\sigma$, the function
$a:=2\pi\sigma1_{\Pi^+}$ has $\nabla
a=j$ on $E(\Kflat_\Omega)$ and is constant on $\Omega$, as $Q$ lies on one side
of $\Pi$ by (A3) of Lemma~\ref{lem:3.5}. Thus $\delta_\Omega(a)=0$.

\emph{(b)} Primal edges are at distance at least $\frac12$ from dual bonds, which gives the first assertion.
Off $C$, $\nabla\Omega_\Xi$ extends continuously across $\Xi$ as the
Biot\textendash Savart field of $C$. Hence $|R^C(b)|\le\frac12\max_{y\in b}J(y)$,
where $J(y):=\oint_C|dl|\,|y-l|^{-2}$. Let $y=y'+h\mathbf e$ with $y'\in\Pi$ and $h\ge0$.
The length of $\{l\in C:|y'-l|\le t\}$ vanishes for $t<d(y',C)$ and is at
most $2\pi t$, since this set lies in the boundary of a convex subset of a disc of radius
$t$. Integration by parts and the substitution $t=h\tan\varphi$ give
$J(y)\le4\pi/d(y,C)$. Finally, $|\Omega_\Xi|<2\pi$, since the radial projection of
the rectangle from a point off $\Pi$ lies in an open hemisphere.

\emph{(c)} For a near cell $\Omega\subseteq Q$, the bonds of $\Kflat_\Omega$
are dual bonds of $\mathbb Z^3$ joining sites of $\hat Q(Q)$, a box of
diameter at most $\sqrt3\,\kbox$ within Euclidean distance $\frac{\sqrt3}2$ of
$C$, so $d(b,C)\le D_{\cN}$. Each $|R^C(b)|^2$ is a fixed nonnegative number
attached to a dual bond of $\mathbb Z^3$, independent of $\Gp$, and passing from $\mathbb Z^3$ to $\Gp$ can only delete
terms of free-space $\ell^2$ sums, since freezing a plaquette deletes a bond and a hole deletes sites and bonds.
The bounds are therefore free-space bounds, independent of the Neumann field and its enhancement at the slit
edges, and with the overlap
$N_{\mathrm{ov}}$ these facts give the first two inequalities. For the last one, we use the
number $\mathcal N(r)$ of dual bonds with $d(b,C)\le r$, which vanishes for
$r<\frac12$. The midpoints of these bonds lie within $r+\frac12$ of $C$ and form three
translates of $\mathbb Z^3$. Hence, with $r_*:=\frac12+\frac{\sqrt3}2$,
$\mathcal N(r)\le3|\{y:d(y,C)\le r+r_*\}|\le3\pi|C|(r+r_*)^2+16\pi(r+r_*)^3$.
By (b) and summation by parts,
$\sum_{d(b,C)\le D}d(b,C)^{-2}=\mathcal N(D)D^{-2}+2\int_{1/2}^D\mathcal N(r)r^{-3}dr$.
The term $3\pi|C|r^2$ gives $6\pi|C|\log D+\Ord(|C|)$, and the other terms give
$\Ord(|C|+D)$. The factor $(2\pi)^2$ of (b) gives $C_{\mathrm{BS}}=(2\pi)^2\cdot6\pi=24\pi^3$.
At $D=D_{\cN}\le2\kbox$ and $|C|\ge8\kbox$ the remaining constants add up to
$C_{\mathrm{sh}}=92$ (Section~\ref{S:G.34}).
\end{proof}

For a polymer $X$ put
\[
B_X(\vartheta):=\tfrac12\mt^2\sum_{x\in X}\vartheta(x)^2
+\frac{2\mt^2}{\eta_H}\sum_{x\in X}(P_X\vartheta)(x)^2,\qquad
T_3:=\Bigl[\int d\mu_{C'_s}(\psi)\,e^{p_3\eta_HB_X(\psi+D)}\Bigr]^{1/p_3}.
\]
With $\gamma=\eta_H\mt^2$ we have
$\frac12p_3\gamma\sum_X\vartheta^2+2p_3\mt^2\sum_X\tilde\delta^2=p_3\eta_HB_X(\vartheta)$.
Hence $T_3$ is the third factor of the H\"older split
(Proposition~\ref{prop:7.37}), and it is the analogue on $\Gp$ of \cite[(9.64)]{BF80}. The
near-cell factor is
\[
\fn'(Y):=2k^2\bGM^{-1}\Bigl[\theta_P\|R^C\|^2_{E(\cN(Y))}
+\frac{4\hat\eta\mt^2}{\eta_H}\sum_{\Omega\subseteq Y\text{ near}}\|P_\Omega f^C\|^2_\Omega\Bigr].
\]
We also use its cellwise version
$\fn''(Y):=2k^2\bGM^{-1}\sum_{\Omega\subseteq Y\text{ near}}\bigl[\frac{\theta_P}{N_{\rm ov}}\|R^C\|^2_{E(\Kflat_\Omega)}
+\frac{4\hat\eta\mt^2}{\eta_H}\|P_\Omega f^C\|^2_\Omega\bigr]$, which is additive over disjoint unions of cells
and satisfies $\fn''\le\fn'$, as each bond of $E(\cN(Y))$ lies in at most $N_{\rm ov}$ of the sets
$E(\Kflat_\Omega)$ (Theorem~\ref{thm:3.20}\,(R6)).

\begin{lemma}\label{lem:7.35}
Assume (L8-1) to (L8-4), and let $c_8$ and $\cnear$ be the constants of Section~\ref{S:G.35}. For every $h$, every polymer $X$ and every $s$,
\[
T_3\le\exp\Bigl\{\frac{c_8|X|+\fn'(X)}{p'}+\frac{\Floc(X)}{p_3}\Bigr\},
\]
and $\fn'=0$ for $k=0$. If
$\min(R,T)\ge R_1(\beta)=(2\rbox+5)L$, then $\fn'(V)\le\cnear|C|$, and uniformly for $\bGM\ge\bGMo$,
\[
\cnear\le0.1264\,\theta_PC_{\mathrm{BS}}+\frac{2\theta_PC_{\mathrm{BS}}C_{\mathrm{sh}}
+1024\pi^2\hat\eta(2\rbox+5)^2c_L^2/(\eta_Hc_1)}{\bGM}.
\]
\end{lemma}

\begin{proof}
\emph{Step 1 (local untranslation identity).} We absorb the shift $D$ by a translation, which turns
$\frac12\langle\psi,C_0^{-1}\psi\rangle+F_2+\Floc$ into one quadratic form of
$\psi+D$. In local form, this is the step of \cite[p.~226]{BF80} in which (7.8) to (7.10) are applied backwards.
Let $Y$ be a union of
elements, let $D_Y$, $w_Y$, $s_Y$ be the sums of $D_\beta$, $w_\beta$, $s_\beta$ over
$S_\beta\subseteq Y$, and let
$\mathcal Q_Y(\vartheta):=\mt^2\|\vartheta\|^2+\|\nabla\vartheta+w_Y\|^2+M_Y^{-2}\|\Delta\vartheta-\nabla^*w_Y\|^2$.
Then for all $\psi\in\mathbb R^V$
\begin{equation}\label{eq:7.6.untr}
\tfrac12\langle\psi,C_0^{-1}\psi\rangle+F_2(Y)+\Floc(Y)=\tfrac12\mathcal Q_Y(\psi+D_Y).
\end{equation}
Indeed, since $C_0^{-1}=\mt^2-\Delta+M_Y^{-2}\Delta^2$ and $-\Delta=\nabla^*\nabla$,
the right side is
$\frac12\langle\psi,C_0^{-1}\psi\rangle+\langle\psi,C_0^{-1}D_Y+s_Y\rangle+\frac12\mathcal Q_Y(D_Y)$
with $s_Y=M_Y^{-2}(-\Delta+M_Y^2)\nabla^*w_Y$. By Proposition~\ref{prop:7.9},
$D_\beta=-\chi_\beta C_0s_\beta$ and $(1-\chi_\beta)s_\beta=0$, so
$C_0^{-1}D_\beta+s_\beta=-[C_0^{-1},\chi_\beta]C_0s_\beta=F_{2,\beta}'$,
supported in $S_\beta$. Therefore the cross term is $F_2(Y)$. Since the functions $D_\beta$,
$\nabla D_\beta$, $w_\beta$, $\Delta D_\beta$, $\nabla^*w_\beta$ are supported in
$S_\beta$ or $E(S_\beta)$, terms of different $\beta$ have disjoint
supports, and
$\frac12\mathcal Q_Y(D_Y)=\sum_{S_\beta\subseteq Y}F_1(S_\beta)=\Floc(Y)$
by Proposition~\ref{prop:7.10} and Corollary~\ref{cor:7.11}.

\emph{Step 2 (the Gaussian integral at fixed $Y$).} We show
\begin{equation}\label{eq:7.6.JY}
J_Y:=\int d\mu_{C_0}(\psi)\exp\bigl\{\hat\eta B_Y(\psi+D)-\Floc(Y)-F_2(Y)\bigr\}
\le\exp\{c_8|Y|+\fn''(Y)\}.
\end{equation}
Each $S_\beta$ lies in $Y$ or is disjoint from it, so $D=D_Y$ on $Y$. Let
$N:=\hat\eta\mt^21_Y+(4\hat\eta\mt^2/\eta_H)P_Y$, so that
$\hat\eta B_Y(\vartheta)=\frac12\langle\vartheta,N\vartheta\rangle$,
since $P_Y=1_YP_Y1_Y$ is an orthogonal projection. By \eqref{eq:7.6.untr} and $\vartheta=\psi+D_Y$, $J_Y$ is the ratio of
the integrals over $\mathbb R^V$ of
$\exp\{-\frac12\mathcal Q_Y(\vartheta)+\frac12\langle\vartheta,N\vartheta\rangle\}$
and of $e^{-\frac12\langle\psi,C_0^{-1}\psi\rangle}$, with quadratic part
$-\frac12\langle\vartheta,(C_0^{-1}-N)\vartheta\rangle$.
For positivity we use \cite[(9.637), (9.640) to (9.642)]{BF80}, with the
cell Poincar\'e inequality summed with overlap $N_{\mathrm{ov}}$, that is,
$\|P_Y\vartheta\|^2\le C_P'L^2\|\nabla\vartheta\|^2$, in place of the Neumann
Laplacian of the cubes in (9.642). We obtain $\langle\vartheta,N\vartheta\rangle\le
\hat\eta\mt^2\|\vartheta\|^2+\theta_P\|\nabla\vartheta\|^2\le
\nu_0\langle\vartheta,C_0^{-1}\vartheta\rangle$, the last step because $C_0^{-1}\ge\mt^2-\Delta$. Thus
$K:=C_0^{1/2}NC_0^{1/2}$ satisfies $0\le K\le\nu_0<1$, and hence
$J_Y=\det(1-K)^{-1/2}\exp\{\sup_\vartheta[-\frac12\mathcal Q_Y(\vartheta)
+\frac12\langle\vartheta,N\vartheta\rangle]\}$.

We first bound the supremum by Lemma~\ref{lem:7.34}, and only near cells leave a remainder. For a far cell we apply Lemma~\ref{lem:7.34}(a) with
$\delta_\Omega(a)=0$ and $\eps\to0$, and for a near cell with $\eps=1$, $a=f^C$
and $\delta_\Omega(f^C)^2\le2C_W^f\kbox^2\|R^C\|^2_{E(\Kflat_\Omega)}+2\|P_\Omega f^C\|^2_\Omega$.
Summing over the cells of $Y$ with overlap $N_{\mathrm{ov}}$ and multiplying
by $2\hat\eta\mt^2/\eta_H$ gives, as
$(2\hat\eta\mt^2/\eta_H)\cdot2C_W^f\kbox^2N_{\mathrm{ov}}=\theta_P$,
$\frac12\langle\vartheta,N\vartheta\rangle\le\frac12\hat\eta\mt^2\|\vartheta\|^2_Y
+\theta_P\|\nabla\vartheta+w_Y\|^2+\fn''(Y)$, the near-cell remainders adding up exactly to $\fn''(Y)$. Since $\mathcal Q_Y(\vartheta)\ge\mt^2\|\vartheta\|^2+\|\nabla\vartheta+w_Y\|^2$,
the exponent is at most
$-\frac12(1-\hat\eta)\mt^2\|\vartheta\|^2-(\frac12-\theta_P)\|\nabla\vartheta+w_Y\|^2+\fn''(Y)\le\fn''(Y)$.

For the determinant, we have $-\frac12\log\det(1-K)\le\operatorname{tr}K/(2(1-\nu_0))$
\cite[Appendix~3]{BF80}, and $\operatorname{tr}K=\operatorname{tr}(NC_0)\le
\hat\eta(1+4/\eta_H)\mt^2K_1M_Y\#Y\le2(1-\nu_0)c_8|Y|$ by $P_Y\le1_Y$,
Theorem~\ref{thm:6.12}, $M_Y=\mt/\lPV$ and $\#Y<C_u\mt^{-3}|Y|$.

\emph{Step 3 (assembly).} We follow \cite[(9.65) to (9.626)]{BF80}, adapted to $\Gp$
as follows. (1) We apply the convexity step \cite[(9.67) to (9.611)]{BF80},
fact S3 of Section~\ref{S:constF}, to the block compressions
$C'_\epsilon=\sum_J1_{Y_J}C'1_{Y_J}$ of \eqref{eq:C.decomp},
$C'_s=\sum_\epsilon\lambda_\epsilon(s)C'_\epsilon$. The factors in this step are finite, because
for $\frac12\langle\vartheta,N_1\vartheta\rangle=p_3\eta_HB_X(\vartheta)$ we have
$\alpha_DN_1\le N$ (as $p_3<p'$), $N$ commutes with each $1_{Y_J}$, and
$C'_\epsilon\le\alpha_DC_{0\epsilon}$, so that
$\|N_1^{1/2}C'_\epsilon N_1^{1/2}\|\le\nu_0<1$ by the positivity step.
(2) In the monotonicity step (9.612) to (9.617) we use Lemma~\ref{lem:7.31}(c), $C'_\epsilon\le\alpha_DC_{0\epsilon}$ with
$C_{0\epsilon}:=\sum_J1_{Y_J}C_01_{Y_J}$, instead of (9.618) to (9.622). (3) After H\"older's inequality (9.625)
with exponents $p'/p_3$, $q'/p_3$, both integrals factor over the sets
$Y_J$, unions of elements (9.626), and the first factor is at most
$\exp\{(c_8|X|+\fn''(X))/p'\}\le\exp\{(c_8|X|+\fn'(X))/p'\}$ by \eqref{eq:7.6.JY}, the additivity of $\fn''$ and $\fn''\le\fn'$. (4) Corollary~\ref{cor:7.20}
with $t_C=1$ and (L8-3) replaces (9.632):
$\bigl(\int d\mu_{C_0}e^{(q'/p')F_2(Y_j)}\bigr)^{1/q'}e^{\Floc(Y_j)/p'}\le e^{\Floc(Y_j)/p_3}$,
as $1/p_3-1/p'=1/q'$. For $k=0$, $\fn'=0$, since $\fn'$ carries the factor $k^2$.

\emph{Step 4 (the perimeter bound).} Lemma~\ref{lem:7.34}(c) and $k\le1$ show that
the first part of $\fn'(V)$ is at most
$2\bGM^{-1}\theta_PC_{\mathrm{BS}}|C|(\log\kbox+C_{\mathrm{sh}})$, as
$|C|=2(R+T)\ge4R_1(\beta)\ge8\kbox$. By Lemma~\ref{lem:7.34}(b),
$\|P_\Omega f^C\|^2_\Omega\le\pi^2|\Omega|\le8\pi^2L^3$. It remains to count the near
cells. A near cell $\Omega\subseteq Q$ lies in a block $Q\subseteq\hat Q(Q)$
all of whose sites are within $\ell^\infty$-distance $\kbox+\frac12$ of $C$.
The $\ell^\infty$-neighbourhood of width $s$ of $C$ has volume at most
$(2s)^2(|C|+8s)$, and for $s=\kbox+2$ and $L\ge4$, $2s\le(2\rbox+5)L$ and
$8s\le(8\rbox+20)L$. Blocks are disjoint with at least $L^3$ sites, and by (R5) a
block contains fewer than $8/c_1$ cells. Hence, as
$|C|\ge4R_1(\beta)=(8\rbox+20)L$,
$N_{\mathrm{near}}\le\frac8{c_1}(2\rbox+5)^2(|C|+(8\rbox+20)L)/L
\le\frac{16}{c_1}(2\rbox+5)^2|C|/L$, and the second part of $\fn'(V)$ is at most
$2\bGM^{-1}\frac{4\hat\eta\mt^2}{\eta_H}8\pi^2L^3N_{\mathrm{near}}
\le\frac{1024\pi^2\hat\eta}{\eta_Hc_1}(2\rbox+5)^2c_L^2\frac{|C|}{\bGM}$, using
$\mt L\le c_L$. Adding the two parts gives $\fn'(V)\le\cnear|C|$. The bound on $\cnear$,
uniform in $\bGM\ge\bGMo$, then follows (Section~\ref{S:G.35}) from
$\log\kbox\le\log(\rbox c_L)+\log(1/\mt)$ and the value of $\mt$.
\end{proof}

By Proposition~\ref{prop:7.8}, $E(\psi)=E'(\psi+g)+\frac12\langle\psi,\nu\psi\rangle$,
where $E'$ collects the Mayer terms of order $s\ge2$ of
Theorem~\ref{thm:6.5}. We write $E(X)$ for the part of $E$ localized in $X$,
as in Section~\ref{sec:7.5}. Both $G_{2,1}$ and $G_{2,2}$ are real, by (S3). Let
$c_\lambda:=\frac14(1-1/p_3)$, $c'_\lambda:=C(e^{-\mu}+\eps_\nu)$,
$c_E:=c'_\lambda+\theta'$,
$G_2^\natural:=G_{2,1}+G_{2,2}+G_{2,3}^{\Dep}+G_{2,3}^\natural$, and assume
\begin{equation}\label{eq:7.6.cl}
2\alpha_Dp_4c_E\le\tfrac12,\qquad
\frac{p_4\alpha_Dc_E^2}{1-2\alpha_Dp_4c_E}+c_E\le c_\lambda .
\end{equation}
For the values of Section~\ref{S:G.36}, \eqref{eq:7.6.cl} follows from
$c_E\le1/70$, which is the case for $\theta_0=0.01$ and $\mu$, $\kappa_S$, $\bGM$
large.

\begin{lemma}\label{lem:7.36}
Assume $\eps_\nu+\theta'\le1/7$ and \eqref{eq:7.6.cl}. For every $h$, every polymer
$X$ and every $s$,
\[
T_4:=\Bigl[\int d\mu_{C'_s}\bigl|e^{E(X)}e^{G_2^\natural(X)}e^{-2\mt^2\sum_X\tilde\delta^2}\bigr|^{p_4}\Bigr]^{1/p_4}
\le\exp\bigl\{c_9|X|+C_{\Dep}\cdep N_{\Dep}(X)+c_\lambda\Floc(X)\bigr\},
\]
with $c_9:=2\alpha_Dc_EK_1C_u/\lPV$.
\end{lemma}

\begin{proof}
The proof of \cite[Lemma~9.9, (9.72) to (9.718)]{BF80} carries over to $\Gp$ with the following changes.
(1) In (9.72), the bracket of $G_{2,1}$ lies in $[0,\frac12y^2]$, where
$y=m\bGM^{1/2}\tilde\delta$, and that of $G_{2,2}$ is at most $y^2$, so by (S4$'$)
$G_{2,1}(X)+G_{2,2}(X)\le\frac32(1+Ce^{-2\kappa_S})\mt^2\sum_X\tilde\delta^2\le2\mt^2\sum_X\tilde\delta^2$.
(2) By Lemma~\ref{lem:7.31}(a), $e^{G_{2,3}^{\Dep}(X)}\le e^{C_{\Dep}\cdep N_{\Dep}(X)}$
and $|G_{2,3}^\natural(X)|\le\frac12\theta'\mt^2\sum_X(\vartheta^2+\psi^2)$. Thus
$c'_\lambda$ of (9.712) becomes $c_E$. (3) On $\Gp$ the bounds (9.73) to (9.712),
$|E(X)|\le\frac12c'_\lambda\mt^2\sum_X(\vartheta^2+\psi^2)$, remain valid.
Indeed, $|\eps_k|\le|m|\bGM^{1/2}|\vartheta|$ as in (9.78), since the phase at $\psi+g$
is $m\bGM^{1/2}(\vartheta+h)$ with $h\in P\mathbb Z$, and (9.711) follows from (S2) of
Theorem~\ref{thm:6.5}, which is uniform in $\Gp$, by the Schur test
\cite[Lemma~9.6]{BF80} together with the bound $\|\nu\|\le\eps_\nu\mt^2$ of (S5). (4) In (9.716) to
(9.718), the bound $C'_s\le\alpha_D\mt^{-2}$ of Lemma~\ref{lem:7.31}(d) takes the place of the lower
bound on $C_s^{-1}$. With it, $K_4:=2p_4c_E\mt^2C_s'^{1/2}1_XC_s'^{1/2}$ has
$\|K_4\|\le2\alpha_Dp_4c_E\le\frac12$, and
$-\frac12\log\det(1-K_4)\le\operatorname{tr}K_4\le p_4c_9|X|$, the last inequality
by Theorem~\ref{thm:6.12}. Condition \eqref{eq:7.6.cl} is (9.718), and the bound
$\frac12\mt^2\|1_XD\|^2\le\Floc(X)$ of Corollary~\ref{cor:7.11} replaces
$\frac12\int f^2\le F_1$.
\end{proof}

Let $\kappa'$ be a term of the expansion of an activity, with the parameters
of \cite[(7.3) to (7.5)]{GM82} referring to $\cU$-cells: as in Lemma~\ref{lem:A.7}, the cubes
$(a_1,\dots,a_t)$ of a Mayer factor are replaced by its number $t$ of points and
its set of $\cU$-cells. We put
$\cP'_G:=e^{\fn'(V)/p'}$ and
\[
T_1:=\Bigl[\int d\mu_{C'_s}\prod_i|\psi(x_i)|^{p_1}\Bigr]^{1/p_1},\qquad
T_2:=\Bigl[\int d\mu_{C'_s}e^{-p_2F_2(X)}\Bigr]^{1/p_2}.
\]

\begin{proposition}[H\"older split]\label{prop:7.37}
Assume (L8-1) to (L8-4), \eqref{eq:7.6.cl} and
\begin{equation}\label{eq:7.6.coll}
p_2\eps(L')\le\tfrac12(1-1/p_3-c_\lambda).
\end{equation}
Fix $h$, $X$, $s$ and a term $\kappa'$.
(a) There is a kernel $k((x_i))\ge0$ such that
\begin{multline*}
\int d\mu_{C'_s}\bigl|e^{E(X,s)}\kappa'e^{G^\natural(X)}e^{-\Floc(X)}e^{-F_2(X)}\bigr|_0\\
\le\int_{x_i\in X}k((x_i))\int d\mu_{C'_s}e^{E}e^{G_2^\natural}e^{-F_2}
\prod_i|\psi(x_i)|\,e^{\frac12\gamma\sum_X\vartheta^2}e^{-\Floc(X)} .
\end{multline*} (b) The right side of (a) is at most
$\int k((x_i))\,T_1T_2\,e^{-\Floc(X)}\,T_3T_4$, with $T_3$ as in
Lemma~\ref{lem:7.35} and $T_4$ as in Lemma~\ref{lem:7.36}. (c) $T_1\le c_T^{\sum_jn_j}\prod_jn_j!$, where $n_j$ is the number of
points $x_i$ in the $j$-th $\cU$-cell and $c_T$ depends only on $p_1$, $r_C$,
$\alpha_DK_1/\lPV$, $K_1''/\lPV$, $c_u$ and $C_{\cU}$; and
$T_2\le e^{p_2\eps(L')\Floc(X)}$. (d) We have
\[
T_2\,e^{-\Floc(X)}\,T_3\,T_4\le\cP'_G\,e^{c_{13}|X|+C_{\Dep}\cdep N_{\Dep}(X)}
e^{-c'\Floc(X)},
\]
with $c_{13}:=c_8/p'+c_9$ and
$c':=1-1/p_3-c_\lambda-p_2\eps(L')\ge c_0':=\frac12(1-1/p_3-c_\lambda)$; at the
values of Section~\ref{S:G.36}, $c_0'=1/16$. For $k=0$, $\cP'_G=1$.
\end{proposition}

\begin{proof}
\emph{(a)} is the analogue on $\Gp$ of \cite[(7.9)]{GM82}, that is, of
\cite[(9.22), (9.23), (9.81) to (9.823)]{BF80}, with the
differentiated factors of Appendix~\ref{app:B}, the rates of
Proposition~\ref{prop:7.38} and Lemma~\ref{lem:7.33} for the derivatives of
$G_1$. By (S3), $E$ and $G_2^\natural$ are real for real $\psi$.
\emph{(b)} follows as in \cite[(7.11)]{GM82}, that is, \cite[(9.25)]{BF80}, since
$(e^{\frac12\gamma\sum_X\vartheta^2}e^{2\mt^2\sum_X\tilde\delta^2})^{p_3}=e^{p_3\eta_HB_X(\vartheta)}$.
\emph{(c)} For $T_1$ we use \cite[(9.31) to (9.33)]{BF80}, with Lemma~\ref{lem:7.31}(d)
in place of the Neumann series of (9.32) and (R7) for the number of
$\cU$-cells within a given distance. The bound on $T_2$ follows from Corollary~\ref{cor:7.20} with
$t_C=\alpha_D\le2$. \emph{(d)} This is \cite[Lemma~13]{GM82} on $\Gp$. By (c) and Lemmas~\ref{lem:7.35},
\ref{lem:7.36}, the exponents of $\Floc(X)$ add up to
$-1+p_2\eps(L')+1/p_3+c_\lambda=-c'$, and \eqref{eq:7.6.coll} gives
$c'\ge c_0'$. Finally, $e^{\fn'(X)/p'}\le\cP'_G$, as $\fn'$ is monotone in $X$.
\end{proof}

\medskip\noindent\emph{Remark.} In these bounds the perimeter is carried by the factor $\cP'_G$ alone.
Far cells contribute nothing, and $E_{\Gp}$ of Corollary~\ref{cor:7.11}\,(d) drops out
of the bounds, since the factor $e^{-k^2E_{\Gp}/2\bGM}$ of Proposition~\ref{prop:7.3} cancels
against $\Floc(V)=F_1+\frac12k^2\bGM^{-1}E_{\Gp}$ and every later use of $\Floc$
is a lower bound. Hence Theorem~\ref{thm:7.61} contains no factor $\log(1+\min(R,T))$,
and $\cP'_G\le e^{2\cnear(R+T)/p'}$.
\medskip

We measure distances between $\cU$-cells in Debye units,
$\dist(A,B):=\mt\,d_{\Gp}(A,B)$. The covariance rate in these units,
$r_C:=r_{\mathrm{cov}}/\mt$, does not depend on $\tau$, by Theorem~\ref{thm:6.12}.
The rate conditions \cite[(9.123), (9.124)]{BF80} require a rate of at least
$(1-\delta_1)/\tilde l_D\ge\frac12$, whereas on $\Gp$ the rate $r_C$ is small. In (9.95) we take the
exponent with the minus sign that (9.124) has.

\begin{proposition}\label{prop:7.38}
Let $0<\delta_2<\bar\delta<r_C$ and put $c'_{\mathrm{rate}}:=\frac12(r_C-\bar\delta)$.
Every step of \cite[Section~9]{BF80} that uses the rate conditions
\cite[(9.123), (9.124)]{BF80}, the covariance bounds \cite[(9.32), (9.34)]{BF80}
or \cite[(9.95)]{BF80} holds on $\Gp$ with the bounds of
Lemma~\ref{lem:7.31}(d) for $C'_s$, with (S2) of Theorem~\ref{thm:6.5} and with any
$\tau$-independent margin, in particular $c'_{\mathrm{rate}}$, in place of a lower bound on the
rates. The pinning factor of
\cite[Lemmas 9.10, 9.11]{BF80} is used with the margin
$c_p:=(r_C-\bar\delta)/18$.
\end{proposition}

\begin{proof}
The argument uses the rate conditions through margins, and the changes are as follows.

(i) \emph{Tree factor} $e^{\bar\delta d}$, $d=\sum_i\dist(\Delta'_i,\Delta''_i)$.
Link $i$ decays at rate $r_C$ by Lemma~\ref{lem:7.31}(d), and pinning brings a factor
$e^{9c_pd}$. The rest $r_C-\bar\delta-9c_p=c'_{\mathrm{rate}}$ is left for the sum over the
positions of $\Delta''_i$, as $d_{\Gp}$ dominates the Euclidean distance and
by (R7) at most $\frac{4\pi}{3}(r+2C_{\cU})^3/c_u$ $\cU$-cells lie within
Debye distance $r$ of a given one.

(ii) \emph{Mayer factors.}
The factors $e^{\bar\delta L_0}$ and $e^{rO_0}$ of \cite[(7.3), (7.4)]{GM82} are absorbed by
the tree decay of (S2) at the Stage~1 rate
$\sqrt3\kappa_YM_Y\ge4\sqrt3\,r_{\mathrm{cov}}>\bar\delta\mt$
of Proposition~\ref{prop:6.11} and by $e^{-\mu(s-1)}$ of (S2). The details are in the \emph{Tree weights} step of the proof of
Proposition~\ref{prop:B.7}.

(iii) \emph{Wick's theorem.} The rate
$\frac12$ is needed only to obtain (9.32) by a Neumann series, whereas
Lemma~\ref{lem:7.31}(d) gives the pointwise bound directly. The
contraction count holds for all rates, with constants depending on $r_C$,
$K_1''/\lPV$, $c_u$, $C_{\cU}$.

(iv) \emph{Clustering.} Since each application has $W\subseteq Z$ or $W\supseteq Z$, we do not use the factor
$e^{-(1-2\delta_1)\dist(X_1,W)}$ of \cite[(9.122)]{BF80}.

(v) We work in the finite box, and the only role of \cite[(A4.25),
(A4.26)]{BF80} is in the infinite-volume limit of
\cite[Lemma~A4.1]{BF80}.
\end{proof}

\paragraph{Seed configurations and the seeded activity.}
A pair $(U,k)$ is \emph{admissible} if $U=V$ and $k\in\{0,1\}$, or
$U=X_0^c:=V\setminus X_0$ with $X_0\subsetneq V$ a union of $\cU$-cells and
$k=0$. A \emph{seed configuration} in $(U,k)$ is a pair $\sigma=(Z,N_Z)$ with
$Z\subseteq U$ a nonempty, possibly disconnected union of $\cU$-cells,
$Z\supseteq X_1$ if $k=1$, and $N_Z$ a wall field on $E(Z)$ whose wall polymers have their closures in $Z$ (so that $Z$ is a union of elements of $N_Z$); for label configurations, $N_Z\in\cN(Z)$. For $X\supseteq Z$,
$\cN_\sigma(X)$ is the set of $N_X\in\cN(X)$ with $N_X|_{E(Z)}=N_Z$ for which
$Z$ is a union of elements of $N_X$. The \emph{seeded activity}
$\cK^{(Z)}_{N_X}(X)$ is the activity of Proposition~\ref{prop:7.25} at $N_X$
with first set $Y_1=Z$, built with $(C',G^\natural)$:
$\cK^{(Z)}_{N_X}(X)=\sum_{\bar y}\int_{[0,1]^{n-1}}ds\int
d\mu_{C'_s}(\psi)\,e^{E(X,s)}\kappa(\bar y,s)\,Q^\natural_{N_X}(X)$, the sum
over $\bar y=(Z,Y_2,\dots,Y_n)$ with the $Y_j$ ($j\ge2$) disjoint unions of
elements and $\bigcup_jY_j=X$, and
$Q^\natural_{N_X}(X):=e^{G^\natural(X)}e^{-\Floc(X)}e^{-F_2(X)}$. We write
$X=Z\sqcup X_\Sigma\sqcup X_o$, where $X_\Sigma$ consists of the non-seed
$\cU$-cells in wall polymers and $X_o$ of the other non-seed $\cU$-cells (the
\emph{lone} cells). A cell is of \emph{class B} if it is one of the $\cU$-cells of
$\bigcup_\beta S_\beta$, and of \emph{class A} otherwise. Since
$\operatorname{supp}D\cup\operatorname{supp}F_2'\subseteq\bigcup_\beta S_\beta$ by Proposition~\ref{prop:7.9}\,(b), (e),
we have $D=F_2'=0$ on every class-A cell.
As in \cite[p.~214]{BF80}, $|\cdot|_0$ is the absolute value taken inside the
sums and integrals that result when all differentiations in $\kappa'$ are
performed.

\begin{lemma}\label{lem:7.39}
Let $(U,k)$ be admissible, $\sigma=(Z,N_Z)$ a seed configuration in $(U,k)$,
$c_1'>0$, $c_4'$ as in Lemma~\ref{lem:7.32} with $c_4=c_1'$, $e^{a''}\ge4C_\varphi$, where
$C_\varphi<\infty$ is the link sum of Lemma~\ref{lem:A.7}, and
$c_2^L:=\log(2eD_{\cU})+4+c_4'+a''+\alpha$. Then
\[
\sum_{\substack{X\supseteq Z\\ X\subseteq U}}\ \sum_{N_X\in\cN_\sigma(X)}
|\cK^{(Z)}_{N_X}(X)|\,e^{\alpha|X|}
\le\sup_{(\cdot)}e^{c_1'\Floc(X,N_X)+c_2^L|X|+\delta_2d}\int d\mu_{C'_s}
\bigl|e^{E(X,s)}\kappa'Q^\natural_{N_X}(X)\bigr|_0 ,
\]
where the supremum runs over $n$, $T$, $\bar y$, the elementary terms,
$s\in[0,1]^{n-1}$ and $N_X\in\cN_\sigma(X)$; $\kappa'$ is the elementary term
without its tree weight $q(T,s)$ (Lemma~\ref{lem:A.1}) and with every Mayer
factor $\mathcal E$ replaced by
$\sum_{\eta^A}e^{c_{10}L_{\eta^A}}e^{c''_{10}t}\mathcal E_{\eta^A}$
(\cite[(9.18), (9.18a), (9.19)]{BF80}); and $d:=\sum_i\dist(\Delta'_i,\Delta''_i)$, the sum over the links $i$ with a covariance line (types (2) to (5), Lemma~\ref{lem:A.1}).
The constant $c_2^L$ depends only on $D_{\cU}$, $c_4'$, $C_\varphi$, $\alpha$,
not on $L'$, and it is bounded uniformly in $\bGM$ and $\tau$ ($c_4'$ depends on them through $A_w$, but it is doubly exponentially small by Lemma~\ref{lem:7.32}). The bound also holds for the sum restricted by
$X\cap W\neq\emptyset$ if $W\subseteq Z$ or $W\supseteq Z$.
\end{lemma}

This lemma is the version on $\Gp$ of \cite[Lemmas 9.1, 9.4]{BF80}, in the distance-zero form, for
seeds that may depend on the configuration, be disconnected, omit $X_1$ when
$k=0$, and lie in $X_0^c$, and we prove it in Proposition~\ref{prop:A.8}. For the last assertion of the lemma, note that every term of the sum then
satisfies the restriction. In the supremum we take
together the elementary terms with the same link data in the sense of Lemma~\ref{lem:A.7},
as in the definition of $\kappa'$ above and in Proposition~\ref{prop:A.8}; that is, $\kappa'$ contains, inside
$|\cdot|_0$, the sums over the assignments of the points of each
$\mathcal E$-factor to the cells of its set of cells, over their positions and
over their charges (and, for $t=2$, over the forms).

\begin{lemma}\label{lem:7.40}
Insert Proposition~\ref{prop:7.37} and the Wick bound into the right side of
Lemma~\ref{lem:7.39}. In the resulting bound, the factor $e^{c_2'|X|}$ of
\cite[(9.823)]{BF80}, with $e^{c_2^L|X|}$ and $e^{c_{13}|X|}$, is at most
$\prod_{\Delta\subseteq X}e^{c_o}\cdot\prod_{\Delta\text{ class B}}e^{c_B(\bD)}
\cdot\prod_{\Delta\cap\Dep\neq\emptyset}e^{C_{\Dep}\cdep}$, the products running over
$\cU$-cells, and the near-cell factor is pulled out globally as $\cP'_G$.
Here $c_o$ does not depend on $L'$ and is bounded uniformly in $\bD$ and $\tau$, and, with the leftover pinning
exponent $q_1=2$ of Lemma~\ref{lem:B.9},
$c_B(\bD)\le2(C_{\mathcal E}^{1/2}+C_G^{1/2})\bD^{-1/2}$, where $C_{\mathcal E}$ and
$C_G$, the constants of Proposition~\ref{prop:B.10}, are independent of $L'$. Every
class-B cell lies in $X_\Sigma$ or in the seed, and every factor on a class-A cell
carries in addition a nonnegative power of $\bD$. The resulting bound depends on the scale $L'$ only
through the set of cells in $X_\Sigma$, through $\eps(L')$, which is decreasing, and
through the monotone hypotheses (L8-3), (L8-4) and \eqref{eq:7.6.coll}.
\end{lemma}

\begin{proposition}\label{prop:7.41}
Insert Proposition~\ref{prop:7.37} and the Wick bound into
Lemma~\ref{lem:7.39} with $c_1'=c'/4$, and distribute $\Floc$ by the
fractions of Remark~\ref{rem:A.10}. Then for $\bGM\ge\bGMo$, in the resulting bound,
every non-seed $\cU$-cell $\Delta$ of $X$ carries a factor at most
$\eps_\Delta$, where
\begin{align*}
\eps_\Delta&:=\Kcell\,e^{\alpha+C_{\Dep}\cdep}(\bD/\lPV)^{1/6}\quad(\Delta\subseteq X_o),\\
\eps_\Delta&:=\Kcell\,e^{\alpha+C_{\Dep}\cdep+c_B(\bD)-(c'/8)\Phi_{\cU}/(1+D_{\cU})}\quad(\Delta\subseteq X_\Sigma),
\end{align*}
and every seed cell carries at most $\Kcell e^{\alpha+C_{\Dep}\cdep}$, times
$e^{c_B(\bD)}$ if it is of class B. Here
$\Kcell:=K_Be^{(c_2^L+c_{13})C_u}\max(1,\lPV^{1/6})$ with $K_B$ independent
of $\bD$, $L'$ and $\tau$, $\Phi_{\cU}$ is the share of
Proposition~\ref{prop:7.22}, and the exponent $p_0=1/6$ is attained.
\end{proposition}

Lemma~\ref{lem:7.40} and Proposition~\ref{prop:7.41} are a quantitative version on $\Gp$ of \cite[Section~9.8,
(9.81) to (9.824)]{BF80} and are proved in
Appendix~\ref{app:B}. With $p_0:=1/6$ we put
\begin{equation}\label{eq:7.6.c16D}
c_{16}^D:=2\Kcell\,e^{\alpha+C_{\Dep}\cdep}\max\Bigl\{(\bD/\lPV)^{p_0},\;
e^{c_B(\bD)-(c'/8)\Phi_{\cU}/(1+D_{\cU})}\Bigr\}.
\end{equation}

\begin{theorem}[Removal of the logarithm]\label{thm:7.42}
Fix $\alpha$. There is $\tau_0$ such that $c_{16}^D\le1$ for $\tau\ge\tau_0$,
and $c_{16}^D\to0$ as $\tau\to\infty$. Consequently the depletion factor
$e^{C_{\Dep}\cdep N_{\Dep}(X)}$, and with it the $\log\tau$ in $\cdep$, is
absorbed by the per-cell smallness of Proposition~\ref{prop:7.41}.
\end{theorem}

\begin{proof}
The first branch of \eqref{eq:7.6.c16D} is at most $1$ if
$\alpha\le p_0\log(\lPV/\bD)-C_{\Dep}\cdep-\log(2\Kcell)$. This holds for
$\tau\ge\tau_0$, because the per-cell gains are exponentially small in $\tau$ and the depletion exponent is
$O(\log\tau)$: we have $\log(1/\bD)\ge\pi^2\vCb(0)\tau-\frac32\log\tau-C$
as $\mt\le2m_D$, $\cdep=O(\log\tau)$ by Proposition~\ref{prop:6.18}, and
$\log\Kcell$ is affine in $\alpha$. In the second branch, Proposition~\ref{prop:7.22} with
$\mt L\ge c_L/2$ and $\mt L'=2\bGM$ gives
$(c'/8)\Phi_{\cU}/(1+D_{\cU})\ge c\,\bD^{-1}\bGM^{-3}$, which dominates the bound
$c_B(\bD)\le C\bD^{-1/2}$ of Lemma~\ref{lem:7.40}. A depleted cell that is lone
or in $X_\Sigma$ has its factor $e^{C_{\Dep}\cdep}$ inside $\eps_\Delta\le\frac12$
(Section~\ref{S:G.42}).
\end{proof}

Let $\tilde\cK(\Delta)$ be the activity of the single element $\Delta$ at
$k=0$ (first set $\Delta$, $n=1$), with no wall on $\Delta$.

\begin{lemma}\label{lem:7.43}
For $\tau\ge\tau_0$ and every wall-free $\cU$-cell $\Delta$, $\tilde\cK(\Delta)\ge\frac12$.
\end{lemma}

\begin{proof}
A single $\cU$-cell that is an element lies in no $S_\beta$, so $D=0$ on
$\Delta$ and $\Floc(\Delta)=F_2(\Delta)=0$. The label is constant on $\Delta$
and, at $k=0$, contributes only a multiple of $2\pi$ to each phase. Hence
$\vartheta=\psi$ on $\Delta$,
$\tilde\cK(\Delta)=\int d\mu_{C'_\Delta}e^{E(\Delta)+G^\natural(\Delta)}$ with
$C'_\Delta$ the $\Delta\times\Delta$ block of $C'$, and the exponent is real by
(S3). Positivity at $k=0$ allows us to use Jensen's inequality, which gives
\[
\tilde\cK(\Delta)\ge\exp\Bigl(-\mathbb E|E(\Delta)|-\mathbb E|G_1(\Delta)|
-\mathbb E|G_{2,1}(\Delta)+G_{2,2}(\Delta)|-\mathbb E\bigl(G_{2,3}^{\Dep}(\Delta)\bigr)_-
-\mathbb E|G_{2,3}^\natural(\Delta)|\Bigr),
\]
expectations under $d\mu_{C'_\Delta}$. A supremum bound does not suffice here, since
$\sup_\psi(-G_{2,3}^{\Dep}(\Delta))$ is only bounded by
$C(e^{-\mu}+e^{-2\kappa_S})\cdep$, which grows like $\log\tau$. Let $x_b:=\alpha_DK_1\bD/\lPV$. Then
$\bGM C'(x,x)\le x_b$ by Lemma~\ref{lem:7.31}(d) and $M_Y=\mt/\lPV$. Moreover,
$\#\Delta<C_u\mt^{-3}$, and $\Delta$ contains at most $C_u/v_-$ cells. As we will show below, each
expectation is at most $C(\bD^{1/2}+\cdep\bD)$, with $C$ depending only on the
constants fixed before $\tau_0$. Since $\cdep\bD=O(\log\tau)e^{-2.494\tau}\to0$,
the sum is then at most $\ln2$ for $\tau\ge\tau_0$.

We begin with $G_{2,3}^{\Dep}$. At $x\in\Dep$ the $x$-term can be negative only because of the
excess $\rho_1(m,x)>\rho^{\mathrm{ref}}(m)$, which is at most
$Ce^{-\mu}\rho^{\mathrm{ref}}(1)$ for $m=\pm1$ by Proposition~\ref{prop:6.6} and
$e^{-\Phi}\le1$, and is bounded by the $|m|\ge2$ moment bound otherwise. With
$1-\cos u\le\frac12u^2$, $\mathbb E\psi(x)^2=C'(x,x)$ and
Proposition~\ref{prop:6.18},
$\mathbb E(G_{2,3}^{\Dep}(\Delta))_-\le C(e^{-\mu}+e^{-2\kappa_S})\mt^2\,\alpha_DK_1M_Y\,\#(\Dep\cap\Delta)
\le C(e^{-\mu}+e^{-2\kappa_S})\alpha_DK_1\cdep\bD/\lPV$.

The other four terms are moment bounds with $\bGM C'(x,x)\le x_b$.
First, $\mathbb E|E(\Delta)|\le C\#\Delta(m_D^2/\bGM)x_b^{3/2}\le C\bD^{1/2}$, as
each term of $E(\Delta)$ has at least three factors $\eps$ or
$\tilde\eps:=\eps-im\bGM^{1/2}\psi$ and the weight
$e^{2\kappa_S(\sum|m_i|-1)}$ of (S2) absorbs the charges.
Second, $\mathbb E|G_1(\Delta)|\le C\bD$, as
$\log\hat R_v(x)=O(v\bD x^4+e^{-\pi^2v/\bD})$ for $|x|\le\bD^{-1/2}$ by Taylor's
formula and the periodization, with a Gaussian tail beyond.
Third, $\mathbb E|G_{2,1}+G_{2,2}|\le C(\bD+\bD^{1/2})$, as the brackets are of
fourth and third order and
$\mathrm{Var}\,\tilde\delta(x)\le2\alpha_D(K_1/\lPV+1/v_-)\mt$. Finally,
$\mathbb E|G_{2,3}^\natural(\Delta)|\le3C_4\theta'C_u(\alpha_DK_1)^2\bD/\lPV^2$
by Lemma~\ref{lem:7.31}(b) with $n=0$ and $\mathbb E\psi^4=3C'(x,x)^2$.
\end{proof}

The bound $\tilde\cK(\Delta)\ge\frac12$ replaces \cite[(A4.16) to (A4.21)]{BF80},
which use translation invariance, whereas here the positivity at $k=0$ of Lemma~\ref{lem:7.44}
and the diagonal bound of Theorem~\ref{thm:6.12} suffice.
}
\providecommand{\bGM}{\beta_{\mathrm{GM}}}
\providecommand{\mt}{\tilde m}
\providecommand{\Floc}{F_1^{\mathrm{loc}}}
\providecommand{\cU}{\mathcal U}
\providecommand{\cK}{\mathcal K}
\providecommand{\cN}{\mathcal N}
\providecommand{\cP}{\mathcal P}
\providecommand{\sF}{\sigma_F}
\providecommand{\epsE}{\varepsilon_E}
\providecommand{\Funit}{F_{\mathrm{unit}}}
\providecommand{\eps}{\varepsilon}
\providecommand{\Ad}{\mathrm{A}}
\providecommand{\Dcell}{D_{\mathrm{cell}}}
\providecommand{\etaseed}{\eta_{\mathrm{seed}}}
\providecommand{\Kflat}{\mathfrak K^{\flat}}
\providecommand{\Glong}{\Gamma_{\mathrm{long}}}
\providecommand{\Nlong}{\mathcal N^{\mathrm{long}}}

\subsection{The Kirkwood\textendash Salsburg system in the class algebra}
\label{sec:7.7}\label{ssec:7.7}

In this subsection we solve the Kirkwood\textendash Salsburg system of Theorem~\ref{thm:7.29}\,(b) by the argument
of \cite[Appendix~4]{BF80} in a weighted $\ell^1$ algebra over $\Glong$, with the area weight of
the classes (Theorem~\ref{thm:7.52}). The area decay of $\mathcal Z_p(X)/Z$ will then follow from that of
$Z_p/Z$, established in Subsection~\ref{sec:7.8}. Activities and outer partition functions are built with the triple
$(C',G^\natural,\fc\cN)$ of Lemma~\ref{lem:7.31}\,(f).

\begin{lemma}[Positivity]\label{lem:7.44}
At $k=0$ and for real $\psi$, the functions $E$, $G$ and $G^\natural$ of
Lemma~\ref{lem:7.wall} are real, and $e^{E(Y)}Q^\natural_N(Y)>0$ for every wall field
$N$ and every union $Y$ of its elements. Hence $\mathcal Z_p(X)\ge0$ for every $X$
and $p$, $Z>0$, and every activity with $n=1$ is positive.
\end{lemma}

\begin{proof}
By (S3) the terms of $m$ and $-m$ in the charge sums of $E$, $G_1$, $G_{2,1}$, $G_{2,2}$ are
complex conjugate, and $G_{2,3}^{\Dep}$, $G_{2,3}^\natural$, $\Floc$, $F_2$ are real. Hence $e^{E(Y)}Q^\natural_N(Y)$ is the
exponential of a real number. The outer partition functions $\mathcal Z_p(X)$ and the activities with $n=1$ are integrals of
such factors, with $\fc\cN>0$, and $Z$ contains $N=0$.
\end{proof}

\begin{lemma}\label{lem:7.46}
For an additive function $p$ on the cycle space of
$\Gp$ let $\Ad(p)$ be the minimum of $\sum_{e\in E}|N(e)|$ over the $N\in\mathbb Z^E$ with
$[N](\gamma)=p(\gamma)$ for every cycle $\gamma$ of $\Gp$, and $\Ad(p):=\infty$ if no such $N$ exists; for
$p\in\Glong$, $\Ad(p)$ is that of the restriction of $p$ to the cycles of $\Gp$. Then
$\Ad(0)=0$, $\Ad(-p)=\Ad(p)$, $\Ad(p+q)\le\Ad(p)+\Ad(q)$, and $\Ad([N])\le\sum_E|N|$ for every
$N\in\mathbb Z^E$; in particular $\Ad<\infty$ on $\Glong$. Fix $\epsE\in(0,c_0'/8]$ and put
$\sigma':=\epsE\sF$ and $\omega(p):=e^{\sigma'\Ad(p)}$. Then $\omega(0)=1\le\omega$ and $\omega(p+q)\le\omega(p)\,\omega(q)$
on $\Glong$, so $\mathcal A_\omega:=\{a:\Glong\to\mathbb C:\|a\|_\omega:=\sum_p|a(p)|\,\omega(p)<\infty\}$, with
$(a*b)(p):=\sum_qa(q)\,b(p-q)$, is a commutative Banach algebra with unit $\delta_0$, and
$\|a*b\|_\omega\le\|a\|_\omega\|b\|_\omega$.
\end{lemma}

\begin{proof}
Sums and negatives of admissible cochains are admissible, $N$ is admissible for
$[N]$, and $\sum_p|(a*b)(p)|\,\omega(p)\le\sum_{p,q}|a(q)|\,\omega(q)\,|b(p-q)|\,\omega(p-q)$.
\end{proof}

\begin{lemma}\label{lem:7.47}
Let $\Omega$ be a cell of the output. For all sites $x,y\in\Omega$,
$d_{\Gp}(x,y)\le\Dcell:=7r_0(2t_1+1)$, where $t_1:=1+\lceil(12k_P/\chi_{\rm N3})\ln k_P^3\rceil$ ($k_P$ in
coarse units). If $c_L\le1.99\cdot10^{-4}\chi_{\rm N3}$, then $\Dcell<L'$ for $\bGM\ge\bGMo$.
\end{lemma}

\begin{proof}[Sketch]
By (N3) of Theorem~\ref{thm:3.20}, there is no $\chi_{\rm N3}$-weak link. Hence, in the component $\mathfrak K$ of
$\Gp\cap\hat Q(Q)$ with $\Omega=Q\cap\mathfrak K$ as in Definition~\ref{def:3.12}, where $|\mathfrak K|\le k_P^3$, balls with $|B_t|\le|\mathfrak K|/2$ grow
by the factor $1+\chi_{\rm N3}/(6k_P)$ per step. Balls of radius $t_1$ about $x$, $y$ therefore meet. A coarse
path of length $\ell$ gives a $\Gp$-path of at most $7r_0(\ell+1)$ fine steps. With $L_c\ge35$
and $c_L\le1.99\cdot10^{-4}\chi_{\rm N3}$, we obtain $\Dcell<L'$ (Section~\ref{S:H1}).
\end{proof}

\begin{lemma}\label{lem:7.prim}
Let $N$ be a long wall field, $w:=2\pi\bGM^{-1/2}N$, and let $\Omega$ be a cell with the set
$\Kflat_\Omega$ of Proposition~\ref{prop:6.10}\,(iv). Then there is $h':\Kflat_\Omega\to P\mathbb Z$, constant on
$\Omega$, with $\nabla h'=w$ on $E(\Kflat_\Omega)$.
\end{lemma}

\begin{proof}
The graph $(\Kflat_\Omega,E(\Kflat_\Omega))$ is
connected, of diameter $d_\flat\le2+0.469\,\bGM/\mt=2+0.235L'$ (Proposition~\ref{prop:6.10}\,(iv), (v)). We define $h'(x)$ as the sum of $w$ along the path from $x_0\in\Omega$
to $x$ in a breadth-first spanning tree. Then $h'$ is $P\mathbb Z$-valued. Such a primitive can fail only on closed walks of about twice the
diameter of $\Kflat_\Omega$, which is less than $L'$. A bond of $E(\Kflat_\Omega)$ off the
tree closes a walk of length at most $2d_\flat+1<2L'$ with two tree paths, on which $N$ has
period $0$. Hence $\nabla h'=w$ on $E(\Kflat_\Omega)$. For $x\in\Omega$, the tree path from $x_0$ to $x$ and the
same-cell step back form a closed m3-walk of length at most
$d_\flat+\Dcell\le2+0.235L'+L+0.964L'<2L'$, by Lemma~\ref{lem:7.47} and Section~\ref{S:H1}, using $L'\ge16$ and
$L\le L'/(4\bGM)$. Therefore $h'(x)=h'(x_0)$ for each such $x$.
\end{proof}

\addtocounter{theorem}{-1}\renewcommand{\thetheorem}{\thesection.\arabic{theorem}a}%
\begin{definition}[Primitive-admissible wall fields]\label{def:7.primadm}
Let $k\in\{0,1\}$, $X$ a union of $\cU$-cells ($X\supseteq X_1$ if $k=1$), and $N$ a wall field on
$E(X)$ whose wall polymers have their closures in $X$. Then $N$ is \emph{primitive-admissible at
coupling $k$} if for every union $Y\subseteq X$ of elements of $N$ and every cell $\Omega\subseteq Y$
there is $h':\Kflat_\Omega\to P\mathbb Z$, constant on $\Omega$, with $\nabla h'=P(N\one_{E(Y)}+k\xi)$ on
$E(\Kflat_\Omega)$.
\end{definition}

\addtocounter{theorem}{-1}\renewcommand{\thetheorem}{\thesection.\arabic{theorem}b}%
\begin{lemma}[Primitives for long and inside fields]\label{lem:7.insideprim}
(a) Every long wall field on $E(X)$ whose wall polymers have their closures in $X$ is
primitive-admissible at $k=0$. (b) If $X\supseteq X_1$ and $N_X$ is an inside field at $k=1$ with
$p_1(N_X)\in\Glong$, then $N_X$ is primitive-admissible at $k=1$.
\end{lemma}

\begin{proof}
(a) For $Y$ a union of elements, $N\one_{E(Y)}$ is long (Lemma~\ref{lem:7.45}\,(a)); apply
Lemma~\ref{lem:7.prim}. (b) Let $Y\subseteq X$ be a union of elements, $N_Y:=N_X\one_{E(Y)}$ and
$\Omega\subseteq Y$ a cell. Each wall of $N_X$ on $E(\Kflat_\Omega)$ lies in $Y$ (first step of the proof
of Lemma~\ref{lem:7.34}\,(a)), and $\xi$ vanishes off $E(X)$. Hence $N_Y+\xi=(N_X+\xi)\one_{E(X)}$ on
$E(\Kflat_\Omega)$, and its period on a closed m3-walk $\gamma$ in $\Kflat_\Omega$ is $-p_1(N_X)(\gamma)$, which
is $0$ if $|\gamma|<2L'$. The breadth-first-tree argument of Lemma~\ref{lem:7.prim}, whose closed walks
(fundamental cycles of length at most $2d_\flat+1$, same-cell closures of length at most $d_\flat+\Dcell$) are
shorter than $2L'$, gives $h'$ with values in $P\mathbb Z$.
\end{proof}
\renewcommand{\thetheorem}{\thesection.\arabic{theorem}}%

\begin{proposition}[Long wall fields]\label{prop:7.long}
At $k=0$, the bounds of Subsection~\ref{ssec:7.6} and of Appendices~\ref{app:A}
and~\ref{app:B}, and Lemma~\ref{lem:7.53}, hold with $\cN(X)$ and $\cN_\sigma(X)$
replaced by $\Nlong(X)$ and
$\Nlong_\sigma(X):=\{N_X\in\Nlong(X):N_X|_{E(Z)}=N_Z,\ Z\text{ a union of elements of }N_X\}$,
for seed configurations $\sigma=(Z,N_Z)$ with $N_Z\in\Nlong(Z)$, with the
integrand of Lemma~\ref{lem:7.wall} and the same constants. In
Lemma~\ref{lem:7.34}\,(a) the label is replaced by the primitive $h'$ of
Lemma~\ref{lem:7.prim}. The same holds at $k=1$ for the inside fields $N_X$ with $p_1(N_X)\in\Glong$, with the
primitive of Lemma~\ref{lem:7.insideprim}\,(b) and $\cP'=\cP'_G$; in general, the bounds hold for every class of
primitive-admissible wall fields (Definition~\ref{def:7.primadm}).
\end{proposition}

\begin{proof}[Sketch]
For the sums over configurations, in Lemma~\ref{lem:7.32}\,(b) and in steps (2), (3) of the proof of
Proposition~\ref{prop:A.9}, the injection $N_X\mapsto(N_I)_{I\subseteq X}$ into integer
interface values is all that is needed, and it is valid for all wall fields. At a fixed configuration the label
occurs in four places: the phases (Lemma~\ref{lem:7.wall}), $w=0$ inside cells, the walls in
\eqref{eq:7.6.untr} and, in Lemma~\ref{lem:7.34}\,(a), a primitive of $w_Y$ on $E(\Kflat_\Omega)$ constant on
$\Omega$. In the last case $N\one_{E(Y)}$ is long by Lemma~\ref{lem:7.45}\,(a), and Lemma~\ref{lem:7.prim} gives such a primitive
(details in Section~\ref{S:H2}). Only the existence of the primitive is used, so the argument applies to every
primitive-admissible field.
\end{proof}

A fixed fraction of the $\Floc$ of each configuration accounts for its walls,
which carry at least the area of their class.

\begin{proposition}[Absolute convergence]\label{prop:7.51}
There is $c<\infty$, depending on $\tau$ and the fixed constants only, such that
$\|\mathcal Z(X)\|_\omega\le\fc\cN\,e^{(c+c_4')|X^c|}$ for every union $X$ of $\cU$-cells, with $c_4'$ from
Lemma~\ref{lem:7.32} at $c_4=c'/2$. Moreover
$\sum_Y\sum_{N_Y}|\cK^{(Z)}_{N_Y}(Y)|\,\omega([N_Y])\,e^{\alpha|Y|}<\infty$ for the seeds $Z$ of
Theorem~\ref{thm:7.29}\,(b), (c), and $\sum_{N_X}|\cK_{N_X}(X)|<\infty$ in Theorem~\ref{thm:7.29}\,(a),
the sum over the inside fields $N_X$ at $k=1$ with $p_1(N_X)\in\Glong$.
Hence the sums of Theorem~\ref{thm:7.29} converge absolutely.
\end{proposition}

\begin{proof}
By H\"older's inequality as in
Proposition~\ref{prop:7.37}\,(b), (d) at $k=0$, with Lemma~\ref{lem:7.33} at $N=0$ and
Proposition~\ref{prop:7.long}, the term of $N$ in $\mathcal Z_p(X)$ is at most $\fc\cN e^{c|X^c|}e^{-c'\Floc(X^c,N)}$.
Moreover, $\omega([N])\le e^{\sigma'\sum|N|}\le e^{(\epsE/2)\Floc(X^c,N)}$, the first inequality by Lemma~\ref{lem:7.46} and the second by Theorem~\ref{thm:7.14}
and $|N|\le N^2$. We also have $\epsE\le c'$, and by Lemma~\ref{lem:7.32}\,(b) the sum of $e^{-(c'/2)\Floc(X^c,N)}$ over the wall fields on $E(X^c)$
is at most $e^{c_4'|X^c|}$. We bound the seeded sums by
Lemma~\ref{lem:7.53}\,(a), (c$'$), with $\eps=\epsE/2$ at $k=0$ and $\eps=0$ at $k=1$. At $k=1$ the
inside fields with $p_1(N_X)\notin\Glong$ contribute $0$ to Theorem~\ref{thm:7.29}\,(a), as
$\mathcal Z_p(X)=0$ for $p\notin\Glong$ by Definition~\ref{def:7.27}(c). For the inside fields at $k=1$ with
$p_1(N_X)\in\Glong$, Lemma~\ref{lem:7.53}\,(a) applies by Lemma~\ref{lem:7.insideprim}\,(b) and
Proposition~\ref{prop:7.long}.
\end{proof}

\paragraph{Constants of the seeded sums.}
The constant $C_9$ bounds the number of interfaces within Euclidean distance $r$ of a given interface
or site by $C_9(r/L+1)^3$. The Euclidean link range is $g:=2L'+4D_E+2$ ($D_E$ of
Proposition~\ref{prop:7.22}). We put $c_2:=c_S+2=O(\log\tau)$ and $c_{16}:=c_{16}^D+\etaseed\to0$ as $\tau\to\infty$,
where $c_{16}^D$ is defined in \eqref{eq:7.6.c16D} and $c_S$, $\etaseed$ in Section~\ref{S:H1}. Let $\cP':=\cP'_G$ if $k=1$ and $\cP':=1$ if $k=0$. For a seed configuration $\sigma=(Z,N_Z)$, $Z^\Sigma$ is the union of its
wall polymers and $Z^o:=Z\setminus Z^\Sigma$. For an anchor $S_0\subseteq U$, that is, a nonempty union of
$\cU$-cells with $S_0\supseteq X_1$ if $k=1$, and for $N_X\in\cN(X)$, $\sat_{N_X}(S_0)$ is the union of the
elements of $N_X$ meeting $S_0$.

\begin{lemma}[Seeded sums at $k=0$ and $k=1$]\label{lem:7.53}
Let $(U,k)$ be admissible, $\bGM\ge\bGMo$, and $\eps\in[0,c'/2]$. (a) \emph{(Variable first set.)} For every anchor $S_0\subseteq U$,
\[
\sum_{S_0\subseteq X\subseteq U}\ \sum_{N_X\in\cN(X)}\bigl|\cK^{(\sat_{N_X}(S_0))}_{N_X}(X)\bigr|\,e^{\alpha|X|+\eps\Floc(X,N_X)}\le\cP'\,e^{c_2|S_0|}.
\]
(b) \emph{(Strict containment.)} The sum in (a) without the term $X=S_0$, $N_X=0$ is at
most $c_{16}\,\cP'\,e^{c_2|S_0|}$. (c$'$) \emph{(Prescribed seed walls, $k=0$.)} For every seed
configuration $\sigma=(Z,N_Z)$ in $(U,0)$ with $Z^o=\emptyset$, every
$\eps\in[0,c'/8]$ and $c_w:=c'/16$,
\[
\sum_{Z\subseteq Y\subseteq U}\ \sum_{N_Y\in\cN_\sigma(Y)}|\cK^{(Z)}_{N_Y}(Y)|\,e^{\alpha|Y|+\eps\Floc(Y,N_Y)}\,e^{c_w\Floc(Z,N_Z)}
\le\prod_{\mathcal S}e^{-(c'/2)\Floc(\mathcal S)},
\]
the product over the wall polymers $\mathcal S$ of $N_Z$.
\end{lemma}

We prove Lemma~\ref{lem:7.53} in Appendix~\ref{app:A}. Proposition~\ref{prop:A.9} treats a fixed seed
configuration, and for (a), (b) the entropy of the first set is controlled by its walls.

\begin{theorem}[{The Kirkwood\textendash Salsburg system in $\mathcal A_\omega$; \cite[Lemma~A4.1]{BF80} on $\Gp$}]\label{thm:7.52}\label{prop:7.52}
Fix a linear order on the $\cU$-cells; for $X\ne\emptyset$ let $\ell(X)$ be its last
$\cU$-cell and $X^*:=X\setminus\ell(X)$. For a $\cU$-cell $\Delta$ and a union $Y\ni\Delta$
of $\cU$-cells let $\cK'_\Delta(Y)\in\mathcal A_\omega$ be the sequence
$q\mapsto\sum\cK^{(\sat(\Delta))}_{N_Y}(Y)$, the sum over the $N_Y$ of
\eqref{eq:7.7.KS} with $[N_Y]=q$ and $(Y,N_Y)\ne(\Delta,0)$. There is exactly one
family $(r(X))_X$ in $\mathcal A_\omega$ with $r(\emptyset)=\delta_0$ and
$\max_{X\ne\emptyset}e^{-\alpha|X|}\|r(X)\|_\omega<\infty$ such that, for every $X\ne\emptyset$,
\begin{equation}\label{eq:7.7.r}
r(X^*)=\tilde\cK(\ell(X))\,r(X)+\sum_{Y\ni\ell(X),\ Y\cap X^*=\emptyset}\cK'_{\ell(X)}(Y)*r(X^*\cup Y),
\end{equation}
and it satisfies, with $\alpha=\ln8$: (a) $\|r(X)\|_\omega\le\frac8{31}e^{\alpha|X|}$ for every $X\ne\emptyset$;
(b) $\mathcal Z(X)=r(X)*\mathcal Z(\emptyset)$ for every $X$, that is, $\mathcal Z_p(X)=\sum_qr(X)(q)\,Z_{p-q}$.
\end{theorem}

\begin{proof}
By Theorem~\ref{thm:7.29}\,(b) the system has the form of \cite[(A4.5)]{BF80} with values in
$\mathcal A_\omega$, and we run the contraction of \cite[(A4.6) to (A4.13)]{BF80} with $\|\cdot\|_\omega$ in place of $|\cdot|$. The
normalization $r(\emptyset)=\delta_0$ replaces the division by $Z$, which is not available for the sequence
$\mathcal Z(\emptyset)$, and (b) follows from uniqueness in the finite box.

We need two bounds. (K1) $\tilde\cK(\Delta)\ge\frac12$, by Lemma~\ref{lem:7.43}, and of the two bounds this is the one that rests on positivity, at $k=0$ in the proof of
Lemma~\ref{lem:7.43}. (K2)
$\sum_{Y\ni\Delta}\|\cK'_\Delta(Y)\|_\omega e^{\alpha|Y|}\le\frac18$. By Lemma~\ref{lem:7.46} and Theorem~\ref{thm:7.14},
$\omega([N_Y])\le e^{\sigma'\sum|N_Y|}\le e^{(\epsE/2)\Floc(Y,N_Y)}$. Hence (K2) follows from Lemma~\ref{lem:7.53}\,(b) with
$S_0=\Delta$, $k=0$ and $\eps=\epsE/2$, over long fields by Proposition~\ref{prop:7.long}, and from the bound
$c_{16}e^{c_2}\le\frac18$ of Lemma~\ref{lem:A.11}.

\emph{Step 1 (the map).} Let $\mathfrak B$ be the Banach space of families
$s=(s(X))_{X\ne\emptyset}$ in $\mathcal A_\omega$ with $|||s|||:=\max_{X\ne\emptyset}e^{-\alpha|X|}\|s(X)\|_\omega$, put $s(\emptyset):=\delta_0$,
and let $(Ts)(X)$ be $\tilde\cK(\ell(X))^{-1}$ times $s(X^*)$ minus the sum in \eqref{eq:7.7.r}.
A family satisfies \eqref{eq:7.7.r} if and only if $Ts=s$.

\emph{Step 2 (contraction).} By $\|a*b\|_\omega\le\|a\|_\omega\|b\|_\omega$, $|X^*\cup Y|=|X|-1+|Y|$,
$\|s(\emptyset)\|_\omega=1$, (K1) and (K2), we obtain
$e^{-\alpha|X|}\|(Ts)(X)\|_\omega\le2e^{-\alpha}[\max(1,|||s|||)+|||s|||/8]=\frac14\max(1,|||s|||)+|||s|||/32$, and in the same way $|||Ts-Ts'|||\le2e^{-\alpha}(1+\frac18)|||s-s'|||=\frac9{32}|||s-s'|||$, since
$(s-s')(\emptyset)=0$. Let $r$ be the fixed point of $T$.

\emph{Step 3 ((a)).} If $|||r|||\ge1$, Step 2 gives $|||r|||\le\frac9{32}|||r|||$, which is
impossible. Therefore $|||r|||\le\frac14+|||r|||/32$, that is, $|||r|||\le\frac8{31}$.

\emph{Step 4 ((b)).} We set $s(X):=r(X)*\mathcal Z(\emptyset)-\mathcal Z(X)$, which has $s(\emptyset)=0$. By associativity and
commutativity, $X\mapsto r(X)*\mathcal Z(\emptyset)$ satisfies \eqref{eq:7.7.r}, and so does $X\mapsto\mathcal Z(X)$, by
\eqref{eq:7.7.KS} with $X_0=X^*$ and $\Delta=\ell(X)$. Consequently $s$ satisfies the linear part of
\eqref{eq:7.7.r} with $s(\emptyset)=0$. Moreover, $|||s|||<\infty$, since there are finitely many $X$ and
$\|\mathcal Z(X)\|_\omega<\infty$ by Proposition~\ref{prop:7.51}. As in Step 2, $|||s|||\le\frac9{32}|||s|||$, and hence $s=0$.
\end{proof}

The bounds in the proof above are uniform in $\Gp$, $\Lambda^\circ$ and $C$. We obtain the invertibility of $\mathcal Z(\emptyset)$
in $\mathcal A_\omega$ as a consequence of Theorem~\ref{thm:7.52}\,(b): $r(V)*\mathcal Z(\emptyset)=\mathcal Z(V)=\fc\cN\,\delta_0$.
}
\providecommand{\bGM}{\beta_{\mathrm{GM}}}
\providecommand{\Floc}{F_1^{\mathrm{loc}}}
\providecommand{\cU}{\mathcal U}
\providecommand{\sF}{\sigma_F}
\providecommand{\epsE}{\varepsilon_E}
\providecommand{\Funit}{F_{\mathrm{unit}}}
\providecommand{\Ad}{\mathrm{A}}
\providecommand{\RNF}{R_{\mathrm{NF}}}
\providecommand{\thNF}{\theta_{\mathrm{NF}}}
\providecommand{\cNF}{c_3^{\mathrm{NF}}}
\providecommand{\fS}{\mathfrak S}
\providecommand{\Glong}{\Gamma_{\mathrm{long}}}
\providecommand{\Nlong}{\mathcal N^{\mathrm{long}}}

\subsection{The area cost of frozen flux: group charge and the count}
\label{sec:7.8}\label{ssec:7.8}

In this subsection we prove that the partition function at fixed class $r$ carries
the area weight of $r$ (Proposition~\ref{prop:7.57}). As announced at the start of Section~\ref{sec:stage2}, the handles of the
insulating set enter the bounds only here; so does (NF$_*$), which holds by (R10) of Theorem~\ref{thm:3.20}. In a configuration of class $r$, a group of wall polymers
realizing $r$ is re-expanded as a seed at $X=\emptyset$, and its entropy is controlled by
Proposition~\ref{prop:7.56}.

\paragraph{Polymers, realizations and the reference realization.}
Fix $r\in\Glong\setminus\{0\}$ with $Z_r\neq0$. For a configuration $N$ of $Z_r$
(Definition~\ref{def:7.27}), $\mathcal P(N)$ is the set of its wall polymers, and we put
$N_{\mathcal S}:=N\one_{E(\mathcal S)}$ for $\mathcal S\in\mathcal P(N)$ and $N_R:=\sum_{\mathcal S\in R}N_{\mathcal S}$, $q_R:=[N_R]$, $|R|:=\|N_R\|_1$ for
$R\subseteq\mathcal P(N)$. By Proposition~\ref{prop:7.9}\,(a), each wall bond lies in exactly one wall polymer.
Hence $\sum_{\mathcal S\in\mathcal P(N)}N_{\mathcal S}=N$ and $q_{\mathcal P(N)}=r$, and $q_R\in\Glong$
by Lemma~\ref{lem:7.45}\,(a). A family $R$ \emph{realizes} $r$ if $q_R=r$, and is
\emph{irreducible} if no nonempty $T\subseteq R$ has $q_T=0$. An inclusion-minimal realizing
subfamily of $\mathcal P(N)$ is irreducible: if $\emptyset\neq T\subseteq R$ had $q_T=0$, then $T\neq R$ as $r\neq0$,
and $R\setminus T$ would realize $r$. Let $\fS(r)$ be the set of families $R$, with their wall
configurations, that occur as irreducible realizations of $r$ in some configuration
of $Z_r$. We fix $R_0\in\fS(r)$ of least area by a deterministic choice, so that
$|R|\ge|R_0|\ge\Ad(r)$ for all $R\in\fS(r)$, the last inequality by Lemma~\ref{lem:7.46}. A polymer is determined by
its wall interfaces and their values, since closures and links are functions of the
walls. Thus $R$ is determined by $N_R$: the wall components of $N_R$ are those of
the members of $R$, closures from different members are not linked (they are distinct wall
polymers of one configuration), so the wall polymers of $N_R$ are exactly the members of $R$. We also set $\nu(R):=\sum_{I\subseteq\supp N_R}N_I^2$, $c_w:=c'/16$,
$\lambda_E:=2c_w\sF=\frac{c'}8\sF$ and
$\Theta_4(r):=\sum_{R\in\fS(r)}e^{-\lambda_E|R|}\prod_{\mathcal S\in R}e^{-(c'/2)\Floc(\mathcal S)}$. The notion of c-connectedness is that of
Definition~\ref{def:3.17}, and distances between bond sets are Euclidean.

\begin{lemma}\label{lem:7.54}
Let $R\in\fS(r)$. Then $N_R-N_{R_0}=\sum_k\eps_k\nabla1_{P_k}$ with $\eps_k\in\{-1,1\}$, where (i) each $P_k$ is
a nonempty c-connected cell set with c-connected complement in its component $\mathfrak K$
of $\Gp$; (ii) $\partial_{\Gp}P_k\subseteq\supp N_R\cup\supp N_{R_0}$; (iii)
$\sum_k|\partial_{\Gp}P_k|=\|N_R-N_{R_0}\|_1\le|R|+|R_0|\le2|R|$; and (iv) every $P_k$ is joined to $\supp N_{R_0}$
by a chain $P_k=P_{k_0},\dots,P_{k_j}$ with $\dist(\partial P_{k_{i-1}},\partial P_{k_i})\le g$ and
$\dist(\partial P_{k_j},\supp N_{R_0})\le g$.
\end{lemma}

\begin{proof}
(1) Since $[N_R-N_{R_0}]=q_R-q_{R_0}=0$, Definition~\ref{def:7.27} gives $N_R-N_{R_0}=\nabla f$ with $f:V\to\mathbb Z$ constant on each cell.

(2) By the coarea formula for the level sets of $f$, $\nabla f=\sum_t\pm\nabla1_{U_t}$ with
$\sum_t|\partial U_t|=\|\nabla f\|_1$. We split each $U_t$ into its c-components $W$ and write
$\nabla1_W=-\sum_i\nabla1_{C_i}$ over the c-components $C_i$ of $\mathfrak K\setminus W$, where $\mathfrak K$ is the component of $W$.
Since each $C_i$ is c-connected with c-connected complement and the cuts partition, (i)
to (iii) hold.

(3) For (iv) we use irreducibility, which forbids groups of pieces
far from $R_0$, since these would be sub-families of $R$ of class zero. We group the pieces into clusters by the relation ``cuts
within distance $g$''. Let $\mathcal V$ be a cluster no cut of which is within $g$ of $\supp N_{R_0}$,
and let $B:=\bigcup_{k\in\mathcal V}\partial P_k$. Within distance $g$ of $B$, $N_{R_0}=0$ and the other pieces
contribute nothing. It follows that $N_R=N_{\mathcal V}:=\sum_{k\in\mathcal V}\eps_k\nabla1_{P_k}$ in this region, and each wall bond of
$N_R$ within distance $g$ of $B$ lies in $B$. Two wall bonds of one polymer are joined by
a chain of wall bonds of that polymer with consecutive gaps at most $g$, since walls of
one $S_\beta$ have gaps less than $2L'+2D_E$ and linked closures add at most $2D_E+2$.
Hence the polymers of $R$ meeting $B$ form a sub-family $T$ with $N_T=N_R1_B=N_{\mathcal V}$, the
gradient of a cell-constant function, and therefore $q_T=0$. By irreducibility $T=\emptyset$, so that
$N_{\mathcal V}=0$. By (iii) the terms of $N_R-N_{R_0}$ do not cancel, and hence neither do those of the
sub-sum: $\|N_{\mathcal V}\|_1=\sum_{k\in\mathcal V}|\partial P_k|$. So $N_{\mathcal V}=0$ forces $\partial P_k=\emptyset$ for every
$k\in\mathcal V$. This is impossible, since by (i) each $P_k$ is a nonempty proper cell set of its
component $\mathfrak K$. Therefore no such cluster exists: every cluster has a cut within $g$ of
$\supp N_{R_0}$, and chaining inside the cluster gives (iv).
\end{proof}

For a finite set $B$ of interfaces and $s>0$, the $s$-clusters of $B$ and their number
$\varphi_s(B)$ are those of Definition~\ref{def:3.17} (for a cut, of its set of interfaces),
and $\mathrm{MST}(B)$ denotes a minimal spanning tree of the complete graph on $B$ with edge
lengths $\ell_e$ the distances of interfaces. Let $j(\ell):=\#\{j\ge0:2^jL<\ell\}$.

\begin{lemma}[Kruskal's identity]\label{lem:7.55}
For every $s>0$, $\#\{e\in\mathrm{MST}(B):\ell_e>s\}=\varphi_s(B)-1$, and
$\sum_{e\in\mathrm{MST}(B)}j(\ell_e)=\sum_{j\ge0}(\varphi_{2^jL}(B)-1)$.
\end{lemma}

\begin{proof}
The edges of $\mathrm{MST}(B)$ of length $\le s$ form a spanning forest of the threshold
graph $\{\ell\le s\}$, since Kruskal's algorithm processes them first. Thus they number
$|B|-\varphi_s(B)$. For the second identity we sum $1[\ell_e>2^jL]$ over $j\ge0$ and $e$.
\end{proof}

Let $\kappa_{\mathrm{NF}}:=4(c_g+4)/c_g^2$, and let $R_{\mathrm{NF}}$, $\thNF$, $\cNF$, $J$ be as in Definition~\ref{def:3.17}.
Applying (R10) in the form of Proposition~\ref{prop:3.19} to the pieces $P_k$ of Lemma~\ref{lem:7.54}, we obtain
\begin{equation}\label{eq:7.8.NF}
J(P_k)\le\thNF\,\Funit\,|\partial_{\Gp}P_k|/L^2 ,
\end{equation}
with the cut measured in fine bonds, of which a coarse face carries at least $r_0^2$. Here $J$ depends
only on the cut.

\begin{proposition}[The count]\label{prop:7.56}
For $\bGM\ge\bGMo$, every $r\in\Glong\setminus\{0\}$ and every $\nu\ge1$,
$\#\{R\in\fS(r):\nu(R)=\nu\}\le\exp((c'/2)\Funit\,\nu)$. Consequently $\Theta_4(r)\le\frac14$.
\end{proposition}

\begin{proof}
By Kruskal's identity, specifying the positions of the cut patches costs, per dyadic scale, the number of
clusters minus one. Below $R_{\mathrm{NF}}$ this is at most $\log_2(R_{\mathrm{NF}}/L)$ per interface, within the
factor $e^{(c'/2)\Funit}$, and above $R_{\mathrm{NF}}$ it is $J$, bounded by the cut \eqref{eq:7.8.NF}.
We first bound the count and then prove the second assertion, $\Theta_4\le\frac14$.

Fix $R$ with $\nu(R)=\nu$. Let $S$, $S_0$ be the interface sets of $\supp N_R$ and
$\supp N_{R_0}$, put $n:=|S|\le\nu$, $n_0:=|S_0|$ and $T:=S\cup S_0$, and let $I_k$ be the interface set of $\partial P_k$.
An interface has between $c_g'L^2$ and $4L^2$ bonds by (R8) and since block sides are less than
$2L$, and $|R_0|\le|R|\le4L^2\nu$. Hence $n_0\le(4/c_g')\nu$, and by Lemma~\ref{lem:7.54}(iii)
$\sum_k|\partial P_k|\le8L^2\nu$, $\sum_k|I_k|\le(8/c_g')\nu\le\kappa_{\mathrm{NF}}\nu$. The last inequality follows from $8/c_g'\le128/(9c_g)\le4(c_g+4)/c_g^2$, which holds since $c_g'\ge\frac9{16}c_g$ by Section~\ref{ssec:D.5}(e) and $c_g\le\frac14$.
Since a bond where $N_R\neq N_{R_0}$ lies on some $\partial P_k$, we have $S\subseteq S_0\cup\bigcup_kI_k\subseteq T$.

\emph{Step 1 (a spanning forest).} For each $k$ let $t_k$ be the least $j$ for which a chain
$P_k=P_{k_0},\dots,P_{k_j}$ as in Lemma~\ref{lem:7.54}(iv) exists. Let $H$ be the graph on $T$ whose edges are the
edges of the $\mathrm{MST}(I_k)$ and, for each $k$, one parent edge at distance at most $2g$: from an
interface of $I_k$ to one of $S_0$ if $t_k=0$, and otherwise to one of $I_{k'}$, where $P_{k'}$ is the
second piece of a chain of length $t_k$, so that $t_{k'}=t_k-1$. Here we use that a bond and its dual coarse face are within distance
$\sqrt3\,r_0\le g/2$. Each node of $T\setminus S_0$ lies in some $I_k$, and $t$ decreases along parent
edges. Hence the parent edges lead to $S_0$. Consequently $H$ contains a spanning forest $F$ of $T$ with roots in $S_0$,
as one sees by adding a virtual node joined to all of $S_0$ and removing it from a spanning
tree. Let
$B:=\sum_{e\in F}j(\ell_e)\le\sum_{e\in H}j(\ell_e)$.

\emph{Step 2 (encoding).} We record the roots (at most $2^{n_0}$ choices), the shape of the
plane forest on $|T|\le n+n_0$ nodes (at most $4^{n+n_0}$), and for each child the number
$j_e$ and the child among the at most $C_9(2^{j_e}+1)^3\le8C_9\,8^{j_e}$ interfaces within
distance $2^{j_e}L$ of its parent. Sequences of $|F|$ nonnegative integers with sum at
most $B_{\max}$ (the bound of Step~3, a function of $\nu$ only) number at most
$2^{B_{\max}+|F|}$. We then record $S\subseteq T$ (at most $2^{n+n_0}$) and the values $N_I$ with
$\sum N_I^2=\nu$ (at most $4^\nu$, counting signs and compositions of $\sum|N_I|\le\nu$). Since $R$ is
determined by $N_R$, we obtain $\ln\#\{R:\nu(R)=\nu\}\le C(n+n_0+\nu)+B_{\max}\ln16$ with $C:=\ln(64C_9)+2$.

\emph{Step 3 (the bound on $B$).} By Lemma~\ref{lem:7.55},
$\sum_{e\in\mathrm{MST}(I_k)}j(\ell_e)=\sum_{j\ge0}(\varphi_{2^jL}(I_k)-1)$. The scales with $2^jL<R_{\mathrm{NF}}$
number at most $j_*:=\cNF\Funit/\ln2+1$, each contributing at most $|I_k|-1$, while the scales
with $2^jL\ge R_{\mathrm{NF}}$ contribute exactly $J(P_k)$, by Definition~\ref{def:3.17}. A parent edge
has $j\le j_g:=\log_2(2g/L)+1=O(\ln\bGM)$, and there are at most $\sum_k|I_k|\le\kappa_{\mathrm{NF}}\nu$
pieces. With \eqref{eq:7.8.NF},
\[
B\le(j_*+j_g)\kappa_{\mathrm{NF}}\nu+\thNF\Funit\sum_k|\partial P_k|/L^2
\le\Bigl(\frac{\cNF\kappa_{\mathrm{NF}}}{\ln2}+8\thNF\Bigr)\Funit\,\nu+(1+j_g)\kappa_{\mathrm{NF}}\nu=:B_{\max}.
\]

\emph{Step 4.} By Definition~\ref{def:3.17}, $4\cNF\kappa_{\mathrm{NF}}=c_0'/4$ and $8\thNF\ln16\le c_0'/8$.
Hence $\ln\#\le\frac38c_0'\Funit\nu+O((1+\ln\bGM)\nu)\le\frac{c'}2\Funit\nu$ for
$\bGM\ge\bGMo$, as $\Funit$ is exponentially large by Section~\ref{S:H1}.

It remains to prove $\Theta_4\le\frac14$, which rests on the factor $2$ of Lemma~\ref{lem:7.32}(a), that is, on the constant $1/18$ of
Theorem~\ref{thm:7.14}, set against the bound of the count. By Lemma~\ref{lem:7.32}(a), $\Floc(\mathcal S)\ge2\Funit\,\nu(\mathcal S)$ for each
wall polymer $\mathcal S$. Hence the summand of $\Theta_4$ is at most $e^{-c'\Funit\nu(R)}$, and
$\Theta_4(r)\le\sum_{\nu\ge1}e^{(c'/2)\Funit\nu}e^{-c'\Funit\nu}\le2e^{-(c'/2)\Funit}\le\frac14$.
\end{proof}

\begin{proposition}[Area cost of frozen flux]\label{prop:7.57}
For every $r\in\Glong$, $Z_r\le e^{-\sigma'\Ad(r)}Z$.
\end{proposition}

\begin{proof}
We extract an irreducible realization from each configuration and
re-expand it as a seed, whose own $\Floc$ covers its cells and shape
and leaves the fraction $c'/2$ over. The outer factors are bounded by
Theorem~\ref{thm:7.52} and by the quantity being bounded, which closes the
argument. We put $C_{\mathrm{FF}}:=\sup_{s\in\Glong}\omega(s)Z_s/Z$. Then $1\le C_{\mathrm{FF}}<\infty$, by taking
$s=0$ and by Proposition~\ref{prop:7.51}. Let $r\neq0$ with $Z_r\neq0$.

(1) \emph{Union bound.} For each configuration we choose one irreducible realization
$R(N)\subseteq\mathcal P(N)$ by a fixed rule. All terms are nonnegative by Lemma~\ref{lem:7.44}. Hence
$Z_r\le\sum_{R\in\fS(r)}Z_r(R)$, with $Z_r(R)$ as in Theorem~\ref{thm:7.29}\,(c). Dropping the
condition $R=R(N)$ only enlarges the right side.

(2) \emph{Re-expansion.} By Theorems~\ref{thm:7.29}\,(c) and~\ref{thm:7.52}\,(b), with
$\sigma:=(\bigcup R,N_R)$,
$Z_r(R)\le\sum_{Y\supseteq\bigcup R}\sum_{N_Y\in\Nlong_\sigma(Y)}|\cK^{(\bigcup R)}_{N_Y}(Y)|\sum_q|r(Y)(q)|\,Z_{r-[N_Y]-q}$.

(3) \emph{Weights.} By Lemma~\ref{lem:7.46}, $\omega(r)\le\omega([N_Y])\,\omega(q)\,\omega(r-[N_Y]-q)$, and
$\omega([N_Y])\le e^{(\epsE/2)\Floc(Y)}$ by Theorem~\ref{thm:7.14}. The definition of $C_{\mathrm{FF}}$ and
Theorem~\ref{thm:7.52}\,(a), which applies as $Y\supseteq\bigcup R\neq\emptyset$, give
$\sum_q|r(Y)(q)|\,\omega(q)\,\omega(r-[N_Y]-q)Z_{r-[N_Y]-q}\le\frac8{31}e^{\alpha|Y|}C_{\mathrm{FF}}Z$.

(4) \emph{Activity.} So that the sum takes the form $\Theta_4(r)$ of Proposition~\ref{prop:7.56}, we insert
$1\le e^{c_w\Floc(\bigcup R)}e^{-\lambda_E|R|}$, which is valid since
$\Floc(\bigcup R)\ge2\sF|R|$ and $2c_w\sF=\lambda_E$. The final estimate of this step holds without this insertion as well, since the proof of $\Theta_4\le\frac14$ bounds $e^{-\lambda_E|R|}$ by $1$ and Lemma~\ref{lem:7.53}\,(c$'$) without the factor $e^{c_w\Floc(Z,N_Z)}\ge1$ follows from (c$'$). Lemma~\ref{lem:7.53}\,(c$'$) over long fields,
which holds by Proposition~\ref{prop:7.long}, applied with the seed configuration $\sigma$, all of whose $\cU$-cells lie
in wall polymers, and with $\eps=\epsE/2\le c'/8$, bounds the sum over $Y$ and $N_Y$, including
$e^{\alpha|Y|}e^{(\epsE/2)\Floc(Y)}e^{c_w\Floc(\bigcup R)}$, by $\prod_{\mathcal S\in R}e^{-(c'/2)\Floc(\mathcal S)}$. Collecting (1)
to (4), we obtain $\omega(r)Z_r/Z\le\frac8{31}C_{\mathrm{FF}}\Theta_4(r)\le\frac2{31}C_{\mathrm{FF}}$ by Proposition~\ref{prop:7.56}.

(5) \emph{The supremum.} With $s=0$, we obtain $C_{\mathrm{FF}}\le\max(1,\frac2{31}C_{\mathrm{FF}})$. Since $C_{\mathrm{FF}}<\infty$, this gives
$C_{\mathrm{FF}}=1$.
\end{proof}

\medskip\noindent\emph{Remark.} The position entropy is part of the count of $\fS(r)$, and so the bound just
proved carries no position factor, such as a sum over a cycle detecting $r$ or $|\gamma|$, even for a long handle with a tapered thin neck, since its cuts of one
class are compared with $R_0$ by Lemma~\ref{lem:7.54}.
\medskip

\begin{lemma}\label{lem:7.58}
Let $\alpha=\ln8$, $\epsE\in(0,c_0'/8]$, $\sigma'=\epsE\sF$ and $\bGM\ge\bGMo$. For every nonempty union
$X$ of $\cU$-cells and every $p\in\Glong$, $0\le\mathcal Z_p(X)/Z\le\frac8{31}e^{\alpha|X|}e^{-\sigma'\Ad(p)}$. The
bound is uniform in $\Gp$, $\Lambda^\circ$ and $C$.
\end{lemma}

\begin{proof}
By Theorem~\ref{thm:7.52}\,(b) and
Lemma~\ref{lem:7.46}, we have $\omega(p)\mathcal Z_p(X)/Z\le\sum_q|r(X)(q)|\,\omega(q)\cdot\omega(p-q)Z_{p-q}/Z$, and this is at most $\|r(X)\|_\omega\le\frac8{31}e^{\alpha|X|}$ by Proposition~\ref{prop:7.57}
and Theorem~\ref{thm:7.52}\,(a). Nonnegativity follows from Lemma~\ref{lem:7.44}.
\end{proof}
}
\providecommand{\bGM}{\beta_{\mathrm{GM}}}
\providecommand{\mt}{\tilde m}
\providecommand{\Gp}{G'}
\providecommand{\cU}{\mathcal U}
\providecommand{\sF}{\sigma_F}
\providecommand{\sV}{\sigma_V}
\providecommand{\epsE}{\varepsilon_E}
\providecommand{\Ad}{\mathrm{A}}
\providecommand{\cnear}{c_{\mathrm{near}}}
\providecommand{\rbox}{r_{\square}}

\subsection{The deterministic area law}
\label{sec:7.9}\label{ssec:7.9}

\providecommand{\Floc}{F_1^{\mathrm{loc}}}
\providecommand{\cP}{\mathcal P}
\providecommand{\fc}{\mathfrak c}
\providecommand{\cN}{\mathcal N}
\providecommand{\cK}{\mathcal K}
\providecommand{\kbox}{k_{\square}}

We now prove the deterministic area law, Theorem~\ref{thm:7.61}.

\begin{lemma}\label{lem:7.59}
Let $X\supseteq X_1$ be a union of $\cU$-cells, $N_X$ an inside field at $k=1$ on $X$, and
$p:=p_1(N_X)$ its outer class (Definition~\ref{def:7.27}). Then
$\sum_{e\in E(X)}|N_X(e)|+\Ad(p)\ge A_{\Gp}(C)$.
\end{lemma}

\begin{proof}
Since linking numbers are additive, the inside walls and a cheapest cochain with the
forced periods form a wall field at $k=1$, and by Theorem~\ref{thm:7.17} this field spans $C$ modulo the frozen complex. Let $N'\in\mathbb Z^E$ realize $\Ad(p)$, which is finite since $-\xi-N_X1_{E(X)}$ has periods $p$. By definition of $p_1$, $N_X1_{E(X)}$ has periods
$-\mathrm{lk}(\cdot,C)-p$, and hence $M:=N_X1_{E(X)}+N'$ has periods $-\mathrm{lk}(\cdot,C)$. Since $\xi$ has periods
$\mathrm{lk}(\cdot,C)$ by Lemma~\ref{lem:7.6}(d), we get $M+\xi=\nabla m$ for some $m:V\to\mathbb Z$, and $M=\nabla m-\xi$ is
the $k=1$ wall field of $m$. By Theorem~\ref{thm:7.17}, $\sum_E|M|\ge\#\{M\neq0\}\ge A_{\Gp}(C)$, and
$\sum_E|M|\le\sum_{E(X)}|N_X|+\sum_E|N'|=\sum_{E(X)}|N_X|+\Ad(p)$.
\end{proof}

\begin{lemma}\label{lem:7.60}
With $C_{\cU}$ and $c_u$ as in (R7) and $C_X:=3C_{\cU}/c_u$,
$|X_1|\le C_X(\mt^2RT+3C_{\cU}\mt(R+T)+9C_{\cU}^2)$.
\end{lemma}

\begin{proof}
The $\cU$-cells of $X_1$ meet bonds crossing $\Xi$, by Definition~\ref{def:7.23}. Thus they lie
within Euclidean distance $D_E+2$ of $\Xi$, where $D_E=C_{\cU}\mt^{-1}\ge4$, in a set of volume at most
$3D_E(R+3D_E)(T+3D_E)$. Each of these cells has at least $c_u\mt^{-3}$ sites by (R7).
\end{proof}

For a rectangle $C$ of sides $R,T$ we put $R_1(\beta):=(2\rbox+5)L$, where $L$ is
the block scale of Definition~\ref{def:3.1} at $\tau=K\beta$.

\begin{theorem}[Deterministic area law]\label{thm:7.61}
Fix the constants of Section~\ref{ssec:D.1} of the supplement. There is $\bGMo$ such that for
$\bGM\ge\bGMo$, for every output $\Gp=\Gp(D)$ of the construction, for every
$D$, and for every rectangle $C$ with $\min(R,T)\ge R_1(\beta)=(2\rbox+5)L$,
\[
W_{\Gp}(C)\le\exp\bigl(-\sV[A_{\Gp}(C)-\eta_SRT]+c_N(R+T)\bigr),
\]
with
\[
\sV=\epsE\sF=\frac{c_g'\,\mt L}{9216}\,\frac{\mt}{\tau}\quad(\epsE=\tfrac1{128}),\qquad
\eta_S=\frac{18c_2C_X}{\epsE\pi^2c_g'\,\mt L}\,\mt\bGM=O(\mt\bGM\log\tau),
\]
and $c_N$ as in Remark~\ref{rem:7.62}. The
degenerate output $\mathbb X=\Lambda^\circ$ ($\Gp=\emptyset$, $W_{\Gp}(C)=1$,
$A_{\Gp}(C)=0$) is included.
\end{theorem}

Section~\ref{S:H1} gives the values of $\sV$ and $\eta_S$. The geometric input of the proof consists of (R2),
(R5) to (R10), Theorems~\ref{thm:6.5} and~\ref{thm:6.9} to \ref{thm:6.12}, Proposition~\ref{prop:6.18}, the frame,
the coarse alignment of the holes and frozen bonds, Lemma~\ref{lem:4.5}, in Lemma~\ref{lem:6.14}, and (N3) of
Theorem~\ref{thm:3.20}, in Lemma~\ref{lem:7.47}. The frame, which does not depend on $D$, is made up of the box
(Lemma~\ref{lem:3.3}), the grid with its windows and removal cells (Definition~\ref{def:3.4}) and the alignment
(A1) to (A3) (Lemma~\ref{lem:3.5}). The holes and frozen bonds are unions of coarse sites and of whole coarse
faces (Section~\ref{ssec:2.5}).

\begin{proof}
We bound the outer factors of the loop expansion by Lemma~\ref{lem:7.58}. If no
bond of $\Gp$ crosses $\Xi$ (for instance if $\Gp=\emptyset$), then $X_1=\emptyset$ and the bound holds
trivially. Indeed, in this case $j=0$ on $E$, which gives $Z(1,\Xi)=Z(0,\Xi)$ and $W_{\Gp}(C)=1$. Moreover, by the proof of Lemma~\ref{lem:7.6}(c), each plaquette of $\Xi$ is then a
face of a hole cube or a frozen face, hence lies in $\FF$, and Definition~\ref{def:2.11} with $\Sigma=\Xi$ gives
$A_{\Gp}(C)=0$. Finally, $\sV\eta_SRT+c_N(R+T)\ge0$.
Otherwise $X_1\neq\emptyset$, and
\begin{align}
W_{\Gp}(C)&=\sum_{X\supseteq X_1}\sum_{N_X}\cK_{N_X}(X)\,\frac{\mathcal Z_{p_1(N_X)}(X)}Z
\le\frac8{31}\sum_{X\supseteq X_1}\sum_{N_X}|\cK_{N_X}(X)|\,e^{\alpha|X|}e^{-\sigma'\Ad(p_1(N_X))}\label{eq:7.9.82}\\
&\le\frac8{31}\,e^{-\epsE\sF A_{\Gp}(C)}\sum_{X\supseteq X_1}\sum_{N_X}|\cK_{N_X}(X)|\,e^{\alpha|X|+\epsE\Floc(X,N_X)}
\le\frac8{31}\,\cP'_G\,e^{c_2|X_1|}e^{-\sV A_{\Gp}(C)}\nonumber\\
&\le\exp\Bigl(-\sV A_{\Gp}(C)+\sV\eta_SRT+\Bigl[\frac{2\cnear}{p'}+3c_2C_XC_{\cU}\mt
+\frac{9c_2C_XC_{\cU}^2}{R_1(\beta)}\Bigr](R+T)\Bigr),\nonumber
\end{align}
and the bracket, $b(\beta)$, is at most $c_N$. The equality follows from Corollary~\ref{cor:7.30} together with
Lemma~\ref{lem:7.31}(f). Its terms with $p_1(N_X)\notin\Glong$ vanish by Definition~\ref{def:7.27}(c).
Consequently, in \eqref{eq:7.9.82}, each sum over $N_X$ runs over the
inside fields at $k=1$ with $p_1(N_X)\in\Glong$. Lemma~\ref{lem:7.58} gives the first inequality. The second uses Lemma~\ref{lem:7.59}, $\sigma'=\epsE\sF$ and $\Floc(X,N_X)\ge2\sF\sum_{E(X)}|N_X|$, which holds by Theorem~\ref{thm:7.14} since $X$ is
a union of elements. The third
follows from Lemma~\ref{lem:7.53}(a) with $k=1$, $U=V$, $S_0=X_1$, $\eps=\epsE\le c'/2$, applied to these fields.
These fields are primitive-admissible at $k=1$ (Lemma~\ref{lem:7.insideprim}\,(b)), so Lemma~\ref{lem:7.53}(a)
applies to them by Proposition~\ref{prop:7.long}. The last inequality uses $\log(8/31)<0$, Lemma~\ref{lem:7.60} with $c_2C_X\mt^2=\sV\eta_S$
and $R+T\ge2R_1(\beta)$, and $\cP'_G=e^{\fn'(V)/p'}\le e^{2\cnear(R+T)/p'}$ (Lemma~\ref{lem:7.35}).
\end{proof}

\begin{remark}[The perimeter constant]\label{rem:7.62}
The proof above obtains $\epsE\sF A_{\Gp}(C)$ from a
fraction $\epsE$ of the wall energy of the seeded polymer, with the area weight, by Lemma~\ref{lem:7.59}; the seed contributes $e^{c_2|X_1|}$, whose bulk part gives $\eta_S$ and whose
remainder is a perimeter term, and the near cells give the perimeter factor $\cP'_G$. Let $c_N:=\sup_{\bGM\ge\bGMo}b(\beta)$, where $b(\beta)$ is the bracket in \eqref{eq:7.9.82} (see also the remark after the proof of Proposition~\ref{prop:7.37}) and $p'=1.3$. It is
finite and depends only on the constants fixed before $K$, since $\cnear$ is bounded uniformly
for $\bGM\ge\bGMo$ by Lemma~\ref{lem:7.35} together with $\log(1/\mt)/\bGM<0.0632$, and the last two terms tend to $0$
(Section~\ref{S:H1}). Since
$\beta$ and $K$ appear in Sections~\ref{sec:3}, \ref{sec:stage1} and~\ref{sec:stage2} only in the combination $\tau=K\beta$,
that is, in $\bGM=4\pi^2\tau$, and every condition on $\bGM$ used there is a lower bound
in terms of items~0 to~14 of Section~\ref{ssec:D.1}, it follows that Theorem~\ref{thm:7.61}, its threshold
$\bGMo$ and $c_N$ use items~0 to~14, not $K$ and $\beta_0(K)$ (items~15 and~16). Section~\ref{sec:8} needs this, since there $c=c_N+10$ must be absolute.
\end{remark}
}
\section{Proof of the main theorem}\label{sec:8}

\subsection{The chain}\label{ssec:8.1}

The deterministic area law holds for each frozen set $D$, and averaged over $D$ it gives
$e^{-\sigV(1-\etaS)RT+c_N(R+T)}\cdot\EE_De^{\sigV(RT-\AG(C))}$. The second factor is the exponential moment of the area deficit
$RT-\AG(C)$, which is exactly what the min-cut lower tail bounds, and the two deficits $\etaS$ and $\etaT$ add. We fix $K>\Kst$ and $\beta\ge\beta_0(K)$ as in Section~\ref{app:D}. With $\tau=K\beta$, the
conditions defining $\beta_0(K)$ include $\beta\ge\pi^{-2}$, $\delta<1$ and the condition \eqref{eq:2.threshold} of Lemma~\ref{lem:2.6}, that is
$\beta\ge\beta_1(K)$, as well as the hypotheses of
Theorems~\ref{thm:5.1} and~\ref{thm:7.61} for $\beta$ and $\bGM=4\pi^2\tau$. They also include
$\etaT+\etaS\le\frac12$ and $\sigV\le1$, which are true for large $\tau$, since $\etaT\to0$
by Theorem~\ref{thm:5.1}, $\etaS=O(\mt\bGM\log\tau)\to0$ by Theorem~\ref{thm:7.61},
and $\sigV=(\cgp\mt L/9216)\mt/\tau\to0$. Let $C$ be a rectangle with
$\min(R,T)\ge\Rone(\beta)=(2\rbox+5)L$ (Definition~\ref{def:3.1}), $\Lamo=\Lamo(C)$
the box of Definition~\ref{def:3.2}, and $\Lam\supseteq C$ any finite box.

\begin{proof}[Proof of Theorem~\ref{thm:1.2}]
We prove the chain \eqref{eq:8.1} to \eqref{eq:8.5}, where $\EE_D$ is the
expectation over $D\sim\Ber(\delta)^{\otimes(\Lamo_2\setminus\bdtwo\Lamo)}$.

\emph{Step \eqref{eq:8.1}.} Lemma~\ref{lem:3.3}(e) gives
$\dist(C,\partial\Lamo)\ge\max(R,T)\ge1$. Hence Proposition~\ref{prop:2.9} applies
with $\Lam'=\Lamo$ and $\delta'=\delta$. By Lemma~\ref{lem:3.3}(a), the box $\Lamo$ is determined by $C$ and the
constants, so it is the same for all $\Lam$ and $D$ and is fixed before
$D$ is drawn.

\emph{Step \eqref{eq:8.2}.} Theorem~\ref{thm:3.20} shows that for each $D\subseteq\Lamo_2\setminus\bdtwo\Lamo$ the
output $(\XX(D),S(D))$ is a deterministic function of $D$ with $D\subseteq\FF(D)$. Hence Proposition~\ref{prop:2.10} applies to each $D$.

\emph{Step \eqref{eq:8.3}.} Theorem~\ref{thm:7.61} applies to each output
$\Gp(D)$ and each rectangle with $\min(R,T)\ge\Rone(\beta)$, pointwise inside
$\EE_D$. We need it for every $D$, because $\Ber(\delta)$
charges each subset of $\Lamo_2\setminus\bdtwo\Lamo$ and the construction is
deterministic. Its functional $A_{\Gp(D)}(C)$ is the min-cut area of
Definition~\ref{def:2.11} for the same output on the same box $\Lamo$.

\emph{Step \eqref{eq:8.4}} is the identity $-\sigV A=-\sigV RT+\sigV(RT-A)$.

\emph{Step \eqref{eq:8.5}.} Theorem~\ref{thm:5.1} at $\sigma=\sigV\in(0,1]$, applied to the same random set $D$, the box $\Lamo$, the output $\Gp(D)$ and the functional $A_{\Gp(D)}(C)$
of Steps \eqref{eq:8.2} and \eqref{eq:8.3}, bounds
the expectation in \eqref{eq:8.4} by $\exp(\sigV\etaT RT+4\abs{C}+4)$.

\emph{Constants.} Since $\abs{C}=2(R+T)$ and $R+T\ge2$, we have $4\abs{C}+4\le10(R+T)$,
and \eqref{eq:8.5} gives \eqref{eq:1.main} with
$\sigma(\beta):=(1-\etaT-\etaS)\sigV$, $\Rone(\beta)=(2\rbox+5)L$, both depending on
$K$ through $\tau=K\beta$, and $c:=c_N+10$. This constant is absolute in the sense of
Theorem~\ref{thm:1.2}, since $c_N$, a supremum over $\bGM\ge\bGMo$, depends only on constants fixed before $K$ (Remark~\ref{rem:7.62}). Using $\etaT+\etaS\le\frac12$ and the bound
$\cgp\mt L\ge\frac{9}{16}\cg\cL$ of Section~\ref{ssec:D.4}, we obtain
\begin{equation}\label{eq:8.sigma}
\sigma(\beta)\ge\frac{\cg\cL\,\mt}{2^{15}\tau}
\ge\frac{(8\pi^2)^{1/2}}{2^{16}}\,\cg\cL\,\tau^{-1/2}e^{-\pi^2\vCb(0)\tau}
\ge1.35\cdot10^{-4}\,\cg\cL\,\tau^{-1/2}e^{-2.49436\,\tau},
\end{equation}
by $\mt\ge\mD/2$ and the formula $\mD=(8\pi^2\tau)^{1/2}e^{-\pi^2\vCb(0)\tau}$
of Section~\ref{sec:stage1}, with the Debye exponent
$\pi^2\vCb(0)=2.4943550871\ldots$ rounded up and the prefactor rounded down
(see Section~\ref{ssec:D.4}). This proves the second assertion of
Theorem~\ref{thm:1.2} for $\beta\ge\beta_0(K)$. Remark~\ref{rem:8.1} gives the case
$\beta<\beta_0(K)$ at the same $K$, and the first assertion follows with one
fixed $K>\Kst$, for instance $K=378$ (any sufficiently large $K$ also works, by the remark after
Lemma~\ref{lem:2.6}).
\qed
\end{proof}

\begin{remark}[Small $\beta$ and the infinite-volume limit]\label{rem:8.1}\label{rem:8.2}
For fixed $K>\Kst$ and $\beta<\beta_0(K)$, Theorem~\ref{thm:2.1}(b), applied pointwise in $\Lam$ and $C$,
gives Theorem~\ref{thm:1.2} at $\beta$ with the three constants
$\sigma(\beta_0(K))$, $c$, $\Rone(\beta_0(K))$. We use monotonicity in $\beta$ on the Wilson side only:
raising $\beta$ raises $\tau$ but lowers $\delta$, and these two changes push the right
side of Theorem~\ref{thm:2.5} in opposite directions. The bound depends on $\Lam$ only
through Proposition~\ref{prop:2.9}, on $D$ only inside $\EE_D$ and on $C$ only
through $R,T$, and by Theorem~\ref{thm:2.1}(c) it holds for the infinite-volume
limit $\ex{\cos\theta_C}^{W}_{\beta}$, which is nondecreasing in $\beta$. Hence the string
tension
$\sigma_\infty(\beta):=\liminf_{\min(R,T)\to\infty}-\frac{1}{RT}\log\ex{\cos\theta_{C_{R,T}}}^{W}_{\beta}\in[0,\infty]$,
where $C_{R,T}$ is a coordinate rectangle of sides $R,T$, is nonincreasing in $\beta$
and at least $\sigma(\beta)$.
\end{remark}

\begin{remark}[Loops below the non-fragmentation scale]\label{rem:8.3}
If $\diam\Lamo\le\RNF$, for instance if both $3\sqrt3\max(R,T)+\sqrt3$ and
$\sqrt3(r_0(\ell+3)+1)$ are at most $\RNF$, then Phase~2 makes no firing by
Proposition~\ref{prop:3.19}, and Proposition~\ref{prop:5.15}, with the deficit
$\etaT^{(1)}$, replaces Theorem~\ref{thm:5.1} in Step~\eqref{eq:8.5}. The string
tension, by contrast, uses Theorem~\ref{thm:5.1} in full.
\end{remark}

\subsection{The order of the constants}\label{app:D}

We fix the constants once, in the order of the list in Section~\ref{ssec:D.1}, which also gives their values and the
statements that use each of them. The absolute constants come first (item~0), among them $\pi^2\vCb(0)$, the threshold
$14.9662$ of Lemma~\ref{lem:2.6} and the constants quoted from \cite{GM82,GM81,BF80} that do not depend
on the parameters of Theorem~\ref{thm:6.5}; those that do are fixed with these parameters. Then come $r_0$, $\cg$ and the other constants of
the construction that precede the scales (items~1 to~10); $\cL$ and the scales $L_c$, $L$, $k_P$, $\ell$, which depend on $\beta$ only through
$\mt$ (item~11); the constants of Stages~1 and~2, among them the parameters $C$, $\delta_{\mathrm{GM}}$, $\mu$ and
$\kappa_S$ of Theorem~\ref{thm:6.5}, and the number $c'_0=\frac1{16}$ (item~12); $\lPV$, and with it $\MY$ and $\mt=\lPV\MY$ (item~13); the cut-rule
constants $\theta_{\mathrm{NF}}$ and $c_3^{\mathrm{NF}}$ of Definition~\ref{def:3.17} (item~14); $K>\Kst$ (item~15); and $\beta_0(K)$, defined by the conditions listed in
Section~\ref{ssec:D.5} (item~16).

Four placements are forced. First, $k_P/L_c$,
$\vartheta$ and $\varepsilon_{\rm th}$ precede $\chi$, because the expansion
constant of Theorem~\ref{thm:4.rest}(c), which bounds $\chi$, is computed from
them. Second, $\chi$ precedes $\cL$, because the condition
$C_P'\cL^2\le(1-4/\pi^2)/8$ that fixes $\cL$ has
$C_P'\le2916\rbox^5C_{\mathrm{bl}}^2/\chi^2$. Since $1-4/\pi^2<\eta_H$ and
$\mt L\le\cL$, the condition that fixes $\cL$ also gives the condition $C_P'(\mt L)^2\le\eta_H/8$ of
Stage~2. Third, $C$, $\delta_{\mathrm{GM}}$, $c_{10}$, $c''_{10}$, $\mu$ and $\kappa_S$
precede $\lPV$, because the Debye condition of Theorem~\ref{thm:6.5} gives
$\lambda_0=\lambda_0(\kappa_S,\mu,C,\delta_{\mathrm{GM}})$ and each earlier
constant uses $\lPV$ only through $\lPV\le\frac14$. Fourth, $\theta_{\mathrm{NF}}$ and
$c_3^{\mathrm{NF}}$ are built from the pure number $c'_0$, and no earlier
constant depends on them.

Each condition defining $\beta_0(K)$ is, at fixed $K$, a lower bound on $\tau=K\beta$ that
persists for larger $\tau$. Those of Section~\ref{sec:2} are $\beta\ge\pi^{-2}$,
$\beta\ge\beta_1(K)$ and $\delta<1$, which \eqref{eq:2.cdelta} gives for all
large $\beta$. In all the others the left side is a fixed function of $\tau$ (built from $\mt$, $\tau$,
$\log\tau$ and the doubly exponential scale $\RNF$) tending monotonically to its
limit. The value $\beta_0(K)$ is the largest of these thresholds divided by $K$. It is very large ($\abs{\log\epspp}\gtrsim10^{14}$ in
Theorem~\ref{thm:5.1}), and the constants are far from sharp, but the proof uses
only that those fixed before $K$ depend on $\tau$ at most through $\mt$, the scales of item~11 and $\cdep$
(Proposition~\ref{prop:6.18}).
}
\section{Corollaries}\label{sec:9}

In this section we prove the two corollaries of the introduction.
The central phases are described by a U(1) Wilson model with configuration-dependent couplings
at most $n\beta$ (Section~\ref{ssec:1.2}), also in the presence of a Higgs term invariant
under the centre. Since the plaquettes of a finite $2$-complex are attached along
closed edge paths, $U_p$ (the ordered product of the links around $p$) and
$\theta_p$ are defined. We write $\Tr$ for the trace of $n\times n$ matrices.

\begin{proposition}[Fr\"ohlich's comparison]\label{prop:9.1}
Let $G\subseteq U(n)$ be a closed subgroup with $\{zI:\abs{z}=1\}\subseteq G$,
$X$ a finite $2$-complex, and
$\mu_{G,\beta}(dU)\propto\prod_p\exp(\beta\ReP\Tr U_p)\prod_edU_e$, $dU_e$ Haar
measure on $G$ (free boundary conditions). For every closed edge path
$\ell$ in $X$,
\begin{equation}\label{eq:9.frohlich}
\Bigl|\Bigl\langle\Tr\prod_{e\in\ell}U_e\Bigr\rangle_{G,\beta}\Bigr|\le2n\,\ex{\cos\theta_\ell}^{W}_{X,2n\beta},
\end{equation}
where the right side is the U(1) Wilson model on the same complex.
\end{proposition}

\begin{proof}
Let $\nu$ be the law on $U(1)^{E(X)}\times G^{E(X)}$ with density proportional to
$\prod_p\exp(\beta\ReP\Tr(\xi_pU_p))$, where $\xi_e=e^{i\phi_e}$ and
$\xi_p=e^{i\phi_p}$ is the product of the $\xi_e^{\pm1}$ around $p$. Since
$\xi_eI\in G$ is central, its image under $(\xi,U)\mapsto\xi U$ is
$\mu_{G,\beta}$, and $\Tr\prod_{e\in\ell}(\xi_eU_e)=e^{i\phi_\ell}\Tr\prod_{e\in\ell}U_e$.
Given $U$, $\xi$ has the law $\ex{\cdot}_U$ with density proportional to
$\prod_p\exp(\beta\abs{\Tr U_p}\cos(\phi_p-\psi_p))$, where $\psi_p:=-\arg\Tr U_p$. Hence
$\abs{\ex{\Tr\prod_{e\in\ell}U_e}_{G,\beta}}\le n\sup_U\sup_\alpha\ex{\cos(\phi_\ell-\alpha)}_U$.
We bound $\ex{\cos(\phi_\ell-\alpha)}_U$ by $\ex{\cos\theta_\ell}^{W}_{X,n\beta}$
using \cite[Proposition~1 and the remark after its proof, p.~21]{MMSP78}, with
couplings $n\beta\ge\beta\abs{\Tr U_p}$ on the plaquette vectors, where terms with the
same vector up to sign are combined. The remark
states this bound (p.~21) for every $\alpha$, as a particular case of Proposition~1, which gives it only for $\alpha$
equal to the phase of the interaction; for each
$\alpha$ it follows from the proof of Proposition~1, since for the edge vector
$M$ of $\ell$ and two copies $\phi,\phi'$ of the angles
$\cos(M\cdot\phi)-\cos(M\cdot\phi'-\alpha)=2\sin(M\cdot\Phi-\frac\alpha2)\sin(M\cdot\Phi'-\frac\alpha2)$,
where $\Phi=\frac12(\phi'+\phi)$ and $\Phi'=\frac12(\phi'-\phi)$. Consequently, the
left side of \eqref{eq:9.frohlich} is at most $n\ex{\cos\theta_\ell}^{W}_{X,n\beta}$,
which is at most the right side, since U(1) Wilson loops are nonnegative and
nondecreasing in the coupling by Ginibre's inequality, on all finite complexes,
as in Theorem~\ref{thm:2.1}(a), (b). The constants
$2n$ and $2n\beta$ in \eqref{eq:9.frohlich} are those of \cite[(23),~(24)]{Fr79}.
\qed
\end{proof}

\begin{corollary}[Corollary~\ref{cor:1.5}]\label{cor:9.5}\label{cor:9.2}
Let $G$ be as in Proposition~\ref{prop:9.1}, with the Wilson action
$\beta\ReP\Tr U_p$ on finite boxes of $\Zthree$, $\rho$ a representation of $G$
with $\rho(zI)=1$ for $\abs{z}=1$, $(\Phi,\rho_0)$ a probability space of Higgs
field values, and $h\ge0$ a measurable function of
$((\rho(U_e))_e,(\varphi_x)_x)$, the Higgs weight, for any values of the Higgs
couplings, with $0<\int h<\infty$ ($h\equiv1$ is the model without Higgs
field). On a finite box $\Lam$ take the measure with density proportional to
$h((\rho(U_e))_e,(\varphi_x)_x)$ times $\prod_p\exp(\beta\ReP\Tr U_p)$ with
respect to $\prod_edU_e\prod_x\rho_0(d\varphi_x)$. Then \eqref{eq:9.frohlich} holds for every closed path $\ell$, and the bound of
Corollary~\ref{cor:1.5} for every rectangle $C$ with
$\min(R,T)\ge\Rone(2n\beta)$ and every finite box $\Lam\supseteq C$, with
$\sigma$, $\Rone$, $c$ of Theorem~\ref{thm:1.2}. In particular the string
tension of $G$ at coupling $\beta$ is at least $\sigma(2n\beta)>0$, also for
$U(2)$ with an adjoint Higgs field, at all gauge and Higgs couplings.
\end{corollary}

\begin{proof}
Since $\rho(\xi_eU_e)=\rho(\xi_eI)\rho(U_e)=\rho(U_e)$, the proof of
Proposition~\ref{prop:9.1} applies to the joint law with the extra factor
$h((\rho(U_e))_e,\varphi)\prod_x\rho_0(d\varphi_x)$, which cancels from the
conditional law of $\xi$. We then apply Theorem~\ref{thm:1.2} at $2n\beta$. The case of
$U(2)$ with an adjoint Higgs field is included, since $\mathrm{Ad}(zU)=\mathrm{Ad}(U)$. (\cite[Corollary~2]{Fr79} states
confinement of static quarks for all $U(n)$ in three dimensions from the
comparison and the results cited there; the area law is new.)
\qed
\end{proof}

\paragraph{The Georgi\textendash Glashow model.} On a finite box $\Lam$ we take
links $U_e\in SU(2)$, $U_{-e}=U_e^{-1}$, a unit adjoint Higgs field
$\mathsf n_x\in S^2$ (the London limit), the adjoint representation
$R:SU(2)\to SO(3)$, $\beta,\kappa>0$, and
\begin{equation}\label{eq:9.GG}
\mu_{\Lam;\beta,\kappa}\propto\exp\Bigl(\beta\sum_p\tfrac12\Tr U_p+\kappa\sum_{e=(x,y)}\mathsf n_x\cdot R(U_e)\mathsf n_y\Bigr)\prod_edU_e\prod_xd\mathsf n_x
\end{equation}
($dU_e$ Haar, $d\mathsf n_x$ uniform on $S^2$). The weight in \eqref{eq:9.GG} is
invariant under $U_e\mapsto g_xU_eg_y^{-1}$, $\mathsf n_x\mapsto R(g_x)\mathsf n_x$
($g\in SU(2)^\Lam$), and $f(U,\mathsf n)$ is \emph{gauge invariant} if it is. Write
$D(\theta)_e:=\mathrm{diag}(e^{i\theta_e},e^{-i\theta_e})$ and $e_3$ for the
constant field $\mathsf n_x=e_3$.

\begin{proposition}[Georgi\textendash Glashow model at infinite Higgs coupling]\label{prop:9.3}
For every finite box $\Lam$, every $\beta>0$ and every bounded continuous gauge
invariant $f$, the limit
$\lim_{\kappa\to\infty}\ex{f}_{\Lam;\beta,\kappa}$ equals
$\ex{f(D(\theta),e_3)}^{W}_{\Lam,\beta}$, the expectation in the U(1) Wilson
model at coupling $\beta$. In particular, for every closed path $C$,
$\ex{\frac12\Tr\prod_{e\in C}U_e}_{\Lam;\beta,\kappa}\to\ex{\cos\theta_C}^{W}_{\Lam,\beta}$.
\end{proposition}

\begin{proof}
We pass to the unitary gauge. Since $R(SU(2))$ is transitive on $S^2$, gauge invariance gives
$\int\!\!\int F\,d\mathsf n\,dU=\int F(U,e_3)\,dU$ for $F=fw$ and $F=w$, where $w$ is the
weight of \eqref{eq:9.GG}. With $U_e$ of diagonal entries $a_e$, $\bar a_e$ and
off-diagonal entries $b_e$, $-\bar b_e$, the hopping factor is
$\prod_ee^{\kappa-2\kappa\abs{b_e}^2}$, Haar measure is $b_e$ uniform on the unit
disc times $\arg a_e=\theta_e$ uniform, and as $\kappa\to\infty$ the weights
$e^{-2\kappa\abs{b_e}^2}$ concentrate at $b=0$, where $U_e=D(\theta)_e$ and
$\frac12\Tr U_p=\cos\theta_p$. The limit is taken in each fixed box, in the
London limit $\abs{\mathsf n_x}=1$, and for $\frac12\Tr$ (the full trace gives
$2\cos\theta_C$).
\qed
\end{proof}

\begin{corollary}[Corollary~\ref{cor:1.4}]\label{cor:9.4}
At $\kappa=\infty$, for every $\beta>0$, every rectangle $C$ with
$\min(R,T)\ge\Rone(\beta)$ and every finite box $\Lam\supseteq C$, the limit
of $\langle\frac12\Tr\prod_{e\in C}U_e\rangle_{\Lam;\beta,\kappa}$ as
$\kappa\to\infty$ has absolute value $\ex{\cos\theta_C}^{W}_{\Lam,\beta}$,
which is at most $\exp(-\sigma(\beta)RT+c(R+T))$.
\end{corollary}

\begin{proof}
This follows from Proposition~\ref{prop:9.3}, Theorem~\ref{thm:2.1}(a) and Theorem~\ref{thm:1.2}.
\qed
\end{proof}

\begin{remark}[Villain action and other models]\label{rem:9.6}\label{rem:9.7}
The theorem of \cite{GM82} holds in the form of Definition~\ref{def:1.1}: for
$\tau\ge K\beta_0(K)$ and $\beta:=\tau/K$, Lemma~\ref{lem:2.3}(c),(d) as in
Proposition~\ref{prop:2.9} give
$\ex{\cos\theta_C}^{V}_{\Lam,\tau}\le\gfix_{\Lamo,\tau}(C;\emptyset)\le\EE_D\gfix_{\Lamo,\tau}(C;D)$,
Steps \eqref{eq:8.2} to \eqref{eq:8.5} give the bound of Theorem~\ref{thm:1.2}
at $\beta$, and Lemma~\ref{lem:2.3}(b), pointwise in $\Lam$, gives smaller
$\tau$. The proof uses $d=3$ throughout. The Georgi\textendash Glashow model at finite Higgs and weak gauge
coupling is not covered: the centre $\ZZ_2$ of $SU(2)$ contains no U(1), and
\cite[Theorem~1]{Fr79} leads to the pure $\ZZ_2$ gauge theory, which confines
only at strong coupling.
\end{remark}
}

\appendix
\providecommand{\Gp}{G'}
\providecommand{\bGM}{\beta_{\mathrm{GM}}}
\providecommand{\mt}{\tilde m}
\providecommand{\eps}{\varepsilon}
\providecommand{\Du}{D_{\mathcal U}}
\providecommand{\Phiu}{\Phi_{\mathcal U}}
\providecommand{\epsE}{\varepsilon_{E}}
\providecommand{\lPV}{\lambda_{\mathrm{PV}}}
\providecommand{\Kcell}{K_{\mathrm{cell}}}
\providecommand{\Floc}{F_{1}^{\mathrm{loc}}}
\providecommand{\Funit}{F_{\mathrm{unit}}}
\providecommand{\supp}{\operatorname{supp}}
\providecommand{\Vol}{\operatorname{Vol}}
\providecommand{\cK}{\mathcal K}
\providecommand{\cN}{\mathcal N}
\providecommand{\cU}{\mathcal U}
\providecommand{\cP}{\mathcal P}
\providecommand{\sat}{\operatorname{sat}}
\providecommand{\bC}{\mathbf C}
\providecommand{\Dep}{\mathfrak{D}}
\providecommand{\bD}{b_{D}}
\providecommand{\NN}{\mathbb{N}}
\providecommand{\RR}{\mathbb{R}}

\section{The activities in tree form: proofs of Lemmas~\ref{lem:7.39} and~\ref{lem:7.53}}
\label{app:A}

In this appendix we prove Lemmas~\ref{lem:7.39} and~\ref{lem:7.53}, carrying \cite[\S9.1]{BF80} with \cite[\S5]{B78} over to $\Gp$. The arguments are given here in condensed form, and Section~\ref{S:constF} lists the structural facts about the expansion that the arguments use.\footnote{Of these facts, S6 is used in the proof of Lemma~\ref{lem:A.1}, S5 in the paragraph on link data and S13 in item~(4) of the proof of Proposition~\ref{prop:A.8}.} Proposition~\ref{prop:A.5} counts the sets with the tree weight $w(T)$ kept outermost, which cancels the moments $d_j!$ that the count produces.

\emph{Conventions.} We assume that $(U,k)$ is admissible and that $\sigma=(Z,N_Z)$ is a seed configuration in $(U,k)$ (Section~\ref{sec:7.6}). By Proposition~\ref{prop:7.long}, at $k=0$ the configuration sets may be those of long wall fields. We write $|Y|$ for the number of $\cU$-cells of $Y$ and set $\dist(A,B):=\mt\,d_{\Gp}(A,B)$. The $\cU$-cell graph joins two $\cU$-cells when a bond of $\Gp$ joins them, and $D_{\cU}$ bounds its degree; by (R7) we may take $D_{\cU}:=\frac{4\pi}3(2C_{\cU}+2)^3/c_u$, a constant, since the $\cU$-cells joined to a given one lie in the ball of radius $2C_{\cU}\mt^{-1}+2$ about one of its sites and each has more than $c_u\mt^{-3}$ sites ($\mt\le1$).

The quantity $\cK^{(Z)}_{N_X}(X)$ is the sum over $\bar y=(Y_1,\dots,Y_n)$, $Y_1=Z$, of the terms of Proposition~\ref{prop:7.25}. For $2\le j\le n$, \emph{link $j$} is the operator $\kappa(j-1)[Y_j]$ of \eqref{eq:C.E20}, which creates $Y_j$, differentiates $s_{j-1}$, and is a sum of elementary terms of the five types listed after \eqref{eq:C.E16}. We also write $X_j:=Y_1\cup\dots\cup Y_j$. A \emph{tree} is a map $T:\{2,\dots,n\}\to\NN$ with $T(j)<j$. We put $d_i(T):=\#\{j:T(j)=i\}$. Thus $d_n(T)=0$ and $\sum_id_i(T)=n-1$. The weight of \cite[(9.115)]{BF80} is
\begin{equation}\label{eq:A.q}
q(T,s):=\prod_{j=2}^n\ \prod_{k=T(j)}^{j-2}s_k,\qquad w(T):=\int_{[0,1]^{n-1}}q(T,s)\,ds
\end{equation}
since $\frac{d}{ds_{j-1}}(s_{T(j)}\cdots s_{j-1})=s_{T(j)}\cdots s_{j-2}$. The variable $s_{n-1}$ does not occur.

\begin{lemma}[Tree form]\label{lem:A.1}
Let $X\supseteq Z$, $N_X\in\cN_\sigma(X)$ and $\bar y=(Z,Y_2,\dots,Y_n)$, $n\ge2$. Expand $\kappa(\bar y,s)Q^\natural_{N_X}(X)$ into elementary terms $\theta=(\theta_2,\dots,\theta_n)$, $\theta_j$ a term of link $j$, and assign to $\theta$ a tree $T=T(\theta)$ and $\cU$-cells $\Delta'_j\subseteq Y_{T(j)}$, $\Delta''_j\subseteq Y_j$: if $\theta_j$ is of type (2) to (5), it contains $\partial_{s_{j-1}}\bC^{\bar y}(x,y)$ with $x\in Y_j$, $y\in Y_a$, $a<j$, and $T(j):=a$, $\Delta''_j$, $\Delta'_j$ the $\cU$-cells of $x$, $y$; if $\theta_j=(\partial_{s_{j-1}}w_s(\tau))e_\tau$ is of type (1), $T(j):=\min\{i:\supp\tau\cap Y_i\ne\emptyset\}$ and $\Delta'_j$, $\Delta''_j$ are the first $\cU$-cells, in a fixed order, of $\supp\tau$ in $Y_{T(j)}$ and in $Y_j$. Then
\begin{equation}\label{eq:A.tree}
(\theta\text{-term})=q(T,s)\,R_\theta(s),
\end{equation}
where $R_\theta(s)$ depends on $s$ only through the weights $w_s(\tau')\in[0,1]$ of $E$-terms, through the covariance $\bC^{\bar y}_s$ and through $e^{E(X,s)}$. Each $\theta$ receives exactly one tree.
\end{lemma}

\begin{proof}
We apply the tree decomposition \cite[(5.1), (5.2)]{B78} to the expansion of Proposition~\ref{prop:7.25}, with the following modifications. (1) We interpolate the covariance as in \cite[App.~E, (E.3)]{GM82}: for $x\in Y_j$, $y\in Y_a$, $\partial_{s_{j-1}}\bC^{\bar y}(x,y)=s_a\cdots s_{j-2}\,\bC(x,y)$. (2) For the weights $w_s(\tau)$ we use the rule of \cite[p.~89]{GM82}, under which $i\in I$ if $1\le i\le n-1$ and $\tau$ has a cube $a_\alpha\subset Y_j$ for some $j>i$ and a cube $a_\beta\subset\bigcup_{\ell\le i}Y_\ell$. Under this rule the expansion is an identity, by the proof of Proposition~\ref{prop:7.25}, and $\partial_{s_{j-1}}w_s(\tau)=\prod_{k=T(j)}^{j-2}s_k$, as among $k\le j-1$ the set $\supp\tau$ meets both $X_k$ and its complement exactly for $k\ge T(j)$. The rule stated in \cite[p.~212]{BF80} has $Y_{i+1}$ in place of $\bigcup_{j>i}Y_j$. Under that rule a term meeting $Y_1$ and $Y_3$ but not $Y_2$ carries no factor $s_1$, and no tree weight dominates it. Finally, link $j$ differentiates $s_{j-1}$ and no other $s_k$, so each $s_k$ is differentiated once, in link $k+1$. \qed
\end{proof}

As the bounds of Section~\ref{sec:7.6} and Appendix~\ref{app:B} on $R_\theta(s)$ are uniform in $s$ (fact S1 of Section~\ref{S:constF}), the weights lie in $[0,1]$ and $\int q(T,s)\,ds=w(T)$, grouping the elementary terms by their link
data $\lambda=(\lambda_j)_{j\ge2}$ (defined below; they determine the tree $T=T_\lambda$) gives
\begin{equation}\label{eq:A.treesum}
|\cK^{(Z)}_{N_X}(X)|\le\sum_{\bar y}\sum_{\lambda}w(T_\lambda)\,\sup_{s}\sum_{\theta:\,\lambda(\theta)=\lambda}
\Bigl|\int d\mu_s\,e^{E(X,s)}R_\theta(s)\,Q^\natural_{N_X}(X)\Bigr|_0 ,
\end{equation}
with the supremum outside the sum over the terms with one link datum, which
Proposition~\ref{prop:B.7} bounds together.

\begin{lemma}\label{lem:A.2}
For $n\ge2$ and $y_1,\dots,y_n\ge0$,
\begin{equation}\label{eq:A.vol}
\sum_T\int_{[0,1]^{n-1}}q(T,s)\prod_{j=2}^ny_{T(j)}\,ds\le e^{y_1+\dots+y_{n-1}} .
\end{equation}
In particular $\sum_Tw(T)\le e^{n-1}$.
\end{lemma}

\begin{proof}
This is \cite[(5.5)]{B78} and \cite[Lemma~5.5]{BF77}, for real weights. The sum over $T$ is $\prod_{m=1}^{n-1}(y_m+P_{m-1}(s))$ with $P_m:=s_m(y_m+P_{m-1})$, $P_0:=0$, and we integrate over $s_{n-2},\dots,s_1$ in turn using $\int_0^1x\,e^{sx}\,ds\le e^x$. \qed
\end{proof}

\begin{lemma}\label{lem:A.3}
For $n\ge2$, $y_1\ge0$ and $b_2,\dots,b_{n-1}\in[0,1)$,
\begin{equation}\label{eq:A.mom}
\sum_Tw(T)\,y_1^{d_1(T)}\prod_{i=2}^{n-1}d_i(T)!\,b_i^{d_i(T)}\le e^{y_1}\prod_{i=2}^{n-1}(1-b_i)^{-1}.
\end{equation}
\end{lemma}

\begin{proof}
We use the identity $d!\,b^d=\int_0^\infty(bt)^de^{-t}\,dt$ for each $i=2,\dots,n-1$. Since $\prod_jy_{T(j)}=\prod_iy_i^{d_i(T)}$ and $d_n(T)=0$, Tonelli's theorem and \eqref{eq:A.vol} with $y_i:=b_it_i$ ($2\le i\le n-1$) bound the left side by $\int_{\RR_+^{n-2}}e^{y_1+\sum_ib_it_i}e^{-\sum_it_i}\,dt=e^{y_1}\prod_{i=2}^{n-1}(1-b_i)^{-1}$. \qed
\end{proof}

For a finite set $A$ of $\cU$-cells, $\mathfrak Y(A)$ is the set of nonempty unions $Y$ of $\cU$-cells each of whose components in the $\cU$-cell graph meets $A$. For $a=\log(2eD_{\cU})+a'+a''$ with $a'\ge4$, $a''\ge0$, and all integers $d\ge0$,
\begin{equation}\label{eq:A.rooted}
\textstyle\sum_{Y\in\mathfrak Y(A)}|Y|^d\,e^{-a|Y|}\le e^{-a''}\,2^{|A|}\,d!\,2^{-d}:
\end{equation}
if $W$ contains each $\cU$-cell independently with probability $p\in[0,1]$, then $p^{|Y|}(1-p)^{|\partial Y|}$, $\partial Y$ the set of cells outside $Y$ adjacent to $Y$, is the probability that $W\supseteq Y$ and $W\cap\partial Y=\emptyset$; for each $W$ at most $2^{|A|}$ sets $Y\in\mathfrak Y(A)$ have this property, since such a $Y$ is the union of the components of $W$ that meet $A\cap Y$. Hence $\sum_{Y\in\mathfrak Y(A)}p^{|Y|}(1-p)^{|\partial Y|}\le2^{|A|}$, and with $|\partial Y|\le D_{\cU}|Y|$ and $p=1/(D_{\cU}+1)$, at most $2^{|A|}(e(D_{\cU}+1))^N\le2^{|A|}(2eD_{\cU})^N$ sets $Y\in\mathfrak Y(A)$ have $|Y|=N$. We conclude with $N^de^{-a'N/2}\le d!\,(2/a')^d$.

\emph{Link data.} The \emph{link datum} $\lambda_j$ of an elementary term $\theta_j$ consists of its type, of the number of points and the set of $\cU$-cells containing them for each of its $E$-terms (at most two), and of the cells $\Delta'_j$, $\Delta''_j$ of Lemma~\ref{lem:A.1}. The set $A(\lambda_j)$ consists of $\Delta''_j$ and the $\cU$-cells of its $E$-terms, and $\varphi\ge0$ is a weight on link data with
\begin{equation}\label{eq:A.Cphi}
C_\varphi:=\sup_{\Delta_0}\sum_{\lambda:\,\Delta'=\Delta_0}\varphi(\lambda)\,2^{|A(\lambda)|}<\infty .
\end{equation}
Each $Y_j$ ($j\ge2$) of a term with $\theta_j\ne0$ lies in $\mathfrak Y(A(\lambda_j))$. Indeed, each element is connected in the $\cU$-cell graph, since a wall component is a component of $\hat\Sigma$ and, by Definition~\ref{def:7.23}, a closure adds bond neighbours and linked closures share a $\cU$-cell. Each element of $Y_j$ also meets $S(\theta_j)\setminus X_{j-1}$, because link $j$ creates $Y_j$ as the union of the elements that meet this set. This set lies in the cells of $A(\lambda_j)$, as the point $y$ of a covariance line lies in $X_{j-1}$.

\begin{proposition}\label{prop:A.5}
Let $a$ be as in \eqref{eq:A.rooted}, with $2C_\varphi e^{-a''}\le\frac12$. For every seed configuration $\sigma=(Z,N_Z)$,
\begin{equation}\label{eq:A.comb}
\sum_{n\ge1}\ \sum_T\ \sum_{(Y_j,\lambda_j)_{j=2}^n}w(T)\prod_{j=2}^n\varphi(\lambda_j)\,e^{-a|Y_j|}\le2e^{|Z|},
\end{equation}
where $Y_1:=Z$, the sum runs over all trees $T$, all link data with $\Delta'_j\subseteq Y_{T(j)}$, and all $Y_j\in\mathfrak Y(A(\lambda_j))$ (the $Y_j$ may overlap, and $Y_j\subseteq U$ need not hold), and the term $n=1$ equals $1$.
\end{proposition}

\begin{proof}
We keep the tree outermost with its weight; each set is rooted at its own link, and each link in an earlier set. The roots in earlier sets produce the moments of Lemma~\ref{lem:A.3}. We fix $n\ge2$ and $T$ and sum in the order $(Y_n,\lambda_n),\dots,(Y_2,\lambda_2)$, innermost first, and inside stage $j$ first over $Y_j$, then over $\lambda_j$ with $\Delta'_j$ fixed, then over $\Delta'_j\subseteq Y_{T(j)}$. In this order every constraint ties a variable only to variables summed further out.

We claim that at stage $j$ the summand depends on $Y_j$ only through $|Y_j|^{d_j(T)}e^{-a|Y_j|}$ and the constraint $Y_j\in\mathfrak Y(A(\lambda_j))$. At stage $j$, \eqref{eq:A.rooted} gives $\sum_{Y_j\in\mathfrak Y(A(\lambda_j))}|Y_j|^{d_j}e^{-a|Y_j|}\le e^{-a''}2^{|A(\lambda_j)|}d_j!\,2^{-d_j}$. The sum over $\lambda_j$ with $\Delta'_j$ fixed then gives at most $C_\varphi e^{-a''}d_j!\,2^{-d_j}$ by \eqref{eq:A.Cphi}, and the remaining sum contributes $\sum_{\Delta'_j\subseteq Y_{T(j)}}1=|Y_{T(j)}|$, a factor that is handled at stage $T(j)<j$. This preserves the claim for the stages that follow.

After stage $2$ the factor $|Z|^{d_1(T)}$ remains, and the sum over $(Y_j,\lambda_j)_{j\ge2}$ is at most $(C_\varphi e^{-a''})^{n-1}|Z|^{d_1(T)}\prod_{j=2}^{n-1}d_j(T)!\,2^{-d_j(T)}$. Multiplying by $w(T)$ and summing over $T$, we obtain the bound $(C_\varphi e^{-a''})^{n-1}e^{|Z|}2^{n-2}$ from Lemma~\ref{lem:A.3} with $y_1=|Z|$ and $b_i=\frac12$. Finally $1+e^{|Z|}\sum_{n\ge2}(C_\varphi e^{-a''})^{n-1}2^{n-2}\le1+\frac12e^{|Z|}\le2e^{|Z|}$. \qed
\end{proof}

\begin{remark}[The printed deduction]\label{rem:A.6}
In \cite[p.~329]{B78} the bound corresponding to Proposition~\ref{prop:A.5}, (5.4), is deduced from (5.5), $\sum_T\int ds\,q(T,s,\bar y)\le e^{|X|}$ (trees at fixed sets), and (5.6) (sets at fixed tree, with the supremum over trees outside). Writing the summand of (5.4) as $f(T,\bar y)g(T,\bar y)$ with $f:=\int ds\,q(T,s,\bar y)e^{-|X|}$, the deduction uses
\[
\sum_{T,\bar y}f(T,\bar y)\,g(T,\bar y)\le\Bigl(\sup_{\bar y}\sum_Tf(T,\bar y)\Bigr)\Bigl(\sup_T\sum_{\bar y}g(T,\bar y)\Bigr),
\]
which fails for general nonnegative $f,g$: for $T,\bar y$ ranging over $\{1,\dots,N\}$ and $f=g=\delta_{T,\bar y}$ the left side is $N$ and the right side $1$. The conclusion holds with a different per-cell constant, by Proposition~\ref{prop:A.5}, or by keeping (5.6) and using $\sum_{T,\bar y}fg\le(\sum_T\sup_{\bar y}fe^{-\epsilon|X|})\sup_T\sum_{\bar y}ge^{\epsilon|X|}$, whose first factor is at most $((1+\epsilon)/\epsilon)^{n-1}$ by Lemma~\ref{lem:A.3} and whose second is (5.6) with $c_5$ raised by $2\epsilon$.

In \cite[p.~215]{BF80} the joint choice for 5$\Sigma$ and 6$\Sigma$ cites ``(5.6) of [1] or Lemma~5.5 of [2]''. The inequality used there is \cite[(5.5)]{B78}, which \cite[Lemma~5.5]{BF77} states at fixed weights, and the conclusion of \cite[Lemma~9.1]{BF80} holds. In \cite[pp.~213--217]{BF80} the sums are listed in the inverse order in which they are performed, so the constraint of 3$\Sigma$, that the set contain one of the cubes `mentioned in 10$\Sigma$', refers to an inner sum; in the count of Proposition~\ref{prop:A.5} each constraint links a variable only to variables outer to it.
\end{remark}

For a link datum $\lambda$, $O_0(\lambda)$ is the number of points of its $E$-terms and $L_0(\lambda)$ the sum over its $E$-terms of half the $\dist$-length of a shortest tree spanning their $\cU$-cells ($L_0=O_0=0$ for type (2)). An $E$-term with cells in two components of $\Gp$ has coefficient $0$ by Proposition~\ref{prop:6.3}(d), and is absent. We put
\begin{equation}\label{eq:A.phi}
\varphi(\lambda):=e^{-\delta_2\one_{\mathrm{line}}(\lambda)\dist(\Delta',\Delta'')}\,e^{-c_{10}L_0(\lambda)}\,e^{-c''_{10}O_0(\lambda)} ,\qquad c_\gamma:=\sup_{\Delta_0}\sum_{\Delta}e^{-\gamma\dist(\Delta_0,\Delta)}\quad(\gamma>0);
\end{equation}
where $\one_{\mathrm{line}}(\lambda)=1$ if $\lambda$ is of a type with a covariance line (types (2) to (5), Lemma~\ref{lem:A.1}) and $\one_{\mathrm{line}}(\lambda)=0$ for type (1); $c_\gamma<\infty$, since by (R7) at most $\frac{4\pi}3(r+2C_{\cU}+1)^3/c_u$ $\cU$-cells lie within distance $r$ of a given one.

\begin{lemma}[Link sum]\label{lem:A.7}
Let $c_{10},\delta_2>0$ and $e^{c''_{10}}\ge16e\max(1,c_{c_{10}/2})$. Then $C_\varphi\le1+8c_{\delta_2}<\infty$. Moreover, for every $E$-term $\mathcal E$ of $\lambda$, with $\mathfrak a$ the set of $\cU$-cells of its points, and every augmented tree $\eta^A$ of \cite[Appendix~1]{BF80} whose marked points lie in the cells of $\mathfrak a$, one in each, the contribution of $\mathcal E$ to $L_0(\lambda)$ is at most $L_{\eta^A}$; so $e^{c_{10}L_0(\lambda)}e^{c''_{10}O_0(\lambda)}$ is dominated, inside $|\cdot|_0$, by replacing each $E$-term $\mathcal E$ by $\sum_{\eta^A}e^{c_{10}L_{\eta^A}}e^{c''_{10}t}\mathcal E_{\eta^A}$, for any decomposition $\mathcal E=\sum_{\eta^A}\mathcal E_{\eta^A}$ (\cite[(9.18), (9.18a)]{BF80}). Which point of $\mathcal E$ lies in which cell of $\mathfrak a$, the positions and the charges are not part of $\lambda$; they are summed inside $|\cdot|_0$.
\end{lemma}

\begin{proof}
The argument is that of 8$\Sigma$ and 10$\Sigma$ of \cite[p.~217]{BF80}, with the choice $B_8B_{10}=e^{c''_{10}O_0}e^{c_{10}L_0}$, together with (9.18) and the lengths \cite[(A1.3)]{BF80}, adapted as follows.

(1) Positions are sets of $\cU$-cells. We count a set of $k$ cells once for its $(k-1)!$ orderings, and the tree sum gives $e^kc_{c_{10}/2}^{k-1}$ (by Cayley's formula, $k^{k-2}$ labelled trees on $k$ vertices, and $k^{k-2}/(k-1)!\le e^k$). The terms with the same link datum differ in the assignment of the points of each $E$-term to the cells of its set, in their positions and in their charges, and for $t=2$ also in the form. Proposition~\ref{prop:B.7} bounds their sum through the bound (M4) of Lemma~\ref{lem:B.1}(d), which holds for the sum over all assignments. This is needed because a set of $k$ cells carries $k!\,S(t,k)$ assignments of $t$ points, where $S$ denotes the Stirling numbers of the second kind, and $t!$ assignments if $k=t$.

(2) The factor $2^{|A(\lambda)|}$ costs $4^k$ per $E$-term with a marked cell. This raises by $\log2$ the constant $c''_{10}$; \cite[p.~217]{BF80} gives no value for it. The cell $\Delta''$ costs $2c_{\delta_2}$ in each of the four types with a covariance line. Hence $C_\varphi\le1+8c_{\delta_2}$.

(3) The length $L_{\eta^A}$ is an infimum over the positions of the marked points in the cells of $\mathfrak a$ and of the auxiliary points, measured with $\dist$, and a tree on given points has at least half the length of a shortest tree spanning them, which dominates one spanning their cells. Hence $L_0\le L_{\eta^A}$, and $e^{c_{10}L_0}|\mathcal E|\le\sum_{\eta^A}e^{c_{10}L_{\eta^A}}|\mathcal E_{\eta^A}|$ as $\mathcal E=\sum_{\eta^A}\mathcal E_{\eta^A}$. Under the conditions $c_{10}\le\frac79r_C$ and $\mu\ge1+2c''_{10}$, whose constants are fixed in the order given in Section~\ref{S:H1}, the proof of Proposition~\ref{prop:B.7} absorbs the factors $e^{c_{10}L_{\eta^A}}e^{c''_{10}t}$ of $\kappa'$ when it bounds the weights of the augmented trees. \qed
\end{proof}

In the count of \cite[(7.3) to (7.5)]{GM82} the Mayer factors $E(a_1,\dots,a_t)$ keep the ordered tuple of cubes $(a_1,\dots,a_t)$ among the summed parameters, with a weight exponential in the least tree length $L(a_1,\dots,a_t)$ of \cite[(1.11b)]{GM82}. This count is that of \cite[10$\Sigma$]{BF80}. There, on \cite[p.~214]{BF80}, the positions of the cubes are summed at a fixed augmented tree $\eta^A$, with the constant weights $b_{\eta^A}$ of \cite[(A1.4)]{BF80}. With the least tree length in place of these weights, the counting step needs a coefficient bound for the sum over all positions of the points with their tree length inside the sum. The derivation of \cite[(4.31)]{GM82} provides it through the pinned norm \cite[(4.12)]{GM82}. Theorem~\ref{thm:6.5} states (S2) in this form, and the count above uses it as explained in Lemma~\ref{lem:A.7} and its proof.

\begin{proposition}[Lemma~\ref{lem:7.39} on $\Gp$]\label{prop:A.8}
Lemma~\ref{lem:7.39} holds, with $C_\varphi$ the constant of Lemma~\ref{lem:A.7}. The same proof gives the bound with $e^{\alpha|X|}$ replaced on the left by $e^{\alpha|X|+\eps\Floc(X,N_X)}$, $\eps\ge0$, and $e^{c_1'\Floc}$ replaced on the right by $e^{(c_1'+\eps)\Floc}$.
\end{proposition}

\begin{proof}
We follow the proof of \cite[Lemma~9.1]{BF80} by the device of \cite[(9.12)]{BF80}, the bound $\sum_v|f(v)|\le(\sum_vB(v)^{-1})\sup_v|B(v)f(v)|$ over the data $v=(n,T,\bar y,(\lambda_j)_{j\ge2},N_X)$ of \eqref{eq:A.treesum}, with $B(v):=e^{a|X\setminus Z|}e^{c_1'\Floc(X,N_X)}\prod_j\varphi(\lambda_j)^{-1}$, $a:=\log(2eD_{\cU})+4+c_4'+a''$, $e^{a''}\ge4C_\varphi$. Terms with the same link data differ only inside $|\cdot|_0$, namely in the maps $l$ of Lemma~\ref{lem:B.2} and in the data of the $E$-terms listed in the proof of Lemma~\ref{lem:A.7}, and their sum is bounded as there. We change the argument at the following points.

(1) For the count of the sets we use Proposition~\ref{prop:A.5} with $a'=4$, after summing $N_X$ innermost by Lemma~\ref{lem:7.32}(b) and keeping only $Y_j\in\mathfrak Y(A(\lambda_j))$. Thus $\sum_vw(T)B(v)^{-1}\le e^{(1+c_4'+\log2)|Z|}\le e^{a|Z|}$, the supremum carries $e^{c_2^L|X|}$, $c_2^L:=a+\alpha$, and Lemma~\ref{lem:A.7} turns $\prod_je^{c_{10}L_0(\lambda_j)+c''_{10}O_0(\lambda_j)}$ and $R_\theta$ into $\kappa'$. (2) The seed is fixed and enters through $|Z|$ and $\cN_\sigma(X)$ alone, not through its connectivity. Seeds depending on the configuration are reduced to fixed ones in Proposition~\ref{prop:A.9}. (3) No step refers to the set $X_1$ of Definition~\ref{def:7.23}(c). (4) For $U=X_0^c$ the activity is independent of the region, since it depends only on the sets of the sequence, the block of the covariance on their union and the configuration there. (5) We use only the distance-zero form ($W\subseteq Z$ or $W\supseteq Z$, where the restriction $X\cap W\ne\emptyset$ is vacuous), since on $\Gp$ a $\cU$-cell can have $d_{\Gp}$-diameter of order $L'$ and the clustering form \cite[(9.122)]{BF80} would therefore cost $e^{O(\mt L')}=e^{O(\tau)}$ per non-seed cell. As $\Floc(X,N_X)$ depends neither on $\psi$ nor on $s$, we may move $e^{\eps\Floc}$ into the supremum. \qed
\end{proof}

For an anchor $S_0$, $\Sigma_{\mathrm{sat}}(S_0)$ is the set of seed configurations $\sigma=(Z,N_Z)$ with $Z\supseteq S_0$ each element of which meets $S_0$. We write $\Floc(\mathcal S,N_{\mathcal S})$ for the sum of $F_1(S_\beta)$ over the wall components of a wall polymer $\mathcal S$, $D_I$ for the largest number of interfaces within Euclidean distance $g$ of a given interface or site, and ($\beta1$), ($\beta3$), ($\beta4$) for the conditions \eqref{eq:A.beta} on $\bGMo$ of Section~\ref{S:H1}.

\begin{proposition}[Core estimate]\label{prop:A.9}
Let $(U,k)$ be admissible, $\bGM\ge\bGMo$ and $\eps\in[0,c'/2]$. For every seed configuration $\sigma=(Z,N_Z)$ in $(U,k)$,
\begin{equation}\label{eq:A.core}
\sum_{\substack{X\supseteq Z\\ X\subseteq U}}\ \sum_{N_X\in\cN_\sigma(X)}|\cK^{(Z)}_{N_X}(X)|\,e^{\alpha|X|+\eps\Floc(X,N_X)}\le\cP'\,e^{c_S|Z^o|}\,e^{-(11c'/16-\eps)\Floc(Z,N_Z)},
\end{equation}
and the sum restricted to $X\neq Z$ is at most $c_{16}^D$ times the right side. Consequently Lemma~\ref{lem:7.53} holds.
\end{proposition}

\begin{proof}
The proof combines the per-cell bound of \cite[\S9.8]{BF80}, applied after Lemma~\ref{lem:7.39} with the factor $e^{-\Floc}$ divided region by region, with a split of the sum by the first set.

(1) \emph{Booking.} Proposition~\ref{prop:A.8}, with $c_1'=c'/4$ and the extraction $e^{\eps\Floc}$, together with Proposition~\ref{prop:7.37}, Lemma~\ref{lem:7.40} and Proposition~\ref{prop:7.41}, leaves $e^{-(3c'/4-\eps)\Floc(X)}$, whose part on $Z$ is constant on $\cN_\sigma(X)$. We divide this factor between the two regions as in Remark~\ref{rem:A.10} of the supplement. On $X\setminus Z$ we use $c'/16$ for $|D|$, $|F_2'|$ and $c'/8$ for the wall polymers, and on $Z$ we use $c'/32$ for $|D|$, $|F_2'|$ and $c'/32$ for the cells of $Z^\Sigma$ under ($\beta1$). Then each non-seed cell carries at most $\eps_\Delta\le\frac12c_{16}^D\le\frac12$ by ($\beta4$), lone seed cells carry at most $e^{c_S}$, and the fraction $c'-c'/4-\eps-c'/32-c'/32=11c'/16-\eps$ remains on the seed. This gives \eqref{eq:A.core}. Terms with $X\ne Z$ contain a non-seed cell and gain $\frac12c_{16}^D$.

(2) \emph{Splitting.} By Lemma~\ref{lem:7.26}, the map $(X,N_X)\mapsto(\sigma(N_X),X,N_X)$, $\sigma(N_X):=(Z,N_X|_{E(Z)})$ with $Z:=\sat_{N_X}(S_0)$, is a bijection onto the triples with $\sigma\in\Sigma_{\mathrm{sat}}(S_0)$, $X\supseteq Z$, $N_X\in\cN_\sigma(X)$, preserving the summands.

(3) \emph{Entropy of the first set.} Section~\ref{S:I2} gives this step in detail. As $11c'/16-\eps\ge c'/8$, the left side of Lemma~\ref{lem:7.53}(a) is at most $\cP'e^{c_S|S_0|}\sum_\sigma\prod_{\mathcal S}e^{-(c'/8)\Floc(\mathcal S,N_{\mathcal S})}$. We assign the wall polymers of $\sigma$ to their first cells in $S_0$ and count them by their wall interfaces and values with Lemmas~\ref{lem:5.12} and~\ref{lem:7.32}(a). Under ($\beta3$) this gives the factor $1+\etaseed$ per cell of $S_0$. From this we obtain (a), then (b), and (c$'$) with $\cP'=1$, $Z^o=\emptyset$. \qed
\end{proof}
}
\providecommand{\bGM}{\beta_{\mathrm{GM}}}
\providecommand{\mt}{\tilde m}
\providecommand{\lPV}{\lambda_{\mathrm{PV}}}
\providecommand{\Gp}{G'}
\providecommand{\Dep}{\mathfrak{D}}
\providecommand{\bD}{b_{D}}
\providecommand{\qpin}{q_{\mathrm{pin}}}
\providecommand{\Kcell}{K_{\mathrm{cell}}}
\providecommand{\cU}{\mathcal{U}}
\providecommand{\eps}{\varepsilon}

\section{Per-cell bounds: proofs of Lemma~\ref{lem:7.40} and Proposition~\ref{prop:7.41}}
\label{app:B}

In this appendix we prove Lemma~\ref{lem:7.40} and Proposition~\ref{prop:7.41}, the bookkeeping of \cite[\S9.8, (9.81) to (9.824)]{BF80} on $\Gp$, again in condensed form. In Debye units ($\bD:=\bGM\mt$) each field derivative carries $\bD^{1/2}$ and each coefficient $1/\bD$. A lone cell is of class A, and by the minimality of \cite[(E.11)]{GM82} it contains a link end or a Mayer cube, which gives it net weight $\ge\frac16$. The cells of $X_{\Sigma}$ are controlled by $F_{1}^{\mathrm{loc}}$, and the factorials by pinning. Since $\Ord((1+r)^{3})$ cells lie within distance $r$ of a cell, a sum of $N$ distances grows like $N^{4/3}$, which grows faster than $\log((N!)^{\qpin})$.

\paragraph{Setting.} Cells are $\cU$-cells, grid cells are denoted $\Omega$, and $\mathrm{vol}_{D}(\Delta):=\mt^{3}\#\Delta\in[c_{u},C_{u}]$. Classes A, B and $X=Z\sqcup X_{\Sigma}\sqcup X_{o}$ are those of Section~\ref{sec:7.6} (seed cells in wall polymers count with $X_{\Sigma}$). On a class-A cell $\vartheta=\psi$ modulo $P\ZZ$ in each phase. In the seeded activity of Section~\ref{sec:7.6}, $X_{i}:=Y_{1}\cup\dots\cup Y_{i}$ and $\kappa(\bar y,s)=\kappa(n-1)\cdots\kappa(1)$ as in \cite[(E.9), (E.10)]{GM82} and Appendix~\ref{app:C}. When its brackets are expanded, $\kappa(i)$ is of one of five \emph{kinds}: K0 $=\partial_{s_{i}}E^{(i)}$, K1 $=\dot C\partial_{x}\partial_{y}$, K2 $=\dot C\partial_{x}(\partial_{y}E)$, K3 $=\dot C(\partial_{x}E)\partial_{y}$ and K4 $=\dot C(\partial_{x}E)(\partial_{y}E)$. Here $\partial_{x}:=\partial/\partial\psi(x)$, $x\in Y_{i+1}$ and $y\in X_{i}$. An \emph{$\mathcal{E}$-factor} is one Mayer term of $E(X,s)$ of the forms \cite[(E.5a) to (E.5d)]{GM82}, with its coefficient $\rho_{t}$ and its $t$ factors. Derivatives act on the factors to the right of the differentiation, as in \cite[(E.13) to (E.16)]{GM82}, which is the rule of \cite[p.~229]{BF80}. The $\psi$-dependent \emph{atoms} of $Q^{\natural}(X)$ are: (a1) $r_{\Omega}$ of \eqref{eq:7.rOmega}, for each grid cell $\Omega\subseteq X$; (a2) $e^{G_{2}(\Omega)}$, $G_{2}(\Omega):=G_{2,1}(\Omega)+G_{2,2}(\Omega)$; (a3) $e^{g_{z}^{\Dep}}$, $z\in X\cap\Dep$; (a4) $e^{g_{z}^{\natural}}$, $z\in X\setminus\Dep$; (a5) $e^{-F_{2}(X)}=e^{-\langle1_{X}\psi,F_{2}'\rangle}$. Here $g_{z}^{\Dep}$, $g_{z}^{\natural}$ are the $z$-terms of $G_{2,3}^{\Dep}$, $G_{2,3}^{\natural}$ and, pairing $m$ with $-m$ by (S3),
\begin{equation}\label{eq:B.G2}
G_{2,1}(\Omega)=\sum_{z\in\Omega}\sum_{m}\rhoref(m)f_{2}(m\bGM^{1/2}\tilde\delta(z)),\qquad G_{2,2}(\Omega)=\sum_{z\in\Omega}\sum_{m}\rhoref(m)v_{m}(\Omega)f_{1}(m\bGM^{1/2}\tilde\delta(z)),
\end{equation}
with $f_{1}(y):=e^{iy}-1-iy$, $f_{2}(y):=\cos y-1+\frac12y^{2}$ and $v_{m}(\Omega):=e^{im\bGM^{1/2}\mathrm{avg}_{\Omega}\vartheta}-1$. Thus $G_{2,1}$ is of fourth order in $\tilde\delta$. The factors of an $\mathcal{E}$-factor are $\eps(\xi)=e^{im\bGM^{1/2}(\vartheta(x)+h(x))}-1$ in (E.5a), $\tilde\eps(\xi):=\eps(\xi)-im\bGM^{1/2}\psi(x)$ in (E.5b) to (E.5d), and the \emph{$\psi$-factor} $im\bGM^{1/2}\psi(x)$ in (E.5c), (E.5d).

\begin{lemma}[Debye units]\label{lem:B.1}
Put $d\hat x:=\mt^{3}\times$ counting measure on $V$, $\hat\psi:=\mt^{-1/2}\psi$, $\delta/\delta\hat\psi(\hat x):=\mt^{-5/2}\partial/\partial\psi(x)$, $\hat C(\hat x,\hat y):=\mt^{-1}C'_{s}(x,y)$ (likewise for $\partial_{s_{i}}C'_{s}$), and $\hat c:=\mt^{-3}c$ for a coefficient $c(x)$ per site. Then (a) $\sum_{x,y}K(x,y)\partial_{\psi(x)}\partial_{\psi(y)}=\iint d\hat x\,d\hat y\,\mt^{-1}K(x,y)\frac{\delta}{\delta\hat\psi(\hat x)}\frac{\delta}{\delta\hat\psi(\hat y)}$; (b) if $F(\psi)=\sum_{z}c(z)f(\bGM^{1/2}\psi(z))$, then $\delta F/\delta\hat\psi(\hat x)=\hat c(\hat x)\,\bD^{1/2}f'(\bD^{1/2}\hat\psi(\hat x))$: each derivative of a function of $\bGM^{1/2}\psi$ produces exactly $\bD^{1/2}$; (c) $|\hat C(\hat x,\hat y)|\le(2K_{1}''/\lPV)e^{-\rcov d(x,y)}$, $\hat C(\hat x,\hat x)\le\alpha_{D}K_{1}/\lPV$ and $\sup_{\hat x}\int d\hat y\,|\hat C(\hat x,\hat y)|e^{\rcov d(x,y)}\le2K_{3}$; (d) for a cell $\Delta$, $k\ge2$ and $C_{k}\le Ck!(2\kappa_{S})^{-k}$: (M1) $\int_{\Delta}d\hat x\sum_{m}|\hat\rho^{\mathrm{ref}}(m)||m|^{k}\le C_{k}\mathrm{vol}_{D}(\Delta)/\bD$; (M2) $\int_{\Dep\cap\Delta}d\hat x\sum_{m}|\hat\rho_{1}-\hat\rho^{\mathrm{ref}}||m|^{k}\le C_{k}\cdep$; (M3) $\int_{\Delta\setminus\Dep}d\hat x\sum_{m}|\hat\rho_{1}-\hat\rho^{\mathrm{ref}}||m|^{k}\le C_{k}\theta'\mathrm{vol}_{D}(\Delta)/\bD$; (M4) for $t\ge2$, a set $\mathfrak a$ of cells and a cell $a_{*}\in\mathfrak a$,
\[
\frac1{t!}\int_{\Xi_{t}(\mathfrak a)}d\hat x_{1}\cdots d\hat x_{t}\sum_{m_{1},\dots,m_{t}}|\hat\rho_{t}|\,e^{2\kappa_{S}\sum_{i}|m_{i}|}\le t\,K_{M4}C_{M4}^{\,t}\,\frac{4\,\mathrm{vol}_{D}(a_{*})}{\bD}\,e^{-\mu(t-1)}e^{-(1-\delta_{\mathrm{GM}})(A/\mt)\hat L(\mathfrak a)},
\]
and hence, for all integers $k_{1},\dots,k_{t}\ge0$, the left side with $\prod_{i}|m_{i}|^{k_{i}}$ in place of $e^{2\kappa_{S}\sum_{i}|m_{i}|}$ is at most $\prod_{i}k_{i}!(2\kappa_{S})^{-k_{i}}$ times the right side. Here $\Xi_{t}(\mathfrak a)$ is the set of $(x_{1},\dots,x_{t})\in V^{t}$ such that $\mathfrak a$ is the set of the cells that contain some $x_{i}$, $A$ is the rate of (S2), $\hat L(\mathfrak a):=\mt L_{\Gp}(\mathfrak a)$ is the tree length of (S2) of the cells of $\mathfrak a$, each listed once, in Debye units, $C_{M4}:=(1-C)^{-1}$ per point and $K_{M4}:=e^{2\kappa_{S}}(1-C)^{-10}$, where $C$ is the parameter of (S2), fixed in $(0,\frac14]$ (item~12 of the list in Section~\ref{ssec:D.1}). The left side is the sum over all assignments of the $t$ points to the cells of $\mathfrak a$ that use every cell; it contains $\int_{a_{1}}\cdots\int_{a_{t}}$ for each $t$-tuple of cells with $\{a_{1},\dots,a_{t}\}=\mathfrak a$.
\end{lemma}

\begin{proof}
Parts (a) and (b) follow from the change of variables, since $\mt^{-5/2}\bGM^{1/2}=\mt^{-3}\bD^{1/2}$. Part (c) is Theorem~\ref{thm:6.12} for $C'\le\alpha_{D}C_{0}$, which holds by Lemma~\ref{lem:7.31}, with $\MY/\mt=1/\lPV$ and entrywise domination of $C'_{s}$. Bound (M1) follows from $\bGM\sum_{m}m^{2}\rhoref(m)=\mt^{2}$ and (S4$'$). Bounds (M2) and (M3) follow from Propositions~\ref{prop:6.18}, \ref{prop:6.6} and \eqref{eq:7.6.mom}. We derive (M4) from (S2). As $e^{2\kappa_{S}\sum_{i}|m_{i}|}=e^{2\kappa_{S}}e^{2\kappa_{S}(\sum_{i}|m_{i}|-1)}$, the left side is $e^{2\kappa_{S}}\frac1{t!}\sum_{x\in\Xi_{t}(\mathfrak a)}\sum_{m}|\rho_{t}|e^{2\kappa_{S}(-1+\sum_{i}|m_{i}|)}$, since in Debye units $\int d\hat x\,|\hat\rho|=\sum_{x}|\rho|$. Each $x\in\Xi_{t}(\mathfrak a)$ has a coordinate in $a_{*}$. As $\rho_{t}$, the weight and $\Xi_{t}(\mathfrak a)$ are symmetric, the sum is at most $t$ times its part with $x_{1}\in a_{*}$. For $x\in\Xi_{t}(\mathfrak a)$, $L_{\Gp}(x_{1},\dots,x_{t})\ge L_{\Gp}(\mathfrak a)$. Indeed, in a tree for $x_{1},\dots,x_{t}$ we keep one marked vertex in each cell of $\mathfrak a$ and declare the others auxiliary. Hence the part is at most $e^{-(1-\delta_{\mathrm{GM}})AL_{\Gp}(\mathfrak a)}$ times the left side of \eqref{eq:6.S2} with $s=t$ and $a_{1}=a_{*}$, and (S2), $(1-C)^{-t-10}=(1-C)^{-10}C_{M4}^{\,t}$ and $|a_{*}|\mD^{2}/\bGM\le4\,\mathrm{vol}_{D}(a_{*})/\bD$, which holds as $\mD\le2\mt$, give (M4). Here the sum over the $|\mathfrak a|!\,S(t,|\mathfrak a|)$ assignments ($S$ as in the proof of Lemma~\ref{lem:A.7}) is a part of the sum over all ordered tuples of points in (S2), and the decay is read off the weight on that part, so the sum over these assignments is bounded by (M4) without a factorial. The form with $\prod_{i}|m_{i}|^{k_{i}}$ follows from $|m|^{k}\le k!(2\kappa_{S})^{-k}e^{2\kappa_{S}|m|}$.
\qed
\end{proof}

Thus the expansion is as in \cite[\S9.8]{BF80}, with $\tilde l_{D}=1$, $\beta\to\bD$, on a fine lattice with a regularized covariance, and infinitely many charges only produce the $k!$ in (d). A factor $f$ has \emph{weight} $w$ if $|f|\le K\bD^{w}\cdot(\text{field monomial})\cdot(\text{numerical factors})$ for all positions in their cells and all fields, with $K$ uniform in $\bD$, $L'$ and $\tau$. Here a field monomial is a product of values $|\hat\psi(\hat x)|$ and averages $\mathrm{avg}_{\Omega}|\hat\psi|$, paid by the Wick bound $T_{1}$ (the Wick bound gives a constant $c_T$, independent of $\bD$, $L'$, $\tau$, per field factor, together with the factorials $\prod_jn_j!$ counted in Lemma~\ref{lem:B.9}). A numerical factor is $|\delta\hat D(\hat x)|:=|\hat D(\hat x)-\mathrm{avg}_{\Omega}\hat D|$ or $|\hat F_{2}'(\hat x)|$, with $\hat D:=\mt^{-1/2}D$, $\hat F_{2}':=\mt^{-5/2}F_{2}'$. The weight of a term is the sum of those of its factors.

\begin{lemma}\label{lem:B.2}
Expand every derivative of $\kappa(\bar y,s)$ by the Leibniz rule, as in \cite[(9.84), (9.85)]{BF80}.
(a) $\kappa(\bar y,s)Q^{\natural}(X)$ is a finite sum of terms, each labelled by the kinds, the $\mathcal{E}$-factors, and a map $l$ that sends each derivative to one atom or one factor of an $\mathcal{E}$-factor which depends on the derivative's point and lies to its right; the term is $[\prod_{i}\partial_{s_{i}}C'(x_{i},y_{i}|s)]\prod_{a}\partial^{l^{-1}(a)}a\prod_{e}[\text{coefficient of }e\cdot\prod_{\phi\in e}\partial^{l^{-1}(\phi)}\phi]$. (b) The multiple derivatives are of the types: i) $\partial^{S}r_{\Omega}$; ii) from an observable (absent here); iii) $\partial^{S}e^{-F_{2}}=(-1)^{|S|}\prod_{j\in S}F_{2}'(x_{j})e^{-F_{2}}$; iv) $\partial^{S}\eps$, $\partial^{S}\tilde\eps$, $\partial^{S}(im\bGM^{1/2}\psi)$; v), vi) the Fa\`a di Bruno blocks $\rhoref(m)\partial^{B}[v_{m}f_{1}]$ and $\rhoref(m)\partial^{B}[f_{2}]$ of $\partial^{S}e^{G_{2}(\Omega)}$; vii), viii) the Fa\`a di Bruno blocks of $\partial^{S}e^{g_{z}^{\Dep}}$ and $\partial^{S}e^{g_{z}^{\natural}}$. (c) With $n_{\alpha}$ link ends, $m_{\alpha}$ factors of $\mathcal{E}$-factors and $w_{\alpha}$ ($=0$ here) factors of an observable in the cell $\Delta_{\alpha}$, the number of maps $l$ is at most $\prod_{\alpha}(m_{\alpha}+w_{\alpha}+5)^{n_{\alpha}}$, and for $c_{p}>0$, $n\ge1$ and integers $m,w\ge0$, $(m+w+5)^{n}\le C_{r}^{n+w}n!\,e^{c_{p}m}$ with $C_{r}:=4e^{5}\max(1,1/c_{p})$.
\end{lemma}

\begin{proof}
The lemma extends \cite[(9.84) to (9.87), Lemma~9.10]{BF80} by two new types, vii) and viii). A bare derivative at $x\in\Delta_{\alpha}$ can hit $r_{\Omega(x)}$, $e^{G_{2}(\Omega(x))}$, $e^{g_{x}^{\Dep}}$ or $e^{g_{x}^{\natural}}$, $e^{-F_{2}}$, or one of the $m_{\alpha}+w_{\alpha}$ factors in $\Delta_{\alpha}$. The count therefore has $+5$ in place of the $+3$ of \cite[(9.86)]{BF80}. The undifferentiated exponentials recombine into $e^{G^{\natural}(X)}$ and $e^{-F_{2}}$, which stay in $T_{4}$ and $T_{2}$ of Proposition~\ref{prop:7.37}. We bound each $r_{\Omega}$ by Lemma~\ref{lem:7.33}, whose factor is included in $T_{3}$. The last inequality follows from $(m+w+5)^{n}\le2^{n}(m^{n}+(w+5)^{n})$, $m^{n}\le n!c_{p}^{-n}e^{c_{p}m}$ and $(w+5)^{n}\le n!e^{w+5}$. The factor $n!$ is controlled by the pinning of Lemma~\ref{lem:B.4}. Within a fixed $l$ the further expansions, namely Fa\`a di Bruno, the Leibniz rule inside $v_{m}f_{1}$, the two parts of $\partial\tilde\delta(z)/\partial\psi(x)$ and the charge sums of Lemma~\ref{lem:B.1}(d), cost at most $C^{n_{\alpha}}(n_{\alpha}!)^{2}$ per cell, and $e^{c_{p}\sum_{\alpha}m_{\alpha}}=\prod_{e}e^{c_{p}t_{e}}$ is a constant per $\mathcal{E}$-factor.
\qed
\end{proof}

\begin{lemma}[Witness]\label{lem:B.3}
(a) \emph{Link localization.} Let $1\le i\le n-1$ and let $Y\in\tilde Y(h)$ with $Y\subseteq Y_{i+1}$. Then $Y$ contains the point $x_{i}$ of the bare derivative of $\kappa(i)$ (kinds K1, K2), or a cube of an $\mathcal{E}$-factor of $\kappa(i)$. (b) Every lone cell is of class A, and $F_{2}'=0$ on it. (c) Let $\Delta\in X_{o}$ with $\Delta\subseteq Y_{i+1}$, $i\ge1$. Then (W1) the bare derivative of $\kappa(i)$ sits at $x_{i}\in\Delta$ and hits an atom of kind (a1) to (a4) of $\Delta$, or (W2) an $\mathcal{E}$-factor of $\kappa(i)$ has a cube equal to $\Delta$.
\end{lemma}

\begin{proof}
(a) By \cite[(E.11)]{GM82}, $Y_{i+1}$ is the smallest union of elements of $\tilde Y(h)$ containing $\mathcal{O}-X_{i}$, where $\mathcal{O}$, the support of the term of $\kappa(i)$, consists of the cubes of its $\mathcal{E}$-factors and the cube of a bare $x_{i}$ if there is one. An element $Y\subseteq Y_{i+1}$ meeting no point of $\mathcal{O}-X_{i}$ would make $Y_{i+1}\setminus Y$ a smaller such union. As cells are $\cU$-cells, ``meets'' is ``contains''.
(b) We have $\mathrm{supp}\,D_{\beta}$, $\mathrm{supp}\,F_{2,\beta}'\subseteq S_{\beta}$ by Proposition~\ref{prop:7.9}, and a lone cell lies in no wall polymer.
(c) By (a), $\Delta$ contains $x_{i}$, or it contains a cube of an $\mathcal{E}$-factor of $\kappa(i)$, which is case (W2). In the first case the bare derivative hits a factor to its right that depends on $\psi(x_{i})$. An atom (a1) to (a4) at $x_{i}$ gives (W1). The atom (a5) and the derivatives of $e^{-F_{2}}$ give $F_{2}'(x_{i})=0$ by (b). A factor at $x_{i}$ of the $\mathcal{E}$-factor of the $y$-bracket of $\kappa(i)$ (kind K2) gives (W2). All other factors are functions of $\psi$ on $X_{i}$, which is disjoint from $Y_{i+1}$.
\qed
\end{proof}

For each $\Delta\in X_{o}$ we fix one witness, (W1) if available and otherwise an $\mathcal{E}$-factor $e$ as in (W2). Let $W(e)$ denote the set of lone cells witnessed by $e$. Each lone cell lies in exactly one $Y_{i+1}$, and by (W2) of Lemma~\ref{lem:B.3}(c) a witness $e$ of a lone cell $\Delta\subseteq Y_{i+1}$ is an $\mathcal{E}$-factor of $\kappa(i)$. It follows that all cells of $W(e)$ lie in $Y_{i(e)+1}$, where $\kappa(i(e))$ is the link of $e$. Each $\kappa(i)$ has at most two $\mathcal{E}$-factors.

The cells of $\Gp$ have polynomial growth: $\#\{\Delta:\mathrm{dist}(\Delta_{0},\Delta)\le r\}\le K_{g}(1+r)^{3}$ in Debye units, for the Euclidean distance between cells and $K_{g}:=(4\pi/3)(1+2C_{\cU})^{3}/c_{u}$, by (R7), $\mathrm{vol}_{D}\ge c_{u}$ and $\mt\le1$ (the sites of the cells lie in a ball of radius $r+2C_{\cU}+\sqrt3\mt/2\le(1+r)(1+2C_{\cU})$). Links have one end in $Y_{i+1}$ and one in $X_{i}$, and the sets $Y_{i+1}$ are pairwise disjoint.

\begin{lemma}\label{lem:B.4}
(a) Let $\Delta_{0}$ be a cell and $\Delta_{1},\dots,\Delta_{N}$ cells, each repeated at most twice, with $r_{k}:=\mathrm{dist}(\Delta_{0},\Delta_{k})$ (Euclidean, in Debye units). For $q\ge1$ and $c_{p}>0$, $(N!)^{q}\le c_{N}^{N}\exp(c_{p}\sum_{k}r_{k})$ with $c_{N}:=e^{c_{p}}\max(1,3q/(c_{p}a_{g}))^{3q}$, $a_{g}:=\frac34(2K_{g})^{-1/3}$. (b) Let $d:=\sum_{i}\mathrm{dist}(\Delta_{i}',\Delta_{i}'')$ be the total length of the links, $\hat L^{E}(e)$ the Euclidean Steiner length, in Debye units, of the cells of the $\mathcal{E}$-factor $e$ (the least Steiner length of points chosen in its cells), and $\hat L_{0}:=\sum_{e}\hat L^{E}(e)$. Pin each $\mathcal{E}$-factor $e$ at a set $\Pi(e)$ of at most four of its cells (chosen in Rule~D, before Theorem~\ref{thm:B.6}), and let $N_{\alpha}:=\#\{e:\Delta_{\alpha}\in\Pi(e)\}$, in either class. Then, with $c=c(q,c_{p},K_{g},C_{\cU})$, $\prod_{\alpha}(n_{\alpha}!)^{q}\le c^{\sum n_{\alpha}}e^{c_{p}d}$ and $\prod_{\alpha}(N_{\alpha}!)^{q}\le c^{\#\mathcal{E}}e^{4c_{p}(\hat L_{0}+2d)}$.
\end{lemma}

\begin{proof}
For (a) we argue as in \cite[Lemma~9.11, (9.820)]{BF80}, with the growth $K_{g}(1+r)^{3}$ in place of the lattice count of \cite[Appendix~2]{BF80}. In increasing order the distances satisfy $r_{(k)}\ge(k/(2K_{g}))^{1/3}-1$. Summing these bounds, we obtain $\sum_{k}r_{k}\ge a_{g}N^{4/3}-N$. Moreover, $qN\log N-c_{p}a_{g}N^{4/3}\le3qN\log^{+}(3q/(c_{p}a_{g}))$.
(b) We first treat the derivatives. Since $\Delta_{\alpha}$ lies in exactly one $Y_{i+1}$, at most one $x$-end lies in $\Delta_{\alpha}$. All other ends in $\Delta_{\alpha}$ are $y$-ends $y_{j}$ of links $j$ with $\Delta_{\alpha}\subseteq X_{j}$, whose partners $x_{j}$ lie in $Y_{j+1}$, disjoint for distinct $j$. We apply (a) to the $y$-ends, with $n!\le2^{n}(n-1)!$. For the $\mathcal{E}$-factors, let $e$ be pinned at $\Delta_{\alpha}$, with link $i=i(e)$. If $\Delta_{\alpha}\subseteq Y_{i+1}$, there are at most two such $e$. Otherwise $\Delta_{\alpha}\subseteq X_{i}$, and $e$ reaches $Y_{i+1}$. For K0 this holds since $\partial_{s_{i}}E^{(i)}$ keeps only Mayer terms with cubes in $X_{i}$ and in $Y_{i+1}$. The $x$-side $\mathcal{E}$-factor of K3, K4 does so through a factor at $x_{i}$, and the $y$-side one of K2, K4 through a factor at $y_{i}$, which is joined to $x_{i}$ by link $i$. Thus $\mathrm{dist}(\Delta_{\alpha},Y_{i+1})\le\hat L^{E}(e)+2C_{\cU}+d_{i}$, as $\hat L^{E}(e)$ is a Steiner length over points of the cubes of $e$ and cells have Euclidean diameter at most $C_{\cU}$ in Debye units. The partner is the cell of $Y_{i+1}$ nearest to $\Delta_{\alpha}$. Since distinct links give disjoint $Y$'s, a partner is used only by $\mathcal{E}$-factors of a single link, hence at most twice (see the paragraph after Lemma~\ref{lem:B.3}). We apply (a). Each $e$ appears at most $|\Pi(e)|\le4$ times and each link at most twice, and the factor $e^{8c_{p}C_{\cU}}$ is absorbed into $c$.
\qed
\end{proof}

With $c_{p}$ of Proposition~\ref{prop:7.38} and $r_{C}=\rcov/\mt$, the link decay $e^{-r_{C}\mathrm{dist}(\Delta_{i}',\Delta_{i}'')}$ of Lemma~\ref{lem:B.1}(c) pays the factor $e^{9c_{p}d}$ and leaves $e^{-(\bar\delta+(r_{C}-\bar\delta)/2)\mathrm{dist}}$. In the proof of Proposition~\ref{prop:B.7}, the Mayer decay of (M4) absorbs the factor $e^{4c_{p}\hat L_{0}}$ together with the factors $e^{c_{10}L_{\eta^{A}}}$ of $\kappa'$. This uses $\hat L\ge3^{-1/2}\hat L^{E}$ and, on outputs, $(1-\delta_{\mathrm{GM}})A/\sqrt3\ge\frac12\kappa_{Y}\MY\ge2r_{C}\mt>4c_{p}\mt$ for $\delta_{\mathrm{GM}}\le\frac12$ (Proposition~\ref{prop:6.11}). Finally, $c^{\#\mathcal{E}}\le c^{2|X|}$ is a constant per cell, as a link has at most two $\mathcal{E}$-factors and there are $n-1<|X|$ links (each $Y_{i+1}$ contains a $\cU$-cell).

In \cite[(9.821)]{BF80}, $N_{\alpha}$ counts distinguished factors $\eps$ in class-A cubes, but the proof in \cite[Appendix~2]{BF80} does not use the class, and (b) pins $\mathcal{E}$-factors at cells of either class, as required by Proposition~\ref{prop:B.10}. As in \cite[(9.822)]{BF80}, we multiply each term by $1=f_{q}\cdot\prod_{\alpha}(n_{\alpha}!)^{q}(N_{\alpha}!)^{q}$, $f_{q}:=\prod_{\alpha}(n_{\alpha}!)^{-q}(N_{\alpha}!)^{-q}$, and bound the second factor by (b).

We state the bounds below in the variables of Lemma~\ref{lem:B.1}, at fixed positions, for $k$ derivatives at points of the factor's cell (for site factors, at the site). Here $\hat\delta:=\mt^{-1/2}\tilde\delta$, $\hat\psi_{\Omega}:=\mathrm{avg}_{\Omega}\hat\psi$, and $v_{\Omega}\in[v_{-},v_{+}]$ is the Debye volume of a grid cell. The chain-rule factors $\delta(\hat\psi_{\Omega})/\delta\hat\psi(\hat x)=1_{\Omega}(\hat x)/v_{\Omega}$ and $\delta\hat\delta(\hat z)/\delta\hat\psi(\hat x)=\hat\delta(\hat x-\hat z)-1_{\Omega}(\hat x)/v_{\Omega}$, $\hat\delta(\hat x-\hat z):=\mt^{-3}1[x=z]$, are not shown. They are absorbed when the line end is integrated, or bounded by $1/v_{-}$. In class B, $|\hat\delta|\le|\hat\delta\hat\psi|+|\delta\hat D|$. A row of Table~\ref{tab:B.5} that carries an equation number of \cite{BF80} restates that bound in Debye units, with the charges and volumes shown, and the other rows are new.

\begin{table}[tbp]
\centering\footnotesize
\begin{tabular}{p{0.30\textwidth}p{0.08\textwidth}p{0.40\textwidth}p{0.10\textwidth}}
\hline\noalign{\smallskip}
factor (\cite{BF80}) & $k$ & bound & weight\\
\noalign{\smallskip}\hline\noalign{\smallskip}
v) $v_{m}(\Omega)$, any class (9.814) & $\ge1$; $0$ & $(|m|\bD^{1/2}/v_{-})^{k}$; $2$ & $k/2$; $0$\\
$v_{m}(\Omega)$, class A & $0$ & $|m|\bD^{1/2}|\hat\psi_{\Omega}|$ & $1/2$\\
vi) $f_{1}(m\bD^{1/2}\hat\delta)$ (9.815) & $0$; $1$; $\ge2$ & $\frac12m^{2}\bD|\hat\delta|^{2}$; $m^{2}\bD|\hat\delta|$; $(|m|\bD^{1/2})^{k}$ & $1$; $1$; $k/2$\\
$f_{1}$, bounded form & $1$ & $2|m|\bD^{1/2}$ & $1/2$\\
vi) $f_{2}(m\bD^{1/2}\hat\delta)$ & $\le4$; $\ge4$ & $m^{4}\bD^{2}|\hat\delta|^{4-k}$; $(|m|\bD^{1/2})^{k}$ & $2$; $k/2$\\
$f_{2}$, bounded form & $2$; $3$ & $2m^{2}\bD$; $|m|^{3}\bD^{3/2}$ & $1$; $3/2$\\
iv) $\tilde\eps$, class A (9.816) & $0$; $1$; $\ge2$ & $\frac12m^{2}\bD|\hat\psi|^{2}$; $m^{2}\bD|\hat\psi|$; $(|m|\bD^{1/2})^{k}$ & $1$; $1$; $k/2$\\
$\tilde\eps$, class B (9.818) & $0$; $1$ & $2+|m|\bD^{1/2}|\hat\psi|$; $2|m|\bD^{1/2}$ & $0$; $1/2$\\
$\psi$-factor, any class (9.816) & $0$; $1$ & $|m|\bD^{1/2}|\hat\psi|$; $|m|\bD^{1/2}$ (zero for $k\ge2$) & $1/2$; $1/2$\\
$\eps$, any class (9.817) & $\ge1$ & $(|m|\bD^{1/2})^{k}$ & $k/2$\\
$\eps$ undistinguished; distinguished, class A (9.819) & $0$ & $2$; $|m|\bD^{1/2}|\hat\psi|$ & $0$; $1/2$\\
vii) $\cos(m\bD^{1/2}\hat\vartheta)-1$ & $\ge1$ & $(|m|\bD^{1/2})^{k}$ & $k/2$\\
viii) $f_{2}(m\bD^{1/2}\hat\psi)$, class A & $\le4$; $\ge4$ & $m^{4}\bD^{2}|\hat\psi|^{4-k}$; $(|m|\bD^{1/2})^{k}$ & $2$; $k/2$\\
viii), class B & $1$; $2$; $\ge3$ & $|m|\bD^{1/2}+m^{2}\bD|\hat\psi|$; $2m^{2}\bD$; $(|m|\bD^{1/2})^{k}$ & $\frac12$; $1$; $k/2$\\
i) $r_{\Omega}$ (Lemma~\ref{lem:7.33}) & $\ge0$ & $c_{7}(c_{7}'\bD^{1/6}/v_{-})^{k}e^{c_{7}''k\log k}e^{\frac12\eta_{H}\mt^{2}\sum_{\Omega}\vartheta^{2}}$ & $k/6$\\
iii) $\hat F_{2}'$, class B only & & numerical factor & $0$\\
\noalign{\smallskip}\hline
\end{tabular}
\caption{Bounds per type (Lemma~\ref{lem:B.5}(a)).}\label{tab:B.5}
\end{table}

\begin{lemma}\label{lem:B.5}
(a) With $C$ numerical and $|m|$ the charge of the factor, the factors satisfy the bounds of Table~\ref{tab:B.5}. (b) \emph{Vertex weights.} A vertex is a block $B$ on a $G_{2}$-, vii)- or viii)-atom with its coefficient, an $r_{\Omega}$ with all its derivatives, an $\mathcal{E}$-factor with its coefficient and factors, or one factor $\hat F_{2}'$. The coefficient weight is $-1$ for one-body, off-$\Dep$ and Mayer coefficients ((M1), (M3), (M4)) and $-\frac16$ for deficit coefficients on $\Dep$ ((M2) with $\cdep\le\bD^{-1/6}$ for $\tau\ge\tau_{0}$). For a block with $k=|B|\ge1$: a $G_{2,1}$-block has weight $\ge1$ in class A, $\ge0$ in class B if $k\ge2$, and for $k=1$ in class B it is a term of weight $0$ equal to $\hat\delta(\hat x)$ at the derivative point plus a term of weight $-\frac12$ with no field or numerical factor; a $G_{2,2}$-block has weight $\ge\frac12$ in class A and $\ge-\frac12$ in class B ($\ge0$ if $k\ge2$), with no numerical factor; a vii)-block has weight $\ge\frac13$; a viii)-block has weight $\ge1$ in class A and $\ge-\frac12$ in class B ($\ge0$ if $k\ge2$); $r_{\Omega}$ with $k$ derivatives has weight $k/6$; $\hat F_{2}'$ has weight $0$. So in class A every non-$\mathcal{E}$ vertex has weight $\ge0$, and $\ge\frac16$ if it carries a derivative; in class B every non-$\mathcal{E}$ vertex has weight $\ge-\frac12$, and is negative only for single-derivative blocks.
\end{lemma}

\begin{proof}
(a) By Lemma~\ref{lem:B.1}(b) each derivative of a function of $\bGM^{1/2}\psi=\bD^{1/2}\hat\psi$ brings $\bD^{1/2}$, and each derivative of $\hat\psi_{\Omega}$ brings $1/v_{\Omega}$. This gives the rows of \cite{BF80}. For the new rows we use the following facts. In class A, $v_{m}(\Omega)=e^{im\bGM^{1/2}\mathrm{avg}_{\Omega}\psi}-1$, since $h\in P\ZZ$, and $|e^{it}-1|\le|t|$. Next, $|f_{1}'|=|e^{iy}-1|\le\min(|y|,2)$. Further, $f_{2}^{(j)}(0)=0$ for $j\le3$ and $|f_{2}^{(4)}|\le1$, hence $|f_{2}^{(k)}(y)|\le|y|^{4-k}/(4-k)!$ for $k\le4$, $|f_{2}^{(k)}|\le1$ for $k\ge4$, $|f_{2}^{(2)}|=|1-\cos y|\le2$ and $|f_{2}^{(3)}|=|\sin y|\le1$. For vii), $|\partial_{t}^{k}e^{imt}|=|m|^{k}$. In class A the factor viii) is $f_{2}(m\bD^{1/2}\hat\psi(\hat z))$, and in class B its first derivative is $-m\bD^{1/2}\sin(\cdot)+m^{2}\bD\hat\psi$ and its second $m^{2}\bD(1-\cos(\cdot))$. Row i) is Lemma~\ref{lem:7.33} with $\partial A_{\Omega}/\partial\psi(x)=\mt^{3}/v_{\Omega}$. For iii), $F_{2}=\int d\hat x\,\hat\psi\hat F_{2}'$ and $\mathrm{supp}\,F_{2}'\subseteq\bigcup S_{\beta}$.
(b) The coefficient weights are those of (M1) to (M4). Moreover, $\cdep=\Ord(\log\tau)$, while $\bD^{-1/6}=e^{\frac16\pi^2\vCb(0)\tau(1+o(1))}$. For $G_{2,1}$ in class A the weight is $-1+2$ for $k\le4$ and $-1+k/2$ for $k\ge4$, while in class B with $k=2,3$ the bounded forms give $-1+1$ and $-1+\frac32$. In class B with $k=1$, as $\sum_{z\in\Omega}\tilde\delta(z)=0$ and $f_{2}'(y)=y-\sin y$, we have $\partial_{x}G_{2,1}(\Omega)=\mt^{2}\tilde\delta(x)-\sum_{m}\rhoref(m)m\bGM^{1/2}[\sin y_{x}-\mathrm{avg}_{\Omega}\sin y]$, whose first term is $(\bD\sum_{m}m^{2}\hat\rho^{\mathrm{ref}}(m))\hat\delta(\hat x)=\hat\delta(\hat x)$ (weight $-1+\frac12+\frac12$) and whose second has weight $-1+\frac12$. For $G_{2,2}$ we split $k=k_{1}+k_{2}$ between $v_{m}$ and $f_{1}$. In class A the weight is $-1+\frac12+\max(1,k_{2}/2)$ if $k_{1}=0$, else $-1+k_{1}/2+\max(1,k_{2}/2)$. In class B we distinguish three cases. If $k_{2}=0$, then $\sum_{z\in\Omega}f_{1}(y_{z})=\sum_{z\in\Omega}(e^{iy_{z}}-1)$ as $\sum_{z}y_{z}=0$, which is bounded by $2\#\Omega$ with no numerical factor, and hence the weight is $-1+k_{1}/2\ge-\frac12$. If $k_{2}=1$, the bounded form gives $-1+k_{1}/2+\frac12$. If $k_{2}\ge2$, the weight is $\ge-1+k_{1}/2+k_{2}/2\ge0$. For a vii)-block the weight is $-\frac16+k/2$. For a viii)-block it is $-1+2$ in class A, and $-1+\frac12$, $-1+1$, $-1+k/2$ in class B.
\qed
\end{proof}

A factor of an $\mathcal{E}$-factor is \emph{gaining} if its weight is at least $\frac12$. This is the case for all differentiated factors, in either class, for all distinguished $\eps$'s (class A), for all $\tilde\eps$'s and $\psi$-factors in class A, and for all $\psi$-factors in class B. Each factor has weight $\ge0$. We select distinguished factors as in \cite[p.~232]{BF80}, with the attribution to cells fixed by Rule D.

\emph{Rule D.} Let $e$ be an $\mathcal{E}$-factor of type (E.5a), $t\ge3$. (D-a) For each $\Delta\in W(e)$ in which $e$ has no differentiated factor, we distinguish one undifferentiated $\eps$ of $e$ in $\Delta$, which exists since $e$ has a cube equal to $\Delta$. (D-b) While $e$ has fewer than three gaining factors, we distinguish another undifferentiated $\eps$ of $e$ in a class-A cell, if there is one. All other undifferentiated $\eps$'s are bounded by $2$. For types (E.5b) to (E.5d) no choice is made. The pinning set $\Pi(e)$ consists of the cells of the factors distinguished in (D-b) (at most three) for (E.5a), or of the cells of $e$ (at most two) for (E.5b) to (E.5d), together with the charged cell defined below, if there is one. Hence $|\Pi(e)|\le4$.

Let $P(e)$ be the weight of $e$ under Rule D. An $\mathcal{E}$-factor gives $\frac16$ to each cell of $W(e)$. Its remainder $\varrho(e):=P(e)-|W(e)|/6$ is assigned to one chosen class-B cube of $e$ (its \emph{charged cell}) if $e$ has a class-B cube, and to any cell of $e$ otherwise. We call $e$ \emph{charged} if $\varrho(e)<0$. Each non-$\mathcal{E}$ vertex is assigned to its own cell.

\begin{theorem}[Net weight per lone cell]\label{thm:B.6}
In every term, after this distribution:
(i) every cell $\Delta\in X_{o}$ receives weight at least $\frac16$; (ii) every class-A cell receives weight at least $0$; (iii) every class-B cell receives weight at least $-N_{\Delta}^{\mathrm{ch}}-\frac12n_{\Delta}^{(1)}$, where $N_{\Delta}^{\mathrm{ch}}$ is the number of $\mathcal{E}$-factors charged to $\Delta$ and $n_{\Delta}^{(1)}\le n_{\Delta}$ the number of single-derivative blocks in $\Delta$.
The value $\frac16$ is attained, both by a lone cell whose only vertex of positive weight is $r_{\Omega}$ with one derivative, and by an $\mathcal{E}$-factor of type (E.5a) with $t=3$ whose three factors are undifferentiated $\eps$'s in three cells of $W(e)$.
\end{theorem}

\begin{proof}
We first treat the $\mathcal{E}$-factors. If all cubes of $e$ are of class A, then $P(e)\ge|W(e)|/6$ and $P(e)\ge0$. In general $P(e)\ge-1+|W(e)|/2$, and hence $\varrho(e)\ge-1$. Indeed, suppose first that all cubes are of class A. For (E.5a), (D-b) stops only when $e$ has $g(e)\ge3$ gaining factors, and the cells of $W(e)$ contain distinct gaining factors by (D-a), hence $g(e)\ge\max(3,|W(e)|)$ and $P(e)\ge-1+g(e)/2\ge g(e)/6\ge|W(e)|/6$. For (E.5b), $P(e)\ge-1+1+1$, and for (E.5c), (E.5d), $P(e)\ge-1+1+\frac12$, both $\ge\frac26$. In general the cells of $W(e)$ are lone, hence of class A by Lemma~\ref{lem:B.3}(b), each contains a gaining factor of $e$, and all other factors have weight $\ge0$.
(ii) A class-A cell receives its non-$\mathcal{E}$ vertices, which have weight $\ge0$ by Lemma~\ref{lem:B.5}(b), credits $\frac16$, and remainders $\varrho(e)\ge0$ of $\mathcal{E}$-factors with no class-B cube.
(i) Let $\Delta\in X_{o}$, with its witness from Lemma~\ref{lem:B.3}(c). In case (W1) a derivative hits an atom of $\Delta$. If this atom is $r_{\Omega}$, its vertex has weight $\ge\frac16$, and otherwise the Fa\`a di Bruno block containing the derivative has weight $\ge\frac13$, since the cell is of class A. In case (W2), $\Delta\in W(e)$ receives $\frac16$ from $e$. All other items received are $\ge0$, as in (ii).
(iii) Class-B non-$\mathcal{E}$ vertices have weight $\ge-\frac12$, with negative weight only for single-derivative blocks. Charged remainders are $\ge-1$, and uncharged ones are $\ge0$.
As for attainment, the first example is (W1) with Lemma~\ref{lem:B.5}(b). In the second, $P(e)=-1+\frac32=3\cdot\frac16$, and the remainder is zero.
\qed
\end{proof}

Each field factor of an $\mathcal{E}$-factor lies in a cell of $\Pi(e)$ or is the single (D-a) factor of a lone cell, and a non-$\mathcal{E}$ vertex carries at most four field factors per derivative. Hence the number $n_{j}$ of field factors in the cell $\Delta_{\alpha}$ satisfies $n_{j}\le1+4(n_{\alpha}+N_{\alpha})$.

\begin{proposition}[Per-cell form of (9.824)]\label{prop:B.7}
Fix the constants of Section~\ref{sec:7.6}, with $c_{10}\le\frac79r_{C}$ in Lemma~\ref{lem:7.39}, $\qpin:=q_{0}+2$ with $q_{0}:=8+c_{7}''$, and $c_{p}$ of Proposition~\ref{prop:7.38}. For every term $\kappa'$ of the activity of a polymer $X$ with seed $Z$ (kinds and link-end cells fixed, and for each $\mathcal{E}$-factor $e$ its number $t_{e}$ of points and the set $\mathfrak a(e)$ of cells containing them; $\kappa'$ contains, inside $|\cdot|_{0}$, the sum over all configurations of the points of $e$ with this set of cells, that is, over the assignments of the $t_{e}$ points to the cells of $\mathfrak a(e)$ that use every cell, over their positions in these cells and over their charges, and over the forms \cite[(E.5b) to (E.5d)]{GM82} if $t_{e}=2$), every $h_{X}$ and every $s$,
\begin{align*}
\int d\mu_{C'_{s}}\bigl|e^{E(X,s)}\kappa'Q^{\natural}(X)\bigr|_{0}&\le\mathcal{P}'_{G}\,e^{c_{13}|X|+C_{\Dep}\cdep N_{\Dep}(X)}\,e^{-c'F_{1}^{\mathrm{loc}}(X)+\frac{c'}{16}F_{1}^{\mathrm{loc}}(X\setminus Z)+\frac{c'}{32}F_{1}^{\mathrm{loc}}(Z)}\\
&\qquad\times\prod_{i}\bigl[K_\ell\,e^{-(\bar\delta+(r_{C}-\bar\delta)/2)\mathrm{dist}(\Delta_{i}',\Delta_{i}'')}\bigr]\prod_{e}\mathfrak{m}_{e}\prod_{\Delta\subseteq X}\Psi_{\Delta},
\end{align*}
where $\prod_i$ runs over the links with a covariance line (types (2) to (5), Lemma~\ref{lem:A.1}), $K_\ell:=2K_{1}''(1+v_{+}/v_{-})^{2}/\lPV$ is the line constant, $\mathfrak{m}_{e}:=e^{-(\mu-1-2c''_{10})(t_{e}-1)}e^{-r_{C}\hat L^{E}(e)}$ is the leftover Mayer weight, $\hat L^{E}(e)$ the Euclidean Steiner length of the cells of $\mathfrak a(e)$ (Lemma~\ref{lem:B.4}(b)) ($\mathfrak{m}_{e}\le1$ for $\mu\ge1+2c''_{10}$), and $\Psi_{\Delta}\le K_{0}^{B}$ for every class-A cell, $\Psi_{\Delta}\le K_{0}^{B}\bD^{1/6}$ for $\Delta\in X_{o}$, and $\Psi_{\Delta}\le K_{0}^{B}e^{c_{B}(\bD)}$ for every class-B cell, with $c_{B}$ as in Proposition~\ref{prop:B.10}. The constant $K_{0}^{B}$ depends only on $\lPV,C,\mu,\kappa_{S},\delta_{\mathrm{GM}},\qpin,c_{p},c'$, the constants of Theorem~\ref{thm:6.12}, Lemma~\ref{lem:7.33} and the Wick bound, and the geometry constants $c_{u},C_{u},C_{\cU},D_{\cU},v_{\pm}$; it does not depend on $\bD$, $L'$ or $\tau$, and it contains neither $K_\ell$ nor $e^{c_{13}}$.
\end{proposition}

We give the proof after Lemma~\ref{lem:B.9}.

The numerical factors $\delta\hat D=\hat D-\mathrm{avg}_{\Omega}\hat D$ and $\hat F_{2}'$ sit at line ends integrated freely over their cells. For a term we put $\mathcal{N}:=\prod_{j}\int_{\Delta_{j}}|N_{j}(\hat x)|\,d\hat x$, the product over its numerical factors $N_{j}$, and $n_{\Delta}^{N}\le n_{\Delta}$ denotes the number of them in the cell $\Delta$.

\begin{lemma}[Numerical factors]\label{lem:B.8}
All numerical factors lie in class-B cells. For every union $Y$ of elements of $\tilde Y(h)$ and every $c>0$, the numerical factors located in $Y$ satisfy
\[
\prod_{j:\,\Delta_{j}\subseteq Y}\int_{\Delta_{j}}|N_{j}|\,d\hat x\le\prod_{\Delta\subseteq Y}\Bigl[\Bigl(\frac{C_{u}}{2c}\Bigr)^{n_{\Delta}^{N}/2}(n_{\Delta}^{N}!)^{1/2}\Bigr]e^{3cF_{1}^{\mathrm{loc}}(Y)}.
\]
In particular, with $3c=c'/16$ on $X\setminus Z$ and $3c=c'/32$ on $Z$, $\mathcal{N}\le\prod_{\Delta}[\cdots]\,e^{\frac{c'}{16}F_{1}^{\mathrm{loc}}(X\setminus Z)+\frac{c'}{32}F_{1}^{\mathrm{loc}}(Z)}$.
\end{lemma}

\begin{proof}
First, $\mathrm{supp}\,D_{\beta}$, $\mathrm{supp}\,F_{2,\beta}'\subseteq S_{\beta}$ by Proposition~\ref{prop:7.9}(b), (e). Hence all numerical factors lie in class-B cells, and $S_{\beta}$ lies in one element of $\tilde Y(h)$. By the Cauchy\textendash Schwarz inequality on each cell and $\|\delta\hat D\|_{L^{2}(\Delta)}\le\|\hat D\|_{L^{2}(\Delta)}$, the left side is at most $\prod_{\Delta}(C_{u}a_{\Delta})^{n_{\Delta}^{N}/2}$ with $a_{\Delta}:=\|\hat D\|^{2}_{L^{2}(\Delta)}+\|\hat F_{2}'\|^{2}_{L^{2}(\Delta)}$, and $a^{n/2}\le(2c)^{-n/2}(n!)^{1/2}e^{ca}$. Finally $\sum_{\Delta\subseteq Y}a_{\Delta}\le(2+\eps(L'))F_{1}^{\mathrm{loc}}(Y)\le3F_{1}^{\mathrm{loc}}(Y)$ by Corollary~\ref{cor:7.11}, Theorem~\ref{thm:7.19} and Corollary~\ref{cor:7.21}.
\qed
\end{proof}

\begin{lemma}\label{lem:B.9}
All factorials produced in a term are at most $\prod_{\alpha}C^{n_{\alpha}+N_{\alpha}}(n_{\alpha}!)^{q_{0}}(N_{\alpha}!)^{q_{0}}$ with $q_{0}:=8+c_{7}''$. Consequently the pinning factor $f_{\qpin}$ with $\qpin=q_{0}+2$ absorbs them and leaves $\prod_{\alpha}(n_{\alpha}!)^{-2}(N_{\alpha}!)^{-2}$.
\end{lemma}

\begin{proof}
We count the factorials per cell, up to factors $C^{n_{\alpha}+N_{\alpha}}$. The Wick bound $T_{1}\le c_{T}^{\sum n_{j}}\prod_{j}n_{j}!$ with $n_{j}\le1+4(n_{\alpha}+N_{\alpha})$ gives $(n_{\alpha}!)^{4}(N_{\alpha}!)^{4}$, the expansions inside the types and Lemma~\ref{lem:B.2}(c) give $(n_{\alpha}!)^{3}$, Lemma~\ref{lem:7.33} gives $(k!)^{c_{7}''}$, since $k\log k\le\log k!+k$, and Lemma~\ref{lem:B.8} gives $(n_{\alpha}!)^{1/2}$. The total exponent is $7.5+c_{7}''\le q_{0}$.
\qed
\end{proof}

\begin{proof}[Proof of Proposition~\ref{prop:B.7}]
We follow the argument of \cite[(9.88) to (9.823)]{BF80} on $\Gp$, make it quantitative, and add the bound on the tree weights of $\kappa'$.

\emph{Assignments.} For each $\mathcal{E}$-factor $e$ we fix its form and a surjection $\pi_{e}:\{1,\dots,t_{e}\}\to\mathfrak a(e)$, the cell of each of its points. For the terms with these assignments we carry out all the steps below except the bound on the tree weights and the collection of the constants, with the maps $l$, Rule~D, the pinning sets $\Pi(e)$, the witnesses and the weights $P(e)$ of these terms. In the integration over the positions we leave out the points of the $\mathcal{E}$-factors. The bounds so obtained are the same for every assignment, because the numbers $m_{\alpha}$, $n_{\alpha}$, $N_{\alpha}$, $n_{j}$ appear only in Lemmas~\ref{lem:B.2}(c), \ref{lem:B.4}(b), \ref{lem:B.9} and in the Wick bound, whose right sides are uniform, and $\hat L^{E}(e)$ depends only on $\mathfrak a(e)$. The positions and charges of the points of $e$ appear only in $|\hat\rho_{t_{e}}|$ and $\prod_{i}|m_{i}|^{k_{i}}$, where $k_{i}$ is the number of derivatives on the $i$th factor. By $|m_{i}|^{k_{i}}\le k_{i}!(2\kappa_{S})^{-k_{i}}e^{2\kappa_{S}|m_{i}|}$, the factorials $k_{i}!$ are among those of Lemma~\ref{lem:B.9}, namely those of the charge sums in the proof of Lemma~\ref{lem:B.2}, and the weight $e^{2\kappa_{S}\sum_{i}|m_{i}|}$ does not depend on the assignment. We therefore bound the sum over the assignments, the positions and the charges of the points of $e$ at once by (M4) with $\mathfrak a=\mathfrak a(e)$. The sum over the forms costs at most a factor $3$.

\emph{Count.} The sum over the maps $l$ and the further expansions is at most the number of terms, which Lemma~\ref{lem:B.2}(c) bounds, times the largest term, the ``Sup'' of \cite[(9.88)]{BF80}.

\emph{Pinning.} We insert $f_{\qpin}$ and bound $\prod_{\alpha}(n_{\alpha}!)^{\qpin}(N_{\alpha}!)^{\qpin}$ by Lemma~\ref{lem:B.4}(b), with the resulting factors absorbed as explained after that lemma. This leaves $e^{-(\bar\delta+(r_{C}-\bar\delta)/2)\mathrm{dist}}$ per link.

\emph{Tree weights.} In the term $\kappa'$ of Lemma~\ref{lem:7.39} an $\mathcal{E}$-factor $e$ appears as $\sum_{\eta^{A}}e^{c_{10}L_{\eta^{A}}}e^{c''_{10}t_{e}}\mathcal{E}_{\eta^{A}}$, where $L_{\eta^{A}}$ is the length \cite[(A1.3)]{BF80}, measured with $\mathrm{dist}$ and with auxiliary points in $V$, of an augmented tree $\eta^{A}$ whose marked points lie in the cells of $\mathfrak a(e)$, one in each, as in Lemma~\ref{lem:A.7}, and $\mathcal{E}=\sum_{\eta^{A}}\mathcal{E}_{\eta^{A}}$. Since only $\sum_{\eta^{A}}\mathcal{E}_{\eta^{A}}=\mathcal{E}$ is used, we may put $\mathcal{E}_{\eta^{A}_{e}}:=\mathcal{E}$ for one such tree $\eta^{A}_{e}$ of least length (a minimizer exists, since $V$ is finite and an auxiliary vertex of degree at most $2$ can be removed without increasing the length) and $\mathcal{E}_{\eta^{A}}:=0$ otherwise. Then $e$ carries the number $e^{c_{10}L^{d}(e)}e^{c''_{10}t_{e}}$, $L^{d}(e):=L_{\eta^{A}_{e}}$, which depends only on $\mathfrak a(e)$, that is, on the link datum. By Lemma~\ref{lem:A.7} and the \emph{Assignments} step, no sum over positions at a fixed $\eta^{A}$ occurs (unlike \cite[p.~214]{BF80}, where the positions are summed at a fixed $\eta^{A}$ with the constant weights $b_{\eta^{A}}$). If $L_{\Gp}(\mathfrak a(e))=\infty$, every configuration with the cells $\mathfrak a(e)$ has points in two components of $\Gp$, so $\rho_{t_{e}}$ vanishes on it by Proposition~\ref{prop:6.3}(d), and $\mathcal{E}=0$. Such a factor is absent, as noted before Lemma~\ref{lem:A.7}, so no product $0\cdot\infty$ arises, and in what follows $L_{\Gp}(\mathfrak a(e))<\infty$. Since $\mathrm{dist}=\mt d_{\Gp}$ and $L_{\Gp}$ is computed with $3^{-1/2}d_{\Gp}$, we have $L^{d}(e)=\sqrt3\,\mt L_{\Gp}(\mathfrak a(e))=\sqrt3\,\hat L(\mathfrak a(e))$. Thus the Mayer decay of (M4) is $e^{-\rho_{S}L^{d}(e)}$ with $\rho_{S}:=(1-\delta_{\mathrm{GM}})A/(\sqrt3\,\mt)\ge2r_{C}$, by the chain of inequalities after Lemma~\ref{lem:B.4}. Also, $L^{d}(e)\ge\hat L^{E}(e)$, since both lengths are minima over points of the cells of $e$ and $d_{\Gp}$ dominates the Euclidean distance. With the pinning factor $e^{4c_{p}\hat L^{E}(e)}$ of Lemma~\ref{lem:B.4}(b) we obtain
\[
e^{c_{10}L^{d}(e)}e^{4c_{p}\hat L^{E}(e)}e^{-\rho_{S}L^{d}(e)}\le e^{-(\rho_{S}-c_{10}-4c_{p})\hat L^{E}(e)}\le e^{-r_{C}\hat L^{E}(e)},
\]
because $\rho_{S}-c_{10}\ge0$ and $\rho_{S}-c_{10}-4c_{p}\ge2r_{C}-\frac79r_{C}-\frac29(r_{C}-\bar\delta)=r_{C}+\frac29\bar\delta$. The bound uses $L^{d}$, not $\hat L^{E}$, because on a slit graph $d_{\Gp}$ is not bounded by a multiple of the Euclidean distance, and hence $e^{-r_{C}\hat L^{E}(e)}$ alone cannot absorb $e^{c_{10}L^{d}(e)}$. Finally, $t_{e}\ge2$ for each $\mathcal{E}$-factor, by \cite[(E.5a) to (E.5d)]{GM82} and Appendix~\ref{app:C}. Hence $e^{c''_{10}t_{e}}\le e^{2c''_{10}(t_{e}-1)}$, which cancels against the factor $e^{-2c''_{10}(t_{e}-1)}$ of $e^{-\mu(t_{e}-1)}$ and leaves $e^{-(\mu-2c''_{10})(t_{e}-1)}$.

\emph{H\"older and positions.} At fixed positions we bound $T_{1}$ by the Wick bound, $T_{2}e^{-F_{1}}T_{3}T_{4}$ by Proposition~\ref{prop:7.37}(d), and the factors $r_{\Omega}$ by Lemma~\ref{lem:B.5}. As $0\le\partial_{s_{i}}p(x,y|s)\le1$ in \cite[(E.3)]{GM82}, $|\partial_{s_{i}}\hat C|\le\mt^{-1}|C'|$. We bound each line by its supremum over its pair of end cells, using Lemma~\ref{lem:B.1}(c), and integrate its ends against a vertex point, a cell average, an $r_{\Omega}$-derivative (cost at most $v_{+}/v_{-}$) or a numerical factor (kept in $\mathcal{N}$). The points of each $\mathcal{E}$-factor $e$, summed over all assignments at once, are integrated against its coefficient by (M4) with $\mathfrak a=\mathfrak a(e)$, and the other internal points by (M1) to (M3) of Lemma~\ref{lem:B.1}(d). This gives at most $\prod_{i}(2K_{1}''/\lPV)e^{-r_{C}\mathrm{dist}(\Delta_{i}',\Delta_{i}'')}\prod_{\text{vertices}}[\text{mass}\cdot\bD^{w}](1+v_{+}/v_{-})^{2(n-1)}\mathcal{N}$.

\emph{Numerical factors and factorials.} We bound these by Lemmas~\ref{lem:B.8} and~\ref{lem:B.9}.

\emph{Weights.} We use Theorem~\ref{thm:B.6} and $\bD\le1$. A cell with weight $\ge\frac16$ gets at most $\bD^{1/6}$, and one with weight $\ge0$ at most $1$. A class-B cell gets $\bD^{-N_{\Delta}^{\mathrm{ch}}-n_{\Delta}^{(1)}/2}$ with $(N_{\alpha}!)^{-2}(n_{\alpha}!)^{-2}$ from Lemma~\ref{lem:B.9} and $N_{\Delta}^{\mathrm{ch}}\le N_{\alpha}$, since a charged $\mathcal{E}$-factor is pinned at its charged cell. Proposition~\ref{prop:B.10} bounds the product by $e^{c_{B}(\bD)}$.

\emph{Constants.} The line constants $(2K_{1}''/\lPV)(1+v_{+}/v_{-})^{2}=K_\ell$ stay in $\prod_iK_\ell$, and $e^{c_{13}|X|}$, the factor of Proposition~\ref{prop:7.37}, stays as it is. Each remaining constant is per cell or per $\mathcal{E}$-factor. Recall that each $\mathcal{E}$-factor that occurs has its cells in one component of $\Gp$. The per-$\mathcal{E}$-factor constants that grow with $t_{e}$ are $2$ for each undistinguished $\eps$, from Rule~D, $C_{M4}$ for each point and the factor $t_{e}$ of (M4), and $e^{c_{p}}$ for each factor, from Lemma~\ref{lem:B.2}(c). Their product is at most $t_{e}(2C_{M4}e^{c_{p}})^{t_{e}}$. As $C\le\frac14$, we have $2C_{M4}\le\frac83$. Moreover, $c_{p}<r_{C}/18<\frac1{540}$, since $r_{C}\le\kappa_{0}\le(\eps_{1}^{\mathrm{HK}})^{1/2}<\frac1{30}$ by Theorem~\ref{thm:6.12}. Therefore $q:=2C_{M4}e^{c_{p}}/e\le0.9829$, $K_{t}:=\sup_{t\ge1}tq^{t-1}\le22$ and $2C_{M4}e^{c_{p}}K_{t}=e\sup_{t\ge1}tq^{t}\le58$. Hence $t_{e}(2C_{M4}e^{c_{p}})^{t_{e}}\le2C_{M4}e^{c_{p}}K_{t}\,e^{t_{e}-1}$, and one factor $e^{-(t_{e}-1)}$ absorbs $e^{t_{e}-1}$. This turns the factor $e^{-(\mu-2c''_{10})(t_{e}-1)}$ obtained above into $e^{-(\mu-1-2c''_{10})(t_{e}-1)}$. The per-$\mathcal{E}$-factor constants independent of $t_{e}$, namely $K_{M4}$, $4C_{u}$, $2C_{M4}e^{c_{p}}K_{t}\le58$, the factor $3$ for the forms when $t_{e}=2$, and the pinning constant of Lemma~\ref{lem:B.4}, amount to a constant per cell, as $\#\mathcal{E}\le2|X|$. Thus the sum over the assignments contributes to $K_{0}^{B}$ only the numerical factor $K_{t}\le22$ per $\mathcal{E}$-factor, that is, at most $22^{2}$ per cell, and the sum over the forms at most $3^{2}$ per cell. In particular, $K_{0}^{B}$ has the dependence stated in the proposition.
\qed
\end{proof}

\begin{proposition}\label{prop:B.10}
Let $q_{1}\ge1$ be the exponent of the leftover pinning factor $\prod_{\alpha}(n_{\alpha}!)^{-q_{1}}(N_{\alpha}!)^{-q_{1}}$ ($q_{1}=2$ for $\qpin=q_{0}+2$). The $\bD$-dependent part of the factor on a class-B cell is at most $e^{c_{B}(\bD)}$, where $c_{B}(\bD):=q_{1}(C_{\mathcal{E}}/\bD)^{1/q_{1}}+q_{1}(C_{G}\bD^{-1/2})^{1/q_{1}}\le q_{1}(C_{\mathcal{E}}^{1/q_{1}}+C_{G}^{1/q_{1}})\bD^{-1/q_{1}}$, $C_{\mathcal{E}}$ being the constant of one charged $\mathcal{E}$-factor and $C_{G}$ that of one single-derivative block; both are independent of $L'$. For $q_{1}=2$, $c_{B}(\bD)\le2(C_{\mathcal{E}}^{1/2}+C_{G}^{1/2})\bD^{-1/2}$. The bound gives $c_{B}(\bD)=o(\Phi_{\cU})$ as $\tau\to\infty$ if and only if $q_{1}>1$.
\end{proposition}

\begin{proof}
By Theorem~\ref{thm:B.6}(iii) and the bound for a class-B cell in the proof of Proposition~\ref{prop:B.7}, the cell carries at most $(C_{\mathcal{E}}/\bD)^{N^{\mathrm{ch}}}(C_{G}\bD^{-1/2})^{n^{(1)}}(N_{\alpha}!)^{-q_{1}}(n_{\alpha}!)^{-q_{1}}$ with $N^{\mathrm{ch}}\le N_{\alpha}$, $n^{(1)}\le n_{\alpha}$. We may assume $C_{\mathcal{E}}/\bD,C_{G}\bD^{-1/2}\ge1$ and raise the exponents to $N_{\alpha}$, $n_{\alpha}$. For $q,a\ge1$, $\sup_{N}a^{N}/(N!)^{q}\le(\sum_{N}a^{N/q}/N!)^{q}=e^{qa^{1/q}}$. For the last assertion, we use $\Phi_{\cU}\asymp\bD^{-1}\bGM^{-3}$, which follows from Proposition~\ref{prop:7.22} as $\mt L'=K_{\mathrm{col}}\bGM$, and $\bD=e^{-\pi^2\vCb(0)\tau(1+o(1))}$. If $q_{1}>1$, then $c_{B}/\Phi_{\cU}\le C\bD^{1-1/q_{1}}\bGM^{3}\to0$. If $q_{1}=1$, the bound is of order $\bD^{-1}$ and the ratio grows like $\bGM^{3}$. As Lemma~\ref{lem:B.4} holds for all $q$, the choice $\qpin=q_{0}+2$ is admissible.
\qed
\end{proof}

The order $\bD^{-1/q_{1}}$ cannot be improved within the bounds of Lemma~\ref{lem:B.5}: a class-B cell can contain an $\mathcal{E}$-factor all of whose factors are undifferentiated class-B $\eps$'s, of weight exactly $-1$, since $\eps=e^{im\bGM^{1/2}(\psi+D+h)}-1$ with $D\neq0$ has no small bound.

\begin{remark}[One or two derivatives on $G_{2}$ in class-A cells]\label{rem:B.11}
On \cite[p.~231]{BF80}, the bounds \cite[(9.814), (9.815)]{BF80} read $(c\beta^{1/2})^{n}\cdot2$ for the $n$th derivative of a term from v), and $c\beta(|\delta|^{2-n}+|g-h|^{2-n})$ if $n<2$, $(c\beta^{1/2})^{n}$ if $n\ge2$, for either term from vi). With these bounds a $G_{2}$-block with one or two derivatives in a class-A cell carries no net power of $\bD$. Indeed, a $G_{2,2}$-block whose single derivative falls on the factor vi) has weight $-1+0+1=0$, and a $G_{2,1}$-block with one or two derivatives has weight $-1+1=0$. A lone cell whose only witness is such a block then receives no power of $\bD$. The class-A bounds of Lemma~\ref{lem:B.5}(a) supply this power. In a class-A cell $|v_{m}(\Omega)|\le|m|\bGM^{1/2}|\mathrm{avg}_{\Omega}\psi|$ as $h\in P\ZZ$, and by charge symmetry $G_{2,1}$ is of fourth order in $\tilde\delta$. By Lemma~\ref{lem:B.5}(b), both give weight at least $\frac12$. Without charge symmetry the third-order form of the second term of vi) already gives $-1+\frac32=\frac12$.
\end{remark}

\begin{proof}[Proof of Lemma~\ref{lem:7.40}]
We insert Proposition~\ref{prop:B.7} into Lemma~\ref{lem:7.39}. With $e^{c_{2}^{L}|X|}$ and the factor $e^{c_{13}|X|}$ of Proposition~\ref{prop:B.7}, its factors are distributed cell by cell as stated, with $c_{o}:=c_{2}^{L}+c_{13}+\log K_{0}^{B}+\log\max(1,K_\ell)$ (link $i$ is mapped injectively to a cell of $Y_{i+1}$, so $\prod_iK_\ell\le\prod_\Delta\max(1,K_\ell)$) and $c_{B}$ of Proposition~\ref{prop:B.10} with $q_{1}=2$. Since a class-B cell meets $\bigcup_{\beta}S_{\beta}$, it lies in a wall polymer, in $X_{\Sigma}$ or in the seed. Factors on class-A cells carry in addition $\bD^{w}$ with $w\ge0$, by Theorem~\ref{thm:B.6}(ii). Their constants are independent of $L'$, and so are $C_{\mathcal{E}}$ and $C_{G}$, which are built from the same quantities (see Section~\ref{S:G.B11}). Indeed, no class-A cell carries $D$, $F_{2}'$ or a quantity depending on $\hat\Sigma$, by Proposition~\ref{prop:7.9}(b), (e). This proves the assertion of the lemma on $L'$, with $\eps(L')$ coming from Lemma~\ref{lem:B.8}.
\qed
\end{proof}

\begin{proof}[Proof of Proposition~\ref{prop:7.41}]
We insert Proposition~\ref{prop:B.7} into Lemma~\ref{lem:7.39} with $c_{1}'=c'/4$. Of $F_{1}^{\mathrm{loc}}$ we use $c'/16$ on $X\setminus Z$ for the numerical factors of Lemma~\ref{lem:B.8} and $c'/8$ there for the wall polymers, and $c'/32$ on $Z$ for the numerical factors. The share $c'/8$ gives each cell of a wall polymer of $X\setminus Z$ the factor $e^{-(c'/8)\Phi_{\cU}/(1+D_{\cU})}$, since $F_{1}(S_{\beta})\ge\Phi_{\cU}\#S_{\beta}$ by Proposition~\ref{prop:7.22}(c), a closure adds at most $D_{\cU}$ cells per cell, and $F_{1}^{\mathrm{loc}}(\mathcal{S})\ge\sum_{S_{\beta}\subseteq\mathcal{S}}F_{1}(S_{\beta})$ by the definition of $F_{1}^{\mathrm{loc}}$ in Corollary~\ref{cor:7.11}. The link decay of Proposition~\ref{prop:B.7} absorbs $e^{\delta_{2}d}$, as $\delta_{2}<\bar\delta$. The Mayer cubes and the per-$\mathcal{E}$-factor constants growing with $t_{e}$ are absorbed inside Proposition~\ref{prop:B.7}, and the leftover Mayer weights of that proposition satisfy $\mathfrak{m}_{e}\le1$, since $\mu\ge1+2c''_{10}$ by item~12 of the list in Section~\ref{ssec:D.1}. With $K_{B}:=K_{0}^{B}\max(1,K_\ell)$, independent of $\bD$, $L'$, $\tau$, the per-cell factors and the line constants (one per link, and the links inject into the cells) give $\eps_{\Delta}$ (Section~\ref{S:G.B11}). The factor $e^{C_{\Dep}\cdep}$ is present only on cells meeting $\Dep$, and the rest of the factor on $X\setminus Z$ is at most $1$. The exponent $p_{0}=\frac16$ is attained by Theorem~\ref{thm:B.6}.
\qed
\end{proof}
}

\providecommand{\bGM}{\beta_{\mathrm{GM}}}
\providecommand{\mt}{\tilde m}
\providecommand{\MY}{M_Y}
\providecommand{\lPV}{\lambda_{\mathrm{PV}}}
\providecommand{\Floc}{F_1^{\mathrm{loc}}}
\providecommand{\UU}{\mathcal{U}}
\providecommand{\Df}{\mathfrak{D}}
\providecommand{\sat}{\operatorname{sat}}
\providecommand{\supp}{\operatorname{supp}}
\providecommand{\one}{\mathbf{1}}
\providecommand{\Z}{\mathbb{Z}}
\providecommand{\R}{\mathbb{R}}
\providecommand{\bC}{\mathbf{C}}
\providecommand{\cM}{\mathcal{M}}

\section{The cluster expansion on slit domains: proof of Proposition~\ref{prop:7.25}}
\label{app:C}

In this appendix we give a condensed proof of Proposition~\ref{prop:7.25}. The expansion of \cite[Appendix~E]{GM82}
is an algebraic identity, built from Bernoulli interpolation of the couplings between consecutive
sets, the fundamental theorem of calculus and Gaussian integration by parts, and its only analytic
input is absolute convergence. On $G'$ absolute convergence needs the Poincaré inequality on cells,
as the Peierls identity of Proposition~\ref{prop:7.4} removes the mass in the cell-fluctuation directions
(Lemma~\ref{lem:C.4}). On block cubes this is the bound of \cite[(9.640) to (9.642)]{BF80}.

\paragraph{Setting.} We fix $h$. Everything below depends on it only through its walls and the integrand of Lemma~\ref{lem:7.24}, and therefore
holds equally for wall fields (Lemma~\ref{lem:7.wall}). The covariance $\bC$ is $C=(C_0^{-1}+\nu)^{-1}$ of Proposition~\ref{prop:7.8} or $C'$
of Lemma~\ref{lem:7.31}, positive definite on $V$ with $\bC\le\alpha C_0$, $\alpha=2$ for $C$ and
$\alpha=\alpha_D\le\frac76$ for $C'$. The integrand is $e^EQ_h$, $Q_h=e^Ge^{-\Floc(V)}e^{-F_2}$, with $G^\natural$
for $G$ when $\bC=C'$. We write $\partial_x:=\partial/\partial\psi(x)$. A \emph{sequence} is $\bar y=(Y_1,\dots,Y_m)$ of nonempty
admissible sets (Definition~\ref{def:7.23}), pairwise disjoint, with the first set $Y_1$
fixed before the expansion starts. Its union is $X(\bar y):=\bigcup_iY_i$.

\paragraph{Interpolation.} For pairwise disjoint admissible $Y_1,\dots,Y_n$ with union
$X$, let $Y_{n+1}:=V\setminus X$, $\chi_i:=\one_{Y_i}$ and, for $s\in[0,1]^n$,
\begin{equation}\label{eq:C.E3}
p(x,y\mid s):=\sum_{1\le i<j\le n+1}s_i\cdots s_{j-1}\bigl[\chi_i(x)\chi_j(y)+\chi_i(y)\chi_j(x)\bigr]
+\sum_{i\le n+1}\chi_i(x)\chi_i(y),
\end{equation}
and $\bC(s)(x,y):=p(x,y\mid s)\bC(x,y)$, as in \cite[(E.2), (E.3)]{GM82}, with $\chi_{n+1}=\one_{V\setminus X}$. For $x\in Y_a$, $y\in Y_b$, $a\le b$,
$p(x,y\mid s)=\prod_{a\le k<b}s_k$, and hence $\partial_{s_k}p\in[0,1]$ vanishes unless $x$, $y$ lie on different
sides of the cut between $\bigcup_{i\le k}Y_i$ and $\bigcup_{i>k}Y_i$. For $\varepsilon\in\{0,1\}^n$ let $\pi(\varepsilon)$ be
the partition of $\{1,\dots,n+1\}$ into the maximal runs of consecutive indices joined by
the $k$ with $\varepsilon_k=1$, $Y_J:=\bigcup_{i\in J}Y_i$ (admissible), and
$\lambda_\varepsilon(s):=\prod_ks_k^{\varepsilon_k}(1-s_k)^{1-\varepsilon_k}$. Then
\begin{equation}\label{eq:C.decomp}
\bC(s)=\sum_\varepsilon\lambda_\varepsilon(s)\,\bC_\varepsilon,\qquad\bC_\varepsilon:=\sum_{J\in\pi(\varepsilon)}\one_{Y_J}\bC\one_{Y_J},
\end{equation}
since $\sum_\varepsilon\lambda_\varepsilon(s)\bC_\varepsilon(x,y)$ is $\bC(x,y)$ times the probability, for independent
Bernoulli$(s_k)$ variables $\varepsilon_k$, that $a$ and $b$ lie in one run. Hence
$\lambda_{\min}(\bC)\le\bC(s)\le\|\bC\|\le\alpha\mt^{-2}$, $\bC(s)(x,x)=\bC(x,x)$, $|\bC(s)(x,y)|\le|\bC(x,y)|$ (so
each entrywise monotone norm of $\bC(s)$ is at most that of $\bC$), and
$\|\cM^{1/2}\bC(s)\cM^{1/2}\|\le\|\cM^{1/2}\bC\cM^{1/2}\|$ for each symmetric $\cM\ge0$ commuting with all
$\chi_i$. For a sequence $\bar y$ of length $n$, we write $\bC^{\bar y}(s,s_n)$ for \eqref{eq:C.E3} built from $\bar y$
($s=(s_1,\dots,s_{n-1})$), $d\mu'^{\,\bar y}_{ss_n}$ for its Gaussian measure and $d\mu^{\bar y}_s$ for the measure at
$s_n=0$.

\paragraph{The weights of $E$.} By (S1) and \cite[(E.5a) to (E.5d)]{GM82}, $E$ is a sum
$\sum_\tau e_\tau$ of terms of two kinds: for $t\ge3$ particles,
$e_\tau=\frac1{t!}\rho_t(\xi_1,\dots,\xi_t)\,\varepsilon_k(\xi_1)\cdots\varepsilon_k(\xi_t)$, and for $t=2$ the three terms $\frac12\rho_2\tilde\varepsilon_k(\xi_1)\tilde\varepsilon_k(\xi_2)$, $\frac12\rho_2\tilde\varepsilon_k(\xi_1)\,im_2\bGM^{1/2}\psi(x_2)$ and
$\frac12\rho_2\,im_1\bGM^{1/2}\psi(x_1)\tilde\varepsilon_k(\xi_2)$. Here $x_i\in V$, $m_i\ne0$,
$\varepsilon_k(\xi)=e^{im(\bGM^{1/2}(\psi+g)(x)+kf(x))}-1$ and $\tilde\varepsilon_k(\xi):=\varepsilon_k(\xi)-im\bGM^{1/2}\psi(x)$. With
$\supp\tau:=\{x_1,\dots,x_t\}$, $I(\tau)$ is the set of $k\le n$ such that $\supp\tau$ meets both
$\bigcup_{i\le k}Y_i$ and $\bigcup_{i>k}Y_i$. We put $w_s(\tau):=\prod_{k\in I(\tau)}s_k$, $E(Y,s):=\sum_{\supp\tau\subseteq Y}w_s(\tau)e_\tau$ and
$E(Y):=E(Y,1)$, and $E^{\bar y}(Y,(s,s_n))$ denotes $E(Y,\cdot)$ built from $\bar y$. These functions agree with $E(X,s)$ of \cite[p.~89]{GM82} and
$E(\Lambda,ss_n)$ of \cite[p.~90, (v)]{GM82}, with sites for the unit cubes $a_i$.

\paragraph{Absolute convergence.} Let $P_V$ be the cell-fluctuation projection,
$(P_V\varphi)|_\Omega:=\varphi|_\Omega-\operatorname{avg}_\Omega\varphi$, and $\|\nu\|_0:=\sup_x\sum_y|\nu(x,y)|$. Summing the fine Poincaré inequality on
cells, given by Theorem~\ref{thm:3.20}\,(R6) and Proposition~\ref{prop:6.10}\,(iv), with the overlap
$N_{\mathrm{ov}}$, we obtain $\|P_V\varphi\|^2\le C_P'L^2\|\nabla\varphi\|^2$, $C_P':=C_W^f(k_\square/L)^2N_{\mathrm{ov}}$, and by (L8-2) of
Subsection~\ref{ssec:7.6}, $C_P'(\mt L)^2\le\eta_H/8<\frac17$. The constants $c_V$, $b_V$, $C_\alpha$ below
serve only to prove convergence and may depend on $G'$.

\begin{lemma}[Growth of $\ln Q_h$]\label{lem:C.3}
Let $\vartheta:=\psi+D$. For every admissible $Y$, every sequence, every $s$ and real $\psi$:
(a) $E(Y,s)$ is real, and $E(Y,s)=B_{Y,s}(\psi)+\frac12\langle\psi,\nu^Y_s\psi\rangle$ with $|B_{Y,s}(\psi)|\le b_V$ and
$\nu^Y_s(x_1,x_2):=\one_Y(x_1)\one_Y(x_2)w_s(x_1,x_2)\nu(x_1,x_2)$, where $w_s(x_1,x_2):=\prod_{a\le k<b}s_k$ for $x_1\in Y_a$,
$x_2\in Y_b$, $a\le b$; moreover $\|\nu^Y_s\|_{2\to2}\le\|\nu\|_0\le\varepsilon_\nu\mt^2$, and $|\partial^\alpha E(Y,s)|\le C_\alpha(1+\|\psi\|)$
for $|\alpha|\ge1$; (b) $G_{2,1}(Y)\le\frac12(1+C_{\mathrm{ref}}e^{-2\kappa_S})\mt^2\|\one_YP_V\vartheta\|^2$; (c) $G_1(Y)\le c_V$,
$|G_{2,2}(Y)|\le c_V(1+\|\psi\|_1)$ and $|G_{2,3}(Y)|\le c_V$; for the terms $G^\Df_{2,3}$, $G^\natural_{2,3}$ of $G^\natural$
(Lemma~\ref{lem:7.31}), $G^\Df_{2,3}(Y)\le c_V$ and $G^\natural_{2,3}(Y)\le c_V+\frac12\theta'\mt^2\|\psi\|^2$; (d) $\Floc(Y)\ge0$ does not
depend on $\psi$, $F_2(Y)$ is linear in $\psi$, and $|\partial^\alpha\ln Q_h(Y)|\le C_\alpha(1+\|\psi\|)$ for $|\alpha|\ge1$.
\end{lemma}

\begin{proof}[Sketch]
(a) By (S3), the terms pair into complex conjugates. With $\varepsilon_k=\tilde\varepsilon_k+im\bGM^{1/2}\psi$ the terms with
$t=2$ add up to $\frac12\rho_2\varepsilon_k\varepsilon_k$ plus a part with charge sum $\frac12\nu$. Using (S2$^+$), we bound $B_{Y,s}$ by
$\sum_{t\ge2}\frac{2^t}{t!}\sum|\rho_t|$, and its $\psi$-derivatives by the same sums with the weights $(\sum_i|m_i|)^n$, and the Schur test with $0\le w_s\le1$ gives the norm bound. (b) Here
$\tilde\delta=P_V\vartheta$, and by (S3) $G_{2,1}$ is a sum of terms $\rho^{\mathrm{ref}}(m)[\cos y-1+\frac12y^2]$ of modulus at most $\frac12|\rho^{\mathrm{ref}}(m)|y^2$, since $0\le\cos y-1+\frac12y^2\le\frac12y^2$, with
$\bGM\sum_mm^2|\rho^{\mathrm{ref}}(m)|\le(1+C_{\mathrm{ref}}e^{-2\kappa_S})\mt^2$. Part (c) is explicit on the finite set $V$, with
Proposition~\ref{prop:7.5} off $\Df$. Part (d) follows from Corollary~\ref{cor:7.11}, Proposition~\ref{prop:7.9}, and the fact that $\ln r_\Omega$ is
smooth and periodic in $A_\Omega$.
\end{proof}

\begin{lemma}[Margin]\label{lem:C.4}
Assume the standing hypothesis \eqref{eq:7.standing}, and let $\kappa_S$ be so large that
$(1+C_{\mathrm{ref}}e^{-2\kappa_S})\eta_H/8\le\frac17$, so that (L8-2) gives $(1+C_{\mathrm{ref}}e^{-2\kappa_S})C_P'(\mt L)^2\le\frac17$. Put
$\cM:=(1+C_{\mathrm{ref}}e^{-2\kappa_S})\mt^2P_V+(\varepsilon_\nu+\theta')\mt^2$ for $\bC=C'$ and
$\cM:=(1+C_{\mathrm{ref}}e^{-2\kappa_S})\mt^2P_V+\varepsilon_\nu\mt^2$ for $\bC=C$. Then $\cM\ge0$ is symmetric, commutes with $\one_\Omega$ for every cell $\Omega$,
$\cM\ge(1+C_{\mathrm{ref}}e^{-2\kappa_S})\mt^2P_V+\|\nu\|_0+\theta'\mt^2\one_{[\bC=C']}$, and
\begin{equation}\label{eq:C.margin}
\langle\varphi,\cM\varphi\rangle\le q_M\langle\varphi,\bC^{-1}\varphi\rangle,\qquad q_M\le\tfrac16\ (\bC=C'),\quad q_M\le\tfrac27\ (\bC=C).
\end{equation}
Moreover $\|\cM^{1/2}\bC(s)\cM^{1/2}\|\le q_M$ for every sequence of admissible sets and every $s$,
and the same holds for $\one_U\bC\one_U$ on $\R^U$ whenever $U$ is a union of cells.
\end{lemma}

\begin{proof}
We control the cell fluctuations by the gradient through the Poincaré
inequality on cells, and the constant part by the mass. Both are dominated by
$C_0^{-1}=\MY^{-2}\Delta^2-\Delta+\mt^2\ge\mt^2-\Delta$, and $C_0^{-1}\le\alpha\bC^{-1}$. The projection $P_V$ is block diagonal over cells,
and thus $\cM$ commutes with each $\one_\Omega$. Moreover, $\|\nu\|_0\le\varepsilon_\nu\mt^2$ by (S5). By the Poincaré inequality
$\|P_V\varphi\|^2\le C_P'L^2\|\nabla\varphi\|^2$,
\[
\langle\varphi,\cM\varphi\rangle\le(1+C_{\mathrm{ref}}e^{-2\kappa_S})C_P'(\mt L)^2\|\nabla\varphi\|^2+(\varepsilon_\nu+\theta')\mt^2\|\varphi\|^2
\le\tfrac17\langle\varphi,C_0^{-1}\varphi\rangle,
\]
which is at most $\frac\alpha7\langle\varphi,\bC^{-1}\varphi\rangle$, with $\alpha=\alpha_D\le\frac76$ for $C'$ and $\alpha=2$ for $C$ (where $\theta'$ is
absent). We now turn to the last statement, for which any $q_M<1$ suffices.
Inequality \eqref{eq:C.margin} says that $\|\cM^{1/2}\bC\cM^{1/2}\|\le q_M$. Each $Y_i$ and $V\setminus X$ is a union of cells, and hence
$\cM$ commutes with each $\chi_i$ and \eqref{eq:C.decomp} applies. Since $\one_U$ commutes with $\cM$, we also get
$\|\cM^{1/2}\one_U\bC\one_U\cM^{1/2}\|\le q_M$, and \eqref{eq:C.decomp} applies on $\R^U$.
\end{proof}

\begin{lemma}\label{lem:C.5}
Call $F:\R^V\to\mathbb C$ \emph{admissible} if $F$ is $C^2$ and, for $|\alpha|\le2$,
$|\partial^\alpha F(\psi)|\le K_F(1+\|\psi\|)^{d_F}\exp(\frac12\langle\psi,\cM\psi\rangle+k_F\|\psi\|_1)$. Under the hypotheses of
Lemma~\ref{lem:C.4}: (a) for every admissible $Y$ and $s'$, and every finite composition $\mathfrak L$
of operators $\partial_x$ and of multiplications by sub-series, with weights in $[0,1]$, of the term expansions of $E(\cdot,s')$, of $\partial_{s_n}E(\cdot,s')$ and of their $\psi$-derivatives (all lie in one polynomial class), the function
$e^{E(Y,s')}\mathfrak LQ_h(Y)$ is admissible, uniformly in $s'$; (b) if $t\mapsto\bC(t)$ is one of the
covariances \eqref{eq:C.E3} with one $s$-variable moving and $F$ is admissible, then
$\int d\mu_{\bC(t)}|F|$ is bounded uniformly in $t$ and in the other variables, $t\mapsto\int d\mu_{\bC(t)}F$ is
$C^1$ on $[0,1]$, and
\begin{equation}\label{eq:C.E14}
\partial_t\int d\mu_{\bC(t)}F=\frac12\sum_{x,y\in V}\partial_t\bC(t)(x,y)\int d\mu_{\bC(t)}\,\partial_x\partial_yF ;
\end{equation}
(c) if $F$ also depends on $t$, is $C^1$ in $t$, and $F$ and $\partial_tF$ are admissible uniformly in $t$,
then $\partial_t\int d\mu_{\bC(t)}F=\int d\mu_{\bC(t)}\partial_tF$ plus the right side of \eqref{eq:C.E14}.
\end{lemma}

\begin{proof}[Sketch]
(a) We collect the exponents of Lemma~\ref{lem:C.3} and use the lower bound on $\cM$ of
Lemma~\ref{lem:C.4}. (b) By Lemma~\ref{lem:C.4} and \eqref{eq:C.decomp}, $\bC(t)^{-1}\ge(\cM+2\epsilon)/q'$ with
$q':=q_M+2\epsilon\|\bC\|<1$ for small $\epsilon>0$. Therefore $|F|g_t$, where $g_t$ is the Gaussian density, is integrable
uniformly in $t$. Since $\partial_tg_t=\frac12\sum_{x,y}\partial_t\bC(t)(x,y)\partial_x\partial_yg_t$, two integrations by parts give
\eqref{eq:C.E14}, the step ``By partial integration'' of \cite[p.~91]{GM82}. (c) This follows from the product rule
under the same domination.
\end{proof}

\paragraph{The operators $\kappa$.} We fix $\bar y$ of length $n$ with $X:=X(\bar y)\subsetneq V$ and write
$E:=E^{\bar y}(V,(s,s_n))=\sum_\tau w(\tau)e_\tau$, $E_x:=\partial_xE$. As $e^{-E}\partial_x\partial_y(e^EF)=(\partial_x+E_x)(\partial_y+E_y)F$ is
symmetric in $x,y$, and $\partial_{s_n}\bC^{\bar y}(x,y)=0$ unless $x,y$ lie on different sides of $X$,
Lemma~\ref{lem:C.5}\,(c) yields
$\partial_{s_n}\int d\mu'^{\,\bar y}_{ss_n}e^EF=\int d\mu'^{\,\bar y}_{ss_n}e^E\tilde\kappa(n)F$ for $F$ not depending on $s_n$ with $e^EF$
admissible, where
\begin{equation}\label{eq:C.E16}
\tilde\kappa(n):=\partial_{s_n}E+\sum_{x\in V\setminus X}\sum_{y\in X}\partial_{s_n}\bC^{\bar y}(x,y)(\partial_x+E_x)(\partial_y+E_y),
\end{equation}
as in \cite[(E.15), (E.16)]{GM82}. As GM note on p.~92, the factor $\frac12$ of (E.16) cancels by symmetry. Inserting $E=\sum_\tau w(\tau)e_\tau$, we write $\tilde\kappa(n)=\sum_\theta\mathfrak t_\theta$ with elementary terms of
five types: (1) $(\partial_{s_n}w(\tau))e_\tau$; (2) $\partial_{s_n}\bC(x,y)\partial_x\partial_y$; (3) $\partial_{s_n}\bC(x,y)w(\tau)(\partial_x\partial_ye_\tau)$; (4)
$\partial_{s_n}\bC(x,y)w(\tau)(\partial_ye_\tau)\partial_x$, respectively $(\partial_xe_\tau)\partial_y$; (5)
$\partial_{s_n}\bC(x,y)w(\tau)w(\tau')(\partial_xe_\tau)(\partial_ye_{\tau'})$. The support $S(\theta)$ is $\supp\tau$ in type (1), and $\{x,y\}$
with the supports of the $\tau$'s involved otherwise. If $\mathfrak t_\theta\ne0$, then $S(\theta)\not\subseteq X$, since in type (1)
$n\in I(\tau)$ and otherwise $x\in V\setminus X$. We put $Y_{n+1}(\theta):=\sat_h(S(\theta)\setminus X)$ and
\begin{equation}\label{eq:C.E20}
\kappa(n)[Y']:=\sum_{\theta:\,Y_{n+1}(\theta)=Y'}\mathfrak t_\theta,\qquad\tilde\kappa(n)=\sum_{Y'}\kappa(n)[Y'],
\end{equation}
as in \cite[(E.11), (E.20)]{GM82}. With $\bar y':=(\bar y,Y')$ and $X':=X\cup Y'$, each $\theta$ in $\kappa(n)[Y']$ has
$S(\theta)\subseteq X'$, coefficients depending on $\psi|_{X'}$ only, derivatives at $x\in Y'$, $y\in X$, and
$E$-terms of $E^{\bar y'}(X',(s,s_n))$ with the same weights. Thus $\kappa(n)[Y']$ depends on $s_1,\dots,s_n$
only. Moreover, $\partial_{s_n}\bC^{\bar y}(x,y)=s_a\cdots s_{n-1}\bC(x,y)=\partial_{s_n}\bC^{\bar y'}(x,y)$ for $x\in Y'$,
$y\in Y_a$. For $\bar y=(Y_1,\dots,Y_m)$ the ordered product $\kappa(\bar y,s):=\kappa(m-1)\circ\cdots\circ\kappa(1)$, with
$\kappa(i):=\kappa(i)[Y_{i+1}]$ built from $(Y_1,\dots,Y_i)$ and each operator acting on everything to its
right, is $\kappa(\bar y,s)$ of \cite[(E.9), (E.10)]{GM82}.

\paragraph{Proof of Proposition~\ref{prop:7.25}.} We follow the proof of \cite[Lemma~E.1,
Proposition~E.2, (E.2) to (E.24)]{GM82}, an induction in $n$. Each step is the fundamental
theorem of calculus \cite[(E.17)]{GM82} in a new variable $s_{n+1}$, started from the
measure and the $E$ of the previous step by the consistency and decoupling relations
\cite[(E.21a), (E.21b)]{GM82}, which hold on $G'$ by \eqref{eq:C.E3} and the definition of
$w_s(\tau)$. Lemma~\ref{lem:C.5} justifies each exchange of derivatives and integrals. The following changes are
needed on $G'$.

(1) \emph{Atoms are sites}, whereas the atoms of \cite[p.~89]{GM82} are unit cubes.
Since each $Y_j$ is admissible, $w_s(\tau)$ depends only on which $Y_j$ contain the points.

(2) \emph{One finite matrix with mass at every site}, and $\chi_{n+1}=\one_{V\setminus X}$, so that the region
outside $\Lambda$ of \cite[p.~92]{GM82} does not occur. Termination occurs at $m\le e_h+1$, since each $Y_j$, $j\ge2$,
contains an element outside $Y_1$ (\cite[p.~93]{GM82}).

(3) \emph{First sets may depend on $h$
and be disconnected}. In \cite[p.~90, (iii)]{GM82} the first set is a fixed union of unit cubes. Here the factorization $Q_h=Q_h(Y_1)Q_h(V\setminus Y_1)$ of Lemma~\ref{lem:7.24} suffices for the
step $n=1$, and the admissibility of each $\bigcup_{j\le i}Y_j$ for the selection rule; every comparison is made at one value of $h$. Type (2) terms
reach the element of $x$ at any distance at which $\bC(x,y)\ne0$, and therefore the sets $X(\bar y)$ are
disconnected in general.

(4) \emph{On $U=X_0^c$ the covariance is $\one_U\bC\one_U$}: marginals of
Gaussian measures are compressions, and hence $\mathcal K_{\bar y}$ involves the covariance only through the block $p(x,y\mid s)\bC(x,y)$,
$x,y\in X(\bar y)$. Lemma~\ref{lem:C.4} covers $\one_U\bC\one_U$.

(5) \emph{The selection rule: $S(\theta)$ contains
the bare point $x$} of each covariance term, so that all derivatives of a term of
$\kappa(n)[Y']$ act in $X\cup Y'$. By contrast, \cite[(E.11)]{GM82} is stated for the supports of the
$E$-terms, and for the term without $E$-terms \cite[p.~91; (E.19), p.~93]{GM82} take $Y_{n+1}\ni x$.

(6) \emph{``$\chi C\chi$'' is an exact block-sum decomposition}: the convex combination of
\cite[p.~88]{GM82} is \eqref{eq:C.decomp}, with compressions to admissible blocks. The bounds on $\bC(s)$ in Lemmas~\ref{lem:C.4}, \ref{lem:C.5} and in Section~\ref{sec:stage2}
(Section~\ref{S:constF}) follow from this decomposition. The integrals $\int d\mu_\bC e^{E(Y)}Q_h(Y)$ are positive, since $E(Y)$, $G$, $\Floc$, $F_2$ are
real by Lemma~\ref{lem:C.3}, and they are finite by Lemma~\ref{lem:C.5}\,(b).
}
\providecommand{\bGM}{\beta_{\mathrm{GM}}}
\providecommand{\mt}{\tilde m}
\providecommand{\MY}{M_{Y}}
\providecommand{\lPV}{\lambda_{\mathrm{PV}}}
\providecommand{\Gp}{G'}
\providecommand{\XX}{\mathbb{X}}
\providecommand{\KK}{\mathfrak{K}}
\providecommand{\RHK}{R_{\mathrm{HK}}}
\providecommand{\rcov}{r_{\mathrm{cov}}}
\providecommand{\etaw}{\eta_{w}}
\providecommand{\eps}{\varepsilon}

\section{Heat kernel, coarse-to-fine transfer and covariance bounds on slit domains}
\label{app:F}

In this appendix we give condensed proofs of Proposition~\ref{prop:6.10} and Theorem~\ref{thm:6.9} and sketch the proofs of Proposition~\ref{prop:6.11}, Theorem~\ref{thm:6.12} and Lemma~\ref{lem:F.7}. Full proofs of the last three are in Section~\ref{S:E2}. Throughout, $\Gp=(V,E)$ is an output, $d=d_{\Gp}$, $\mathcal{E}(f):=\sum_{\{x,y\}\in E}(f(x)-f(y))^{2}$, and $\partial U=\partial_{\Gp}U$ is the set of bonds of $E$ with exactly one end in $U$.

\subsection{Coarse-to-fine transfer: proof of Proposition~\ref{prop:6.10}}
\label{ssec:F.1}

A large fine set either fills many coarse cells, whose boundary (R2) controls, or is spread thinly inside cells, where box isoperimetry applies (canonical paths and Loomis\textendash Whitney). Counting faces combines the two estimates. The coarse cells, the sites $c\in V_{c}$ of the coarse slit graph $\Gamma=(V_{c},E_{c})$ of \S\ref{ssec:6.3}, are boxes $B_{c}$ with sides in $[r_{0},2r_{0}-1]$ by Definition~\ref{def:3.4}, and by \S\ref{ssec:2.5} the box graph of $B_{c}\cup B_{c'}$, $\{c,c'\}\in E_{c}$, lies in $\Gp$. Thus a coarse bond between a \emph{heavy} cell ($|A\cap B_{c}|\ge|B_{c}|/2$) and a light one carries at least $r_{0}^{2}/4$ bonds of $\partial A$ in $B_{c}\cup B_{c'}$, and $|\partial A|\ge(r_{0}^{2}/24)|\partial_{\Gamma}H|$ for the set $H$ of heavy cells. Combined with (R2), in the form of Theorem~\ref{thm:4.7}(i), (ii), this proves (i), and the same argument with the expansion of $K$ in place of (R2) proves (iii). We deduce (ii) from (i) as in Theorem~\ref{thm:4.7}(iv), and from Theorem~\ref{thm:4.7}(ii) and $r_{0}^{3}V_{*}=2V_{*}^{f}$. Part (iv) is (iii) for the blow-up of the set $\KK^{\flat}_{c}$ of (R6) (Lemma~\ref{lem:4.16}) with Cheeger's inequality, the overlap count inherited and $|\Omega|\ge c_{1}L^{3}$ from (R5). Finally, (v) holds because balls in a connected graph of degree at most $6$ and expansion $h\le1$ grow by the factor $1+h/6$ per step up to half the vertices, so the diameter is at most $2+(24/h)\log|\KK^{\flat}|<L'$ under (L8-2). The constants are those of \S\ref{S:E1.ctf}.

\subsection{Nash's inequality from Faber\textendash Krahn, and the on-diagonal bound}
\label{ssec:F.2}

\begin{lemma}[Nash inequality]\label{lem:F.3}
For $f\ge0$ on $V$ with $\sum f=1$, and every $r\ge1$ with $2r^{3}\le V_{*}^{f}$, $\|f\|_{2}^{2}\le A_{N}r^{2}\mathcal{E}(f)+r^{-3}$, where $A_{N}:=24\cdot2^{2/3}/c_{F}^{2}$.
\end{lemma}

\begin{proof}
We put $s:=1/(2r^{3})$. Then $f^{2}\le2(f-s)_{+}^{2}+2\min(f,s)^{2}$ and $\sum\min(f,s)^{2}\le s$. The function $(f-s)_{+}$ vanishes off $\Omega_{s}:=\{f>s\}$, which has $|\Omega_{s}|<1/s=2r^{3}\le V_{*}^{f}$, and $\mathcal{E}((f-s)_{+})\le\mathcal{E}(f)$. By Proposition~\ref{prop:6.10}(ii), $\sum(f-s)_{+}^{2}\le(12/c_{F}^{2})(2r^{3})^{2/3}\mathcal{E}(f)$.
\qed
\end{proof}

\begin{proof}[Proof of (HK0) and (HK1)]
We obtain (HK0) from the degree bound alone, by Davies' perturbation argument. For $\psi:=-\alpha d(x,\cdot)$,
\[
\langle g,e^{\psi}\Delta_{\Gp}e^{-\psi}g\rangle=-\mathcal{E}(g)+2\sum_{\{x,y\}\in E}(\cosh(\psi(x)-\psi(y))-1)g(x)g(y)\le6(\cosh\alpha-1)\|g\|^{2},
\]
so $\|e^{\psi}P_{t}e^{-\psi}\|_{2\to2}\le e^{6(\cosh\alpha-1)t}$ and $p_{t}(x,y)=e^{\psi(y)-\psi(x)}\langle\delta_{x},e^{\psi}P_{t}e^{-\psi}\delta_{y}\rangle\le e^{-\alpha d+6(\cosh\alpha-1)t}$. At $\alpha=\frac14$, $6(\cosh\frac14-1)<\frac14-\frac1{17}$. The number $N_{t}$ of jumps up to time $t$ is dominated by a Poisson variable of mean $6t$, and $d(x,X_{t})\le N_{t}$.

For (HK1), the function $\psi(t):=\|P_{t}\delta_{x}\|^{2}=p_{2t}(x,x)$ is nonincreasing, $\psi\le1$ and $\psi'=-2\mathcal{E}(P_{t}\delta_{x})$. With $r(t)$ given by $r^{-3}=\psi/2$, the condition $2r^{3}\le V_{*}^{f}$ becomes $\psi\ge4/V_{*}^{f}$, which holds on an interval $[0,t_{*}]$. On this interval Lemma~\ref{lem:F.3} implies $-\psi'\ge\psi^{5/3}/(2^{2/3}A_{N})=c_{N}\psi^{5/3}$, hence $\psi(t)\le(\frac23c_{N}t)^{-3/2}$, while $\psi<4/V_{*}^{f}$ after $t_{*}$. Hence $p_{s}(x,x)\le\max(C_{H}s^{-3/2},4/V_{*}^{f})$, $p_{s}(x,y)\le(p_{s}(x,x)p_{s}(y,y))^{1/2}$, and $4/V_{*}^{f}=\frac{64}{27}\RHK^{-3}$ (\S\ref{S:E1.scales}). The zero modes lie below the plateau, since $p_{t}(x,x)\to1/|\KK_{x}|<1/(2V_{*}^{f})$, the last inequality by Proposition~\ref{prop:6.10}(ii).
\qed
\end{proof}

\subsection{Barlow's argument: proof of Theorem~\ref{thm:6.9}}
\label{ssec:F.3}

From now on we assume $\bGM\ge\bGMo$, so that $\mt\le K_{0}(2C_{H})^{-1/3}$, $\RHK\ge K_{0}/\mt\ge(2C_{H})^{1/3}$, and $p_{t}\le C_{H}t^{-3/2}$ on $[1,\RHK^{2}]$ by (HK1). We obtain (HK1$'$) as in \cite[Lemma~3.3(b), Prop.~3.4]{Ba04}, (HK2) as in \cite[Lemmas~3.5, 3.6, Prop.~3.7]{Ba04}, and (HK3) as the fixed-time form of \cite[Thm~3.8]{Ba04}, with the following adjustments, each forced by $\Gp$:
\par(a) \cite[Prop.~3.1, Cor.~3.2]{Ba04}, the on-diagonal decay in a very good ball for $t\le cR^{2}/\log R$, are replaced by (HK1), which holds at all centres. Every ball is a very good ball with $N_{B}=1$ \cite[Def.~1.7]{Ba04}, there is no localization and no $\log R$ loss, and the window is $t\le\eps_{1}^{\mathrm{HK}}\RHK^{2}$. The exit-time bound $\frac12+4C_{\mathrm{mo}}^{2}t/r^{2}$ of \cite[Lemma~3.5]{Ba04} holds at all centres for $t\le\RHK^{2}/2$, and exceeds $1$ for larger $t$ when $r\le2C_{\mathrm{mo}}\RHK$.
\par(b) The walk is the variable-speed walk with counting measure on a finite, possibly disconnected graph, and \cite[(3.8) to (3.10)]{Ba04} hold without change, with $a_{xy}=1$, $\mu\equiv1$, as finite sums.
\par(c) The volume bound \cite[(1.1)]{Ba04} enters only \cite[Lemma~3.3(b)]{Ba04}, through $\sum_{y}e^{-a\,d(x,y)}\le c_{4}a^{-3}$ for $0<a\le2$, which follows from $d_{\Gp}\ge\|\cdot\|_{1}$: $\sum_{y}e^{-a\,d(x,y)}\le\sum_{z\in\Zthree}e^{-a\|z\|_{1}}=\coth(a/2)^{3}\le(1+2/a)^{3}\le64a^{-3}$.
\par(d) \cite[Lemma~1.1]{Ba04} is replaced by (HK0), (HK1) and the Poisson domination of $N_{t}$. Let $M(t):=\mathbb{E}_{x}d(x,X_{t})$. By (HK1), $M(t)\ge\mathbb{P}_{x}(X_{t}\ne x)\ge1-C_{H}t^{-3/2}\ge\frac12$ for $t\ge(2C_{H})^{2/3}$, and the bound by $N_{t}$ gives $M(t)\le6t$ at earlier times. For $\rho>A_{1}t/4$ ($A_{1}$ of \S\ref{S:E1.HK}), the event $\tau(x,\rho)<t$ forces $N_{t}>\rho$.
\par(e) (HK4) is not in \cite{Ba04}, but it follows from (HK2). With $D:=d(w,X_{s})$, we split the sum in $\mathbb{E}e^{\theta D}-1\le\theta\sum_{k\ge1}e^{\theta k}\mathbb{P}(D\ge k)$ at $k=s$ into a Gaussian part, summed with $\theta\le s^{-1/2}$, and a Poisson part, summed with $\theta\le c_{\mathrm{ex}}/2$. For $s<1$, $D\le N_{s}$.
\par For (HK3), Chapman\textendash Kolmogorov gives $p_{t}(x,y)\le2\sup_{z,w}p_{t/2}(z,w)\sup_{w}\mathbb{P}_{w}(d(w,X_{t/2})\ge d/2)$, and (HK1), (HK2) bound the two factors. The constants $C_{\mathrm{mo}},c_{\mathrm{ex}},A_{1},C_{\mathrm{gs}},c_{\mathrm{gs}},C_{M}$ of \S\ref{S:E1.HK} are functions of $C_{H}$.

\subsection{Resolvent decay and covariance bounds}
\label{ssec:F.4}\label{ssec:F.5}

For Proposition~\ref{prop:6.11}, (HK4) gives $\mathbb{E}_{w}e^{\theta_{M}d(w,X_{s})}\le\frac32$ for $s\le T_{1}:=\min(M^{-2},\eps_{1}^{\mathrm{HK}}\RHK^{2})$. Using blocks of length $T_{1}$, we obtain $\mathbb{E}_{x}e^{\theta_{M}d(x,X_{t})}\le\frac32e^{M^{2}t/2}$, and the factor $e^{-M^{2}t}$ controls this growth in $\sum_{y}v^{(M)}_{\Gp}(x,y)e^{\theta_{M}d(x,y)}=\int_{0}^{\infty}e^{-M^{2}t}\mathbb{E}_{x}e^{\theta_{M}d(x,X_{t})}dt$. For Theorem~\ref{thm:6.12} we write $C_{0}=\int_{0}^{\infty}k(t)P_{t}\,dt$ with $0\le k(t)\le\min(\MY^{2}t,\varpi)e^{-r_{-}t}$ (\S\ref{S:E1.cov}). With this representation, (H1), (H1$'$), (H2) follow from (HK0), (HK1), (HK3), \eqref{eq:6.LW} and the blocking bound by splitting the time integral, and (H$_{C}$) follows from the Schur test and a Neumann series in $\|\cdot\|_{\rcov}$.

\subsection{Bounds on the lattice Yukawa kernel}
\label{ssec:F.6}

\begin{lemma}[Lattice Yukawa kernel]\label{lem:F.7}
Let $0<M\le\frac15$ and $v^{(M)}:=(-\Delta_{\Zthree}+M^{2})^{-1}$. There is an absolute constant $C_{L}$ such that for all $x\in\Zthree$ and all unit vectors $e$,
\begin{equation}\label{eq:F.L1L2}
v^{(M)}(x)\le C_{L}(1+|x|)^{-1}e^{-3M|x|/4},\qquad|v^{(M)}(x+e)-v^{(M)}(x)|\le C_{L}(1+|x|)^{-2}e^{-3M|x|/4}.
\end{equation}
If $\Lambda'=\prod_{\mu}[1,n_{\mu}]$ is a box with $\min_{\mu}n_{\mu}\ge M^{-1}$ and $u_{x}:=v^{(M)}_{\Lambda'}(\cdot,x)$ is the kernel of the box with all its bonds, then for $x,y,y+e\in\Lambda'$, with $C_{L}'$ absolute,
\begin{gather*}
\mathrm{(L1)}\ \ u_{x}(y)\le C_{L}'(1+|y-x|)^{-1}e^{-M|y-x|/2},\\
\mathrm{(L2)}\ \ |u_{x}(y+e)-u_{x}(y)|\le C_{L}'(1+|y-x|)^{-2}e^{-M|y-x|/2}.
\end{gather*}
\end{lemma}

\noindent\emph{Sketch.} Exponential tilts of the three one-dimensional heat kernels and the unimodality of the Poisson law give \eqref{eq:F.L1L2} after the time integral, the rate $3M/4$ coming from $M^{2}t+\frac{9}{64}|x|^{2}/t\ge\frac34M|x|$. The box kernel is the image sum $v^{(M)}_{\Lambda'}(y,x)=\sum_{g\in\mathcal{G}}v^{(M)}(y-gx)$ over the group $\mathcal{G}$ generated by the reflections in the face planes. Since $\min_{\mu}n_{\mu}\ge M^{-1}$, the images cost a factor $(1+M|y-x|)^{3}$, which the loss of rate from $3M/4$ to $M/2$ absorbs.
}
\providecommand{\ZZ}{\mathbb{Z}}
\providecommand{\RR}{\mathbb{R}}
\providecommand{\one}{\mathbf{1}}
\providecommand{\dist}{\operatorname{dist}}
\providecommand{\ex}[1]{\langle #1\rangle}
\providecommand{\abs}[1]{\lvert #1\rvert}
\providecommand{\vt}{\tilde v}
\providecommand{\omb}{\omega_{\beta}}
\providecommand{\Kst}{K_{*}}

\section{Comparing Wilson and Villain loops by correlation inequalities}
\label{app:G}

Numbers \ref{app:G}.$n$ refer to the statements and subsections of this appendix, while equation numbers (E.$n$), and statement numbers cited with \cite{GM82}, refer to Appendix~E of \cite{GM82}.

Results for the Wilson action are often obtained by comparison with the Villain action, or by redoing the Villain analysis with additional estimates \cite[Section~6 and Appendix~B]{FS81KT}, \cite{FS82}. For upper bounds on Wilson loops, which is the direction needed for confinement, Theorem~\ref{thm:2.5} bounds the Wilson loop by an average of Villain loops in which a sparse random set of plaquettes is frozen. Freezing turns the problem into a Villain model on a domain with insulating walls, and this is where most of the length of this paper lies. It is natural to ask whether a better comparison avoids frozen plaquettes.

Two facts are known. First, the character coefficients $I_k(\beta)/I_0(\beta)$ of the Wilson weight, with $I_k(\beta)\sim(\beta/2)^k/k!$ as $k\to\infty$, decay more slowly than those of any Villain weight, $e^{-k^2/(2t)}$, so no Villain plaquette weight dominates the Wilson weight coefficient by coefficient (Section~\ref{ssec:1.3}). Second, Ginibre's inequality \cite{Gi70} transfers to Villain weights through complexes that realize them as limits: cable graphs for spin systems, credited to Bellissard in \cite[Section~3]{FS81} and developed in \cite[Appendix~A]{AHPS} and \cite{DG25,DvEG}, and carpet graphs for lattice gauge theory \cite{CG25}. Such arguments give inequalities for all characters at once.

To our knowledge, the following box version of the first fact is new. On every finite box, an inequality between Wilson and Villain loops that holds for all charges, even with a perimeter factor, forces the Villain side to contain arbitrarily large couplings (Corollary~\ref{cor:G.4} and Proposition~\ref{prop:G.6}). The proof pairs a lower bound from Ginibre's inequality with the spin-wave upper bound for Villain loops, which we extend to Villain mixtures (Theorem~\ref{thm:G.2} and Remark~\ref{rem:G.5}). We complement it with three related results. Wilson loops are not monotone in the character coefficients of the plaquette weights (Theorem~\ref{thm:G.8}, proved by an exact computation). If a character coefficient of a weight exceeds that of the Villain weight $\vt_t$, every representation of the weight as a mixture of Villain weights carries an explicit minimal mass of couplings above $t$ (Proposition~\ref{prop:G.10}). Freezing plaquettes has increasing returns: the effect of freezing two plaquettes can exceed the product of the effects of freezing each of them (Proposition~\ref{prop:G.12}).

Section numbers of \cite{FS81KT} and \cite{FS81} refer to their preprints IHES/P/81/25 and IHES/P/80/10.

\subsection{Setting}\label{ssec:G.1}

\paragraph{Complexes and loops.} We use the notation of Section~\ref{sec:2} for a finite $2$-dimensional cell complex $X$: $E(X)$, $P(X)$, $[p:e]$, $\theta_p$, $\partial$, $Z_2(X)$ and $\theta_C$. For $n\in\ZZ^{P(X)}$ we have $\theta_{\partial n}=\sum_pn_p\theta_p$, and we set $\abs{\partial n}=\sum_e\abs{(\partial n)_e}$. We let $L(X)=\ker(\partial\otimes\RR)\subseteq\RR^{P(X)}$, with the standard inner product of $\RR^{P(X)}$. Since $\partial$ is an integer matrix, $Z_2(X)$ is a lattice of full rank in $L(X)$. We write $\delta_p\in\ZZ^{P(X)}$ for the indicator of $p$. In a box of $\Zthree$ every integer $1$-cycle is a boundary $\partial n$.

\paragraph{Weights.} A \emph{plaquette weight} is an even continuous function $w\colon\mathbb{T}\to[0,\infty)$ whose Fourier coefficients $c_k(w)=\int_{\mathbb{T}}w(\varphi)e^{-ik\varphi}\,d\varphi/2\pi$ are nonnegative and summable, with $c_0(w)=1$. The Wilson weight is $w^{W}_{\beta}(\varphi)=e^{\beta\cos\varphi}/I_0(\beta)$, with $c_k(w^{W}_{\beta})=I_k(\beta)/I_0(\beta)$. The Villain weight $\vt_t$ of Section~\ref{sec:2} has $c_k(\vt_t)=e^{-k^2/(2t)}$ for $t>0$, and $\vt_0=1$ is a removed plaquette. A weight is of \emph{Ginibre type} if it is proportional to $e^{f}$, where $f(\varphi)=\sum_{j=1}^J\lambda_j\cos j\varphi$ with all $\lambda_j\ge0$. The weight $w^{W}_{\beta}$ is of Ginibre type. A \emph{Villain mixture with couplings at most $\tau$} is $w=\int_{[0,\tau]}\vt_t\,\nu(dt)$ for a probability measure $\nu$ on $[0,\tau]$. Its coefficients satisfy $c_k(w)\le e^{-k^2/(2\tau)}$ for $k\neq0$.

\paragraph{Gibbs measures.} Given weights $w=(w_p)_{p\in P(X)}$, let $\mu_{X,w}$ be the probability measure on $\mathbb{T}^{E(X)}$ proportional to $\prod_pw_p(\theta_p)\prod_ed\theta_e/2\pi$, with expectation $\ex{\cdot}_{X,w}$. If $w_p=w^{W}_{\beta}$ for all $p$, this is the Wilson expectation $\ex{\cdot}^{W}_{X,\beta}$ of Section~\ref{sec:2}. For $n\in\ZZ^{P(X)}$ the Wilson loop of $\partial n$ is $\cos\theta_{\partial n}$. We use one monotonicity property: if every $w_p$ is of Ginibre type, with coefficients $\lambda_{p,j}$, then $\ex{\cos\theta_{\partial n}}_{X,w}$ is nondecreasing in every $\lambda_{p,j}$, by Ginibre's inequality \eqref{eq:2.ginibre}.

\emph{Villain monotonicity} states that $\ex{\cos\theta_{\partial n}}_{X,w}$ is nondecreasing in every $t_p$ if $w_p=\vt_{t_p}$. It is proved on the lattice in \cite[Lemma~C.1]{Gu80}, on the cubes $\{x\in\ZZ^d:\abs{x}_\infty\le L\}$ in \cite[Corollary~1.6]{CG25}, and on finite two-dimensional complexes in Lemma~\ref{lem:2.3}(b). We do not use it in this appendix.

\begin{lemma}\label{lem:G.1}
For weights $w$ and $n\in\ZZ^{P(X)}$,
\[
\ex{\cos\theta_{\partial n}}_{X,w}=\frac{\hat Z(n)}{\hat Z(0)},\qquad \hat Z(n)=\sum_{m\in Z_2(X)}\prod_{p\in P(X)}c_{\abs{n_p+m_p}}(w_p).
\]
\end{lemma}

\begin{proof}
Each series $w_p(\theta_p)=\sum_kc_k(w_p)e^{ik\theta_p}$ converges absolutely and uniformly. We expand the product of the weights times $e^{i\theta_{\partial n}}$ and integrate term by term. The term indexed by $l\in\ZZ^{P(X)}$ contributes $\prod_pc_{\abs{l_p}}(w_p)$ if $\partial(l+n)=0$, and nothing otherwise. Writing $l=m-n$ with $m\in Z_2(X)$ and replacing $m$ by $-m$ gives $\hat Z(n)$. The substitution $\theta\mapsto-\theta$ shows that $\ex{e^{i\theta_{\partial n}}}_{X,w}=\ex{\cos\theta_{\partial n}}_{X,w}$. The term $m=0$ gives $\hat Z(0)\ge1$.
\qed
\end{proof}

For Villain weights on a box, Lemma~\ref{lem:G.1} is Lemma~\ref{lem:2.7}.

\subsection{All-charge comparisons}\label{ssec:G.2}

\begin{theorem}\label{thm:G.2}
Let $n_0\in\ZZ^{P(X)}$ with $\partial n_0\neq0$, and put $\kappa=\dist(n_0,L(X))^2$. Then $\kappa>0$, and the following hold.
\begin{itemize}
\item[(a)] If every $w_p$ is a Villain mixture with couplings at most $\tau$, then $\ex{\cos k\theta_{\partial n_0}}_{X,w}\le e^{-\kappa k^2/(2\tau)}$ for every $k\in\ZZ$.
\item[(b)] If every $w_p$ is of Ginibre type and $\partial\delta_{p_0}\neq0$, then $\ex{\cos k\theta_{p_0}}_{X,w}\ge c_k(w_{p_0})$ for every $k\ge0$. In particular $\ex{\cos k\theta_{p_0}}_{X,w}\ge I_k(\beta)/I_0(\beta)$ if $w_{p_0}=w^{W}_{\beta}$.
\end{itemize}
\end{theorem}

The proof of (a) uses the following elementary fact.

\begin{lemma}\label{lem:G.3}
Let $\Gamma$ be a lattice of full rank in a finite-dimensional real inner product space $L$ with norm $\|\cdot\|$, and let $x\in L$. Then $\sum_{m\in\Gamma}e^{-\|m+x\|^2/2}\le\sum_{m\in\Gamma}e^{-\|m\|^2/2}$.
\end{lemma}

\begin{proof}
By Poisson summation the left side equals $c\sum_{\xi\in\Gamma^*}e^{-2\pi^2\|\xi\|^2}e^{2\pi i\langle\xi,x\rangle}$, where $\Gamma^*$ is the dual lattice and $c>0$ depends only on $\Gamma$. All coefficients are positive, so the sum is largest at $x=0$.
\qed
\end{proof}

\begin{proof}[Proof of Theorem~\ref{thm:G.2}]
Since $\partial n_0\neq0$, the vector $n_0$ does not lie in $L(X)$, so $\kappa>0$.

(a) Write $w_p=\int\vt_t\,\nu_p(dt)$ and $\nu=\bigotimes_p\nu_p$. We have $c_j(\vt_t)=e^{-j^2/(2t)}$, read as $\one_{j=0}$ when $t=0$, and $c_j(w_p)=\int c_j(\vt_t)\,\nu_p(dt)$ by Fubini's theorem. By Lemma~\ref{lem:G.1} and Tonelli's theorem, $\hat Z(n)=\int\hat Z_t(n)\,\nu(dt)$ for $n\in\ZZ^{P(X)}$, where
\[
\hat Z_t(n)=\sum_{m\in Z_2(X)}\prod_{p\in P(X)}e^{-(n_p+m_p)^2/(2t_p)}.
\]
It suffices to show $\hat Z_t(kn_0)\le e^{-\kappa k^2/(2\tau)}\hat Z_t(0)$ for every $t\in[0,\tau]^{P(X)}$. Let first $t\in(0,\tau]^{P(X)}$. We equip $\RR^{P(X)}$ with the inner product $\langle x,y\rangle_t=\sum_px_py_p/t_p$ and its norm $\|\cdot\|_t$, so that $\hat Z_t(n)=\sum_{m\in Z_2(X)}e^{-\|n+m\|_t^2/2}$. Let $\Pi_t$ be the orthogonal projection onto $L(X)$ for this inner product, and $\kappa_t=\|n_0-\Pi_tn_0\|_t^2$. For $m\in Z_2(X)$ we have $\|kn_0+m\|_t^2=k^2\kappa_t+\|k\Pi_tn_0+m\|_t^2$. Lemma~\ref{lem:G.3} with $\Gamma=Z_2(X)$ and $x=k\Pi_tn_0$ gives $\hat Z_t(kn_0)\le e^{-k^2\kappa_t/2}\hat Z_t(0)$. Since $t_p\le\tau$, we have $\|x\|_t^2\ge\abs{x}^2/\tau$ for all $x$, and hence $\kappa_t\ge\kappa/\tau$. If some coordinates of $t$ vanish, we replace them by $s\in(0,\tau]$. As $s\downarrow0$, every term of the resulting sums decreases to the corresponding term for $t$, and the inequality passes to the limit by dominated convergence.

(b) By Ginibre's inequality, $\ex{\cos k\theta_{p_0}}_{X,w}$ decreases when all coefficients $\lambda_{p,j}$ with $p\neq p_0$ are lowered to $0$. Then $\mu_{X,w}$ is proportional to $w_{p_0}(\theta_{p_0})\prod_ed\theta_e/2\pi$. Since $\partial\delta_{p_0}$ is a nonzero integer vector, $\theta\mapsto\theta_{p_0}$ is a continuous homomorphism of $\mathbb{T}^{E(X)}$ onto $\mathbb{T}$, so it pushes Haar measure to Haar measure. Hence the expectation equals $\int\cos(k\varphi)w_{p_0}(\varphi)\,d\varphi/2\pi=c_k(w_{p_0})$.
\qed
\end{proof}

\begin{corollary}\label{cor:G.4}
Let $p_0\in P(X)$ with $\partial\delta_{p_0}\neq0$, and let $\beta>0$. Let $v=(v_p)_{p\in P(X)}$ be Villain mixtures with couplings at most $\tau$. For all constants $a,b\ge0$ there is $k\ge1$ with
\[
\ex{\cos k\theta_{p_0}}^{W}_{X,\beta}>ae^{bk\abs{\partial\delta_{p_0}}}\,\ex{\cos k\theta_{p_0}}_{X,v}.
\]
In particular no inequality $\ex{\cos\theta_{\partial n}}^{W}_{X,\beta}\le ae^{b\abs{\partial n}}\ex{\cos\theta_{\partial n}}_{X,v}$ holds for all $n\in\ZZ^{P(X)}$.
\end{corollary}

\begin{proof}
Let $\kappa_{p_0}=\dist(\delta_{p_0},L(X))^2>0$. By Theorem~\ref{thm:G.2} and $I_k(\beta)\ge(\beta/2)^k/k!\ge(\beta/2k)^k$, the left side is at least $(\beta/2k)^k/I_0(\beta)$ and the right side is at most $ae^{bk\abs{\partial\delta_{p_0}}-\kappa_{p_0}k^2/(2\tau)}$. The ratio of the two bounds tends to infinity as $k\to\infty$.
\qed
\end{proof}

\begin{remark}\label{rem:G.5}
(i) For $w_p=\vt_\tau$ on every plaquette, part (a) is the classical spin-wave upper bound: the Villain loop is at most the value $e^{-\kappa k^2/(2\tau)}$ of the same observable for the Gaussian spin wave. It follows from the decoupling of the Villain model into a spin wave and a Coulomb gas \cite[Proposition~5.2]{GS23}, and in three dimensions it gives the bound of Glimm and Jaffe \cite{GJ77}, see \cite[Corollary~7.3]{GS23}. In the proof above, Lemma~\ref{lem:G.3} bounds the Coulomb-gas factor by $1$. Part (a) adds uniformity over Villain mixtures with couplings at most $\tau$. (ii) For a box $\Lam$ of $\Zthree$, $L(\Lam)$ is spanned by the boundaries of the cubes of $\Lam$. These are the stars of the dual graph of Section~\ref{ssec:2.4}, whose sites are the cubes and the exterior ground site, as a network with unit conductances. The matrix of the orthogonal projection onto their span is the transfer-current matrix $Y$ \cite[Section~2.4]{LP16}. Hence $\dist(\delta_p,L(\Lam))^2=1-Y(p,p)$, and $Y(p,p)$ is the effective resistance across $p$ \cite[Section~4.2]{LP16}. For a single cube, $\kappa=5/6$.
\end{remark}

Comparisons are often made through auxiliary complexes. Let $G$ be a finite two-dimensional complex with plaquette weights $w$, and let $\gamma\in\ZZ^{P(G)}$ with $\partial\gamma\neq0$. The \emph{weight realized by $(G,\gamma,w)$} is the density $w_G$ of the law of $\theta_{\partial\gamma}$ under $\mu_{G,w}$ with respect to $d\varphi/2\pi$. Its coefficients are $c_k(w_G)=\ex{\cos k\theta_{\partial\gamma}}_{G,w}$. If a copy of $G$ is glued into a plaquette $p$ of $X$ along $\partial\gamma=\partial\delta_p$ in place of $w_p$, then integrating out the new edges replaces $w_p$ by $w_G$. Cable graphs and carpet graphs are of this kind.

\begin{proposition}\label{prop:G.6}
In this setting the following hold.
\begin{itemize}
\item[(a)] If every plaquette $p$ of $G$ carries a weight of Ginibre type, with coefficients $\lambda_{p,j}$, then each $c_k(w_G)$ is nondecreasing in every $\lambda_{p,j}$.
\item[(b)] If every plaquette of $G$ carries a Villain mixture with couplings at most $\tau$, then $c_k(w_G)\le e^{-\kappa_Gk^2/(2\tau)}$ with $\kappa_G=\dist(\gamma,L(G))^2>0$.
\end{itemize}
In particular, let $\beta>0$. If $(G,\gamma,w)$ realizes $w^{W}_{\beta}$ and every weight of $G$ is of Ginibre type, then raising any of the coefficients $\lambda_{p,j}$ produces a weight $w'$ with $c_k(w')\ge I_k(\beta)/I_0(\beta)$ for all $k$, and no complex whose plaquettes carry Villain mixtures with bounded couplings realizes such a weight.
\end{proposition}

\begin{proof}
Part (a) is Ginibre's inequality applied to $\cos k\theta_{\partial\gamma}$ on $G$, and part (b) is Theorem~\ref{thm:G.2}(a) on $G$ with $n_0=\gamma$. The last assertion follows, since $(\beta/2)^k/(k!\,I_0(\beta))$ is not bounded by $e^{-\kappa_Gk^2/(2\tau)}$ for large $k$.
\qed
\end{proof}

\begin{remark}\label{rem:G.7}
Ginibre's inequality, Villain monotonicity, complexes as in Proposition~\ref{prop:G.6}, and the annealed Hartman\textendash Watson representation with monotone rounding of the couplings (Theorem~\ref{thm:2.5}) all give inequalities for all characters at once. By Corollary~\ref{cor:G.4} none of them bounds Wilson loops by Villain loops with bounded couplings, even on one box. If the Villain side of such a comparison is a Villain mixture, its mixing measure gives positive mass to arbitrarily large couplings, and we bound the mass of large couplings from below in Section~\ref{ssec:G.4}. In the comparison of Theorem~\ref{thm:2.5}, which holds for every $\eps>0$, the plaquettes that carry these large couplings, those above $1/\eps$, form a random set of frozen plaquettes: their couplings are raised to infinity.
\end{remark}

\subsection{Monotonicity in character coefficients}\label{ssec:G.3}

Ginibre's inequality gives monotonicity in the Fourier coefficients of the logarithm of a weight, and Villain monotonicity raises all coefficients $c_k(\vt_t)$ at once. Wilson loops are not monotone in each coefficient $c_k(w_p)$ of the weights themselves.

Let $X$ be the boundary of the unit cube, with its $8$ vertices, $12$ edges and $6$ faces oriented by the outward normal, and fix a face $p_0$. For a weight $w$ and $x\ge0$ let $w_x=w+2(x-c_1(w))\cos\varphi$, so that $c_1(w_x)=x$ and $c_k(w_x)=c_k(w)$ for $k\neq\pm1$. Put $w_x$ on $p_0$ and $w$ on the other five faces, and let $\mathcal{W}(x)=\hat Z(\delta_{p_0})/\hat Z(0)$ as in Lemma~\ref{lem:G.1}. When $w_x\ge0$, this is $\ex{\cos\theta_{p_0}}$.

\begin{theorem}\label{thm:G.8}
Let $w$ be either the Wilson weight $w^{W}_{8}$ or the Villain weight $\vt_t$ with $e^{-1/(2t)}=15/16$. Then $\mathcal{W}$ is strictly decreasing on $[0,\infty)$. In particular, for $0<s\le s_w=\tfrac12\min_{\mathbb{T}}w$ the function $w+2s\cos\varphi$ is a plaquette weight with $c_1$ larger than that of $w$ and all other coefficients equal, and its Wilson loop is strictly smaller. Here $s_w=e^{-8}/(2I_0(8))$ for $w=w^{W}_{8}$, and $s_w>0$ for $w=\vt_t$. Moreover, if every face carries $w+2s\cos\varphi$, then $\frac{d}{ds}\ex{\cos\theta_{p_0}}$ is negative at $s=0$.
\end{theorem}

\begin{proof}
The complex $X$ is a sphere, so $Z_2(X)=\ZZ\one_{P(X)}$, where $\one_{P(X)}=\sum_p\delta_p$. Write $c_k=c_k(w)$. Lemma~\ref{lem:G.1} with $n=\delta_{p_0}$ and $m=-h\one_{P(X)}$ gives $\hat Z(\delta_{p_0})=\sum_{h\in\ZZ}c'_{\abs{h-1}}c_{\abs{h}}^5$ and $\hat Z(0)=\sum_hc'_{\abs{h}}c_{\abs{h}}^5$, where $c'$ denotes the coefficients of the weight on $p_0$. Collecting the terms that contain $c'_1=x$ we get
\[
\mathcal{W}(x)=\frac{A_C+xB_C}{A_0+xB_0},\qquad
\begin{aligned}
A_0&=1+2\textstyle\sum_{h\ge2}c_h^6, & B_0&=2c_1^5,\\
A_C&=c_1^5+\textstyle\sum_{h\ge3}c_{h-1}c_h^5+\sum_{h\ge1}c_{h+1}c_h^5, & B_C&=1+c_2^5.
\end{aligned}
\]
Hence $\mathcal{W}'(x)=S/(A_0+xB_0)^2$ with $S=A_0B_C-A_CB_0$, a number independent of $x$. It remains to show $S<0$. If every face carries $w+2s\cos\varphi$, the derivative at $s=0$ has the sign of $U=Z_C'Z_0-Z_CZ_0'$, where $Z_C=A_C+c_1B_C$, $Z_0=A_0+c_1B_0$, $Z_C'=B_C+5c_1^4(1+c_2)$ and $Z_0'=12c_1^5$.

We evaluate $S$ and $U$ in exact rational arithmetic with intervals. Both are homogeneous of degree $11$ in $(c_k)$, so for $w^{W}_{8}$ we may replace $c_k$ by $a_k=I_k(8)=\sum_{m\ge0}4^{2m+k}/(m!\,(m+k)!)$. The partial sums with $m<80$ are rational lower bounds. Consecutive terms have ratio at most $16/(81(81+k))$ beyond $m=80$, so we bound the remainder by a geometric series. Comparing the series term by term gives $I_{k+1}(8)\le 4I_k(8)/(k+1)$, and with it we bound the parts of the sums over $h$ with $\abs{h}>40$ by geometric series. For $\vt_t$ the coefficients $c_k=(15/16)^{k^2}$ are rational, and we bound the parts with $\abs{h}>40$ again by geometric series. All operations are exact, with interval endpoints rounded outward. This computation, carried out by a program of the scripts archive of the paper (see the Data availability statement), gives $S<0$ and $U<0$ in both cases, and the Lean kernel decides the same inequalities in the formalization \cite{TomLean}.

Finally, $\min w^{W}_{8}=e^{-8}/I_0(8)$ is attained at $\varphi=\pi$, and $\vt_t>0$ is continuous. For $0<s\le s_w$ the function $w+2s\cos\varphi$ is nonnegative, even and continuous, with nonnegative summable coefficients.
\qed
\end{proof}

The enclosures are $S=-2.2085\ldots\cdot10^{28}$ and $U=-2.5854\ldots\cdot10^{28}$ for $w^{W}_{8}$, with $c_k$ replaced by $I_k(8)$, and $S=-0.2776\ldots$ and $U=-0.3255\ldots$ for $\vt_t$.
At $x=c_1$ the program also encloses the loop and its derivative: $\ex{\cos\theta_{p_0}}=0.9464\ldots$ and
$\mathcal{W}'(c_1)=-0.0321\ldots$ for $w^{W}_{8}$, and $\ex{\cos\theta_{p_0}}=0.9476\ldots$ and $\mathcal{W}'(c_1)=-0.0342\ldots$
for $\vt_t$.

\begin{remark}\label{rem:G.9}
(i) The weight $w^{W}_{8}$ is of Ginibre type, so by Ginibre's inequality its Wilson loops increase with every coefficient of $\log w^{W}_{8}$. Raising $c_1$ alone changes every Fourier coefficient of the logarithm, not all in the same direction, and Theorem~\ref{thm:G.8} shows that this can lower the loop. (ii) Villain monotonicity, Lemma~\ref{lem:2.3}(b), is therefore not an instance of a general monotonicity in character coefficients, in line with the observation of \cite[Remark~1.7]{CG25} that $\log\vt_t$ is not positive definite. (iii) The phenomenon comes from light tails and large $\beta$ or $t$. The Poisson kernel $P_r(\varphi)=\sum_kr^{\abs{k}}e^{ik\varphi}$, $0<r<1$, has $\log P_r$ with nonnegative Fourier coefficients $r^{\abs{k}}/\abs{k}$ for $k\neq0$, so Ginibre's inequality applies to it by approximation. For $P_r$ on all faces, summing geometric series in the proof of Theorem~\ref{thm:G.8} gives $S=1-r^{10}>0$. Heuristically, for $w^{W}_{\beta}$ or $\vt_t$ the dual variable $h$ of the proof is close to a centered Gaussian of variance about $\beta/6$ or $t/6$. On $p_0$ the loop replaces $h$ by $h-1$, whose law is close to a Gaussian with the same variance and mean $-5/6$. When the variance exceeds $1$, shifting a Gaussian lowers the probability of $\{\abs{h}=1\}$ to second order.
\end{remark}

\subsection{Villain-mixture representations}\label{ssec:G.4}

A \emph{Villain-mixture representation} of a plaquette weight $w$ is a probability measure $\omega$ on $(0,\infty)$ with $c_k(w)=\int e^{-k^2u/2}\,\omega(du)$ for all $k\in\ZZ$. The variable $u$ is the inverse coupling, so that $w=\int\vt_{1/u}\,\omega(du)$. For $w^{W}_{\beta}$ the Hartman\textendash Watson law $\omb$ of Lemma~\ref{lem:2.2} is such a representation \cite[Theorem~1.3]{HW74}. Yor \cite{Yo80} described its density, and Patie \cite[Remark~2.7]{Pat09} relates it to the infinitely divisible laws of Hartman \cite{Har76}. It is used in comparisons in \cite[(9.10)]{AHPS} and in Section~\ref{sec:2}. By Lemma~\ref{lem:2.2} it is the only probability measure that reproduces $I_\nu(\beta)/I_0(\beta)$ for every real $\nu\ge0$.

\begin{proposition}\label{prop:G.10}
Let $\omega$ be a Villain-mixture representation of a weight $w$ with coefficients $c_k$, let $\eps>0$, and let $1\le k_1<k_2$ be integers with $c_{k_1}>e^{-k_1^2\eps/2}$. Then
\[
\omega((0,\eps))\ \ge\ \bigl(c_{k_1}-e^{-k_1^2\eps/2}\bigr)^{k_2^2/(k_2^2-k_1^2)}\,c_{k_2}^{-k_1^2/(k_2^2-k_1^2)}.
\]
Moreover $\omega((0,\eps))\le e^{k^2\eps/2}c_k$ for every $k$.
\end{proposition}

In terms of couplings, $\omega((0,\eps))$ is the mass of couplings above $1/\eps$. For $\omega=\omb$ the last bound is the tail bound of Lemma~\ref{lem:2.2}.

\begin{proof}
Let $F(s)=\int_{(0,\eps)}e^{-su}\,\omega(du)$ for $s\ge0$. By H\"older's inequality $\log F$ is convex. Put $s_i=k_i^2/2$. Then $F(s_2)\le c_{k_2}$, and $F(s_1)\ge c_{k_1}-\int_{[\eps,\infty)}e^{-s_1u}\omega(du)\ge c_{k_1}-e^{-s_1\eps}>0$. Convexity on $[0,s_2]$ gives $\log F(s_1)\le(1-s_1/s_2)\log F(0)+(s_1/s_2)\log F(s_2)$, that is
\[
\log F(0)\ge\frac{s_2\log F(s_1)-s_1\log F(s_2)}{s_2-s_1},
\]
and the bounds on $F(s_1)$ and $F(s_2)$ give the claim. The upper bound is the Chernoff bound $\one_{u<\eps}\le e^{k^2(\eps-u)/2}$.
\qed
\end{proof}

\begin{remark}\label{rem:G.11}
(i) Only integer frequencies constrain $\omega$, and the representation is then not unique in general. Suppose some representation has a density bounded below by a positive constant on an interval $[u_1,u_2]\subset(0,\infty)$. In the variable $x=e^{-u/2}$ the constraints read $\int x^{k^2}\,\tilde\omega(dx)=c_k$. Let $[a,b]$ be a closed interval inside $(e^{-u_2/2},e^{-u_1/2})$. Since $\sum_{k\ge1}k^{-2}<\infty$, the functions $x^{k^2}$, $k\ge0$, do not span a dense subspace of $C[a,b]$ \cite{CE43}, \cite[Theorem~29]{Alm07}. Hence some signed measure $\zeta\neq0$ on $[a,b]$ annihilates all of them. The image of $\zeta$ under a dilation $x\mapsto sx$ annihilates them too, since $\int(sx)^{k^2}\zeta(dx)=s^{k^2}\int x^{k^2}\zeta(dx)=0$. We average these images against a continuous probability density in $s$ supported near $1$. The average is $h(x)\,dx$ with $h$ continuous and supported in $[e^{-u_2/2},e^{-u_1/2}]$, and it annihilates all $x^{k^2}$. As the support of the density shrinks to $\{1\}$, the averages converge weakly to $\zeta$, so $h\neq0$ for a small enough support. Adding a small multiple of $h\,dx$ gives another representation. Proposition~\ref{prop:G.10} applies to all of them. (ii) Numerical. For $w^{W}_{\beta}$ with $\eps=1/(K\beta)$ and $\beta\to\infty$, we choose $k_1$ just above the crossing of $I_k(\beta)/I_0(\beta)$ and $e^{-k^2\eps/2}$ and $k_2=k_1+1$. At leading order Proposition~\ref{prop:G.10} then gives mass at least $e^{-63.02\tau}$ above the coupling $\tau=K\beta$ when $K=\Kst$, while the upper bound gives $e^{-29.93\tau}$, the order of the density $\delta$ of frozen plaquettes in Lemma~\ref{lem:2.6}. Hence, at leading order, the analogue of the condition \eqref{eq:2.threshold} on the density of frozen plaquettes fails for every representation when $K\le45.3$.
\end{remark}

\subsection{Increasing returns of freezing}\label{ssec:G.5}

Let $X$ and $p_0$ be as in Theorem~\ref{thm:G.8}, let $\tau>0$, and let $D$ be a set of faces with $p_0\notin D$. Put $\vt_\tau$ on the faces outside $D$ and $\vt_s$ on those in $D$, and let $g(D)=\lim_{s\to\infty}\ex{\cos\theta_{p_0}}$. By monotone convergence in Lemma~\ref{lem:G.1} the limit exists and is given by the dual formula with $c_k=1$ on $D$, which is finite since $p_0\notin D$. In the notation of Section~\ref{sec:2}, $g(D)$ is the frozen expectation $g_{X,\tau}(\partial\delta_{p_0};D)$. The effect of a frozen set is not bounded by the product of the effects of its plaquettes.

\begin{proposition}\label{prop:G.12}
For $\tau\ge3$ and distinct faces $p,p'\neq p_0$,
\[
g(\{p,p'\})\,g(\emptyset)>g(\{p\})\,g(\{p'\}).
\]
Hence $D\mapsto\log g(D)$ is not submodular, and $g(D)/g(\emptyset)\le\prod_{q\in D}g(\{q\})/g(\emptyset)$ fails for $D=\{p,p'\}$.
\end{proposition}

\begin{proof}
By Lemma~\ref{lem:G.1} the value $g(D)$ depends only on $d=\abs{D}$. Write $g_d$ for it. The dual formula gives $g_d=\sum_he^{-((h-1)^2+(5-d)h^2)/(2\tau)}\big/\sum_he^{-(6-d)h^2/(2\tau)}$. Completing the square,
\[
\log g_d=-\frac{5-d}{2\tau(6-d)}+\log R_d,\qquad R_d=\frac{\Theta_d(1/(6-d))}{\Theta_d(0)},\quad \Theta_d(y)=\sum_{h\in\ZZ}e^{-(6-d)(h-y)^2/(2\tau)}.
\]
By Poisson summation $\Theta_d(y)$ is a positive multiple of $1+2\sum_{m\ge1}e^{-2\pi^2\tau m^2/(6-d)}\cos(2\pi my)$. This gives $(1-2\varrho)/(1+2\varrho)\le R_d\le1$ with $\varrho=e^{-\pi^2\tau/3}/(1-e^{-\pi^2\tau/3})$. The claim is $\log g_2-2\log g_1+\log g_0>0$. The leading terms contribute $\frac1\tau(-\frac38+\frac45-\frac5{12})=\frac1{120\tau}$, and the terms $\log R_d$ contribute at least $-2\lambda$ with $\lambda=\log\frac{1+2\varrho}{1-2\varrho}$. For $\tau\ge3$ we have $\varrho<5.2\cdot10^{-5}$ and $\lambda<2.1\cdot10^{-4}$, so $240\tau\lambda<1$ at $\tau=3$. Since $\tau\lambda$ decreases in $\tau$ for $\tau\ge3$, the claim follows.
\qed
\end{proof}
}

\section*{Acknowledgements}
Generative AI tools were used to write code, run analyses and draft text; the author directed the work and takes full
responsibility for its content.

\section*{Declarations}
\noindent\textbf{Conflict of interest.} The author declares no conflict of
interest.

\noindent\textbf{Data availability.} No data were generated or analysed. The file \texttt{scripts.zip} of the
\mbox{Zenodo} record of this paper, \url{https://doi.org/10.5281/zenodo.23445737}, holds the program
\texttt{cube\_amonotone.py}, which carries out the exact computation in the proof of Theorem~\ref{thm:G.8}, the
scripts behind the threshold values of Section~\ref{ssec:D.3}, and scripts that check several statements of the
paper numerically; its README describes each of them. Of the proofs in the paper, only that of
Theorem~\ref{thm:G.8} depends on these scripts.

\bibliographystyle{alpha}
\bibliography{refs}
\label{page:endmain}

\clearpage
\setcounter{section}{0}
\renewcommand{\thesection}{S\arabic{section}}
\begin{center}{\Large\bfseries Supplementary material}\end{center}
\section{Constants}\label{S:constants}
\subsection{Constants of Sections~\ref{sec:2} and~\ref{sec:8}: the ordered list, the threshold $K_*$ and the conditions defining $\beta_0(K)$}\label{S:constA}

Here we give the values of the constants; the order in which they are fixed, and the reasons for it, are explained in Section~\ref{app:D}. The values were evaluated by computer in high-precision arithmetic, but the proofs need only the inequalities stated.

\subsubsection{The ordered list}\label{ssec:D.1}

Each constant is fixed after all constants above it, and the constants in items~1 to~14 are independent of $\beta$, $\tau$, $K$, the loop $C$ and $D$, except that $\bar r_0$ of item~1 is $r_0$ or $2r_0-1$ according to $\Grid(C)$, and that the depletion constant $\cdep$ of item~12 depends on $\tau$. The scales of item~11 depend on $\beta$ through $\tilde m$ alone, and the derived quantities $c_g'$ and $c_u$, which lie between two constants of this kind, depend on $\beta$ only through the size of $L_c$ ($c_g'$) and through its rounding ($c_u$). Constants built from $c_u$, such as $D_{\cU}$ of Appendix~\ref{app:A} and $K_0^B$ of Appendix~\ref{app:B}, likewise lie between two constants of this kind, and we treat them as constants. Illustrative values are given at $(c_g,c_s,r_0)=(1/4,1,1)$. We use coarse units for the regularization and fine units in Sections~\ref{sec:stage1} and~\ref{sec:stage2}.

\begin{enumerate}
\item[0.] \emph{Absolute constants.} Among them are $c_I=2^{2/3}$ and the Debye exponent $\pi^2v_{\mathrm{Cb}}(0)=2.49435508\ldots$, where $v_{\mathrm{Cb}}(0)=(-\Delta_{\mathbb Z^3})^{-1}(0,0)=\frac{\sqrt6}{192\pi^3}\Gamma(\frac1{24})\Gamma(\frac5{24})\Gamma(\frac7{24})\Gamma(\frac{11}{24})=0.25273101\ldots$ is half of Watson's integral \cite{Wat39} in the normalization of \cite{Zu11}, whose closed form is \cite[(3.6)]{Zu11}, and $\log(1/m_D)=\pi^2v_{\mathrm{Cb}}(0)\tau-\frac12\log(8\pi^2\tau)$. The rounding rule for the Debye exponent is in Section~\ref{ssec:D.4}. The other absolute constants are $C_{\mathrm{bl}}=384$ (Proposition~\ref{prop:6.10}), the window count $151^3$ (Definition~\ref{def:3.4}), $\alpha=\ln8$ and $8/31$ (Theorem~\ref{thm:7.52} and Lemma~\ref{lem:7.58}), $c_T=2+\log96$ and the threshold $14.9662>6\pi^2v_{\mathrm{Cb}}(0)=14.96613\ldots$ (Lemma~\ref{lem:2.6} and Section~\ref{sec:mincut}), and the absolute constants among those of Göpfert\textendash Mack, Brydges\textendash Federbush and Stage~1 quoted in Sections~\ref{sec:stage1}, \ref{sec:stage2} and Appendices~\ref{app:A} to~\ref{app:C}. The constants of these authors and of Stage~1 that depend on the parameters $C$, $\delta_{\mathrm{GM}}$, $\mu$, $\kappa_S$, $\lambda_{\mathrm{PV}}$ of Theorem~\ref{thm:6.5} are fixed together with these parameters, in items~12 and~13.
\item[1.] $r_0\ge1$, an integer, and $\bar r_0:=2r_0-1$, or $\bar r_0:=r_0$ if $\Grid(C)$ has no exceptional layer; they are used in Definition~\ref{def:3.4}, Lemma~\ref{lem:5.11} and $\eta_T$.
\item[2.] $c_g:=1/4$, the constant of the property (S-face) of Definition~\ref{def:3.14}, which the construction provides for all $c_g\le1-134\vartheta$ by Theorem~\ref{thm:4.rest}. The proofs hold under $0<c_g\le1/4$, so every $c_g\in(0,1/4]$ is admissible.
\item[3.] $\eta_w=1/10$, the hull-weight deficit of the frozen bonds (Definition~\ref{def:3.8}).
\item[4.] $c_\rho=1/12$, that is, $\rho_\lambda=L_c/12$ and $\lambda=12/L_c$, the equality case of Lemma~\ref{lem:4.8}.
\item[5.] Before $\chi$ and $c_L$ we fix $k_P/L_c=304$, so that the windows of Definition~\ref{def:3.4} have sides less than $151\cdot2L_c<k_P$, as in (G4), and we set $r_\square:=(k_P+3)/L_c\le305$.
\item[6.] $c_1:=\min(c_g/2,((1-\eta_w)c_Ic_\rho/2)^3)$, which equals $2.109\cdot10^{-4}$ for $c_g\ge4.22\cdot10^{-4}$. We set $c_{\mathrm{sh}}:=c_1/216$; Table~\ref{tab:S:B1.hull} gives their uses.
\item[7.] $c_s\in[3.3\cdot10^{-7},10]$ (illustration $c_s=1$), with $w_*:=(1+c_s)|\partial\Delta|$ (Definition~\ref{def:3.11}) and $\lambda_U:=2c_s/\ell$ (Definition~\ref{def:3.forced}); then $2c_s\ell^2>1$, as $\ell\ge308L_c\ge1232$ (used in Theorem~\ref{thm:4.42}).
\item[8.] The forcing threshold $\vartheta=3\cdot10^{-5}$ and the penalty $\mu_{\rm B}=25$ of Definition~\ref{def:3.forced}, and the thin-set constant $\varepsilon_{\rm th}=0.008$ of Section~\ref{ssec:4.3}. They satisfy $24.6<\mu_{\rm B}$ (Lemma~\ref{lem:4.forced}), $128\vartheta<\varepsilon_{\rm th}$, $2\varepsilon_{\rm th}+12\vartheta\le0.0164$ and $2/\sqrt{0.0164}>14$ (Lemma~\ref{lem:4.thin}), and $254.4\vartheta\le\varepsilon_{\rm th}$ (Theorem~\ref{thm:4.rest}); the arithmetic is in Section~\ref{S:B1.cells}.
\item[9.] The weak-link constant $\chi:=2.0\cdot10^{-8}$ and $\chi_{\rm N3}:=\chi/3=6.67\cdot10^{-9}$ of Definition~\ref{def:3.13}. Its value depends only on items~3, 5 and~8, item~3 through the factor $12/\eta_w$ of Proposition~\ref{prop:4.sealcount} in the expansion bound, and this value is constrained by Theorem~\ref{thm:4.rest}(c), which provides the expansion $h_\star=2.2\cdot10^{-11}/L_c$ on each component of $\Gp\cap\hat Q$ at rest. Since $3k_Ph_\star=2.006\cdot10^{-8}\ge\chi$ (Section~\ref{S:B1.cells}), (N3) holds with $\chi_{\rm N3}$ and (R6) with $\chi$.
\item[10.] \emph{Consequences, fixed with the constants above and independent of $L_c$ and $\ell$.} $C_1:=7+27\cdot24.6/\vartheta=2.214\cdot10^7$, $C_{\rm seal}:=726$, $\Cst:=C_1+2C_{\rm seal}=2.2141\cdot10^7$ and $C_{\mathrm{free}}:=\Cst(2+1/(2c_s)+1/3696)=5.536\cdot10^7$ (Propositions~\ref{prop:4.sealcount} and~\ref{prop:4.39}); $C_a:=4362+5808/c_s$, $C_b:=2+4/c_s$ and $C_c:=1+2/c_s$, equal to $10170$, $6$, $3$ (Lemma~\ref{lem:5.31}), and the response load $C_{\mathrm{resp}}:=375C_a+125C_c\Cst=8.307\cdot10^9$ of Section~\ref{ssec:5.6}; finally, $C_{\mathrm{rm}}=20$, $C_{\mathfrak B}:=30\Cst+170C_{\mathrm{resp}}+30=1.413\cdot10^{12}$ and $C_3:=C_{\mathrm{free}}+20(2\Cst+4C_{\mathrm{resp}}+1)=6.655\cdot10^{11}$ (Proposition~\ref{prop:5.36}, Theorem~\ref{thm:5.1}).
\item[11.] The constant $c_L$ is subject to $C_P'c_L^2\le(1-4/\pi^2)/8$ and $c_L\le\frac12$. The first condition holds if $2916r_\square^5C_{\mathrm{bl}}^2c_L^2/\chi^2\le(1-4/\pi^2)/8$, since $C_P'=C_W^fr_\square^2N_{\mathrm{ov}}\le2916r_\square^5C_{\mathrm{bl}}^2/\chi^2$, where $C_W^f=108C_{\mathrm{bl}}^2/\chi^2$ is the constant of Proposition~\ref{prop:6.10}(iv) and $N_{\mathrm{ov}}\le27r_\square^3$. At $r_\square=305$ this yields the bound $8.09\cdot10^{-12}\chi$, and the admissible choice is $c_L\le8.0\cdot10^{-12}\chi$, that is, $c_L=1.6\cdot10^{-19}$ at $\chi=2.0\cdot10^{-8}$. This choice implies $\tilde mL\le\frac12$ for Lemma~\ref{lem:4.19} and $c_L\le1.99\cdot10^{-4}\chi_{\rm N3}=1.33\cdot10^{-12}$ for Lemma~\ref{lem:7.47}. The scales of Definition~\ref{def:3.1} are then $L_c:=c_L\tilde m^{-1}/r_0$ rounded down to an even integer, $L:=r_0L_c$, $k_P:=304L_c$, $\ell_D\approx\tilde m^{-1}$, a multiple of $L$, and $\ell:=\ell_D/r_0\ge k_P+4L_c$. The derived quantities are $c_g'=c_g(1+1/L_c)^{-2}$ and $c_u:=r_0^3V_*\tilde m^3=((1-\eta_w)c_I\tilde mL/24)^3$, which lies in $[((1-\eta_w)c_Ic_L/48)^3,((1-\eta_w)c_Ic_L/24)^3]$ once $c_L\tilde m^{-1}\ge4r_0$ (Section~\ref{S:B1.scales}). Further derived quantities are $R_{\mathrm{HK}}=(1-\eta_w)c_Ic_\rho L/3$ and the constant $K_0:=(1-\eta_w)c_Ic_\rho c_L/6=0.01984\,c_L$, whose uses are listed in Section~\ref{S:E1.scales}. Since $c_L/2\le\tilde mL\le c_L$, we have $K_0\le\tilde mR_{\mathrm{HK}}\le2K_0$.
\item[12.] \emph{Stage~1 and Stage~2 constants.} These are the constants of Sections~\ref{sec:stage1} and~\ref{sec:stage2} and Appendices~\ref{app:A} to~\ref{app:C}. We first fix the parameters $C\in(0,\frac14]$ and $\delta_{\mathrm{GM}}\in(0,\frac12]$ of Theorem~\ref{thm:6.5}, where this $C$ is not the loop; the bound $\delta_{\mathrm{GM}}\le\frac12$ is used for (S5) on outputs and in Lemma~\ref{lem:B.4} and Proposition~\ref{prop:B.7}. The bound $C\le\frac14$ on this parameter implies $2C_{M4}=2/(1-C)\le\frac83$, hence $q:=2C_{M4}e^{c_p}/e\le0.9829$ since $c_p<\frac1{540}$. In Proposition~\ref{prop:B.7}, therefore, one factor $e^{-(t_e-1)}$ absorbs the per-$\mathcal E$-factor constants $t_e(2C_{M4}e^{c_p})^{t_e-1}$ up to the factor $\sup_{t\ge1}tq^{t-1}\le22$, which contributes one factor per cell; here $t_e$ is the factor of (M4), Lemma~\ref{lem:B.1}(d), which bounds the sum over all assignments of the points to their cells.

Next come $\eta_H=0.62$, $(p_3,p',q',p_1,p_2,p_4)=(1.2,1.3,15.6,20,20,15)$, $c_\lambda=1/24$, $\theta_0=0.01$ and the link-sum rate $c_{10}$ with $0<c_{10}\le\frac79r_C$, where $r_C$ is evaluated at $\tilde mL=c_L/2$. We use this bound in Proposition~\ref{prop:B.7}, where, for $\delta_{\mathrm{GM}}\le\frac12$, the rate of (S2) in Debye units is at least $2r_C$ for all $\lambda_{\mathrm{PV}}\le\frac14$, by Proposition~\ref{prop:6.11}, and pinning uses $4c_p<\frac29r_C$ of this rate for every $\bar\delta$. The quantity $r_C=\min(\kappa_0/2,(2c_{\mathrm{gs}})^{1/2})$, as in Theorem~\ref{thm:6.12}, is computed with the constant $K_0$ of item~11, the value of $\tilde mR_{\mathrm{HK}}$ at $\tilde mL=c_L/2$, which is the worst case of $\tilde mL\in[c_L/2,c_L]$. It is therefore determined by items~0 to~11, and hence $c_{10}$ is fixed independently of $\beta$, $\lambda_{\mathrm{PV}}$ and $\bar\delta$.

We then fix $c''_{10}$ with $e^{c''_{10}}\ge16e\max(1,c_{c_{10}/2})$ (Lemma~\ref{lem:A.7}), and then the parameters $\mu$ and $\kappa_S$ of Theorem~\ref{thm:6.5}, both large: $\mu\ge1+2c''_{10}$, after $c_{10}$ and $c''_{10}$ (the factor $e^{-\mu(s-1)}$ of (S2) pays $e^{c''_{10}t}\le e^{2c''_{10}(t-1)}$, Proposition~\ref{prop:B.7}), and $\kappa_S$ so large that $(1+C_{\mathrm{ref}}e^{-2\kappa_S})\eta_H/8\le\frac17$ (Lemma~\ref{lem:C.4}), where $C_{\mathrm{ref}}:=8C$ with $C$ the constant of \eqref{eq:6.S4c} on $\Zthree$, so that $\bGM\sum_mm^2|\rho^{\mathrm{ref}}(m)|\le(1+C_{\mathrm{ref}}e^{-2\kappa_S})\mt^2$ (as $\mD\le2\mt$). This constant $C$ comes from the tail bound \eqref{eq:6.S4c}, that is, from (S2) at $s=1$ with $a_1=\{x\}$ and from \eqref{eq:6.lead}, whose right sides involve the parameter $C$ and $\mu$ but not $\kappa_S$. Hence the constant $C$ of the tail bound can be chosen independently of $\kappa_S$ and $\lambda_{\mathrm{PV}}$, and this last condition can be met. At $C=\frac14$, for instance, $c''_{10}=2200$, $\mu=4401$ and $\kappa_S\ge5.8$ are admissible; the binding condition on $\kappa_S$ is \eqref{eq:7.6.cl}.

Further, we require $\varepsilon_\nu+\theta'\le1/7$ and $c_E\le1/70$. The depletion constant $\cdep:=A_0^D+A_1^D\log(C'\tau/\theta_0)$ of Proposition~\ref{prop:6.18} depends on $\tau$, but all its occurrences are in lower bounds on $\tau$ (Theorem~\ref{thm:7.42}). Next we fix $K_{\mathrm{col}}=2>27\cdot2.4944/(4\pi^2)=1.706$ and $\varepsilon_0=0.003125$ (Corollary~\ref{cor:7.21}), and $0<\delta_2<\bar\delta<r_C$ and $c'_{\mathrm{rate}}=\frac12(r_C-\bar\delta)$ (Proposition~\ref{prop:7.38}). Last come the pure number $c'_0:=\frac12(1-1/p_3-c_\lambda)=1/16$, a lower bound for the fraction $c'$ of Proposition~\ref{prop:7.37}, then $\varepsilon_E=c'_0/8=1/128$, $c_w=c'/16$ and $\alpha=\ln8$.
\item[13.] $\lambda_{\mathrm{PV}}\le\min\bigl(\frac14,\lambda_0\bigr)$, chosen after $c_L$ and after $\mu$, $\kappa_S$ of item~12, where $\lambda_0=\lambda_0(\kappa_S,\mu,C,\delta_{\mathrm{GM}})$ of Theorem~\ref{thm:6.5} comes from the Debye condition in its proof. Here $M_Y$ is the smallest fixed point of \eqref{eq:6.mass} and $\tilde m:=\lambda_{\mathrm{PV}}M_Y$, as in Section~\ref{ssec:6.2}. The constants above are the same for every $\lambda_{\mathrm{PV}}$ allowed by the a priori bound $\lambda_{\mathrm{PV}}\le\frac14$, whereas those of Göpfert\textendash Mack, Brydges\textendash Federbush and Stage~1 that depend on $(\lambda_{\mathrm{PV}},\kappa_S,\mu,C,\delta_{\mathrm{GM}})$, among them the threshold $\bGMo$ of Theorem~\ref{thm:6.5}, are fixed here, after $\lambda_{\mathrm{PV}}$.
\item[14.] \emph{The cut-rule constants.} Those of Definition~\ref{def:3.17} are $\theta_{\mathrm{NF}}:=c'_0c_g/(90(c_g+4))=4.08\cdot10^{-5}$, $c_3^{\mathrm{NF}}:=c'_0c_g^2/(64(c_g+4))=1.44\cdot10^{-5}$ and $\kappa_{\mathrm{NF}}:=4(c_g+4)/c_g^2=272$ at $c_g=1/4$. They belong to the regularization but come after item~12, because they are built from the pure number $c'_0$, and the constants above are independent of $\theta_{\mathrm{NF}}$ and $c_3^{\mathrm{NF}}$.
\item[15.] $K>K_*=377.38$, which implies $c_\delta^{\mathrm{sp}}(K)/(2K)>14.9662$ (see Section~\ref{ssec:D.3}). Proposition~\ref{prop:5.15} and Theorem~\ref{thm:5.1} use this threshold.
\item[16.] $\beta_0(K)$, determined by the conditions of Section~\ref{ssec:D.5}.
\end{enumerate}

\noindent Among the scales determined by these constants are $\tilde m^{-1}$, with $\tilde m\in[m_D/2,2m_D]$, and $L_c$, $L$, $k_P$, $\ell$ as in item~11. These constants further determine $V_*=(1-\eta_w)^3\rho_\lambda^3/2$, $V_{\mathcal U}=\tilde m^{-3}$, $R_*=(J+1)w'+6L_c=O((\ell/c_s)\log(\ell/c_s))$ of Theorem~\ref{thm:4.42} together with $r'=2R_*+2$, and $R_{\mathrm{resp}}=10\ell$ together with $R'=R_*+2R_{\mathrm{resp}}$. The remaining scales are $L'=2(\beta_{\mathrm{GM}}/a)\tilde m^{-1}$, $R_1(\beta)=(2r_\square+5)L$, $F_{\mathrm{unit}}=\sigma_FL^2=(\pi^2c_g'/18)(\tilde mL)^3/(\tilde m\beta_{\mathrm{GM}})$ of Definition~\ref{def:3.17}, $R_{\mathrm{NF}}=Le^{c_3^{\mathrm{NF}}F_{\mathrm{unit}}}$, and the box $\Lamo$ of side $M_\Lambda'\in\{M_\Lambda,M_\Lambda+1\}$, where $M_\Lambda=\max(3\max(R,T),r_0(\ell+3))$.

\subsubsection{The threshold $K_*$ of Lemma~\ref{lem:2.6}}\label{ssec:D.3}

Lemma~\ref{lem:2.6} gives, with $a_*$ the solution of $a=K\log(2a)$,
\begin{equation}\label{eq:D.csp}
c_\delta(K)=\frac{|\log\delta|}{\beta}\ge c_\delta^{\mathrm{sp}}(K)-\frac1{4a_*}-O\Big(\frac{\log\beta}\beta\Big),\qquad c_\delta^{\mathrm{sp}}(K):=a_*\Big(\tfrac12\log(2a_*)-1\Big)+1 .
\end{equation}
The function $K\mapsto c_\delta^{\mathrm{sp}}(K)/(2K)$ is increasing. Indeed $da_*/dK=a_*\log(2a_*)/(a_*-K)$ and $a_*-K=K(\log(2a_*)-1)$, which yield
\[
\frac{dc_\delta^{\mathrm{sp}}}{dK}=\Big(\tfrac12\log(2a_*)-\tfrac12\Big)\frac{da_*}{dK}=\tfrac12\log^2(2a_*),
\]
and $K\cdot\frac12\log^2(2a_*)-c_\delta^{\mathrm{sp}}(K)=K\log(2a_*)-1>0$. Hence $\frac{d}{dK}\frac{c_\delta^{\mathrm{sp}}(K)}{K}>0$. The threshold of Lemma~\ref{lem:2.6} and \eqref{eq:2.threshold} is $14.9662$, that is, $6\pi^2v_{\mathrm{Cb}}(0)=14.96613\ldots$ rounded up. The equation $c_\delta^{\mathrm{sp}}(K)/(2K)=14.9662$ has the unique root $K=377.3635$, and $K_*:=377.38$ lies above it. Hence $c_\delta^{\mathrm{sp}}(K)/(2K)>14.9662$ for all $K\ge K_*$. Numerically, at $K=K_*$,
\[
\frac{c_\delta^{\mathrm{sp}}(K_*)}{2K_*}=14.96639,\qquad a_*=3321.43,\qquad \frac1{4a_*}=7.5\cdot10^{-5}.
\]

\emph{The computation behind \eqref{eq:2.cdelta}} (proof of Lemma~\ref{lem:2.6}). We put $\nu:=\astar\beta$ in Lemma~\ref{lem:2.4}(iv). Then
$\nu^2\eps/2=\astar^2\beta/(2K)$. Let $h(\mu):=(\beta/2)^\mu/\Gamma(\mu+1)$ and
$\mu_n:=(n^2+\nu^2)^{1/2}\ge\nu$. From the series
$I_\mu(\beta)=\sum_{k\ge0}(\beta/2)^{\mu+2k}/(k!\,\Gamma(\mu+k+1))$ and
$\Gamma(\mu+k+1)=\Gamma(\mu+1)\prod_{i=1}^k(\mu+i)\ge\Gamma(\mu+1)(\mu+1)^k$,
we get $I_\mu(\beta)\le h(\mu)\sum_{k\ge0}(\beta^2/(4(\mu+1)))^k/k!=h(\mu)e^{\beta^2/(4(\mu+1))}\le h(\mu)e^{\beta/(4\astar)}$ for
$\mu\ge\nu$. Since $\log\Gamma$ is convex, its difference quotients increase.
Thus for $\mu\ge\nu$
\[
\log\Gamma(\mu+1)-\log\Gamma(\nu+1)\ge(\mu-\nu)\bigl(\log\Gamma(\nu+1)-\log\Gamma(\nu)\bigr)=(\mu-\nu)\log\nu ,
\]
hence $h(\mu)\le h(\nu)(\beta/(2\nu))^{\mu-\nu}\le h(\nu)2^{-(\mu-\nu)}$, because
$\nu=\astar\beta\ge\beta$. As $\mu_n-\nu\ge0$ for $\abs{n}\le\nu$ and
$\mu_n-\nu\ge\abs{n}-\nu$ otherwise,
\[
\sum_{n\in\ZZ}I_{\mu_n}(\beta)\le e^{\beta/(4\astar)}h(\nu)\Bigl(2\nu+1+2\sum_{m>\nu}2^{-(m-\nu)}\Bigr)\le(2\astar\beta+5)\,e^{\beta/(4\astar)}h(\nu).
\]
Stirling's inequality $\Gamma(x+1)\ge(x/e)^x$ gives
$\log h(\nu)\le-\astar\beta(\log2\astar-1)$. Finally, for $\beta\ge\pi^{-2}$,
using $\cos\theta\ge1-\theta^2/2$ on $[0,\beta^{-1/2}]\subseteq[0,\pi]$, we obtain
$I_0(\beta)=\frac1\pi\int_0^\pi e^{\beta\cos\theta}d\theta\ge\frac{1}{\pi\sqrt\beta}e^{\beta-1/2}$.
Collecting these bounds, we obtain
\[
\log\delta\le-\beta\Bigl[\astar(\log2\astar-1)-\frac{\astar^2}{2K}+1-\frac{1}{4\astar}\Bigr]+\log(2\astar\beta+5)+\log(\pi\sqrt\beta)+\frac12 ,
\]
and $\astar^2/(2K)=\frac12\astar\log2\astar$ by the definition of $\astar$, which
gives \eqref{eq:2.cdelta}, with exactly the function $c_\delta^{\mathrm{sp}}(K)$ and the
correction $1/(4a_*)$ of \eqref{eq:D.csp}.

The term $O(\log\beta/\beta)$ is absorbed in the term $O(\log\tau+\log r_0)$ of Lemma~\ref{lem:5.16}. The margins $c_\delta^{\mathrm{sp}}(K)/(2K)-14.9662$ that appear in $\eta_T$ are:
\[
\begin{array}{c|cccc}
K & 377.38 & 10^3 & 3\cdot10^3 & 10^4\\ \hline
c_\delta^{\mathrm{sp}}(K)/(2K) & 14.9664 & 19.521 & 25.289 & 32.372\\
\text{margin} & 1.9\cdot10^{-4} & 4.55 & 10.32 & 17.41
\end{array}
\]

\emph{Numerical values.} To more digits, $a_*(K_*)=3321.434$ and $c_\delta^{\mathrm{sp}}(K_*)/(2K_*)=14.966392$; moreover $1/(8K_*a_*)=9.97\cdot10^{-8}$, and the root is $377.363524$, or $377.363533$ with the correction $-1/(4a_*)$. The correction $1/(4a_*)$ in \eqref{eq:D.csp} moves $c_\delta/(2K)$ by $10^{-7}$ and the root by less than $10^{-4}$. The entropy exponent of Lemma~\ref{lem:5.16} is $6\pi^2v_{\mathrm{Cb}}(0)<6\cdot2.4943551=14.96613\ldots<14.9662$, so the threshold is computed without the rounding of the Debye exponent up to $2.4944$ used elsewhere; with the rounded exponent the threshold is $14.9664$, whose root $377.3807$ exceeds $K_*$.

\subsubsection{The string tension and the rounding rule}\label{ssec:D.4}

The slit-domain tension of Theorem~\ref{thm:7.61} is $\sigma_V=\varepsilon_E\sigma_F=(c_g'\tilde mL/9216)\tilde m/\tau$, with $c_g'=c_g(1+1/L_c)^{-2}$ and with $\tilde mL>c_L-2r_0\tilde m$ by Section~\ref{ssec:3.1}. For $L_c\ge35$, one of the conditions of Section~\ref{ssec:D.5}(e), we have $c_L\tilde m^{-1}\ge35r_0$, and hence $c_g'\tilde mL\ge(36/35)^{-2}(1-2/35)\,c_gc_L\ge0.89\,c_gc_L\ge(9/16)c_gc_L$. In \eqref{eq:8.sigma}, $18432\cdot16/9=32768=2^{15}$ exactly and $(8\pi^2)^{1/2}/2^{16}=1.3559\cdot10^{-4}$. We take $c_g=1/4$ and $c_L=8.0\cdot10^{-12}\cdot2.0\cdot10^{-8}=1.6\cdot10^{-19}$, as in items~9 and~11, although any smaller admissible $c_L$ also works; at these values \eqref{eq:8.sigma} gives $\sigma(\beta)\ge5.4\cdot10^{-24}(K\beta)^{-1/2}e^{-2.49436K\beta}$.

Throughout, and in particular in Section~\ref{ssec:8.1}, we round as follows. The Debye exponent is $\pi^2v_{\mathrm{Cb}}(0)=2.49435508\ldots<2.4943551$. In a lower bound on $\tilde m$, and hence on a tension, the exponent is rounded up, to $2.49436$, and prefactors are rounded down, to $1.35\cdot10^{-4}$. In an upper bound on a scale the exponent is rounded up to $2.4944$: since $\tilde m\ge m_D/2$, $\log(1/\tilde m)\le2.4943551\tau-\frac12\log(8\pi^2\tau)+\log2\le2.4944\tau$ for all $\tau\ge1$. We use this form for $K_{\mathrm{col}}>27\cdot2.4944/(4\pi^2)$, in the logarithmic step of the proof of Lemma~\ref{lem:6.19}, and in the bound $\log(1/\tilde m)/\beta_{\mathrm{GM}}\le2.4944/(4\pi^2)=0.063184<0.0632$ of Remark~\ref{rem:7.62}. The bounds that determine $K_*$ use the exponent without rounding. There the entropy $6\log R_*$ of a root and a part in Lemma~\ref{lem:5.16}, with $\ell\approx\tilde m^{-1}/r_0$ and with $R_*$, $r'$ and $R'$ of Section~\ref{sec:mincut}, is bounded with $\pi^2v_{\mathrm{Cb}}(0)<2.4943551$, as in Sections~\ref{ssec:D.3} and~\ref{S:const5} (for the threshold with the rounded exponent, see the end of Section~\ref{ssec:D.3}). Where the exponent serves as a lower bound on $\log(1/\tilde m)$, it is rounded down: the hypothesis of Lemma~\ref{lem:6.19} has $2.494\tau$, because it is used to bound $\tilde m\le2m_D\le17.8\,\tau^{1/2}e^{-2.494\tau}$ from above.

\subsubsection{The conditions defining $\beta_0(K)$}\label{ssec:D.5}

We fix the constants of items~0 to~15. As explained in Section~\ref{app:D}, each condition below is a lower bound on $\tau=K\beta$. Then $\beta_0(K)$ is $\tau_0/K$ for the largest of the thresholds $\tau_0$.
\begin{enumerate}
\item[(a)] \emph{Section~\ref{sec:mincut}.} The conditions used in Proposition~\ref{prop:5.13}, Lemma~\ref{lem:5.14}, Proposition~\ref{prop:5.15}, Lemma~\ref{lem:5.25}, Proposition~\ref{prop:5.27}, Lemma~\ref{lem:5.35}, Proposition~\ref{prop:5.36}, Lemmas~\ref{lem:5.37}, \ref{lem:5.38} and Theorem~\ref{thm:5.1} are
\begin{gather*}
\varepsilon_c\le\tfrac12,\qquad\eta_T^{(1)}\le1,\qquad\varepsilon''\le\tfrac12,\qquad\eta_T\le1,\qquad 2^{29}\cdot26e\cdot\varepsilon''^{1/(3C_{\mathfrak B})}\le\tfrac12,\\
2\big(2\lceil R'/L_c\rceil+5\big)^3\,\varepsilon''^{\,1/(48\varepsilon_D(1+8L_c)(\Cst+3C_{\mathrm{resp}}+6))}\le\tfrac12,\\
8\varepsilon_D(1+8L_c)(\Cst+3C_{\mathrm{resp}}+6)(48C_{\mathrm{resp}}+2)\le1\quad\text{(that is, \eqref{eq:5.16})},\\
R_{\mathrm{NF}}\ge\max\big(3456\bar r_0L_c,\ 2048\bar r_0L_c/(c_{\mathrm{sh}}\theta_{\mathrm{NF}}F_{\mathrm{unit}})\big).
\end{gather*}
Here $\varepsilon_D=3.2768\cdot10^7\bar r_0L_c/(c_{\mathrm{sh}}\theta_{\mathrm{NF}}F_{\mathrm{unit}}R_{\mathrm{NF}})$ is doubly exponentially small. The factor $2^{29}$ contains the factor $2^b$ for the sizes of the components in Lemma~\ref{lem:5.38}. The binding condition is the first of \eqref{eq:5.17}, $|\log\varepsilon''|\ge75.16\,C_{\mathfrak B}\approx1.1\cdot10^{14}$, while $\eta_T\le1$ needs about $1.7\cdot10^{13}$ and $\eta_T^{(1)}\le1$ needs about $3.6\cdot10^8$; all three values are computed in Section~\ref{S:const5}.
\item[(b)] \emph{Section~\ref{sec:stage1}.} The conditions are $\tilde m\le K_0(2C_H)^{-1/3}$; $M_Y\le m_W$ and $\tilde m\le\min(1/68,m_W/2)$ for Theorem~\ref{thm:6.12}, with $m_W=\min(1,c_{\mathrm{ex}}/2)$ of Proposition~\ref{prop:6.11}; $M_Y\le\frac15$, for Lemma~\ref{lem:F.7} on $\Lambda^\circ$ in the proof of Theorem~\ref{thm:6.16}; $m_D/2\le\tilde m\le2m_D$; the coupling condition of Lemma~\ref{lem:6.19}, $\tau\ge2$ and $2.494\tau\ge\ln(C^{\mathrm{dep}}_*/\theta_0)+2.5\ln\tau$, whose exponent is rounded as in Section~\ref{ssec:D.4}; and the threshold $\beta_{\mathrm{GM}}\ge\bGMo(\lambda_{\mathrm{PV}},\kappa_S,\mu,C,\delta_{\mathrm{GM}})$ of Theorem~\ref{thm:6.5}.
\item[(c)] \emph{Section~\ref{sec:stage2} and Appendices~\ref{app:A} to~\ref{app:C}.} The conditions are the slack in $\varepsilon_\nu+\theta'\le1/7$ and $c_E\le1/70$, which needs the terms $\beta_{\mathrm{GM}}M_Y$ and $e^{-\mu}$, $e^{-2\kappa_S}$ to be small; $\varepsilon(L')\le\varepsilon_0$ of Corollary~\ref{cor:7.21}; ($\beta1$) and ($\beta3$) of \eqref{eq:A.beta}; the cell-diameter conditions of Lemma~\ref{lem:7.47} and Proposition~\ref{prop:6.10}, of the form $2+C(k_\square/\chi)\log k_\square<L'$; the conditions of Proposition~\ref{prop:7.41} and Theorem~\ref{thm:7.42}, the latter a condition $0.4157\,\tau\ge C_{\mathfrak D}A_1^D\log(C'\tau/\theta_0)+\alpha+O(\log\tau)$; $c_{16}e^{c_2}\le1/8$ of Lemma~\ref{lem:A.11}; the conditions of Lemma~\ref{lem:7.53}; $\Theta_4\le\frac14$ of Proposition~\ref{prop:7.57}; and $F_{\mathrm{unit}}\ge(8C'/c')(1+\ln\beta_{\mathrm{GM}})$ of Proposition~\ref{prop:7.56}.
\item[(d)] \emph{Section~\ref{sec:8}.} The conditions are $\eta_T+\eta_S\le\frac12$ and $\sigma_V\le1$. Both deficits are explicit upper bounds that are eventually decreasing in $\tau$: $\eta_T=O_K(1/\tau)$, because, by Lemma~\ref{lem:5.16}, $|\log\varepsilon''|\ge\gamma_T(K)\tau-O(\log\tau+\log r_0)$ with $\gamma_T(K)>0$ for $K>K_*$, and $\eta_S=O(\tilde m\beta_{\mathrm{GM}}\log\tau)$.
\end{enumerate}
\begin{enumerate}
\item[(e)] \emph{Rounding of $L_c$ in Section~\ref{sec:3}.} The conditions are $L_c\ge4$, which implies $c_g'\ge(16/25)c_g\ge(9/16)c_g$ for Lemma~\ref{lem:4.17}; $L_c\ge35$, used for $\lambda=12/L_c\le1$ in Proposition~\ref{prop:4.39}, in Lemma~\ref{lem:7.47} and for $c_g'\tilde mL\ge(9/16)c_gc_L$ in Section~\ref{ssec:D.4} and Section~\ref{ssec:8.1}; and $c_L\tilde m^{-1}\ge4r_0$, which gives $\tilde mL\ge c_L/2$ (the bounds on $c_u$ and on $C_{\mathcal U}$, Lemma~\ref{lem:4.19}).
\item[(f)] \emph{Section~\ref{sec:2}.} The conditions are $\beta\ge\pi^{-2}$, for Lemma~\ref{lem:2.6}; \eqref{eq:2.threshold}, that is, $\beta\ge\beta_1(K)$; and $\delta<1$, which follows from \eqref{eq:2.cdelta} for all large $\beta$ and is used in Step~\eqref{eq:8.1}, with $\delta'=\delta$ in Proposition~\ref{prop:2.9}.
\end{enumerate}
In addition, the range of Remark~\ref{rem:8.3}, where Phase~2 is empty, is nonempty once $\sqrt3(r_0(\ell+3)+1)\le R_{\mathrm{NF}}$. This condition belongs to that remark rather than to Theorem~\ref{thm:1.2}, and it holds for all large $\tau$ because $r_0\ell\approx\tilde m^{-1}$ is exponential and $R_{\mathrm{NF}}$ is doubly exponential in $\tau$.

All these thresholds are finite but very large (Section~\ref{app:D}).

\subsection{Constants of Section~\ref{sec:3} and Subsections~\ref{ssec:4.1} and~\ref{ssec:4.3}}\label{S:B1}

We group the constants by the main-text statement in which they appear.

\paragraph{Definition~\ref{def:3.1}: the scales.}\label{S:B1.scales} The screening mass satisfies $m_D/2\le\tilde m\le2m_D$, so that $\log(1/\tilde m)=\pi^2v_{\rm Cb}(0)\,\tau+O(\log\tau)$ with $\pi^2v_{\rm Cb}(0)=2.49435508\ldots$, and $\tilde m^{-1}$ is the Debye length. The scales are fixed at item~11 of the ordered list in Section~\ref{ssec:D.1}. Since $c_L\le1/308$, \eqref{eq:3.Hell} follows from $\ell\ge\tilde m^{-1}/r_0\ge L_c/c_L$. Of the conditions $L_c\ge4$ and $c_L\tilde m^{-1}\ge4r_0$ of Section~\ref{ssec:D.5}(e), the first implies $c_g':=c_g(1+1/L_c)^{-2}\ge\frac9{16}c_g$, and by the second, $L>c_L\tilde m^{-1}-2r_0\ge c_L\tilde m^{-1}/2$, that is, $\tilde mL\ge c_L/2$.

\paragraph{Definition~\ref{def:3.4}: windows.} A window is the union of the blocks at block distance at most $75$ from a block. It therefore has at most $151$ blocks per direction, each of coarse side at most $2L_c-1$, and its sides are at most $151(2L_c-1)<302L_c<k_P$. A block lies in the windows of the at most $151^3=3.44\cdot10^6$ blocks at block distance at most $75$ from it, and $151^3<27\cdot304^3=7.6\cdot10^8=27(k_P/L_c)^3$, the overlap bound of (R6).

\paragraph{Definition~\ref{def:3.forced}, Lemma~\ref{lem:4.forced}, Theorem~\ref{thm:4.30}: the cost of a fill.}\label{S:B1.forced} A block $Q$ has $|\partial Q|\le6(2L_c-1)^2<24L_c^2$ and $|Q|<8L_c^3$. Hence, with $\lambda_U=2c_s/\ell$, $c_s\le10$ and $\ell\ge308L_c$, we have $\lambda_U|Q|<16c_sL_c^3/\ell\le0.52L_c^2$ and $|\partial Q|+\lambda_U|Q|<24.6L_c^2<\mu_{\rm B}L_c^2=25L_c^2$. Also $\lambda_U=2c_s/\ell\le20/(308L_c)<\lambda=12/L_c$. A removal cell $\Delta$ with sides $a,b,c\in[\ell,3\ell)$ has $|\Delta|/|\partial\Delta|=abc/(2(ab+bc+ca))\le(abc)^{1/3}/6<\ell/2$ by the arithmetic-geometric mean inequality, so $\lambda_U|\Delta|<c_s|\partial\Delta|$ and $|\partial\Delta|+\lambda_U|\Delta|<w_*$. The number of forced blocks is at most $27|\Mset|/(\vartheta L_c^2)$, since a mark lies in at most $27$ sets $3Q$. Hence filling them costs at most $27\cdot24.6/\vartheta=664.2/\vartheta$ per mark.

\paragraph{Lemma~\ref{lem:4.5}: constants that depend on the hull weight.}\label{S:B1.hull} Frozen bonds have the hull weight $2-\eta_w$ of Definition~\ref{def:3.8}, and the hull weight affects the constants in two ways. Through Lemma~\ref{lem:4.5} it determines $(a_\eta,b_\eta)=(1+1/(1-\eta_w),1/(1-\eta_w))$ and the factor $1/(1-\eta_w)$ in the bound on $N_2(W)$. Table~\ref{tab:S:B1.hull} lists the constants that depend on $\eta_w$ through Lemma~\ref{lem:4.5}, with their formula in $\eta_w$ and their value at $\eta_w=1/10$; the other constants, among them the constant $c_0$ of the density bound, the bound on $c_L$, $K_*$, $\sigma(\beta)$ and $R_1(\beta)$, do not depend on $\eta_w$ through Lemma~\ref{lem:4.5}. In the cell analysis the hull weight appears, by Lemma~\ref{lem:4.11}(b), as the deficit $\eta_w$ of a short segment through a frozen bond, which yields the factor $120=12/\eta_w$ of Proposition~\ref{prop:4.sealcount}. Through this factor, $C_{\rm seal}=726$, the number $134$ of non-good lines per unit of $m_Q+m_{Q'}$ and the $158$ in the expansion bound behind $h_\star$ depend on $\eta_w$ (see the arithmetic in Section~\ref{S:B1.cells}), and so do the constants built from them, such as $\Cst$, $C_{\mathrm{free}}$ and $C_{\mathrm{resp}}$, and the constraint on the weak-link constant $\chi$ (items~9 and~10 of Section~\ref{ssec:D.1}).

\begin{table}[ht]
\caption{Constants depending on the hull weight; values at $\eta_w=1/10$, $c_I=2^{2/3}$, $\rho_\lambda=L_c/12$.}\label{tab:S:B1.hull}
\centering\footnotesize
\begin{tabular}{p{2.9cm}p{3.5cm}p{3.3cm}p{2.3cm}}
\hline
quantity & formula & value at $\eta_w=1/10$ & used in\\
\hline
$(a_\eta,b_\eta)$ & $(1+1/(1-\eta_w),\,1/(1-\eta_w))$ & $(19/9,10/9)$ & Lemma~\ref{lem:4.5}, (R2)\\
isoperimetry for $|W|\le V_*$ & $(c_I/(2a_\eta))|W|^{2/3}$ & $(9c_I/38)|W|^{2/3}=0.376|W|^{2/3}$ & Theorem~\ref{thm:4.7}(i)\\
$V_*$ & $(1-\eta_w)^3\rho_\lambda^3/2$ & $0.3645\rho_\lambda^3$ & Definition~\ref{def:3.1}, (R2)\\
volume growth constant & $(c_I/(48a_\eta))^3$ & $3.844\cdot10^{-6}$ & Theorem~\ref{thm:4.7}(iii)\\
saturation radius of (iii) & $24a_\eta(1-\eta_w)\rho_\lambda$ & $45.6\rho_\lambda$ & Theorem~\ref{thm:4.7}(iii)\\
Faber\textendash Krahn constant & $c_I^2/(48a_\eta^2)$ & $c_I^2/213.93$ & Theorem~\ref{thm:4.7}(iv)\\
$N_2(W)$ factor & $1/(1-\eta_w)$ & $10/9$ & Lemmas~\ref{lem:4.5}, \ref{lem:6.14}\\
$c_1$ (second entry) & $((1-\eta_w)c_I/24)^3$ & $2.109\cdot10^{-4}$ & (R5)\\
$c_u$ & $((1-\eta_w)c_I\tilde mL/24)^3$ & $(0.0595\,\tilde mL)^3$ & (R7)\\
$R_{\mathrm{HK}}$, $K_0$ & $(1-\eta_w)c_I\rho_\lambda/3$, $K_0=(1-\eta_w)c_Ic_\rho c_L/6$ & $\tilde mR_{\mathrm{HK}}=0.0397\tilde mL\in[K_0,2K_0]$, $K_0=0.01984c_L$ & Theorem~\ref{thm:6.9}\\
$c_F$ (fine isoperimetry) & $(c_I/(2a_\eta))/152.4$ & $2.467\cdot10^{-3}$ & Proposition~\ref{prop:6.10}\\
$C_{\mathcal U}$, $N_{\mathrm{near}}$, $A_2$ & through $1/c_1$ and $1/c_u$, at most the factor $(1-\eta_w)^{-3}$ & factor at most $1.372$ & (R7), Section~\ref{sec:stage2}\\
$c_{\mathrm{sh}}$ & $c_1/216$ & $9.77\cdot10^{-7}$ & Lemma~\ref{lem:5.22}\\
\hline
\end{tabular}
\end{table}

\noindent At $\eta_w=1/10$, $c_1$ equals its second entry, with the factor $(1-\eta_w)^3=0.729$, when $c_g\ge4.22\cdot10^{-4}$, and $c_1=c_g/2$ otherwise. Hence the factor $(1-\eta_w)^{-3}=1.372$ in the table is an upper bound. The Nash constant $A_N=24\cdot2^{2/3}/c_F^2$ of Appendix~\ref{app:F} depends on $\eta_w$ only through $c_F$.

\paragraph{Theorem~\ref{thm:4.7}.}\label{S:B1.geom} We have $V_*=(c_I\rho_\lambda/(2b_\eta))^3$, since $(c_I/2)^3=\frac12$. Hence $b_\eta\lambda V_*^{1/3}=c_I/2$. In (iii), $\alpha:=c_I/(12a_\eta)=0.0626\ldots\le\frac14$, and the two expressions in the minimum are equal at $r=24(a_\eta/b_\eta)\rho_\lambda=45.6\rho_\lambda$.

\paragraph{Lemma~\ref{lem:4.19}.}\label{S:B1.U} We set
\[
C_{\Uc}:=8\sqrt3\,(2+6/c_g)/(c_1c_L^2),
\]
which bounds $2\sqrt3(2+6/c_g)/(c_1(\tilde mL)^2)$ when $\tilde mL\ge c_L/2$.

\paragraph{Proposition~\ref{prop:4.22}: the constant $c_0$.}\label{S:B1.c0} The fine plaquettes dual to the insulating bonds with an end in a removal cell $\Delta$, those of frozen bonds counted twice, number at most $\bar r_0^{\,2}(3+c_s)|\partial\Delta|$, since a coarse bond is dual to at most $\bar r_0^{\,2}$ fine plaquettes. Hence they number at most $c_0/(z\tau)$ with $c_0:=z\tau\,\bar r_0^{\,2}(3+c_s)\max_\Delta|\partial\Delta|$, where $z$ is the monopole activity of Section~\ref{sec:stage1}, $m_D^2=2\beta_{\rm GM}z$ and $z\tau=m_D^2/(8\pi^2)$. Using $\bar r_0<2r_0$, $r_0\ell=\ell_D<\tilde m^{-1}+L\le(1+c_L)\tilde m^{-1}$, $m_D\le2\tilde m$, the bound $|\partial\Delta|<54\ell^2$ of (G3) and $c_L\le\frac12$, we obtain
\[
c_0<\frac{4\tilde m^2}{8\pi^2}\cdot4\cdot54\,(3+c_s)(1+c_L)^2\tilde m^{-2}\le\frac{243}{\pi^2}(3+c_s)<25(3+c_s),
\]
independently of $\beta$. On a grid without exceptional layers and with cubic removal cells, $c_0=6(3+c_s)(m_D\ell_D)^2/(8\pi^2)$.

\paragraph{Lemma~\ref{lem:4.runs}, Proposition~\ref{prop:4.giant}, Lemmas~\ref{lem:4.thin}, \ref{lem:4.aux}, Proposition~\ref{prop:4.sealcount}, Theorem~\ref{thm:4.rest}: the cell analysis.}\label{S:B1.cells} With $\vartheta=3\cdot10^{-5}$ and $\varepsilon_{\rm th}=0.008$:
\begin{itemize}
\item \emph{Giant.} $12+6.4\cdot9\cdot2=127.2<128$, and $(1-3\vartheta)L_c^3>384\vartheta L_c^3=3\cdot128\vartheta L_c^3$ since $384\vartheta=0.01152$.
\item \emph{Thin sets.} $2\varepsilon_{\rm th}+12\vartheta=0.01636\le0.0164\le\frac14$, and $2/\sqrt{0.0164}=15.62\ge15.6>14$. Non-giant components are thin since $128\vartheta=0.00384<\varepsilon_{\rm th}$.
\item \emph{Seal.} Lemma~\ref{lem:4.11}(b) with $\mathrm{supp}\le6(m_Q+m_{Q'})$ and $\eta_w=1/10$ gives $120(m_Q+m_{Q'})$ lines, plus $m_Q+m_{Q'}$ lines with a mark, in total $121$ per unit of $m_Q+m_{Q'}$, and $C_{\rm seal}=6\cdot121=726$.
\item \emph{Interfaces.} $1+121+12=134$ non-good lines per unit of $m_Q+m_{Q'}<\vartheta L_c^2$, so $\good(I)\ge(1-134\vartheta)|F'|=0.99598|F'|\ge c_g|F'|$.
\item \emph{Expansion.} For two large sides, $\varepsilon_{\rm th}-127.2\vartheta=0.004184\ge\varepsilon_{\rm th}/2$, so $\mathrm{cut}\ge\varepsilon_{\rm th}L_c^2/25.6=3.125\cdot10^{-4}L_c^2$. For adjacent in/out blocks, $\mathrm{cut}\ge(1-158\vartheta-2\varepsilon_{\rm th})L_c^2/3=0.32642L_c^2$. Both are at least $3.1\cdot10^{-4}L_c^2$, and $|A|<k_P^3/2=1.4047\cdot10^7L_c^3$. Hence $h_\star\ge3.1\cdot10^{-4}/(1.4047\cdot10^7L_c)=2.207\cdot10^{-11}/L_c\ge2.2\cdot10^{-11}/L_c$. Then $3k_Ph_\star=912\cdot2.2\cdot10^{-11}=2.006\cdot10^{-8}\ge\chi=2.0\cdot10^{-8}$, and $\chi_{\rm N3}=\chi/3=6.67\cdot10^{-9}\le k_Ph_\star=6.69\cdot10^{-9}$.
\end{itemize}
Theorem~\ref{thm:4.rest} shows that the construction provides $c_1$-type volume $(1-3\vartheta)L_c^3$ per cell, one interface per block face and six neighbouring cells per cell. These facts imply (R5) and (R8) with the constants $c_1$, $1/c_g$ and $6/c_g$, which we keep.

\subsection{Constants of Sections~\ref{ssec:4.9} and~\ref{ssec:4.4}}\label{S:C1}

We first list the sizes on the loop-adapted grid and then the constants, grouped by main-text statement.

\paragraph{Sizes on the loop-adapted grid.} Table~\ref{tab:4.1} lists the sizes on $\Grid(C)$ and, for orientation only, their values on the uniform grid, in which all coarse cells have side $r_0$ and all block spacings are $L_c$; the constant $c_1$ in the table is that of item~6 of Section~\ref{ssec:D.1}. The opacity property (S-face) of Definition~\ref{def:3.14} is normalized per face, with the threshold $c_g|F(I)|$, and hence the good fraction and the cell degree are independent of the face area.

\begin{table}[ht]
\caption{Sizes on the loop-adapted grid $\Grid(C)$, in coarse units unless stated, with the values on the uniform grid for orientation and the statements in which each size is used.}\label{tab:4.1}
\centering\footnotesize
\begin{tabular}{p{3.0cm}p{2.9cm}p{2.2cm}p{2.7cm}}
\hline
quantity & on $\Grid(C)$ & uniform grid & used in\\
\hline
block side & $[L_c,2L_c-1]$ & $L_c$ & (G1)\\
block and cell volume & $<8L_c^3$ & $L_c^3$ & (R5); Lemma~\ref{lem:4.19} ($\tilde mL\le\frac12$)\\
face area $|F'|$ & $[L_c^2,4L_c^2)$ & $L_c^2$ & (S-face), Lemma~\ref{lem:4.forced}\\
block diameter & $<2\sqrt3L_c$ & $\sqrt3L_c$ & Cor.~\ref{cor:4.10}\\
window side; windows per site & $<k_P$; $\le151^3$ & $151L_c$; $151^3$ & (G4), (R6)\\
removal cell side & $[\ell,3\ell)$ & $\ell$ & Lemma~\ref{lem:5.32}\\
$|\partial\Delta|$ & $[6\ell^2,54\ell^2)$ & $6\ell^2$ & (R9), $c_0$ of Section~\ref{S:B1.c0}\\
reach $R_*$ & Theorem~\ref{thm:4.42} & same & Theorem~\ref{thm:3.21}\\
$k_P$ & $304L_c$ & same & (G4), (R6)\\
$c_1$; good fraction; cell degree & $2.109\cdot10^{-4}$; $c_g$; $6/c_g$ & same & (R5), (R8)\\
fine good fraction & $c_g(1+1/L_c)^{-2}$ & $c_g$ & Lemma~\ref{lem:4.17}\\
$q_m$ & $3(\bar r_0+1)^3\delta^{1/2}\le24r_0^3\delta^{1/2}$ & $3(r_0+1)^3\delta^{1/2}$ & Lemma~\ref{lem:5.11}\\
fine area of a coarse face & $\le\bar r_0^2<4r_0^2$ & $r_0^2$ & Section~\ref{sec:5} ($\bar r_0^2$ in $\eta_T$)\\
$c_I$ for $|W|\le V_*$ & $2^{2/3}$ & $2^{2/3}$ & Theorem~\ref{thm:4.7}\\
\hline
\end{tabular}
\end{table}

\paragraph{Lemma~\ref{lem:4.density}(c) and Theorem~\ref{thm:4.42}: the reach.} In Lemma~\ref{lem:4.density}(c), $\lambda^{-1}+\lambda_U^{-1}=L_c/12+\ell/(2c_s)\le\ell/c_s$ because $c_s\le10<6\ell/L_c$, and $w\ge120\ell/c_s$ gives the factor $30\ell/(c_sw)\le\frac14$. In Theorem~\ref{thm:4.42}, $w'\le121\ell/c_s+2L_c\le122\ell/c_s$, since $2L_c\le\ell/c_s$, and for $\rho\ge4L_c$ we have $g(\rho)\le54\rho^2+54c_s\rho^3/\ell$. Therefore $g((J+1)w')\le10^8(J+1)^3(\ell/c_s)^2$, and $J$ is at most the least integer with $5^J>10^8(J+1)^3(\ell/c_s)^2$, that is, $J=2\log_5(\ell/c_s)+O(\log\log(\ell/c_s))$. Hence
\[
R_*=(J+1)w'+6L_c\le122(J+1)\ell/c_s+6L_c=O\bigl((\ell/c_s)\log(\ell/c_s)\bigr);
\]
with $\log\ell=\pi^2\vCb(0)\,\tau+O(\log\tau)$ this is $R_*\approx380\,\tau\,\ell/c_s$, and $\log R_*=\log(1/\tilde m)+O(\log\tau)$, the only property of $R_*$ that the threshold $K_*$ uses, in Lemma~\ref{lem:5.16}. In the clearing-out step, each site of $\XX^1$ weighs at least $\lambda_U$ in $f$, because $\lambda\ge\lambda_U$. Further, $|B_{w'}(x)|\ge\ell^3$ since $w'\ge\ell$ and each side of $\Lamo$ is at least $\ell$, and $\lambda_U\ell^3=2c_s\ell^2>1$.

\paragraph{Proposition~\ref{prop:4.39}: the area per part.} The competitor costs at most $(6+\lambda)N(P)$ for the marks, $24.6L_c^2$ for each of at most $27N(P)/(\vartheta L_c^2)$ forced blocks, and $2$ for each of at most $C_{\rm seal}N(P)=726N(P)$ seal bonds. We therefore set
\[
C_1:=7+27\cdot24.6/\vartheta=7+664.2/\vartheta=2.214\cdot10^7,\qquad\Cst:=C_1+2C_{\rm seal}=C_1+1452=2.2141\cdot10^7
\]
at $\vartheta=3\cdot10^{-5}$, and $\Cfree:=\Cst(2+1/(2c_s)+1/3696)=5.536\cdot10^7$ at $c_s=1$, where $1/(\lambda\ell)=L_c/(12\ell)\le1/3696$ by \eqref{eq:3.Hell}. The marks of a set $3Q$ are within distance $3(2L_c-1)<6L_c$ of each other, and a bond of $\bar S^1$ has both ends within $4L_c-3$ of a mark of its two blocks, so $R_*+4L_c<r'=2R_*+2$.

\paragraph{Lemma~\ref{lem:4.21}.} The comparison base is $\XX^1\cup R_t$ rather than the base $B_t=\MM\cup\Uf^1\cup R_t$. A bond from a block of $\Uf^1$ to hull material of $\XX^1$ lies in $\partial\Uf^1$ but not in $\mathrm{ins}^1=\partial\XX^1$, and the marks do not bound such bonds, so only the comparison with $\XX^1\cup R_t$ gives a bound by $\mathrm{ins}^1$.

\subsection{Constants of Section~\ref{sec:5}}\label{S:const5}

\paragraph{Theorem~\ref{thm:5.1}.} The constants of \eqref{eq:5.2} are
\[
c_T'':=1+\log(64\kpl),\qquad \kpl:=12e,\qquad C_3:=\Cfree+20(2C_\star'+4\Cresp+1)\quad(\text{Proposition~\ref{prop:5.36}}),
\]
with $\Cst$, $\Cfree$ and $\Cresp$ as in item~10 of Section~\ref{ssec:D.1}. The quantity $\eps''$ is defined in \eqref{eq:5.epspp}, with $R'=R_*+2\Rresp=R_*+20\ell$ and $r'=2R_*+2$. The conditions used in the proof are those of item~(a) of Section~\ref{ssec:D.5}, which hold for large $\tau$ because $|\log\eps''|\ge\gamma_T(K)\tau-O(\log\tau+\log r_0)$ with $\gamma_T(K)>0$ for $K>K_*$ (Lemma~\ref{lem:5.16}), the constants $C_3$ and $C_{\Bfr}$ are free of $\beta$ and $K$, the quantities $\epsD$, $M_m$ and $\RNF^{-1}$ are doubly exponentially small in $\tau$, and $\log d_0=O(\log\tau)$. The constants $\Cresp$, $C_{\Bfr}$ and $C_3$ are fixed before $K$, and hence they affect $\beta_0(K)$ but not $K_*$. As for the order of $\etaT$, we have $\Cfree=O(C_\star'(1+1/c_s))$, hence $C_3=O((C_\star'+\Cresp)(1+1/c_s))$, and Lemma~\ref{lem:2.6} shows that $\gamma_T(K)$ is of the order of $\log^2K$ as $K\to\infty$.

\paragraph{Lemma~\ref{lem:5.16}.} Since $R_*\ge w'\ge12\ell$ by Theorem~\ref{thm:4.42}, we have $2R_*+1\le3R_*$ and $2r'+1=4R_*+5\le5R_*$. With the bound $q_m\le24r_0^3\delta^{1/2}$ of Lemma~\ref{lem:5.11} we obtain
\begin{align*}
\log\epsc&\le\log(8e)+3\log3+3\log5+\log24+6\log R_*+3\log r_0-\tfrac12|\log\delta|\\
&\le14.4+6\log R_*+3\log r_0-\tfrac12|\log\delta| .
\end{align*}
By Theorem~\ref{thm:4.42}, $R_*=O((\ell/c_s)\log(\ell/c_s))$ (Section~\ref{S:C1}), and $\ell\le\ell_D=O(\mt^{-1})$ by Definition~\ref{def:3.1}. With $\log(1/\mt)=\pi^2\vCb(0)\,\tau+O(\log\tau)$ and $\pi^2\vCb(0)<2.4943551$, these bounds and the bound on the entropy exponent in Section~\ref{ssec:D.3} give $6\log R_*\le14.9662\,\tau+O(\log\tau)$. Also $\tfrac12|\log\delta|=c_\delta(K)\beta/2=\tau c_\delta(K)/(2K)$. This is the first inequality of \eqref{eq:5.9}. By Lemma~\ref{lem:2.6}, $c_\delta(K)\ge c_\delta^{\mathrm{sp}}(K)-1/(4a_*)-O(\log\beta/\beta)$. The function $\gamma_T(K):=(c_\delta^{\mathrm{sp}}(K)-1/(4a_*))/(2K)-14.9662$ is increasing in $K$, and its zero is given, together with the margins, in Section~\ref{ssec:D.3}. Hence $\gamma_T(K)>0$ for $K>K_*$. The $O(\log\beta/\beta)$ term contributes $O(\log\tau)$ after multiplication by $\tau/(2K)$. For $\eps''$ we use $\Rresp=10\ell$ from Lemma~\ref{lem:5.32}, $L_c\le\ell$ and $R_*=O(\ell\tau)$. Hence $\log(2R'+4L_c+1)\le\log(1/\mt)+O(\log\tau)$, and the same computation applies.

\paragraph{Lemma~\ref{lem:5.25}.} In the proof of (c), with $\rho=\rho(s)$, we have $|x_c-x_Q|_\infty\le\lceil\rho\rceil/2\le\rho/2+1$. In (ii), the ends of the other cut bond are at distance more than $73\rho-1$ from $x_c$ by (U2), hence more than $73\rho-\rho/2-2>9\rho$ from $x_Q$. In (iii), two bonds of $B_\infty(x_c,10\rho)$ have ends within coarse distance $20\rho$, and $\sqrt3\rbar(20\rho+1)<128\rbar\rho=s$, so (U1) applies. For the last claim of (c), the two chosen bonds have ends within coarse distance $\rho+2$, and $\sqrt3\rbar(\rho+3)\le s$. The range $27L_c\le\rho\le\mathrm{side}_{\min}/24$ holds. The lower bound follows from $\RNF\ge3456\rbar L_c$. For the upper bound, $\varphi_s\ge2$ forces $s$ below the lattice diameter of $\Lc$, which is at most $\sqrt3\rbar\,\mathrm{side}_{\max}\le2\sqrt3\rbar\,\mathrm{side}_{\min}$, the last inequality by Lemma~\ref{lem:3.3}(d), and hence $\rho<2\sqrt3\,\mathrm{side}_{\min}/128$.

\paragraph{Lemmas~\ref{lem:5.31} and~\ref{lem:5.32}.} In the proof of Lemma~\ref{lem:5.31}, summing $121(m_q+m_{q'})$ over the six faces of each block $q$ of $\Delta_0$, we count at most $121(6m(\Delta_0)+6m(U))\le1452\,m$ seal bonds with an end in $\Delta_0$ and, summing over the blocks of $U$, at most $121(6m(U)+6m(U_2(\Delta_0)))\le1452\,m$ seal bonds with both ends in $U$. With $u_3\le c_s^{-1}(i+2904m+2n)$ we obtain
\[
\mathrm{units}(\Delta_0)\le1452m+2u_3+i+2\cdot1452m+2n+6m\le\Bigl(4362+\frac{5808}{c_s}\Bigr)m+\Bigl(2+\frac4{c_s}\Bigr)n+\Bigl(1+\frac2{c_s}\Bigr)i,
\]
since $1452+2904+6=4362$ and $2\cdot2904=5808$. At $c_s=1$, $(C_a,C_b,C_c)=(10170,6,3)$, and the loads of Lemma~\ref{lem:5.32} are $125C_a=1.27\cdot10^6$, $125C_b=750$ and $125C_c=375$. Thus
\[
\Cresp:=375C_a+125C_c\Cst=3.81\cdot10^6+375\cdot2.2141\cdot10^7=8.307\cdot10^9 ;
\]
the term $125C_c\Cst$, the load of the Phase-1 insulation owned by a part, dominates. The range is $\Rresp=10\ell\ge9\ell+1$.

\paragraph{Proposition~\ref{prop:5.34}.} A touched $Y_\Delta$ costs at most $2|\partial\Delta|+2|\mathrm{ins}^1\cap E(\Delta)|+4.5|\partial\Delta|=6.5|\partial\Delta|+2|\mathrm{ins}^1\cap E(\Delta)|$. A touched $Y=Y_{\Delta'}$ costs $2|\partial Y|+|Y|/\ell$, which by the bound on $|\partial Y|$ is at most $4\,\#\{\text{bonds of types (a) to (c) with an end in }\Delta'\}+13|Y|/\ell$, and $13|Y|/\ell\le(26L_c/\ell)m(\Delta')$. Since a bond has two ends, it is charged by at most two cells of the third kind, at most $4$ each. A unit $(\Delta,b)$ is charged at most $6.5$ by $Y_\Delta$ and at most $4$ by the cell of the other end of $b$. A bond of $\mathrm{ins}^1$ inside a removed cell is charged at most $2$ by $Y_\Delta$. All coefficients are at most $20$.

\paragraph{Lemma~\ref{lem:5.35} and Proposition~\ref{prop:5.36}.} The constants are
\begin{gather*}
C_{\Bfr}:=30C_\star'+170\Cresp+30,\qquad M_m:=16\epsD(1+8L_c)(C_\star'+3\Cresp+6),\\
C_3:=\Cfree+20(2C_\star'+4\Cresp+1).
\end{gather*}

To obtain $C_{\Bfr}$, we consider the proof of Lemma~\ref{lem:5.35}(c). There the bonds of $\mathrm{ins}^1$ number at most $C_\star'N_{\mathcal Q}$, as in (b), the cut bonds number at most $A_{\mathcal F}$, and the units number at most $2\Cresp(N_{\mathcal Q}+A_{\mathcal F})$ by (a), each of these being charged by at most two blocks. The marks involved, at most $N_{\mathcal Q}$, belong to parts of $\Qatt$, each charged by at most seven blocks. Summing, we obtain
\[
|\Bfr_{\mathcal F}|\le2\bigl(C_\star'N_{\mathcal Q}+A_{\mathcal F}+2\Cresp(N_{\mathcal Q}+A_{\mathcal F})\bigr)+7N_{\mathcal Q}\le(2C_\star'+4\Cresp+10)N_{\mathcal Q}
\]
by Lemma~\ref{lem:5.35}(b), and $2C_\star'+4\Cresp+10\le C_{\Bfr}$.

For $M_m$, we have $m\le2A_{\mathcal F}\le16\epsD(1+8L_c)(C_\star'+3\Cresp+6)N_{\mathcal Q}=M_mN_{\mathcal Q}$ by Lemma~\ref{lem:5.35}(d) and~(b).

\emph{The size of $\Cresp$.} We have $\Cresp\ge125C_c\Cst\ge125\Cst$, in particular $\Cresp\ge C_\star'+20$.

For $C_3$, we note that $20(u_{\mathrm{touch}}+c_{\mathrm{touch}}+i_{\mathrm{touch}})\le20(\Cresp+C_\star')N_{\mathcal Q}+20(\Cresp+1)A_{\mathcal F}$ in the proof of Proposition~\ref{prop:5.36}, and that $26L_c/\ell\le26/308<1/11$. With $A_{\mathcal F}\le N_{\mathcal Q}/(48\Cresp+2)$, all terms with $A_{\mathcal F}$ together are at most $N_{\mathcal Q}$, and
\[
\#(\Xi'\cap\Ffr)/\rbar^{2}\le\bigl(\Cfree+20C_\star'+20\Cresp+2\bigr)N_{\mathcal Q}\le C_3N_{\mathcal Q}.
\]

\emph{The binding condition.} The first condition of \eqref{eq:5.17} reads $|\log\eps''|\ge3C_{\Bfr}\ln(2^{30}\cdot26e)=75.16\,C_{\Bfr}$. At the illustrative values $(c_g,c_s,r_0)=(1/4,1,1)$, the constants of item~10 are $\Cst=2.2141\cdot10^7$, $\Cfree=5.536\cdot10^7$ and $\Cresp=8.307\cdot10^9$, hence $C_3=6.655\cdot10^{11}$ and $C_{\Bfr}=1.413\cdot10^{12}$, and this condition becomes $|\log\eps''|\ge1.06\cdot10^{14}$. The condition $\etaT\le1$ needs $|\log\eps''|\ge3c_T''\rbar^{2}C_3=1.73\cdot10^{13}\rbar^{2}$, with $c_T''=8.644$, and $\etaT^{(1)}\le1$ needs $|\log\epsc|\ge c_T\rbar^{2}\Cfree=3.63\cdot10^8\rbar^{2}$. Each of these is a lower bound on $\tau$, by Lemma~\ref{lem:5.16}.

\subsection{Constants of Section~\ref{sec:stage1} and Appendix~\ref{app:F}}\label{S:E1}

\paragraph{Masses (Section~\ref{sec:stage1}).}\label{S:E1.mass}
The Coulomb kernel at the origin is
\[
\vCb(0)=\tfrac12W_{S}=(\sqrt6/(192\pi^{3}))\Gamma(\tfrac1{24})\Gamma(\tfrac5{24})\Gamma(\tfrac7{24})\Gamma(\tfrac{11}{24})=0.2527310098\ldots,
\]
where $W_{S}=\pi^{-3}\int_{[0,\pi]^{3}}(3-\sum_{i}\cos x_{i})^{-1}dx$ is Watson's integral \cite{Wat39}, in the Gamma-function form of \cite[(3.6)]{Zu11}. Numerically, $\pi^{2}\vCb(0)=2.4943550871\ldots$. Since
$\mD=(8\pi^{2}\tau)^{1/2}e^{-\pi^{2}\vCb(0)\tau}$ and $2(8\pi^{2})^{1/2}=17.77153\ldots$, it follows that
\begin{equation}\label{S:E1.mt}
\mt\le2\mD\le17.7716\,\tau^{1/2}e^{-2.494\tau}\qquad(\mD/2\le\mt\le2\mD).
\end{equation}
Here $2.494$ is $\pi^{2}\vCb(0)$ rounded down to a lower bound by the rule of Section~\ref{ssec:D.4}, and the hypothesis of Lemma~\ref{lem:6.19} has the same $2.494$ on its left.

\paragraph{Theorem~\ref{thm:6.5}: the Debye condition.}\label{S:E1.debye}
In \cite[(4.24)]{GM82}, at $A=(1-\delta)M$, \cite{GM82} use $\|v^{2}\|_{A}\le2(\delta M)^{-2}$ \cite[Lemma~5, (3.26)]{GM82}, and the Debye condition \cite[p.~34]{GM82} reads $1\le\frac14\delta^{2}C^{3}(1-C)^{8}e^{-2\kappa'}M^{2}/\mD^{2}$, where $\kappa'=\kappa_{S}+\frac12\mu$ by the choice made on \cite[p.~35]{GM82}. On $\Gp$ the bound $\|v^{2}_{\Gp}\|_{A}\le3\MY^{-2}$ holds by Proposition~\ref{prop:6.11} on outputs for $A\le\sqrt3\,\kappa_{Y}\MY$, and by $\|v^{2}_{\Gp}\|_{0}\le\MY^{-2}$ on all slit graphs. It replaces $2(\delta M)^{-2}$, that is, $\frac12\delta^{2}$ by $\frac13$, and hence $\frac14\delta^{2}$ becomes $\frac16$:
\[
1\le\tfrac16C^{3}(1-C)^{8}e^{-2\kappa_{S}-\mu}\MY^{2}/\mD^{2}.
\]
Since $\MY\ge\mD/(2\lPV)$ on $I_{Y}$, $\MY^{2}/\mD^{2}\ge1/(4\lPV^{2})$, and the condition holds for all $\MY\in I_{Y}$ once
\begin{equation}\label{S:E1.lPV}
\lPV^{2}\le C^{3}(1-C)^{8}e^{-2\kappa_{S}-\mu}/24 .
\end{equation}
In \cite[(4.23)]{GM82} the bound $\|v^{1}_{\Gp}\|_{A}\le3M_{1}^{-2}$, with $M_{1}=-4\pi\bGM^{-1}\ln(1-C)$, is a power of $\bGM$, which $e^{\bGM\eps_{1}}$ absorbs for $\bGM\ge\bGMo$. Neither condition depends on $\Gp$.

\paragraph{The scales of \S\ref{ssec:6.3}: the rounding of $L$.}\label{S:E1.scales}
By Definition~\ref{def:3.1}, $L=r_{0}L_{c}$ with $L_{c}$ the largest even integer $\le c_{L}\mt^{-1}/r_{0}$, and $c_{L}\mt^{-1}\ge4r_{0}$. Hence
\begin{equation}\label{S:E1.L}
c_{L}/(2\mt)\le L\le c_{L}/\mt .
\end{equation}
In fine units $r_{0}\rho_{\lambda}=L/12=c_{\rho}L$, where $c_{\rho}=1/12$, while $c_{I}=2^{2/3}$, $V_{*}=(1-\etaw)^{3}\rho_{\lambda}^{3}/2$ in coarse units and $V_{*}^{f}=r_{0}^{3}V_{*}/2$, so $\RHK=(1-\etaw)c_{I}r_{0}\rho_{\lambda}/3$ and $4\RHK^{3}/V_{*}^{f}=64/27$ exactly. The constant
\[
K_{0}:=(1-\etaw)c_{I}c_{\rho}c_{L}/6\qquad(=0.2381\,c_{\rho}c_{L}=0.01984\,c_{L}\ \text{at}\ \etaw=0.1)
\]
satisfies $K_{0}\le\mt\RHK\le2K_{0}$ by \eqref{S:E1.L}. We use $K_{0}$ only through the lower bound $\mt\RHK\ge K_{0}$, in the following places.
\begin{itemize}
\item[(a)] In Proposition~\ref{prop:6.11}, $(\eps_{1}^{\mathrm{HK}})^{1/2}\MY\RHK\ge(\eps_{1}^{\mathrm{HK}})^{1/2}K_{0}/\lPV$, so the rate of \eqref{eq:6.LW} at $M=\MY$ is at least $\kappa_{Y}\MY$.
\item[(b)] In (H2) of Theorem~\ref{thm:6.12}, at the mass $r_{-}^{1/2}\ge\mt$ the rate of \eqref{eq:6.LW} is at least $\kappa_{W}\mt\min(1,(\eps_{1}^{\mathrm{HK}})^{1/2}\mt\RHK)\ge\kappa_{0}\mt$, using $(\eps_{1}^{\mathrm{HK}})^{1/2}K_{0}<1$.
\item[(c)] In (H1) and (H1$'$) of Theorem~\ref{thm:6.12}, $\RHK^{-3}r_{-}^{-1}\le\mt^{-2}\RHK^{-3}\le\mt K_{0}^{-3}=\lPV K_{0}^{-3}\MY$.
\item[(d)] In Lemma~\ref{lem:6.15}, $\RHK^{-1}+\MY^{-2}\RHK^{-3}\le\mt K_{0}^{-1}+\lPV^{2}\mt K_{0}^{-3}=\eta_{G}$.
\item[(e)] In Appendix~\ref{app:F}, $\mt\le K_{0}(2C_{H})^{-1/3}$ gives $\RHK\ge(2C_{H})^{1/3}$.
\end{itemize}
By \eqref{S:E1.L}, $\MY r_{0}\rho_{\lambda}=c_{\rho}\MY L\in[c_{\rho}c_{L}/(2\lPV),\,c_{\rho}c_{L}/\lPV]$. We will use the lower bound in (f) to (h) below.

\paragraph{Theorem~\ref{thm:6.9}.}\label{S:E1.HK}
We set $c_{N}:=c_{F}^{2}/(24\cdot2^{4/3})=1.006\cdot10^{-7}$, $C_{H}:=(c_{N}/3)^{-3/2}=1.628\cdot10^{11}$ and $\eps_{1}^{\mathrm{HK}}:=(\log2)^{2}/512=9.38\cdot10^{-4}$. In the proof, given in Appendix~\ref{app:F}, the constant of Lemma~\ref{lem:F.3} is $A_{N}:=24\cdot2^{2/3}/c_{F}^{2}$, so that $c_{N}=1/(2^{2/3}A_{N})$, and (HK0) uses $6(\cosh\frac14-1)=0.1885<\frac14-\frac1{17}=0.1912$. The constants $C_{\mathrm{mo}}\ge1$ of the moment bound, $c_{\mathrm{ex}}:=\min((\log2)/(2A_{1}),\frac12)$ with $A_{1}:=512C_{\mathrm{mo}}^{2}/(\log2)^{2}$ (so $\eps_{1}^{\mathrm{HK}}=C_{\mathrm{mo}}^{2}/A_{1}$), $C_{\mathrm{gs}}:=8C_{H}e^{c_{\mathrm{ex}}/2}$, $c_{\mathrm{gs}}:=c_{\mathrm{ex}}/8$ and $C_{M}$, which contains the factor $e^{1/(2c_{\mathrm{ex}})}$, depend only on $C_{H}$. Hence $\kappa_{W}=\min(1,1/(2C_{M}))$ of Proposition~\ref{prop:6.11} is extremely small and fixed before $\bGM$; what matters is that it is a constant independent of $\tau$, $K$, $C$ and $D$.

\paragraph{Proposition~\ref{prop:6.10}.}\label{S:E1.ctf}
The box inequalities of the sketch in \S\ref{ssec:F.1} are as follows. For a box graph $B$ with sides in $[n_{\min},n_{\max}]$, $\varrho:=n_{\max}/n_{\min}$, and $U\subseteq B$, canonical paths give $|\partial_{B}U|\ge4|U||B\setminus U|/(n_{\max}|B|)\ge2\min(|U|,|B\setminus U|)/n_{\max}$, and the Loomis\textendash Whitney inequality gives $|\partial_{B}U|\ge3|U|^{2/3}-3|U|/n_{\min}$. Hence $|\partial_{B}U|\ge c_{\mathrm{box}}(\varrho)|U|^{2/3}$ for $|U|\le|B|/2$, with $c_{\mathrm{box}}(\varrho):=6/(3\varrho+2)$, as one sees by minimizing over $|U|^{1/3}=t\,n_{\min}$. In the face count, for a heavy cell $c$ and a light neighbour $c'$, the set $D:=B_{c}\cup B_{c'}$ has $n_{\max}(D)<4r_{0}$ and $|U||D\setminus U|/|D|\ge r_{0}^{3}/4$ for $U:=A\cap D$, so $D$ carries at least $\gamma_{D}r_{0}^{2}$ bonds of $\partial A$, $\gamma_{D}=\frac14$. Summing over $\partial_{\Gamma}H$, where $\Gamma=(V_{c},E_{c})$ is the coarse slit graph of \S\ref{ssec:6.3}, counts the bonds inside a coarse cell at most six times. Hence $|\partial A|\ge(\gamma_{D}r_{0}^{2}/6)|\partial_{\Gamma}H|$. With the constant $a:=9c_{I}/38$ of (R2), and using $c_{\mathrm{box}}(\varrho)>\frac34$ for $\varrho<2$ and $\gamma_{D}=\frac14$, the two cases of (i) give $|\partial A|\ge2^{-2/3}c_{\mathrm{box}}|A|^{2/3}\ge0.47|A|^{2/3}$ and $|\partial A|\ge(\gamma_{D}a/6)16^{-2/3}|A|^{2/3}=(a/152.39\ldots)|A|^{2/3}$, so $c_{F}:=a/152.4=2.467\cdot10^{-3}$. The constant $C_{\mathrm{bl}}:=384$ comes from $\gamma_{D}r_{0}^{3}/(12v_{\max})\ge1/384$ with $v_{\max}<8r_{0}^{3}$. Further, $C_{W}^{f}:=108C_{\mathrm{bl}}^{2}/\chi^{2}$, $k_{\square}:=r_{0}(k_{P}+3)$, $h^{f}:=\chi/(3C_{\mathrm{bl}}k_{\square})$, $12/(h^{f})^{2}=C_{W}^{f}k_{\square}^{2}$ and $N_{\mathrm{ov}}\le27(k_{P}/L_{c})^{3}$. For (v), (L8-2) reads $C_{W}^{f}r_{\square}^{2}N_{\mathrm{ov}}(\mt L)^{2}\le\eta_{H}/8$ and $C_{W}^{f}r_{\square}^{2}(\mt L)^{2}=12\mt^{2}/(h^{f})^{2}$, so $1/h^{f}\le96^{-1/2}\mt^{-1}\le0.103\,\mt^{-1}$, since $\eta_{H}<1\le N_{\mathrm{ov}}$. Moreover $\log k_{\square}\le\log(1/\mt)\le2.4944\,\tau=0.0632\,\bGM$. Also $|\KK^{\flat}|\le|V\cap\hat Q(Q)|\le k_{\square}^{3}$: a window has fewer than $k_P$ coarse sites per direction (G4), all of fine side $r_0$ except at most one exceptional layer per special interval (Section~\ref{S:B2.grid}), of fine side at most $2r_0-1$, and it meets at most three special intervals per direction, so its fine sides are at most $r_0k_P+3(r_0-1)<k_{\square}$. Hence the diameter bound $2+(24/h^{f})\cdot3\log k_{\square}\le2+24\cdot0.103\cdot3\cdot0.0632\,\bGM/\mt<2\bGM/\mt=L'$ holds.

\paragraph{Theorem~\ref{thm:6.12}.}\label{S:E1.cov}
Let $\varpi:=(1-4\lPV^{2})^{-1/2}\in[1,1.16]$ and $r_{\pm}:=\frac12\MY^{2}(1\pm(1-4\lPV^{2})^{1/2})$, so that $r_{-}=2\mt^{2}/(1+(1-4\lPV^{2})^{1/2})\in[\mt^{2},1.08\,\mt^{2}]$ for $\lPV\le\frac14$. The constants, which depend only on $(c_{F},c_{\rho},c_{L},\lPV)$, are
\begin{align*}
K_{1}&:=\tfrac12+2C_{H}(1+\varpi)+\varpi C_{H}\lPV K_{0}^{-3},\qquad K_{3}:=3\varpi\mt^{2}/r_{-}\le3.5,\\
K_{1}'&:=\max\bigl\{eK_{1},\ \varpi\bigl(313+C_{\mathrm{gs}}(2\pi/c_{\mathrm{gs}})^{1/2}+4C_{H}(\eps_{1}^{\mathrm{HK}})^{-3/2}\lPV K_{0}^{-3}\bigr)\bigr\},\\
K_{1}''&:=\max(eK_{1},K_{1}').
\end{align*}
The bound $de^{-3d/68}\le\MY d^{2}e^{-3d/68}\le(136/3)^{2}e^{-2}\MY<279\,\MY$, valid for $d\ge\MY^{-1}$, together with $279\le313$, accounts for the number $313$.

\paragraph{Section~\ref{ssec:6.4}: the depletion constants.}\label{S:E1.dep}
Both $A_{0}^{\mathrm{AD}}$ and $C_{F}:=C_{\mathrm{Gr}}C_{L}'$ are upper-bound constants, taken $\ge1$; we also take $C_{\mathrm{Gr}}\ge1$ and $C_{L}'\ge C_{\rm im}$, the constant of the far images in step (iv) of the proof of Theorem~\ref{thm:6.17}, so that $C_{F}\ge C_{\rm im}$. Let $\eps_{0}:=\theta_{0}/(2\pi^{2}\tau)$, $\eps_{F}:=\eps_{0}/(4C_{F})$,
\[
L_{g}:=3+\log_{2}^{+}\bigl(8\lPV/(c_{\rho}c_{L})\bigr),\qquad C_{R}:=256\pi^{2}A_{0}^{\mathrm{AD}}C_{F}L_{g},\qquad C_{S}:=CA_{0}^{\mathrm{AD}}\bigl(1+\lPV/(c_{\rho}c_{L})\bigr),
\]
so that $32A_{0}^{\mathrm{AD}}L_{g}/\eps_{F}=C_{R}\tau/\theta_{0}$ and $\Rdep=\lceil C_{R}\tau/\theta_{0}\rceil$. Moreover, $2/\eps_{F}=16\pi^{2}C_{F}\tau/\theta_{0}$ and $\Rdep\le2C_{R}\tau/\theta_{0}$. Since $A_{0}^{\mathrm{AD}}\ge1$ and $L_{g}\ge3$, $C_{R}\ge768\pi^{2}C_{F}\ge24\pi^{2}C_{\rm im}$.
\begin{itemize}
\item[(f)] In Theorem~\ref{thm:6.16}, the shells $2^{k}\le r(y)<2^{k+1}$ with $2^{k+1}\le r_{0}\rho_{\lambda}$ contribute at most $8A_{0}^{\mathrm{AD}}$ each to $S_{2}(x)$ by Lemma~\ref{lem:6.14}, and hence at most $CA_{0}^{\mathrm{AD}}\log(2/\MY)$ up to $r\le\MY^{-1}$. The shells beyond $\MY^{-1}$ are summed with $e^{-\MY r/2}$, and the growth term gives $\sum_{k}(16A_{0}^{\mathrm{AD}}/(r_{0}\rho_{\lambda}))2^{k}e^{-\MY2^{k}/2}\le CA_{0}^{\mathrm{AD}}/(\MY r_{0}\rho_{\lambda})\le2CA_{0}^{\mathrm{AD}}\lPV/(c_{\rho}c_{L})$. Since $\log(2/\MY)\ge1$, $S_{2}(x)\le C_{S}\log(2/\MY)$.
\item[(g)] In step (ii) of the proof of Theorem~\ref{thm:6.17}, with $R\le r_{0}\rho_{\lambda}$, the far pairs contribute at most $16A_{0}^{\mathrm{AD}}/R+(16A_{0}^{\mathrm{AD}}/(r_{0}\rho_{\lambda}))(2+\log_{2}^{+}(4/(\MY r_{0}\rho_{\lambda})))\le(16A_{0}^{\mathrm{AD}}/R)(3+\log_{2}^{+}(8\lPV/(c_{\rho}c_{L})))=16A_{0}^{\mathrm{AD}}L_{g}/R$, which is at most $\eps_{F}/2$ for $R=\Rdep$. In the second sum, the shells with $2^{k+1}R>r_{0}\rho_{\lambda}$ give $\sum_{j\ge0}e^{-a2^{j}}$ with $a>\MY r_{0}\rho_{\lambda}/4$, which is at most $2+\log_{2}^{+}(1/a)$.
\item[(h)] For Lemma~\ref{lem:6.19}, see the proof below.
\end{itemize}
As a function of $\Rdep$, the covering number $N_{\mathrm{cov}}$ of Proposition~\ref{prop:6.18} takes the same value for all $\Rdep\le\ell_{D}$, so it does not depend on $\theta_{0}$. It is a very large constant, and it raises $\bGMo$ through a single condition, that of Theorem~\ref{thm:7.42}.

\paragraph{Proof of Lemma~\ref{lem:6.19}.}\label{S:E1.619}
The hypothesis says $C_{*}^{\mathrm{dep}}\tau^{5/2}e^{-2.494\tau}\le\theta_{0}$, and we use the bound \eqref{S:E1.mt} throughout.

We first prove $\Rdep\le r_{0}\rho_{\lambda}$. Since $A_{0}^{\mathrm{AD}},C_{F}\ge1$ and $L_{g}\ge3$, $C_{R}\ge768\pi^{2}>7.5\cdot10^{3}$. Hence $y:=C_{R}\tau/\theta_{0}\ge1.5\cdot10^{6}$ and $\Rdep=\lceil y\rceil\le(1+10^{-6})y$. By \eqref{S:E1.L}, $r_{0}\rho_{\lambda}=c_{\rho}L\ge c_{\rho}c_{L}/(2\mt)$, so it suffices that $2(1+10^{-6})C_{R}\tau\mt\le c_{\rho}c_{L}\theta_{0}$. Using $\tau\ge2$,
\begin{align*}
2(1+10^{-6})C_{R}\tau\mt&\le2(1+10^{-6})\,17.7716\,C_{R}\tau^{3/2}e^{-2.494\tau}\le17.8\,C_{R}\tau^{5/2}e^{-2.494\tau}\\
&\le c_{\rho}c_{L}C_{*}^{\mathrm{dep}}\tau^{5/2}e^{-2.494\tau}\le c_{\rho}c_{L}\theta_{0}.
\end{align*}

Next, $\Rdep\le\ell_{D}$ because $r_{0}\rho_{\lambda}=c_{\rho}L<L\le\ell_{D}$.

Finally, for (F), $\log(2/\MY)=\ln(2\lPV/\mt)\le\ln(1/\mD)=\pi^{2}\vCb(0)\tau-\frac12\ln(8\pi^{2}\tau)\le2.4944\,\tau$, using $\lPV\le\frac14$ and $\mt\ge\mD/2$. With $\eta_{G}=\mt(K_{0}^{-1}+\lPV^{2}K_{0}^{-3})$,
\[
C_{F}\eta_{G}C_{S}\log(2/\MY)\le C_{F}C_{S}(K_{0}^{-1}+\lPV^{2}K_{0}^{-3})\cdot17.7716\,\tau^{1/2}e^{-2.494\tau}\cdot2.4944\,\tau,
\]
and this is at most $\eps_{0}/4=\theta_{0}/(8\pi^{2}\tau)$ because $8\pi^{2}\cdot17.7716\cdot2.4944<3500.2\le3.6\cdot10^{3}$ and $C_{*}^{\mathrm{dep}}\ge3.6\cdot10^{3}C_{F}C_{S}(K_{0}^{-1}+\lPV^{2}K_{0}^{-3})$.
\qed

\subsection{Constants of Subsections~\ref{ssec:7.0} to~\ref{ssec:7.5} and Appendix~\ref{app:C}}\label{S:constF}

\paragraph{Proposition~\ref{prop:7.5}.} Off $\Df$, (S4$'$) gives
$|\rho_1(\pm1,x)-\rho^{\mathrm{ref}}(1)|\le\rho^{\mathrm{ref}}(1)(1-e^{-\theta_0}+Ce^{-\mu})$, and
$\sum_{|m|\ge2}(|\rho_1(m,x)|+|\rho^{\mathrm{ref}}(m)|)m^2\le2Ce^{-2\kappa_S}m_D^2/\bGM$. Since
$2\bGM\rho^{\mathrm{ref}}(1)=\mt^2-\bGM\sum_{|m|\ge2}m^2\rho^{\mathrm{ref}}(m)\le(1+Ce^{-2\kappa_S})\mt^2$
and $m_D\le2\mt$, we obtain, after enlarging $C$,
\[
\sum_m|\rho_1(m,x)-\rho^{\mathrm{ref}}(m)|\,m^2\le\theta'\,\mt^2/\bGM\qquad(x\notin\Df),
\]
with $\theta'$ of \eqref{eq:7.thetaprime}. On $\Df$ we have $|\rho_1(\pm1,x)|\le\rho^{\mathrm{ref}}(1)(1+Ce^{-\mu})$ and, since $e^{-\Phi}\le1$,
$\rho_1(\pm1,x)-\rho^{\mathrm{ref}}(1)\le Ce^{-\mu}\rho^{\mathrm{ref}}(1)$. These bounds, the same tails and
$2\bGM\rho^{\mathrm{ref}}(1)\le2\mt^2$ give the constants of part (b). At $\theta_0=0.01$ we have
$1-e^{-\theta_0}=0.00995\ldots$, and hence the standing hypothesis \eqref{eq:7.standing} holds
once $\varepsilon_\nu+C(e^{-\mu}+e^{-2\kappa_S})\le0.1329$.

\paragraph{The partial fractions \eqref{eq:7.partialfrac} and Lemma~\ref{lem:7.18}.}
If $\lPV\le\frac14$, then $\varpi=(1-4\lPV^2)^{-1/2}\in[1,1.155]$, $c_+\le\varpi$,
$c_-\le1.26\,\lPV^2$, $\mt^2\le r_-\le1.08\,\mt^2$ and $r_+\ge0.932\,\MY^2$.
In part (b), with $\alpha=\mt/3$, the bound $\cosh x-1\le\frac{x^2}2\cosh x\le0.5281\,x^2$
for $x\le\frac13$ gives $6(\cosh\alpha-1)\le0.3521\,\mt^2$. With this $\alpha$, part (a) gives
$\|e^\rho R_{r_-}e^{-\rho}\|\le(r_--0.3521\mt^2)^{-1}\le1.544\,\mt^{-2}$ and
$\|e^\rho R_{r_+}e^{-\rho}\|\le(r_+-0.3521\mt^2)^{-1}\le1.1\,\MY^{-2}=1.1\,\lPV^2\mt^{-2}$, and hence
$\|e^\rho\mathcal Ge^{-\rho}\|\le(1.155\cdot1.544+1.26\cdot1.1\,\lPV^4)\mt^{-2}\le1.80\,\mt^{-2}$.
The matrix $e^\rho\nabla^*e^{-\rho}$ has entries $\pm e^{\rho(x)-\rho(b)}$ with
$|\rho(x)-\rho(b)|\le\frac13$, at most $6$ per row and $2$ per column. Hence its
norm is at most $\sqrt{12}\,e^{1/3}\le4.835$, and $1.80\cdot4.835\le8.71$.
In part (c), $\|[\Delta,\chi]\|\le6\Lambda_\chi$ and $\|\Delta\|\le12$, so
$\|[C_0^{-1},\chi]\|\le\MY^{-2}\cdot2\cdot12\cdot6\,\Lambda_\chi+6\Lambda_\chi=\Lambda_\chi(144\MY^{-2}+6)\le150\,\Lambda_\chi\MY^{-2}$, since $\MY\le1$.

\paragraph{Theorem~\ref{thm:7.19}.} The constant is $C_\varepsilon:=3.8\cdot10^9$. With
$d(A',V_B)\ge L'/2-2$, $\Lambda_\chi=4/L'$, $\MY^{-4}=\lPV^4\mt^{-4}$ and $\mt\le1$,
Lemma~\ref{lem:7.18}\,(b), (c) give
$\|F'_{2,\beta}\|^2\le(150\cdot\frac4{L'}\MY^{-2})^2(8.71\,\mt^{-2})^2e^{4/3}e^{-\mt L'/3}\|w_\beta\|^2
\le1.71\cdot10^6\cdot16\,e^{4/3}\,\mt^2(\mt L')^{-2}\lPV^4\mt^{-8}e^{-\mt L'/3}\|w_\beta\|^2$.
Theorem~\ref{thm:7.14} and $c_g'\ge\frac9{16}c_g$ give
\[
\sum_{\beta:S_\beta\subseteq X}\|w_\beta\|^2\le18\,\Floc(X)/(c_g'\mt^2L/2)\le32\,\Floc(X)/(c_g\mt^2L/2).
\]
The product of the numerical factors is $1.71\cdot10^6\cdot16\cdot32\,e^{4/3}\le3.4\cdot10^9\le C_\varepsilon$.

\paragraph{Corollary~\ref{cor:7.21}.} The Hölder exponents and the constant of
Subsection~\ref{ssec:7.6} are $p_2=20$, $p_3=1.2$, $p'=1.3$, $q'=15.6$ and
$c_\lambda=\frac1{24}$, so
$\varepsilon_0=\min\{(1-\frac1{1.2}-\frac1{24})/40,\,2(1.3/15.6)^2\}=\min\{0.003125,\,0.0139\}=0.003125$.
Dropping $(\mt L')^{-2}\le1$ in \eqref{eq:7.epsL}, we obtain $\varepsilon(L')\le\varepsilon_0$
as soon as
\[
\mt L'\ge27\log(1/\mt)+3\log\bigl(C_\varepsilon\lPV^4/(c_g\varepsilon_0\,\mt L/2)\bigr).
\]
The second term is bounded, since $c_L/2\le\mt L\le c_L$ by Definition~\ref{def:3.1}.
For the first, Definition~\ref{def:3.1} and Section~\ref{S:G.35} give $\log(1/\mt)\le2.4944\,\tau=(2.4944/4\pi^2)\bGM$
for $\tau\ge\tau_0$, so the right side is at most $1.706\,\bGM+O(1)$, while
$\mt L'=\Kcol\bGM=2\bGM$.

\paragraph{Proposition~\ref{prop:7.22}.} The constant of part (c) is
\[
\Phi_\UU:=\frac{\pi c_gc_u}{24}\,\frac{(\mt L)^3}{(\mt\bGM)(\mt L')^3(11/10)^3}.
\]
Dividing (a) by (b) gives the share $\frac{\pi c_g'c_u}{12}(\mt L)^3/((\mt\bGM)(\mt L')^3(1+\eta_{\mathrm{col}})^3)$,
which is at least $\Phi_\UU$ because $c_g'\ge\frac9{16}c_g\ge\frac12c_g$ and
$\eta_{\mathrm{col}}\le\frac1{10}$ for $\bGM\ge\bGMo$. The latter holds since $L$ and $D_E$ are $O(\mt^{-1})$,
while $L'=2\bGM\mt^{-1}$. With $\mt L\ge c_L/2$ and $\mt L'=2\bGM$,
$\Phi_\UU\ge\pi c_gc_uc_L^3/(24\cdot64\cdot1.331\,\mt\bGM^4)$, and
$\mt\bGM^4\to0$ because $\log(1/\mt)=\pi^2\vCb(0)\tau+O(\log\tau)$ and $\bGM=4\pi^2\tau$.

\paragraph{Structural facts of the expansion (Appendix~\ref{app:C}).} The bounds on
the activities in Subsections~\ref{ssec:7.6} and~\ref{ssec:7.7} and in Appendices~\ref{app:A}
and~\ref{app:B} use the expansion of Proposition~\ref{prop:7.25} only through
the following facts S1 to S14, and none of them uses translation invariance, an
infinite lattice, or the shape of the atoms. Each fact is followed by the place where it is
proved; ``change~$n$'' is item~($n$) of the changes in the proof of Proposition~\ref{prop:7.25}
in Appendix~\ref{app:C}. These facts are not the properties (S1) to (S5) of Theorem~\ref{thm:6.5}.
\begin{itemize}
\item[S1] $\bC_s(x,x)=\bC(x,x)$ and $|\bC_s(x,y)|\le|\bC(x,y)|$; the weighted norms and
$\|\bC_s\|_{2\to2}$ are bounded by those of $\bC$ (change~6).
\item[S2] $\bC_s=\sum_\varepsilon\lambda_\varepsilon\bC_\varepsilon$, where each $\bC_\varepsilon$ is
block diagonal with admissible blocks (change~6).
\item[S3] $\log\int d\mu_\bC\,e^{\frac12\langle\psi+D,M(\psi+D)\rangle}$ is convex in $\bC$ for
$M\ge0$ (\cite[(9.67) to (9.611)]{BF80}; Lemma~\ref{lem:7.35}, Step~3).
\item[S4] The integrand of $\mathcal K_{\bar y}$ depends on $\psi|_{X(\bar y)}$ only, and covariance
derivatives have both ends in $X(\bar y)$ (Proposition~\ref{prop:7.25}).
\item[S5] Each $Y_{i+1}$ is a union of elements, each meeting $S(\theta_i)\setminus X_i$, the part of
the support of its link outside $X_i$ (selection rule \eqref{eq:C.E20}).
\item[S6] Link $i$ carries $\prod_{k=T(i+1)}^{i-1}s_k$ (covariance link from $Y_{T(i+1)}$) or
$\prod_{k=j_{\min}}^{i-1}s_k$ ($E$-term link); all other $s$-dependence is through weights in
$[0,1]$ and through $\bC_s$ (\eqref{eq:C.E3}, \eqref{eq:C.E16}).
\item[S7] There are five types of terms per link (\eqref{eq:C.E16}).
\item[S8] The link ends $x\in Y_{i+1}$ and $y\in Y_{T(i+1)}$ are counted in $\UU$-cells (selection
rule \eqref{eq:C.E20}).
\item[S9] The sequences consist of disjoint admissible sets, and there are finitely many of them
(change~2).
\item[S10] $Z(Y)>0$, and the one-set activities are positive (Proposition~\ref{prop:7.25}).
\item[S11] The first set may depend on $h$ (change~3).
\item[S12] The first set may be disconnected (change~3).
\item[S13] The expansion is on $U=X_0^c$ with $\one_U\bC\one_U$, and the activities do not depend
on $U$ (change~4).
\item[S14] Every term converges absolutely (Lemmas~\ref{lem:C.4} and~\ref{lem:C.5}).
\end{itemize}
The facts are used as follows: S1 in Lemma~\ref{lem:7.31} and Appendices~\ref{app:A} and~\ref{app:B}; S2 and S3 in
Lemma~\ref{lem:7.35}; S4 in Lemma~\ref{lem:7.39} and Theorem~\ref{thm:7.29}(b), (c); S5 and S6 in
Lemma~\ref{lem:7.39} (Appendix~\ref{app:A}); S7 and S8 in Appendix~\ref{app:A}; S9 in Lemma~\ref{lem:7.39}
and Proposition~\ref{prop:7.51}; S10 in Lemmas~\ref{lem:7.43} and~\ref{lem:7.44} and
Proposition~\ref{prop:7.57}; S11 in Theorem~\ref{thm:7.29} and Lemma~\ref{lem:7.53}; S12 in
Theorem~\ref{thm:7.29}(c) and Lemma~\ref{lem:7.53}; S13 in Theorems~\ref{thm:7.29}(b) and~\ref{thm:7.52} and in Lemma~\ref{lem:7.39} (Appendix~\ref{app:A}); and S14 throughout.

\subsection{Section~\ref{sec:7.6} and Appendix~\ref{app:B}}\label{S:G}

\subsubsection{The H\"older exponents and (L8-1)}\label{S:G.L8}
Section~\ref{ssec:D.1} fixes $\eta_H=0.62$, $p_1=p_2=20$, $p_3=1.2$, $p_4=15$,
$p'=1.3$ and $q'=15.6$. These values satisfy $\sum_i1/p_i=1$ and $1/p'+1/q'=1/p_3$. With the bound
$\alpha_D\le7/6$ of Lemma~\ref{lem:7.31}(c) we get
$\hat\eta=\alpha_Dp'\eta_H\le\frac76\cdot1.3\cdot0.62=0.94033\ldots<0.941$, so
(L8-1) holds. Moreover, $\eta_*\le\bar\eta\le1-4/\pi^2+2C_{\hat\rho}e^{-2\kappa_S}<0.62$ for $\kappa_S$
so large that $C_{\hat\rho}e^{-2\kappa_S}<0.0126$, since $1-4/\pi^2=0.5947\ldots$.

\subsubsection{Lemma~\ref{lem:7.34}(c): the shell constant $C_{\mathrm{sh}}=92$}\label{S:G.34}
With $r_*=\frac12+\frac{\sqrt3}2$, $\int_{1/2}^D(r+r_*)^2r^{-3}dr\le\log(2D)+4r_*+2r_*^2$
and $(r+r_*)/r\le1+2r_*$ for $r\ge\frac12$, the shell count of the proof gives,
for $D\ge1$,
\[
\sum_{d(b,C)\le D}d(b,C)^{-2}
\le6\pi|C|\Bigl[\log(2D)+4r_*+2r_*^2+\tfrac12(1+r_*)^2\Bigr]
+16\pi\bigl[(1+r_*)^3+2(1+2r_*)^3\bigr]D .
\]
The first bracket is at most $\log(2D)+12$, since $4r_*+2r_*^2+\frac12(1+r_*)^2=11.9951\ldots$, and
$16\pi[(1+r_*)^3+2(1+2r_*)^3]=16\pi\cdot117.207\ldots=5891.4\ldots\le5892$. At
$D=D_{\cN}=\sqrt3\,\kbox+1\le2\kbox$ and $|C|\ge8\kbox\ge4D_{\cN}$ the sum is at most
$6\pi|C|\bigl(\log\kbox+\log4+12+1473/(6\pi)\bigr)\le6\pi|C|(\log\kbox+92)$,
since $\log4+12+1473/(6\pi)=91.53\ldots$. Multiplying by $(2\pi)^2$ gives
$S(D_{\cN})\le24\pi^3|C|(\log\kbox+92)$.

\subsubsection{Lemma~\ref{lem:7.35}: the constants $c_8$ and $\cnear$}\label{S:G.35}
We set
\[
c_8:=\frac{\hat\eta(1+4/\eta_H)K_1C_u}{2\lPV(1-\nu_0)},\qquad
\cnear:=\frac{2\theta_PC_{\mathrm{BS}}(\log\kbox+C_{\mathrm{sh}})}{\bGM}
+\frac{1024\pi^2\hat\eta}{\eta_Hc_1}\,\frac{(2\rbox+5)^2c_L^2}{\bGM}.
\]

\emph{Uniformity of $\cnear$.} Since $\kbox=\rbox L\le\rbox c_L/\mt$,
$\log\kbox\le\log(\rbox c_L)+\log(1/\mt)$ with $\log(\rbox c_L)<0$. With
$\mt\ge m_D/2$, $m_D=(8\pi^2\tau)^{1/2}e^{-\pi^2\vCb(0)\tau}$,
$\pi^2\vCb(0)<2.4943551$ and $\bGM=4\pi^2\tau$,
\[
\frac{\log(1/\mt)}{\bGM}\le\frac{2.4943551}{4\pi^2}
-\frac{\frac12\log(8\pi^2\tau)-\log2}{4\pi^2\tau}<0.0632\qquad(\tau\ge1).
\]
Hence $2\theta_PC_{\mathrm{BS}}\log\kbox/\bGM<0.1264\,\theta_PC_{\mathrm{BS}}$, which gives the bound
on $\cnear$ in Lemma~\ref{lem:7.35}. The coefficient $\theta_P$ in that bound satisfies
$\theta_P\le(4\hat\eta/\eta_H)C_P'c_L^2\le\frac12$ by (L8-2), uniformly in $\bGM$.

\subsubsection{Lemma~\ref{lem:7.36}: the condition \eqref{eq:7.6.cl}}\label{S:G.36}
At $p_4=15$ and $p_3=1.2$, so that $c_\lambda=\frac14(1-1/p_3)=\frac1{24}$, and with $\alpha_D\le\frac76$
and $c_E\le\frac1{70}$, we have $2\alpha_Dp_4c_E\le2\cdot\frac76\cdot\frac{15}{70}=\frac12$ and
$p_4\alpha_Dc_E^2/(1-2\alpha_Dp_4c_E)+c_E\le\frac{15\cdot(7/6)}{4900\cdot(1/2)}+\frac1{70}
=0.00714\ldots+0.01428\ldots=0.02142\ldots\le\frac1{24}$.
In Proposition~\ref{prop:7.37}(d), $c_0'=\frac12(1-1/p_3-c_\lambda)=\frac12(\frac16-\frac1{24})=\frac1{16}$.

\subsubsection{Theorem~\ref{thm:7.42}: the threshold $\tau_0$}\label{S:G.42}
With $\pi^2\vCb(0)=2.49435508\ldots$ and $p_0=\frac16$, the lone-cell condition
in the proof of Theorem~\ref{thm:7.42} reads
\[
0.4157\,\tau\ge C_{\Dep}A_1^D\log(C'\tau/\theta_0)+\alpha+O(\log\tau),
\]
where $\cdep=A_0^D+A_1^D\log(C'\tau/\theta_0)$ is the depletion constant of Proposition~\ref{prop:6.18}.
The threshold $\tau_0$ is explicit, and it is large because $A_1^D$ is large, as $\theta_0$ is
small and the covering number of Proposition~\ref{prop:6.18} is large. We need only that
this threshold exists, and the same holds for the threshold of Lemma~\ref{lem:A.11}. For
the wall-polymer cells, Proposition~\ref{prop:7.22} gives
$\Phi_{\cU}\ge(\pi c_gc_uc_L^3/24)/(64\bGM^3(11/10)^3\bD)$.

\subsubsection{Lemma~\ref{lem:7.40} and Proposition~\ref{prop:7.41}: constants and fractions}\label{S:G.B11}
\emph{Constants on class-A cells.} They, and the constants $C_{\mathcal{E}}$, $C_{G}$ of Proposition~\ref{prop:B.10}, are independent of $L'$ and depend only on the following quantities:
\begin{itemize}
\item $c_{7},c_{7}',c_{7}''$ of Lemma~\ref{lem:7.33} (through $v_{\pm}$, $\eta_{H}$, $\kappa_{S}$, that is, on the grid scale $L$);
\item $C_{k}$, $\theta'$, $\cdep$ of Lemma~\ref{lem:B.1}(d);
\item $\mu,\kappa_{S},\delta_{\mathrm{GM}},C$ and the Mayer decay;
\item $K_{1}'',K_{3},\alpha_{D}K_{1},\lPV,r_{C}$ of the lines;
\item $C_{r}(c_{p})$ and $c(\qpin,c_{p},K_{g},C_{\cU})$ of the count and the pinning;
\item $c_{T}$ of the Wick bound;
\item $c_{13}$ and $C_{\Dep}\cdep$ of the H\"older split;
\item the geometry constants $c_{u},C_{u},C_{\cU},D_{\cU},v_{\pm},K_{g}$.
\end{itemize}

\emph{Fractions of $F_{1}^{\mathrm{loc}}$.} They are those of Remark~\ref{rem:A.10} in Section~\ref{S:H1}. Proposition~\ref{prop:7.41} uses the fractions for the numerical factors, which are $c'/16$ on $X\setminus Z$ (Lemma~\ref{lem:B.8}) and $c'/32$ on $Z$, and the fraction $c'/8$ on $X\setminus Z$ for the wall polymers.

\emph{Cells.} Put $K_{B}:=K_{0}^{B}\max(1,K_\ell)$, with the line constant $K_\ell$ of Proposition~\ref{prop:B.7}. Each cell carries the per-cell constants $e^{c_{2}^{L}+c_{13}}$ and at most the factor $\max(1,K_\ell)$ of the line constants, as the links inject into the cells. Since $C_{u}\ge1$, $c_{2}^{L}+c_{13}\ge0$ and $\alpha>0$, it follows that $e^{c_{2}^{L}+c_{13}}\le e^{\alpha}e^{(c_{2}^{L}+c_{13})C_{u}}$. With $\Kcell=K_{B}e^{(c_{2}^{L}+c_{13})C_{u}}\max(1,\lPV^{1/6})$ and $\bD^{1/6}=\lPV^{1/6}(\bD/\lPV)^{1/6}$, Proposition~\ref{prop:B.7} shows that a non-seed lone cell carries at most $\Kcell e^{\alpha+C_{\Dep}\cdep}(\bD/\lPV)^{1/6}$, that a non-seed cell of a wall polymer carries at most $\Kcell e^{\alpha+C_{\Dep}\cdep+c_{B}(\bD)}e^{-(c'/8)\Phi_{\cU}/(1+D_{\cU})}$, and that a seed cell carries at most $\Kcell e^{\alpha+C_{\Dep}\cdep}$, times $e^{c_{B}(\bD)}$ if it is of class B.

\subsection{Constants of Sections~\ref{sec:7.7} to~\ref{sec:7.9} and Appendix~\ref{app:A}}\label{S:H1}

\paragraph{Lemma~\ref{lem:7.47}.}
Since $t_1\le2+(36k_P/\chi_{\rm N3})\ln k_P$, we have $\Dcell\le35r_0+504\,r_0k_P\ln k_P/\chi_{\rm N3}$. In
fine units $r_0k_P=304L\le304c_L/\mt$ and $k_P<1/\mt$, and therefore
$\ln k_P\le\ln(1/\mt)<0.0632\,\bGM$, as in the proof of Lemma~\ref{lem:7.35}.
Hence
$\Dcell\le35r_0+504\cdot304\cdot0.0632\,(c_L/\chi_{\rm N3})\,\bGM/\mt\le35r_0+1.927\,\bGM/\mt$
for $c_L\le1.99\cdot10^{-4}\chi_{\rm N3}$, which is less than $L'=2\bGM/\mt$ for
$\bGM\ge\bGMo$, since $L_c\ge35$ gives $35r_0\le L$. The choice of $c_L$
through $C_P'c_L^2\le(1-4/\pi^2)/8$ places it far below the threshold
$1.99\cdot10^{-4}\chi_{\rm N3}$.

\paragraph{Constants of the seeded sums (before Lemma~\ref{lem:7.53}).}
The constant $C_9$ exists because blocks have side at least $L$, a block has
fewer than $8/c_1$ cells and a cell has at most $6/c_g$ interfaces. We
set\footnote{The summand $1+c_4'+\log2$ of $c_S$ is already contained in $\log\Kcell$: in the proof of
Proposition~\ref{prop:A.8} the factor $e^{(1+c_4'+\log2)|Z|}$ is bounded by $e^{a|Z|}$, which enters
$c_2^L=a+\alpha$ and hence $\Kcell$. Counting it twice only enlarges $c_S$, which the later arguments allow.}
\[
c_S:=\log\Kcell+\alpha+C_{\Dep}\cdep+1+c_4'+\log2,\qquad
\etaseed:=6D_Ie^{-(c'/4)\Funit}.
\]
Since the constants $\log\Kcell$, $D_{\cU}$ and $\alpha$ are fixed numbers and
$\cdep=O(\log\tau)$ by Proposition~\ref{prop:6.18}, we have $c_S=O(\log\tau)$ and
$c_2=c_S+2=O(\log\tau)$. Moreover $\log D_I=O(\log\tau)$, as $g/L=O(\tau)$, while
$\Funit$ is exponentially large in $\tau$, so that $\etaseed$ is doubly
exponentially small.

\paragraph{Proposition~\ref{prop:7.56} (the count), Step~4.}
By Definition~\ref{def:3.17}, $c_3^{\mathrm{NF}}\kappa_{\mathrm{NF}}=c_0'/16$
and $\thNF=c_0'c_g/(90(c_g+4))$, and hence
\[
\frac{\ln16}{\ln2}\,c_3^{\mathrm{NF}}\kappa_{\mathrm{NF}}=4c_3^{\mathrm{NF}}\kappa_{\mathrm{NF}}=\frac{c_0'}4,\qquad
8\thNF\ln16=\frac{8\ln16}{90}\,\frac{c_g}{c_g+4}\,c_0'\le\frac{c_0'}8 .
\]
With $n\le\nu$, $n_0\le(4/c_g')\nu$ and $j_g=O(\ln\bGM)$, the terms
$C(n+n_0+\nu)$ and $(1+j_g)\kappa_{\mathrm{NF}}\nu\ln16$ of Steps~2 and~3 are at
most $C'(1+\ln\bGM)\nu$ with $C'$ depending only on $C_9$, $c_g$, $c_g'$, $c_L$, $C_{\cU}$ and
$\kappa_{\mathrm{NF}}$ (constants of Section~\ref{ssec:D.1}; $c_g'$ through $M\le(1+4/c_g')\nu$, and $c_L$, $C_{\cU}$ through
$j_g$). Hence
$\ln\#\le\frac38c_0'\Funit\nu+C'(1+\ln\bGM)\nu\le\frac{c'}2\Funit\nu$ as soon as
$\Funit\ge(8C'/c_0')(1+\ln\bGM)$, which holds for $\bGM\ge\bGMo$ because
$\Funit=(\pi^2c_g'/18)(\mt L)^3/(\mt\bGM)\ge(\pi^2c_g'/144)c_L^3/(\mt\bGM)$
($\mt L\ge c_L/2$) is exponentially large. In addition, $c'\ge c_0'$.

\paragraph{Lemma~\ref{lem:7.60} and Theorem~\ref{thm:7.61}.}
Under the weight $2-\eta_w$ of the hull functional the constant $c_u$ of (R7)
is the value fixed in Section~\ref{ssec:D.1}, and only the value of $C_X=3C_{\cU}/c_u$
depends on it. Since small components of $\Gp$ are $\cU$-cells of Debye volume at
least $c_u$, they are counted in Lemma~\ref{lem:7.60}. In
Theorem~\ref{thm:7.61}, with $\sF=(\pi^2c_g'/18)(\mt L)\mt/\bGM$
from Definition~\ref{def:3.17} and with $\bGM=4\pi^2\tau$,
\[
\sV=\epsE\,\frac{\pi^2c_g'\mt L}{18}\,\frac{\mt}{\bGM}=\epsE\,\frac{c_g'\mt L}{72}\,\frac{\mt}{\tau}
=\frac{c_g'\mt L}{9216}\,\frac{\mt}{\tau}\quad\text{at }\epsE=\tfrac1{128}=\tfrac{c_0'}8,
\qquad\eta_S=\frac{c_2C_X\mt^2}{\sV},
\]
which is the displayed expression, and it is $O(\mt\bGM\log\tau)$ because
$c_2=O(\log\tau)$.

\paragraph{Remark~\ref{rem:7.62} (the perimeter constant).}
By Lemma~\ref{lem:7.35}, $\cnear\le0.1264\,\theta_PC_{\mathrm{BS}}+O(1/\bGM)$ with
$C_{\mathrm{BS}}=24\pi^3$, uniformly for $\bGM\ge\bGMo$. The coefficient
$0.1264\,\theta_PC_{\mathrm{BS}}$ decreases as $c_L$ decreases, since $\theta_P\propto c_L^2$. Since
$c_2=O(\log\tau)$, $\mt=e^{-\pi^2\vCb(0)\tau(1+o(1))}$ and
$R_1(\beta)=(2\rbox+5)L\ge(2\rbox+5)c_L/(2\mt)$, the terms $3c_2C_XC_{\cU}\mt$ and
$9c_2C_XC_{\cU}^2/R_1(\beta)$ of $c_N$ tend to $0$.

\paragraph{Lemma~\ref{lem:A.7} and the order of $c_{10}$, $c''_{10}$, $\mu$.}
We fix $c_{10}$ first, independently of
$\bar\delta$, with $c_{10}\le\frac79r_C$. Then we fix $c''_{10}$, which depends only on $c_{10}$ and on the growth constants
$C_{\cU}$, $c_u$, with
$e^{c''_{10}}\ge16e\max(1,c_{c_{10}/2})$, and finally we fix the Mayer parameter $\mu$ with
$\mu\ge1+2c''_{10}$.

\paragraph{Proposition~\ref{prop:A.8} (the constant $c_2^L$).}
The constant $c_2^L=a+\alpha$, where $a=\log(2eD_{\cU})+4+c_4'+a''$ with
$e^{a''}\ge4C_\varphi$, is built from $D_{\cU}$, $c_4'$, $C_\varphi$ and
$\alpha$. Of these, $c_4'$ depends on $\tau$, but it is doubly exponentially small and decreasing in
$\tau$, hence bounded uniformly for $\tau\ge\tau_0$, and in this sense $c_2^L$ is uniform in $L'$,
$\bGM$ and $\tau$. Lemma~\ref{lem:A.11} provides a finite threshold $\tau_0$, which grows with $c_2^L$.

\paragraph{Proposition~\ref{prop:A.9}: the conditions on $\bGMo$.}
The conditions on $\bGMo$ used in Proposition~\ref{prop:A.9} are
\begin{equation}\label{eq:A.beta}
\begin{split}
&(\beta1)\ \ \tfrac{c'}{32}\Phi_{\cU}\ge(1+D_{\cU})\bigl(c_S+c_B(\bD)\bigr),\\
&(\beta3)\ \ 3eD_Ie^{-(c'/4)\Funit}\le\tfrac12,\qquad(\beta4)\ \ c_{16}^D\le1 .
\end{split}
\end{equation}

\paragraph{Proposition~\ref{prop:A.9}: the booking of $\Floc$.}
In the proof of the core estimate \eqref{eq:A.core}, we book the factor
$e^{-(3c'/4-\eps)\Floc(X)}$ left by Proposition~\ref{prop:7.37} and the
extraction according to Remark~\ref{rem:A.10}, as follows.
\begin{itemize}
\item \emph{Type-iii factors, per region.} The factors $|D|$ and $|F_2'|$ at derivatives in a region involve $D$ and $F_2'$ on that region only, because $\supp D_\beta$ lies in the element of $S_\beta$. For a wall component with $N$ factors $|D|$, $n_\Delta$ of them in the cell $\Delta$, let $a_\Delta:=\|D\|_{L^2(\Delta)}$, so that $\sum_\Delta a_\Delta^2\le2\Floc/\mt^2$. The weighted arithmetic-geometric mean inequality and the identity $\sup_{F>0}(2F/(\mt^2N))^{N/2}e^{-\delta F}=(e\delta\mt^2)^{-N/2}$ give, for every $\delta>0$,
\[
\prod_\Delta a_\Delta^{n_\Delta}\,e^{-\delta\Floc}\le(\delta\mt^2)^{-N/2}\prod_\Delta(n_\Delta!)^{1/2},
\]
and $F_2'$ is handled in the same way. These constants enter $\Kcell$ and the pinning factor of Proposition~\ref{prop:B.7} and Lemma~\ref{lem:B.8}. We take $\delta=c'/16$ on $X\setminus Z$ and $\delta=c'/32$ on $Z$.
\item \emph{(M-$\Sigma$) on $X\setminus Z$.} We use $c'/8$ of $\Floc(X\setminus Z)$. A wall polymer $\mathcal S$ is the union of the closures of its components, $|\operatorname{cl}(S_\beta)|\le(1+D_{\cU})|S_\beta|$, and $F_1(S_\beta)\ge\Phi_{\cU}|S_\beta|$ by Proposition~\ref{prop:7.22}(c). Hence $\Floc(\mathcal S)\ge\Phi_{\cU}|\mathcal S|/(1+D_{\cU})$, and each $\cU$-cell of a wall polymer of $X\setminus Z$ receives the factor $e^{-(c'/8)\Phi_{\cU}/(1+D_{\cU})}$ of $\eps_\Delta$ in Proposition~\ref{prop:7.41}. Lone cells of $X\setminus Z$ carry $\eps_\Delta=\Kcell e^{\alpha+C_{\Dep}\cdep}(\bD/\lPV)^{1/6}$. The remaining fraction $9c'/16-\eps\ge c'/16$ of $\Floc(X\setminus Z)$ contributes a factor bounded by $1$.
\item \emph{Seed.} We use $c'/32$ of $\Floc(Z,N_Z)$ for the $\cU$-cells of $Z^\Sigma$. Each carries at most $\Kcell e^{\alpha+C_{\Dep}\cdep+c_B(\bD)}\le e^{c_S+c_B(\bD)}$ by Proposition~\ref{prop:7.41}, and $\Floc(Z^\Sigma)\ge\Phi_{\cU}|Z^\Sigma|/(1+D_{\cU})$ pays it under ($\beta1$). Lone seed cells carry at most $\Kcell e^{\alpha+C_{\Dep}\cdep}\le e^{c_S}$. What remains is
\[
e^{-(c'-c'/4-\eps-c'/32-c'/32)\Floc(Z,N_Z)}=e^{-(11c'/16-\eps)\Floc(Z,N_Z)}.
\]
\end{itemize}
Each non-seed cell carries at most $\eps_\Delta\le\frac12c_{16}^D\le\frac12$ by ($\beta4$).
In (a) and (b) of Lemma~\ref{lem:7.53}, $c'/8$ of the fraction $11c'/16-\eps$ left on the seed is used for the entropy of the first set, and at least $c'/16$ is left unused. In (c$'$) of that lemma this fraction pays $\eps\le c'/8$ and $c_w=c'/16$ and leaves at least $c'/2$.

\begin{remark}[The division of $F_1$]\label{rem:A.10}
The proofs of Lemma~\ref{lem:7.53} and Proposition~\ref{prop:7.41} divide $\Floc$ as follows. On the non-seed elements ($X\setminus Z$) they use the fraction $1-c'$ for the H\"older split and \cite[Lemma~13]{GM82} ($T_2,T_3,T_4$) in Proposition~\ref{prop:7.37}, $c'/4$ for the $h$-entropy $c_1'$ of Proposition~\ref{prop:A.8}, $\eps$ for the extraction of Lemma~\ref{lem:7.53} ($\eps\le c'/2$; the chain uses $\eps\le\epsE$), $c'/16$ for the type-iii factors $|D|$, $|F_2'|$ of Proposition~\ref{prop:A.9}, and $c'/8$ for the cells of wall polymers, (M-$\Sigma$) in Proposition~\ref{prop:7.41}; the remaining $9c'/16-\eps\ge c'/16$ is discarded. On the seed elements ($Z$) they use $1-c'$, $c'/4$ and $\eps$ in the same way, $c'/32$ for the type-iii factors and $c'/32$ for the seed cells of $Z^\Sigma$ under ($\beta1$); the remaining $11c'/16-\eps$ is retained. In each region the fractions add up to $1$, and no fraction is used twice. Since the walls on $Z$ are fixed on $\cN_\sigma(X)$ and no wall lies on an interface between $Z$ and $X\setminus Z$, the weight $e^{c_1'\Floc(X\setminus Z)}$ suffices in the sum over $N_X$ in the proof of Proposition~\ref{prop:A.8}, and the fraction for the $h$-entropy on $Z$ is optional. We keep it, at the cost of $c'/4$ on the seed that would otherwise be left unused, so that Lemma~\ref{lem:7.39} can be stated with $e^{c_1'\Floc(X,N_X)}$.
\end{remark}

\begin{lemma}[Constant check]\label{lem:A.11}
There is $\tau_0$ such that for $\tau\ge\tau_0$ the conditions \eqref{eq:A.beta} hold and $c_{16}e^{c_2}\le\frac18$.
\end{lemma}

\begin{proof}
Each condition compares an energy that is exponentially large in $\tau$ with a cost that is logarithmic in $\tau$ or of smaller exponential order. Since $\mt\le2\mD$ and $\mD=(8\pi^2\tau)^{1/2}e^{-\pi^2\vCb(0)\tau}$, $\log(1/\bD)\ge\pi^2\vCb(0)\,\tau-\frac32\log\tau-C$, with $\pi^2\vCb(0)=2.49435508\ldots$. For the constants of the seeded sums we showed (Section~\ref{S:H1}) that $\log D_I=O(\log\tau)$ and, by Proposition~\ref{prop:6.18}, that $c_S=O(\log\tau)$ and $c_2=c_S+2=O(\log\tau)$. By Proposition~\ref{prop:7.22}, $\Phi_{\cU}\ge c\,\bD^{-1}\bGM^{-3}$ with $c>0$ fixed, while $c_B(\bD)\le C\bD^{-1/2}$ by Lemma~\ref{lem:7.40}. Moreover, since $\mt L\ge c_L/2$, we have $\Funit=(\pi^2c_g'/18)(\mt L)^3(\mt\bGM)^{-1}\ge(\pi^2c_g'/144)c_L^3(\mt\bGM)^{-1}$. Hence ($\beta1$) and ($\beta3$) hold for $\tau\ge\tau_0$, and ($\beta4$) is the content of Theorem~\ref{thm:7.42}. For the product, the first branch of $c_{16}^D$ gives
\[
\log\bigl(2\Kcell e^{\alpha+C_{\Dep}\cdep}(\bD/\lPV)^{1/6}e^{c_2}\bigr)=O(\log\tau)-\tfrac16\pi^2\vCb(0)\,\tau ,
\]
and the second gives $O(\log\tau)+c_B(\bD)-(c'/8)\Phi_{\cU}/(1+D_{\cU})$, which tends to $-\infty$ faster. Consequently, $c_{16}^De^{c_2}=\exp(O(\log\tau)-\pi^2\vCb(0)p_0\tau(1+o(1)))$ with $p_0=1/6$. Also $\etaseed e^{c_2}=6D_Ie^{c_S+2}e^{-(c'/4)\Funit}$ is doubly exponentially small. Both are at most $\frac1{16}$ for $\tau\ge\tau_0$. \qed
\end{proof}

\InputIfFileExists{supp/S1_I}{}{}
\section{Full proofs of steps sketched in the main text}\label{S:proofs}
Part~S4 gives more detailed versions of some of these proofs: those of Section~\ref{S:proofs5} in
Section~\ref{S4-S:S4.2}, those of Section~\ref{S:E2} in Section~\ref{S4-S:S4.3}, those of
Section~\ref{S:F2} in Sections~\ref{S4-S:S4.4} and~\ref{S4-S:S4.5}, those of Section~\ref{S:H2}
in Section~\ref{S4-S:S4.8}, in a version that includes the bounds for every primitive-admissible wall field, and
those of Section~\ref{S:I2} also in Section~\ref{S4-S:S4.8}.

\subsection{Full proofs for Section~\ref{sec:2}}\label{S:proofsA}

\paragraph{Proof of Lemma~\ref{lem:2.3}.}\label{S:lem2.3}
We verify Ginibre's hypothesis \cite{Gi70} for the Villain weights directly. A continuous
function $K$ on $\mathbb T^m\times\mathbb T^m$ is a \emph{positive kernel} if it is a uniform limit
of finite sums $\sum_iu_i(\alpha)u_i(\beta)$ with continuous real $u_i$. Positive kernels form a
cone closed under products and under pullback by homomorphisms $\mathbb T^{m'}\to\mathbb T^m$, and
$\iint s(\alpha)s(\beta)K(\alpha,\beta)\,d\alpha\,d\beta\ge0$ for every continuous real $s$. A
continuous $w:\mathbb T\to\RR$ is \emph{admissible} if $(a,b)\mapsto w(a+b)w(a-b)$ is a positive
kernel, and we write $w_1\preceq w_2$ if $(a,b)\mapsto w_1(a+b)w_2(a-b)-w_1(a-b)w_2(a+b)$ is one.

\emph{Step 1 (Ginibre's argument).} Let $w_p>0$ be admissible, $\mu\propto\prod_pw_p(\theta_p)\,d\theta$,
and $B,B'$ integer vectors. The map $(\alpha,\beta)\mapsto(\alpha+\beta,\alpha-\beta)$ of
$\mathbb T^{E}\times\mathbb T^{E}$ onto itself preserves Haar measure, and
$\prod_pw_p(\theta_p(\alpha)+\theta_p(\beta))\,w_p(\theta_p(\alpha)-\theta_p(\beta))$ is a positive
kernel in $(\alpha,\beta)$. Doubling the variables and using
$\cos(x+y)\pm\cos(x-y)\in\{2\cos x\cos y,\,-2\sin x\sin y\}$ gives $\ex{\cos B\cdot\theta}_\mu\ge0$ and
$\ex{\cos B\cdot\theta\,\cos B'\cdot\theta}_\mu\ge\ex{\cos B\cdot\theta}_\mu\ex{\cos B'\cdot\theta}_\mu$,
as in \cite[Propositions~2, 3]{Gi70}. If moreover $w_q\preceq w_q'$ for one plaquette $q$, with
$w_q'>0$, then the same change of variables in
$\iint(\cos B\cdot\theta'-\cos B\cdot\theta)\,w_q(\theta_q)w_q'(\theta_q')\prod_{p\ne q}w_p(\theta_p)w_p(\theta_p')\,d\theta\,d\theta'$,
symmetrized under $\beta\mapsto-\beta$, shows that replacing $w_q$ by $w_q'$ does not lower
$\ex{\cos B\cdot\theta}$.

\emph{Step 2 (the Villain weights).} Let $t,t_1,t_2>0$. Expanding $\vt_t(a\pm b)$ in Fourier series
and putting $p=n+m$, $r=n-m$ (so $p\equiv r$ mod $2$ and $n^2+m^2=\frac12(p^2+r^2)$) gives
\begin{gather*}
\vt_t(a+b)\vt_t(a-b)=\tfrac12\bigl(c(a)c(b)+\tilde c(a)\tilde c(b)\bigr),\\
c(a):=\sum_pe^{-p^2/4t}\cos pa,\qquad\tilde c(a):=\sum_p(-1)^pe^{-p^2/4t}\cos pa ,
\end{gather*}
so $\vt_t$ is admissible. For $t_1\le t_2$, with $\sigma:=\frac1{8t_1}+\frac1{8t_2}$ and
$\kappa:=\frac1{4t_1}-\frac1{4t_2}\ge0$,
\[
\vt_{t_1}(a+b)\vt_{t_2}(a-b)-\vt_{t_1}(a-b)\vt_{t_2}(a+b)
=2\sum_{p\equiv r\ (2)}e^{-\sigma(p^2+r^2)}\sinh(\kappa pr)\,\sin pa\,\sin rb ,
\]
a positive kernel: expand $\sinh(\kappa pr)=\sum_{j\ \mathrm{odd}}\kappa^jp^jr^j/j!$ and
$\one_{p\equiv r}=\frac12(1+(-1)^{p+r})$; the series converges absolutely, as $\sigma-\kappa/2=1/(4t_2)>0$.
Hence $\vt_{t_1}\preceq\vt_{t_2}$. Finally $\vt_t(a-b)-\vt_t(a+b)=2\sum_ne^{-n^2/2t}\sin na\,\sin nb$, so
$1\preceq\vt_t$, and the constant $1$ (a removed plaquette) is admissible.

\emph{Step 3 (conclusion).} Step~1 with $w_p=\vt_{t_p}$ gives (a), with $\ex{\sin\theta_C}=0$ by
$\theta\mapsto-\theta$. Step~1 with $\vt_{t_1}\preceq\vt_{t_2}$, one plaquette at a time, gives (b), and
with $1\preceq\vt_t$ and the admissibility of $1$ it gives (c).

\emph{Step 4 (frozen limits).} For fixed couplings off $D$, the expectation is a
nondecreasing function of the couplings on $D$ with values in $[0,1]$, and
therefore has a limit. This limit inherits monotonicity in the other
couplings. If $D\subseteq D'$ and $s\ge\tau$, then
$t^{\tau,D,s}\le t^{\tau,D',s}$ coordinatewise, so
$\ex{\cos\theta_C}^{V}_{t^{\tau,D,s}}\le\ex{\cos\theta_C}^{V}_{t^{\tau,D',s}}\le g_{X,\tau}(C;D')$,
and $g_{X,\tau}(C;D)\le g_{X,\tau}(C;D')$. Monotonicity in $\tau$ passes to the
limit in the same way.
\qed

\paragraph{Proof of Lemma~\ref{lem:2.8}.}\label{S:lem2.8}
Let $Q_s(k):=\frac1{2\tau}\abs{\one_{F^c}(n_0+\partial k)}^2+\frac1{2s}\abs{\one_F(n_0+\partial k)}^2$,
so that by Lemma~\ref{lem:2.7} the left side of \eqref{eq:2.frozen} at
finite $s$ is $N_s/D_s$ with $N_s:=\sum_ke^{-Q_s(k)}$ and $D_s$ the same sum
for $n_0=0$. We write $k=w'+w$ with $w'\in W'$, $w=\sum_\KK a_\KK\one_\KK\in W$.
Since $\one_{F^c}\partial w=0$,
\[
N_s=\sum_{w'\in W'}e^{-\frac1{2\tau}\abs{\one_{F^c}(n_0+\partial w')}^2}S_s(w'),\qquad
S_s(w'):=\sum_{a\in\ZZ^{\mathcal K_F}}e^{-\frac1{2s}\abs{b+Ba}^2},
\]
with $b:=\one_F(n_0+\partial w')\in\RR^F$ and $Ba:=\partial(\sum_\KK a_\KK\one_\KK)\in\RR^F$.
The matrix $M:=B^{\mathsf T}B$ is positive definite because $\partial$ is
injective. Completing the square, we get $\abs{b+Ba}^2=(a-a_0)\cdot M(a-a_0)+\abs{Pb}^2$
with $a_0:=-M^{-1}B^{\mathsf T}b$ and $P:=1-BM^{-1}B^{\mathsf T}$ the orthogonal
projection onto $(\operatorname{im}B)^\perp$. Poisson summation in $a$ gives
\begin{gather*}
S_s(w')=(2\pi s)^{\abs{\mathcal K_F}/2}(\det M)^{-1/2}\,e^{-\abs{Pb}^2/(2s)}\bigl(1+r_s(w')\bigr),\\
\abs{r_s(w')}\le\rho_s:=\sum_{0\ne m\in\ZZ^{\mathcal K_F}}e^{-2\pi^2s\abs{m}^2/\norm{M}},
\end{gather*}
uniformly in $w'$, and $\rho_s\to0$ exponentially. The prefactor is the same
in $N_s$ and $D_s$ and cancels. In the remaining sums the factor
$e^{-\abs{Pb}^2/(2s)}(1+r_s)$ tends to $1$, is bounded by $2$ for large $s$,
and differs from $1$ by at most $\abs{\one_F(n_0+\partial w')}^2/(2s)+\rho_s$.
If $\one_{F^c}\partial w'=0$ for some $w'\in W'\otimes\RR$, then $w'$ is constant
on each component of the unfrozen graph and vanishes on the component of the
ground, so $w'\in(W\otimes\RR)\cap(W'\otimes\RR)=\{0\}$. Hence
$\abs{\one_{F^c}(n_0+\partial w')}^2\ge c\abs{w'}^2-C$ for some $c>0$, and
$e^{-\frac1{2\tau}\abs{\one_{F^c}(n_0+\partial w')}^2}$ times any polynomial in
$w'$ is summable over $W'$. Dominated convergence gives \eqref{eq:2.frozen},
and the bound on the factor gives the rate $O(1/s)$. Since the summands of
\eqref{eq:2.frozen} are $W$-invariant functions of $k\in\ZZ^{\Lam'_3}$, the
sums do not depend on the complement.
\qed

\subsection{Section~\ref{sec:3}: the loop-adapted grid}\label{S:B2}

\paragraph{Construction of $\mathfrak G(C)$ (Definition~\ref{def:3.4}).}\label{S:B2.grid} With $\Lbox=\prod_\mu[a_\mu,b_\mu]$ as in Definition~\ref{def:3.2}, we construct the objects in the following order.
\begin{enumerate}
\item \emph{Special planes.} In direction $\mu$ the special planes are the lattice planes $\{x_\mu=s\}$, $s\in\mathcal S_\mu$, with $\mathcal S_1=\{a_1,r_1,r_1+R,b_1\}$, $\mathcal S_2=\{a_2,t_1,t_1+T,b_2\}$, $\mathcal S_3=\{a_3,z_0,b_3\}$. A \emph{special interval} is the fine interval between two consecutive special planes of one direction.
\item \emph{Sites.} A special interval of fine length $D'$ is divided into $\lfloor D'/r_0\rfloor$ consecutive intervals: $\lfloor D'/r_0\rfloor-1$ of length $r_0$, followed, at the end with the larger coordinate, by one of length $r_0+(D'\bmod r_0)\in[r_0,\bar r_0]$. The sites (coarse cells) are the boxes $\prod_\mu I_\mu$ with $I_\mu$ such intervals. Their fine sides lie in $[r_0,\bar r_0]$, and equal $r_0$ outside at most one exceptional layer per special interval. The sites form, combinatorially, a box of $\mathbb Z^3$. We denote by $E$ its nearest-neighbour bonds, and the coarse face dual to a bond is the common face of its two sites.
\item \emph{Blocks.} In coarse index units, a special interval of coarse length $n$ is divided into $p:=\lfloor n/L_c\rfloor$ consecutive intervals, and with $r:=n-pL_c$ the first $(r\bmod p)$ of them have length $L_c+\lceil r/p\rceil$ and the others $L_c+\lfloor r/p\rfloor$. Their endpoints are the \emph{block planes}. Blocks are the boxes between consecutive block planes of all three directions, and a \emph{block face} is the common face of two adjacent blocks.
\item \emph{Removal cells.} In each direction the block intervals, in increasing order, are grouped greedily into consecutive groups of total coarse length $\ge\ell$, and a final remainder of length $<\ell$ is merged into the last group. Removal cells $\Delta$ are the boxes $\prod_\mu J_\mu$ with $J_\mu$ such groups, and $|\partial\Delta|$ is the number of bonds of $\mathbb Z^3$ (in coarse index units) with exactly one end in $\Delta$, bonds leaving $\Lbox$ included.
\item \emph{Windows.} We index the blocks by their position in each direction. The block distance is the $\ell^\infty$ distance of these indices. For each block $Q$, $\hat Q(Q)$ is the union of the blocks at block distance at most $75$ from $Q$, a block-aligned box, and $\Qc$ is the set of windows.
\item \emph{Tube zones and tubes.} Let $F'$ be a block face between blocks $Q$, $Q'$, in the plane $P'$. Its tube zone $Z(F')$ is the set of sites of $Q\cup Q'$ among the $L_c/2$ layers nearest to $P'$ on each side. For a bond $b$ across $F'$, its tube $\tau_b$ is the segment of the line of $b$ formed by the $L_c$ sites of that line in $Z(F')$. We write $|F'|$ for the number of coarse bonds across $F'$.
\end{enumerate}
Where the rules leave a choice, we put the exceptional coarse layer at the end with the larger coordinate and the longer block intervals first. These conventions depend only on $C$ and the constants, and any others with this property would do.

\paragraph{Proof of Lemma~\ref{lem:3.3}(a) to (d).}\label{S:B2.box}
(a) The quantity $M_\Lambda$ is determined by $R$, $T$, $r_0$ and $\ell$, and the centre is determined by $C$. The grid is built from $\Lbox$ and the planes of $C$ by rules that involve only the constants.

(b) By (e) and $m\ge\min(R,T)\ge R_1\ge L$, the special planes $a_\mu$, $b_\mu$ are at fine distance $\ge L$ from the planes of $C$, and $R,T\ge R_1\ge L$.

(c) Step~2 cuts a special interval of fine length $D'$ into $\lfloor D'/r_0\rfloor$ coarse intervals. Hence a side of $\Lbox$ made of at most three special intervals has at least $M'_\Lambda/r_0-3\ge\ell$ coarse cells. Step~4 tiles each side of coarse length $\ge\ell$ by groups of block intervals, so $\Lbox$ is a union of removal cells. Removal cells are unions of blocks by construction. The bound $\ell\ge k_P+4L_c$ is \eqref{eq:3.Hell}.

(d) Each $n_\mu$ lies in $[M'_\Lambda/r_0-3,\ M'_\Lambda/r_0]$, and $M'_\Lambda/r_0\ge\ell+3\ge6$. Hence $\max_\mu n_\mu\le M'_\Lambda/r_0\le2(M'_\Lambda/r_0-3)\le2\min_\mu n_\mu$.

\paragraph{Proof of (G1) to (G5).}\label{S:B2.G}
(G1) By Lemma~\ref{lem:3.3}(b), a special interval has fine length $D'\ge L=r_0L_c$, hence coarse length $n=\lfloor D'/r_0\rfloor\ge L_c$ and $p\ge1$. Since $r<L_c$, each block interval has length in $[L_c,2L_c-1]$. A block lies in one special interval in each direction, so it meets at most one exceptional layer per direction, and its fine side lies in $[r_0L_c,(2L_c-1)r_0+(r_0-1)]=[L,2L-1]$.

(G2) This holds because the construction is a tensor product and the endpoints of special intervals are block planes.

(G3) A greedy group closes as soon as its length reaches $\ell$, and a block interval has length at most $2L_c-1$. Hence a group has length in $[\ell,\ell+2L_c-2]$. Adding a final remainder of length at most $\ell-1$ keeps it at most $2\ell+2L_c-3<3\ell$, by \eqref{eq:3.Hell}. Every side of $\Lbox$ is $\ge\ell$ by Lemma~\ref{lem:3.3}(c), so there is at least one group per direction. Then $|\partial\Delta|=2(s_1s_2+s_2s_3+s_3s_1)$ with $s_\mu\in[\ell,3\ell)$.

(G4) A window has at most $151$ blocks per direction, each of coarse side at most $2L_c-1$ by (G1), so its sides are at most $151(2L_c-1)<k_P$. If $u\in Q$ and $|y-u|_\infty\le75L_c$, the segment from $u$ to $y$ in each direction crosses at most $1+\lfloor(75L_c-1)/L_c\rfloor=75$ block planes, since consecutive parallel block planes are at least $L_c$ apart. Thus the block of $y$ is at block distance at most $75$ from $Q$, and $Q_\infty(u,75L_c)\cap\Lbox\subseteq\hat Q(Q)$. An adjacent block is at block distance $1$ and therefore lies in the window. Finally, a block lies in $\hat Q(Q)$ only for the at most $151^3$ blocks $Q$ at block distance at most $75$ from it.

(G5) This holds because blocks have at least $L_c$ layers and parallel block planes are at least $L_c$ apart.


\subsection{Section~\ref{sec:5}}\label{S:proofs5}

\paragraph{The dual paths of Lemma~\ref{lem:5.3}.} In fine coordinates with $\Xi=[0,R]\times[0,T]\times\{0\}$, the dual paths are the closed paths of fine cubes in the planes $y=j+\frac12$ around the rectangles of cube centres $[i+\frac12,2R-i-\frac12]\times[i-R+\frac12,R-i-\frac12]$ in $(x,z)$, $0\le j<T$, $0\le i<R$. These paths are disjoint, and in these coordinates the path of $(i,j)$ crosses $\Xi$ once, at $x=i+\frac12$.

\paragraph{Proof of Lemma~\ref{lem:5.12}.} By \cite[Lemma~2.1 and the sentence after it, p.~141]{GK04}, the first number is at most the number $\binom{Dn}{n}/((D-1)n+1)=\binom{Dn}{n-1}/n$ of $n$-vertex subtrees containing the root of the infinite rooted $D$-ary tree, $D:=D_{\max}$, and $\binom{Dn}{n-1}/n\le D^{n-1}n^{n-1}/n!\le(eD)^{n-1}$ as $n!\ge e(n/e)^n$. For the second, we choose the first plaquette at an edge of $C$ of each component among at most $4|C|$ plaquettes ($2^{4|C|}$ choices), the sizes of the components in the order of these roots ($2^a$), and each component by the first count with $D_{\max}=12$.

\paragraph{Proof of Lemma~\ref{lem:5.14}.} We put $\lambda:=\log(2^{k+1}\kpl)$ and $\omega:=|\log\eps|/F_*\ge c_T(k)/\eta\ge c_T(k)=1+\lambda$. By Lemma~\ref{lem:5.12}, $\PD(E)\le2^{4|C|+j}\sum_{a\ge RT}e^{-a(\omega-\lambda)+\omega(1-\eta)RT}\le2^{4|C|+j}\frac{e}{e-1}e^{RT(\lambda-\eta\omega)}\le2^{4|C|+j+2}e^{-RT}$, as $\omega-\lambda\ge1$. Since $j\le3$, $\ED e^{\sigma X}\le e^{\sigma\eta RT}+e^{\sigma RT}\PD(E)\le e^{\sigma\eta RT}(1+2^{4|C|+5})\le e^{\sigma\eta RT+4|C|+4}$.

\paragraph{Proof of Lemma~\ref{lem:5.18}(a), (c) to (e).} We fix the cut $j$ and write $U:=U_j$, $U':=U_j'$, with $U^\alpha$ and $\mathcal W(\alpha)$ as in Definition~\ref{def:5.17}.

(a) Let $(u,v)$ be a bond of $\Sep^*_j$ with $u\in U^\alpha$ and $v\in W$. If $v\in U$, then $v$ is box-adjacent to $u$ and lies in $U^\alpha$, which is impossible. Hence $v\notin U$. If $(u,v)$ is a $\Gp_{t_j}$-bond, then $v\in\KK_j$ because $\KK_j$ is a component, so $v\in U'$ and the bond is a cut bond. Otherwise $v\in\XX_{t_j}$ (a hole face) or the bond is frozen.

(c) Let $W\in\mathcal W(\alpha)$ and $z\in W\cap U'$. We take a $\Gp_{t_j}$-path in $\KK_j$ from $z$ to $U^\alpha$ and let $y$ be the site just before its first entry into $U^\alpha$. The segment from $z$ to $y$ is a box path avoiding $U^\alpha$, so $y\in W$. Also $y\in\KK_j$, and $y\notin U$ because a site of $U$ box-adjacent to $U^\alpha$ lies in $U^\alpha$. Thus $y\in U'$, and the bond from $y$ into $U^\alpha$ is a cut bond in $\partial(U^\alpha,W)$. A bond $(u,v)$ with $u\in U$, $v\notin U$ determines $\alpha$ (the box-component of $u$ in $U$) and $W$ (the box-component of $v$ in $\Lc\setminus U^\alpha$), so distinct pieces are disjoint. Each piece contains one of the $A_j$ cut bonds.

(d) For a cut bond $(u,z)$ with $u\in U^\alpha$, $z\in U'$, the box-component of $z$ in $\Lc\setminus U^\alpha$ belongs to $\mathcal W(\alpha)$. The same argument with $U$ and $U'$ exchanged gives the piece of $\Sep^{**}_j$.

(e) Let $u^*$ be the last site of $U$ on the path, $u^*\in U^{\alpha^*}$. The rest of the path lies in $\Lc\setminus U\subseteq\Lc\setminus U^{\alpha^*}$ and is box-connected. Hence it lies in one box-component $W^*$ of $\Lc\setminus U^{\alpha^*}$, and $W^*\in\mathcal W(\alpha^*)$ because the path ends in $U'$. The bond leaving $u^*$ lies in $\partial(U^{\alpha^*},W^*)$.

\paragraph{Proof of Lemma~\ref{lem:5.22}.} \emph{Shells.} We partition $\Lc$ into \emph{shell boxes} as a tensor product, cutting each coordinate axis of $\Lc$ into intervals of lengths in $[6L_c,12L_c)$ (possible since every coarse side of $\Lc$ is at least $\ell\ge308L_c$, Lemma~\ref{lem:3.3}(c)). We index the boxes by $\iota\in\prod_\mu[0,n^{\mathrm{sh}}_\mu)$ and put $\delta(b,b'):=|\iota-\iota'|_\infty$ and $n_{\min}:=\min_\mu n^{\mathrm{sh}}_\mu$. For a box $b_0$ and $i\ge1$ the \emph{shell} $\mathrm{Sh}_i(b_0)$ is the union of the boxes $b$ with $\delta(b,b_0)\in\{3i,3i+1,3i+2\}$.

\emph{Shell claim: if $\mathrm{Sh}=\mathrm{Sh}_i(b_0)$ with $3i+2\le(n_{\min}-1)/2$ contains a whole cell of $U_j$ and a whole cell of $U_j'$ (cells at time $t_j$), then each of $\Sep^*_j$ and $\Sep^{**}_j$ has at least $(c_1/24)L_c^2$ bonds with both ends in $\mathrm{Sh}$.} We write $U:=U_j$. Let $A'$ be the set of sites joined to $U$ by a box path with no bond of $\Sep^*_j$, and $B':=\Lc\setminus A'$. Then $U_j'\subseteq B'$ by Lemma~\ref{lem:5.18}(e), and every box bond between $A'$ and $B'$ lies in $\Sep^*_j$. We put $W:=A'\cap\mathrm{Sh}$. The sets $W$ and $\mathrm{Sh}\setminus W$ contain a cell each, hence at least $c_1L_c^3$ sites each, and it suffices to find $(c_1/24)L_c^2$ box bonds inside $\mathrm{Sh}$ between them. The boxes of $\mathrm{Sh}$ form a face-connected family: since $3i+2\le(n_{\min}-1)/2$, each index sphere of radius $3i$, $3i+1$, $3i+2$ around $b_0$ keeps, in each direction, at least one of its two faces inside the index box, so its clipped faces form a connected surface, and consecutive spheres are face-adjacent. Put $f_b:=|W\cap b|/|b|$.

\emph{Case 1: $f_b\le\frac12$ for all boxes $b$ of $\mathrm{Sh}$.} The box Cheeger inequality of the proof of Lemma~\ref{lem:5.22} in each box (sides less than $12L_c$) gives at least $|W|/(24L_c)\ge c_1L_c^2/24$ bonds.

\emph{Case 2: $f_b\ge\frac12$ for all $b$.} The same argument applies to $\mathrm{Sh}\setminus W$.

\emph{Case 3: otherwise.} There are face-adjacent boxes $b_1,b_2$ with $f_{b_1}\ge\frac12>f_{b_2}$. In the box $b_1\cup b_2$ (sides less than $24L_c$) both $W$ and its complement have at least $\frac12(6L_c)^3=108L_c^3$ sites, and the same inequality gives at least $108L_c^3/(48L_c)>2L_c^2>(c_1/24)L_c^2$ bonds.

The same argument with $U$ and $U_j'$ exchanged gives the bound for $\Sep^{**}_j$.

\emph{Counting shells.} Let $b_0$ be the shell box of $x$. The $U_j$-end of the cut bond of Lemma~\ref{lem:5.22}(i) lies within $\rho/2+2$ of $x$, hence at index distance less than $\rho/(12L_c)+2$ from $b_0$, and the far site of (ii) lies at index distance more than $3\rho/(4L_c)-1$. A c-path in $U_j$ from the former to the latter changes the index distance by at most $1$ per step, and cells lie in one block, of side less than the box sides. Hence each shell $\mathrm{Sh}_i(b_0)$, $i\in I(\rho):=\{i\in\ZZ:\rho/(36L_c)+1\le i\le\rho/(4L_c)-1\}$, contains a whole cell of $U_j$, and likewise of $U_j'$. These shells are disjoint, lie within $9\rho$ of $x$, satisfy $3i+2\le(n_{\min}-1)/2$, and number $|I(\rho)|\ge\rho/(9L_c)$ for $\rho\ge27L_c$. In this range, the shell claim applied in each shell gives $(\rho/(9L_c))(c_1/24)L_c^2=c_{\mathrm{sh}}\rho L_c$ bonds of $\Sep^*_j$, and as many of $\Sep^{**}_j$.

\paragraph{Proof of Lemma~\ref{lem:5.38}(b).} We take the breadth-first spanning tree from $o$ of the graph defining anchoring, with fixed orders. Along this tree, the block set is determined by the following data:
\begin{itemize}
\item the sizes of the components ($2^b$ choices);
\item the anchors of the children of $o$ in $Z(\Xi')$, $|Z(\Xi')|\le2a$, and of the children of each component $C$ among its blocks ($2^{2a+k_o}$ and $2^{|C|+k_C}$ choices, $k_o+\sum_Ck_C=m$);
\item the positions of the root blocks relative to their anchors ($(2d_0+1)^3$ choices each);
\item the components, $*$-connected sets of given sizes containing their root blocks ($(26e)^{|C|}$ choices by Lemma~\ref{lem:5.12} with $D_{\max}=26$).
\end{itemize}
With $\sum_C|C|=b$ the product is $2^b\cdot2^{2a+b+m}(2d_0+1)^{3m}(26e)^b$.

\subsection{Full proofs for Section~\ref{sec:stage1} and Appendix~\ref{app:F}}\label{S:E2}

We use the conventions of Appendix~\ref{app:F}.

\paragraph{Proof of Proposition~\ref{prop:6.11}.}\label{S:E2.611}
Diffusive displacement is needed only up to the time $T_{1}=\min(M^{-2},\eps_{1}^{\mathrm{HK}}\RHK^{2})$. Beyond this time the walk is controlled by blocks of length $T_{1}$, in each of which it moves a distance of order $T_{1}^{1/2}$; the resulting bound grows like a ballistic motion at speed of order $T_{1}^{-1/2}$, and the mass term controls this growth.

We write $\sum_{y}v^{(M)}_{\Gp}(x,y)e^{\theta d(x,y)}=\int_{0}^{\infty}e^{-M^{2}t}\,\mathbb{E}_{x}e^{\theta d(x,X_{t})}dt$. If $T_{1}=M^{-2}$ then $\theta_{M}=\kappa_{W}M=\kappa_{W}T_{1}^{-1/2}$. If $T_{1}=\eps_{1}^{\mathrm{HK}}\RHK^{2}<M^{-2}$ then $\theta_{M}=\kappa_{W}M^{2}T_{1}^{1/2}$. In both cases $\theta_{M}\le\min(T_{1}^{-1/2},c_{\mathrm{ex}}/2)$, as $M\le m_{W}$, and $C_{M}\theta_{M}T_{1}^{1/2}\le C_{M}\kappa_{W}M^{2}T_{1}\le\frac12M^{2}T_{1}\le\frac12$. Hence (HK4) gives $\sup_{w}\mathbb{E}_{w}e^{\theta_{M}d(w,X_{s})}\le q:=1+C_{M}\theta_{M}T_{1}^{1/2}\le\frac32$ for $s\le T_{1}$. For $t=kT_{1}+s$ with $0\le s<T_{1}$, $d(x,X_{t})\le\sum_{i\le k}d(X_{(i-1)T_{1}},X_{iT_{1}})+d(X_{kT_{1}},X_{t})$, and the Markov property gives
\[
\mathbb{E}_{x}e^{\theta_{M}d(x,X_{t})}\le q^{k+1}\le q\exp\bigl(C_{M}\theta_{M}t/T_{1}^{1/2}\bigr)\le q\,e^{M^{2}t/2}.
\]
Hence the integral is at most $2q/M^{2}\le3M^{-2}$, which is \eqref{eq:6.LW}.

At $M=\MY$, $(\eps_{1}^{\mathrm{HK}})^{1/2}\MY\RHK\ge(\eps_{1}^{\mathrm{HK}})^{1/2}K_{0}/\lPV$ by \eqref{eq:6.scales}, which gives \eqref{eq:6.CY}. The bound for $v^{2}_{\Gp}$ follows from the inequalities $0\le v^{2}_{\Gp}\le v_{\Gp}$ of Lemma~\ref{lem:6.2}(iii), and that for $v^{1}_{\Gp}$ follows from \eqref{eq:6.LW} at $M=M_{1}$, since $M\mapsto\theta_{M}$ is nondecreasing. Finally, $\kappa_{0}\le\kappa_{W}$ because $(\eps_{1}^{\mathrm{HK}})^{1/2}K_{0}<1$. If $(\eps_{1}^{\mathrm{HK}})^{1/2}K_{0}\ge\lPV$, then $\kappa_{Y}\MY=\kappa_{W}\mt/\lPV\ge\kappa_{0}\mt/\lPV$. Otherwise $\kappa_{Y}\MY=\kappa_{0}\mt/\lPV^{2}$. In both cases $\kappa_{Y}\MY\ge4\kappa_{0}\mt$, because $\lPV\le\frac14$.
\qed

\paragraph{Proof of Theorem~\ref{thm:6.12}.}\label{S:E2.612}
Since $C_{0}$ is a difference of two resolvents, and hence a positive mixture of heat kernels, every bound on it reduces to the heat kernel. With $r_{\pm}:=\frac12\MY^{2}(1\pm(1-4\lPV^{2})^{1/2})$, the roots of $s^{2}+\MY^{2}s+\MY^{2}\mt^{2}$ are $-r_{\pm}$, and partial fractions give
\begin{gather*}
C_{0}=F(-\Delta_{\Gp}),\qquad F(s)=\frac{\MY^{2}}{(s+r_{-})(s+r_{+})}=\varpi\Bigl(\frac1{s+r_{-}}-\frac1{s+r_{+}}\Bigr),\\
C_{0}=\varpi\bigl(v^{(r_{-}^{1/2})}_{\Gp}-v^{(r_{+}^{1/2})}_{\Gp}\bigr)=\int_{0}^{\infty}k(t)P_{t}\,dt,
\end{gather*}
with $k(t)=\varpi(e^{-r_{-}t}-e^{-r_{+}t})$ and $0\le k(t)\le\min(\MY^{2}t,\varpi)e^{-r_{-}t}$. Since $F(0)=\mt^{-2}$, $C_{0}1_{\KK}=\mt^{-2}1_{\KK}$, that is, the zero modes are part of $C_{0}$. We put $R:=\RHK$, so that $R^{-3}r_{-}^{-1}\le\mt^{-2}R^{-3}\le\mt K_{0}^{-3}=\lPV K_{0}^{-3}\MY$ by \eqref{eq:6.scales}.

(H1). Since $p_{t}\le1$ on $[0,1]$ and $p_{t}\le C_{H}t^{-3/2}+C_{H}R^{-3}$ on $[1,\infty)$ by (HK1), we have
\begin{align*}
C_{0}(x,x)&\le\int_{0}^{1}\MY^{2}t\,dt+C_{H}\int_{1}^{\infty}\min(\MY^{2}t,\varpi)t^{-3/2}dt+C_{H}R^{-3}\int_{0}^{\infty}\varpi e^{-r_{-}t}dt\\
&\le\tfrac12\MY+2C_{H}(1+\varpi)\MY+\varpi C_{H}\lPV K_{0}^{-3}\MY,
\end{align*}
where the middle integral is split at $t=\MY^{-2}$, whatever the order of $\MY^{-2}$ and $R^{2}$.

(H2). Entrywise, $C_{0}\le\varpi v^{(r_{-}^{1/2})}_{\Gp}$, because $v^{(r_{+}^{1/2})}_{\Gp}\ge0$ by Lemma~\ref{lem:6.2}. Since $\mt\le m_{W}/2$, the mass $r_{-}^{1/2}\le1.04\,\mt$ is at most $m_{W}$, and \eqref{eq:6.LW} at this mass has $\theta\ge\kappa_{W}\mt\min(1,(\eps_{1}^{\mathrm{HK}})^{1/2}\mt R)\ge\kappa_{0}\mt$, since $\mt R\ge K_{0}$ and $(\eps_{1}^{\mathrm{HK}})^{1/2}K_{0}<1$. Hence the row sums of $C_{0}e^{\kappa_{0}\mt d}$ are at most $3\varpi/r_{-}=K_{3}\mt^{-2}$, and $C_{0}$ is symmetric.

(H1$'$). First, $C_{0}(x,y)\ge0$ because $k\ge0$ and $p_{t}\ge0$. If $d:=d(x,y)<\MY^{-1}$, then $C_{0}(x,y)\le(C_{0}(x,x)C_{0}(y,y))^{1/2}\le K_{1}\MY\le eK_{1}\MY e^{-\kappa_{0}'\mt d}$, since $\kappa_{0}'\mt d\le\lPV\le1$. If $d\ge\MY^{-1}$, we use $C_{0}(x,y)\le\varpi\int_{0}^{\infty}e^{-r_{-}t}p_{t}(x,y)\,dt$ and split the time axis. (i) For $t\le d$, by (HK0) the contribution is at most $\varpi de^{-d/17}\le313\varpi\MY e^{-d/68}\le313\varpi\MY e^{-\kappa_{0}'\mt d}$, because $de^{-3d/68}\le\MY d^{2}e^{-3d/68}<279\,\MY$ for $d\ge\MY^{-1}$, and $\kappa_{0}'\mt\le1/68$. (ii) For $d\le t\le T_{2}:=2\eps_{1}^{\mathrm{HK}}R^{2}$ we have $t\ge d\ge1$, and (HK3) together with $r_{-}t+c_{\mathrm{gs}}d^{2}/(2t)\ge(2c_{\mathrm{gs}}r_{-})^{1/2}d$ bounds the contribution by
\begin{align*}
\varpi C_{\mathrm{gs}}e^{-(2c_{\mathrm{gs}}r_{-})^{1/2}d}\int_{0}^{\infty}t^{-3/2}e^{-c_{\mathrm{gs}}d^{2}/(2t)}dt&=\varpi C_{\mathrm{gs}}(2\pi/c_{\mathrm{gs}})^{1/2}d^{-1}e^{-(2c_{\mathrm{gs}}r_{-})^{1/2}d}\\
&\le\varpi C_{\mathrm{gs}}(2\pi/c_{\mathrm{gs}})^{1/2}\MY e^{-\kappa_{0}'\mt d}.
\end{align*}
(iii) For $t\ge T_{2}$, as in the proof of (HK3), $p_{t}(x,y)\le2\sup p_{t/2}\cdot\sup_{w}\mathbb{P}_{w}(d(w,X_{t/2})\ge d/2)$. Here $\sup p_{t/2}\le C_{H}(\eps_{1}^{\mathrm{HK}})^{-3/2}R^{-3}$ by (HK1), and Markov's inequality, combined with the blocking bound from the proof of Proposition~\ref{prop:6.11} in \S\ref{S:E2.611} at the mass $r_{-}^{1/2}$ and with its rate $\theta\ge\kappa_{0}\mt$, gives $\mathbb{P}_{w}(d(w,X_{t/2})\ge d/2)\le q\,e^{-\theta d/2}e^{r_{-}t/4}$. Thus the contribution is at most $4\varpi C_{H}(\eps_{1}^{\mathrm{HK}})^{-3/2}R^{-3}r_{-}^{-1}e^{-\kappa_{0}\mt d/2}\le4\varpi C_{H}(\eps_{1}^{\mathrm{HK}})^{-3/2}\lPV K_{0}^{-3}\MY e^{-\kappa_{0}'\mt d}$. The three parts give $K_{1}'$.

(H$_{C}$). By the Schur test, $\|\nu\|_{2\to2}\le\|\nu\|_{0}\le\eps_{\nu}\mt^{2}$, and $C_{0}^{-1}\ge\mt^{2}$. Hence $C_{0}^{-1}+\nu\ge(1-\eps_{\nu})C_{0}^{-1}>0$, $C$ exists, is positive definite, and $C\le(1-\eps_{\nu})^{-1}C_{0}$, which gives $C(x,x)\le2K_{1}\MY$ because $\eps_{\nu}\le1/(2K_{3})<\frac12$. Since $\rcov\le\kappa_{0}\mt$, $\|C_{0}\|_{\rcov}\le K_{3}\mt^{-2}$, and the Neumann series $C=\sum_{n\ge0}(-1)^{n}C_{0}(\nu C_{0})^{n}$ converges in $\|\cdot\|_{\rcov}$ with $\|C\|_{\rcov}\le K_{3}\mt^{-2}\sum_{n}(K_{3}\eps_{\nu})^{n}\le2K_{3}\mt^{-2}$. For the pointwise bound we write $C=C_{0}-C_{0}\nu C$ and use $e^{\rcov d(x,y)}\le e^{\rcov d(x,z)}e^{\rcov d(z,w)}e^{\rcov d(w,y)}$ and the bound $\sup_{z}C_{0}(x,z)e^{\rcov d(x,z)}\le K_{1}''\MY$, which follows from (H1), (H1$'$) and $\rcov\le\kappa_{0}'\mt$. This gives $|C(x,y)|e^{\rcov d(x,y)}\le K_{1}''\MY(1+\|\nu\|_{\rcov}\|C\|_{\rcov})\le K_{1}''\MY(1+2K_{3}\eps_{\nu})\le2K_{1}''\MY$. A matrix $C_{s}$ dominated entrywise by $|C|$, with $C_{s}(x,x)\le C(x,x)$, inherits the three bounds.
\qed

\paragraph{Proof of Lemma~\ref{lem:F.7}.}\label{S:E2.F7}
The heat kernel of $\Zthree$ factorizes into three one-dimensional kernels, and an exponential tilt of each turns the Gaussian bound into a bound with exponential decay. The time integral then produces the factor $(1+|x|)^{-1}$ from times of order $|x|^{2}$ and the decay from the mass.

Let $q_{t}(k):=e^{-2t}I_{k}(2t)$ be the transition probability of the continuous-time walk on $\ZZ$ that jumps to each neighbour at rate $1$, so that the kernel of $e^{t\Delta_{\Zthree}}$ is $p_{t}(x)=\prod_{\mu}q_{t}(x_{\mu})$ and $v^{(M)}(x)=\int_{0}^{\infty}e^{-M^{2}t}p_{t}(x)\,dt$. For $\lambda\ge0$, $q^{\lambda}_{t}(k):=q_{t}(k)e^{\lambda k-2t(\cosh\lambda-1)}$ is the law at time $t$ of $N_{+}-N_{-}$ with independent Poisson variables of means $te^{\lambda}$ and $te^{-\lambda}$. Conditioning on $N_{-}$, we get $\sup_{k}q^{\lambda}_{t}(k)\le\sup_{j}\mathbb{P}(N_{+}=j)\le\sqrt2(1+t)^{-1/2}$, by Stirling's formula. Conditioning on $N_{-}$, $q^{\lambda}_{t}$ is a mixture of translates of the Poisson law $\pi$ of mean $te^{\lambda}$, which is unimodal ($\pi(j)/\pi(j-1)=te^{\lambda}/j$); hence $\sum_{k}|q^{\lambda}_{t}(k+1)-q^{\lambda}_{t}(k)|\le\sum_{k}|\pi(k+1)-\pi(k)|=2\max_{j}\pi(j)\le2\sqrt2(1+t)^{-1/2}$. Writing $q^{\lambda}_{t}=q^{\lambda}_{t/2}*q^{\lambda}_{t/2}$, we obtain $\sup_{k}|q^{\lambda}_{t}(k+1)-q^{\lambda}_{t}(k)|\le8(1+t)^{-1}$. Hence, for $k\ge0$,
\begin{align*}
q_{t}(k)&\le\sqrt2(1+t)^{-1/2}e^{-\lambda k+2t(\cosh\lambda-1)},\\
|q_{t}(k+1)-q_{t}(k)|&\le\bigl(8(1+t)^{-1}+\sqrt2\lambda(1+t)^{-1/2}\bigr)e^{-\lambda k+2t(\cosh\lambda-1)},
\end{align*}
the second because $q_{t}(k+1)-q_{t}(k)=e^{-\lambda k+2t(\cosh\lambda-1)}[e^{-\lambda}q^{\lambda}_{t}(k+1)-q^{\lambda}_{t}(k)]$ and $1-e^{-\lambda}\le\lambda$. For negative $k$ the bounds follow by symmetry, at the cost of a factor $e^{\lambda}$ in the difference bound.

We let $r:=|x|\ge1$, choose $\lambda_{\mu}:=s|x_{\mu}|/r$ with $s\in[0,\frac12]$, and use $\sum_{\mu}2(\cosh\lambda_{\mu}-1)\le\sum_{\mu}\lambda_{\mu}^{2}\cosh\lambda_{\mu}\le s^{2}\cosh s$ and $\sum_{\mu}\lambda_{\mu}|x_{\mu}|=sr$. Then $p_{t}(x)\le2^{3/2}(1+t)^{-3/2}\exp(-sr+ts^{2}\cosh s)$. For $t\le r$ the choice $s=\frac12$ makes the exponent at most $-r/2+r\cosh(\frac12)/4\le-0.218\,r$. For $t\ge r$ the choice $s=r/(2t)\le\frac12$ makes the exponent at most $-r^{2}/(4t)\,(2-\cosh\frac12)\le-0.218\,r^{2}/t$, and
\[
M^{2}t+0.218\,r^{2}/t\ge\bigl(M^{2}t+\tfrac{9}{64}r^{2}/t\bigr)+0.077\,r^{2}/t\ge\tfrac34Mr+0.077\,r^{2}/t.
\]
Hence
\begin{align*}
v^{(M)}(x)&\le2^{3/2}e^{-0.218r}\int_{0}^{r}(1+t)^{-3/2}dt+2^{3/2}e^{-3Mr/4}\int_{0}^{\infty}t^{-3/2}e^{-0.077r^{2}/t}dt\\
&\le2^{5/2}e^{-0.218r}+\frac{2^{3/2}\sqrt\pi}{\sqrt{0.077}\,r}e^{-3Mr/4},
\end{align*}
and $e^{-0.218r}\le e^{-3Mr/4}e^{-0.068r}$ because $\frac34M\le0.15$. This gives the first bound in \eqref{eq:F.L1L2}. For the difference in direction $e=\pm e_{\mu}$ only the factor $q_{t}(x_{\mu})$ changes, and the same choices of $\lambda$ give $|p_{t}(x+e)-p_{t}(x)|\le C(1+t)^{-3/2}\bigl((1+t)^{-1/2}+s\bigr)\exp(-sr+ts^{2}\cosh s)$. For $t\le r$ this is bounded as before. For $t\ge r$ we take $s=r/(2t)$ and use $\int_{0}^{\infty}(t^{-2}+rt^{-5/2})e^{-0.077r^{2}/t}dt\le Cr^{-2}$. For $r<1$ both bounds are trivial.

For the box, the reflections in the six face planes (at half-integer positions) generate a group $\mathcal{G}$ that acts coordinatewise, and $v^{(M)}_{\Lambda'}(y,x)=\sum_{g\in\mathcal{G}}v^{(M)}(y-gx)$. Indeed, the right side is $\mathcal{G}$-invariant, so across each face its two values agree and the Neumann condition holds, and the orbit of $x$ meets $\Lambda'$ only at $x$. Coordinatewise $|y_{\mu}-(gx)_{\mu}|\ge|y_{\mu}-x_{\mu}|$ for $x_{\mu},y_{\mu}\in[1,n_{\mu}]$, so $|y-gx|\ge r_{yx}:=|y-x|$, and $\#\{g:|y-gx|\le\rho\}\le8(\rho M+1)^{3}$ because $n_{\mu}\ge M^{-1}$. For a decreasing $f$, summation by parts gives $\sum_{g}f(|y-gx|)\le8\bigl[(1+Mr_{yx})^{3}f(r_{yx})+\int_{r_{yx}}^{\infty}3M(1+M\rho)^{2}f(\rho)\,d\rho\bigr]$. With $f(\rho)=C_{L}(1+\rho)^{-j}e^{-3M\rho/4}$, $j\in\{1,2\}$, this is at most $C(1+Mr_{yx})^{3}(1+r_{yx})^{-j}e^{-3Mr_{yx}/4}\le C'(1+r_{yx})^{-j}e^{-Mr_{yx}/2}$. Since the differences $u_{x}(y+e)-u_{x}(y)$ are the sums of the differences of the images, (L1) and (L2) follow.
\qed

\providecommand{\Floc}{F_1^{\mathrm{loc}}}
\providecommand{\UU}{\mathcal U}
\providecommand{\one}{\mathbf 1}
\providecommand{\bGM}{\beta_{\mathrm{GM}}}
\providecommand{\mt}{\tilde m}
\providecommand{\MY}{M_Y}
\subsection{Full proofs for Subsection~\ref{ssec:7.5}}\label{S:F2}

\paragraph{Proof of Lemma~\ref{lem:7.wall}.}
The integrand of Proposition~\ref{prop:7.8} depends on the configuration only
through the field $\psi+g+k\bGM^{-1/2}f$, read modulo $P\Z$ at each site and in
each cell average, and through the walls. We write $\vartheta:=\psi+D$.

(a) Let $h\in\mathcal H$ and $k\in\{0,1\}$. By \eqref{eq:7.shift},
$g+k\bGM^{-1/2}f=h+D$. Since $h$ takes values in $P\Z=2\pi\bGM^{-1/2}\Z$,
$e^{im\bGM^{1/2}h(x)}=1$, so each phase $e^{im(\bGM^{1/2}(\psi+g)(x)+kf(x))}$ of
$E'$, $\varepsilon_k$, $G_{2,2}$ and $G_{2,3}$ equals $e^{im\bGM^{1/2}\vartheta(x)}$. In
$G_1=\sum_\Omega\ln r_\Omega(A_\Omega)$ the function $r_\Omega$ of \eqref{eq:7.rOmega}
depends on $A_\Omega$ only through $A_\Omega+k\bGM^{-1/2}\bar f_\Omega$ and is $P$-periodic
in it. This quantity is $\operatorname{avg}_\Omega\vartheta+h|_\Omega$, as $h$ is constant
on $\Omega$. Hence $r_\Omega(A_\Omega)$ is $r_\Omega$ at $k=0$ evaluated at
$\operatorname{avg}_\Omega\vartheta$. Likewise
$\tilde\delta=(\psi+g+k\bGM^{-1/2}f)-\operatorname{avg}_\Omega(\psi+g+k\bGM^{-1/2}f)=\vartheta-\operatorname{avg}_\Omega\vartheta$
on $\Omega$. The terms added in $G^\natural$ (Lemma~\ref{lem:7.31}) are functions of
$\vartheta$ and $\psi$ through the same phases. The term $F_1(S_\beta)$ is given by
\eqref{eq:7.F1Sbeta}, and $F_2(Y)=\langle\one_Y\psi,F_2'\rangle$ with
$F_2'=\sum_\beta F'_{2,\beta}$ and $F'_{2,\beta}=-[C_0^{-1},\chi_\beta]C_0s_\beta$
as in Proposition~\ref{prop:7.9}\,(e), both functions of $w$ alone. Hence the two
definitions of $Q_N(Y)$ and of $E(Y,s)$ agree.

(b) After the replacement, the action is a sum of three kinds of terms.
The cell terms are $\ln r_\Omega(\operatorname{avg}_\Omega\vartheta)$. The site terms are the
terms of $G_{2,1}$, $G_{2,2}$, $G_{2,3}$, of $G^\natural$ and of $E$, functions of
$\psi$ and $\vartheta$ at the sites of their support (the $E$-terms with weights
$w_s(\tau)$). The wall-component terms $F_1(S_\beta)$ and
$\langle\psi,F'_{2,\beta}\rangle$ are supported in $S_\beta$
by Proposition~\ref{prop:7.9}\,(e). On a union $Y$ of wall components and of
$\UU$-cells outside $\hat\Sigma$, Proposition~\ref{prop:7.9}\,(f) shows that $D|_Y=\sum_{S_\beta\subseteq Y}D_\beta|_Y$ depends on $N$
only through $N|_{E(Y)}$. The proof of Lemma~\ref{lem:7.24} needs no more than
these facts. An admissible set contains
each cell and each wall component entirely or not at all, which gives the
factorization and the locality for every wall field. Definition~\ref{def:7.23}
and the proof of Lemma~\ref{lem:7.26} use only the wall sites, the wall
components and Proposition~\ref{prop:7.9}\,(a), all of which are defined from
$N$. On $U\ne V$ we take closures in $U$, which is consistent because the
wall polymers of $N$ have their closures in $U$.

Proposition~\ref{prop:7.25} is
proved in Appendix~\ref{app:C} from the factorization, the admissibility of the
sets of a sequence and absolute convergence. Absolute convergence rests on
Lemma~\ref{lem:C.3}, whose bounds are pointwise in $(\psi,\vartheta)$, and on the Poincar\'e inequality on cells
of Lemma~\ref{lem:C.4}, which does not involve the configuration. Here Lemma~\ref{lem:C.3}\,(b) is a bound
on $G_{2,1}$ by $\|\one_YP_V\vartheta\|^2$.

In
Theorem~\ref{thm:7.14}, the bound \eqref{eq:7.H6p} of
Corollary~\ref{cor:7.11}\,(c) with $g-H=D$ and $\nabla g+k\bGM^{-1/2}R_f=w+\nabla D$ reads
$\Floc(Y)\ge\frac12\mt^2\|D\|^2_Y+\frac12\|w+\nabla D\|^2_{E(Y)}$, which follows from
\eqref{eq:7.F1Sbeta} by dropping the term with $\MY^{-2}$, the supports of
different $\beta$ being disjoint. In the same theorem, the tube argument uses only that $w$ vanishes
on the bonds inside cells and is constant on interfaces
(Lemma~\ref{lem:7.13}). The proof of Proposition~\ref{prop:7.5} bounds
$\cos(mu)-1$ at each site, for an arbitrary real number $u$. Here
$u=\bGM^{1/2}(\varphi+k\bGM^{-1/2}f-h)(x)$, which after the translation
$\varphi=\psi+g$ equals $\bGM^{1/2}\vartheta(x)$. \qed

\InputIfFileExists{supp/S2_G}{}{}
\providecommand{\Nlong}{\mathcal N^{\mathrm{long}}}
\providecommand{\Kflat}{\mathfrak K^{\flat}}
\providecommand{\kbox}{k_{\square}}
\providecommand{\cN}{\mathcal N}
\providecommand{\Floc}{F_1^{\mathrm{loc}}}
\providecommand{\Funit}{F_{\mathrm{unit}}}
\subsection{Full proofs for Subsections~\ref{sec:7.7} to~\ref{sec:7.9}}\label{S:H2}

\paragraph{Proof of Proposition~\ref{prop:7.long}.}
The bounds of Subsection~\ref{ssec:7.6} and Appendices~\ref{app:A},
\ref{app:B} are stated for the configurations $\cN(X)$, $\cN_\sigma(X)$ produced by
labels. We go through the uses of the label in their proofs and show that
each of them survives for long wall fields. To this end we fix $k=0$ and a configuration $N_X\in\Nlong(X)$ (or
$\Nlong_\sigma(X)$), with the integrand of Lemma~\ref{lem:7.wall}.

(i) \emph{Sums over configurations.} They enter only through the following.
Lemma~\ref{lem:7.32}\,(b) maps $N_X$ to its interface values
$(N_I)_{I\subseteq X}\in\mathbb Z^{n_I(X)}$ and drops the realizability constraints. A wall
field is constant on interfaces and vanishes off them. Hence the map is
injective on $\Nlong(X)$, and the bound holds there as well. In the proof of
Proposition~\ref{prop:A.9}, step (2) splits by the first set
$\sat_{N_X}(S_0)$ through Lemma~\ref{lem:7.26}, which holds in wall form
by Lemma~\ref{lem:7.wall}\,(b), and the seed configuration it produces is long
by Lemma~\ref{lem:7.45}\,(a). Step (3) of that proof uses Lemma~\ref{lem:5.12} to count wall polymers with their walls
through their wall interfaces and values at the Euclidean link scale $g$.
This is again a count of integer functions on interfaces, combined
with $\Floc(\mathcal S)\ge2\Funit\sum_IN_I^2$ from Lemma~\ref{lem:7.32}\,(a). The sum over
link data and sets in Propositions~\ref{prop:A.5} and~\ref{prop:A.8} does not
involve the configuration beyond these.

(ii) \emph{Bounds at a fixed configuration.} Each is a bound on a Gaussian
integral whose integrand is a function of $\psi$, $D$, $w$ and $F_2'$, that
is, of $\psi$ and $N_X$, by Lemma~\ref{lem:7.wall}. The label enters these proofs
only in the following places.
\begin{itemize}
\item Appendix~\ref{app:B} uses the label in the phases of class-A cells, through
$e^{im\bGM^{1/2}h(x)}=1$. Under Lemma~\ref{lem:7.wall} the phase is
$e^{im\bGM^{1/2}\vartheta(x)}$ by definition. The same holds for the cell function
$\hat R_v$ of Lemma~\ref{lem:7.33}, evaluated at $\operatorname{avg}_\Omega(\psi+D)$, and
for Lemma~\ref{lem:7.43}, where $D=0$ on the wall-free cell $\Delta$.
\item Theorem~\ref{thm:7.14} enters through Lemma~\ref{lem:7.32}\,(a), and it holds for
every wall field by Lemma~\ref{lem:7.wall}\,(b).
\item In Step~1 of Lemma~\ref{lem:7.35}, the identity \eqref{eq:7.6.untr} is stated
in terms of $D_Y$, $w_Y$ and $s_Y$ for unions of elements and involves no
extension of $h$. At $k=0$ the near-cell term vanishes.
\item The functions $G^\natural$ of Lemma~\ref{lem:7.31} are functions of
$\vartheta=\psi+D$ and $\psi$, bounded pointwise.
\item Lemma~\ref{lem:7.34}\,(a) is used in Step~2 of Lemma~\ref{lem:7.35}. Its proof
first shows that $w=w_Y$ on $E(\Kflat_\Omega)$, using only (L8-4) and
Proposition~\ref{prop:7.9}, and then puts $a_\Omega:=h$ on $\Kflat_\Omega$, using
$P_\Omega h=0$ and $\nabla h=w$ on $E(\Kflat_\Omega)$. For a long configuration, Lemma~\ref{lem:7.45}\,(a) shows that the
field $N_X\one_{E(Y)}$ on the union $Y$ of elements is long, and Lemma~\ref{lem:7.prim} gives $h':\Kflat_\Omega\to P\mathbb Z$,
constant on $\Omega$, with $\nabla h'=w_Y$ on $E(\Kflat_\Omega)$. We apply
Proposition~\ref{prop:6.10}\,(iv) to $\vartheta+h'$. Since $\nabla(\vartheta+h')=\nabla\vartheta+w_Y$ on
$E(\Kflat_\Omega)$ and $P_\Omega(\vartheta+h')=P_\Omega\vartheta$, we obtain
$\|P_\Omega\vartheta\|^2\le C_W^f\kbox^2\|\nabla\vartheta+w_Y\|^2_{E(\Kflat_\Omega)}$, which is (a)
at $k=0$. Step~2 of Lemma~\ref{lem:7.35} then goes through without change.
\end{itemize}
Absolute convergence at a fixed configuration (Proposition~\ref{prop:7.25})
follows as in Section~\ref{S:F2}, the estimates used there being
pointwise in $(\psi,D)$. Lemma~\ref{lem:7.34}\,(a) is the only place where the
bounds use that the configuration is long: for a wall field with a nonzero
period on a short cycle of $\Kflat_\Omega$, or on a same-cell closure through
$\Omega$, no primitive of $w_Y$ constant on $\Omega$ exists. With (i) and (ii), the
proofs of Propositions~\ref{prop:7.37}, \ref{prop:7.41}, Lemmas~\ref{lem:7.39},
\ref{lem:7.40} and Proposition~\ref{prop:A.9} hold with the same constants, and
so does Lemma~\ref{lem:7.53}. \qed

\subsection{Full proofs for Appendix~\ref{app:A}}\label{S:I2}

\paragraph{Proof of Proposition~\ref{prop:A.9}, step (3).}
We use the notation of Lemma~\ref{lem:7.53} and Proposition~\ref{prop:A.9}, and we put
$w(\mathcal S):=e^{-(c'/8)\Floc(\mathcal S,N_{\mathcal S})}$.
Since $Z^o\subseteq S_0$ and $11c'/16-\eps\ge c'/8$, steps (1)
and (2) show that the left side of Lemma~\ref{lem:7.53}(a) is at most
$\cP'e^{c_S|S_0|}\sum_\sigma\prod_{\mathcal S}w(\mathcal S)$, the sum over
$\sigma\in\Sigma_{\mathrm{sat}}(S_0)$ and the product over the wall polymers
$\mathcal S$ of $\sigma$. Assigning the disjoint wall polymers of $\sigma$ to
their first cells in $S_0$ bounds the sum by
$\prod_{\Delta\subseteq S_0}\bigl(1+\sum_{\mathcal S\ni\Delta}w(\mathcal S)\bigr)$,
and
\[
\sum_{\mathcal S\ni\Delta}w(\mathcal S)\le6D_Ie^{-(c'/4)\Funit}=\etaseed
\]
under ($\beta3$). Indeed, Lemmas~\ref{lem:7.6}(a) and~\ref{lem:7.26} show that a wall polymer with its walls is determined by its wall
interfaces and their values. These interfaces form a set
connected at Euclidean scale $g$ containing one of the $D_I$ interfaces
within distance $g$ of $\Delta$, so the lattice-animal count of
Lemma~\ref{lem:5.12}, together with $\Floc(\mathcal S)\ge2\Funit\sum_IN_I^2$
from Lemma~\ref{lem:7.32}(a), gives
$\sum_{n\ge1}D_I(eD_I)^{n-1}(3e^{-(c'/4)\Funit})^n$. This gives (a).

In (b),
the terms with $\sigma=(S_0,0)$ and $X\ne S_0$ gain $c_{16}^D$, and those
whose $\sigma$ contains a wall polymer contribute
$(1+\etaseed)^{|S_0|}-1\le\etaseed e^{2|S_0|}$, so the sum without the term
$X=S_0$, $N_X=0$ is at most
$(c_{16}^D+\etaseed)\cP'e^{(c_S+2)|S_0|}=c_{16}\cP'e^{c_2|S_0|}$.

In (c$'$) we have
$\cP'=1$ and $Z^o=\emptyset$, and \eqref{eq:A.core} with $\eps\le c'/8$ and
$c_w=c'/16$ leaves $e^{-(c'/2)\Floc(Z,N_Z)}$.

\label{page:end}

\end{document}